\pdfoutput=1
\documentclass[a4paper,12pt,oneside]{report}
\usepackage[top=2.5cm, bottom=2.5cm, left=4cm, right=2cm, headheight=15pt]{geometry}

\newcommand{\linespacing}{1}
\renewcommand{\baselinestretch}{\linespacing}
\usepackage[british]{babel}

\usepackage{graphicx}
\usepackage{color}
\usepackage{tikz}
\usepackage{enumerate} 
\usepackage{rotating} 
\usepackage{fancyhdr}
\fancypagestyle{plain}{
	\fancyhf{} 
	\fancyhead[C]{\thepage} 
	\chead{\thepage}
	
}
\usepackage{amssymb}
\usepackage{physics} 
\usepackage{bm}        
\usepackage{verbatim}

\usepackage[sort&compress,numbers]{natbib}
\usepackage[colorlinks=true, linkcolor=black!60!blue, citecolor=black!60!blue, urlcolor=black!60!blue]{hyperref} 
\newcommand*{\boxcolor}{black!60!blue}

\makeatletter
\renewcommand{\boxed}[1]{\textcolor{\boxcolor}{
\tikz[baseline={([yshift=-1ex]current bounding box.center)}] \node [rectangle, minimum width=2ex,rounded corners,draw] {\normalcolor\m@th$\displaystyle#1$};}}
\makeatother

\usepackage{ragged2e} 
\usepackage{lmodern}
\newlength\longest

\usepackage{afterpage}

\usepackage{listings}

\definecolor{dkgreen}{rgb}{0,0.6,0}
\definecolor{gray}{rgb}{0.5,0.5,0.5}
\definecolor{mauve}{rgb}{0.58,0,0.82}

\definecolor{coolblack}{rgb}{0.0, 0.18, 0.39}  

\lstnewenvironment{mathematicathesis}
  {\lstset{
  tabsize=3,
  frame=tblr, 
  language=Mathematica,
  basicstyle=\small\ttfamily,
  keywordstyle=\color{black},
  commentstyle=\color{coolblack},
  showstringspaces=false,
  mathescape=true}}
{} 
  
\graphicspath{{./pictures/}}
\usepackage{subcaption}
\usepackage{diagbox}
\usepackage{csquotes}
\usepackage{mathtools}
\usepackage{cleveref}
\usepackage{framed}
\usepackage{placeins}

\makeatletter
\newcommand*{\transpose}{%
  {\mathpalette\@transpose{}}%
}
\newcommand*{\@transpose}[2]{%
  \raisebox{\depth}{$\m@th#1\intercal$}%
}
\makeatother

\begin{document}

\pagenumbering{roman}

\thispagestyle{empty}
\begin{flushright}
\includegraphics[width=6cm]{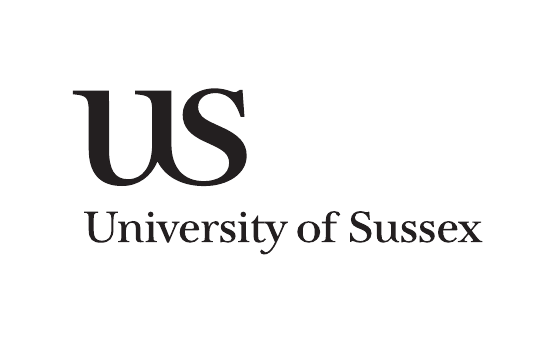}
\end{flushright}	
\vskip40mm
\begin{center}

\LARGE\textbf{Non-asymptotic quantum metrology}
\vskip2mm

\Large\textit{Extracting maximum information from limited data}
\vskip10mm

\Large
\textbf{Jes\'{u}s Rubio Jim\'{e}nez}
\normalsize

\large
\vskip30mm
Department of Physics \& Astronomy \\
\vskip4mm
University of Sussex \\
\end{center}
\vfill
\begin{flushleft}
\large

Submitted for the degree of Doctor of Philosophy \\

July 2019
\end{flushleft}		


\chapter*{\vspace{-1cm}Declaration}
I hereby declare that, unless otherwise indicated, the results presented in this thesis are the product of my own independent research, and that, to the best of my knowledge, all the relevant sources have been appropriately acknowledged. In addition, this thesis has not been and will not be submitted in whole or in part to another University for the award of any other degree.

Some of these results have already been published or are available in a preprint:
\begin{enumerate}
\item \emph{Non-asymptotic analysis of quantum metrology protocols beyond the Cram\'{e}r-Rao bound}, \underline{Jes\'{u}s Rubio}, Paul Knott and Jacob Dunningham, J. Phys. Commun. 2 015027 (2018).
\item \emph{Quantum metrology in the presence of limited data}, \underline{Jes\'{u}s Rubio} and Jacob Dunningham, New J. Phys. 21 043037 (2019).
\item \emph{Designing quantum experiments with a genetic algorithm}, Rosanna Nichols, Lana Mineh, \underline{Jes\'{u}s Rubio}, Jonathan C. F. Matthews and Paul A. Knott, Quantum Sci. Technol. 4 045012 (2019).
\item \emph{Bayesian multi-parameter quantum metrology with limited data}, \underline{Jes\'{u}s Rubio} and Jacob Dunningham, arXiv:1906.04123 (2019).
\end{enumerate}
I confirm that I am responsible for the calculations and contents in works $1$, $2$ and $4$, where the co-authors provided invaluable support and guidance by discussing the results, verifying the methods and proposing changes to enhance the quality of the manuscripts. Regarding work $3$, my contribution was twofold. On the one hand, I prepared a Matlab module with a Bayesian fitness function to be included within the AdaQuantum algorithm. On the other hand, I performed the Bayesian analysis of some of the experimental designs produced by this algorithm. 

Finally, some of the asymptotic results for sensor-symmetric networks in chapter \ref{chap:networks} were partially obtained in collaboration with Paul Knott, Timothy Proctor and Jacob Dunningham, while I was fully responsible for the Bayesian calculations. 

\vskip10mm

\raggedleft Jes\'{u}s Rubio Jim\'{e}nez


%
%
%
%
%
%
%
\thispagestyle{empty}
\newpage
\null\vskip10mm
\begin{center}
\large
\underline{UNIVERSITY OF SUSSEX}
\vskip20mm

\textsc{Jes\'{u}s Rubio Jim\'{e}nez, Doctor of Philosophy}
\vskip20mm

\underline{\textsc{Non-asymptotic quantum metrology}}
\vskip0mm

\underline{\textsc{Extracting maximum information from limited data}}
\vskip20mm
\underline{\textsc{Summary}}
\vskip2mm
\end{center}

\renewcommand{\baselinestretch}{1.0}
\small\normalsize

\justify{Science relies on our practical ability to extract information from reality, since processing this information is essential for developing theories that explain our world. This thesis is precisely the study of how to extract and process information using quantum systems when a constrained amount of resources means that the available data is limited. The natural framework for this task is quantum metrology, a set of tools to model and design quantum measurement strategies. Equipped with this theory, we advocate a Bayesian approach as the appropriate formalism to study systems with a finite amount of resources, which is a non-asymptotic problem, and we propose a methodology for non-asymptotic quantum metrology. To start with, we show the consistency of taking those solutions that are optimal in the asymptotic regime of many trials as a guide to calculate a generalised measure of uncertainty in the Bayesian framework. This provides an approximate but useful way of studying the non-asymptotic regime whenever a direct Bayesian optimisation is intractable, and it avoids non-physical results that can arise when only the asymptotic theory is employed. Secondly, we construct a new non-asymptotic Bayesian bound without relying on the previous approximation by first selecting the optimal quantum strategy for a single shot, and then simulating a sequence of repetitions of this scheme, which is suitable for experiments where we do not wish or cannot correlate different trials. These methods are applied to a Mach-Zehnder interferometer, which is a single-parameter problem, and to quantum sensing networks where the nodes are either qubits or optical modes, which are multi-parameter protocols. Our results provide a detailed characterisation of how the interplay between prior information, entanglement and a limited amount of data affects the performance of quantum metrology protocols, which has important implications for the analysis of theory and experiments in this field.}


\clearpage

\thispagestyle{empty}

\vspace*{5cm} 
\begin{quote} 
\centering 

\parbox{10cm}{
  {\center\normalsize\itshape
To those who dedicate their lives to pursue the path of knowledge with kindness and honesty, and without ever ceasing to seek the best approximation to the truth.
  \par\bigskip}
  \par} 
  
\end{quote}
\vspace*{\fill}

\clearpage


\chapter*{Acknowledgements}
\renewcommand{\baselinestretch}{\linespacing}
\small\normalsize

Those of us who find meaning in creating and understanding often tend to enjoy the pleasure of a good challenge, and a doctoral thesis has proven to be one of the best. I feel very grateful for this fantastic opportunity, which would not have been possible without my supervisor Jacob Dunningham. He not only gave me the chance to live in another country and join his group at Sussex, but he has also provided excellent guidance, full of insights and ideas. From our many discussions I have learned a very natural and intuitive way of doing physics, and it has been a pleasure to work in such friendly company. 

It has also been fantastic working and discussing physics with the rest of the group - Paul, Simon, Anthony, James, Michail and Sam. Paul's guidance, always full of optimism, was invaluable during his time as my second supervisor, and I have benefited enormously from our collaborations (my current appreciation of numerical techniques would have not happened without his help!).

I am indebted to collaborators and friends - Alfredo, Tim, George, Dominic, Matthew, Jasminder, Christos, Francesco and many others - who through many interesting discussions have helped me to refine my ideas, and to develop them further. I would also like to thank Dimitrios for his help with the mathematical literature of probability theory, and Petr for introducing me to Jaynes's work; my thesis would have been very different without it. These interactions have been possible thanks to attending many interesting conferences, which I was able to do thanks to the generous financial support provided by the University of Sussex and the South East Physics Network (SEPnet) to complete my PhD.

Working towards my PhD at Sussex has been a great experience. Its friendly staff are always ready to be helpful, and its library is superb. In fact, the library has been an invaluable source of material during development of many sections of this thesis. It has been a great pleasure meeting many fascinating people, and to have had the opportunity of developing many activities. There was always time to chat with Luc\'{i}a and let the idealists inside us to run free, to have entertaining discussions and share experiences with Harry, Nick and Jean-Marc, or to run teaching workshops with Reece. And I enormously enjoyed every session of our AMO Theory Club - Kathryn, Pedro, Germ\'{a}n, Samuel, Graham, Charlie and Michail - which was always an excellent opportunity to learn something new beyond our thesis topics. 

In addition, I am very grateful for all the support and encouragement that I have received from Kathryn during this journey. She has been a continuous source of inspiration to explore new ways of thinking about the world, and to better appreciate the wonders of life. In addition, I would like to thank her for having read some of the preliminary drafts and having provided invaluable feedback.  

Last but not least, I would like to thank all the people who have also supported my project both from home and from many other parts of the world: Isabel and Alfredo, for our fantastic meetings when I visit Madrid; In\'{e}s and Eli, for a friendship that never ends; H\'{e}ctor, because indeed great minds think alike; Anah\'{i}, for always being so trustworthy; Ana, for always being open to the world; Pablo, for our conversations about life; and my family, for your support no matter where we are.  

Thanks to all of you.



\newpage
\pdfbookmark[0]{Contents}{contents_bookmark}
\tableofcontents 
\listoftables
\phantomsection
\addcontentsline{toc}{chapter}{List of Tables}
\listoffigures
\phantomsection
\addcontentsline{toc}{chapter}{List of Figures}


\newpage
\pagenumbering{arabic}

\chapter{Introduction}
\label{chap:intro}

The primary empirical idea that underlies science in general and physics in particular is the existence of a part of the world that is independent of the mental activity of human beings. When this external world affects us and is affected by our actions in any way that we can perceive, we say that an interaction has taken place, and that this happens is precisely what gives us the opportunity of enquiring about the fundamental nature of reality.

To understand how this is possible, first we note that, to some extent, we have the ability to divide our perception of the external world as if it were a collection of different entities that can be categorised. In addition, by noticing that human beings are just another part of the world, we can extend the idea of interaction to apply between any two or more of these entities. Therefore, we can always choose a small set of entities that is delimited in space and time, which we call a system, and we can generate a chain of local interactions between them in order to study how the system behaves in a controlled way. This is what an experiment does.

For this process to be successful, it is crucial that we can recognise how the system changes under the influence of our actions, and the appearance of these changes is what we call events. A careful exposition of the central role that events play in physics can be found, for instance, in the treatments of Englert \cite{englert2013} and Haag \cite{haag1996, haag2016}. Crucially, the chain of actions that ends in the creation of an event does not need to involve a human observer directly manipulating the system of interest at every stage. For example, we can consider an interaction between the system and some artificial object that only interacts with us at a later stage. As a consequence, while it is true that which events we observe and how we do it depend on the actions that we choose to implement in the laboratory, it may be considered that the physical nature of a given event does not rely on the observer, and in that sense we can say that events give rise to objective facts. An example that illustrates well this point is that of atomic spectra, which are independent of who is performing the experiments that create the events that determine them.

The study of the events that we can observe allows us to speak about natural phenomena in terms of relationships between physical properties, and this provides us with a basis to construct the theories that help us to explain the world around us. Although as a first approximation we can describe such properties using verbal statements, a greater precision (and often clarity) is achieved by capturing their essence with mathematical objects and relationships. Of particular importance are quantitative representations, which associate numbers with properties. One of the two fundamental procedures to accomplish this task is to measure. 

Measurements can be regarded as collections of interactions between the system and some device, such that the properties that emerge from the events created in this way can be associated with magnitudes by comparison with respect to standard references or units \cite{vim2012}. Depending on the experiment, properties are sometimes quantified using integers or rational numbers, and in general they can be characterised by a set of real numbers, each of them lying within some interval whose width is related to the experimental conditions of the measurement. The use of real numbers is particularly useful in this context, since the fact that they form a complete ordered field\footnote{Here \emph{field} refers to the mathematical concept (see, e.g., \cite{spivak1980}), and not to the physical object that will appear in chapter \ref{chap:conceptual}.} \cite{spivak1980} provides us with a powerful tool for the practical necessity of comparing any two given magnitudes that may vary in a smooth way. The part of science that studies the design and implementation of measurements is metrology \cite{vim2012}.

Unfortunately, not all physical properties can be associated with a measurement procedure in such direct fashion. In fact, most of them require more complicated schemes. This is the case, for example, when we need to assign a value to a parameter that represents the difference of optical phases, which is one of the central scenarios that we will study in this thesis. In that case, our measurement is based on the detection of light \cite{helstrom1976, rafal2015, dowling2014}, and a difference of phases can only be quantified by means of a theoretical relationship that connects such parameter with the outcomes of the measurement. We then say that the parameter has to be estimated. 

The estimation of physical parameters has a wide range of applications. These include the important case of generating an educated guess for the value of those properties without a direct measurement scheme, or for which a measurement scheme is difficult to implement, and also the task of connecting parameters of a theoretical nature with experimental procedures, which is crucial to assigning values to the constants that characterise a theory. We see then that an estimation procedure is the second fundamental way of assigning numerical values to physical properties, and the formal framework that studies how to construct useful estimation schemes is estimation theory \cite{helstrom1976, jaynes2003, kay1993}. Note, however, that while estimation theory opens the door to quantify a larger set of properties, it still relies on the direct measurement of more primitive quantities that are to be related to the property of interest. 

When we succeed in assigning numerical values to some property, we say that we have extracted information from the natural world. Sometimes this assignment is unique, but in most cases we can only find a set of possible values that are compatible with the property that we are studying, and a collection of weights indicating how likely each value is on the basis of what we know about the situation at hand. When this happens, the information that we gain is partial in the sense that it does not lead to a unique answer. This ambiguity is captured by the notion of uncertainty, and the partial information that is available is modelled via the concept of probability \cite{jaynes2003}. Importantly, note that partial information as defined here does not necessarily imply a notion of incompleteness, since partial information may be all that a given system is able to offer for the property that we are trying to associate with it. 

The previous discussion highlights the crucial importance of our ability to extract information in the generation, development and testing of our scientific knowledge, and this motivates searching for new ways of enhancing both our measurement and estimation techniques. This is, in a general sense, our key motivation for addressing the research question that we develop in the following paragraphs.  

One of the most powerful known ways of improving how we measure and make estimates is to exploit the fundamental role that quantum mechanics - which is one of our fundamental theories about the universe - plays in the technological revolution known as \emph{the second quantum revolution} \cite{dowling2003}, whose aim is to exploit the physical principles of quantum theory for the development of new technologies in the areas of sensing, computation and communication \cite{degen2017, browne2017, barnett2017, acin2018}. These include important applications such as gravitational wave detection \cite{aasi2013, pitkin2011}; medical and biological imaging \cite{taylor2013, taylor2015, taylor2016}; measurements for other fragile systems such as atoms, molecules or spin ensembles \cite{eckert2007, pototschnig2011, carlton2010, wolfgramm2013, PaulProctor2016}; magnetic sensing \cite{baumgart2016}; quantum radar \cite{shabir2015, kebei2013, lanzagorta2012} and lidar \cite{lanzagorta2012, wang2016, zhuang2017}; navigation \cite{Dowling1998}; and quantum networks for distributed sensing \cite{proctor2017networked, proctor2017networkedshort, ge2018, eldredge2018, altenburg2018, qian2019} and for satellite-to-ground cryptography \cite{liao2017}, among others. In general, quantum technologies can be classified depending on whether information is extracted, processed, transferred or stored \cite{samuel2018}. 

Despite this broad scope, the common denominator underlying all quantum technologies is precisely the possibility of performing high-precision measurements, since as Dowling and Milburn argued in \cite{dowling2003}, that is a crucial requirement for the success of any technology. This gives rise to the application of quantum mechanics to the enhancement of our measurement techniques, and this process initiates a very useful feedback loop whose logic was particularly well captured by the reasoning advanced by Dunningham in \cite{dunningham2006}: measurements provide the basic information to create theories, and those theories allow us to find better measurement techniques that, in turn, might give rise to a whole new theoretical framework. The result of this is a theory to design and implement measurements by exploiting the quantum properties of light and matter. In other words, we have a \emph{quantum metrology}.   

From a formal perspective, quantum metrology can be seen as a collection of techniques that rely on quantum mechanics in order to extract information about unknown physical quantities from the outcomes of experiments \cite{giovanetti2006review, dunningham2006, paris2009, rafal2015}. Expressed in this way, its final aim is to find the strategy that can extract information with the greatest possible precision for a given amount of physical resources, and, as a consequence, it sets an optimisation problem. To solve it, first we need to define some mathematical quantity that acts as a figure of merit and informs us about the uncertainty of the estimation process, and then we can minimise such quantity with respect to the features that we can typically control, which include the details of the preparation of the experiment, the measurement scheme itself and the statistical functions employed in the analysis of the experimental data to generate an estimate. 

In practice the quality of the information extracted in this way is restricted by factors such as the number of probes, measurements or repetitions of the experiment, or by the energy that the experimental arrangement can employ. The latter constraint is particularly relevant for cases where we are interested in studying fragile systems \cite{eckert2007, pototschnig2011, carlton2010, taylor2013, taylor2015, taylor2016, PaulProctor2016}. On the other hand, the number of times that we can interact with the system under study by performing several measurements is always finite and potentially small. This is a possibility that could arise, for instance, in tracking scenarios where a scheme for remote sensing can only have access to a few observations before the object of interest is out of reach \cite{shabir2015, kebei2013, lanzagorta2012,wang2016,zhuang2017}.

It is also important to appreciate that the measurement data is not the only source of information that we can use to make estimates. To formulate an estimation problem we normally need to further specify certain details such as the instructions that we must follow in order to implement some metrology scheme in the laboratory, or any other piece of information that we may have about the unknown parameters whose values we wish to learn, the origin of such information being different from the measurement data. An example of the latter is the range of possible values that such parameters could take. This kind of information is said to be known a priori, in the sense that it precedes the initialisation of the experiment.

Given this state of affairs, it is crucial to observe that quantum metrology protocols are typically designed around the assumption that we have an abundance of measurement data (see, e.g., \cite{braunstein_gaussian1992, rafal2015, braun2018, tsang2016, liu2016, lumino2017}), and this clearly excludes real-world scenarios with very limited data such as those mentioned above. Moreover, it is frequent to find studies where it is assumed that we are working either in the high prior information regime \cite{paris2009, rafal2015}, or in the presence of complete ignorance \cite{ariano1998, chiara2003, chiribella2005, holevo2011, rafal2015}, which is the other extreme, while we may expect a realistic amount of prior knowledge to normally be moderate. In addition, these assumptions are also unsatisfactory from a theoretical point of view. Indeed, we will see that the mathematical consequence of assuming a large number of data is that the framework derived from such premise is generally valid only in an asymptotic sense, and that imposing a very large amount of prior knowledge effectively restricts the validity of our estimates to a local region of the parameter space. On the contrary, an approach of a more fundamental nature should ideally not rely on an asymptotic approximation to be relevant and useful, and it should allow for the possibility of accessing any regime of prior knowledge. 

Therefore, there is an unmet need of developing methods to study and design metrology protocols that operate in this largely unexplored and more realistic regime. This directly leads us to our thesis, which we now enunciate as follows:
\begin{framed}
\justify{~\\[-20pt]The number of times that we can access a physical system to extract information via quantum metrology and estimation theory is \emph{always} finite, and possibly small, and a realistic amount of prior information will typically be moderate. As a consequence, theoretical consistency demands a quantum metrology methodology that can depart from both asymptotic approximations and restricted parameter locations, while practical convenience requires that such methodology is also sufficiently flexible and easy to use in applications where the amount of data is limited. We submit that this methodology can and should be built on the Bayesian framework of probability theory, and that its construction can be carried out and adapted for both single and multi-parameter schemes, including important models such as the Mach-Zehnder interferometer and quantum sensing networks. Finally, we advance that this methodology generates a wealth of new results characterising an interesting interplay between different amounts of data, the prior information and quantum correlations. In other words, we propose, construct, explore and exploit a \emph{non-asymptotic quantum metrology}.}
\end{framed}

The first step to accomplish our goals will be the introduction of the fundamental concepts that we need to develop our ideas, a task that will be carried out in chapter \ref{chap:conceptual}. We will start by arguing that a version of probability theory where the focus lies on the information content of our probability models is the most suitable choice for studying the regime of limited data, and we will review the basic elements of this approach. Once we have the means to model information in terms of probabilities, we will proceed to study how such information can be encoded in quantum systems, and we will revisit the formalism of quantum mechanics to verify that using the version of the Bayesian framework that we consider in this thesis is compatible with the usual notions in quantum theory. We then complete our conceptual framework presenting the characterisation of a generic Mach-Zehnder interferometer and the quantum sensing network model introduced by Proctor \emph{et al.} \cite{proctor2017networked}, where the latter is the type of multi-parameter scheme that we will examine. 

At this point we will have all the ingredients to formulate the problem of this thesis in a formal way, which will happen in chapter \ref{chap:methodology}. We will focus our attention on experiments that are repeated a certain number of times, and upon introducing a notion of resources that is relevant for our purposes, we will define the regime of limited data in terms of a low number of trials. Then we will carry out a detailed analysis of different measures of uncertainty that we could consider as the figure of merit to be optimised, and a measure of uncertainty that is appropriate for designing inference schemes from theoretical considerations will be selected. 

Equipped with this uncertainty, we will review the fundamental equations for the optimal quantum strategy in a Bayesian context \cite{helstrom1976, helstrom1974, holevo1973b, holevo1973}, and also a set of bounds (including the widely used Cram\'{e}r-Rao bound \cite{paris2009, rafal2015}) that are often employed and that can be useful due to the general difficulties to solve the previous equations, and we will highlight the assumptions that go into the construction of these tools. Furthermore, we will revisit some known results in the single-shot regime, and we will demonstrate that a new way of understanding the latter is possible by explicitly separating the quantum and classical steps during the process of optimising the single-shot uncertainty. This will be one of our first novel insights. 

The analysis of the advantages and potential drawbacks of different tools will prepare the ground for the construction of our non-asymptotic methodology, which will emerge as a useful method for quantum metrology that is more general than simply using bounds, and while it is not as general as solving the fundamental equations for the optimal strategy, our methods will be associated with calculations that are more tractable than those in the latter approach. 

Two key approaches will constitute the basis of this methodology. We will always select the estimator that is optimal for any number of repetitions, such that this part of the problem is always exact in all our calculations, and we propose two different methods to select the quantum strategy, i.e., how the system is to be prepared and which measurement scheme should be selected. On the one hand, we propose to employ the known asymptotic theory as a guide, and to choose a quantum strategy that is guaranteed to be optimal as the number of repetitions grows, even when the analysis of the scheme is done with an uncertainty that also works in the non-asymptotic regime. That is, if we were not sure a priori about how many times the experiment is to be repeated, a weak condition that we could impose on the optimisation would be that the performance of the scheme should not break in the long run. While this does not guarantee that the solutions arising from this method will be optimal for a low number of trials, we will see that this is a useful approximation that will allow us to extract some information about the non-asymptotic regime of our metrology schemes.

As for the second method, we will go a step further and construct a fully Bayesian approach based on selecting the quantum strategy that is optimal for a single shot, so that this scheme is then repeated as many times as the application at hand demands or allows for. In this way only the resources that are needed will be optimised, and the necessity of relying on tools that assume a large amount of data will be completely eliminated from our calculations. Hence, chapter \ref{chap:methodology} will lay a bridge between the current state of the art and the novel ideas that our work introduces. 

The next four chapters will be dedicated to developing the theory associated with our non-asymptotic methodology by dividing this process in four different steps, one per chapter, and we will demonstrate these ideas explicitly with specific metrology schemes. The first step is to construct the hybrid method (exact estimation plus asymptotically optimal quantum strategy) for single-parameter schemes, a task that will be carried out in chapter \ref{chap:nonasymptotic}. Then we will demonstrate its usefulness in the context of a Mach-Zehnder interferometer that has been prepared using current techniques in optical interferometry, and we will use our method to address two questions: when does the Cram\'{e}r-Rao bound stop being valid, since in general it is only meaningful in an asymptotic sense, and how the validity of predictions of such tool change when the experiment is operating in the non-asymptotic regime. Our results will verify that the number of repetitions and the minimum amount of prior information needed to recover the asymptotic behaviour crucially depend on the state of the system. In addition, we will propose a simple analytical relation to identify and prevent the appearance of states for which the number of trials required to match the asymptotic uncertainty grows unbounded, while, at the same time, almost no information is gained for a low number of repetitions.

Our study of the Mach-Zehnder interferometer will continue in chapter \ref{chap:limited}, but this time we will implement our second method, that is, the optimisation of the uncertainty in a shot-by-shot fashion. We will see that this technique generates bounds on the estimation error that can be tight both for a single shot (by construction) and for a large number of them, since the predictions of the Cram\'{e}r-Rao bound are sometimes recovered as a limiting case within our approach. This partially fundamental character will further allow us to provide the first rigorous characterisation of the interplay between the amount of data, the prior information and the photon correlations associated with the interferometer, fulfilling in this way one of the main claims of our thesis for single-parameter protocols. Remarkably, we have found evidence of the potential existence of a trade-off between the asymptotic and non-asymptotic performances that is associated with the photon correlations within each optical mode. More concretely, while a large amount of the latter is beneficial asymptotically, sometimes it appears to be detrimental for a low amount of data. Moreover, our bounds provide us with a new benchmark to study whether certain practical measurements are actually optimal in the regime of limited data, and we have shown that the bounds that emerge from our technique are superior to other alternatives in the literature such as the quantum Ziv-Zakai and Weiss-Weinstein bounds \cite{tsang2012, tsang2016} whenever we restrict our attention to identical and independent experiments. As a final demonstration of the power of our single-parameter approach we have combined our methods with a genetic algorithm for state engineering that has been developed by our colleagues at the University of Nottingham, and we have shown that our Bayesian methodology can predict schemes that not only supersede standard benchmarks, but that have the potential to be experimentally feasible. 

The transition from single to multi-parameter estimation problems is made in chapter \ref{chap:networks}. Here we return to the hybrid method where the quantum strategy is asymptotically selected and we extend it to cover cases with several parameters. Once this step has been achieved, we proceed to apply it to a collection of sensors that are spatially distributed, which can be modelled with the framework for quantum sensing networks developed by Proctor \emph{et al.} \cite{proctor2017networked, proctor2017networkedshort}. The presence of several parameters opens the door to a vast set of new possibilities to enhance our estimation protocols, and for that reason it is useful to introduce some definitions that help us to identify in a transparent way what our final goal is. To that end we define, on the one hand, the notion of natural or primary properties of the network, and, on the other hand, the concept of derived or secondary properties, where the former refers to the original parameters of the system and the latter to functions of them. Our task in this chapter is then to determine which role the correlations between sensors play in the estimation of global properties, where these are understood as linear functions that depend non-trivially on several parameters that were originally encoded in a local way. Assuming a network where each node is a qubit, we first solve this problem asymptotically to extract the solutions that will serve us as a guide at a latter point. In particular, we will uncover the link between the geometry of the vectors associated with the components of the linear functions and the amount of inter-sensor correlations that are needed for achieving the asymptotically optimal error, and we will show that how much entanglement is required for a given geometry crucially varies with the number of repetitions of the experiment, which is a result fully compatible with our findings in the non-asymptotic study of the Mach-Zehnder interferometer. 

The final step of our methodology, which is implemented in chapter \ref{chap:multibayes}, will focus on generalising the shot-by-shot method to the multi-parameter regime. To achieve this goal we will first derive a new multi-parameter single-shot quantum bound, and we will show under which circumstances it can be saturated. This is perhaps one of the most important results that we report in this thesis. We will calculate this bound both for the qubit network studied in chapter \ref{chap:networks} and for a discrete model for phase imaging, and we will show that entanglement is not needed for the estimation of the original parameters of the network when the experiment operates in the regime of moderate prior knowledge and limited data. The crucial importance of this finding stems from the fact that an analogous result had only been established in a clear way in terms of the asymptotic theory \cite{knott2016local, proctor2017networked, proctor2017networkedshort}. 

In chapter \ref{chap:future} we will identify some of the limitations of our current approach and will discuss some ideas to overcome them, as well as potential ideas for the future of non-asymptotic quantum metrology, while chapter \ref{chap:conclusions} will be dedicated to the analysis of the unified perspective that will emerge from the findings and conclusions presented in previous chapters. 

Finally, we would like to draw attention to the fact that in appendices \ref{app:numsingle} and \ref{app:multinum} we provide a comprehensive numerical toolbox for optical interferometry and two-parameter estimation problems that is based on MATLAB and Mathematica algorithms. Hence, the interested readers will have the opportunity of either reproducing our results or adapting our algorithms to their specific problems. The relative simplicity and efficiency of these algorithms might help to overcome the extended perception that Bayesian techniques, while often conceptually clearer, are somehow less accessible due to the numerical character of the associated calculations. For the details of some of our analytical calculations and extended discussions about our methods, see appendix \ref{app:supplemental}.
\chapter{Conceptual framework}
\label{chap:conceptual}

As a first step we review the tools and concepts needed for our discussion, which will rest on three fundamental pillars: how we handle information, how quantum systems are described and which type of schemes are useful for quantum metrology. 

\section{Fundamentals I: probability theory}
\label{sec:probability}

Our main aim is to study how quantum metrology protocols are to be designed when the finite character of the number of observations is explicitly taken into account, with a particular emphasis on the regime of limited data. In this context it is natural to employ a formulation of probability theory where the central focus lies on the information content of our probabilities, and this is precisely the path that we will follow. The formal elements of this approach, which can be seen as a part of the Bayesian paradigm \cite{jaynes2003}, are briefly reviewed in the following sections\footnote{The Bayesian framework can also be constructed using bets, profit and degrees of belief \cite{definetti1990, bernardo1994}. Here we do not follow this subjectivist approach because metrology rests on the study of natural phenomena, and this is an impersonal enterprise. One may also work with a purely measure-theoretic version of the theory \cite{rosenthal2006} if the problem can be recast in the language of sets, although this is not always possible when the prior information acquires an important role \cite{jaynes2003}. Finally, note that a definition of probabilities in terms of relative frequencies that arise in a repeated experiment is not appropriate for us, since the regime of limited data includes, by definition, scenarios where the number of trials is low, or where some events happen only once. Moreover, note that it can be consistently argued that probabilities and relative frequencies are conceptually different quantities, where the former are assigned by us or by our theory and the latter are empirical facts (see \cite{jaynes2003, jeffreys1961, jiangwei2014}, and also our discussion in section \ref{subsec:lln}).}. 

\subsection{Calculus of probabilities}
\label{subsec:calprob}

Following the expositions given by Ballentine \cite{ballentine1998}, Van Horn \cite{vanhorn2003} and Jaynes \cite{jaynes2003}, we can capture the rules of probability theory using the following axioms:
\begin{enumerate}
\item $0 \leqslant P(A|B) \leqslant 1$,
\item $P(A|B) = 1$ when it can be concluded that $A$ is true on the basis of $B$,
\item $P(\neg A | B) = 1 - P(A | B)$, and
\item $P(A \land B | C) = P(A | C) P(B |A \land C)$,
\end{enumerate}
where $A$, $B$ and $C$ are propositions, and the symbols $\neg$ and $\land$ are the connectives for negation and conjunction, respectively \cite{nidditch1962, copi2016}. The probability $P(A|B)$ is to be understood as the degree of plausibility for $A$ to be true given $B$, and it can be seen as a carrier of information. More concretely, $B$ encodes either what is known about the real world or hypotheses about it, i.e., it represents a state of information \cite{jaynes2003, vanhorn2003}, and the logical analysis of this information is what determines the plausibility associated with the proposition $A$, which is the object of our enquiry. The two extremal values of the scale of plausibility correspond to the most informative scenarios, which recover as particular cases the two truth values found in propositional logic \cite{jaynes2003, nidditch1962, copi2016}, while any other intermediate plausibility will be associated with an uncertain scenario. Thus probability theory is seen as an extension of propositional calculus that allows us to encode and manipulate information in uncertain situations\footnote{There has been some debate as whether the word \emph{plausibility} should be employed as it is done here \cite{shafer2004, vanhorn2004}, since this word is used in a different sense in the theory of belief functions \cite{shafer2004}. Nevertheless, it can be argued that, historically, its use in our context is older \cite{vanhorn2004}, and we have found that, in practice, it is particularly convenient for studying estimation problems, which is our topic of discussion.} \cite{jaynes2003}. 

This way of understanding probabilities can be justified via Cox's work \cite{cox1946, cox1961}, provided that his assumptions for a reasonable measure of plausability are accepted. There is a rich literature about the validity, scope and limitations of this approach \cite{paris1994, vanhorn2003, colyvan2004, clayton2017, vanhorn2017}, but for our purposes suffice it to say that: i) there exists a rigorous treatment of Cox's ideas (see \cite{paris1994}), and (ii) in practice it can be successfully applied to a wide range of real-world problems, as the work of authors such as Jeffreys \cite{jeffreys1961} and Jaynes \cite{jaynes2003} demonstrates.

Two important concepts are those of mutual exclusivity and independence \cite{jaynes2003}. Mutually exclusive propositions satisfy that $P(A_1 \lor \cdots \lor A_s | I_0) = \sum_{i=1}^s P(A_i | I_0)$, where $\lor$ indicates disjunction \cite{nidditch1962, copi2016}, and we have that $\sum_{i=1}^s P(A_i | I_0) = 1$ if $\lbrace A_i \rbrace$ are also exhaustive. In addition, independence is expressed as $P(B_1 \land \cdots \land B_r | I_0) = \prod_{j=1}^r P(B_j | I_0)$. Beyond these notions, for us the key result that can be derived from these axioms is Bayes theorem: 
\begin{equation}
P(A | B \land I_0) = \frac{P(A | I_0) P(B | A \land I_0)}{P(B | I_0)},
\label{bayestheorem1}
\end{equation}
where $P(B | I_0) = P(A | I_0 ) P(B |A  \land I_0 ) + P(\neg A | I_0 ) P[B |(\neg A) \land I_0]$. 

To understand equation (\ref{bayestheorem1}), suppose we take $A$ to be a proposition about theoretical parameters, and imagine that the experimental outcomes are encoded in $B$. Furthermore, $I_0$ represents our initial state of information, which in this case includes the conditions under which the experiment is performed and the possible ranges for parameters and outcomes\footnote{In quantum metrology we can think of the prior information $I_0$ as a formal representation of the operational information that indicates how the experiment is to be arranged and performed, which in general is a collection of instructions expressed in the language of experimental physics.}. Then we can see that, according to equation (\ref{bayestheorem1}), the prior probability $P(A | I_0)$ is updated using the new information about $A$ provided by the empirical evidence $B$, which is encoded in the likelihood $P(B | A \land I_0)$, and the denominator acts as a normalisation constant. The overall result is the construction of the posterior probability $P(A | B \land I_0)$, which gives us the plausibility for $A$ to be true given the prior information $I_0$ and the empirical data $B$. 

When the propositions refer to variables, as it is the case of $A$ and $B$ in the previous example, probabilities are defined in term of certain probability functions that act as mathematical models for the information about the situation under analysis. For example, 
\begin{equation}
P(\Delta_{\theta'}| I_0) \equiv P(\theta \in \Delta_{\theta'}| I_0) = \int_{\Delta_{\theta'}} d\theta'' ~p(\theta'') 
\label{cont1}
\end{equation}
is the probability that $\theta$ lies in an interval of boundaries $\theta'$ and $\theta' + \Delta_{\theta'}$, where we have introduced the probability density function $p(\theta)$. In a similar way,
\begin{equation}
P( \Delta_{\theta'} \land \Delta_{m'} | I_0 ) = \int_{\Delta_{\theta'}} d\theta'' \int_{\Delta_{m'}} dm''~p(\theta'', m''),
\label{cont2}
\end{equation}
where $p(\theta, m)$ is a joint density. Note that while probabilities are dimensionless numbers, probability densities can have units.  

For the conditional density we may use equations (\ref{cont1}) and (\ref{cont2}), assume that $\Delta_{m'} \ll 1$ and $\Delta_{\theta'} \ll 1$, such that 
\begin{equation}
P( \Delta_{m'} | \Delta_{\theta'} \land I_0 )= \frac{P( \Delta_{\theta'} \land \Delta_{m'} | I_0 )}{P(\Delta_{\theta'}| I_0)} \rightarrow \frac{p(\theta', m')}{p(\theta')}\Delta_{m'},
\end{equation}
and take $p(m|\theta) = p(\theta, m)/p(\theta)$. The linear approximation in the last step can be found by integrating the Taylor expansions of the density functions $p(\theta'', m'')$ and $p(\theta'')$ around $\theta'$ and $m'$. Since this procedure also applies to $P(\Delta_{\theta'} |\Delta_{m'} \land I_0 )$, we have that $p(\theta|m) = p(\theta) p(m|\theta)/p(m)$, with $p(m)=\int d\theta ~p(\theta) p(m|\theta)$. That is, we have a version of Bayes theorem in terms of densities, and the same idea is valid when we consider vector variables. We note that in this thesis we follow the convention of omitting integration limits of general expressions where the integration is taken over the complete parameter domain, as it is the case in the latter integral. 

Sometimes it is useful to employ the more compact notation $P(d\theta| I_0 ) = p(\theta)d\theta$ and $P(d\theta \land dm | I_0) = p(\theta, m)d\theta dm$, which arises from equations (\ref{cont1}) and (\ref{cont2}) by taking infinitesimally small intervals. In general we will use the language of continuous variables because this also includes discrete cases when we allow the densities under our integration symbols to involve sums of Dirac deltas \citep{jaynes2003, breuer2002}. In those cases where an explicitly discrete treatment is more convenient, we will use the notation $P(n = n' | I_0 ) \equiv p(n')$, where $p(n)$ is a probability mass function. 

The previous description is suitable for those variables for which only uncertain information is available. Common reasons for this situation to arise are lack of knowledge, lack of control in an experiment and the existence of fundamental limits that nature imposes. The latter scenario is best illustrated by quantum systems. 

Finally, instead of working with the variables themselves, we often wish to consider some function of them. In that case, a useful quantity to have an idea of the magnitude of such function is the average. For instance, we could have $\bar{f} = \int d\theta dm~ p(m,\theta) f(m, \theta)$. 

\subsection{Law of large numbers}
\label{subsec:lln}

Another important result that we will exploit is the law of large numbers. Given the proposition $B$ representing a physical event generated in an experiment specified in $I_0$, the weak version of this law is \cite{ballentine2016, rosenthal2006} 
\begin{equation}
\lim_{\mu \to \infty} P(\abs{f_\mu - P(B|I_0)} \geqslant \varepsilon~ |I_0) = 0,
\label{lawlargenumbers}
\end{equation}
where $f_\mu = n_B(\mu)/\mu$ is the relative frequency of $B$ after performing $\mu$ independent repetitions of the experiment in $I_0$, and $\varepsilon$ is a positive number. We say that $f_\mu$ converges in probability to $P(B|I_0)$ \cite{rosenthal2006}.

The importance of this result is that it offers an empirical link between probabilities and relative frequencies, since the latter are quantities that we measure in the laboratory. To see why, we propose the following argument. First we recall that $P(B|I_0)$ is the plausibility for the event $B$ to happen when the experiment in $I_0$ is run once. Presumably, $I_0$ encodes the procedure that generates the events, and it also contains the fact that our actions do not produce the same event in each new repetition. The key observation is that $f_\mu$ in equation (\ref{lawlargenumbers}) is based on the same $I_0$, in the sense that we could perform a simulation where $P(B|I_0)$ and $P(\neg B|I_0)$ are used for generating $\mu$ outcomes, and calculate the relative frequency $f_\mu$ for the event $B$ from them. In view of this, equation (\ref{lawlargenumbers}) expresses the intuitive idea that if in each run some outcomes are more (less) likely to appear, then as we increase the number of repetitions it is also likely that the largest (lowest) values for relative frequencies correspond to the largest (lowest) single-shot probabilities.

The frequency $f_\mu$ in the previous discussion is not yet factual, but a prediction made on the basis of $I_0$ and our model $P(B|I_0)$. If we now perform the actual experiment and we observe that the experimental frequencies after a very large number of trials are compatible with those that come from the model, we may think of it as a good representation of the available information about the physical phenomenon that gave rise to the experiment. Moreover, since it is likely that $f_\mu$ and $P(B|I_0)$ are close in the long run, we may also imagine that the experimental frequencies are to be compared to the probability $P(B|I_0)$ directly.

This idea becomes even more meaningful when we consider the strong version of the law, which instead states that $f_\mu \rightarrow P(B|I_0)$ almost surely as $\mu$ increases \cite{rosenthal2006}. Crucially, comparing probabilities and relative frequencies is precisely what is done in practice with quantum experiments. A good example can be found in the results of \cite{baumgart2016} for the implementation of a magnetometer. In particular, this work shows a good agreement between the quantum-mechanical probabilities and the measured frequencies, which is fully compatible with our rationale above. As a consequence, this way of looking at the law of large numbers provides a clear link with experiments while probabilities are still seen as mathematical models that encode information, and that are qualitatively different from the concept of relative frequency. 

\section{Fundamentals II: quantum mechanics}
\label{sec:qmech}

In his celebrated work of 1925 (page 261 of \cite{waerden1967}), Heisenberg offered an insight that would eventually lead to the modern formalism of quantum mechanics. Starting by representing the dynamical variables\footnote{The dynamical variables represent elementary properties that we can use to describe a physical system, and also its variation in time. Examples of these include the positions and momenta of an ensemble of particles, the components of their spin or the amplitude of a field.} with Fourier terms, his key idea was to modify these terms such that the experimental facts of the atomic realm could be accommodated, while still retaining the form of the classical laws of dynamics. As a result, later work built on this premise produced a new formalism broad enough to generate probability models that can capture the behaviour of the phenomenology of quantum systems, whose nonclassical features stem from the discreteness associated with the quantum of action $h$. We turn now our attention to how this theory describes the physical systems that we use in quantum metrology.  

\subsection{Elements of the theory}
\label{subsec:elements}

A useful way of looking at the theory is to decompose it in three parts:
\begin{enumerate}
\item Each dynamical variable is represented with a Hermitian operator whose spectrum contains the real numbers that such variable can take, or the intervals in which it can lie. Physical systems are then characterised by a set $\boldsymbol{Z}(t) = \lbrace Z_1(t), Z_2(t), \dots\rbrace$ of these Hermitian operators, for which
\begin{equation}
\left[Z_i(t),Z_j(t)\right] = Z_i(t)Z_j(t) - Z_j(t)Z_i(t) \neq 0
\label{commutator}
\end{equation}
for at least some of the cases where $i \neq j$. That the Hermitian operators for different dynamical variables may not commute is a mathematical representation of the fundamental limits associated with the existence of $h$.
\item To model more complex aspects of the quantum realm we can construct general functions $f(\boldsymbol{Z}(t), t)$, and we can consider their evolution in time, which for closed systems is given by Heisenberg's equation of motion \cite{englert2013, schwinger2001}
\begin{equation}
\frac{d}{d t} f(\boldsymbol{Z}(t), t) = \frac{\partial}{\partial t}f(\boldsymbol{Z}(t), t) + \frac{1}{i\hbar} \left[f(\boldsymbol{Z}(t), t),H(\boldsymbol{Z}(t), t)\right],
\label{heisenbergeq}
\end{equation}
with initial condition $f(\boldsymbol{Z}(t_0), t_0) = f_0$. The function $H(\boldsymbol{Z}(t), t)$ is the Hamiltonian, a Hermitian operator that generates the temporal displacement, and $\hbar = h/(2\pi)$ is the reduced Planck constant. Note that equation (\ref{heisenbergeq}) also gives the evolution of the dynamical variables themselves when $f(\boldsymbol{Z}(t), t) = Z_i(t)$, for any $i$. Among all the functions $f(\boldsymbol{Z}(t), t)$, two of them play a crucial role in the theory:
\begin{enumerate}
\item[2.i.] The density operator $\rho(\boldsymbol{Z}(t), t)$ is a positive semi-definite Hermitian operator satisfying $\mathrm{Tr}[\rho(\boldsymbol{Z}(t),t)] = 1$, and it represents how the system is prepared at some moment in time \cite{englert2013,ballentine1998}. We call this the state preparation procedure \cite{ballentine1998}, or simply state. When the system is closed, the details of the initial preparation are preserved as time passes, and as such we have that $d\rho(\boldsymbol{Z}(t),t)/dt = 0$ \cite{englert2013, schwinger2001}. Inserting this fact into equation (\ref{heisenbergeq}) we find von Neumann's equation
\begin{equation}
\frac{\partial}{\partial t}\rho(\boldsymbol{Z}(t), t) =  \frac{1}{i\hbar} \left[H(\boldsymbol{Z}(t), t), \rho(\boldsymbol{Z}(t), t)\right],
\label{vonneumann}
\end{equation}
with initial condition $\rho(\boldsymbol{Z}(t_0), t_0) = \rho_0$.
\item[2.ii.] The probability operator
\begin{equation}
E(\Delta_{m'}, \boldsymbol{Z}(t), t) = \int_{\Delta_{m'}} dm''~E(m'', \boldsymbol{Z}(t), t),
\end{equation}
also positive semi-definite and Hermitian, represents a measurement device or instrument \cite{helstrom1976} that interacts with a system described by $\boldsymbol{Z}(t)$, generating an event characterised by an outcome $m$ that lies in some subinterval of width $\Delta_{m'}$. We say that $E(m, \boldsymbol{Z}(t), t)$ generates a probability-operator measurement (POM)\footnote{Also known as positive operator-valued measure (POVM) \cite{englert2013, jiangwei2014, barnett2014}.}, such that the identity is resolved as
\begin{equation}
\int dm~E(m, \boldsymbol{Z}(t), t) = \mathbb{I}.
\end{equation}
\end{enumerate}
\item The Born rule establishes that the probability density for observing the outcome $m$ at time $t$ is given by
\begin{equation}
p(m|t) = \mathrm{Tr}\left[E(m, \boldsymbol{Z}(t), t) \rho(\boldsymbol{Z}(t), t) \right],
\label{bornrule}
\end{equation}
and it provides the link between theory and experiment. In particular, the probability model $p(m|t)$ can be used to predict the relative frequencies that we measure in the laboratory, following the rationale that we discussed in section \ref{subsec:lln} in connection with the law of large numbers.
\end{enumerate}

According to the Born rule, quantum probabilities can be seen as depending on two different types of information. On the one hand, they depend on our choices for the functions $\rho(\cdot)$ and $E(\cdot)$. Following current practice, we will employ rank-one operators such as pure states and projective measurements when, for all practical purposes, the preparation of systems and instruments involves a degree of control so high that can be thought of as to provide maximum information. Otherwise, mixed states (i.e., density operators for which $\mathrm{Tr}[\rho(t)^2]\neq \mathrm{Tr}[\rho(t)] = 1$ \cite{breuer2002}) and more general POMs are to be employed. Note that projective measurements originate in the idea of quantum observable. In particular, an observable is a physical quantity represented by a Hermitian operator whose eigenvectors give rise to a measurement scheme. For example, upon calculating the spectral decomposition $Z(t) = \int dz\,z \ketbra{z,t}$ for the dynamical variable $Z(t)$, we can implement its measurement using the projectors $\ketbra{z,t} = E(z,t)$, for which $ E(z,t)dz E(z',t) dz' = \delta(z - z')E(z, t)dz dz' $. 
 
On the other hand, quantum probabilities also incorporate the nonclassicality that emerges from $h$ via the commutation relations for the dynamical variables, and also through the law of evolution in equation (\ref{heisenbergeq}). That $p(m|t)$ takes into account the relevant role of $h$ in quantum physics offers, in fact, a good way of understanding the success of quantum technologies. Probabilities in classical physics are less constrained because their only source of uncertainty is the lack of knowledge about an over-idealised initial state of affairs, and thus some of the models that they admit do not correspond with reality. On the contrary, quantum protocols built using equation (\ref{bornrule}) are based on a more realistic description of natural phenomena, and as such they give us a superior framework to explore which are the best technologies that nature allows.

This way of breaking the theory into what experimenters can freely modify (states and measurements) and a physical law (the existence of $h$) provides an heuristic intuition that is extremely useful to encode and manipulate information in quantum systems, which is crucial to design metrology protocols. Beyond that, one can simply focus on the probability models that emerge from equation (\ref{bornrule}) as the physically meaningful quantities encoding information about quantum systems, and we can generally regard the operators that appear in the theory as abstract tools.

The previous perspective suggests that probabilities for classical and quantum systems differ in the origin of the uncertain information that they encode, which in turn affects how they are mathematically constructed\footnote{In \cite{isham1995} Isham points out that the key difference between classical and quantum probabilities is that while the former are based on ratios of volumes, the latter come from a version of the Pythagorean theorem with complex numbers.}, but not necessarily on what probability as a concept is. This is further supported by the fact that it is possible to show that no formal contradiction emerges between probability theory and quantum mechanics when the former is properly applied \cite{ballentine1986}. This includes cases where a single event is involved \cite{ballentine1998, ballentine1986}, and also joint probabilities for events associated with commuting POMs\footnote{For example, given the POMs generated by $E(m, Z(t_0),t) \equiv E(m,t)$, $F(k, Z(t_0),t) \equiv F(k,t)$ and the state $\rho(Z(t_0),t) \equiv \rho(t)$ in the Schr\"{o}dinger picture, if $[E(m,t), F(k,t)] = 0$, then 
\begin{align}
p(m,k|t) &= \mathrm{Tr}\left[E(m,t)F(k,t)\rho(t)\right]
\nonumber \\
&= \mathrm{Tr}\left[\sqrt{F(k,t)}E(m,t)\sqrt{F(k,t)}\rho(t)\right]
\nonumber \\
&= \mathrm{Tr}\left[F(k,t)\rho(t)\right]\mathrm{Tr}\left[E(m,t)\frac{\sqrt{F(k,t)}\rho(t)\sqrt{F(k,t)}}{\mathrm{Tr}\left[F(k,t)\rho(t)\right]}\right]
\nonumber \\
&= p(k|t) p(m|k, t), 
\nonumber
\end{align}
which is the product rule (axiom 4) of probability theory.} \cite{breuer2002, ballentine1998}. A proper probability model for the joint occurrence of events with non-commuting POMs cannot be constructed on the basis of such POMs, but there may be other POMs that provide less precise information about those events in a joint manner (see section 3.6 of \cite{holevo2011}), which again would give a probability compatible with the usual rules. Therefore, we conclude that we can safely exploit the Bayesian framework that we described in section \ref{sec:probability} for the design of quantum metrology protocols\footnote{The use of different probability systems in quantum mechanics has been previously explored in the literature, including our current approach (see, e.g., \cite{ballentine1986, ballentine1998, ballentine2016}). Nevertheless, we are not aware of other works that follow the same argumentation that we propose here, and thus we consider our presentation to be an important step to enhance the conceptual understanding of the role of quantum theory in metrology. Importantly, despite our joint use of quantum mechanics and Bayesian probabilities, our approach is not related to \emph{QBism} \cite{fuchs2017}, since the latter interprets the quantum formalism using de Finetti's personalist philosophy, which, as we previously mentioned, leads to an alternative formulation of Bayesian theory that we do not use here.}.

\subsection{Light, atoms and quantum information}
\label{subsec:qapp}

The applications of quantum mechanics range from the fundamental description of natural entities to the pragmatic aspects of encoding information in quantum systems. Here we collect both types of result in order to prepare the ground for the metrological protocols in the next sections. 

We start with the description of electromagnetic radiation in free space. Given the Hermitian operators $\boldsymbol{E}(\boldsymbol{x},t)$ and $\boldsymbol{B}(\boldsymbol{x},t)$ associated with the electric and magnetic fields at position $\boldsymbol{x}$, we may decompose them as
\begin{equation}
\boldsymbol{E}(\boldsymbol{x},t) = \sum_{\boldsymbol{k}, \sigma} \boldsymbol{E}_{\boldsymbol{k}, \sigma}(\boldsymbol{x},t), ~~ \boldsymbol{B}(\boldsymbol{x},t) = \sum_{\boldsymbol{k}, \sigma} \boldsymbol{B}_{\boldsymbol{k}, \sigma}(\boldsymbol{x},t)
\end{equation}
in a portion of space of volume $\mathcal{V} = L^3$ and periodic boundary conditions, where the form of the modes $\boldsymbol{E}_{\boldsymbol{k}, \sigma}(\boldsymbol{x},t)$ and $\boldsymbol{B}_{\boldsymbol{k}, \sigma}(\boldsymbol{x},t)$ is \cite{barnett2002, ballentine1998}
\begin{equation}
\boldsymbol{E}_{\boldsymbol{k}, \sigma}(\boldsymbol{x},t) = i \sqrt{\frac{\hbar \omega_k}{2\varepsilon_0 \mathcal{V}}} \left[\boldsymbol{\varepsilon}_{\boldsymbol{k},\sigma}a_{\boldsymbol{k},\sigma}\mathrm{e}^{i(\boldsymbol{k}\cdot \boldsymbol{x}-\omega_k t)} - \boldsymbol{\varepsilon}_{\boldsymbol{k},\sigma}^{*} a_{\boldsymbol{k},\sigma}^\dagger\mathrm{e}^{-i(\boldsymbol{k}\cdot \boldsymbol{x}-\omega_k t)} \right]
\end{equation}
and $\boldsymbol{B}_{\boldsymbol{k}, \sigma}(\boldsymbol{x},t) = [\boldsymbol{k}\times\boldsymbol{E}_{\boldsymbol{k}, \sigma}(\boldsymbol{x},t)]/\omega_k$. In addition, $\boldsymbol{k} = k\boldsymbol{\hat{u}}_k = 2\pi(n_x, n_y, n_z)/L$ is a wavevector, $\sigma$ is an index with two values, $\omega_k = c k$ is an angular frequency, $c = 1/\sqrt{\varepsilon_0 \mu_0}$ is the speed of light, $\varepsilon_0$ is the vacuum permittivity and $\mu_0$ is the vacuum permeability. The operators $a_{\boldsymbol{k},\sigma}$ and $a_{\boldsymbol{k},\sigma}^\dagger$ satisfy the commutation relations 
\begin{equation}
[a_{\boldsymbol{k},\sigma}, a_{\boldsymbol{k}',\sigma'}] = [a_{\boldsymbol{k},\sigma}^\dagger, a_{\boldsymbol{k}',\sigma'}^\dagger] = 0, ~~ [a_{\boldsymbol{k},\sigma}, a_{\boldsymbol{k}',\sigma'}^\dagger] = \delta_{\boldsymbol{k},\boldsymbol{k}'}\delta_{\sigma, \sigma'},
\label{lightcomm}
\end{equation}
while we have that $\boldsymbol{k}\cdot \boldsymbol{\varepsilon}_{\boldsymbol{k},\sigma} = 0$, $\boldsymbol{\varepsilon}_{\boldsymbol{k},\sigma}\boldsymbol{\varepsilon}_{\boldsymbol{k},\sigma'}^{*} = \delta_{\sigma, \sigma'}$ and $\sum_\sigma (\boldsymbol{\varepsilon}_{\boldsymbol{k},\sigma})_\alpha (\boldsymbol{\varepsilon}_{\boldsymbol{k},\sigma}^{*})_\beta = \delta_{\alpha, \beta} - k_\alpha k_\beta /k^2$ for the polarization vector $\boldsymbol{\varepsilon}_{\boldsymbol{k},\sigma}$, where $\alpha$, $\beta$ are vector components.

If we further use $\boldsymbol{E}_{\boldsymbol{k}, \sigma}(\boldsymbol{x},t)$ and $\boldsymbol{B}_{\boldsymbol{k}, \sigma}(\boldsymbol{x},t)$ to calculate the operator
\begin{align}
H' &= \frac{1}{2} \sum_{\boldsymbol{k} \boldsymbol{k}', \sigma \sigma'} \int_{\mathcal{V}} d\boldsymbol{x} \left[\varepsilon_0 \boldsymbol{E}_{\boldsymbol{k}, \sigma}(\boldsymbol{x},t) \cdot \boldsymbol{E}_{\boldsymbol{k}', \sigma'}(\boldsymbol{x},t) + \frac{\boldsymbol{B}_{\boldsymbol{k}, \sigma}(\boldsymbol{x},t) \cdot \boldsymbol{B}_{\boldsymbol{k}', \sigma'}(\boldsymbol{x},t)}{\mu_0}  \right]
\nonumber \\
&= \sum_{\boldsymbol{k}, \sigma} \hbar \omega_k \left( a_{\boldsymbol{k}, \sigma}^\dagger a_{\boldsymbol{k}, \sigma} + \frac{1}{2} \right) \equiv \sum_{\boldsymbol{k}, \sigma} H_{\boldsymbol{k}, \sigma}',
\end{align}
then we can construct the single-mode Hamiltonian
\begin{equation}
H_{\boldsymbol{k}, \sigma} = H'_{\boldsymbol{k}, \sigma} - H_{0,k} = \hbar \omega_k a_{\boldsymbol{k}, \sigma}^\dagger a_{\boldsymbol{k}, \sigma} \equiv \hbar \omega_i a_i a_i^\dagger = H_i
\label{lighthamil}
\end{equation}
where $H_{0,k} = \hbar \omega_k/2$ and we have introduced the notational change $(\boldsymbol{k}, \sigma) \rightarrow i$ for simplicity. Note that $\boldsymbol{E}_{\boldsymbol{k},\sigma}(\boldsymbol{x},t)$ and $\boldsymbol{B}_{\boldsymbol{k},\sigma}(\boldsymbol{x},t)$ satisfy Heisenberg's equation of motion in equation (\ref{heisenbergeq}) when we treat them as dynamical variables labelled by $\boldsymbol{x}$ and we use the Hamiltonian in equation (\ref{lighthamil}). Furthermore, the new notation implies that the relations in equation (\ref{lightcomm}) become
\begin{equation}
\comm*{\hat{a}_i}{\hat{a}_j} = \comm*{\hat{a}_i^{\dagger}}{\hat{a}_j^{\dagger}} = 0, ~~\comm*{\hat{a}_i}{\hat{a}_j^{\dagger}} = \delta_{ij}.
\label{commlightsimple}
\end{equation}

Upon diagonalising equation (\ref{lighthamil}) we find that $H_i = \hbar \omega_i \sum_{n_i} n_i \ketbra{n_i}$, with $n_i = 0, 1, 2, \dots$ and $\langle n_i | n_i' \rangle = \delta_{n_i,n_i'}$. The eigenvector $\ket{n_i}$ is seen as the state for $n_i$ quanta of light, or photons, each of them with energy $\hbar \omega_i$ and characterised by the properties indicated in $i$. Since $a_i^\dagger \ket{n_i} = \sqrt{n_i + 1}\ket{n_i + 1}$ and $a_i \ket{n_i} = \sqrt{n_i}\ket{n_i - 1}$ \cite{rafal2015}, we can interpret $a_i^\dagger$ and $a_i$ as creation and annihilation operators, respectively. Moreover, $\ket{0}$ is a state without photons, i.e., the vacuum, and $N_i \ket{n_i} = n_i \ket{n_i}$, $N_i \equiv a_i^\dagger a_i$ being the number operator. For $j$ independent modes we have that the most general initial density operator that we can construct is \cite{rafal2015}
\begin{equation}
\rho_0 = \sum_{\boldsymbol{n}, \boldsymbol{n}} c_{\boldsymbol{n}, \boldsymbol{n}'} \ketbra{\boldsymbol{n}}{\boldsymbol{n}'},
\label{genlightstate}
\end{equation}
where $\boldsymbol{n} = (n_1, \dots, n_j)$ and $\ket{\boldsymbol{n}} = \ket{n_1, \dots, n_j} = \ket{n_1}\otimes \cdots \otimes\ket{n_j}$. Equation (\ref{genlightstate}) and the operators $a_i^\dagger$, $a_i$, for $i = 1, \dots, j$, are sufficient to describe the optical systems that we will employ. 

On the other hand, we may also consider sensors built with atoms at low energies. Suppose we have an atom with two energy levels, $\hbar \omega_0$ and $\hbar \omega_1$, where the former is associated with the ground state $\ket{0}$ and the latter with the excited state $\ket{1}$. This is one of the possible ways of implementing the notion of qubit in a real system \cite{nielsen2010}. Its most general initial state is \cite{barnett2002}
\begin{equation}
\rho_0 = \frac{1}{2}\left(\mathbb{I} + \boldsymbol{\sigma}\cdot \boldsymbol{\hat{r}}\right) = \sum_{i, i'=0}^1 c_{i,i'} \ketbra{i}{i'},
\end{equation}
where $\boldsymbol{\hat{r}}$ is some real unit vector, $\mathbb{I}$ is the identity matrix, $\boldsymbol{\sigma} = (\sigma_x, \sigma_y, \sigma_z)$ with components
\begin{equation}
\sigma_x = \begin{pmatrix}
0 & 1 \\
1 & 0
\end{pmatrix}, ~~
\sigma_y = \begin{pmatrix}
0 & -i \\
i & 0
\end{pmatrix}, ~~
\sigma_z = \begin{pmatrix}
1 & 0 \\
0 & -1
\end{pmatrix},
\label{paulimatrices}
\end{equation}
which are the Pauli matrices, and we have assumed the convention $\bra{0} = (1, 0)$ and $\bra{1} = (0, 1)$. We notice that, given that the operators in equation (\ref{paulimatrices}) can be associated with physical quantities that are observable\footnote{In particular, $\sigma_x$ and $\sigma_y$ represent the real and complex parts of the complex dipole moment, while the atomic inversion is given by $\sigma_z$ \cite{barnett2002}.} \cite{barnett2002}, we may use them as dynamical variables that are expressed in the Schr\"{o}dinger picture. Those parts of our study based on this description will be focused on a Hamiltonian with the form $H \propto \sigma_z$, for some proportionality constant with units of energy, and, as we will see, the generalisation to several independent two-level systems is analogous to the case for independent optical modes. 

Let us imagine now that we wish to find the time-evolved state $\rho(t) \equiv \rho(\boldsymbol{Z}(0),t)$ for some time-independent Hamiltonian $H \equiv H(\boldsymbol{Z}(0))$ such as those found in the previous discussion, where we have assumed that $t_0 = 0$. In that case, von Neumann's equation (\ref{vonneumann}) implies that $\rho(t) = U(t) \rho_0 U(t)^\dagger = \mathrm{e}^{-i H t/\hbar} \rho_0 \mathrm{e}^{i H t/\hbar}$, which is an example of unitary evolution because $U(t)U(t)^\dagger = U(t)^\dagger U(t) = \mathbb{I}$. We typically look at $\rho(t)$ as representing the state of the system at time $t$. However, an alternative possibility is to imagine that the elapsed time $t$ is an unknown parameter to be found by performing measurements on a system that has evolved (i.e., it has changed) and where $\rho_0$ and $H$ are known. Given this point of view, it would be desirable to examine whether the same idea can be exploited for more general parameters representing other changes in the system. In that case, we could use quantum systems to encode and manipulate information. 

It turns out that the answer is in the affirmative. A general parameter $\theta$ can be encoded in a probe with initial state $\rho_0$ via a generator $K$, which is a Hermitian operator. For parameter-independent generators we can mimic the time evolution and take the encoding to be formally expressed as $\rho(\theta) = \mathrm{e}^{-i K \theta} \rho_0 \mathrm{e}^{i K \theta}$, which is consistent with the fact that any unitary transformation of a density operator produces a new valid quantum state \cite{dunningham2018}, and we may recast this as the solution to the operator differential equation
\begin{equation}
\frac{d\rho(\theta)}{d\theta} = i[\rho(\theta),K],
\label{genuniencoding}
\end{equation}
with initial condition $\rho(0)=\rho_0$. This equation, which is valid for generic parameters \cite{kok2010}, leads us to a more abstract formalism where the focus is shifted to the information represented by $\theta$, while the mechanical description in terms of dynamical variables is no longer explicit. As a consequence, Hermitian operators such as $\rho(\theta)$ are seen as only depending on the general parameter $\theta$, which justifies the use of the total derivative in equation (\ref{genuniencoding}). This contrasts with the partial derivative in von Neumann's equation (\ref{vonneumann}), where states were treated as functions of both the dynamical variables and the parametric time.  

The departure from the mechanical picture becomes even more apparent in scenarios with several unknown parameters $\boldsymbol{\theta} = (\theta_1, \dots, \theta_d)$. If these are encoded by means of a set of parameter-independent commuting generators $\boldsymbol{K} = (K_1, \dots, K_d)$ (i.e., $[K_i, K_j] = 0$ for all $i$, $j$), then we can trivially upgrade equation (\ref{genuniencoding}) to the vector equation 
\begin{equation}
\boldsymbol{\nabla} \rho(\boldsymbol{\theta}) = i \left[ \rho(\boldsymbol{\theta}), \boldsymbol{K} \right],
\label{vecunienc}
\end{equation}
with initial condition $\rho(\boldsymbol{0}) = \rho_0$, and its solution is $\rho(\boldsymbol{\theta}) = \mathrm{e}^{-i \boldsymbol{K} \cdot \boldsymbol{\theta}} \rho_0 \mathrm{e}^{i \boldsymbol{K}\cdot \boldsymbol{\theta}}$.

The protocols in this thesis are based on the class of schemes where the information is encoded employing either equation (\ref{genuniencoding}) or equation (\ref{vecunienc})\footnote{We leave the study of schemes with more general unitary transformations, non-unitary encodings or non-commuting generators \cite{Szczykulska2016, baumgratz2016} for future work.}. Once this has been achieved, the next step is to extract that information by performing a statistical analysis that involves the probabilities generated via the Born rule in equation (\ref{bornrule}), which for general parameters can be expressed as
$p(m|\boldsymbol{\theta})=\mathrm{Tr}\left[E(m)\rho(\boldsymbol{\theta})\right]$. It is clear that the efficiency of these information-processing techniques will depend on the characteristics of states, measurements and generators, among which entanglement and other correlations deserve special attention. 

Given a system whose space of operators is partitioned in terms of $j$ subsystems, we say that a generic ket $\ket{a}$ is \emph{entangled} when it cannot be expressed as
\begin{equation}
\ket{a} = |a^{(1)}\rangle\otimes\cdots\otimes|a^{(j)}\rangle, 
\end{equation}
which applies to both pure states and projective measurements. More generally,
we say that a generic operator $A_x$ (e.g., density operators or general POMs) depending on some variable $x$ is entangled with respect to the chosen partition if it cannot be written as \cite{dunningham2018}
\begin{equation}
A_x = \int dx~p(x)A_x^{(1)}\otimes\cdots\otimes A_x^{(j)},
\label{genericseparable}
\end{equation}
for some probability density $p(x)$. However, note that while an operator of the form of equation (\ref{genericseparable}) is not entangled, in general it will still be correlated. A complete lack of correlations between subsystems would imply that
\begin{equation}
A_x = A_x^{(1)}\otimes\cdots\otimes A_x^{(j)}.
\end{equation}
Providing new insights to understand the role of correlations in quantum metrology is one of our main goals.

\section{Fundamentals III: quantum schemes}
\label{sec:qmetrology}

Sections \ref{sec:probability} and \ref{sec:qmech} have provided us with the necessary tools to model information in general and to encode it in physical systems whose quantum aspects are relevant in particular. We conclude this chapter by presenting the schemes that will serve as the basis of the quantum protocols in later chapters.

\subsection{The Mach-Zehnder interferometer}
\label{subsec:optint}

The Mach-Zehnder interferometer is an optical system formed by two electromagnetic modes with the same frequency and a series of passive, lossless and linear elements \cite{rafal2015, yurke1986}. This arrangement lies at the heart of quantum metrology as a paradigmatic protocol for phase estimation. 

An elegant and economical way of describing this scheme is to employ the Jordan-Schwinger map \cite{yurke1986, rafal2015}
\begin{equation}
J_x = \frac{1}{2}\left(a_1^\dagger a_2 + a_1 a_2^\dagger \right), ~~ J_y = \frac{i}{2}\left(a_1 a_2^\dagger - a_1^\dagger a_2 \right), ~~ J_z = \frac{1}{2}\left(a_1^\dagger a_1 - a_2^\dagger a_2 \right),
\end{equation}
where $J_x$, $J_y$, $J_z$ are angular momentum operators satisfying the commutation relations $[J_i, J_j] = i\epsilon_{ijk}J_k$, with $i, j, z = x, y, z$, for the Lie algebra of the $SU(2)$ group \cite{yurke1986}. This arises from the commutation relations in equation (\ref{commlightsimple}) that are satisfied by the creation and annihilation operators $a_1^\dagger$, $a_2^\dagger$, $a_1$ and $a_2$.

\begin{figure} [t]
    \centering
    \includegraphics[trim={0cm 0cm 0cm 0.25cm},clip,width=12cm]{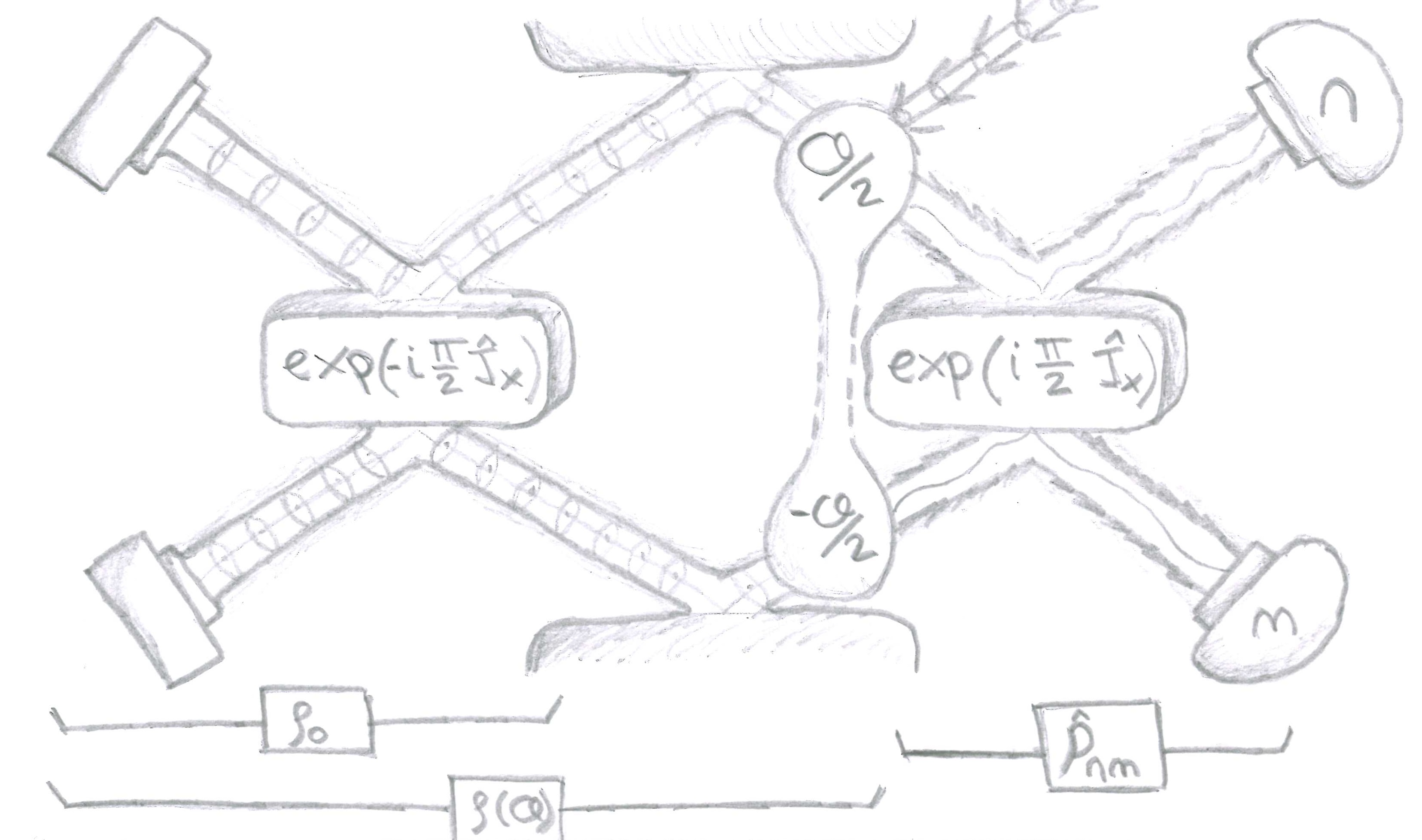}
    \caption[Artistic representation of the Mach-Zehnder interferometer]{Artistic representation of the Mach-Zehnder interferometer. A probe state $\rho_0$ is first prepared by mixing two light beams with a $50$:$50$ beam splitter. Then the probe interacts with an external entity whose properties we wish to study, so that an unknown parameter $\theta$ that is related to them is encoded as $\rho(\theta)$. Finally, the light beams are recombined with a second beam splitter and the number of clicks of each detector are measured. The information about the parameter can be extracted by processing these data.}
    \label{mzinterferometer}
\end{figure}

Within this framework we can capture the action of a $50$:$50$ beam splitter with the unitary operator $U_{\mathrm{BS}} = \mathrm{exp}(-i \frac{\pi}{2}J_x)$, while a difference of phase shifts $\theta$ is modelled with $U(\theta) = \mathrm{exp}(-i J_z\theta)$ \cite{yurke1986, rafal2015}. If the initial state for the two ports is $\rho_0'$ and $\lbrace \ketbra{n_1, n_2}\rbrace$ is a POM that counts the number of photons at the end of each port, then the standard configuration of this interferometer can be implemented with the sequence of operations
\begin{equation}
\rho_0' \rightarrow U_{\mathrm{BS}} \rightarrow U(\theta) \rightarrow U_{\mathrm{BS}}^\dagger \rightarrow \ketbra{n_1, n_2},
\label{mzsequence}
\end{equation}
which is visually represented in figure \ref{mzinterferometer}. More concretely, we have a protocol with three steps: i) preparation of the state $U_\mathrm{BS} \rho_0' U_\mathrm{BS}^\dagger$, (ii) encoding of the unknown parameter $\theta$ in the transformed probe $U(\theta)U_\mathrm{BS} \rho_0' U_\mathrm{BS}^\dagger U(\theta)^\dagger$, and (iii) recording of the read-out $(n_1, n_2)$ that has been produced by the measurement scheme $U_\mathrm{BS}\ketbra{n_1, n_2}U_\mathrm{BS}^\dagger$. Note that, as we mentioned in section \ref{subsec:qapp}, any unitary transformation of a quantum state produces a new density operator. 

As a generalisation of the previous idea we can consider two-mode interferometry protocols based on the sequence
\begin{equation}
\rho_0 \rightarrow \mathrm{e}^{-iJ_z\theta} \rightarrow E(m),
\label{mzgeneral}
\end{equation}
for some state $\rho_0$ and POM $E(m)$. This generalised Mach-Zehnder interferometer\footnote{Importantly, the model in equation (\ref{mzgeneral}) is a direct representation of realistic experiments when either $[\rho_0, N_T] = 0$ or $[E(m), N_T] = 0$, or both, where $N_T = a_1^\dagger a_1 + a_2^\dagger a_2$ is the total number operator \cite{yurke1986, pezze2015}. If this is not the case, then the $SU(2)$ symmetry is not satisfied in practice because a second parameter is imprinted by $N_T$ in the transformed probe \cite{pezze2015}. Here we assume that the experiment has been calibrated such that this parameter can be set to zero whenever the previous conditions are not fulfilled, and we consider that only the resources that enter into the scheme once it has been calibrated are relevant \cite{proctor2017networked}.} is the model that we will employ in chapters \ref{chap:nonasymptotic} and \ref{chap:limited} to study single-parameter estimation problems. 

Although the full development of estimation techniques will be carried in the next chapter, let us consider here a simple estimation strategy to illustrate how an interferometer can be used to extract information about $\theta$, as well as to introduce those concepts that play a crucial role in optical interferometry. In particular, suppose that the scheme in equation (\ref{mzgeneral}) is initialised in the pure state $ \rho_0 = \ketbra{\psi_0}$ and that we measure the observable $M = \int dm\,m \ketbra{m}$, recording the outcome $m$. If we repeat this protocol $\mu$ times and $\mu \gg 1$, then in practice we may assume that 
\begin{equation}
\frac{1}{\mu}\sum_{i=1}^\mu m_i \approx \langle \psi_0 | U(\theta)^\dagger M U(\theta) |\psi_0 \rangle = \langle \psi_0 | M(\theta) |\psi_0 \rangle = \langle  M(\theta) \rangle
\end{equation}
due to the law of large numbers, where $M(\theta) = U(\theta)^\dagger M U(\theta)$ and we have introduced the notation $\langle \psi_0 |\Box | \psi_0 \rangle = \langle \Box \rangle$.

Next we observe that while $\mu^{-1}\sum_{i=1}^\mu m_i$ is empirically determined, $\langle M(\theta)\rangle$ is a function of $\theta$ that can be calculated from the theory. Let us further imagine that, according to our prior information, $\theta$ is very close to some known value $\theta'$. In that case we can calculate the Taylor expansion of $\langle M(\theta) \rangle$ and assume that
\begin{equation}
\frac{1}{\mu}\sum_{i=1}^\mu m_i \approx \langle M(\theta') \rangle + \frac{d \langle M(\theta') \rangle}{d\theta}(\theta - \theta'),
\end{equation}
which gives us an estimate for the unknown value if we solve it as an equation for $\theta$.

Finally, given that the relationship between the average $\langle M(\theta)\rangle$ and the parameter $\theta$ is approximately linear, to a good approximation we can connect their uncertainties via the error propagation formula \citep{yurke1986, rafal2015, HofmannHolger2009}
\begin{equation}
\Delta \theta^2 \approx \frac{\Delta M(\theta)^2}{\abs{d\langle M(\theta) \rangle/d\theta}^2},
\label{errorprop}
\end{equation}
where $\Delta M(\theta)^2 = \langle  M(\theta)^2 \rangle - \langle M(\theta)  \rangle^2$. Thus we can use equation (\ref{errorprop}) to quantify the quality of our estimation. 

This simple estimation technique shows in a particularly transparent way how the assumptions of having an abundance of measurement data and a good prior knowledge can enter in quantum metrology protocols. In chapter \ref{chap:methodology} we perform a detailed analysis of these restrictions, and the results in this thesis will demonstrate that the methods that we have proposed open the door to design practical schemes beyond such limitations. 

Despite these difficulties, equation (\ref{errorprop}) can still be useful. On the one hand, this uncertainty can always be accessed experimentally and employed as a measure of sensitivity (see, e.g., \cite{baumgart2016}), since the law of large numbers also implies that
\begin{equation}
\Delta M(\theta)^2 \approx \frac{1}{\mu}\sum_{i=1}^\mu m_i^2 - \left(\frac{1}{\mu}\sum_{i=1}^\mu m_i \right)^2
\end{equation} 
when $\mu$ is large, and $d\langle M(\theta) \rangle/d\theta$ can be approximated by a known constant in the regime that we are considering. On the other hand, equation (\ref{errorprop}) provides a theoretical method to compare the efficiency of different quantum strategies. 

Furthermore, by combining equation (\ref{errorprop}) with the generalised Mandelstam-Tamm uncertainty relation $2 \Delta J_z \Delta M(\theta) \geqslant \abs{d \langle M(\theta)\rangle/d \theta}$ \cite{HofmannHolger2009} we find the quantum Cram\'{e}r-Rao bound for pure states \cite{HofmannHolger2009}
\begin{equation}
\frac{\Delta M(\theta)^2}{\abs{d\langle M(\theta) \rangle/d\theta}^2} \geqslant \frac{1}{4\Delta J_z^2}.
\label{simpleqcrb}
\end{equation}
This result indicates that the sensitivity improves when $4\Delta J_z^2$ increases, and how this is to be achieved can be revealed if we rewrite such quantity as \cite{sahota2015}
\begin{equation}
4\Delta J_z^2 = \bar{n}_1\left( 1 + \mathcal{Q}_1 \right) + \bar{n}_2\left( 1 + \mathcal{Q}_2 \right) - 2\mathcal{J}\sqrt{\bar{n}_1\bar{n}_2\left( 1 + \mathcal{Q}_1 \right)\left( 1 + \mathcal{Q}_2 \right)},
\label{fishercorrelations}
\end{equation}
where $\bar{n}_i = \langle a_i^\dagger a_i \rangle$ is the mean number of quanta sent through the $i$-th port, 
\begin{equation}
\mathcal{Q}_i = \frac{1}{\bar{n}_i}\left[\langle (a_i^\dagger a_i)^2 \rangle - \bar{n}_i\left(\bar{n}_i+ 1\right)\right]
\end{equation}
is the Mandel $\mathcal{Q}$-parameter that quantifies the photon correlations within the $i$-th mode, and
\begin{equation}
\mathcal{J} = \frac{\langle a_1^\dagger a_1 a_2^\dagger a_2 \rangle - \bar{n}_1\bar{n}_2}{\Delta(a_1^\dagger a_1) \Delta(a_2^\dagger a_2)}
\end{equation}
is a parameter quantifying the correlations between modes. Indeed equation (\ref{fishercorrelations}) shows that we can enhance the sensitivity by either increasing $\mathcal{Q}_i$ or making $\mathcal{J}$ negative, or both. 

These expressions can be further simplified when we consider the important family of path-symmetric probes introduced by Hofmann \cite{HofmannHolger2009}, which is precisely the class of schemes that we will exploit. Following the characterisation given by Sahota and Quesada \cite{sahota2015}, we say that a state is path-symmetric when the conditions
\begin{equation}
\bar{n}_1 = \bar{n}_2 \equiv \bar{n}/2, ~~ \langle (a_1^\dagger a_1)^2\rangle = \langle (a_2^\dagger a_2)^2\rangle
\end{equation}
are satisfied. In that case we have that 
\begin{equation}
4\Delta J_z^2 = \bar{n}(1+\mathcal{Q})(1-\mathcal{J}),
\label{fishercorrpathsym}
\end{equation}
where the parameters quantifying correlations have become
\begin{align}
\mathcal{Q} &= \frac{1}{2\bar{n}}\left[4\langle (a_1^\dagger a_1)^2 \rangle - \bar{n}\left(\bar{n} + 2\right)\right] = \frac{1}{2\bar{n}}\left[4\langle (a_2^\dagger a_2)^2 \rangle - \bar{n}\left(\bar{n} + 2\right)\right],
\nonumber \\
\mathcal{J} &= \frac{\langle a_1^\dagger a_1 a_2^\dagger a_2 \rangle - \bar{n}^2/4}{\Delta(a_1^\dagger a_1)^2}  = \frac{\langle a_1^\dagger a_1 a_2^\dagger a_2 \rangle - \bar{n}^2/4}{\Delta(a_2^\dagger a_2)^2}.
\label{correlationsintro}
\end{align}

Given the nature of $\mathcal{Q}$ and $\mathcal{J}$, from now on we will refer to the former as the amount of \emph{intra-mode correlations}, while the latter will be understood as quantifying \emph{inter-mode correlations} \cite{knott2016local, proctor2017networked}. These are the type of correlations that will be relevant for our analysis of interferometric schemes. 

\subsection{Multi-parameter protocols}
\label{subsec:multischemes}

Single-parameter protocols such as the Mach-Zehnder interferometer provide a simple and intuitive formalism to understand the fundamental limits that nature imposes on our estimation strategies. However, real-world applications typically give rise to estimation problems with several unknown pieces of information. For instance, we may need to determine the range and velocity of a moving object \cite{zhuang2017}, quantify phases and phase diffusion \cite{vidrighin2014, szczykulska2017}, reconstruct an image \cite{humphreys2013, knott2016local, zhang_lu2017}, estimate the components of a field \cite{baumgratz2016}, assess the spatial deformations of a grid of sources \cite{jasminder2016, jasminder2018} or implement distributed sensing protocols using quantum networks \cite{proctor2017networked, proctor2017networkedshort, ge2018, eldredge2018, altenburg2018, qian2019}. For that reason, the second part of this thesis will be dedicated to the study of multi-parameter schemes. 

Our starting point is the general framework for \emph{quantum sensing networks} introduced by Proctor \emph{et al.} \cite{proctor2017networked}. This is a model for spatially distributed sensing, where in general there will be several sets of unknown parameters, and each set will be encoded locally in a sensor. The importance of this configuration is that it allows us to investigate whether the estimation of such parameters can be enhanced by exploiting inter-sensor and intra-sensor correlations \cite{proctor2017networked, proctor2017networkedshort, sammy2016compatibility, eldredge2018, ge2018}, which in a sense generalise the analogous notions for the Mach-Zehnder interferometer \cite{proctor2017networked, knott2016local}. 

Here we focus on a particular subset of the problems that this formalism can accommodate. In particular, we will consider schemes with a single parameter encoded in each sensor, and possibly including an ancillary system. The fact that the sensors are spatially distributed is imposed by means of the local unitary encoding
\begin{equation}
U(\boldsymbol{\theta}) = \mathbb{I}\otimes U_1(\theta_1)\otimes \cdots \otimes U_d(\theta_d),
\label{distsensing}
\end{equation}
while both the state $\rho_0$ and the measurement $E(m)$, which are defined on the same space that $U(\boldsymbol{\theta})$ is, can be correlated with respect to different sensors. Within this framework, we examine two cases. 

In chapters \ref{chap:networks} and \ref{chap:multibayes} we explore the role of inter-sensor correlations in a network designed to estimate local or global properties in the presence of different amounts of data. Each sensor is modelled as a qubit and no ancillary system is assumed, which implies that, in this case, the first component of the partition in equation (\ref{distsensing}) is absent. This scheme could be implemented, for example, with atoms (see section \ref{subsec:qapp} and \cite{proctor2017networked}).

\begin{figure}[t]
\centering
\includegraphics[trim={0cm 0.5cm 0cm 0.25cm},clip,width=14cm]{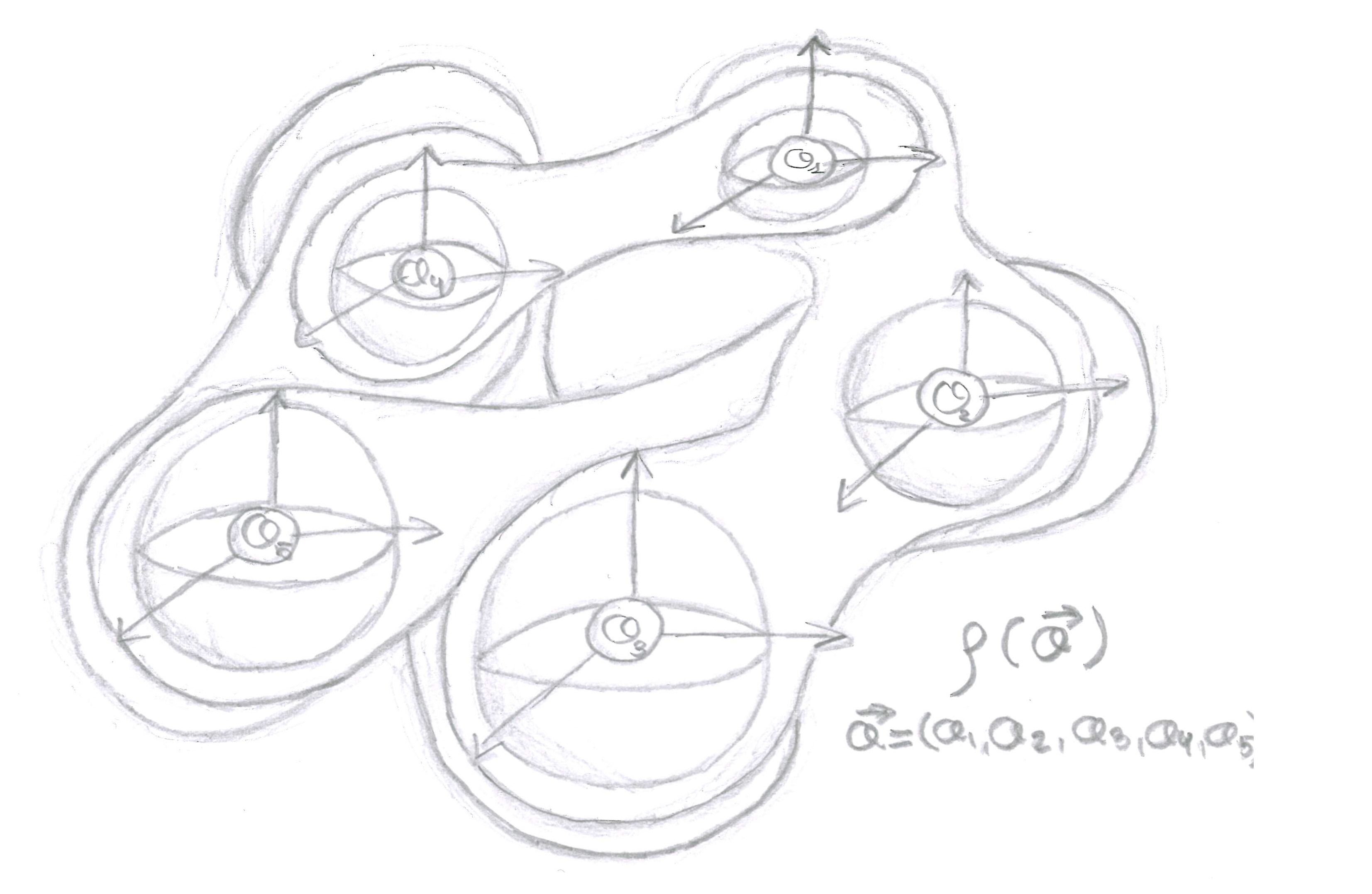}
\caption[Artistic representation of a quantum sensing network]{Artistic representation of a quantum sensing network. Several qubit sensors are entangled to estimate a set of unknown parameters, which are encoded in the transformed probe $\rho(\boldsymbol{\theta}) = U(\boldsymbol{\theta}) \rho_0 U(\boldsymbol{\theta})^\dagger$.}
\end{figure}

On the other hand, in chapter \ref{chap:multibayes} we also examine a network where each sensor is an optical mode encoding an unknown phase shift that we wish to determine, including an extra mode that acts as a phase reference and whose phase is assumed to be known from the calibration of the experiment\footnote{Counting the resources of this extra beam is motivated when we wish to entangle it with the rest of the network because entangled states are generally difficult to prepare in the laboratory \cite{proctor2017networked}. As with the Mach-Zehnder interferometer, we assume any extra resources that may be needed to calibrate the experiment such that this model applicable in practice \cite{jarzyna2012, proctor2017networked}.} \cite{proctor2017networked}. Therefore, this also fulfils the condition for distributed sensing in equation (\ref{distsensing}). This arrangement encompasses an important imagining protocol that has been extensively studied both with Bayesian \cite{chiara2003, spagnolo2012} and non-Bayesian tools \cite{spagnolo2012, humphreys2013}, and that has produced a rich literature about the potential enhancement that the global estimation of several parameters might provide, or the lack of it \cite{humphreys2013, knott2016local, proctor2017networked, proctor2017networkedshort, altenburg2018}. This protocol is one of the possible ways in which we can generalise two-mode interferometry to the multi-parameter case. For an alternative generalisation where each sensor is a full interferometer, see \cite{knott2016local}.

It is important to note that existing literature employs the notions of \emph{local} and \emph{global} estimation strategies in two different ways. Within the context of this section, a strategy is said to be local if neither the state nor the measurement present correlations with respect to the partition in equation (\ref{distsensing}), and global otherwise. That is, a local strategy is uncorrelated according to the definitions in section \ref{subsec:qapp}. However, if instead we focus on the more general context of estimation theory, local strategies are those that rely on a high amount of prior knowledge about the parameters of interest, while a global estimation refers to situations where we are almost completely ignorant about them. In the next chapter we will see that our protocols will be designed to operate in the intermediate regime between the two latter extremes. The exact meaning that the terms \emph{local} and \emph{global} have in each situation will be clear from the context.   

\section{Chapter summary}

The formalism reviewed in this chapter enables us to perform three tasks that are crucial for the development of quantum metrology: representing information, encoding it in quantum systems, and manipulating those systems to extract such information, which is the final goal.

We have seen that the uncertain information associated with natural phenomena can be suitably modelled with the objective version of Bayesian probability theory. The importance of this fact will become apparent once we have introduced the condition of limited data in our estimation protocols. Moreover, this perspective has offered a clear intuition to understand the empirical link between our probability models and the relative frequencies that we measure in the laboratory, as well as their conceptual differences. Crucially, this link plays a fundamental role in verifying the validity of quantum predictions.

On the other hand, we have revisited the foundations of quantum mechanics to understand the reasons behind the success of quantum technologies in general, and of quantum metrology in particular. The key observation is that while the existence of the quantum of action constrains the probability models that we can construct, we also have freedom to choose the functions of dynamical variables that give rise to states and measurements. Hence, the combination of both features provides us with a framework that is more realistic than the classical paradigm and flexible enough to develop new applications on the basis of quantum entities.

As a final step we have described two important classes of quantum schemes: a generalised version of the Mach-Zehnder interferometer and a quantum sensing network model for distributed sensing, and we have identified the types of correlations that may be present in these systems as one of the fundamental features to be exploited in our study of quantum metrology protocols. 
\chapter{Towards a non-asymptotic methodology}
\label{chap:methodology}

In this chapter we present the strategy that we will follow to construct our non-asymptotic methodology for quantum metrology. Some of the results that we use to motivate and justify such strategy are new, although at this stage we still need to rely on known ideas that are already available in the literature.

\section{Formulation of the problem}
\label{sec:problem}

The basic information needed for the estimation problems that we find in quantum metrology is encoded in the joint probability
\begin{eqnarray}
P\left[E(d\boldsymbol{m}) \land \varrho(d\boldsymbol{\theta}) | I_0\right] = p(\boldsymbol{\theta})\mathrm{Tr}\left[E(\boldsymbol{m})\varrho(\boldsymbol{\theta}) \right] d\boldsymbol{\theta} d\boldsymbol{m},
\label{basicinfo}
\end{eqnarray}
where $p(\boldsymbol{\theta})$ is the prior probability and we have used both the Born rule $p(\boldsymbol{m}|\boldsymbol{\theta}) = \mathrm{Tr}\left[E(\boldsymbol{m})\varrho(\boldsymbol{\theta})\right]$ and the fact that $p(\boldsymbol{\theta}, \boldsymbol{m}) = p(\boldsymbol{\theta}) p(\boldsymbol{m}|\boldsymbol{\theta})$. More concretely, we have the following propositions:
\begin{itemize}
\item[i)] $E(d\boldsymbol{m})$, with $\boldsymbol{m} = (m_1, \dots, m_\mu)$, indicates that the measurement scheme $E$ generates $\mu$ observations with outcomes lying within $\boldsymbol{m}$ and $\boldsymbol{m} + d\boldsymbol{m}$. The set of outcomes $\boldsymbol{m}$ is the empirical data. 
\item [ii)] $\varrho(d\boldsymbol{\theta})$, with $\boldsymbol{\theta} = (\theta_1, \dots, \theta_d)$, represents the fact that the interaction between an external system that we wish to study and the initial probe $\varrho_0$ encodes in our scheme $d$ parameters with values lying within $\boldsymbol{\theta}$ and $\boldsymbol{\theta} + d\boldsymbol{\theta}$. The number of unknown parameters is the dimension of the estimation problem\footnote{Notice that this is different from the dimension of the space where the operators that represent states and measurements are defined.}. 
\item [iii)] $I_0$ is the prior information. We can split it as $I_0 = \Omega_{\boldsymbol{m}} \land \Omega_{\boldsymbol{\theta}} \land I_{\mathrm{exp}}$, where
\begin{itemize}
\item[a)] $\Omega_{\boldsymbol{m}}$ and $\Omega_{\boldsymbol{\theta}}$ indicate the values that the experimental outcomes and parameters can take when we only know the physical nature of the quantities that they represent.
\item[b)] $I_{\mathrm{exp}}$ represents the operational information that maps our actions in the laboratory with the mathematical representation for states, measurements and the mechanism that encodes the parameters in the probe, as well as any further prior knowledge that we may have about the external system that we wish to sense and whose properties are represented by the parameters\footnote{The proposition $I_0$ is normally omitted in quantum metrology discussions. However, the information that it encodes is crucial to implement theoretical protocols in real life, and in practice it is always implicitly taken into account. By making it explicit and giving it a formal representation inside the theory we can keep track of the assumptions that go into our calculations in an economical way. Although a study of the practical implementation of our results is beyond the scope of this thesis, we will still use $I_0$ to make the assignment of the prior probability $p(\boldsymbol{\theta})$ more transparent.}.
\end{itemize}
\end{itemize}
In short, the likelihood function $p(\boldsymbol{m}|\boldsymbol{\theta})$ encodes the information about the process that generates the outcomes and its relationship with the parameters, while the prior $p(\boldsymbol{\theta})$ includes what is known about $\boldsymbol{\theta}$ before the experiment is performed. It is important to observe that while the likelihood model is given by the laws of quantum mechanics, the prior probability must be assigned by other means. A practically motivated procedure to select priors will be developed in chapter \ref{chap:nonasymptotic}.

While the previous description is completely general, our study will be focused on identical and independent experiments. Hence, our calculations will be based on a specific partition of the space of operators where 
\begin{equation}
\varrho(\boldsymbol{\theta})=\smash[b]{ \underbrace{\rho(\boldsymbol{\theta}) \otimes \cdots \otimes \rho(\boldsymbol{\theta})\,}_\text{$\mu$ times}}
\label{icopies}
\end{equation}
and 
\begin{equation}
E(\boldsymbol{m}) = E(m_1)\otimes \cdots \otimes E(m_\mu).
\label{iidpom}
\end{equation}
As a consequence, the likelihood function becomes
\begin{equation}
p(\boldsymbol{m}|\boldsymbol{\theta}) = \prod_{i=1}^{\mu}p(m_i|\boldsymbol{\theta}) =\prod_{i=1}^{\mu} \mathrm{Tr}\left[E(m_i)\rho(\boldsymbol{\theta})\right].
\label{iidlikelihood}
\end{equation}
This configuration can model either repetitions of a given experiment, or a collection of $\mu$ copies of some system where independent and identical measurements are performed. We will normally describe our results in terms of the first picture, although both are mathematically equivalent. Note that states and measurements describing a single repetition or copy can still be entangled with respect to its internal structure, as we saw in section \ref{sec:qmetrology}. 

The motivation to select the class of schemes specified in equations (\ref{icopies} - \ref{iidlikelihood}) is twofold. On the practical side, strategies where the same scheme is repeated several times are relevant for any experimental arrangement where we cannot or do not wish to correlate different runs, which include a wide range of practical scenarios. On the other hand, having a sequence of repetitions is an intuitive way of examining the transition from the regime of limited data that we wish to explore, to the asymptotic regime of many trials, and this transition is crucial to define the non-asymptotic regime. In addition, this assumption greatly simplifies the complex numerical calculations involved in Bayesian scenarios. Admittedly, there are other interesting practical possibilities that emerge when we allow for POMs that do not satisfy the constraint in equation (\ref{iidpom}), or for a collection of states that are not identical. For instance, we could consider adaptive schemes, where the strategy for each new trial is selected taking into account the information provided by the previous outcome \citep{berry2000, esteban2017, lumino2017}, and they could be a better choice in some scenarios. These techniques are beyond the scope of this work, although in chapter \ref{chap:limited} we briefly explore the case of general collective measurements \cite{kolodynski2014, jarzyna2016thesis} that are implemented on $\mu = 10$ copies of a NOON state.

We will also assume that in each repetition the parameters are encoded via the unitary transformation $U(\boldsymbol{\theta})=\mathrm{exp}(-i \boldsymbol{K}\cdot \boldsymbol{\theta})$, where $\boldsymbol{K} = (K_1, \dots, K_d)$ are commuting generators. Thus we can define the following quantum metrology protocol: 
\begin{enumerate}
\item A probe state $\rho_0$ is prepared.
\item The interaction with an external system transforms the probe as $\rho_0 \rightarrow \rho(\boldsymbol{\theta}) = U(\boldsymbol{\theta})\rho_0 U^\dagger(\boldsymbol{\theta})$,
\item A measurement scheme with elements $\lbrace E(m_i)\rbrace$ is performed to extract the information about $\boldsymbol{\theta}$.
\item The process is repeated $\mu$ times, which generates the data $\boldsymbol{m}=(m_1, \dots, m_\mu)$.
\end{enumerate}

To make the protocol more realistic, we typically define some notion of resources that allows us to capture further constraints that different applications may impose in practice. To this end, Proctor \emph{et al.} introduced in \cite{proctor2017networked} the resource operator $R$, which is Hermitian, and defined the average amount of resources for a single shot as $\langle R \rangle = \mathrm{Tr}(\rho_0 R)$. In addition, they imposed that $\comm*{R}{U(\boldsymbol{\theta})} = 0$, since this implies that $\mathrm{Tr}[\rho(\boldsymbol{\theta}) R] = \mathrm{Tr}(\rho_0 R) = \langle R \rangle$. In words, the resources are conserved during the interaction between the probe and the external system, a condition that guarantees that the resource counting will not depend on the unknown parameters. 

Following this approach, we have that the total amount of resources consumed on average by the protocol described above is $\mu \langle R \rangle$. In turn, we can now formally define the regime of limited data as the regime where $\mu$ is low. In principle we could consider any realistic value for $\langle R \rangle$ without leaving this regime. However, the nature and scope of our study imposes two constraints on $\langle R \rangle$. On the one hand, one of our main goals is to identify the novel effects that emerge directly from having different amounts of data, and, as such, we have chosen $\langle R \rangle$ to be sufficiently low to guarantee that $\mu$ is the dominant contribution to the total resources. Importantly, this condition means that our results are particularly useful for the study of fragile systems \cite{eckert2007, pototschnig2011, carlton2010, taylor2013, taylor2015, taylor2016, PaulProctor2016}. On the other hand, $\langle R \rangle$ needs to be large enough to allow for certain quantum enhancements to be relevant, which in the majority of our schemes can be achieved simply by requiring that $\langle R \rangle > 1$\footnote{An illustrative example to justify this choice is the case of a Mach-Zehnder interferometer with a single photon, since in that scenario we have that the sensitivity of coherent and NOON states is the same, in spite of the fact that the coherent state is classical-like (see section \ref{subsec:commoninter}).}. Crucially, this configuration will allow us to explore the interplay between a small amount of data and the usefulness of exploiting quantum features such as squeezing and sensor entanglement. 

\section{Uncertainty and estimation}
\label{sec:uncertainty}

Once the outcomes ${\boldsymbol{m}}=(m_1, \dots, m_\mu)$ have been generated with the protocol that we have described, the next step is to develop a technique to extract information from them. We already saw a first approximation of the type of procedure that is suitable for this task when we revisited the Mach-Zehnder interferometer in section \ref{sec:qmetrology}, where we were able to identify two key elements: a way of making estimates that inform us about the true values of the parameters, and a second type of quantity that represents the quality of this process. The former are formally encoded in a vector estimator $\boldsymbol{g}(\boldsymbol{m}) = ( g_1(\boldsymbol{m}), \dots, g_d (\boldsymbol{m}))$, which is a function of the experimental outcomes, while the latter is a measure of uncertainty. We then seek the estimators and the quantum protocol for which the uncertainty is minimal, and this search can be rephrased as an optimisation problem \cite{jaynes2003}.

To construct the measure of uncertainty, first we introduce the error or deviation function $\mathcal{D} [\boldsymbol{g}({\boldsymbol{m}}),\boldsymbol{\theta}]$, which quantifies the deviation of our estimates $\boldsymbol{g}(\boldsymbol{m})$ when the parameters happened to be $\boldsymbol{\theta}$. Its choice relies on the nature of the variables that we wish to estimate. In our case, these will be either optical phases, differences of optical phases or simply periodic parameters; consequently, the deviation function should respect their periodic character \cite{helstrom1976, holevo2011, berry2015, kolodynski2014}. 

One of the simplest options that satisfy this requirement for a single parameter is the sine error \cite{demkowicz2011, kolodynski2014, rafal2015}
\begin{equation}
\mathcal{D} [g({\boldsymbol{m}}),\theta] = 4~\mathrm{sin}^2\left\lbrace\left[g(\boldsymbol{m})-\theta\right]/2\right\rbrace.
\label{sinerror}
\end{equation}
In principle we could base our analysis of single-parameter schemes on equation (\ref{sinerror}), since Demkowicz-Dobrza\ifmmode \acute{n}\else \'{n}\fi{}ski found in \cite{demkowicz2011} a completely analytical solution to the problem of phase estimation for $\mu = 1$ using this type of error. However, the extension of this result to the case where many repetitions are considered is still numerically challenging. Instead, here we argue that the characteristics of the regime of limited data motivate an important simplification. 

If the empirical data is limited, then the prior information about the unknown parameters included in $I_0$ will generally play an active role in their estimation. Therefore, a natural regime to study situations where the number of measurements is small is the regime of moderate prior knowledge. This is an intermediate case between complete ignorance and an amount of prior knowledge so high that the problem can be recast in a local form \cite{durkin2007, demkowicz2011}. Then we may say that, in a certain sense, the quantity $\abs{g(\boldsymbol{m})-\theta}/2$ will be moderately small, so that it is meaningful to approximate equation (\ref{sinerror}) as 
\begin{equation}
\mathcal{D} [g({\boldsymbol{m}}),\theta] \approx \left[g(\boldsymbol{m}) - \theta\right]^2.
\label{squarerror}
\end{equation}
In appendix \ref{prior_sinapprox_appendix} we evaluate the error in the truncation of the Taylor expansion that leads to equation (\ref{squarerror}) in different ways, and we show that the main conclusions of our results for single-parameter protocols, which operate in the regime of moderate prior knowledge, are not affected by it. Therefore, we can safely exploit the mathematical simplicity of the square error in the context of phase estimation. 

Furthermore, given the relative freedom to choose deviation functions, we can apply the same logic to multi-parameter problems, and we can require that any reasonable error that we may use for several periodic parameters also approaches its squared version in the intermediate regime of prior information. That is, 
\begin{equation}
\mathcal{D} [\boldsymbol{g}({\boldsymbol{m}}),\boldsymbol{\theta}, \mathcal{W}] \approx  \mathrm{Tr}\left\lbrace \mathcal{W} \left[\boldsymbol{g}(\boldsymbol{m}) - \boldsymbol{\theta}\right] \left[\boldsymbol{g}(\boldsymbol{m}) - \boldsymbol{\theta}\right]^\transpose  \right\rbrace,
\label{multideviation}
\end{equation}
where $\mathcal{W} = \mathrm{diag}(w_1, \dots, w_d)$, $\mathrm{Tr}(\mathcal{W}) = 1$ and $w_i \geqslant 0$ indicates the relative importance of estimating the $i$-th parameter \cite{proctor2017networked}. In that way, the optimal strategy will produce the smallest errors for the most relevant parameters.

Using the chosen $\mathcal{D} [\boldsymbol{g}({\boldsymbol{m}}),\boldsymbol{\theta}, \mathcal{W}]$ as a basis, it is possible to construct different types of uncertainty depending on which information is assumed to be exactly known and which information is only partial \cite{jaynes2003}. As a result, different authors often base their analysis of metrology protocols on different quantities \cite{rafal2015, li2018}. 

To simplify this state of affairs, here we propose a progressive construction of different measures of uncertainty, using as a guide the physical requirements imposed by the three basic situations that we could face in an scenario with unknown parameters: a real experiment that is performed in the laboratory, the simulation of a hypothetical experiment, and the theoretical study of a real or hypothetical experiment. Moreover, we show that this method gives a clear physical meaning to the figure of merit that is suitable to design protocols from theory. While the calculations in this thesis will be based on the square errors in equations (\ref{squarerror}) and (\ref{multideviation}), we draw attention to the fact that the following discussion is also applicable to more general deviation functions. For a discussion about the relation of our strategy with other approaches in the literature, see appendix \ref{sec:otheruncertainty}. 

Let us first recall that the raison d'\^{e}tre of any experiment is to produce outcomes. Since in the first scenario these are known, the measure of uncertainty employed by an experimentalist will depend on $\boldsymbol{m}$. On the contrary, it is clear that it should not depend on $\boldsymbol{\theta}$ because the parameters are what we seek. Therefore, we need a probability function with information about the parameters for a given set of outcomes, which is precisely what the posterior $p(\boldsymbol{\theta}|\boldsymbol{m})$ provides, and the error of the estimation is
\begin{equation}
\epsilon(\boldsymbol{m}) = \int d\boldsymbol{\theta} ~p(\boldsymbol{\theta}|\boldsymbol{m}) ~\mathcal{D}[\boldsymbol{g}(\boldsymbol{m}),\boldsymbol{\theta}, \mathcal{W}].
\label{errexp}
\end{equation}
This is the uncertainty that arises from gathering and processing data in a real experiment, both in quantum \cite{PaulProctor2016} and classical \cite{jaynes2003} scenarios\footnote{Technically, if we use the square error in the laboratory then we need to calculate the square root of equation (\ref{errexp}), so that both the estimates and the uncertainty have the same units. However, the present form is more convenient for studies of a theoretical nature.}. As we saw in the previous chapter, the posterior probability is the result of applying Bayes theorem (see equation \ref{bayestheorem1}), and it can be calculated as $p(\boldsymbol{\theta}|\boldsymbol{m}) \propto p(\boldsymbol{\theta}) p(\boldsymbol{m}|\boldsymbol{\theta})$. Importantly, in quantum metrology we can assume that the likelihood models are a good representation of reality because, so far, the quantum framework has passed all experimental tests. In addition, the prior knowledge stored in $p(\boldsymbol{\theta})$ will typically include the multivariate domain in which we can expect to find the parameters, a piece of information that can be given, for instance, by the results of past experiments. 

Apart from analysing a specific experiment, usually we also want to enhance its design in order to improve the precision of the estimation protocol. This study will often occur outside of the laboratory, in which case we no longer have access to specific measurement outcomes. In turn, a measure of uncertainty that is useful for designing experiments cannot depend on $\boldsymbol{m}$. Since equation (\ref{errexp}) already gives us the experimental error, now we need a probability function with information about the possible experimental outcomes that the configuration under analysis could produce. One possibility is to employ $p(\boldsymbol{m}|\boldsymbol{\theta}')$, where $\boldsymbol{\theta}'$ is our simulation of the true values, and calculate the average of the errors for all the possible experimental outcomes associated with $\boldsymbol{\theta}'$ weighted by their likelihood, i.e.,
\begin{equation}
\epsilon(\boldsymbol{\theta}') = \int d\boldsymbol{m}~p(\boldsymbol{m}|\boldsymbol{\theta}')~\epsilon(\boldsymbol{m}).
\label{errsim}
\end{equation}
This is the appropriate quantity if our aim is to simulate experiments and study their performance on average, as it is the case, for example, in \cite{PaulProctor2016}.

The previous uncertainty still depends on the specific values $\boldsymbol{\theta}'$ of the simulation. If we instead follow a purely theoretical approach, then we need to take into account the fact that both outcomes and true values for the parameters are unknown to the theorist. In that case, the relevant information about the possible outcomes is $p(\boldsymbol{m}) = \int d\boldsymbol{\theta}' p(\boldsymbol{\theta}') p(\boldsymbol{m}|\boldsymbol{\theta}')$, and by taking either the average $\int d\boldsymbol{m} ~p(\boldsymbol{m})\epsilon(\boldsymbol{m}) = \bar{\epsilon}$ weighted over $p(\boldsymbol{m})$, or the average $\int d\boldsymbol{\theta}' p(\boldsymbol{\theta}')\epsilon(\boldsymbol{\theta}') = \bar{\epsilon}$ weighted over our prior knowledge of $\boldsymbol{\theta}'$, and using that $p(\boldsymbol{m}) p(\boldsymbol{\theta}|\boldsymbol{m}) = p(\boldsymbol{\theta}, \boldsymbol{m})$, we finally obtain the error
\begin{equation}
\bar{\epsilon} = \int d\boldsymbol{\theta} d\boldsymbol{m} ~p(\boldsymbol{\theta}, \boldsymbol{m}) ~\mathcal{D}[\boldsymbol{g}(\boldsymbol{m}),\boldsymbol{\theta}, \mathcal{W}],
\label{errthe}
\end{equation}
which is independent of the values of parameters and outcomes. Following the previous discussion, $\bar{\epsilon}$ represents the uncertainty on average about the knowledge that we can acquire in principle with the experimental configuration that is being studied. As such, this is the suitable figure of merit to design experiments from theory in order to make optimal inferences, and we will make use of it from now on. 

\section{Quantum estimation and metrology}
\label{sec:qestimation}

\subsection{The fundamental equations of the optimal strategy}
\label{subsec:fundeq}

Using the joint probability for quantum systems in equation (\ref{basicinfo}) and the error in equation (\ref{errthe}), we have that the uncertainty is
\begin{equation}
\bar{\epsilon} = \int d\boldsymbol{\theta} d\boldsymbol{m} ~p(\boldsymbol{\theta})\mathrm{Tr}\left[E(\boldsymbol{m})\varrho(\boldsymbol{\theta})\right]\mathcal{D}[\boldsymbol{g}(\boldsymbol{m}),\boldsymbol{\theta}, \mathcal{W}].
\label{qerrgen}
\end{equation}
Assuming that the prior has been assigned, let us first consider a case where both $\varrho_0$ and the details of its transformation $\varrho_0 \rightarrow \varrho(\boldsymbol{\theta})$ are known. Then, the optimisation of the protocol is achieved by minimising equation (\ref{qerrgen}) with respect to the vector estimator $\boldsymbol{g}(\boldsymbol{m})$ and the measurement scheme $E(\boldsymbol{m})$. 

To simplify the problem we can combine $\boldsymbol{g}(\boldsymbol{m})$ and $E(\boldsymbol{m})$ into a single object by labelling the POM elements with the estimates as $E(\boldsymbol{g}) = \int d\boldsymbol{m}~\delta\left(\boldsymbol{g}(\boldsymbol{m})-\boldsymbol{g}\right) E(\boldsymbol{m})$ \cite{rafal2015}. As a result, equation (\ref{qerrgen}) can be recast in the form
\begin{equation}
\bar{\epsilon} = \int d\boldsymbol{\theta} d\boldsymbol{g} ~p(\boldsymbol{\theta})\mathrm{Tr}\left[E(\boldsymbol{g})\varrho(\boldsymbol{\theta})\right]\mathcal{D}(\boldsymbol{g},\boldsymbol{\theta}, \mathcal{W}).
\label{errhelstrom} 
\end{equation} 
How equation (\ref{errhelstrom}) is to be optimised has been known since the works of Helstrom \cite{helstrom1976, helstrom1974} and Holevo \cite{holevo1973b, holevo1973}. Following their expositions in \cite{helstrom1976} and \cite{holevo1973b}, respectively, first we rewrite equation (\ref{errhelstrom}) as $\bar{\epsilon} = \int  d\boldsymbol{g} ~\mathrm{Tr}\left[E(\boldsymbol{g})Q(\boldsymbol{g})\right]$, with $Q(\boldsymbol{g}) = \int d\boldsymbol{\theta}  ~p(\boldsymbol{\theta})\varrho(\boldsymbol{\theta})\mathcal{D}(\boldsymbol{g},\boldsymbol{\theta}, \mathcal{W})$. If $E_\mathrm{opt}(\boldsymbol{g})$ is the optimal strategy, then there exists a Hermitian operator $Y$ satisfying that 
\begin{equation}
\begin{cases}
Y = \int d\boldsymbol{g} ~Q(\boldsymbol{g}) E_\mathrm{opt}(\boldsymbol{g}) = \int d\boldsymbol{g} E_\mathrm{opt}(\boldsymbol{g}) Q(\boldsymbol{g}), \\
Q(\boldsymbol{g}) - Y \geqslant 0,
\end{cases}
\label{optequations}
\end{equation}
and we have that $\bar{\epsilon} \geqslant \bar{\epsilon}_{\mathrm{min}} = \mathrm{Tr}(Y)$.

The operator inequality is to be understood as $\langle u| Q(\boldsymbol{g}) |u \rangle \geqslant \langle u| Y |u \rangle $ for any $\ket{u}$. In addition, the conditions in equation (\ref{optequations}), together with the closure relation $\int d\boldsymbol{g} E_\mathrm{opt}(\boldsymbol{g}) = \mathbb{I}$, imply that 
\begin{equation}
\left[Q(\boldsymbol{g}) - Y \right]E_\mathrm{opt}(\boldsymbol{g})d\boldsymbol{g} = 0.
\label{practicalcondition}
\end{equation}
Therefore, if we can find the Hermitian operator $Y$ that satisfies the previous inequality and gives us the minimum value for $\mathrm{Tr}(Y)$, then we may use equation (\ref{practicalcondition}) to construct the optimal strategy\footnote{When the operators are represented by matrices we can find the estimates by imposing that the determinant of $\left[Q(\boldsymbol{g}) - Y \right]$ vanishes, and construct the optimal POM elements from its null space. We recall that $\ket{v}$ belongs to the null space of $A$ when $A \ket{v} = 0 $ \cite{mathematics2004}.}. When this is not possible, the conditions in equation (\ref{optequations}) offer at least a way to verify whether a given measurement is optimal. Helstrom applied the latter approach to several examples in \cite{helstrom1976}. 

A second possibility is to assume that the POM is known and minimise equation (\ref{errhelstrom}) with respect to the initial probe state $\varrho_0$. Macieszczak \emph{et al.} studied this problem in \cite{macieszczak2014bayesian} for a single parameter and the square error, and proposed an heuristic algorithm to find the state and measurement scheme that are simultaneously optimal\footnote{We may think of the optimisation of the measurement scheme as the goal of quantum metrology, and of the full optimisation as the aim of quantum estimation theory.}. We now adapt their arguments to the general case of this section. 

If we express the parameter encoding as $\varrho(\boldsymbol{\theta})=\Lambda_{\boldsymbol{\theta}} (\varrho_0)$, then we can define a dual map $\Lambda_{\boldsymbol{\theta}}^{*}$ for which $\mathrm{Tr}\left[B \Lambda_{\boldsymbol{\theta}}(C) \right] = \mathrm{Tr}\left[\Lambda_{\boldsymbol{\theta}}^{*}(B) C \right]$ \cite{macieszczak2014bayesian}. As a consequence, equation (\ref{errhelstrom}) is equivalent to
\begin{equation}
\bar{\epsilon} = \mathrm{Tr}\left\lbrace \varrho_0 \int d\boldsymbol{\theta} ~p(\boldsymbol{\theta})  \Lambda_{\boldsymbol{\theta}}^{*}\left[ \int d\boldsymbol{g}~E(\boldsymbol{g})\mathcal{D}(\boldsymbol{g},\boldsymbol{\theta}, \mathcal{W}) \right] \right\rbrace \equiv \mathrm{Tr}\left(\varrho_0 \Gamma \right),
\label{errmacieszczak} 
\end{equation}
and the optimal probe is a pure state given by the eigenvector of $\Gamma$ with the minimum eigenvalue \cite{macieszczak2014bayesian}. In addition, we can constrain the minimisation with further conditions such as a fixed amount of resources $\langle R \rangle$ \cite{ariano1998}. Then, by combining this procedure with Helstrom and Holevo's approach we may be able to construct the general optimal solution. In particular, we can calculate the optimal measurement $E^{\text{\tiny{(0)}}}_{\boldsymbol{g}}$ for an initial seed $\varrho_0^{\text{\tiny{(0)}}}$, introduce this POM in equation (\ref{errmacieszczak}) to find its optimal state $\varrho_0^{\text{\tiny{(1)}}}$, and repeat the process until the solutions converge \cite{macieszczak2014bayesian}.

The work of Macieszczak \emph{et al.} \cite{macieszczak2014bayesian} demonstrates that the previous strategy succeeds at least for the square error. This is a crucial result, since the framework in this section can provide general solutions to a wide range of estimation problems. Unfortunately, these results present two important difficulties. One of them is that, except for a few cases such as those that admit covariant measurements \cite{ariano1998, chiara2003, chiribella2005, holevo2011, rafal2015}, deriving exact solutions from this formalism is known to be challenging \cite{helstrom1976}, and this makes it difficult to exploit it in many practical scenarios. 

As for the second difficulty, the fundamental equations can predict optimal strategies that do not represent the repetitions of an experiment, which is the physical model that we have assumed in our definition for the regime of limited data. We can see why this is the case by noticing that, according to equation (\ref{optequations}), the optimal POM for our initial state $\varrho_0 = \rho_0 \otimes \rho_0 \otimes \cdots$ with $\mu$ copies of $\rho_0$ can be collective \cite{jarzyna2016thesis}, and this would contradict our requirement of independent measurements in equation (\ref{iidpom})\footnote{Remarkably, in section \ref{measurements_section} we demonstrate that a collective POM is not better than independent measurements for NOON states in a Mach-Zehnder interferometer.}. Similarly, the optimal state that arises from equation (\ref{errmacieszczak}) for independent and identical measurements might entangle different copies of the probe. 

Given these difficulties, an alternative path that is commonly followed in quantum metrology is to bound the estimation error \cite{rafal2015, li2018, tsang2012, tsang2016, liu2016}. The key advantage of this method is that it sometimes produces tight bounds, and in many situations this is sufficient to extract useful information about the fundamental precision that a given scheme could achieve \cite{rafal2015, Szczykulska2016, haase2018jul}, which in turn provides a mathematically simpler way of finding optimal solutions without relying on the exact theory. This is a common feature, for instance, of the asymptotic regime of many repetitions \cite{rafal2015, jarzyna2015true, haase2018jul}. Note, however, that in general the latter is only appropriate for applications with an abundance of measurement data, which is precisely the requirement that we wish to weaken in this thesis, while those bounds that can be applied with a low number of trials contain less information about fundamental limits because they tend to be loose in such regime \cite{tsang2012, tsang2016}.

The fact that the reliance on bounds is a type of approximation that will generally introduce limitations on the applicability of the results cannot be overstated. Indeed, the misapplication of bounds can often lead to paradoxes \cite{li2018}. An important example of these that we will revisit in later sections is that of the states that appear to provide an infinite precision \cite{rivas2012, zhang2013}, even when in practice they do not and are, in fact, inefficient \cite{tsang2012, giovannetti2012subheisenberg, berry2012infinite, pezze2013, rafal2015} (although not completely useless \cite{alfredo2017}). The key observation is that such paradoxes appear when the assumptions that go into the construction of such bounds are not appropriately taken into account, and in principle there is no reason to think that there is a problem with the physical scheme itself unless the paradox also arises in the exact theory. 

This state of affairs best highlights the importance of developing a non-asymptotic methodology and its potential usefulness in applications: we seek a formulation able to provide more information about the regime of limited data than what current techniques can do, and yet make it tractable enough to be useful in real-world applications. To achieve this, we will perform two preliminary analyses. Firstly, we will examine which elements of the strategy based on bounds can be exploited for our purposes. Secondly, we will explore how Helstrom and Holevo's approach can be adapted to study repetitions. The foundation of our methodology will emerge from the practical combination of both, and while our solutions will not be as general as what the exact theory could offer, we will show that they give us access to a regime largely unexplored and required for further practical progress in the field of quantum metrology and sensing. 

\subsection{Cram\'{e}r-Rao bounds}
\label{subsec:crb}

Let us make our choice of deviation function in section (\ref{sec:uncertainty}) explicit, such that the uncertainty of the estimation is approximately given by the mean square error
\begin{equation}
\bar{\epsilon} \approx \bar{\epsilon}_{\mathrm{mse}} = \int d\boldsymbol{\theta} d\boldsymbol{m} ~p(\boldsymbol{\theta}, \boldsymbol{m})~\mathrm{Tr}\left\lbrace \mathcal{W} \left[\boldsymbol{g}(\boldsymbol{m}) - \boldsymbol{\theta}\right] \left[\boldsymbol{g}(\boldsymbol{m}) - \boldsymbol{\theta}\right]^\transpose \right\rbrace.
\label{msethesis}
\end{equation}
In addition, we notice that, according to the condition in equation (\ref{iidlikelihood}) for independent and identical experiments, $p(\boldsymbol{\theta}, \boldsymbol{m}) = p(\boldsymbol{\theta})\prod_{i=1}^\mu p(m_i|\boldsymbol{\theta})$.

A widely used method to compare estimation schemes consists in optimising equation (\ref{msethesis}) by approaching the Cram\'{e}r-Rao bound \cite{rafal2015, kay1993, paris2009}. If we rewrite the mean square error as
\begin{equation}
\bar{\epsilon}_{\mathrm{mse}} = \int d\boldsymbol{\theta} ~p(\boldsymbol{\theta}) ~\mathrm{Tr}\left\lbrace \mathcal{W}\left[C(\boldsymbol{\theta}) + \boldsymbol{b}(\boldsymbol{\theta})\boldsymbol{b}(\boldsymbol{\theta})^\transpose \right] \right\rbrace,
\label{msedecomposition}
\end{equation}
where we have introduced the covariance matrix of the vector estimator
\begin{align}
C(\boldsymbol{\theta}) = &\int d\boldsymbol{m}~p(\boldsymbol{m}|\boldsymbol{\theta})\boldsymbol{g}(\boldsymbol{m})\boldsymbol{g}(\boldsymbol{m})^\transpose 
\nonumber \\
& - \int d\boldsymbol{m}~p(\boldsymbol{m}|\boldsymbol{\theta})\boldsymbol{g}(\boldsymbol{m})\int d\boldsymbol{m}~p(\boldsymbol{m}|\boldsymbol{\theta})\boldsymbol{g}(\boldsymbol{m})^\transpose 
\end{align}
and its vector bias $\boldsymbol{b}(\boldsymbol{\theta}) = \int d\boldsymbol{m}~ p(\boldsymbol{m}|\boldsymbol{\theta}) \left[ \boldsymbol{g}(\boldsymbol{m}) - \boldsymbol{\theta}\right]$, then the classical version of the Cram\'{e}r-Rao bound is \cite{kay1993, jaynes2003, bayesbounds2007}
\begin{align}
\bar{\epsilon}_{\mathrm{mse}} \geqslant  &\int d\boldsymbol{\theta}~p(\boldsymbol{\theta})~\mathrm{Tr}\left( \mathcal{W}\left\lbrace\left[\mathbb{I} + \frac{\partial \boldsymbol{b}(\boldsymbol{\theta})}{\partial \boldsymbol{\theta}} \right] \frac{F(\boldsymbol{\theta})^{-1}}{\mu} \left[\mathbb{I} + \frac{\partial \boldsymbol{b}(\boldsymbol{\theta})}{\partial \boldsymbol{\theta}} \right]^\transpose \right\rbrace \right)
\nonumber \\
& + \int d\boldsymbol{\theta}~p(\boldsymbol{\theta})~\mathrm{Tr}\left[ \mathcal{W} \boldsymbol{b}(\boldsymbol{\theta})\boldsymbol{b}(\boldsymbol{\theta})^\transpose \right],
\label{ccrb}
\end{align}
which is given in terms of the Fisher information matrix\footnote{Notice that the multidimensional integral in equation (\ref{fim}) is equivalent to $\mu$ times the integral over a single observation $m$ because we are assuming independent and identical trials.}
\begin{align}
F(\boldsymbol{\theta}) &= \frac{1}{\mu}\int \frac{d\boldsymbol{m}}{p(\boldsymbol{m}|\boldsymbol{\theta})} \left[\frac{\partial p(\boldsymbol{m}|\boldsymbol{\theta})}{\partial \boldsymbol{\theta}} \right] \left[\frac{\partial p(\boldsymbol{m}|\boldsymbol{\theta})}{\partial \boldsymbol{\theta}} \right]^\transpose
\nonumber \\
&= \int \frac{dm}{p(m|\boldsymbol{\theta})} \left[\frac{\partial p(m|\boldsymbol{\theta})}{\partial \boldsymbol{\theta}} \right] \left[\frac{\partial p(m|\boldsymbol{\theta})}{\partial \boldsymbol{\theta}} \right]^\transpose.
\label{fim}
\end{align}

A bound is particularly useful when it can be saturated. In our case, the necessary and sufficient condition for the saturation of the previous inequality is \cite{kay1993, jaynes2003}
\begin{equation}
\left[ \boldsymbol{g}(\boldsymbol{m}) - \boldsymbol{\theta} \right] =  \left[\mathbb{I} + \frac{\partial \boldsymbol{b}(\boldsymbol{\theta})}{\partial \boldsymbol{\theta}} \right] \frac{F(\boldsymbol{\theta})^{-1}}{p(\boldsymbol{m}|\boldsymbol{\theta})\mu}  \frac{\partial p(\boldsymbol{m}|\boldsymbol{\theta})}{\partial \boldsymbol{\theta}} + \boldsymbol{b}(\boldsymbol{\theta}).
\label{crbcondition}
\end{equation}
For a given vector estimator and likelihood function, we can always use equation (\ref{crbcondition}) to verify whether our scheme achieves the bound. However, ideally we would like to exploit the result in equation (\ref{ccrb}) to find optimal strategies in a more general fashion. This can be done in two steps. 

The first step is to employ the so-called maximum likelihood estimator, which is defined as $\boldsymbol{g}(\boldsymbol{m}) = \mathrm{max}_{\boldsymbol{\theta}}\left[ p(\boldsymbol{m}| \boldsymbol{\theta}) \right]$ \cite{rafal2015, kay1993}. The key advantage of this tool lies on its asymptotic properties in the limit $\mu \rightarrow \infty$. In particular, the maximum likelihood is asymptotically unbiased \cite{rafal2015, kay1993}, i.e., $\boldsymbol{b}(\boldsymbol{\theta}) \rightarrow \boldsymbol{0}$ as $\mu$ grows, and it satisfies the saturation condition in equation (\ref{crbcondition}) in such limit, which implies that it is also asymptotically optimal \cite{kay1993, vaart1998}. Thus, we can always approach the Cram\'{e}r-Rao bound in the regime of many repetitions $\mu \gg 1$, where it becomes
\begin{equation}
\bar{\epsilon}_{\mathrm{mse}} \approx \frac{1}{\mu}\int d\boldsymbol{\theta}~p(\boldsymbol{\theta}) ~\mathrm{Tr}\left[\mathcal{W} F(\boldsymbol{\theta})^{-1}\right].
\label{ccrbmulti}
\end{equation} 
If we further assume that the Fisher information does not depend on the parameters, so that $F(\boldsymbol{\theta})=F$ for all $\boldsymbol{\theta}$, then we conclude that $\bar{\epsilon}_{\mathrm{mse}} \approx \mathrm{Tr}(\mathcal{W} F^{-1})/\mu$.

The true usefulness of this method is revealed when we further consider the quantum aspect of the problem, which is the second step. According to equation (\ref{fim}), the Fisher information matrix only depends on the likelihood function, which is constructed out of the measurement scheme and the transformed state. In the single-parameter case, this matrix is reduced to the scalar
\begin{equation}
F(\theta) = \int \frac{dm}{p(m|\theta)}\left[\frac{\partial p(m|\theta)}{\partial \theta}\right]^2,
\label{fishersingleparameter}
\end{equation} 
and by maximizing it over all the POMs, Braunstein and Caves \cite{BraunsteinCaves1994} proved the inequality $F(\theta) \leqslant F_q(\theta) = \mathrm{Tr}\left[ \rho(\theta) L(\theta) \right]$, where $F_q(\theta)$ is the quantum Fisher information originally introduced by Helstrom in \cite{helstrom1967mmse}. The symmetric logarithmic derivative $L(\theta)$ is obtained by solving $L(\theta)\rho(\theta) + \rho(\theta)L(\theta) = 2 \partial \rho(\theta)/\partial \theta$, and the bound on the Fisher information may be saturated with a measurement scheme based on the projections onto the eigenstates of $L(\theta)$ \cite{BraunsteinCaves1994, genoni2008}. Moreover, $F_q$ does not depend on $\theta$ when the transformation is a unitary that takes the form $U(\theta)=\mathrm{exp}(-i K \theta)$ \cite{pezze2014, rafal2015}. Therefore, $\bar{\epsilon}_{\mathrm{mse}} \approx \bar{\epsilon}_{cr} = 1/\left(\mu F_q\right)$ given a POM for which $F(\theta) = F_q$, which is a quantum version of the scalar Cram\'{e}r-Rao bound.  

From this we conclude that the asymptotically optimal precision for a single parameter is a function of $\rho(\theta)$ alone, and that to find optimal probes in this regime we just need to maximize the quantum Fisher information. Since we also know an estimator and a POM that can achieve this precision, we have completed our search of the optimal strategy for many trials. Furthermore, an alternative and simpler procedure that is very useful in practice is to consider a collection of relevant probe states and compare their performances using $1/\left(\mu F_q\right)$, a quantity that is independent of both the estimator and the measurement scheme.

To extend this idea to the multi-parameter case, we may construct a quantum version of the Fisher information matrix with components \cite{helstrom1967mmse, helstrom1976, Szczykulska2016, sammy2016compatibility, proctor2017networked}
\begin{equation}
\left[F_q(\boldsymbol{\theta})\right]_{ij} = \frac{1}{2}\mathrm{Tr}\left\lbrace \rho(\boldsymbol{\theta}) \left[L_i(\boldsymbol{\theta}) L_j(\boldsymbol{\theta}) + L_j(\boldsymbol{\theta}) L_i(\boldsymbol{\theta}) \right] \right\rbrace,
\label{qfimgen}
\end{equation} 
where $L_i(\boldsymbol{\theta})$ is the symmetric logarithm derivative of the $i$-th parameter. Assuming that the state is pure, it can be shown that there is an individual measurement on a single copy of the system for which $F(\boldsymbol{\theta}) = F_q(\boldsymbol{\theta})$ if and only if \cite{sammy2016compatibility, pezze2017simultaneous}
\begin{equation}
\langle \psi(\boldsymbol{\theta})|[L_i(\boldsymbol{\theta}), L_j(\boldsymbol{\theta})]|\psi(\boldsymbol{\theta})\rangle = 0,
\label{weakcommutation}
\end{equation}
for all $i$, $j$. Moreover, if the parameters are encoded with the unitary $U(\boldsymbol{\theta}) = \mathrm{exp}(-i \boldsymbol{K}\cdot \boldsymbol{\theta})$, then the weak commutation condition in equation (\ref{weakcommutation}) is satisfied when $[K_i, K_j] = 0$, for all $i$, $j$ \cite{sammy2016compatibility, pezze2017simultaneous}, and the components of such matrix are
\begin{equation}
(F_q)_{ij} =  4\left( \langle  \psi_0 | K_i K_j |  \psi_0 \rangle - \langle  \psi_0 | K_i |  \psi_0 \rangle \langle  \psi_0 | K_j |  \psi_0 \rangle \right),
\label{fimpur}
\end{equation}
so that the optimal asymptotic error is $\bar{\epsilon}_{cr} = \mathrm{Tr}(\mathcal{W} F_q^{-1})/\mu$ when $F(\boldsymbol{\theta})=F_q$. Importantly, this is again a function of $\rho(\boldsymbol{\theta})$ alone. It is interesting to note that from equation (\ref{fimpur}) we recover a notion of sensitivity similar to what we found via the error propagation formula in section \ref{subsec:optint}, since for a single parameter it becomes $F_q = 4 \left(\langle \psi_0 | K^2 |\psi_0 \rangle -  \langle \psi_0 | K |\psi_0 \rangle^2\right)$ and, in that case, $\bar{\epsilon}_{cr} = 1/(4\mu  \Delta K^2)$\footnote{Nevertheless, in section \ref{subsec:optint} we used the quantity $1/(4\Delta K^2)$ as an approximation to the experimental error that was measured in the laboratory after many trials, while in this context it plays the role of a quantity that gives information about the precision for each shot in the regime of many of them.}. 

The restrictive conditions required to exploit the multi-parameter result contrast with the generality of the scalar case. Fortunately, all the multi-parameter schemes in this thesis are based on pure states and commuting generators, and as such the condition in equation (\ref{weakcommutation}) is satisfied. In those cases where this is not true, one can still study bounds based on the right logarithmic derivatives that arise from $\rho(\boldsymbol{\theta})R_i(\boldsymbol{\theta}) = \partial \rho(\boldsymbol{\theta})/\partial \theta_i$, which can be tighter than the bounds based on equation (\ref{qfimgen}) \cite{Szczykulska2016, helstrom1976}, or rely on the Holevo Cram\'{e}r-Rao bound \cite{holevo2011, sammy2016compatibility, albarelli2019}. The latter is a more general result and can be asymptotically approached if we allow for collective measurements within the framework of quantum local asymptotic normality \cite{sammy2016compatibility, gill2011}. However, note that as in the case of Helstrom and Holevo's Bayesian approach, these collective measurements would no longer represent repetitions of an experiment, which is the practical case that we wish to investigate. 

It is also important to observe that the classical Cram\'{e}r-Rao bound in equation (\ref{ccrbmulti}) that acts as the basis of the previous method can be also obtained without taking the limit $\mu \rightarrow \infty$ if we assume unbiased or locally unbiased estimators\footnote{An estimator is unbiased if $\boldsymbol{b}(\boldsymbol{\theta}) = \boldsymbol{0}$ for all $\boldsymbol{\theta}$, while it is locally unbiased at $\boldsymbol{\theta}_0$ if $\boldsymbol{b}(\boldsymbol{\theta}_0) = \boldsymbol{0}$ and $\int d\boldsymbol{m} ~ g_i(\boldsymbol{m})~ \partial p(\boldsymbol{m}|\boldsymbol{\theta}_0)/\partial \theta_j = \delta_{ij}$ \cite{fraser1964}.} \cite{kay1993, rafal2015, kolodynski2014, hall2012}. Nevertheless, here we are interested in considering a more general set of scenarios where these restrictions do not necessarily apply, and this generality means that in most cases we can only approach the Cram\'{e}r-Rao bound asymptotically.

From this discussion we can readily extract an important conclusion that will prove to be useful for our methodology. Since the quantum Cram\'{e}r-Rao error can be approached asymptotically when the appropriate conditions are fulfilled, the bounds derived with this tool are fundamental in the regime of many trials. As a consequence, we may see the Bayesian uncertainty in equation (\ref{msethesis}) as the true underlying theory (for a moderate amount of prior knowledge) that can give us the optimum in general and the Cram\'{e}r-Rao bound as an approximation to it that works and is recovered in certain situations. While the formal framework to exploit this idea in a consistent way is given by the theory of local asymptotic normality \cite{lecam1986, vaart1998, gill2011}, for our purposes it suffices to follow some known heuristic arguments that will be revisited in chapters \ref{chap:nonasymptotic} and \ref{chap:networks}, and to combine them with the numerical simulations of practical schemes that constitute a part of our results. 

\subsection{Other quantum bounds}
\label{subsec:alternativebounds}

Although the Cram\'{e}r-Rao bound generates fundamental limits once we have collected enough data, there is no reason to expect that these results will be valid out of this regime, and this motivates the search of quantum bounds that are valid for all $\mu$. This idea was precisely explored by Tsang \cite{tsang2012} and Lu and Tsang \cite{tsang2016}, where the two families of Bayesian bounds \cite{bayesbounds2007} were extended to the quantum regime. 

According to their results, the single-parameter quantum Ziv-Zakai bound for a flat prior of width $W_0$ is \cite{tsang2012}
\begin{equation}
\bar{\epsilon}_{\mathrm{mse}} \geqslant \frac{1}{2} \int_{0}^{W_0} d\theta \theta \left(1 - \frac{\theta}{W_0} \right) \left[1 - \sqrt{1 - \abs{f(\theta)}^{2\mu}} ~\right],
\label{qzzb}
\end{equation}
where $f(\theta) = \bra{\psi_0}\ket{\psi(\theta)}$, $\ket{\psi_0}$ is a pure state and $\ket{\psi(\theta)}$ is the result of having encoded the parameter with a unitary transformation. Note that the parameter domain of the integral in equation (\ref{qzzb}) is always $[0, W_0]$, independently of where the uniform prior is centred. Equation (\ref{qzzb}) can be derived by reinterpreting the expression for the mean square error as a binary hypothesis problem \citep{bayesbounds2007, tsang2012}. 

Furthermore, the quantum Weiss-Weinstein bound, which belongs to the second Bayesian family and is based on the covariance inequality\footnote{The covariance inequality is $\int d\boldsymbol{\theta}d\boldsymbol{m} ~ p(\boldsymbol{\theta}, \boldsymbol{m})~f(\boldsymbol{\theta}, \boldsymbol{m}) f(\boldsymbol{\theta}, \boldsymbol{m})^\transpose \geqslant \mathcal{T} \mathcal{G}^{-1} \mathcal{T}$, where $\mathcal{T} = \int d\boldsymbol{\theta}d\boldsymbol{m} ~ p(\boldsymbol{\theta}, \boldsymbol{m})~f(\boldsymbol{\theta}, \boldsymbol{m}) g(\boldsymbol{\theta}, \boldsymbol{m})^\transpose$ and $\mathcal{G} = \int d\boldsymbol{\theta}d\boldsymbol{m} ~ p(\boldsymbol{\theta}, \boldsymbol{m})~ g(\boldsymbol{\theta}, \boldsymbol{m}) g(\boldsymbol{\theta}, \boldsymbol{m})^\transpose$ \cite{bayesbounds2007}.}, establishes that \cite{tsang2016}
\begin{equation}
\bar{\epsilon}_{\mathrm{mse}} \geqslant \sup_{0 < s < 1} \frac{\theta^2 f_c(s,\theta)^2 \abs{f(\theta)}^{4\mu}}{h_c\left(s, \theta\right)\abs{f(\theta)}^{2\mu} - 2f_c(s,2\theta)\mathrm{Re}\left\lbrace \left[f(\theta)^{2} {f(2\theta)}^{*}\right ]^\mu \right\rbrace},
\label{qwwb}
\end{equation}
where 
\begin{equation}
h_c\left(s, \theta\right) = f_c(2s,\theta) + f_c(2-2s,\theta),
\end{equation}
\begin{equation}
f_c(s,\theta) = \int_{\lbrace \theta'\hspace{-0.1em},\hspace{0.2em} p(\theta')\neq 0 \rbrace} d\theta' p(\theta' + \theta)^s p(\theta')^{1-s}.
\label{qwwbpriorterm}
\end{equation} 

Interestingly, these tools share the simplicity of the Cram\'{e}r-Rao bound to some extent, since the quantities in equations (\ref{qzzb} - \ref{qwwbpriorterm}) do not depend on either the estimator or the POM. Thanks to this we can derive lower bounds for a given transformed state $\rho(\theta)$ and any desired number of copies. Moreover, while the Cram\'{e}r-Rao bound is a local quantity that depends on the derivatives of the likelihood function, the Ziv-Zakai and Weiss-Weinstein bounds are able to access the global topology of the parameter domain. This is particularly transparent when we observe that equations (\ref{qzzb}) and (\ref{qwwb}) are given in terms of the fidelity $|f(\theta)|^2$.  

These bounds can provide useful information about the non-asymptotic regime where the number of repetitions is low, but they also present important limitations. For example, the quantum Ziv-Zakai bound can recover the asymptotic scaling given by the Cram\'{e}r-Rao bound, but it is not tight in general \cite{tsang2012}. The situation improves with the quantum Weiss-Weinstein bound, since it is asymptotically tight. However, it is not guaranteed that we can saturate this bound in the regime with a finite number of measurements \cite{tsang2016}. 

Similar problems arise when we consider other quantum bounds. For instance, Liu and Yuan \cite{liu2016} introduced the quantum optimal-bias bound by optimising the trade-off between bias and variance in the scalar version of equation (\ref{msedecomposition}). This result is also valid for all $\mu$, but it is lower than the Cram\'{e}r-Rao bound by construction and, as we will see, the latter is sometimes lower than the optimal error when it is applied out of its regime of applicability. 

In addition, there exists a Bayesian version of the Cram\'{e}r-Rao bound based on the van Trees inequality \cite{gill1995}. Unfortunately, its derivation requires that the prior probability satisfies the boundary conditions $p(a) \rightarrow 0$ and $p(b) \rightarrow 0$, and this excludes the case of a flat prior between $a$ and $b$. 

Some of these caveats are also inherited by the multi-parameter generalisation of these results, as the multi-parameter Ziv-Zakai bound introduced by Zhang and Fan \cite{zhang2014} exemplifies. Worse, in some cases we can even lose the computational advantage provided by the use of bounds, which is precisely what happens with the multi-parameter Weiss-Weinstein bound \cite{tsang2016}. 

Due to these difficulties, we will not employ these types of bounds to derive our results. Nonetheless, in chapter \ref{chap:limited} we compare the single-parameter Ziv-Zakai and Weiss-Weinstein bounds in equations (\ref{qzzb} - \ref{qwwbpriorterm}) with the results that arise from our proposed strategy, and we show that our methodology is a superior choice to study the non-asymptotic regime of the practical situations under consideration. 

\subsection{The single-shot paradigm}
\label{subsec:singleshotparadigm}

Instead of addressing a multi-parameter experiment that is repeated $\mu$ times directly, let us change our strategy and focus our attention on a simpler scenario first: a single shot of a system with one parameter. Since $\mu = 1$, collective measurements do not arise, and thus the optimal strategy that satisfies Helstrom and Holevo's condition in equation (\ref{optequations}) presents no practical difficulty. Moreover, the solution for the squared deviation function is known \cite{personick1971, helstrom1976, macieszczak2014bayesian}. We dedicate the rest of this section to review this result. 

Assuming that the probe state $\rho_0$ and the unitary operator $U(\theta)$ are known, we wish to optimise the mean square error
\begin{equation}
\bar{\epsilon}_{\mathrm{mse}} = \int d\theta dm~p(\theta,m) \left[g(m)-\theta\right]^2,
\label{singlemse}
\end{equation}
which arises from equation (\ref{msethesis}) when $\mu = 1$ and $d = 1$, over all possible measurement schemes and estimators. Following Macieszczak \emph{et al.} \cite{macieszczak2014bayesian}, first we rewrite equation (\ref{singlemse}) as 
\begin{equation}
\bar{\epsilon}_{\mathrm{mse}} = \int d\theta p(\theta) \theta^2 + \mathrm{Tr}(S_2 \rho - 2 S \bar{\rho}),
\label{mseqform}
\end{equation}
where $\rho = \int d\theta p(\theta)\rho(\theta)$ and $\bar{\rho} = \int d\theta p(\theta)\rho(\theta) \theta$ are state moments, and having the POM and estimator inside the operators $S = \int dm~ g(m)E(m)$ and $S_2 = \int dm~ g(m)^2E(m)$. 

The measurement scheme $E(m)$ is completely general. However,  Macieszczak \emph{et al.} \cite{macieszczak2014bayesian} proved that restricting the possible POMs to the class of projective measurements does not lead to a loss of optimality. To see it, let us rewrite the operators $S$ and $S_2$ as
\begin{equation}
S = \int dg~g E(g), ~~S_2 = \int dg~g^2 E(g),
\end{equation}
where we have used $E(g) = \int dm~\delta(g(m)-g) E(m)$ to relabel the POM elements with the estimates, and notice that 
\begin{equation}
S_2 - S^2 = \int dg~g^2 E(g) - \left[\int dg~g E(g)\right]^2 \geqslant 0
\end{equation}
due to the operator version of Jensen's inequality\footnote{Given a convex function $f(t)$, Jensen's operator inequality establishes that $\int dt~E(t)f(t) \geqslant f\left[\int dt~E(t)t\right] $ \cite{hansen2003, macieszczak2014bayesian}.} \cite{hansen2003, macieszczak2014bayesian}. This implies that $\mathrm{Tr}(S_2 \rho) \geqslant \mathrm{Tr}(S^2 \rho)$, which can be saturated by choosing a projective measurement; consequently, in the latter case we have that
\begin{equation}
\bar{\epsilon}_{\mathrm{mse}} = \int d\theta p(\theta) \theta^2 + \mathrm{Tr}\left(\rho S^2 - 2\bar{\rho}S\right).
\label{quantumse}
\end{equation}

The final step to find the optimum is to minimise the latter equation with respect to $S$, arriving at \cite{personick1971, macieszczak2014bayesian}
\begin{equation}
\bar{\epsilon}_{\mathrm{mse}} \geqslant \int d\theta p(\theta)\theta^2 - \mathrm{Tr}\left(\bar{\rho}S\right),
\label{singleshot_bound}
\end{equation}
where now $S$ satisfies $S\rho + \rho S  = 2\bar{\rho}$. This is the minimum uncertainty.

A key advantage of this result is that the single-shot optimal strategy can be explicitly constructed from 
\begin{equation}
S = \int ds~s E(s) = \int ds~s \ketbra{s},
\label{singleshot_strategy}
\end{equation}
since the inequality in equation (\ref{singleshot_bound}) is saturated when the projectors $\lbrace \ket{s}\rbrace$ associated with the estimates $\lbrace s\rbrace$ are used as the measurement scheme. In fact, the eigenvalues $\lbrace s\rbrace$ are precisely the estimates given by the mean of the posterior density $p(\theta|s)\propto p(\theta) p(s|\theta)$ \cite{personick1971}, which is the classical solution for the optimal estimator \cite{jaynes2003, jesus2017, rafal2015}, and for that reason we will refer to the observable $S$ as the optimal quantum estimator.

Since equation (\ref{singleshot_strategy}) provides the optimal strategy, it must fulfil Helstrom and Holevo's condition in equation (\ref{optequations}). That this is indeed the case was shown by Helstrom in \cite{helstrom1976}. Following his discussion, we start by noticing that, in this case, $Q(g) = \int dg ~ p(\theta) \rho(\theta)(g-\theta)^2 =  g^2\rho - 2 g \bar{\rho} + \rho_2$, where $\rho_2 = \int d\theta p(\theta) \rho(\theta)\theta^2$. We may now use this and the optimal POM $\lbrace \ketbra{s} \rbrace$ to construct the operator $Y$ given in equation (\ref{optequations}), finding that
\begin{equation}
Y = S^2\rho - 2 S \bar{\rho} + \rho_2 = \rho S^2 - 2 \bar{\rho} S + \rho_2.
\label{optoperator}
\end{equation}
The goal is then to show that $Q(g) - Y$ is semi-definitive positive. To achieve this, let us rewrite $Y$ as $Y = \rho_2 - S \rho S$ by using the two forms in equation (\ref{optoperator}) and the equation satisfied by the optimal quantum estimator $S$ \cite{helstrom1976}. Similarly, $Q(g) = \rho_2 + g \rho g -  g \rho S -  S \rho g$. Hence, $Q(g) - Y = (S - g\mathbb{I})\rho(S - g\mathbb{I})$ and \cite{helstrom1976}
\begin{equation}
\langle u|\left[ Q(g) - Y \right] |u\rangle = \langle u| (S - g\mathbb{I})\rho(S - g\mathbb{I}) |u\rangle \geqslant 0,
\end{equation}
as required, since $|u\rangle$ can be any state and we may choose $|u\rangle = (S - g\mathbb{I})^{-1} |\bar{u}\rangle$, with $|\bar{u}\rangle$ arbitrary \cite{helstrom1976}.

Equation (\ref{singleshot_bound}) was originally discovered by Personick \cite{personick1969thesis, personick1971} and explored by him and others in the context of communication theory \cite{personick1969thesis, personick1971, helstrom1976}, and it has been more recently used to study a depolarizing channel \cite{mashide2002}, for frequency estimation \cite{macieszczak2014bayesian}, for magnetic sensing \cite{sekatski2017} and to estimate the coupling strengh of an optomechanical system \cite{bernad2018}. Moreover, a formally similar result emerges in the construction of the quantum Allan variance \cite{chabuda2016allanvariance}. Nevertheless, this result does not appear to have been fully exploited to study phase estimation in the regime of limited data and an intermediate prior that we are considering here.

As with the Cram\'{e}r-Rao bound, the results derived from the optimal single-shot mean square error will be fundamental and achievable, and there is an algorithm to calculate the optimal strategy explicitly. One of the central ideas in this thesis is the proposal of a practical way of exploiting this result for $\mu \neq 1$ and $d\neq 1$. 

\subsection{A new derivation of the optimal single-shot mean square error for a single parameter}
\label{subsec:originalderivation}

The final idea that we need to construct our non-asymptotic framework, which we discuss now, emerges in a rather surprising way when we derive the optimal single-shot mean square error using a different method. To the best of our knowledge, the path that we follow in this section has not been considered in existing literature. 

Given the single-parameter error in equation (\ref{singlemse}), our first step is to perform a classical optimisation over all the possible estimators. If we look at $\bar{\epsilon}_{\mathrm{mse}}$ in equation (\ref{singlemse}) as a functional of $g(m)$, then we can formulate the variational problem \cite{jaynes2003}
\begin{equation}
\delta \bar{\epsilon}_{\mathrm{mse}}\left[g(m)\right] = \delta \int dm~\mathcal{L}\left[m, g(m) \right] = 0,
\label{variationprob}
\end{equation}
where we have defined the object $\mathcal{L}\left[m, g(m) \right] = \int d\theta p(\theta, m)\left[g(m) - \theta\right]^2$, and, mathematically, equation (\ref{variationprob}) is equivalent to requiring that \cite{mathematics2004}
\begin{equation}
\frac{d \bar{\epsilon}_{\mathrm{mse}}\left[g(m)+\beta h(m)\right]}{d\beta} \bigg\rvert_{\beta = 0} = 0,~~\text{for~all}~~h(m). 
\end{equation}
In our case we have that
\begin{align}
\frac{d \bar{\epsilon}_{\mathrm{mse}}\left[g(m)+\beta h(m)\right]}{d\beta} &= \frac{d}{d\beta} \int dm~\mathcal{L}\left[m, g(m) + \beta h(m) \right] 
\nonumber \\
&= 2 \int d\theta dm~p(\theta, m) \left[ g(m) + \beta h(m) - \theta \right] h(m),
\label{firstvariation}
\end{align}
which means that the requirement to find the extrema of the error is
\begin{equation}
\frac{d \bar{\epsilon}_{\mathrm{mse}}\left[g(m)+\beta h(m)\right]}{d\beta} \bigg\rvert_{\beta = 0} = 2 \int d\theta dm~p(\theta, m) \left[ g(m) - \theta \right] h(m) = 0,
\label{variationcondition}
\end{equation}
and this implies that $\int d\theta p(\theta, m) [g(m) - \theta] = 0$ if equation (\ref{variationcondition}) it is to be satisfied by an arbitrary $h(m)$. By decomposing the joint probability as $p(\theta, m) = p(m) p(\theta|m)$, where the posterior satisfies that $p(\theta|m) \propto p(\theta)p(m|\theta)$, we see that the solution $g(m) = \int d\theta p(\theta|m) \theta$ makes the error $\bar{\epsilon}_{\mathrm{mse}}\left[g(m)\right]$ in equation (\ref{singlemse}) extremal, which is a well-known result in probability theory\footnote{Interestingly, we may say that $\mathcal{L} = \int d\theta p(\theta, m)\left[g(m) - \theta\right]^2$, $\int d\theta p(\theta, m) [g(m) - \theta] = 0$ and the estimator $g(m) = \int d\theta p(\theta|m) \theta$ are to Bayesian estimation theory what the Lagrangian, the Euler-Lagrange equations and the trajectory of the system are to analytical mechanics, respectively.} \cite{jaynes2003}.

To verify that this is a minimum we can use the functional version of the second derivative test. Calculating the second variation from equation (\ref{firstvariation}) we see that
\begin{equation}
\frac{d^2 \bar{\epsilon}_{\mathrm{mse}}\left[g(m)+\beta h(m)\right]}{d\beta^2} \bigg\rvert_{\beta = 0} = 2 \int d\theta dm~p(\theta, m) h(m)^2 > 0
\end{equation}
for non-trivial variations; consequently, choosing the estimator $g(m) = \int d\theta p(\theta|m) \theta$ gives the minimum mean square error. 

Upon introducing $g(m) = \int d\theta p(\theta|m) \theta$ in equation (\ref{singlemse}) we thus find the bound
\begin{align}
\bar{\epsilon}_{\mathrm{mse}} \geqslant ~& \epsilon_{\mathrm{opt}}^c = \int dm~p(m) \left\lbrace \int d\theta p(\theta|m) \theta^2 -  \left[\int d\theta p(\theta|m) \theta \right]^2 \right\rbrace
\nonumber \\
& = \int d\theta p(\theta) \theta^2  - \int \frac{dm}{\int d\theta p(\theta)p(m|\theta)}\left[\int d\theta p(\theta)p(m|\theta) \theta \right]^2,
\label{classicalbound}
\end{align}
where the second line can be obtained by noticing that $\int dm~p(m) p(\theta|m) = p(\theta)$ and using Bayes theorem (see section \ref{sec:probability}). 

The first line of equation (\ref{classicalbound}) is the familiar expression for the variance of the posterior probability averaged over the probability $p(m)$, which represents the theoretical information about the possible values for the outcomes. The second expression, on the other hand, contains a term that displays a remarkable similarity with the expression for the classical Fisher information in equation (\ref{fishersingleparameter}), and this formal analogy becomes even more apparent when we further consider the quantum part of the problem. In particular, by inserting $p(m|\theta) = \mathrm{Tr}[E(m) \rho(\theta)]$ in equation (\ref{classicalbound}) we find that
\begin{equation}
\epsilon_{\mathrm{opt}}^c = \int d\theta p(\theta)\theta^2 - \int dm  \frac{\mathrm{Tr}\left[ E(m) \bar{\rho} \right]^2}{\mathrm{Tr}\left[ E(m) \rho \right]},
\label{bayesanalogygen}
\end{equation}
with $\rho = \int d\theta p(\theta) \rho(\theta)$ and $\bar{\rho} = \int d\theta p(\theta) \rho(\theta) \theta$. This suggests that it may be possible to bound this term with a procedure similar to the proof proposed by Braunstein and Caves \cite{BraunsteinCaves1994} to derive the quantum Cram\'{e}r-Rao bound.

Following this analogy we can introduce the Bayesian counterpart of the equation for the symmetric logarithmic derivative in section \ref{subsec:crb}, that is, $S \rho + \rho S = 2\bar{\rho}$\footnote{However, note that $\bar{\rho}$ is not a derivative.}. This allows us to manipulate the second term in the right hand side of equation (\ref{bayesanalogygen}) as 
\begin{align}
\int dm  \frac{\mathrm{Tr}\left[ E(m) \bar{\rho}_u \right]^2}{\mathrm{Tr}\left[ E(m) \rho \right]} &= \int dm\left(\frac{\mathrm{Re}\left\lbrace\mathrm{Tr}\left[E(m) S\rho\right]\right\rbrace}{\sqrt{\mathrm{Tr}\left[ E(m) \rho\right]}}\right)^2 
\nonumber \\
&\leqslant  \int dm~\abs{\frac{\mathrm{Tr}\left[E(m) S\rho\right]}{\sqrt{\mathrm{Tr}\left[ E(m) \rho\right]}}}^2
\nonumber \\
&=\int dm~ \abs{\mathrm{Tr}\left[\frac{\rho^{\frac{1}{2}}E(m)^{\frac{1}{2}}}{\sqrt{\mathrm{Tr}\left[ E(m) \rho \right]}} E(m)^{\frac{1}{2}}S \rho^{\frac{1}{2}}\right]}^2 
\nonumber \\
&\leqslant  \int dm~ \mathrm{Tr}\left[ E(m) S \rho S \right] = \mathrm{Tr}(\rho S^2) = \mathrm{Tr}\left(\bar{\rho} S \right) ,
\label{qinequalities}
\end{align}
where we have used the Cauchy-Schwarz inequality 
\begin{equation}
|\mathrm{Tr}[X^\dagger Y ]|^2 \leqslant \mathrm{Tr}[ X^\dagger X] \mathrm{Tr}[Y^\dagger Y]
\end{equation}
with $X = E(m)^{\frac{1}{2}} \rho^{\frac{1}{2}}/\sqrt{\mathrm{Tr}\left[ E(m) \rho \right]}$, $Y = E(m)^{\frac{1}{2}} S \rho^{\frac{1}{2}}$. As expected, the operations performed in equation (\ref{qinequalities}) are formally identical to those appearing in the proof of the Braunstein-Caves inequality \cite{BraunsteinCaves1994, genoni2008}. 

The combination of equations (\ref{classicalbound} - \ref{qinequalities}) finally gives us the chain of inequalities
\begin{equation}
\bar{\epsilon}_{\mathrm{mse}} \geqslant \epsilon_{\mathrm{opt}}^c \geqslant \epsilon_{\mathrm{opt}}^q = \int d\theta p(\theta) \theta^2 - \mathrm{Tr}(\bar{\rho} S),
\label{chain}
\end{equation}
where the second term in $\epsilon_{\mathrm{opt}}^q$ can be seen as a Bayesian counterpart of the quantum Fisher information. 

From our discussion of the classical optimisation we see that the first inequality in equation (\ref{chain}) is saturated when the estimator is given by the average over the posterior. On the other hand, the quantum bound in equation (\ref{qinequalities}) relies on two inequalities. The first of them is saturated when $\mathrm{Tr}[E(m)S\rho]$ is real, while the Cauchy-Schwarz inequality is saturated if and only if $X \propto Y$ for some proportionality constant \cite{helstrom1968multiparameter}. In our case this implies that
\begin{equation}
\frac{E(m)^{\frac{1}{2}}\rho^{\frac{1}{2}}}{\mathrm{Tr}\left[E(m) \rho \right]} = \frac{E(m)^{\frac{1}{2}}S\rho^{\frac{1}{2}}}{\mathrm{Tr}\left[E(m) S \rho \right]}.
\end{equation}

These conditions are fulfilled by constructing a POM based on the projections onto the eigenstates of $S$. To verify this, let us first consider the eigendecomposition $S = \int ds~s \ketbra{s}$. Then, by using the POM $E(s) =  \ketbra{s}$ we find that
\begin{align}
\int ds  \frac{\mathrm{Tr}\left[ E(s) \bar{\rho} \right]^2}{\mathrm{Tr}\left[ E(s) \rho \right]} &= \int ds\left(\frac{\mathrm{Re}\left\lbrace\mathrm{Tr}\left(\ketbra{s} S\rho\right)\right\rbrace}{\sqrt{\mathrm{Tr}\left( \ketbra{s} \rho\right)}}\right)^2
\nonumber \\
&= \int ds~ s^2 \mathrm{Tr}\left(\ketbra{s} \rho\right]) = \mathrm{Tr}(\rho S^2),
\end{align}
as required, since $\mathrm{Tr}(\rho S^2) = \mathrm{Tr}(\bar{\rho} S)$. Therefore, we have recovered the result for the single-shot mean square error reviewed in section \ref{subsec:singleshotparadigm}. Moreover, further intuition can be gained by noticing that $\mathrm{Tr}(\rho S) = \int d\theta p(\theta) \theta$, so that we can rewrite the quantum bound in equation (\ref{chain}) as
\begin{equation}
\bar{\epsilon}_{\mathrm{mse}}\geqslant\Delta \theta^2_p - \Delta S^2_{\rho},
\label{myexpression}
\end{equation}
where we have defined the prior uncertainty as
\begin{equation}
\Delta \theta^2_p = \int d\theta p(\theta) \theta^2 - \left[\int d\theta p(\theta) \theta \right]^2
\end{equation}
and $\Delta S^2_\rho = \mathrm{Tr} \left(S^2 \rho \right) - \mathrm{Tr}\left(S \rho \right)^2$. In words, the uncertainty of our estimation is lower bounded by the difference between the prior variance and the variance of the optimal quantum estimator.

One could be tempted to argue that by introducing $S \rho + \rho S = 2 \bar{\rho}$ into the derivation we are somehow assuming the answer, as this is indeed the formal solution that arises from the direct optimisation in section \ref{subsec:singleshotparadigm} that combines classical and quantum elements. However, note that this equation is introduced here as a redefinition of $\bar{\rho}$ that allows us to derive a bound, and whose form is imposed by exploiting the formal analogy with the Fisher information. In addition, there is a way to see how $S \rho + \rho S = 2 \bar{\rho}$ emerges without performing the quantum optimisation. If we combine Bayes theorem and the Born rule as
\begin{equation}
p(\theta|m) = \frac{p(\theta)\mathrm{Tr}[E(m)\rho(\theta)]}{\int d\theta p(\theta)\mathrm{Tr}\left[E(m)\rho(\theta)\right]},
\end{equation}
then we have that $g(m) = \int d\theta p(\theta|m)\theta = \mathrm{Tr}[E(m)\bar{\rho}]/\mathrm{Tr}[E(m)\rho]$ for the optimal estimator. By further rearranging its terms we find that 
\begin{align}
0 &= \mathrm{Tr}\left[g(m)E(m) \rho - \bar{\rho}E(m)  \right],
\nonumber \\
0 &= \mathrm{Tr}\left[g(m)E(m) \rho + \rho g(m)E(m) - 2\bar{\rho}E(m) \right],
\nonumber \\
0 &= \mathrm{Tr}\left[\rho g(m)E(m) - \bar{\rho}E(m) \right],
\end{align}
which at this stage are fully equivalent. However, we can rewrite them as
\begin{eqnarray}
\mathrm{Tr}\left[L \rho - \bar{\rho}  \right] = 0, ~~ \mathrm{Tr}\left[S \rho + \rho S - 2\bar{\rho} \right]  = 0, ~~ \mathrm{Tr}\left[\rho R - \bar{\rho} \right] = 0,
\end{eqnarray}
respectively, after taking the integral over the experimental outcomes, so that they are satisfied when $L \rho = \bar{\rho}$, $S\rho + \rho S = 2\bar{\rho}$ and $\rho R = \bar{\rho}$. Note that these equations are not equivalent, and since we have implicitly assumed that the unknown parameter is a real quantity, we need a quantum estimator that is Hermitian, which in general is only given by $S$. Remarkably, this way of looking at the problem is reminiscent of the quantization rule to upgrade expressions with real quantities from their classical form to their quantum version by replacing c-numbers with q-numbers.

Thus we conclude that while the proofs by Personick \cite{personick1971}, Helstrom \cite{helstrom1976} and Macieszczak \emph{et al.} \cite{macieszczak2014bayesian} are preferred from a mathematical point of view, our presentation provides physically useful insights. Certainly, its major strength is that it clearly separates the classical optimisation from the manipulations associated with the quantum part of the problem, in complete analogy with the original derivation by Braunstein and Caves for the Fisher information \cite{braunstein1996}. This separation between classical and quantum contributions to the process of optimising the error is precisely the insight that we were looking for, and it will provide us with a powerful heuristic intuition to understand why the method proposed in the next section works. 

\section{Constructing a non-asymptotic metrology}
\label{subsec:constructingmethod}

We finally have all the pieces that we need in order to formulate the central method that we propose in this thesis:
\begin{enumerate}
\item A common way of extracting information from reality is to repeat a given experiment a certain number of times while we also exploit quantum features such as squeezing or entanglement in each shot. The number of trials is always finite in practice, and potentially small. Moreover, a realistic amount of prior knowledge will typically be moderate.
\item According to our discussion in section \ref{sec:uncertainty}, a suitable figure of merit to study this experimental arrangement is the Bayesian square error in equation (\ref{msethesis}). 
\item Our first step is to consider the optimisation of the estimator and that of the quantum strategy in a separate fashion. While this separation is commonly exploited in the context of the Cram\'{e}r-Rao bound (section \ref{subsec:crb}), both minimisations are usually merged when the Bayesian approach is employed (sections \ref{subsec:fundeq} and \ref{subsec:singleshotparadigm}). Our discussion in section \ref{subsec:originalderivation} demonstrates that splitting the problem in this way is also meaningful within the Bayesian framework.
\item We will choose the Bayesian estimator that is optimal with respect to the square error criterion for any number of repetitions. Therefore, this part of the problem will always be exact in all our calculations. 
\item The quantum strategy will be selected in two ways. One of them is to use the asymptotic regime as a guide and consider quantum schemes that are asymptotically optimal according to the Cram\'{e}r-Rao bound. The solutions that emerge from this method, which is based on a direct analysis of the non-asymptotic uncertainties that we calculate, provide less general but useful information about the non-asymptotic regime\footnote{Note that the optimal Bayesian estimator can approach the Cram\'{e}r-Rao bound asymptotically in the same way that the maximum of the likelihood does. We will recall the arguments showing that this is the case in chapters \ref{chap:nonasymptotic} and \ref{chap:networks}.}.
\item A different possibility is to select the quantum scheme that is optimal for a single shot of the experiment, and then repeat this strategy as many times as the application at hand demands or allows for. To achieve this, we will exploit the quantum square error in sections \ref{subsec:singleshotparadigm} and \ref{subsec:originalderivation}, and in chapter \ref{chap:multibayes} we will generalise this result to cover the multi-parameter regime. This procedure generates uncertainties that have been optimised in a shot-by-shot fashion and, as such, we are only optimising the resources that we need.  
\end{enumerate}

One may see the first procedure to select the quantum strategy as analogous to a semiclassical approximation, where the Fisher information and the Cram\'{e}r-Rao bound are to the Bayesian uncertainty what the part of the problem that is modelled by classical mechanics is to the quantum degrees of freedom. This contrasts with the different logic that is followed by the shot-by-shot method, which instead of optimising the protocol assuming that we will have many trials, it selects a strategy with a good performance for the experiments that will actually be performed. 

Other techniques could be more general. However, our method provides a direct link with the reality of experimental practice, while, at the same time, we will see that it relies on computations that are generally tractable. We recall that we do not apply Helstrom and Holevo's method in its most general way (section \ref{subsec:fundeq}), neither shall we use the Holevo Cram\'{e}r-Rao bound directly\footnote{Although if we can saturate the multi-parameter quantum Cram\'{e}r-Rao bound, then we are also saturating the version given by Holevo \cite{sammy2016compatibility}.}, because our model represents repetitions and, as such, it excludes the possibility of considering collective measurements. Furthermore, we have also excluded the application of lower bounds for which it is not clear whether they are tight in the regime of limited data, as is the case in general for the proposals considered in section \ref{subsec:alternativebounds}.

The rest of this work is dedicated to implementing this programme, and its potential application shall be illustrated by analysing the performance predicted by our methodology for optical interferometers and quantum sensing networks. 

\section{Summary of results and conclusions}

In this chapter we have laid a bridge between the results that are available in the literature and those that we intend to derive in the following chapters. We have reviewed the fundamental equations that the optimal quantum strategy needs to satisfy in the Bayesian framework, identifying its strengths in the single-shot regime, and acknowledging its practical limitations when collective measurements are allowed. We have also discussed the useful aspects of an approach based on bounding the estimation error, with a particular emphasis on the potentially fundamental character of the Cram\'{e}r-Rao bound in the regime of many repetitions.  

Furthermore, we have offered a new perspective on the derivation of the quantum strategy that makes the mean square error optimal for a single shot. A key novelty is the explicit separation of the classical and quantum contributions to the optimisation of the uncertainty, in analogy with Braunstein and Caves's original derivation of the inequality for the Fisher information \cite{BraunsteinCaves1994}. This formal connection between Bayesian quantities and those associated with Fisher methods has appeared in \cite{jesus2019b}
\begin{displayquote}
\emph{Bayesian multi-parameter quantum metrology with limited data}, \underline{Jes\'{u}s Rubio} and Jacob Dunningham, arXiv:1906.04123 (2019).
\end{displayquote}

From the analysis of these results we propose a strategy to study and design experiments that takes into account the challenges faced in practice, focusing our attention on limited amounts of measurement data and moderate prior knowledge. In particular, it is argued that the classical part of the problem can always be treated exactly, while the quantum part can be approximated to make the optimisation more tractable. This is the case if we employ the asymptotic regime given by the Cram\'{e}r-Rao bound as a guide. Alternatively, a more powerful approach is to repeat the quantum strategy that is optimal for a single shot. The philosophy of both methods, being completely different, can complement each other, since the former assumes in advance that a large number of experiments will eventually be performed while the latter only focus on those that will actually happen. 

Finally, we have carried out a detailed analysis of how different measures of uncertainty should be used in metrology, and the quantity that is relevant for our purposes has been identified, which provides our results with a strong conceptual and physically rigorous foundation. The three-step construction based on the characteristics of experiments, simulations and theoretical studies was included in \cite{jesus2017}
\begin{displayquote}
\emph{Non-asymptotic analysis of quantum metrology protocols beyond the Cram\'{e}r-Rao bound}, \underline{Jes\'{u}s Rubio}, Paul Knott and Jacob Dunningham, J. Phys. Commun. 2 015027 (2018).
\end{displayquote}

\chapter{Non-asymptotic analysis of single-parameter protocols}
\label{chap:nonasymptotic}

\section{Goals for the first stage of our methodology}

We start by considering an experiment that has been repeated $\mu$ times and where there is an unknown parameter $\theta$. Given that configuration, the main aim of this chapter is to analyse the non-asymptotic performance of metrology protocols that have been optimised as if the asymptotic theory were valid, and to explore the structure of the non-asymptotic regime with concrete examples. This is achieved by utilising a versatile numerical framework  that combines the exact optimal estimator with the asymptotically optimal quantum strategy, implementing in this way the first version of our methodology in chapter \ref{chap:methodology}. 

A crucial advantage of this approach is that it provides the means to investigate the regime of validity of the quantum Cram\'{e}r-Rao bound for specific strategies. Moreover, it allows us to understand what happens in practice with the conclusions extracted from this bound in the regime where it is not a valid approximation. We will address these questions as a first application of our methods, having chosen a selection of schemes among those that are commonly employed in the context of optical quantum metrology \cite{yurke1986, berry2000, durkin2007, dowling2008, HofmannHolger2009, gerry2010, chiruvelli2011, rafal2015, PaulProctor2016, dowling2014, sahota2015, sahota2016, lee2016}.

The fundamental importance of determining when this bound should be employed becomes apparent if we take into account that many protocols are designed by simply maximising the quantum Fisher information \cite{rafal2015, PaulProctor2016}, and that the assumptions that go into the construction of this tool are often not explicitly taken into account \cite{PaulProctor2016, proctor2017networked}. For example, in chapter \ref{chap:methodology} we saw that this technique normally requires many repetitions to be useful, and that this is an important drawback to study realistic physical systems. Since in general it is not possible to foresee when and how the Cram\'{e}r-Rao bound is going to fail in a concrete practical scenario from the asymptotic theory itself, a closer analysis of those schemes that are asymptotically optimal is needed. 

This problem has been widely acknowledged in the literature, both before \cite{braunstein_gaussian1992, hall2012, tsang2012, tsang2016, rafal2015, liu2016} and after \cite{smirne2018, lumino2017, braun2018, haase2018may} the appeareance of our results in this chapter (which were published in \cite{jesus2017}), and several solutions have been proposed. The direct approach based on choosing some general measure of uncertainty and estimating how many measurements are needed such that the results of the asymptotic theory are valid has been implemented numerically \cite{braunstein_maxlikelihood1992, braunstein_gaussian1992}. The early proposal in \cite{braunstein_gaussian1992} provides, in addition, an analytical estimate of this number, a result that can be derived by generalising the likelihood equation. More recently, Tsang \cite{tsang2012} succeeded in capturing the effect of the prior information with his quantum Ziv-Zakai bound (see section \ref{subsec:alternativebounds}), and the works based on the quantum Weiss-Weinstein \cite{tsang2016} and optimal-bias bounds \cite{liu2016} also included repetitions.  

From our discussion in chapter \ref{chap:methodology} we can readily see the advantages of our methodology over previous ideas. The numerical nature of the proposal in this chapter is shared with the work in \cite{braunstein_maxlikelihood1992}. However, the latter does not take into account the prior information, while our method is based on Bayesian techniques. On the other hand, the fact that we are modelling the prior knowledge rigorously is shared with the alternative quantum bounds in \cite{tsang2012, tsang2016, liu2016}. Nevertheless, we have seen that these are not tight in general. On the contrary, we calculate the actual uncertainties associated with the schemes under analysis, which implies that, by construction, we know how to generate them in a hypothetical experiment. Thus our pragmatic approach is both more general in the sense that it does not ignore important information, and also useful in practice because it relies on real uncertainties. 

The results in this chapter show that, once we have fixed the measurement strategy, both the number of trials and the minimum prior knowledge needed to reach the asymptotic regime are state-dependent, so that the conclusions about the relative performance of different optical schemes change in the non-asymptotic regime. As a result, in general we can say that maximizing the Fisher information alone does not always guarantee the best precision for experiments with a limited number of observations, a conclusion with important implications for the analysis of theory and experiments in quantum metrology.

\section{Methodology (part A)} \label{method}

\subsection{The asymptotic regime as a guide}
\label{theory}

The uncertainty in equation (\ref{msethesis}) is reduced to
\begin{equation}
\bar{\epsilon}_{\mathrm{mse}} = \int d\theta d\boldsymbol{m} ~p(\theta, \boldsymbol{m}) \left[g(\boldsymbol{m}) - \theta \right]^2
\label{errwork}
\end{equation}
for $\mu$ repetitions and a single parameter (i.e., $d = 1$), and the estimator $g(\boldsymbol{m})$ that makes this error minimum can be found by solving the variational problem \cite{jaynes2003} 
\begin{align}
\delta \bar{\epsilon}_{\mathrm{mse}}\left[g(\boldsymbol{m}) \right] = \delta \int d\boldsymbol{m}~ \mathcal{L}\left[\boldsymbol{m},g(\boldsymbol{m})\right]= 0,
\end{align}
where $\mathcal{L}\left[\boldsymbol{m}, g(\boldsymbol{m})\right] = \int d\theta p(\boldsymbol{m},\theta) \left[g(\boldsymbol{m}) - \theta\right]^2$. This problem is formally identical to the analogous case for a single repetition that we examined in section \ref{subsec:originalderivation}. Consequently, we know that the optimal estimator is $g(\boldsymbol{m}) = \int d\theta p(\theta|\boldsymbol{m})\theta$, with $p(\theta|\boldsymbol{m}) \propto p(\theta)p(\boldsymbol{m}|\theta)$, and that equation (\ref{errwork}) becomes 
\begin{equation}
\bar{\epsilon}_{\mathrm{mse}} = \int d\boldsymbol{m} p(\boldsymbol{m})\epsilon(\boldsymbol{m}),
\label{erropt}
\end{equation}
where $p(\boldsymbol{m}) = \int d\theta p(\theta)p(\boldsymbol{m}|\theta)$ and  
\begin{align}
\epsilon(\boldsymbol{m}) = \int d\theta p(\theta|\boldsymbol{m}) \theta^2- \left[\int d\theta p(\theta|\boldsymbol{m}) \theta \right]^2.
\label{errbayes}
\end{align}
Note that equation (\ref{errbayes}) is the experimental error identified in section \ref{sec:uncertainty}. The uncertainty in equation (\ref{erropt}) is a function of the number of repetitions $\mu$ and, as such, it is the quantity that we will employ to study the low-$\mu$ regime.

Once we have selected $g(\boldsymbol{m})$, we wish to choose some quantum protocol that is asymptotically optimal, but for our purposes this is only meaningful if we can treat the Cram\'{e}r-Rao bound as an approximation to equation (\ref{erropt}). Fortunately, it is known that this is the case not only for the maximum likelihood estimator reviewed in section \ref{subsec:crb}, but also for equation (\ref{erropt}) \cite{bernardo1994, cox2000}. We now recall an heuristic version of the standard argument that leads to this result (this can be found, e.g., in \cite{jaynes2003, cox2000, bernardo1994}), so that we can identify the key assumptions and the nature of the approximation that we intend to exploit.

Let us imagine a hypothetical scenario where the likelihood $p(\boldsymbol{m}|\theta)$ as a function of $\theta$ becomes narrower and concentrated around a maximum $\theta_{\boldsymbol{m}}$ when $\mu \gg 1$ \cite{cox2000}, where the observations $\boldsymbol{m}$ were originated from an unknown parameter $\theta'$. In addition, the prior knowledge is enough to identify a region of the parameter domain that contains $\theta'$ and in which such maximum is absolute and unique, although the experimental information dominates in this regime. This can be captured by a prior probability that is approximately flat in that region, whose width we can express as $(b-a)$. Hence, $p(\theta) \approx 1/(b-a)$ when $\theta \in [a,b]$, and zero otherwise. 

If we express the likelihood formally as $p(\boldsymbol{m}|\theta) = \mathrm{exp}{\left\lbrace \mathrm{log} \left[p(\boldsymbol{m}|\theta)\right]\right\rbrace}$, then the first step is to calculate the Taylor expansion
\begin{eqnarray}
\mathrm{log} \left[p(\boldsymbol{m}|\theta)\right] \approx \mathrm{log} \left[p(\boldsymbol{m}|\theta_{\boldsymbol{m}})\right] + \frac{1}{2} \frac{\partial^2 \mathrm{log} \left[p(\boldsymbol{m}|\theta_{\boldsymbol{m}})\right]}{\partial \theta^2} (\theta - \theta_{\boldsymbol{m}})^2,
\end{eqnarray}
where the first order term has vanished because $\theta_{\boldsymbol{m}}$ represents a maximum. Additionally, by the law of large numbers (section \ref{subsec:lln})
\begin{eqnarray}
\frac{\partial^2\mathrm{log} \left[p(\boldsymbol{m}|\theta_{\boldsymbol{m}})\right]}{\partial \theta^2} = \sum_{i=1}^{\mu} \frac{\partial^2 \mathrm{log} \left[p(m_i|\theta_{\boldsymbol{m}})\right]}{\partial \theta^2}\approx  \mu\int dm\hspace{0.1em} p(m|\theta')\frac{\partial^2 \mathrm{log} \left[p(m|\theta')\right]}{\partial \theta^2},
\label{lln}
\end{eqnarray}
where we have also used that $\theta_{\boldsymbol{m}} \approx \theta'$ due to the consistency of the maximum of the likelihood \cite{cox2000, kay1993, pezze2014}. Therefore,
\begin{align}
p(\boldsymbol{m}|\theta) \approx p(\boldsymbol{m}|\theta')\mathrm{exp} \left[ -\frac{\mu F(\theta')}{2}\left(\theta - \theta'\right)^2 \right],
\label{gaussian_likelihood}
\end{align}
where $F(\theta')$ is the classical Fisher information in equation (\ref{fishersingleparameter}) that arises from expanding the derivative of equation (\ref{lln}). 

By performing the calculation\footnote{The details of both this calculation and those in equation (\ref{gaussiansingle}) can be found in appendix \ref{sec:multigaussian}.}
\begin{eqnarray}
\int_{a}^b d\theta p(\theta)p(\boldsymbol{m}|\theta)  \approx \frac{p(\boldsymbol{m}|\theta')}{b-a}\int_{-\infty}^\infty d\theta\hspace{0.15em} \mathrm{e}^{ - \frac{\mu F(\theta')}{2}(\theta - \theta')^2} = \frac{p(\boldsymbol{m}|\theta')}{b-a} \sqrt{\frac{2 \pi}{\mu F(\theta')}},
\label{norm_asy}
\end{eqnarray}
where the approximation of the infinite limits holds due to the concentration of $p(\boldsymbol{m}|\theta)$ around a single point, we can approximate the posterior probability by a Gaussian density as
\begin{equation}
p(\theta|\boldsymbol{m}) = \frac{p(\theta)p(\boldsymbol{m}|\theta)}{\int d\theta p(\theta)p(\boldsymbol{m}|\theta)}\approx \sqrt{\frac{\mu F(\theta')}{2\pi}} \mathrm{exp} \left[ -\frac{\mu F(\theta')}{2}(\theta - \theta')^2 \right].
\label{gaussian}
\end{equation}
In turn we can now calculate the Gaussian integrals
\begin{align}
\int_{a}^b d\theta p(\theta|\boldsymbol{m})\theta &\approx   \sqrt{\frac{\mu F(\theta')}{2 \pi}}\int_{-\infty}^\infty d\theta\hspace{0.15em}\mathrm{e}^{ - \frac{\mu F(\theta')}{2}(\theta - \theta')^2}\theta = \theta',
\nonumber \\
\int_{a}^b d\theta p(\theta|\boldsymbol{m})\theta^2 &\approx   \sqrt{\frac{\mu F(\theta')}{2 \pi}}\int_{-\infty}^\infty d\theta\hspace{0.15em} \mathrm{e}^{- \frac{\mu F(\theta')}{2}(\theta - \theta')^2}\theta^2 =  (\theta')^2 + \frac{1}{\mu F(\theta')}
\label{gaussiansingle}
\end{align}
and introduce them in equation (\ref{errbayes}), finding that $\epsilon(\boldsymbol{m}) \approx 1/[\mu F(\theta')]$ for the variance of the posterior. 

Finally, by integrating the approximation for $\epsilon(\boldsymbol{m})$ with respect to the outcomes we conclude that the error in equation (\ref{erropt}) can be approximated as
\begin{equation}
\bar{\epsilon}_{\mathrm{mse}} \approx \int d\theta' p(\theta')  \int d\boldsymbol{m} ~\frac{p(\boldsymbol{m}|\theta')}{\mu F(\theta')} = \int d\theta' \frac{p(\theta') }{\mu F(\theta')},
\label{asymapproxmse}
\end{equation}
which is the single-parameter version of the classical Cram\'{e}r-Rao bound in equation (\ref{ccrbmulti}). As a consequence, if the measurement scheme is given by the projections onto the eigenspaces of the symmetric logarithmic derivative $L(\theta)$, or by some POM consistent with these, and we have that $F(\theta) = F_q$, then $\bar{\epsilon}_{\mathrm{mse}} \approx 1/(\mu F_q)$, where we recall that $F_q = \mathrm{Tr}[\rho(\theta)L(\theta)]$ is the quantum Fisher information and that this quantity does not depend explicitly on the parameter when the latter is encoded as $\rho(\theta)=\mathrm{e}^{-i K \theta}\rho_0 \mathrm{e}^{i K \theta}$, where $\rho(\theta)$ is the transformed state\footnote{Even if the Fisher information (classical or quantum) depends explicitly on the parameter, we can still derive a lower bound on equation (\ref{asymapproxmse}). In effect, let us define the functions $u(\theta) = \sqrt{p(\theta)}/\sqrt{\mu F(\theta)}$, $v(\theta) = \sqrt{p(\theta)\mu F(\theta)}$, and apply the Cauchy-Schwarz inequality \cite{mathematics2004}
\begin{equation}
\int d\theta \abs{u(\theta)}^2  \int d\theta \abs{v(\theta)}^2 \geqslant \abs{\int d\theta u(\theta) \overline{v(\theta)}}^2; 
\nonumber
\end{equation}
then equation (\ref{asymapproxmse}) satisfies 
\begin{equation}
\bar{\epsilon}_{\mathrm{mse}} \approx \int d\theta \frac{p(\theta)}{\mu F(\theta)} \geqslant \frac{1}{\int d\theta p(\theta)\mu F(\theta)}.
\nonumber
\label{priorind}
\end{equation}
The inequality can be saturated when $u(\theta) \propto v(\theta)$, with a constant of proportionality that does not depend on the parameter, and this is fulfilled if and only if the Fisher information $F$ is a constant. This result can also be derived using Jensen's inequality \cite{kolodynski2014}.}. 

As we announced, this discussion demonstrates that the Bayesian uncertainty in equation (\ref{erropt}) can be approximated by the quantum Cram\'{e}r-Rao bound under certain circumstances. Therefore, in our work this result plays the role of an asymptotic approximation, instead of being employed as a proper bound that is generally valid. This perspective\footnote{While in physics it is very natural (and useful) to explore how the limiting cases of some theories recover the results given by less general theories, in estimation theory it is still common to make a distinction between local and global approaches \cite{paris2009, rafal2015, li2018}, where the former is associated with the Fisher information and the frequentist interpretation of probabilities and the latter with Bayesian techniques. However, the view exploited here is conceptually simpler, and as we argue in appendix \ref{sec:otheruncertainty}, it can be applied to most cases that we may find in practice, which cast doubts on the necessity of introducing different frameworks. The results in this thesis constitute an explicit piece of evidence in favour of our perspective.} allows us to define the asymptotic regime by two basic properties:
\begin{itemize}
\item[i)] the number of trials $\mu$ is sufficiently large, and
\item[ii)] the prior information is enough to localise the relevant domain,
\end{itemize}
while the non-asymptotic emerges when these requirements are not fulfilled\footnote{Note that we are implicitly assuming that the probability densities are regular enough to perform the operations in the previous calculations. That this is the case will become apparent in the outputs of our numerical simulations.}. It is clear that whenever two strategies are being compared in terms of the quantum Cram\'{e}r-Rao bound, in general it is also necessary to indicate how large $\mu$ needs to be such that $\bar{\epsilon}_{\mathrm{mse}} \approx 1/(\mu F_q)$ is a good approximation. Moreover, if the likelihood reaches its maximum for several values of the parameter, then we need enough prior knowledge to select a single peak. 

The verification of the fulfilment of these restrictions is often not done in the literature, a problem that can be overcome by using the framework in the next sections. Once we have identified the boundary separating the asymptotic and non-asymptotic regimes, we can proceed to also analyse the low-$\mu$ performance of our protocols. The intuition behind this idea can be understood as follows. Suppose that a given scheme is designed to be run a certain number of times. In many situations we might not know in advance the amount of data that will be generated, and  one of the weakest conditions that we can impose in those cases is that the scheme is optimal after many repetitions. If the experiment happens to produce a low amount of data, our protocol may not be optimal, but at least we will always be certain that its performance in the long run will not break. A useful analogy is to imagine a function $f(x)$ for which the only known piece of information is that $f(x) \rightarrow a$ as $x \rightarrow \infty$. In general the limit cannot select a unique solution, but it will constrain the search of $f(x)$ to some extent. Similarly, we can think of the Cram\'{e}r-Rao bound as an asymptotic guide for the quantum strategy to be employed in the absence of a better solution even when $\mu$ is low.

\subsection{Experimental configuration and prior knowledge}
\label{experiment_prior}

Consider an experiment where a system described by $\rho(\theta) = \mathrm{e}^{-i K\theta}\rho_0 \mathrm{e}^{i K\theta}$ is measured with a scheme that is optimal with respect to the quantum Cram\'{e}r-Rao bound (i.e., where the classical and quantum Fisher information coincide), and that this configuration is summarised in $p(\boldsymbol{m}|\theta)=\prod_{i=1}^\mu\mathrm{Tr}[E(m_i)\rho(\theta)]$, with $\theta$ unknown. In this section we discuss our method to select a prior that is suitable for this arrangement and compatible with the idea of using the asymptotic regime as a guide.

For many practical cases such as those that we will consider, it is reasonable to assume that we know a priori that the parameter is localised somewhere within a domain of width $W_0$, and that this domain is centred around the value $\bar{\theta}$, a state of information that can be represented by the uniform density
\begin{equation}
p(\theta) = 1/W_0,~\mathrm{for}~\theta \in [\bar{\theta}-W_0/2, \bar{\theta}+W_0/2],
\label{prior_probability}
\end{equation}
and $p(\theta)=0$ otherwise\footnote{It could be argued that a more realistic way of capturing this state of knowledge is to use a probability function such as $p(\theta)=\gamma \hspace{0.1em} \mathrm{exp}\left[-(\theta-\bar{\theta})^{2\gamma}\right]/\Gamma[1/(2\gamma)]$, which is a box-like Gaussian density with a flat peak \cite{braunstein_gaussian1992}. Nevertheless, the idealisation in equation (\ref{prior_probability}) suffices for our purposes, since, as we discussed in section \ref{sec:uncertainty}, in general the use of the square error as the measure of uncertainty is already an approximation.}. Importantly, we have seen that $W_0$ must be sufficiently small to guarantee that the square error is an appropriate deviation function for the estimation of periodic parameters, and this constraint on $W_0$ implies that, in general, the prior knowledge represented in equation (\ref{prior_probability}) will be moderate. A first rough estimate of this threshold is $W_0 \leqslant \pi$, as our calculation in appendix \ref{prior_sinapprox_appendix} for the schemes of this chapter shows. This estimate will be refined in chapter \ref{chap:limited}. 

We may justify equation (\ref{prior_probability}) from first principles with Jaynes's method of transformation groups \cite{jaynes1968, jaynes2003, toussaint2011}, whose key idea is to express the consequences of the propositions that constitute our prior information $I_0$ as mathematical statements and to impose these conditions to construct the density $p(\theta)$. To see how this is possible, first we notice that the periodic nature of the magnitude for an optical phase implies that in principle $0 \leqslant \theta < 2\pi$. If we are completely ignorant about such magnitude, then our state of information does not change when we rotate the phase by some arbitrary angle, so that it can be treated as a location parameter. Formally, this means that our state of knowledge is invariant under the transformation $\theta \rightarrow \theta' = \theta + c$, for some constant $c$ and taking it to be modulo $2\pi$. As a result, the problems associated with the estimation of $\theta$ and $\theta'$ are equivalent, which amounts to imposing that $p(\theta) d\theta = p(\theta') d\theta'=p(\theta + c) d\theta$ \cite{jaynes2003}, that is, $p(\theta) = p(\theta + c)$. This functional equation is satisfied when $p(\theta) \propto 1$, and upon its normalisation we conclude that $p(\theta) = 1/(2\pi)$ for $0 \leqslant \theta < 2\pi$, which is precisely the probability measure employed by authors such as Helstrom \cite{helstrom1976} and Holevo \cite{holevo2011} for the study of an angular variable in the absence of prior knowledge.

Now we observe that, as Jaynes notes in \cite{jaynes2003}, a transformation group is an idealisation that can only be approximate in real-world problems. However, we may consider the argument above as valid for some region of the parameter domain where the invariance is approximately satisfied. If that region is of width $W_0$, then we recover the prior in equation (\ref{prior_probability}). 

Although this argument is conceptually appealing and a flat prior simplifies the calculations, other authors have successfully employed different prior densities in the context of phase estimation\footnote{For example, a prior emerging from a diffusive evolution offers an elegant transition from a high amount of prior information to a state of complete ignorance \cite{demkowicz2011}, and a popular choice is to use a Gaussian prior \cite{macieszczak2014bayesian, friis2017}, which according to the principle of maximum entropy \cite{jaynes2003} amounts to assuming that the prior knowledge is given as the first two moments of some prior density.}. Moreover, there are other techniques to choose a prior probability from formal arguments \cite{kass1996}, and we could also imagine that our experiment was previously carried by a different team and that we have a summary of their findings encoded in $p(\theta)$. However, this malleable character of the prior probability is not a problem for us, since we can always treat $W_0$ and $\bar{\theta}$ as referring to the parameter domain itself and take other prior measures on that region.

While the values for $W_0$ and $\bar{\theta}$ will be given in practice by the prior information about $\theta$, our discussion in the previous section suggests that $W_0$ has to fulfil certain requirements associated with the likelihood model if the scheme that we intend to implement is to be useful. In particular, we have seen that the likelihood function needs to be concentrated around its highest peak in order to be able to use the approximation $\bar{\epsilon}_{\mathrm{mse}} \approx 1/(\mu F_q)$. This local behaviour implies that, for a given scheme, the width of the parameter domain must be such that the solution to the problem $\partial p(\boldsymbol{m}|\theta)/\partial \theta = 0$ includes an asymptotically unique absolute maximum. Hence, we introduce the quantity $W_{\mathrm{int}}$, which we call \emph{intrinsic width}, and we define it as the width that fulfils the above criterion on average. Moreover, the prior probability should not modify the information of the likelihood in the region where it becomes narrower, which is already satisfied by equation (\ref{prior_probability}).

We will see that different states are associated with a different $W_{\mathrm{int}}$, and, as a consequence, only those states with a value for $W_{\mathrm{int}}$ that is greater than or equal to the width imposed by the experiment would be useful in a real scenario. In fact, if $W_0 > W_{\mathrm{int}}$, then the experiment cannot distinguish between two or more equally likely values, and the error tends to a constant when $\mu \gg 1$\footnote{This is an example of how the asymptotic approximation might fail, and new cases will arise when we extend these ideas to multi-parameter scenarios in chapter \ref{chap:networks}.}. When $W_0 = W_{\mathrm{int}}$, we will refer to equation (\ref{prior_probability}) as the \emph{intrinsic prior} of our particular strategy.

To find $W_{\mathrm{int}}$ we can plot the posterior probability $p(\theta| \boldsymbol{m})$ as a function of $\theta$ directly, since its relative extremes coincide with those of the likelihood when the prior is flat. This procedure depends on the simulation of several random outcomes $\boldsymbol{m}$ for different values of the parameter, and thus the solution is necessarily probabilistic. However, this is enough for our purposes because our analysis only requires that this is satisfied in the asymptotic regime, where $\mu$ is large. Furthermore, we will see that in some simple cases it is possible to relate the conclusions extracted from the numerical study with the analysis of the symmetries of the likelihood.

Recalling that the use of probability densities is a shorthand for
\begin{equation}
P(d\theta, d\boldsymbol{m} | I_0) = p(\theta)\prod_{i=1}^\mu\mathrm{Tr}[E(m_i)\rho(\theta)] d\theta d\boldsymbol{m},
\end{equation}
(see chapter \ref{chap:conceptual}), as a final observation we notice that we can think of the method in this section as if we had introduced an extra piece of information in $I_0$ indicating that the experiment is well designed, in the sense that it can give potentially partial but unambiguous information about $\theta$.  

\subsection{Asymptotic approximation threshold}
\label{subsec:asymsatu}

Once we have a method to choose the prior probability, the next step is to devise a procedure that allows us to identify the boundary between the asymptotic regime, which as we have seen is dominated by the experimental data, and the non-asymptotic regime, where the prior information plays a central role. This can be achieved by introducing the relative error 
\begin{equation}
\varepsilon_\tau = \frac{|\bar{\epsilon}_{\mathrm{mse}}(\mu_{\tau}) - \bar{\epsilon}_{\mathrm{cr}}(\mu_{\tau})|}{\epsilon_{\mathrm{mse}}(\mu_{\tau})} = \frac{|\bar{\epsilon}_{\mathrm{mse}}(\mu_{\tau}) - 1/(\mu_{\tau} F_q)|}{\epsilon_{\mathrm{mse}}(\mu_{\tau})},
\label{saturation}
\end{equation}
for $\bar{\epsilon}_{\mathrm{mse}} \neq 0$, which is a simple but effective way of quantifying the deviation of the quantum Cram\'{e}r-Rao bound $\bar{\epsilon}_{\mathrm{cr}} = 1/(\mu F_q)$ with respect to the exact Bayesian error $\bar{\epsilon}_{\mathrm{mse}}$ in equation (\ref{erropt}). Equation (\ref{saturation}) will give us the minimum number of observations $\mu_{\tau}$ that is needed such that the approximation $\bar{\epsilon}_{\mathrm{mse}} \approx 1/(\mu F_q)$ is valid for a given threshold $\varepsilon_{\tau}$, and such threshold needs to be chosen according to the requirements of the specific experimental configuration that is being analysed.

\subsection{Numerical mean square error}
\label{subsec:numalgorithm}

The final ingredient is an algorithm for the exact calculation of our central quantity, that is, the mean square error in equation (\ref{erropt}), where the optimal estimator $g(\boldsymbol{m}) = \int d\theta p(\theta|\boldsymbol{m})\theta$ has already been selected. Since this integral has $(\mu+1)$ dimensions and we are interested in studying its behaviour as $\mu$ increases, in general we can only compute it numerically. Although this is a purely numerical problem that is well known in the Bayesian literature \cite{jaynes2003, rafal2015} and can be treated with standard numerical techniques \cite{mathematics2004, numerics2014matlab}, we would like to highlight our specific calculation scheme because it has proven to be relatively straightforward, very efficient for a reasonable number of trials and robust against small variations of several numerical parameters\footnote{We have found that the average run-time of our algorithm using a standard workstation is no more than two days for any graph of $\bar{\epsilon}_{\mathrm{mse}}$ in any of the figures of this thesis. The single-parameter version of the algorithm can be found in appendix \ref{sec:msematlab}.}. In particular, we have followed a three-step method:
\begin{enumerate}
\item We sample a collection of $\mu$ experimental outcomes $\boldsymbol{m}$ from $p(m|\theta')$, and we use them to update the prior in equation (\ref{prior_probability}) via Bayes theorem until we have generated the posterior probability  $p(\theta|\boldsymbol{m})$, so that we can calculate its variance in equation (\ref{errbayes}) and obtain $\epsilon(\boldsymbol{m})$. The integral that defines this quantity can be calculated with a standard deterministic method. 

\item The previous $\mu$-trial process is repeated many times and the variances emerging from them are averaged. Therefore, by virtue of the law of large numbers we can construct the quantity
\begin{equation}
\epsilon(\theta') = \int d\boldsymbol{m}p(\boldsymbol{m}|\theta')\epsilon(\boldsymbol{m}).
\label{simulation}
\end{equation}
This is an instance of the Monte Carlo techniques employed for multidimensional integrals \cite{mathematics2004, numerics2014matlab}. 

\item By approximating the domain of $p(\theta')$ with a discrete counterpart, implementing the first two steps for each value of $\theta'$ and taking the average 
\begin{eqnarray}
\int d\theta' p(\theta')\epsilon(\theta')=\bar{\epsilon}_{\mathrm{mse}}
\label{finalmse}
\end{eqnarray}
weighted by $p(\theta')$ we finally arrive at the mean square error. For the integral over $\theta'$ we can again use a deterministic numerical method.
\end{enumerate}

Surprisingly, the decomposition of the mean square error that provides us with an efficient algorithm parallels in a perfect way our discussion about different measures of uncertainty in section \ref{sec:uncertainty}, which highlights a remarkable connection between the design of the algorithm and the physical principles that are relevant for the present situation. The numerical details, as well as the code to implement it in MATLAB, are provided in appendix \ref{sec:msematlab}. 

\section{Our methodology in action: results and discussion} 
\label{results}

The methodology that we have described is general enough to accommodate a wide range of estimation problems, and in this section we explore its application to phase estimation in optical interferometry \cite{rafal2015, yurke1986}. 

\subsection{Common states in two-mode interferometry}
\label{subsec:commoninter}

Suppose we are working with the Mach-Zehnder interferometer that we reviewed in chapter \ref{chap:conceptual}, so that the parameter $\theta$ is encoded as a difference of phase shifts by means of the unitary operator $\mathrm{exp}(-i K\theta) = \mathrm{exp}(-i J_z\theta)$, with $J_z = (a_1^{\dagger}a_1 - a_2^{\dagger}a_2)/2$ and where $a_i,a_i^{\dagger}$ are the creation and annihilation operators for the modes $i = 1,2$. Here we focus on a collection of states that together represent techniques commonly employed in optical quantum metrology \cite{rafal2015, sahota2015, PaulProctor2016, jarzyna2012}. Concretely, we consider:
\begin{enumerate}
\item Coherent states 
\begin{equation}
\ket{\psi_0} = U_{\mathrm{BS}} D_1(\alpha) \ket{0, 0} = |\alpha/\sqrt{2},-i\alpha/\sqrt{2}\rangle,
\label{coherentcommon}
\end{equation}
where $D_1(\alpha) = \mathrm{exp}(\alpha a_1^{\dagger} - \alpha^{*}a_1)$ is the displacement operator and we recall that $U_{\mathrm{BS}} = \mathrm{exp}(-i\frac{\pi}{2}J_x)$, with $J_x = (a_1^{\dagger}a_2 + a_2^{\dagger}a_1)/2$, is the 50:50 beam splitter. See appendix \ref{sec:optrans} for the calculation of the second equality in equation (\ref{coherentcommon}).  
\item NOON states
\begin{equation}
\ket{\psi_0} = \frac{1}{\sqrt{2}}(|N, 0\rangle + |0, N\rangle).
\label{noon}
\end{equation}
\item Twin squeezed vacuum 
\begin{equation}
\ket{\psi_0} = S_1(r) S_2(r)|0, 0\rangle = |r,r\rangle,
\end{equation}
where $S_i(r) = \mathrm{exp}\{[r^{*}a_i^2 - r (a_i^{\dagger})^2]/2\}$, for $i = 1, 2$, are squeezing operators. 
\item Squeezed entangled states 
\begin{equation}
\ket{\psi_0} = \mathcal{N} (|r,0\rangle +|0,r\rangle)
\end{equation}
where $\mathcal{N} = \left[2 + 2/\mathrm{cosh}(\abs{r})\right]^{-1/2}$.
\end{enumerate}

Coherent states present no photon correlations, since as Sahota and Quesada \cite{sahota2015} showed, in that case $\mathcal{Q} = \mathcal{J} = 0$\footnote{Note that in this chapter $\mathcal{Q}$ and $\mathcal{J}$ denote the intra-mode and inter-mode correlations for a Mach-Zehnder interferometer that we reviewed in section \ref{subsec:optint}.}, and this means that, according to equation (\ref{fishercorrpathsym}), the quantum Fisher information for these probes is $F_q = \bar{n}$, where $\bar{n} = \langle \psi_0 |(N_1 + N_2)| \psi_0\rangle$ is the total mean number of quanta per trial, $N_i = a_i^\dagger a_i$ and $R = N_1 + N_2$ is our resource operator for the interferometer. As such, we say that their precision is asymptotically given by the \emph{standard quantum limit}. Contrarily, for NOON states we have that $\mathcal{Q} = \bar{n}/2-1$ and $\mathcal{J} = -1$ \cite{sahota2015}, so that $F_q = \bar{n}^2$, and this $o(\bar{n})$ enhancement is commonly denominated \emph{Heisenberg limit}\footnote{While the standard quantum limit and the Heisenberg limit are useful notions, it is important to observe that their definition depends in a crucial way on which measure of uncertainty is chosen \cite{li2018, pezze2015, rafal2015}, and in general there are not unique definitions even when the latter is fixed \cite{braun2018}. Moreover, what we ultimately want is to enhance the overall performance for a given amount of resources, which does not necessarily involve an analysis of the scaling with $\bar{n}$. In fact, in chapter \ref{chap:multibayes} we will see that the scaling with the number of parameters in a multi-parameter protocol is sometimes more relevant than other contributions to the uncertainty. For a discussion about the relative importance of studying the scaling of the error see \cite{PaulProctor2016, braun2018}.} \cite{dowling2008}. Despite the fact that NOON states present both intra-mode and inter-mode correlations, it can be shown that only intra-mode correlations are necessary to achieve a Heisenberg scaling, as the twin squeezed vacuum with $\mathcal{Q} = \bar{n} + 1$, $\mathcal{J} = 0$ and $F_q = \bar{n}^2 + 2\bar{n}$ demonstrates \cite{rafal2015, sahota2015}, although squeezed entangled states, which have both types of correlations, constitute a precision improvement over the previous probes \cite{PaulProctor2016}. Note that although we have selected pure probes for simplicity, the ideas that form the basis for our methods are also applicable to mixed states.

\begin{figure}[t]
\centering
\includegraphics[trim={1.6cm 0.1cm 1.5cm 0.5cm},clip,width=15.5cm]{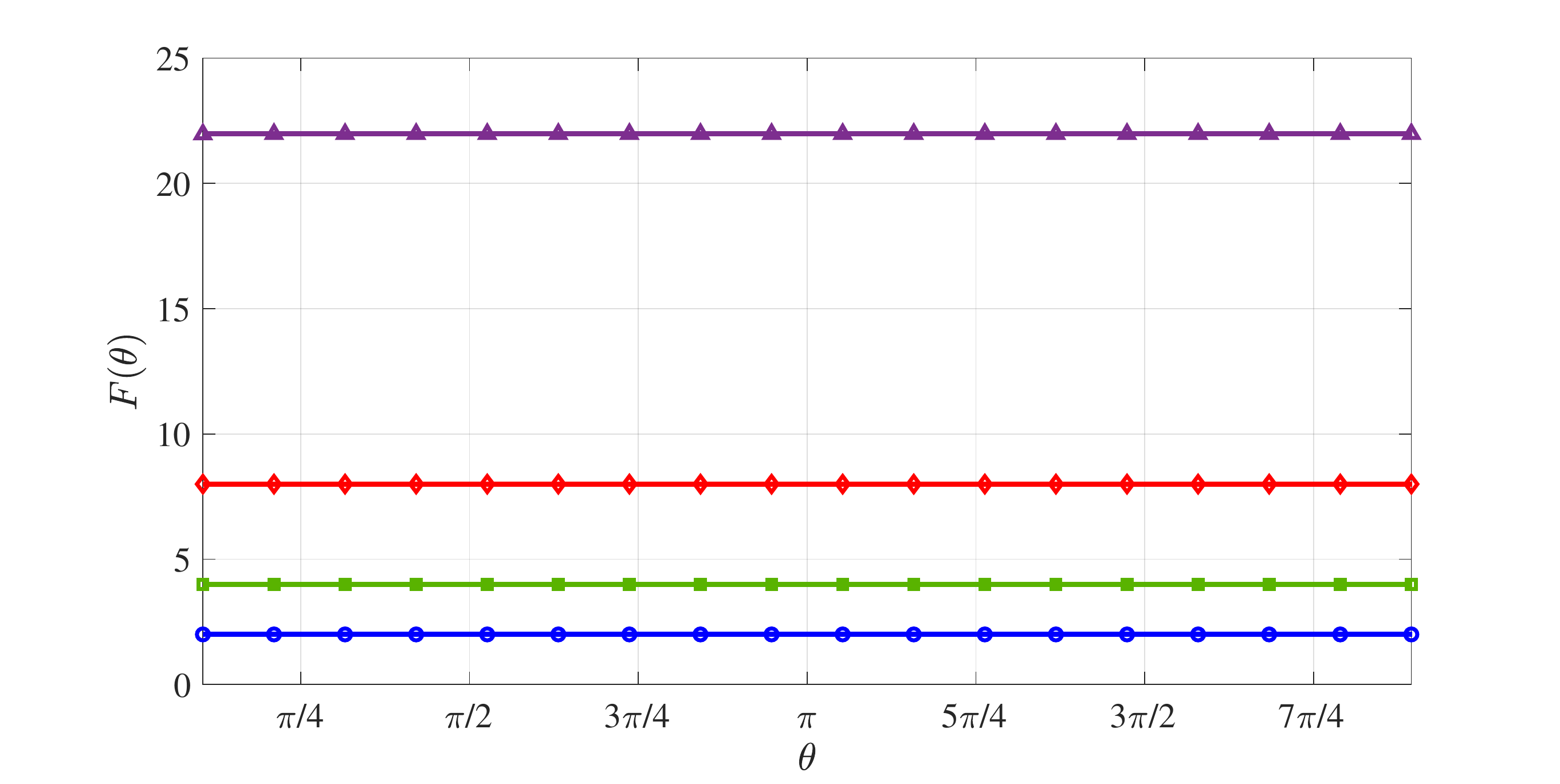}
	\caption[Saturation of the Braunstein-Caves inequality for optical schemes]{Quantum Fisher information (solid lines) and its classical counterpart (symbols) for coherent (circles), NOON (squares), twin squeezed vacuum (diamonds) and squeezed entangled (triangles) states, with $\bar{n} = 2$ and a photon counting measurement that has been implemented after the action of a $50$:$50$ beam splitter $U_{\mathrm{BS}} = \mathrm{exp}(-i\frac{\pi}{2}J_x)$. This numerical calculation illustrates the saturation of the Braunstein-Caves inequality for path-symmetric states in equation (\ref{bcavesine}).}
\label{braunsteincaves}
\end{figure}

A common property of these configurations is that they belong to the family of path-symmetric states that we reviewed in section \ref{subsec:optint}, so that each mode is associated with the same mean number of photons and the same photon number variance. For this class of probes Hofmann showed that \cite{HofmannHolger2009}
\begin{align}
F(\theta) &= \int \frac{dm}{p(m|\theta)}\left[\frac{\partial p(m|\theta)}{\partial \theta}\right]^2 = \mathrm{Tr}\left[\rho(\theta) L(\theta) \right]
\nonumber \\
&= 4 \left(\langle \psi_0 |J_z^2| \psi_0\rangle - \langle \psi_0 |J_z| \psi_0\rangle^2 \right) = F_q
\label{bcavesine}
\end{align}
if we implement a photon-counting measurement after the action of a 50:50 beam splitter, the POM elements of this scheme being
\begin{equation}
\left\lbrace \mathrm{exp}\left(-i\frac{\pi}{2}J_x \right) \ketbra{k}\mathrm{exp}\left(i\frac{\pi}{2}J_x\right) \right\rbrace_k,
\label{photonpom}
\end{equation}
with $N_1\otimes N_2 = \int dk~k\ketbra{k}$. That is, the classical Fisher information for path-symmetric probes reaches the bound imposed by the Braunstein-Caves inequality \cite{BraunsteinCaves1994}. In figure \ref{braunsteincaves} we show an explicit calculation illustrating this fact for $\bar{n} = 2$. 

Crucially, this implies that any discrepancy between the mean square error $\bar{\epsilon}_{\mathrm{mse}}$ in equation (\ref{erropt}) and the Cram\'{e}r-Rao bound $\bar{\epsilon}_{\mathrm{cr}} = 1/(\mu F_q)$ must necessarily come from the asymptotic approximation that we discussed in section \ref{theory}.  

\subsection{Prior information analysis}
\label{subsec:prioranalysis}

\begin{figure}[t]
\centering
\includegraphics[trim={0cm 0.1cm 0cm 0cm},clip,width=7.85cm]{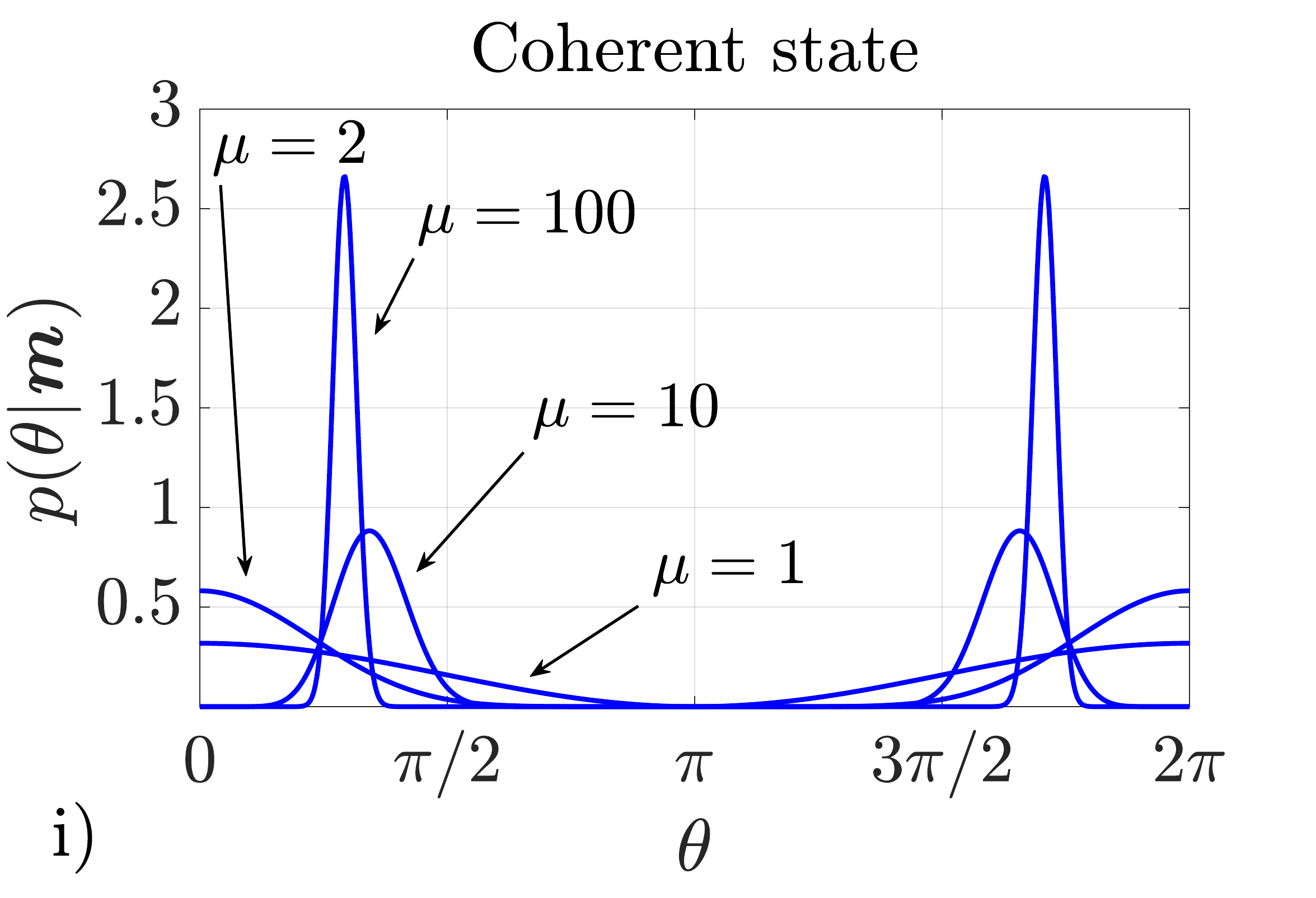}\includegraphics[trim={0cm 0.1cm 0cm 0cm},clip,width=7.85cm]{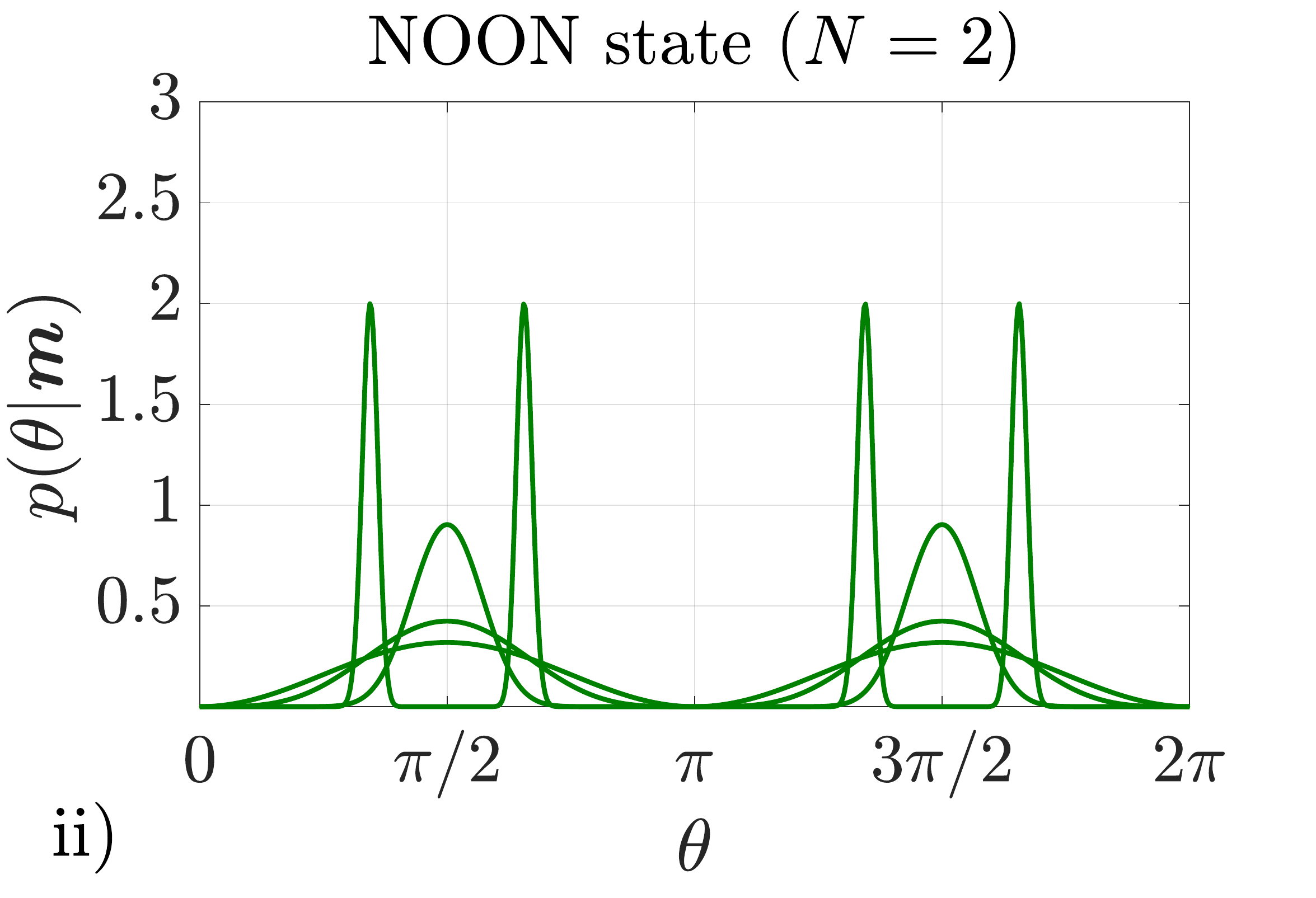}
\includegraphics[trim={0cm 0.1cm 0cm 0cm},clip,width=7.85cm]{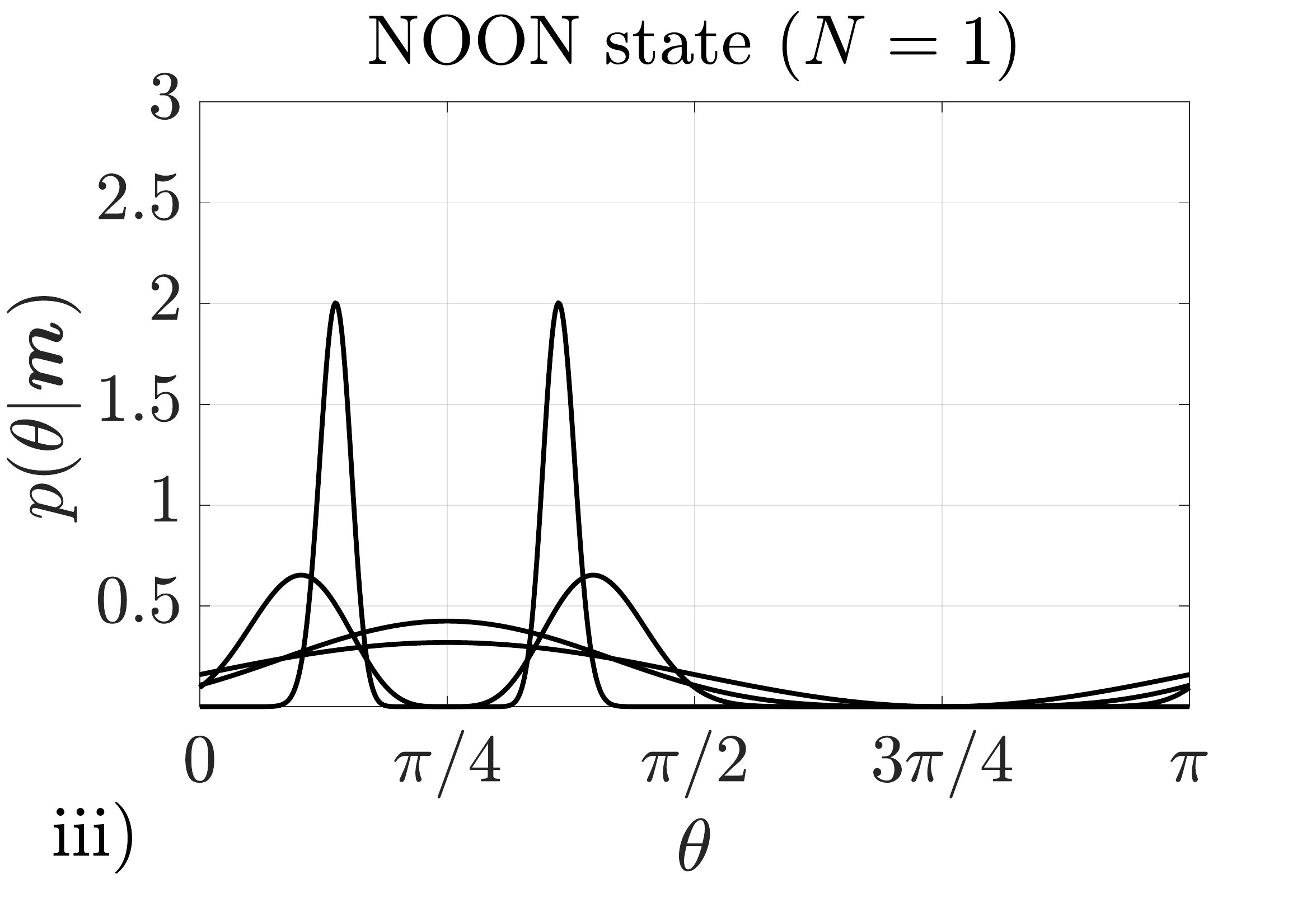}\includegraphics[trim={0cm 0.1cm 0cm 0cm},clip,width=7.85cm]{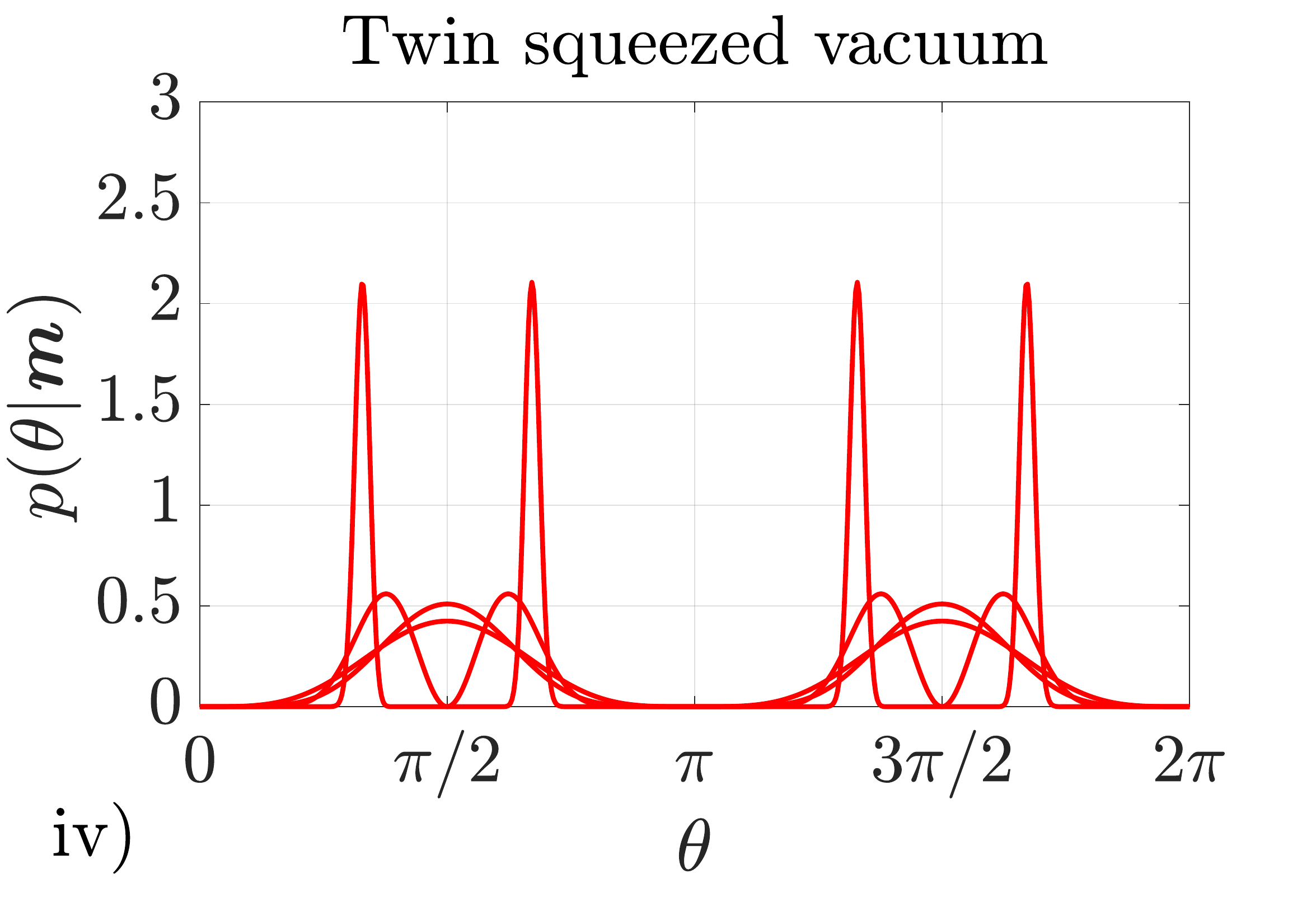}
	\caption[Prior information analysis of a single-parameter scheme]{Posterior density functions for random simulations of 1, 2, 10 and 100 trials, a flat prior and a photon-counting measurement implemented after the action of a 50:50 beam splitter. The initial probes are (i) coherent state with $\bar{n} = 2$, (ii) NOON state with $\bar{n} = 2$, (iii) NOON state with $\bar{n} = 1$, and (iv) twin squeezed vacuum with $\bar{n} = 2$. We draw attention to the fact that these configurations cannot distinguish a unique value  when the initial prior is set to $W_0 = 2\pi$, even if we are in the asymptotic regime with $\mu \gg 1$.}
\label{priornonasymtptotic}
\end{figure}

The first step to apply our numerical strategy is to identify the intrinsic width $W_{\mathrm{int}}$ of each state for a given $\bar{n}$ and the POM in equation (\ref{photonpom}), recalling that we have defined $W_{\mathrm{int}}$ as the largest value that the intrinsic width $W_0$ can take such that the likelihood $p(\boldsymbol{m}|\boldsymbol{\theta})$ is concentrated around a single absolute maximum. Some of the random simulations that are required to achieve this goal are shown in figure \ref{priornonasymtptotic}, which allow us to deduce the size of the maximum width by direct examination. The algorithm employed to generate them can be found in appendix \ref{sec:priormatlab}.

For a twin squeezed vacuum and a squeezed entangled state we have found that $W_{\mathrm{int}} = \pi/2$, while coherent states have $W_{\mathrm{int}} = \pi$. Note that those results hold for any $\bar{n}$. On the contrary, with NOON states we have that $W_{\mathrm{int}} = \pi/\bar{n}$ or $W_{\mathrm{int}} = \pi/(2\bar{n})$ depending on whether the value for $N = \bar{n}$ in equation (\ref{noon}) is even or odd, provided that we choose a prior centred around $\bar{\theta} = W_{\mathrm{int}}/2$. 

The value of $W_{\mathrm{int}}$ for coherent states was also determined in \cite{kolodynski2014} by examining the regions where the single-shot likelihood function $p(m|\theta)$ increases or decreases monotonically as a function of $\theta$. This motivates the search of an alternative way of determining $W_{\mathrm{int}}$ by studying the form of $p(m|\theta)$.

From figure (\ref{priornonasymtptotic}) we observe that the posterior $p(\theta|m) \propto p(m|\theta)$ of our schemes presents two types of symmetry: the periodicity of an imaginary envelope and an axis of symmetry within each period. We can formalise these by imposing
\begin{align}
p(m|\theta) &= p(m|\theta + \mathcal{T}),
\nonumber \\
p(m|\mathcal{S} - \theta) &= p(m|\mathcal{S} + \theta)
\label{symconstr}
\end{align}
when the prior is flat, where $\mathcal{T}$ is the period of the envelope and $\mathcal{S}$ is the position of the axis of symmetry. In addition, a form of the second condition that is more useful in calculations is
\begin{equation}
p(m|\theta) = p(m|2\mathcal{S}-\theta),
\label{easiersym}
\end{equation}
which is found after introducing the change of variables $\theta \rightarrow \mathcal{S} - \theta$.

Let us check whether the previous idea allows us to recover the intrinsic width that our numerical method associates with NOON states. The single-shot likelihood function of this scheme is
\begin{equation}
p(n_1, n_2 | \theta) = || \langle n_1, n_2 | \mathrm{e}^{-i\frac{\pi}{2}J_x} \mathrm{e}^{-i J_z \theta}| \psi_0 \rangle ||^2,
\end{equation}
where $\ket{\psi_0}$ is the NOON state in equation (\ref{noon}) and we have changed the notation as $m \rightarrow (n_1,n_2)$ to make the fact that each port has its own output explicit. A lengthy but straightforward calculation that we include in appendix \ref{sec:optrans} shows that
\begin{equation}
p(n_1, n_2 | \theta) = p(n,N-n|\theta) = \frac{2 N! \hspace{0.2em} \mathrm{cos}^2\left[N\theta/2 + (2n-N)\pi/4\right]}{2^N n! (N-n)!}.
\end{equation}
Introducing this probability in the periodicity condition of equation (\ref{symconstr}), and recalling that $\mathrm{cos}^2(x)=[1+\mathrm{cos}(2x)]/2$, we have that 
\begin{eqnarray}
\mathrm{cos}\left(x_{n,N}\right) = \mathrm{cos}\left(x_{n,N} + N\mathcal{T}\right) = \mathrm{cos}\left(x_{n,N}\right)\mathrm{cos}\left(N \mathcal{T} \right)  - \mathrm{sin}\left(x_{n,N}\right)\mathrm{sin}\left(N \mathcal{T} \right),
\end{eqnarray}
with $x_{n,N} = N\theta + (2n-N)\pi/2$, and this implies that
\begin{equation}
\mathcal{T} = \frac{2\pi k}{N}, ~~ \text{with}~~k=0, \pm 1, \pm 2, \dots.
\label{periodicity}
\end{equation}
Similarly, from the condition in equation (\ref{easiersym}) we find
\begin{align}
\mathrm{cos}\left(x_{n,N}\right) = &~\mathrm{cos}\left[2\mathcal{S}N + \left(2n - N\right)\pi - x_{n, N}\right]
\nonumber \\
= &~\mathrm{cos}\left(x_{n,N}\right)\mathrm{cos}\left[2\mathcal{S}N + \left(2n - N\right)\pi \right]  
\nonumber \\
& + \mathrm{sin}\left(x_{n,N}\right)\mathrm{sin}\left[2\mathcal{S}N + \left(2n - N\right)\pi \right],
\end{align}
so that
\begin{equation}
\mathcal{S} = \frac{\pi\left(k-n\right)}{N} + \frac{\pi}{2},  ~~ \text{with}~~k=0, \pm 1, \pm 2, \dots.
\label{mirrorsymmetry}
\end{equation}

To guarantee that the products of single-shot likelihoods do not generate symmetric absolute maxima we need an interval where the likelihood of a single trial does not contain redundant information. Equation (\ref{periodicity}) means that the width of such interval must be  equal to or less than the period $2\pi/N$. In addition, given that
\begin{equation}
\mathcal{S}(k+1) - \mathcal{S}(k) = \frac{\pi}{N}
\end{equation}
for the points of symmetry in equation (\ref{mirrorsymmetry}), we see that, actually, the width cannot be larger than $\pi/N$. We also notice that the axes of symmetry in equation (\ref{mirrorsymmetry}) only contain the point $\theta = 0$ when $2(n-k) = N$, which can only happen when $N$ is even. If $N$ is odd, then we may find $\theta = 0$ as the middle point between axes, since
\begin{equation}
\mathcal{S} \pm \frac{\pi}{2N} = \frac{\pi\left[2\left(k-n\right) \pm 1 \right]}{2N} + \frac{\pi}{2} = 0 ~~\Rightarrow~~ N \pm 1 =2(n-k),
\end{equation}
and the latter condition can be satisfied for odd $N$. As a consequence, if the parameter domain is $[0, W_\mathrm{int}]$, as it is the case in this chapter, then we conclude that $W_\mathrm{int} = \pi/N = \pi/\bar{n}$ when $\bar{n}$ is even, and $W_\mathrm{int} = \pi/(2N) = \pi/(2\bar{n})$ when $\bar{n}$ is odd, since in the latter case only half of the width free of redundancies is included in the domain. We thus arrive in this way at the same result found in the simulations.

This demonstrates that if the analytical formula for the likelihood function is known, then it is sometimes possible to derive the intrinsic width explicitly by analysing the symmetries of the quantum probability for a single shot. Thus our method complements the previous proposal in \cite{kolodynski2014} based on studying the monotonicity of the likelihood. Moreover, our numerical approach provides the means to find $W_\mathrm{int}$ even if the analytical expression for $p(m|\theta)$ is not available, which is sometimes the situation for more complicated states. This is usually the case, for example, when the quantum circuit is designed using state engineering algorithms \cite{knott2016}.

\begin{figure}[t]
\centering
\includegraphics[trim={0.1cm 0.1cm 0.5cm 0.5cm},clip,width=7.7cm]{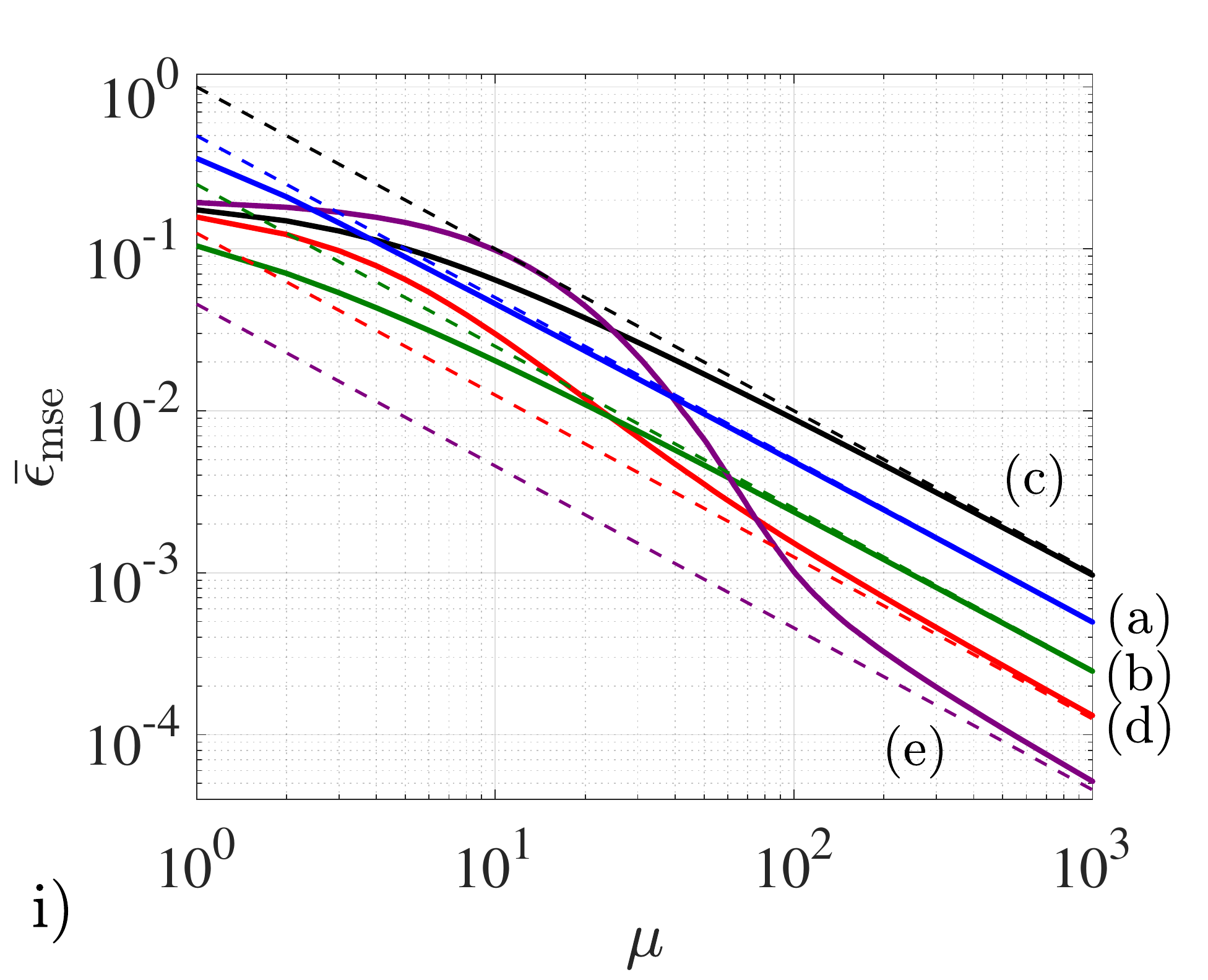}\includegraphics[trim={0.1cm 0.1cm 0.5cm 0.5cm},clip,width=7.7cm]{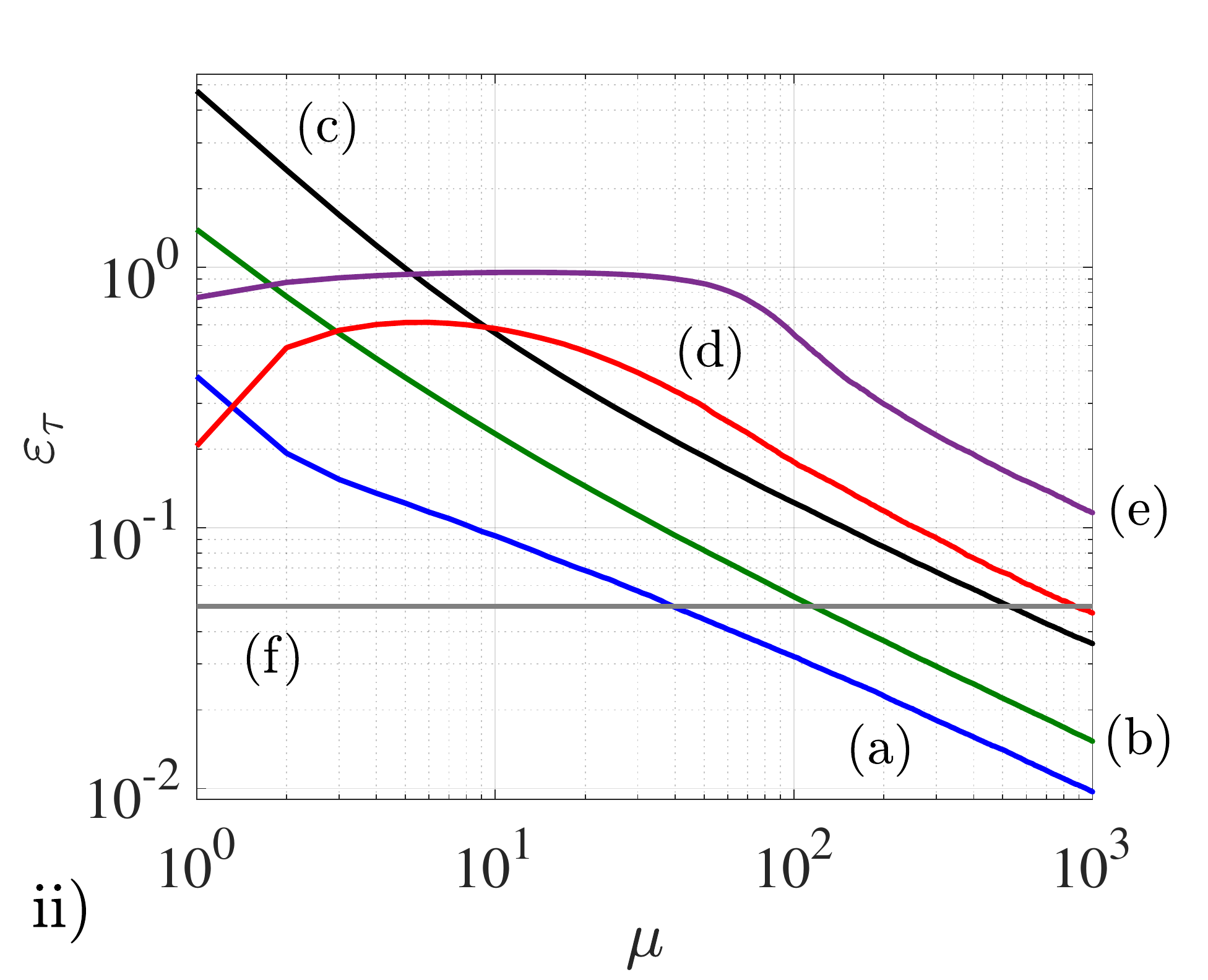} 
\includegraphics[trim={0.1cm 0.1cm 0.5cm 0.5cm},clip,width=7.7cm]{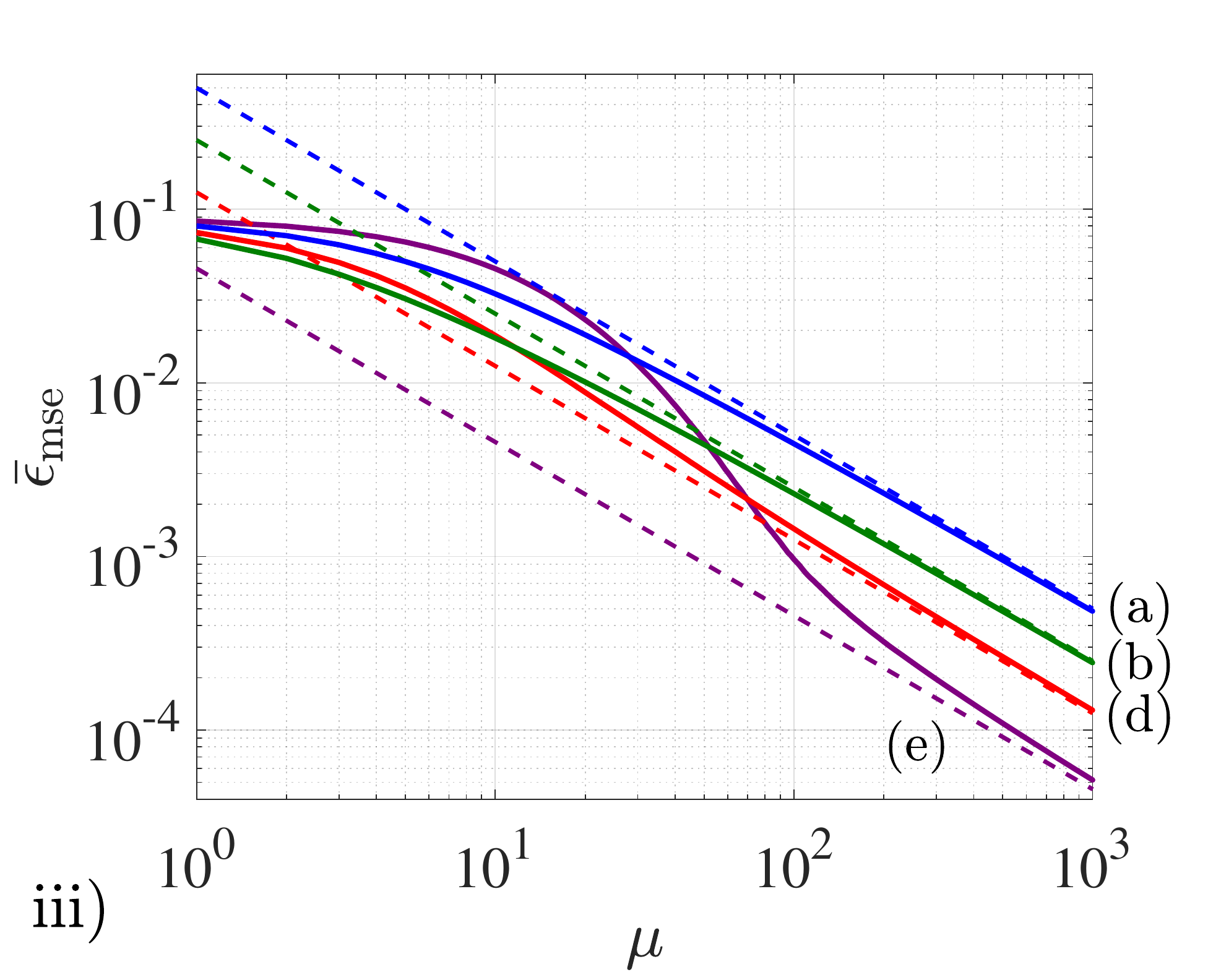}\includegraphics[trim={0.1cm 0.1cm 0.5cm 0.5cm},clip,width=7.7cm]{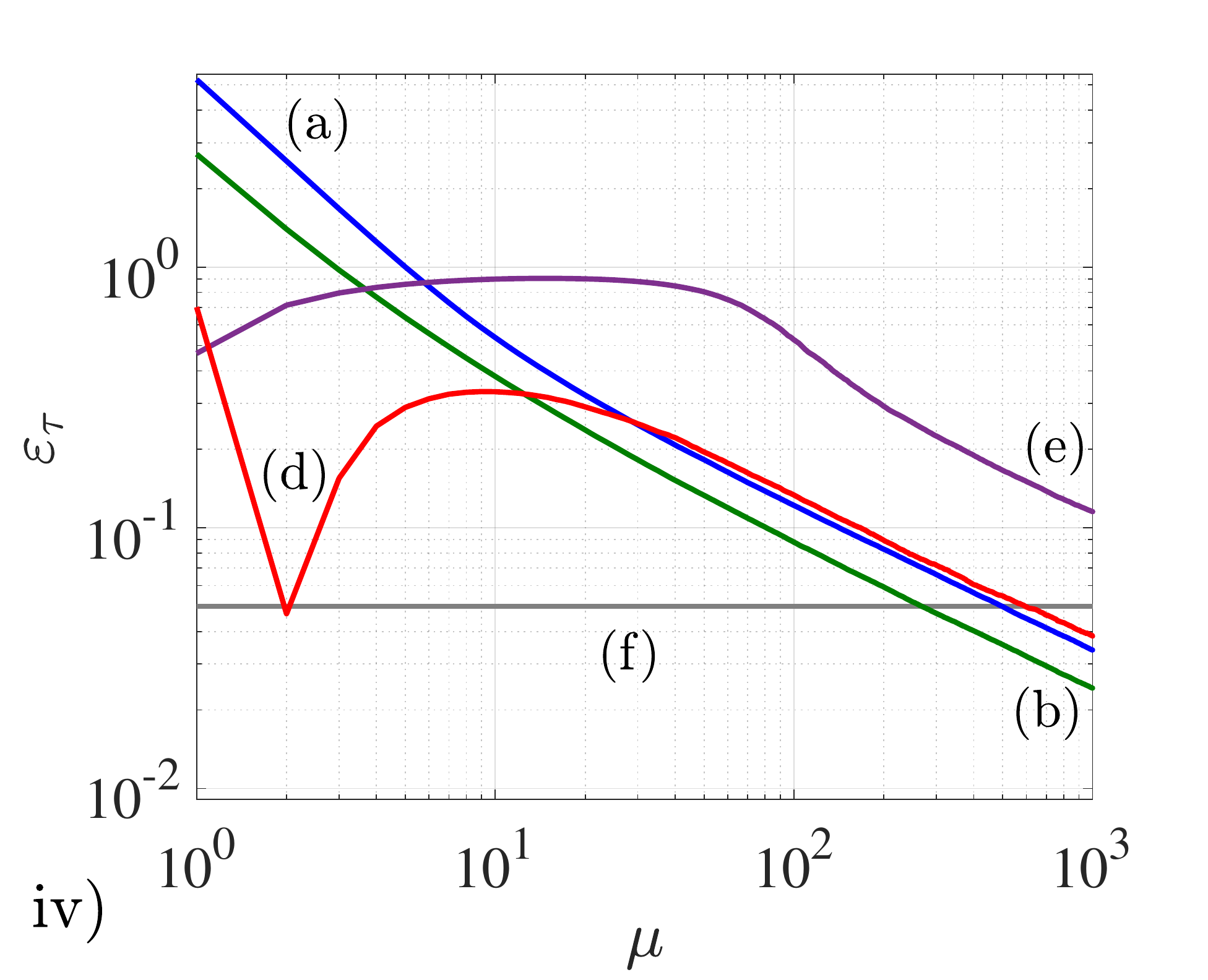}
\caption[Non-asymptotic analysis of the Mach-Zehnder interferometer]{i) Mean square error (solid lines) for the POM in equation (\ref{photonpom}) and quantum Cram\'{e}r-Rao bound (dashed lines) for (a) a coherent state with $\bar{n} = 2$ and $W_{\mathrm{int}} = \pi$, (b) a NOON state with $\bar{n} = 2$ and $W_{\mathrm{int}} = \pi/2$, (c) a NOON state with $\bar{n} = 1$ and $W_{\mathrm{int}} = \pi/2$, (d) a twin squeezed vacuum with $\bar{n} = 2$ and $W_{\mathrm{int}} = \pi/2$, and (e) a squeezed entangled state with $\bar{n} = 2$ and $W_{\mathrm{int}} = \pi/2$, where $\bar{n}$ is the mean number of quanta per trial and $W_{\mathrm{int}}$ is the intrinsic width; (ii) relative error defined by equation (\ref{saturation}) with (f) a threshold $\varepsilon_\tau = 0.05$ for the states considered in figure \ref{mainresult}.i; (iii) repetition of the calculation performed in figure \ref{mainresult}.i with a common prior width $W_0 = \pi/3$ and the same values for $\bar{n}$; and (iv) relative error for the states considered in figure \ref{mainresult}.iii. These results are examined in the main text.}
\label{mainresult}
\end{figure}

The important observation is that none of these states allows us to uniquely identify the relative phase shift when we have no information about its possible values, that is, if $W_0= 2\pi$. We conclude then that the scheme that we are employing introduces some limitations to the estimation protocol, in spite of the fact that the measurement is optimal according to the quantum Cram\'{e}r-Rao bound criterion. 

\subsection{Uncertainty as a function of the number of trials}
\label{subsec:uncertaintynonasymptotic}

Once $W_{\mathrm{int}}$ is known we can use the uniform prior in equation (\ref{prior_probability}) with $W_0 = W_{\mathrm{int}}$ and $\bar{\theta} = W_{\mathrm{int}}/2$ to perform the numerical calculation of the mean square error $\bar{\epsilon}_{\mathrm{mse}}$ in equation (\ref{erropt}), which can be achieved by means of the algorithm described in section \ref{subsec:numalgorithm} (see also appendix \ref{sec:msematlab}). In addition, the algorithm in appendix \ref{subsec:qcrbmatlab} gives the quantum Cram\'{e}r-Rao bound $\bar{\epsilon}_{\mathrm{cr}} = 1/(\mu F_q)$, and the combination of $\bar{\epsilon}_{\mathrm{mse}}$ and $\bar{\epsilon}_{\mathrm{cr}}$ allows us to obtain the relative error $\varepsilon_{\tau}$ in equation (\ref{saturation}). While the explicit form of the optimal estimator $g(\boldsymbol{m}) = \int d\theta p(\theta|\boldsymbol{m})\theta$ will not be provided, note that this is already included within the numerical calculation of $\bar{\epsilon}_{\mathrm{mse}}$.

The results of these operations are shown in figure \ref{mainresult}.i and figure \ref{mainresult}.ii, where we have assumed that the experiment can only be repeated $\mu = 10^3$ times as an extra constraint. For this number of observations, the mean square error of coherent, NOON and twin squeezed vacuum states is close enough to the result predicted by the quantum Cram\'{e}r-Rao bound. In particular, their relative error is smaller than the selected threshold $\varepsilon_\tau = 0.05$. However, the minimum number of observations that are needed in order to reach that threshold is different for different states, and the squeezed entangled state does not even reach it in the regime that we are studying. This state-dependent phenomenon, whose concrete values are indicated in table \ref{table_summary}, has important consequences. 

If we consider first the comparison between a NOON state and a twin squeezed vacuum with $\bar{n} = 2$, $W_{\mathrm{int}} = \pi/2$, we can see that the latter is a better choice according to the Fisher information, but its error is higher for $\mu < 20$. Even if we focus on the results of the asymptotic regime, the twin squeezed vacuum requires $\mu \sim 10^3$ observations to achieve it, while the NOON state only needs $\mu \sim 10^2$. Thus a state whose Fisher information is maximum with respect to other probes can still produce a larger error if the experiment is operating outside of the asymptotic regime. Moreover, although it was shown that only the intra-mode correlations are crucial to surpass the standard quantum limit in the regime where the Fisher approach is valid \cite{sahota2015, sahota2016, proctor2017networked}, this comparison between a NOON state, which includes both types of correlations, and a twin squeezed vacuum, that has intra-mode correlations only, suggests that the role of photon correlations in metrology should be revisited for the non-asymptotic regime. This study will be carried out in detail in section \ref{correlations_section} using a more sophisticated approach.

On the other hand, a coherent state with $\bar{n} = 2$, $W_{\mathrm{int}} = \pi$ is less precise than a NOON state with $\bar{n} = 1$, $W_{\mathrm{int}} = \pi/2$ when $\mu \sim 1$. This implies that there is a region in which a probe with fewer resources can still beat a scheme with more photons if the prior knowledge of the former is higher. By combining these observations with those extracted from the previous probes we conclude that the Cram\'{e}r-Rao bound can both overestimate and underestimate the precision outside of its regime of validity. It is particularly relevant to draw attention to the latter case, since the fact that NOON and coherent states display a mean square error which is lower than their respective Cram\'{e}r-Rao bounds for low values of $\mu$ demonstrates that the unbiased estimators of the local theory are not always optimal.

The analysis of the squeezed entangled state provides further details of the properties of the non-asymptotic regime. In particular, its performance is worse than all the previous cases for $\mu \sim 10$, and it only becomes the best choice when the number of repetitions is greater than $\mu \sim 10^2$. Surprisingly, this result is showing that while states with an indefinite number of photons can do better than the optimal choice for a finite number of quanta, NOON states have the best absolute precision among the cases that we have studied if the number of observations is less than $\mu \sim 10$.

\begin{table} [t]
\centering
{\renewcommand{\arraystretch}{1.05}
\begin{tabular}{|l|c|c|c|c|}
\hline
Probe state & $\bar{n}$ & $W_{\mathrm{int}}$ & $\mu_{\tau} (W_{\mathrm{int}})$ & $\mu_{\tau} (W_0=\pi/3)$\\
\hline
\hline
$|\alpha/\sqrt{2},-i\alpha/\sqrt{2}\rangle$ & $2$ & $\pi$ & $3.9\cdot 10$ & $4.97\cdot 10^2$\\
NOON state (even $N$) & $2$ & $\pi/2$ & $1.15\cdot 10^2$ &$2.67\cdot 10^2$\\
NOON state (odd $N$)& $1$ & $\pi/2$ & $5.26\cdot 10^2$ & - \\
$S_1(r)S_2(r)\ket{0,0}$ & $2$ & $\pi/2$ & $8.74\cdot 10^2$ & $5.95\cdot 10^2$\\
$\mathcal{N}(\ket{r,0}+\ket{0,r})$ & $2$ & $\pi/2$ & $>10^3$ &$ >10^3$\\
\hline
\end{tabular}}
\caption[Conditions to reach the asymptotic regime]{Numerical values of $W_{\mathrm{int}}$ and $\mu_{\tau}$ obtained in figure \ref{priornonasymtptotic} and figure \ref{mainresult}, respectively, for an asymptotically optimal strategy and a threshold $\varepsilon_\tau = 0.05$. The representation of the posterior probability $p(\theta|\boldsymbol{m})$ for the squeezed entangled state that provides the value of its intrinsic width was very similar to that of the twin squeezed vacuum, and therefore it has been omitted in figure \ref{priornonasymtptotic} for brevity. In addition, note that we have chosen $\bar{n} = 2$ for most of our schemes in order to detect a significant improvement over the standard quantum limit.}
\label{table_summary}
\end{table}

To have a fairer comparison, we have repeated the calculation with a common width $W_0 = \pi/3$ and $\bar{n} = 2$. Figures \ref{mainresult}.iii and \ref{mainresult}.iv show that, while the numerical values are different, the qualitative conclusions are the same. Nonetheless, there is an important difference given that the prior knowledge is now higher. For the NOON and coherent states, $\mu_{\tau}$ has increased with respect to the previous calculation, since the starting difference between the error and the bound is now greater. On the other hand, for the twin squeezed vacuum there is a point where now the mean square error crosses the Cram\'{e}r-Rao bound before a stable saturation is reached. This happens because for $W_0 = W_{\mathrm{int}}$ the error approached the bound from above, while for $W_0 = \pi/3$ the error begins below the bound and then crosses it to achieve the asymptotic regime from above. This suggests that if we keep increasing our prior information and we make the width of the parameter domain very small, then the number of observations needed to approach the Cram\'{e}r-Rao bound will grow. 

\subsection{A practical relation to prevent \emph{infinite-precision} states}
\label{subsec:infiniteprecision}

It is possible to formalise the previous phenomenon and derive an intuitive and informative relation that detects states that are not well-behaved. Firstly, we note that the uncertainty of an estimation that is made before we perform the experiment is represented by the variance of the prior probability 
\begin{equation}
\bar{\epsilon}_{\mathrm{mse}}({\mu = 0}) = \Delta \theta^2_p = \int d\theta p(\theta) \theta^2 - \left[\int d\theta p(\theta) \theta \right]^2
\end{equation}
that we introduced in section \ref{subsec:originalderivation}, which for the flat density in equation (\ref{prior_probability}) is
\begin{equation}
\Delta \theta^2_p = \left(\bar{\theta}^2 + \frac{W_0^2}{12}\right) - \bar{\theta}^2 = \frac{W_0^2}{12}.
\end{equation}
On the other hand, we know that the precision is given by the Fisher information when $\mu \gg 1$; consequently, an estimation protocol is only worthwhile when
\begin{equation}
\Delta \theta^2_p(\rho) > \frac{1}{\mu(\rho) F_q(\rho)}
\label{criterion}
\end{equation}
is asymptotically satisfied, where we have made explicit the dependence on the state to indicate that the values of $\mu$ and $\Delta \theta^2_p$ guarantee that the Cram\'{e}r-Rao regime can be reached. If equation (\ref{criterion}) were not fulfilled, then the experiment would not be telling us more than what we already knew. By reorganizing the terms we  arrive at
\begin{equation}
\mu(\rho) > \frac{1}{\Delta \theta^2_p(\rho) F_q(\rho)},
\label{fundamental}
\end{equation}
which is a constraint based on practical requirements.

According to equation (\ref{fundamental}), the number of required observations will increase when the Fisher information is fixed and the prior knowledge is improved, which is consistent with the results of figure \ref{mainresult}. Furthermore, we have seen that the prior width cannot be arbitrarily large if we want to employ certain states in an experiment. Thus, if we maximise the Fisher information at the expense of decreasing the maximum prior uncertainty, and the latter phenomenon is faster, then the number of observations will tend to infinity\footnote{It is important to note that equation (\ref{fundamental}) only helps to predict cases where $\mu(\rho)$ grows indefinitely. Any other finite result will constitute a necessary but not sufficient condition that the value of the number of observations needed to reach the asymptotic regime must satisfy.}.

This is precisely the case of the family of one-mode states
\begin{equation}
\ket{\psi_0} = \sqrt{1 - \delta}\ket{0} + \sqrt{\delta}\ket{N/\delta}
\label{infinite}
\end{equation}
that was considered, e.g., in \cite{alfredo2017}, where $0 < \delta < 1$ and $N/\delta$ is an integer. To see it, 
first we perform an analysis of the periodicity associated with the unitary transformation 
\begin{equation}
\ket{\psi(\theta)} = \sqrt{1-\delta}\ket{0} + \sqrt{\delta}\mathrm{e}^{-i\theta N/\delta}\ket{N/\delta}. 
\end{equation}
By imposing $\ket{\psi(\theta)} = \ket{\psi(\theta + \mathcal{T})}$ we find that
\begin{equation}
\mathrm{exp}\left(-i N\theta/\delta\right)=\mathrm{exp}\left(-i N\theta/\delta\right)\mathrm{exp}\left(-i N\mathcal{T}/\delta\right),
\end{equation}
which implies that $\mathcal{T} = 2\pi k\delta/N$, with $k = 0, \pm 1, \pm 2, \dots$. This indicates that $W_{\mathrm{int}} \leqslant 2\pi\delta/N$, so that $\Delta \theta^2_p \leqslant \pi^2 \delta^2/(3N^2)$ when $W_0 = W_\mathrm{int}$. Furthermore, the Fisher information is
\begin{eqnarray}
F_q = 4[ \langle \psi_0 | (a^\dagger a )^2 | \psi_0 \rangle - \langle \psi_0 | a^\dagger a  | \psi_0 \rangle^2] = 4\left(\frac{N^2}{\delta} - N^2 \right) = \frac{4N^2(1-\delta)}{\delta}.
\end{eqnarray}
Hence, from equation (\ref{fundamental}) we have that
\begin{equation}
\mu(\delta) > \frac{3}{4\pi^2 \delta (1- \delta)}.
\label{infinitesolution}
\end{equation}
The Fisher information suggests that we can get an infinite precision in the limit $\delta \rightarrow 0$ for a fixed number of resources per trial $\bar{n} = N$, but equation (\ref{infinitesolution}) shows that this conclusion only holds if the total number of resources is actually infinite, which is consistent with the analyses of sub-Heisenberg strategies in the literature \cite{tsang2012, berry2012infinite, hall2012}. From a physical point of view we conclude that it is not advantageous to use states for which the majority of our resources have to be employed in making our scheme as sensitive as the prior uncertainty that we already had.

\section{Summary of results and conclusions}

The first step of our methodology, which combines the optimal Bayes estimator with a quantum strategy that is asymptotically optimal, has been implemented in a numerical fashion. This process has involved a rigorous analysis of the prior knowledge required by given state and POM and the estimation of the number of repetitions that are needed to reach the asymptotic regime. This has allowed us to explore the limitations of approximating the Bayesian mean square error by the quantum Cram\'{e}r-Rao bound for practical scenarios that are relevant in quantum metrology, to characterise the boundary that separates the asymptotic and non-asymptotic regimes, and to perform a first analysis of the non-asymptotic regime. 
 
We have applied this strategy to coherent, NOON, twin squeezed vacuum and squeezed entangled states for the estimation of phase shifts in optical interferometry, finding that the conditions for approaching the Cram\'{e}r-Rao bound crucially vary with the state of the system once the POM has been fixed. Moreover, we have proposed a simple and practical criterion to detect states that may require an infinite amount of trials before they provide useful information beyond the prior knowledge.

From the results of our simulations we can conclude that maximizing the Fisher information alone is not always enough to find the best precision in general. For instance, while a twin squeezed vacuum outperforms NOON states according to the Fisher information, we have found that this conclusion may not hold when the number of observations is low. Similarly, a squeezed entangled state is asymptotically better than the previous examples, but it is the worst choice for small values of $\mu$ among the schemes that we have examined. In fact, a coherent state with no correlations and a NOON state with less photons per observation outperform it when $\mu \sim 10$. An additional lesson extracted from section \ref{results} is that the role of inter-mode and intra-more correlations and the use of states with an indefinite number of quanta to enhance the precision needs to be revisited in the non-asymptotic regime.

As a consequence, for a real experiment either we need to perform a fully Bayesian analysis or we must estimate explicitly the number of observations that are required to guarantee that we are operating in the asymptotic regime if we want to follow the path of the Fisher information. This practice will improve the quality and fairness of the comparisons between strategies, helping us to understand the fundamental limits of estimation theory and aiding the design of quantum sensing protocols for quantum technologies. A clear demonstration of the latter can be found in \cite{jesus2018dec}
\begin{displayquote}
\emph{Designing quantum experiments with a genetic algorithm}, Rosanna Nichols, Lana Mineh, \underline{Jes\'{u}s Rubio}, Jonathan C. F. Matthews and Paul A. Knott, Quantum Sci. Technol. 4 045012 (2019).
\end{displayquote}
where in collaboration with the University of Nottingham and the University of Bristol we succeeded in combining the methods of this chapter with a genetic algorithm in order to design optical experiments that can be accessed with current technology. This proposal will be examined in section \ref{sec:genetic} once we have completed our methodology for single-parameter estimation protocols. 

Finally, we notice that the contents in this chapter have been published in \cite{jesus2017}
\begin{displayquote}
\emph{Non-asymptotic analysis of quantum metrology protocols beyond the Cram\'{e}r-Rao bound}, \underline{Jes\'{u}s Rubio}, Paul Knott and Jacob Dunningham, J. Phys. Commun. 2 015027 (2018).
\end{displayquote}
\chapter{Quantum metrology in the presence of limited data}
\label{chap:limited}

\section{Goals for the second stage of our methodology}

The results obtained so far have provided us with a first quantitative characterisation of how much information our schemes can extract in the presence of limited data. However, while the approach that we have followed is more general than simply maximising the Fisher information, our method still relies on the Cram\'{e}r-Rao bound to select the quantum strategy that is asymptotically optimal. The main goal of this chapter is to go a step further and construct a strategy where the quantum optimisation is directly performed in the non-asymptotic regime. 

This is precisely where the second version of our methodology in chapter \ref{chap:methodology} enters the scene. The key idea is to find the measurement scheme predicted by the optimal single-shot mean square error in sections \ref{subsec:singleshotparadigm} and \ref{subsec:originalderivation}, and use that measurement in a sequence of repeated experiments. As such, the new method that we propose here combines analytical and numerical techniques, and we will demonstrate its potential using a Mach-Zehnder interferometer that operates in the regime of limited data and moderate prior knowledge. 

We will show that the bounds that arise from this technique are tight and can be approached in principle both for a single shot (by construction) and in the asymptotic regime of many measurements, since the results predicted by the Fisher information are recovered in the latter case. Admittedly, this does not guarantee that our solution will be generally optimal for a few trials (in that case an adaptive scheme could be better than repeating the same measurement). Nevertheless, we will see that having an error that is a function of the number of repetitions where the first point is already tight, and that also tends towards the asymptotically optimal solution as the number of shots grows, is enough to draw conclusions to important questions such as the role of photon number correlations or the performance of experimentally feasible measurements in the non-asymptotic regime. 

For instance, we have found an example where the correlations between the paths of the Mach-Zehnder interferometer appear to be particularly useful in this regime, and we have demonstrated that while measuring quadratures and counting photons after the action of a beam splitter are asymptotically equivalent in an ideal scenario, the former measurement scheme is better for a low number of repeated experiments. In addition, we will show that the combination of both the methods in this chapter and those in chapter \ref{chap:nonasymptotic} with a genetic algorithm allows us to design quantum experiments for engineering optical states that supersede certain benchmarks in the regime that we are studying. These findings may prove to be important in the development of quantum enhanced metrology applications where practical considerations mean that we are limited to a small number of trials.

It is interesting to note that two related approaches were proposed during the development of the work in this chapter. On the one hand, Lumino \emph{et al.} \cite{lumino2017} exploited techniques from machine learning to optimise an experimental implementation with a low number of shots. On the other hand, Mart\'{i}nez-Vargas \emph{et al.} \cite{esteban2017} presented a modification of the quantum van Trees inequality and used it to construct an adaptive strategy based on a parameter-independent single-shot measurement scheme. The experimental nature of Lumino \emph{et al.} \cite{lumino2017} and the adaptive character of Mart\'{i}nez-Vargas \emph{et al.} \cite{esteban2017} are aspects that are not covered here. On the other hand, the advantage of our proposal is its fundamental character, in the sense that it is built on the true optimum for a single shot. In fact, our results can be seen as a non-trivial generalisation with respect to those that are obtained when the Fisher information is used instead. Thus our work and those mentioned above are complementary.

\section{Methodology (part B)}\label{sec:methodb}

\subsection{Shot-by-shot strategy}
\label{subsec:shotbyshot}

Our starting point is the single-parameter configuration exploited in chapter \ref{chap:nonasymptotic}, where we had a quantum probe with statistical properties described by the density matrix $\rho_0$, and an unknown parameter $\theta$ that was encoded in the probe state through the unitary transformation $\rho(\theta) = \mathrm{e}^{-i K\theta}\rho_0 \mathrm{e}^{i K\theta}$. 

The fundamental difference is that now we perform the measurement described by the POM elements $\lbrace \ketbra{s}\rbrace$, which are the eigenstates of the optimal quantum estimator $S = \int ds~s\ketbra{s}$ that satisfies the equation $S\rho + \rho S = 2\bar{\rho}$, with $\rho = \int d\theta p(\theta)\rho(\theta)$, $\bar{\rho} = \int d\theta p(\theta)\rho(\theta)\theta$ and $p(\theta)$ being the prior that is to be updated after a single shot. As we saw in chapter \ref{chap:methodology}, this is the quantum strategy that minimises the single-shot mean square error. 

Furthermore, the prior knowledge is again represented by the uniform density of width $W_0$ and centred around $\bar{\theta}$ that was introduced in equation (\ref{prior_probability}). While in chapter \ref{chap:nonasymptotic} we found that the square error was a reasonable approximation for periodic parameters when $W_0 \leqslant \pi$, a more powerful analysis based on the schemes of this chapter  reveals that a better estimate of that threshold is $W_0 \lesssim 2$\footnote{Note that most schemes in chapter \ref{chap:nonasymptotic} do still fulfil this requirement, the only exception being the coherent state with $W_0 = \pi$.}. The discussion that leads to this conclusion can be found in appendix \ref{prior_sinapprox_appendix}. In addition, in section \ref{prior_section} we will see that the local regime of prior information is not properly recovered until the prior width is $W_0 = 0.1$ or smaller. Hence, we will work in the regime of moderate prior information with $0.1<W_0<2$.

Once we have calculated the projectors of $S$, we proceed to repeat the same optimal experiment $\mu$ times, so that the uncertainty associated with the overall experience is given by
\begin{equation}
\bar{\epsilon}_{\mathrm{mse}} = \int d\boldsymbol{s}~p(\boldsymbol{s}) \left\lbrace \int d\theta p(\theta|\boldsymbol{s}) \theta^2 - \left[\int d\theta p(\theta|\boldsymbol{s}) \theta \right]^2 \right\rbrace.
\label{shotbyshotmse}
\end{equation}
This is the single-parameter error in equation (\ref{erropt}) after having selected the optimal estimator $g(\boldsymbol{s}) = \int d\theta p(\theta|\boldsymbol{s}) \theta$, where in this case the posterior is $p(\theta|\boldsymbol{s})= p(\theta) p(\boldsymbol{s}|\theta)/p(\boldsymbol{s})$, the likelihood is $p(\boldsymbol{s}|\theta) = \prod_{i=1}^\mu \langle s_i | \rho(\theta) |s_i \rangle$ and $p(\boldsymbol{s}) = \int d\theta p(\theta)p(\boldsymbol{s}|\theta)$.   

\begin{figure}
\centering
\includegraphics[trim={0.1cm 9.25cm 0.5cm 0cm},clip,width=15cm]{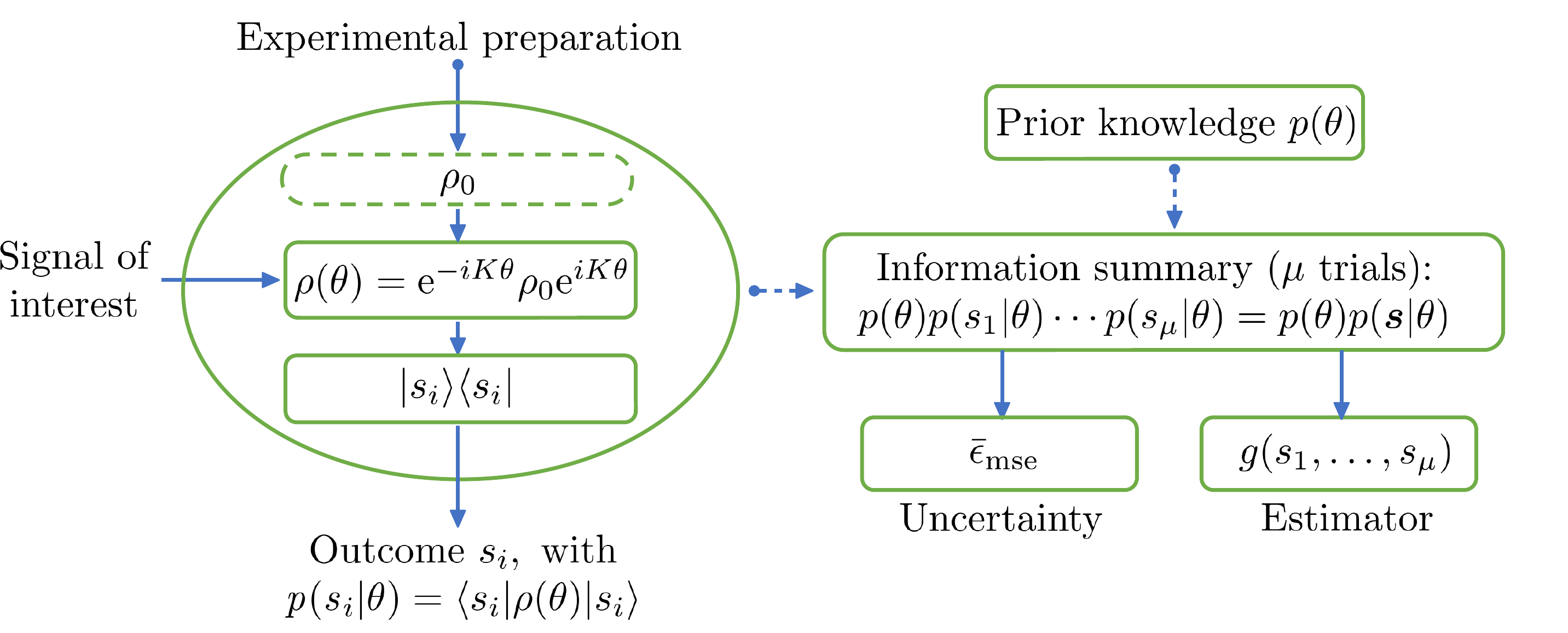}
	\caption[Extraction of information from a quantum sensor]{Representation of the extraction of information from a quantum sensor following the shot-by-shot strategy. This process consists of three stages: preparation of the probe state $\rho_0$, parameter encoding $\mathrm{exp}(-i K\theta)$ and measurement scheme $\ketbra{s_i}$. The statistics of the outcome $s_i$ is given by the Born rule, and the protocol is repeated $\mu$ times. Taking also into account any prior information that we may have we can construct an estimator $g(s_1,\dots,s_\mu)$ as a function of the experimental outcomes, and assess its performance using the measure of uncertainty $\bar{\epsilon}_\mathrm{mse}$.}
\label{singleshotvisual}
\end{figure}

Therefore, this approach combines numerical simulations with the rigorous foundation provided by an analytical and potentially reachable quantum bound. A visual representation of this strategy can be found in figure \ref{singleshotvisual}. 

\subsection{Calculation scheme for the optimal single-shot strategy}
\label{numcal}

The error in equation (\ref{shotbyshotmse}) can be numerically calculated as a function of $\mu$ following the same three-step algorithm that we discussed in section \ref{subsec:numalgorithm}. On the other hand, to obtain the eigendecomposition of the quantum estimator $S$ that gives rise to the POM $\ketbra{s_i}$ we need to solve $S \rho + \rho S = 2 \bar{\rho}$ for given $\rho_0$, $K$ and $p(\theta)$. 

By expanding $\rho$ in the basis of its eigenvectors, that is, $\rho = \sum_i p_i \ketbra{\phi_i}$, and inserting it into $S \rho + \rho S = 2 \bar{\rho}$, we find that
\begin{equation}
S\rho + \rho S - 2\bar{\rho} = \sum_{ij}\left[\left(p_i + p_j\right)\bra{\phi_i}S\ket{\phi_j} - 2\bra{\phi_i} \bar{\rho} \ket{\phi_j} \right]\ketbra{\phi_i}{\phi_j} = 0,
\end{equation}
so that we can formally express the solution for $S$ as
\begin{equation}
S = 2\sum_{ij} \frac{\bra{\phi_i} \bar{\rho} \ket{\phi_j}}{p_i+p_j}\ketbra{\phi_i}{\phi_j}.
\label{quantumestimator_rho}
\end{equation}
Importantly, equation (\ref{quantumestimator_rho}) is only defined on the support of $\rho$, since $S \rho + \rho S = 2 \bar{\rho}$ is a Sylvester equation and, as such, it only has a unique solution in the subspace where the spectra of $\rho$ and $-\rho$ are disjoint (see pages 203, 204 of \cite{bhatia1997}). However, this is not a problem, since the quantity $\mathrm{Tr}(\bar{\rho}S) = \mathrm{Tr}(\rho S^2)$ appearing in the single-shot bound in equation (\ref{singleshot_bound}) only depends on the terms associated with the support of $\rho$, that is, 
\begin{equation}
\mathrm{Tr}(\rho S^2) = \sum_{i} p_i \bra{\phi_i} S^2 \ket{\phi_i} = \sum\limits_{\substack{\lbrace i,\hspace{0.2em} p_i\neq 0 \rbrace}} p_i \bra{\phi_i} S^2 \ket{\phi_i}.
\end{equation}

Unfortunately, the analytical calculation of equation (\ref{quantumestimator_rho}) is challenging for indefinite photon number states, since they belong to a space whose dimension is infinite. For that reason, we have employed a hybrid method where $\rho$ and $\bar{\rho}$ are calculated analytically and $S$ is obtained numerically from equation (\ref{quantumestimator_rho}) with the algorithm in appendix \ref{sec:singleshotalgorithm}. 

To find the analytical expressions of $\rho$ and $\bar{\rho}$, first we recall that the generator for the Mach-Zehnder interferometer that we are analysing is $K = J_z$. By expanding the transformed pure state $\ket{\psi(\theta)} = \mathrm{e}^{-i J_z\theta}\ket{\psi_0}$ in the number basis as $\ket{\psi(\theta)} = \sum_{nm}\mathrm{e}^{-i (n-m)\theta/2}c_{nm}\ket{nm}$, where $c_{nm}$ are the components of the initial state $\ket{\psi_0}$, we can construct the density matrix
\begin{equation}
\rho(\theta) = \ketbra{\psi(\theta)} =\sum_{nmlk} \mathrm{e}^{-i (n-m)\theta/2} \mathrm{e}^{i (k-l)\theta/2} c_{nm}c^{*}_{kl}\ketbra{nm}{kl},
\end{equation}
with $c_{nm}c_{kl}^{*}=\left(\rho_{0}\right)_{nmkl}$. Then, given that $p(\theta)=1/W_0$ when $\theta$ lies between $\bar{\theta} - W_0/2$ and $\bar{\theta} + W_0/2$, we have that
\begin{equation}
\rho = \int d\theta p(\theta) \rho(\theta)= \sum_{nmkl} \mathcal{K}_{nmkl} c_{nm}c^{*}_{kl}\ketbra{nm}{kl}
\label{zeroth_qmoment_code}
\end{equation}
and
\begin{equation}
\bar{\rho} = \int d\theta p(\theta)\rho(\theta) \theta = \sum_{nmkl} \mathcal{L}_{nmkl} c_{nm}c^{*}_{kl}\ketbra{nm}{kl},
\label{first_qmoment_code}
\end{equation}
where
\begin{equation}
\mathcal{K}_{nmkl} = \frac{1}{W_0}\int_{\bar{\theta}-W_0/2}^{\bar{\theta}+W_0/2}d\theta \mathrm{e}^{-ix_{nmkl}\theta/2},
\label{k_semianalytical_int}
\end{equation}
\begin{equation}
\mathcal{L}_{nmlk} = \frac{1}{W_0}\int_{\bar{\theta}-W_0/2}^{\bar{\theta}+W_0/2} d\theta \theta \mathrm{e}^{-ix_{nmkl}\theta/2}
\end{equation}
and $x_{nmkl}=n-m+l-k$. These integrals can be computed directly, finding that
\begin{equation}
\mathcal{K}_{nmkl} = \frac{4}{W_0}\frac{A_{nmkl} B_{nmkl}}{x_{nmkl}},
\label{k_semianalytical_sol}
\end{equation}
and
\begin{equation}
\mathcal{L}_{nmkl} = \frac{2 A_{nmkl}}{x_{nmkl}} \left(\frac{2 B_{nmkl} D_{nmkl}}{W_0} + iC_{nmkl}\right),
\end{equation}
where we have defined
\begin{align}
A_{nmkl}&=\mathrm{exp}\left(-i x_{nmkl}\bar{\theta}/2\right),
\nonumber \\
B_{nmkl}&=\mathrm{sin}\left(x_{nmkl}W_0/4\right),
\nonumber \\
C_{nmkl}&=\mathrm{cos}\left(x_{nmkl}W_0/4\right)~\mathrm{and}~
\nonumber \\
D_{nmkl}&=\bar{\theta}-2 i/x_{nmkl}.
\label{d_semianalytical_auxiliar}
\end{align}
Note that all the elements $\mathcal{K}_{nmkl}$ and $\mathcal{L}_{nmkl}$ are well defined except when $x_{nmkl}$ vanishes, in which case we have an indetermination. In those cases we need to take the limits
\begin{equation}
\underset{x_{nmlk}\rightarrow 0}{\mathrm{lim}} \mathcal{K}_{nmkl}=1 ,~~\underset{x_{nmkl}\rightarrow 0}{\mathrm{lim}} \mathcal{L}_{nmkl}=\bar{\theta}.
\label{singularcases}
\end{equation}
Since $\mathcal{K}_{nmkl}$, $\mathcal{L}_{nmkl}$ and $c_{nm}c^{*}_{kl}$ can be seen as $(nm \times kl)$ matrices, we can finally rewrite equations (\ref{zeroth_qmoment_code}) and (\ref{first_qmoment_code}) as $\rho = \rho_0 \circ \mathcal{K}$ and $\rho = \rho_0 \circ \mathcal{L}$, where we are using the entrywise product of matrices defined as $X \circ Y = \sum_{ij} X_{ij} Y_{ij}\ketbra{i}{j}$ \cite{horn1985}.

\section{Our methodology in action: results and discussion}
\label{sec:methodlimited}

\subsection{Highly-sensitive states in two-mode interferometry}
\label{employed_states}

Previously we saw that the coherent state $|\alpha/\sqrt{2},-i\alpha/\sqrt{2}\rangle$ is a natural benchmark to evaluate the enhancement derived from quantum resources such as entanglement or squeezing, while the NOON state $(\ket{N,0} + \ket{0,N})/\sqrt{2}$ is an intuitive example of a definite photon number state that reaches the Heisenberg limit \cite{dowling2008} when enough prior knowledge is available (see \cite{berry2012infinite, hall2012} and its analysis in section \ref{subsec:prioranalysis}). This justifies their use in this chapter mainly as a reference, although we will also highlight those features related to the regime of limited data\footnote{Other aspects of these two states have been extensively studied in previous works. See, e.g., \cite{kolodynski2014, jarzyna2016thesis, hall2012, berry2012infinite}.}.

The principal analysis will instead be dedicated to states that are experimentally feasible and whose quantum Fisher information is large with respect to the two previous benchmarks; i.e., we wish to optimise the non-asymptotic regime of states with a great sensitivity. According to the work by Knott \emph{et al.} \cite{PaulProctor2016}, this is precisely the case of the other two probes that we examined in chapter \ref{chap:nonasymptotic}: the twin squeezed vacuum state $\ket{r,r}=S_1(r)S_2(r)\ket{0,0}$, where $S_i(r) = \mathrm{exp}\lbrace[r^{*}a_i^2-r(a_i^{\dagger})^2]/2\rbrace$, and the squeezed entangled state $\mathcal{N}_{ses}\left(\ket{r,0}+\ket{0,r}\right)$, where $\mathcal{N}_{ses} = [2+2/\mathrm{cosh}(|r|)]^{-1/2}$. Additionally, this is also true for the twin squeezed cat state $\mathcal{N}_{tscs}\left[S(r)\left(\ket{\alpha}+\ket{-\alpha}\right)\right]^{\otimes 2}$, with $\mathcal{N}_{tscs}=(2+2\mathrm{exp}(-2|\alpha|^2)^{-1/2}$ and $\ket{\alpha} = D(\alpha)\ket{0}$. 

In order to have a fair comparison, the parameters that define the previous states have been chosen such that, on average, all the strategies utilise the same amount of resources (see the third column in table \ref{tablesummary}). In particular, $\bar{n} = \bra{\psi_0} R \ket{\psi_0} = \bra{\psi_0} (N_1 + N_2) \ket{\psi_0} = 2$ for all $\ket{\psi_0}$. This energy constraint fixes the parameters of all the states except those of the twin squeezed cat state; the parameters of the latter case will be chosen such that the quantum Fisher information is maximum in all the sections of this work except in sections \ref{correlations_section} and \ref{prior_section}, where we also consider an intermediate scenario. Note that the fact that $\bar{n} = 2$ for all our protocols implies that we are working in the low photon number regime \cite{PaulProctor2016}.

\begin{sidewaystable}
\renewcommand{\baselinestretch}{1.0}
\centering
{\renewcommand{\arraystretch}{1.2}
\begin{tabular}{|l|c|c|c|c|c|c|}
\hline
Probe state & $\ket{\psi_0}$  & State parameters  & $\mathcal{Q}$ & $\mathcal{J}$ & $F_q$ & $\mu_{\tau} (\rho)$\\
\hline
\hline
Twin squeezed vacuum state & $S_1(r)S_2(r)\ket{0,0}$ & $r=\mathrm{asinh}\left( 1 \right)$ & $3$ & $0$ & $8$ & $5$ \\ 
\begin{tabular}{@{}l@{}}Twin squeezed cat state (intermediate) \\ Twin squeezed cat state (optimal)\end{tabular} & $\mathcal{N}_{\mathrm{tscs}}\left[S(r)\left(\ket{\alpha}+\ket{-\alpha}\right)\right]^{\otimes 2}$& \begin{tabular}{@{}c@{}}$r=1.103$, $\alpha=1.090$ \\ $r=1.215$, $\alpha=0.9601$\end{tabular} & \begin{tabular}{@{}c@{}}$10.00$ \\ $11.75$\end{tabular}& \begin{tabular}{@{}c@{}}$0$ \\ $0$\end{tabular} & \begin{tabular}{@{}c@{}} $22.00$ \\ $25.49$\end{tabular} & \begin{tabular}{@{}c@{}}$42$ \\ $66$\end{tabular} \\
Squeezed entangled state & $\mathcal{N}_{\mathrm{ses}}\left(\ket{r,0}+\ket{0,r}\right)$ & $r=\log\left(2+\sqrt{3}\right)$ & $9$ & $-0.1$ & $22$ & $45$ \\ 
NOON state & $\left(\ket{N 0}+\ket{0 N}\right)/\sqrt{2}$& $N=2$ &$0$ & $-1$ & $4$ & $116$ \\
Coherent state & $|\alpha/\sqrt{2},-i\alpha/\sqrt{2}\rangle$ & $\alpha = \sqrt{2}$ & $0$ & $0$ & $2$ & $282$ \\ 
\hline
\end{tabular}}
\caption[Properties of several probe states for optical interferometry]{Properties of the probe states considered in the main text. The state parameters have been chosen such that the mean number of photons is $\bar{n} = 2$. Furthermore, $\mathcal{Q}$ and $\mathcal{J}$ represent the amount of intra-mode and inter-mode correlations in the interferometer defined in \cite{sahota2015} and section \ref{subsec:optint}. Finally, $\mu_{\tau}(\rho)$ indicates the state-dependent number of repetitions that are required for the quantum Cram\'{e}r-Rao bound to be a good approximation to the bounds based on the optimal single-shot strategy in figure \ref{bounds_results}, according to the methodology discussed in chapter \ref{chap:nonasymptotic} with relative error $\varepsilon_\tau = 0.05$, prior mean $\bar{\theta}=0$ and prior width $W_0 = \pi/2$. Note that $\rho$ includes the information of the initial probe, the encoding of the signal and the prior knowledge.}
\label{tablesummary}
~\\[15pt]
\centering
{\renewcommand{\arraystretch}{1.2}
\begin{tabular}{|l|c|c|c|c|c|}
\hline
\diagbox{Probe state ~~~~~~~~~~}{$\mu\cdot\bar{\epsilon}_{\mathrm{mse}}(\mu, W_0)$} & $\mu=1$, $W_0=\pi/2$ & $\mu=1$, $W_0=\pi/3$ & $\mu=1$, $W_0=\pi/4$ & $\mu=1$, $W_0=0.1$ & $\mu \gg 1$ \\
\hline
\hline 
Twin squeezed vacuum state & $9.93\cdot 10^{-2}$ & $5.83\cdot 10^{-2}$ & $3.81\cdot 10^{-2}$ & $8.28\cdot 10^{-4}$ & $1.25\cdot 10^{-1}$  \\ 
Twin squeezed cat state (intermediate) & $1.50\cdot 10^{-1}$ & $6.48\cdot 10^{-2}$ & $3.61\cdot 10^{-2}$ & $8.19\cdot 10^{-4}$ & $4.55\cdot 10^{-2}$  \\ 
Twin squeezed cat state (optimal) & $1.42\cdot 10^{-1}$ & $7.10\cdot 10^{-2}$ & $4.11\cdot 10^{-2}$ & $8.17\cdot 10^{-4}$ & $3.92\cdot 10^{-2}$  \\ 
Squeezed entangled state & $1.12\cdot 10^{-1}$ & $5.61\cdot 10^{-2}$ & $3.47\cdot 10^{-2}$ & $8.19\cdot 10^{-4}$ & $4.55\cdot 10^{-2}$   \\ 
NOON state & $1.04\cdot 10^{-1}$ & $6.47\cdot 10^{-2}$ & $4.21\cdot 10^{-2}$ & $8.31\cdot 10^{-4}$ & $2.5\cdot 10^{-1}$  \\
Coherent state & $1.44\cdot 10^{-1}$ & $7.71\cdot 10^{-2}$ & $4.66\cdot 10^{-2}$ & $8.33\cdot 10^{-4}$ & $5\cdot 10^{-1}$   \\ 
\hline
\end{tabular}}
\caption[Local regimes: single-shot with informative prior, and asymptotic]{Optimal single-shot mean square error with different prior widths for the states considered in the main text and asymptotic performance for $\mu\gg1$. The state parameters are those indicated in table \ref{tablesummary}. We notice that the asymptotic ordering of probe states and the ordering for $W_0=0.1$ and a single shot are identical, which implies that the local regime is achieved for such a prior width.}
\label{prior_effect_summary}
\end{sidewaystable}

Finally, we will assume that the prior width is $W_0 = \pi/2 < 2$ and that the prior mean is $\bar{\theta} = 0$. The former is consistent with our findings in chapter \ref{chap:nonasymptotic}, while the latter is justified by the fact that, as we will see, the fundamental bounds generated by the method in section \ref{sec:methodb} do no depend on $\bar{\theta}$, and $\bar{\theta} = 0$ is more natural than $\bar{\theta} = W_0/2$ in optical interferometry \cite{demkowicz2011}. Moreover, the results that arise from $\bar{\theta} = 0$ and those associated with $\bar{\theta} = W_0/2$ can be related with controllable phase shifts as part of the POM. This will be demonstrated in section \ref{measurements_section}.

\subsection{Quantum bounds in the presence of limited data}
\label{main_results}

\begin{figure}[t]
\centering
\begin{tabular}{l l}
\begin{tabular}{@{}c@{}}\includegraphics[trim={0.2cm 0.1cm 1cm 1.2cm},clip,width=9.65cm]{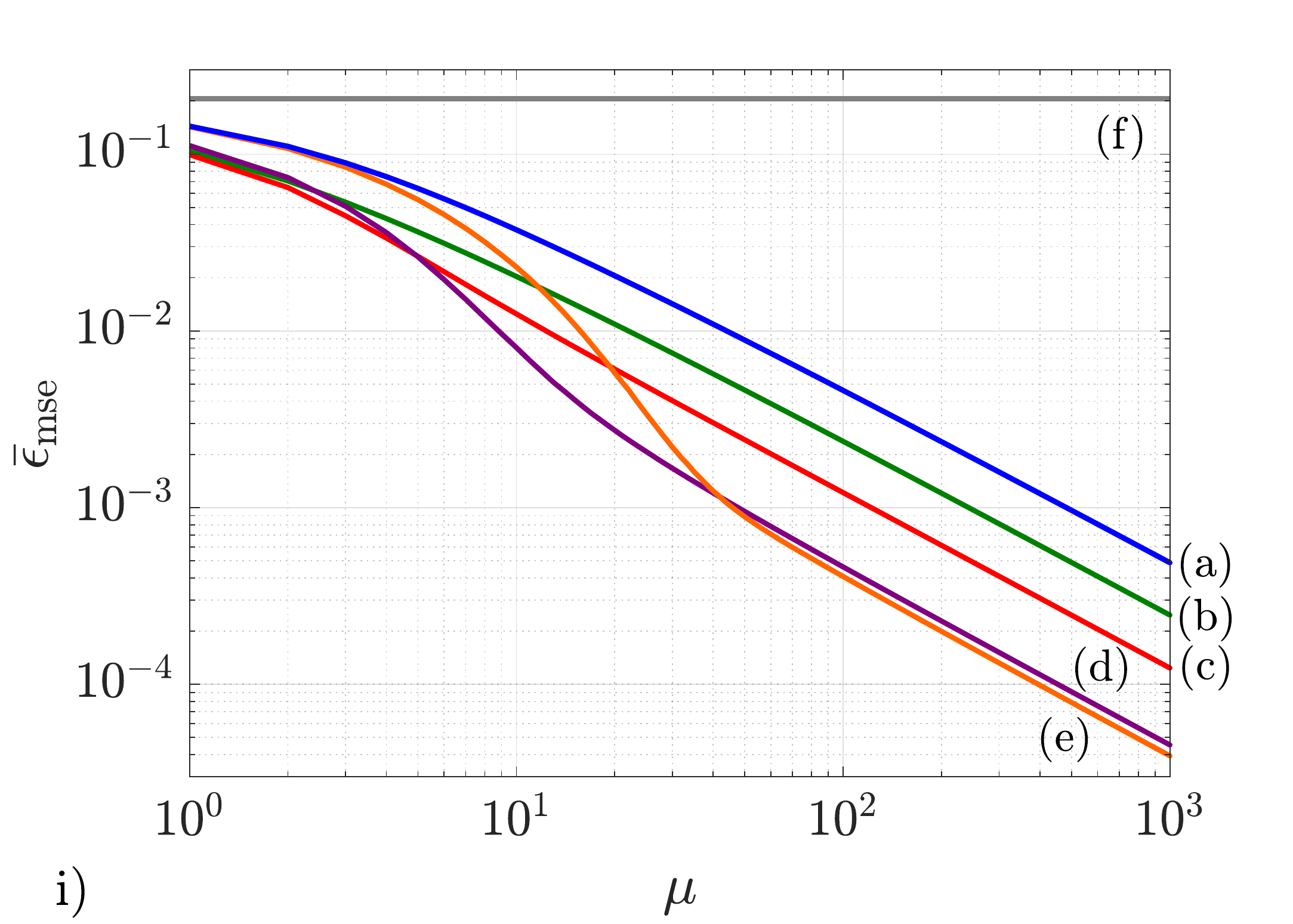}\end{tabular} & \begin{tabular}{@{}c@{}} \includegraphics[trim={0.35cm 0cm 1.1cm 1cm},clip,width=4.85cm]{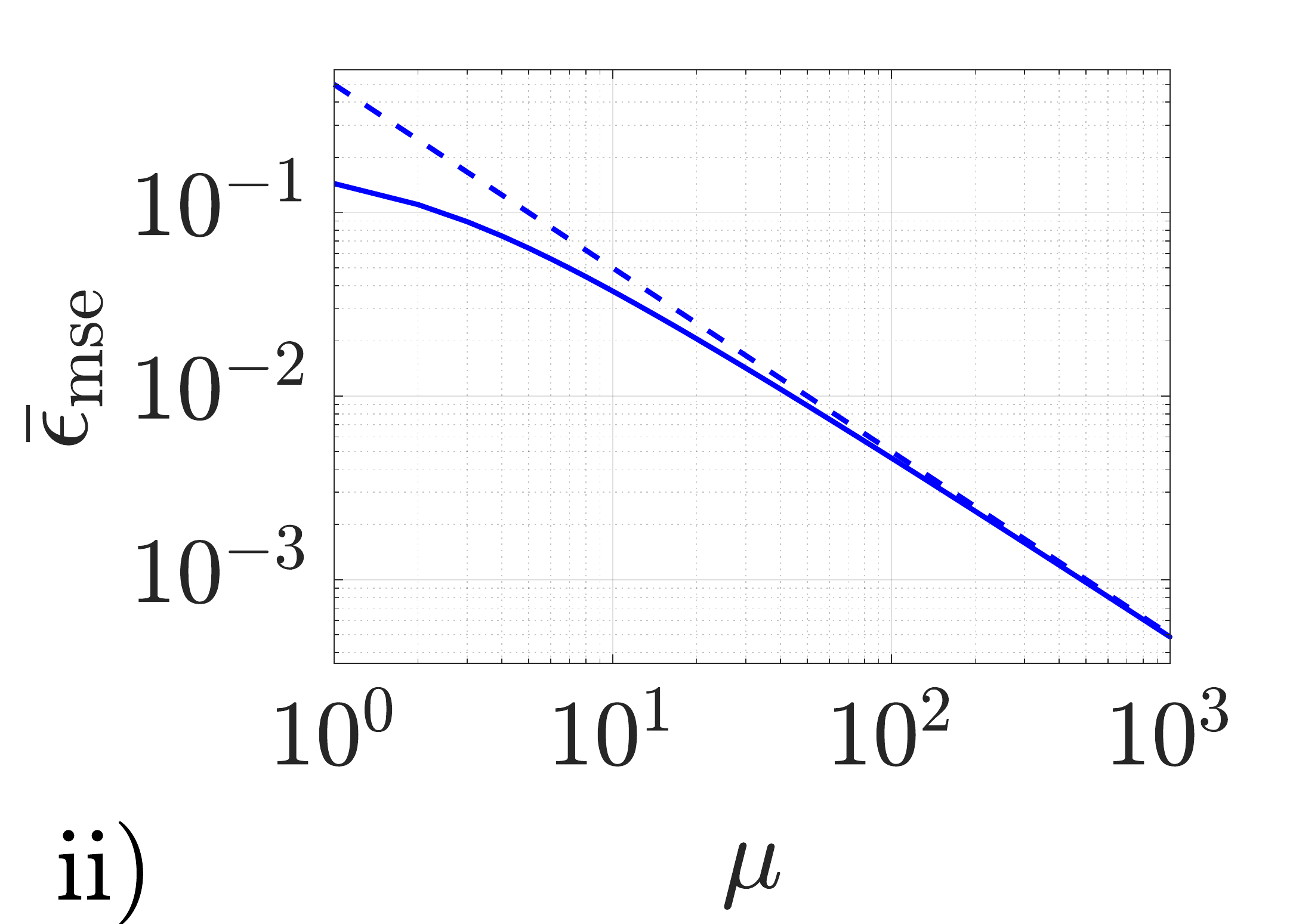} \\ \includegraphics[trim={0.35cm 0cm 1.1cm 1cm},clip,width=4.85cm]{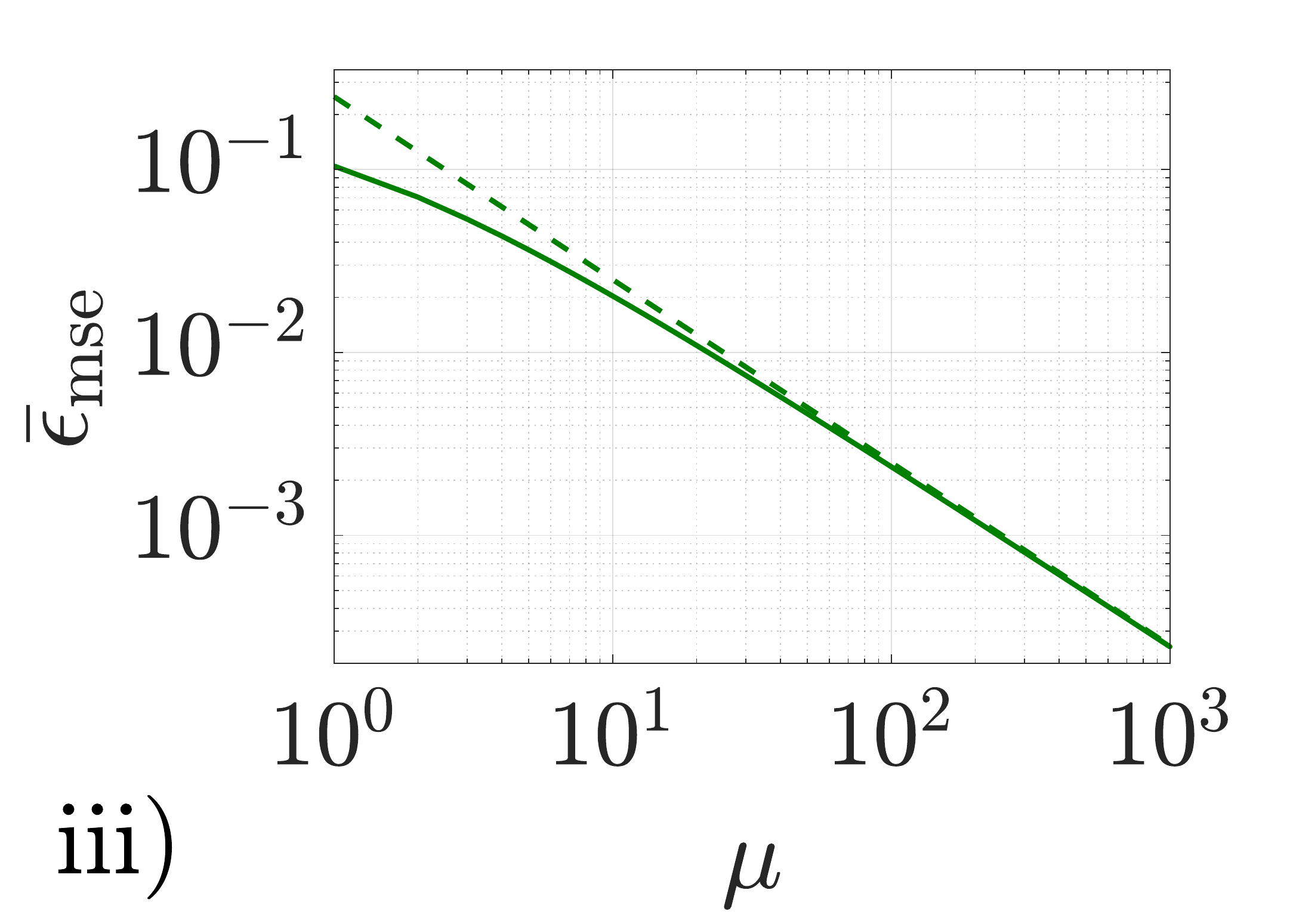}\end{tabular} \\[0pt]
\includegraphics[trim={0.35cm 0cm 1.1cm 1cm},clip,width=4.85cm]{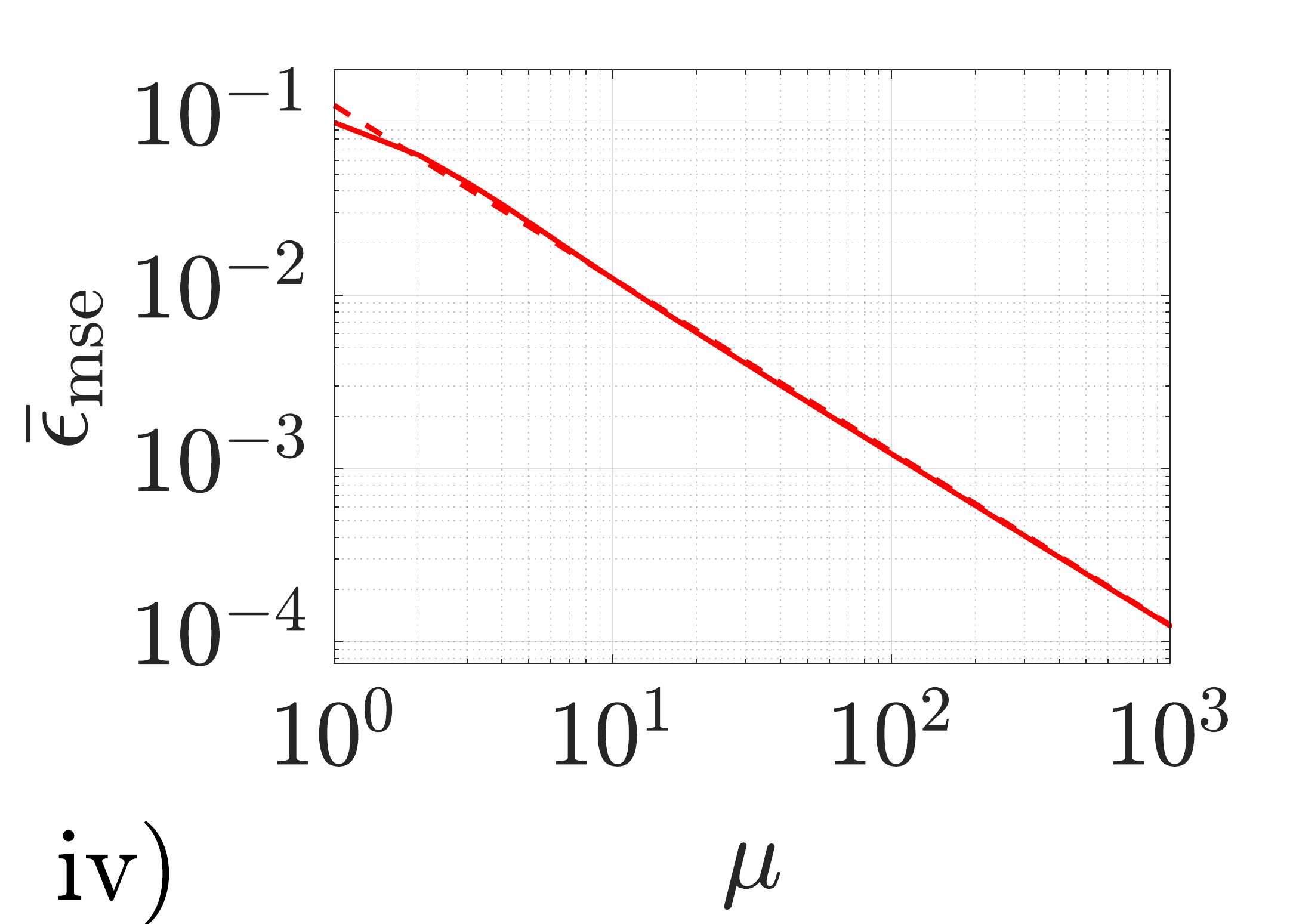}\includegraphics[trim={0.35cm 0cm 1.1cm 1cm},clip,width=4.85cm]{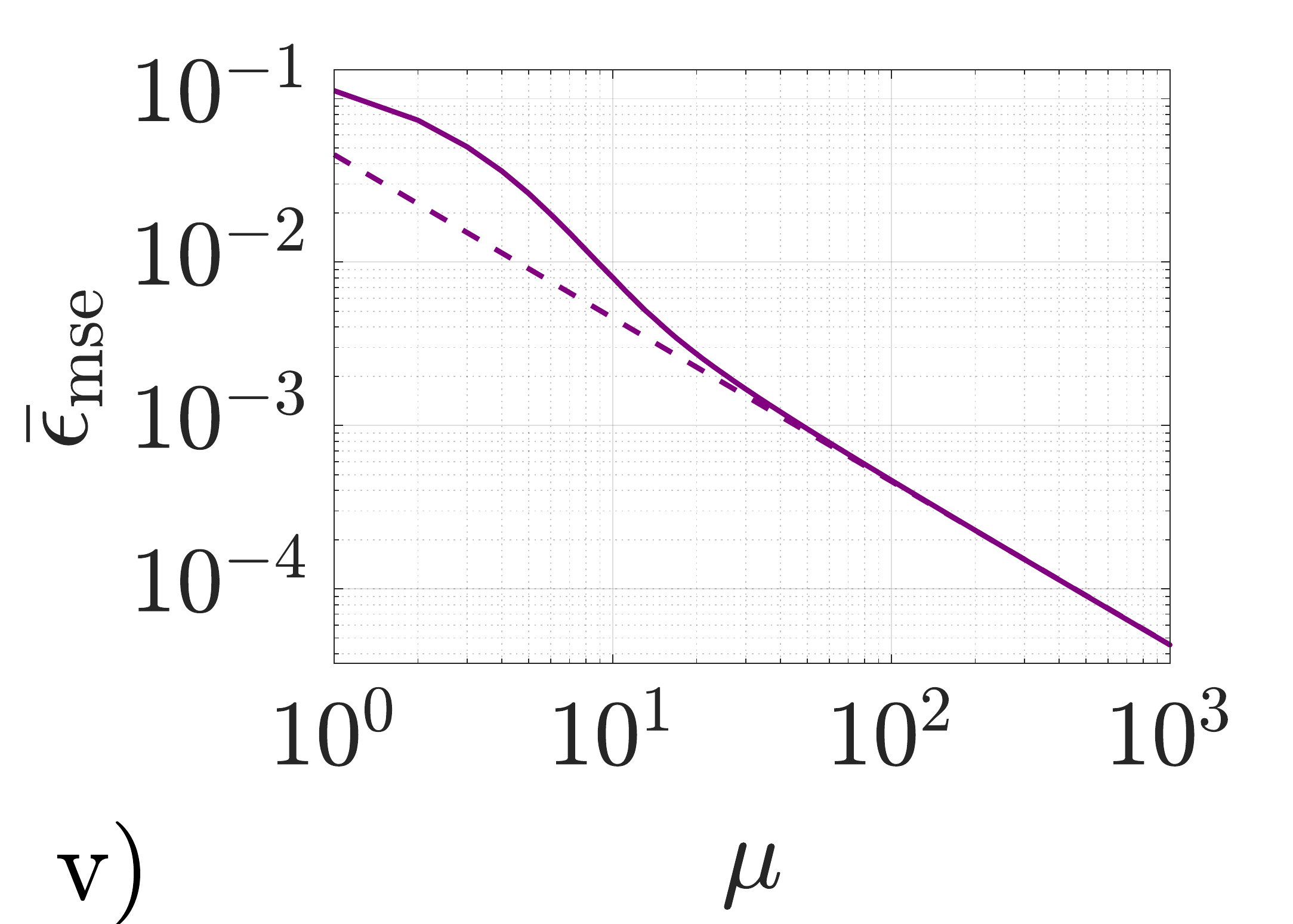} & \includegraphics[trim={0.35cm 0cm 1.1cm 1cm},clip,width=4.85cm]{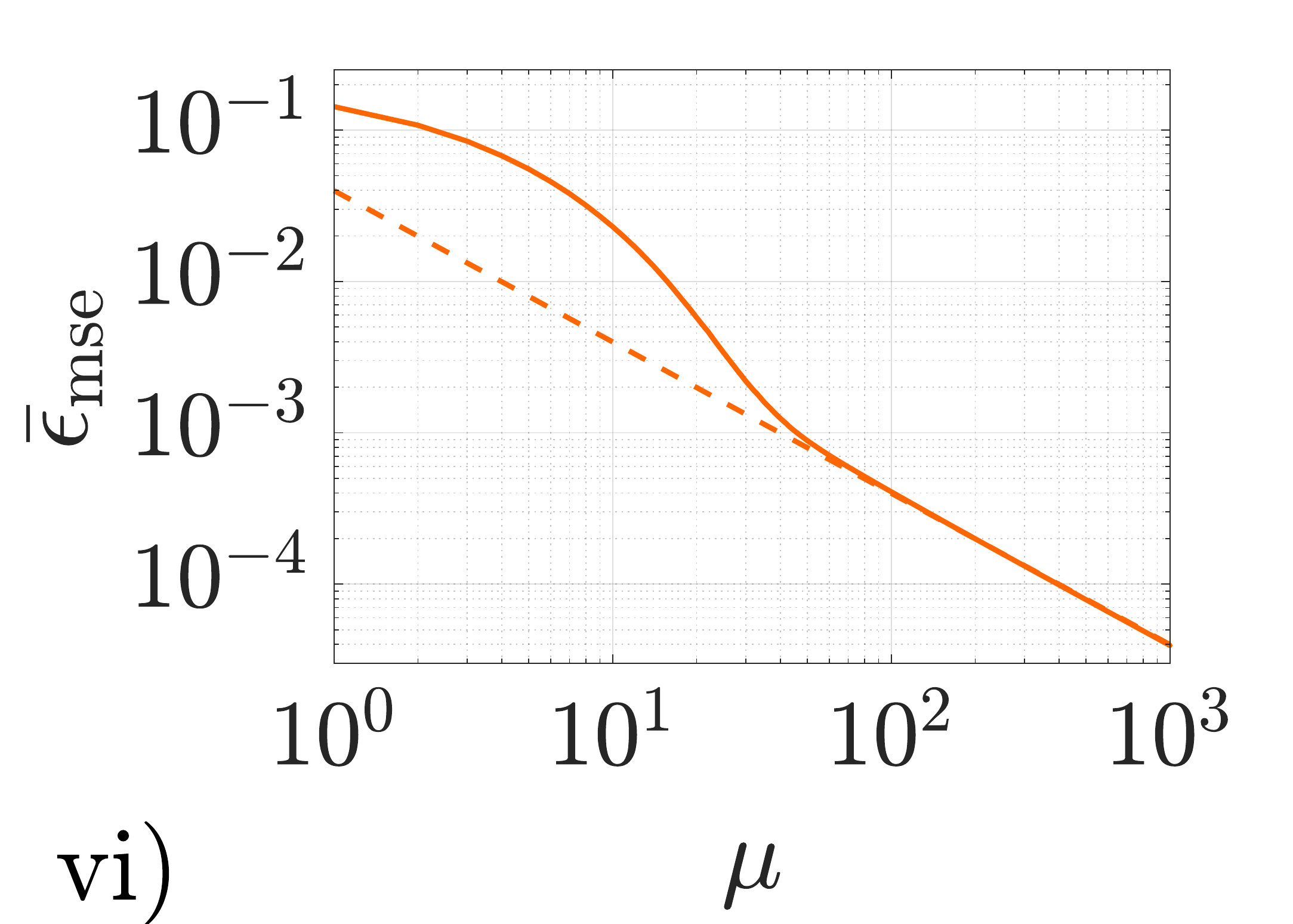}  \\[0pt]
\end{tabular}
\caption[Shot-by-shot quantum bounds for the Mach-Zehnder interferometer]{i) Mean square error as a function of the number of repetitions using the optimal single-shot strategy for (a) the coherent state, (b) the NOON state, (c) the twin squeezed vacuum state, (d) the squeezed entangled state, and (e) the twin squeezed cat state, with mean number of photons $\bar{n}=2$, prior mean $\bar{\theta} = 0$ and prior width $W_0 = \pi/2$, while (f) represents the variance of the prior probability; (ii) mean square error based on the optimal single-shot strategy (solid line) and quantum Cram\'{e}r-Rao bound (dashed line) for the same coherent state, (iii) NOON state, (iv) twin squeezed vacuum state, (v) squeezed entangled state and (vi) twin squeezed cat state considered in (i). These graphs constitute the main results of section \ref{main_results}, and their consequences are analysed in the main text.}
\label{bounds_results}
\end{figure}

The application of the method described in section \ref{sec:methodb} to interferometric configurations leads to the results shown in figure \ref{bounds_results}.i, where the mean square error in equation (\ref{shotbyshotmse}) is plotted as a function of the number of repetitions for the optical probes in section \ref{employed_states}: (a) coherent state, (b) NOON state, (c) twin squeezed vacuum state, (d) squeezed entangled state and (e) twin squeezed cat state. Let us proceed to analyse the consequences of these graphs.

To start with, figure \ref{bounds_results}.i presents two different regimes. On the one hand, the performance of all the states becomes linear with the number of repetitions in the logarithmic scale when $\mu \gtrsim 10^2$. This is precisely the behaviour that we would expect in the asymptotic regime $\mu \gg 1$, since in that case the mean square error can be approximated by the Cram\'{e}r-Rao bound as $\bar{\epsilon}_{\mathrm{mse}}\approx 1/(\mu F)$, and as such $\mathrm{log}(\bar{\epsilon}_{\mathrm{mse}}) \approx - \mathrm{log}(\mu) - \mathrm{log}(F)$. In this regime we can observe that the graphs of different states do not intersect each other. This property allows us to identify the twin squeezed cat state as the best asymptotic choice, followed by the squeezed entangled state, the twin squeezed vacuum state, the NOON state and, finally, the coherent state, whose performance is the worst. We notice that this is consistent with the findings in \cite{PaulProctor2016}.

On the other hand, the graphs deviate from this logarithmic linear approximation when $1 \leqslant \mu \lesssim 10^2$ and, as a consequence, a non-trivial structure emerges in this part of the plot. This is the non-asymptotic regime of limited data for the schemes in this chapter. Since the graphs no longer follow straight lines, they intersect each other, and this implies that the ordering of the states in terms of their performance depends on the number of repetitions. For instance, the twin squeezed vacuum state produces the lowest uncertainty when $1\leqslant\mu<5$, while the squeezed entangled state is the best option when $5<\mu<40$. In addition, the twin squeezed cat state is recovered as the best probe when $\mu>40$, although it practically has the same performance as the coherent state when  $\mu = 1, 2 , 3$. Interestingly, the coherent state is also associated with the largest uncertainty for a low number of trials.

That the strategy leading to the lowest uncertainty can depend on the number of repetitions in a crucial way was already demonstrated in chapter \ref{chap:nonasymptotic}. However, our previous results were based on a specific measurement scheme (counting photons after the action of a $50$:$50$ beam splitter), while now the bounds are constructed by repeating a single-shot strategy that has been optimised over all possible POMs. Thus, the results in figure \ref{bounds_results}.i generalise those in chapter \ref{chap:nonasymptotic} and put the state-dependence behaviour of the non-asymptotic regime on a more solid basis.

For these results to be useful, we need to understand the optimality and saturability of the bounds. The uncertainty for $\mu = 1$ is already optimal by construction and can always be reached in principle for any given state using the single-shot POM in equation (\ref{singleshot_strategy}). This means that other tools such as the quantum Ziv-Zakai bound \cite{tsang2012} and the quantum Weiss-Weinstein bound \cite{tsang2016} will necessarily produce less tight single-shot results whenever their value is different from the solution found here. The demonstration of this fact is provided in section \ref{sec:alternativevssingleshot}.

Furthermore, figures \ref{bounds_results}.ii - \ref{bounds_results}.vi show how our results for each state approach the quantum Cram\'{e}r-Rao bound asymptotically, that is, $\bar{\epsilon}_{\mathrm{mse}} \approx 1/(\mu F_q)$  when $\mu \gg 1$. Taking into account that the bounds for a large number of trials that can be constructed using the quantum Cram\'{e}r-Rao bound are fundamental, we conclude that our bounds are also optimal in this limit. As a result, if we work in the regime of intermediate prior knowledge and $\rho(\theta)$ and $p(\theta)$ are given, then the scheme developed in section \ref{sec:methodb} is optimal both for a single shot and a large number of trials. Moreover, it is also optimal for any number of trials if we exclude the possibility of having adaptive measurements and focus on identical and independent experiments.

To quantify the number of repetitions that are needed to reach this asymptotic regime where our methods are no longer required we can follow section \ref{subsec:asymsatu}, construct the relative error $\varepsilon_\tau = |\bar{\epsilon}_{\mathrm{mse}}(\mu_{\tau}) - 1/(\mu_{\tau} F_q)|/\epsilon_{\mathrm{mse}}(\mu_{\tau})$ in equation (\ref{saturation}) and select $\mu_{\tau}$ after imposing that $\varepsilon_\tau \approx 0.05$ for each state. According to the results of this calculation, which are summarised in the last column of table \ref{tablesummary}, the uncertainty for the twin squeezed vacuum state agrees with the prediction of the quantum Cram\'{e}r-Rao bound when the number of trials is as low as $\mu_{\tau} = 5$. Therefore, in this case the asymptotic theory mostly gives the right answer. However, the squeezed entangled state and the twin squeezed cat state require $\mu_{\tau}=45$ and $\mu_{\tau}=66$, respectively, and the quantum Cram\'{e}r-Rao bound overestimates the performance of these probes in the regime of limited data because the graphs of our bounds are higher (figures \ref{bounds_results}.v and \ref{bounds_results}.vi). We note that it is in scenarios of this type where we could not extract useful information from the quantum optimal-bias bound derived in \cite{liu2016}, since for a flat prior this quantity is always lower than the quantum Cram\'{e}r-Rao bound by construction. Finally, the NOON state needs $\mu_{\tau}=116$ and the coherent state requires $\mu_{\tau}=282$, but the Cram\'{e}r-Rao bound prediction underestimates the precision of these protocols when $\mu$ is low. It is interesting to observe that the chosen probes exemplify the three basic behaviours that we could expect to find in the non-asymptotic regime, that is, that the Cram\'{e}r-Rao bound is lower, higher or approximately equal to the Bayesian mean square error.

It is possible to perform the explicit calculation of the optimal measurement scheme that generates the results associated with the NOON state. Using the notation introduced in section \ref{numcal}, its initial density matrix is
\begin{equation}
\rho_0 = 
\begin{pmatrix}
\left(\rho_0\right)_{2020} & \left(\rho_0\right)_{2002} \\
\left(\rho_0\right)_{0220} & \left(\rho_0\right)_{0202}
\end{pmatrix} = \frac{1}{2}
\begin{pmatrix}
1 & 1 \\
1 & 1 \\
\end{pmatrix},
\end{equation} 
while from equations (\ref{k_semianalytical_sol} - \ref{singularcases}) we have that
\begin{equation}
\mathcal{K} = 
\begin{pmatrix}
\mathcal{K}_{2020} & \mathcal{K}_{2002} \\
\mathcal{K}_{0220} & \mathcal{K}_{0202}
\end{pmatrix} = \frac{1}{\pi}
\begin{pmatrix}
\pi & 2 \\
2 & \pi
\end{pmatrix}
\end{equation}
and
\begin{equation}
\mathcal{L} = 
\begin{pmatrix}
\mathcal{L}_{2020} & \mathcal{L}_{2002} \\
\mathcal{L}_{0220} & \mathcal{L}_{0202}
\end{pmatrix} = \frac{i}{\pi}
\begin{pmatrix}
0 & -1 \\
1 & 0  
\end{pmatrix}
\end{equation}
Therefore, 
\begin{equation}
\rho = \rho_0 \circ \mathcal{K} = \frac{1}{2\pi}
\begin{pmatrix}
\pi & 2 \\
2 & \pi
\end{pmatrix} =
\frac{\mathbb{I}}{2} + \frac{\sigma_x}{\pi}
\label{rhonoon}
\end{equation}
and
\begin{equation}
\bar{\rho} = \rho_0 \circ \mathcal{L} = \frac{i}{2\pi}
\begin{pmatrix}
0 & -1 \\
1 & 0
\end{pmatrix} = \frac{\sigma_y}{2\pi}.
\label{rhobarnoon}
\end{equation}
Inserting equations (\ref{rhonoon}) and (\ref{rhobarnoon}) in $S \rho + \rho S = 2 \bar{\rho}$ we find that the equation to be solved is 
\begin{equation}
S + \frac{1}{\pi}\left\lbrace S, \sigma_x \right\rbrace = \frac{\sigma_y}{\pi},
\end{equation}
where $\left\lbrace X, Y \right\rbrace = X Y + Y X$. Recalling that the anticommutator for the Pauli matrices is $\left\lbrace \sigma_i, \sigma_j \right\rbrace = 2\delta_{ij}$, for $i, j, k = x, y, z$, by inspection we conclude that the solution is $S = \sigma_y/\pi$. This implies that the optimal single-shot POM is given by the eigenvectors
\begin{align}
\ket{s_1} &= \frac{1}{\sqrt{2}} 
\begin{pmatrix} 
i  \\
1  
\end{pmatrix} = \frac{1}{\sqrt{2}} (i\ket{2,0}+\ket{0,2}), 
\nonumber \\
\ket{s_2} &= \frac{1}{\sqrt{2}}
\begin{pmatrix} 
1 \\
i 
\end{pmatrix} = \frac{1}{\sqrt{2}} (\ket{2,0}+i\ket{0,2}),
\label{noon_projectors}
\end{align}
and that the Bayesian estimates that the NOON state predicts for $\theta$ are given by the eigenvalues $s_1=-1/\pi$ and $s_2=1/\pi$. In section \ref{measurements_section} we will construct physical measurements that realise these projectors exactly. In addition, it is important to note that, while this spectrum of estimates is discrete and the difference of phase shifts $\theta$ is a continuous variable, Luis and Pe{\v{r}}ina \cite{alfredo1996} showed that this behaviour is not contradictory due to the existence of an ultimate quantum limit to the uncertainty in phase estimation.

\begin{figure}[t]
\centering
\includegraphics[trim={2.25cm 0.25cm 1.3cm 0.5cm},clip,width=15.5cm]{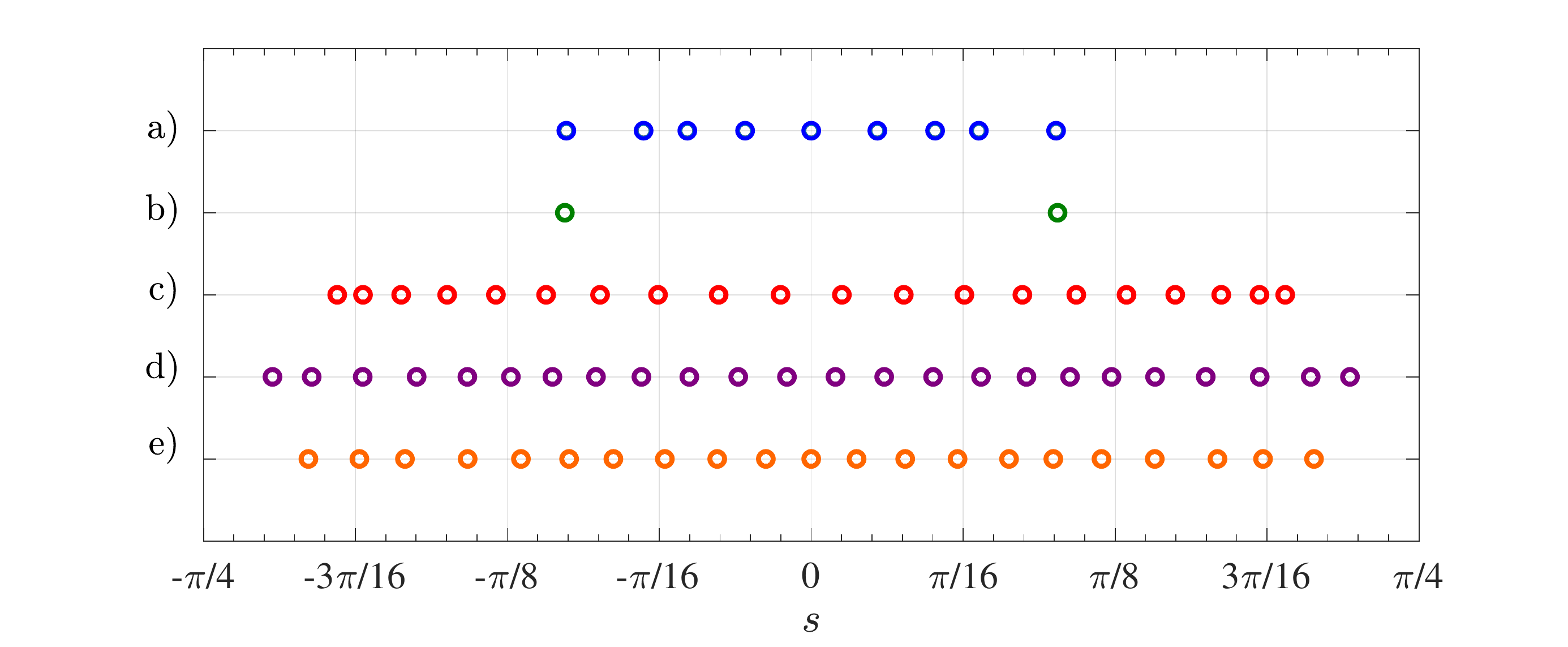}
	\caption[Spectra of the optimal quantum estimators]{Spectrum of the optimal quantum estimator $S$ for (a) the coherent state, (b) the NOON state, (c) the twin squeezed vacuum state, (d) the squeezed entangled state, and (e) the twin squeezed cat state, with $\bar{n}=2$, $\bar{\theta} = 0$ and $W_0 = \pi/2$. The details of this calculation can be found in appendix \ref{sec:singleshotalgorithm}.}
\label{bayes_spectra}
\end{figure}

Although the numerical character of the projectors for the indefinite photon number states makes it difficult to visualise their structure, we can still provide a partial characterisation of these single-shot strategies through the spectra of $S$. A numerical approximation of these spectra has been represented in figure \ref{bayes_spectra} for the coherent state, the twin squeezed vacuum state, the squeezed entangled state and the twin squeezed cat states, which shows their Bayesian estimates distributed within the parameter domain $[\bar{\theta}-W_0/2,\bar{\theta}+W_0/2]$\footnote{We draw attention to the fact that the particular number of estimates represented in the approximated spectra of figure \ref{bayes_spectra} for each indefinite photon number state depends on the numerical truncation of the support where $S$ is defined (see section \ref{numcal}), which in our case assumes that an eigenvalue of $\rho$ is non-zero when its value is higher than $\sim 10^{-12}$. See appendix \ref{sec:singleshotalgorithm} for more details about the numerical approximations employed in this chapter.}.

We finish this analysis by noting that both the projectors $\lbrace \ket{s} \rbrace$ and the estimates $\lbrace s \rbrace$ depend on the specific shape of the prior probability $p(\theta)$. Interestingly, in our case we have verified numerically that while the results change with $W_0$, they do not depend on $\bar{\theta}$. Nonetheless, in section \ref{measurements_section} we will see that this is no longer true for measurement schemes different from the optimal single-shot strategy.

\subsection{The role of intra-mode and inter-mode correlations for a low number of repetitions}\label{correlations_section}

Following our discussion in section \ref{subsec:optint}, there are two types of correlations that are relevant for optical metrology: the intra-mode correlations quantified by the Mandel $\mathcal{Q}$-parameter, and the inter-mode correlations quantified by $\mathcal{J}$. We recall that these quantities were defined in equation (\ref{correlationsintro}) for path-symmetric states, which is the family of probes to which the states in our analysis belong \cite{PaulProctor2016, sahota2015, HofmannHolger2009}. 

These quantities play a crucial role in the regime where $\bar{\epsilon}_{\mathrm{mse}} \approx 1 /(\mu F_q)$ because the quantum Fisher information for path-symmetric pure states can be rewritten as $F_q = 4\Delta J_z^2 = \bar{n}(1+\mathcal{Q})(1-\mathcal{J})$ (\cite{sahota2015, knott2016local} and section \ref{subsec:optint}). Therefore, we can control the asymptotic performance by changing $\mathcal{Q}$ and $\mathcal{J}$. Recalling that $-1\leqslant\mathcal{Q}<\infty$ and $-1\leqslant\mathcal{J}\leqslant 1$, optimising the performance amounts to increasing the intra-mode correlations as much as possible, since path entanglement can only improve the precision by a factor of $2$ at most. To verify that the asymptotic part of figure \ref{bounds_results}.i is consistent with this way of proceeding we have calculated the amount of intra-mode and inter-mode correlations and the quantum Fisher information for each state\footnote{An analytical calculation of these quantities for coherent, NOON and twin squeezed vacuum states is available in \cite{sahota2015} and in section \ref{subsec:commoninter}, while the results for squeezed entangled and twin squeezed cat states can be found in \cite{PaulProctor2016}.}, and the results can be found in the fourth, fifth and sixth columns of table \ref{tablesummary}, respectively. As expected, the twin squeezed cat state, which was found to be the asymptotically optimal choice, has the largest values for $F_q$ and $\mathcal{Q}$ among the states that we are studying.

On the other hand, we have also demonstrated that this state is not better than a coherent state when $\mu \sim 1$, in spite of the fact that for the coherent state we have $\mathcal{Q} = 0$ and $\mathcal{J}=0$, and that the other three probes perform better in the low trial number regime. This already supports the idea that the clear role that photon number correlations play asymptotically is not preserved when $\mu$ is low, something that was suggested by the results in section \ref{subsec:uncertaintynonasymptotic} using a specific POM. While it is not currently possible to find a rigorous relationship between uncertainty and correlations that is also valid in the regime of limited data because an analytical expression for $\bar{\epsilon}_{\mathrm{mse}}(\mu)$ is not available, we can still exploit the methodology introduced in section \ref{sec:methodb} to further explore this idea.

First we note that the twin squeezed cat state can be seen as a family of states defined in terms of the parameters $r$ and $\alpha$. Since this state is separable with respect to the arms of the interferometer, $\mathcal{J} = 0$, and as such we are free to choose different combinations of $r$ and $\alpha$ to control the Mandel $\mathcal{Q}$-parameter while keeping $\bar{n}=2$ and $W_0=\pi/2$ unchanged. The particular instance of the twin squeezed cat family with $\mathcal{Q} = 11.75$ and $F_q = 25.49$ considered until now is the optimal choice after maximising $F_q$ numerically\footnote{The optimisation has been performed using the analytical expression for the quantum Fisher information of the twin squeezed vacuum state that is provided in the appendices of \cite{PaulProctor2016}. Notice that our result is consistent with the equivalent optimisation that was carried out there.}. A second example with $\mathcal{Q} = 10.00$ and $F_q = 22.00$ has been included in table \ref{tablesummary} to represent the intermediate case. In addition, the twin squeezed vacuum state is recovered within the twin squeezed cat family when we choose $\alpha = 0$ \cite{PaulProctor2016}, and for this state we have that $\mathcal{Q} = 3$ and $F_q = 8$. 

\begin{figure}[t]
\centering
\includegraphics[trim={1cm 0.1cm 1.3cm 1cm},clip,width=14.75cm]{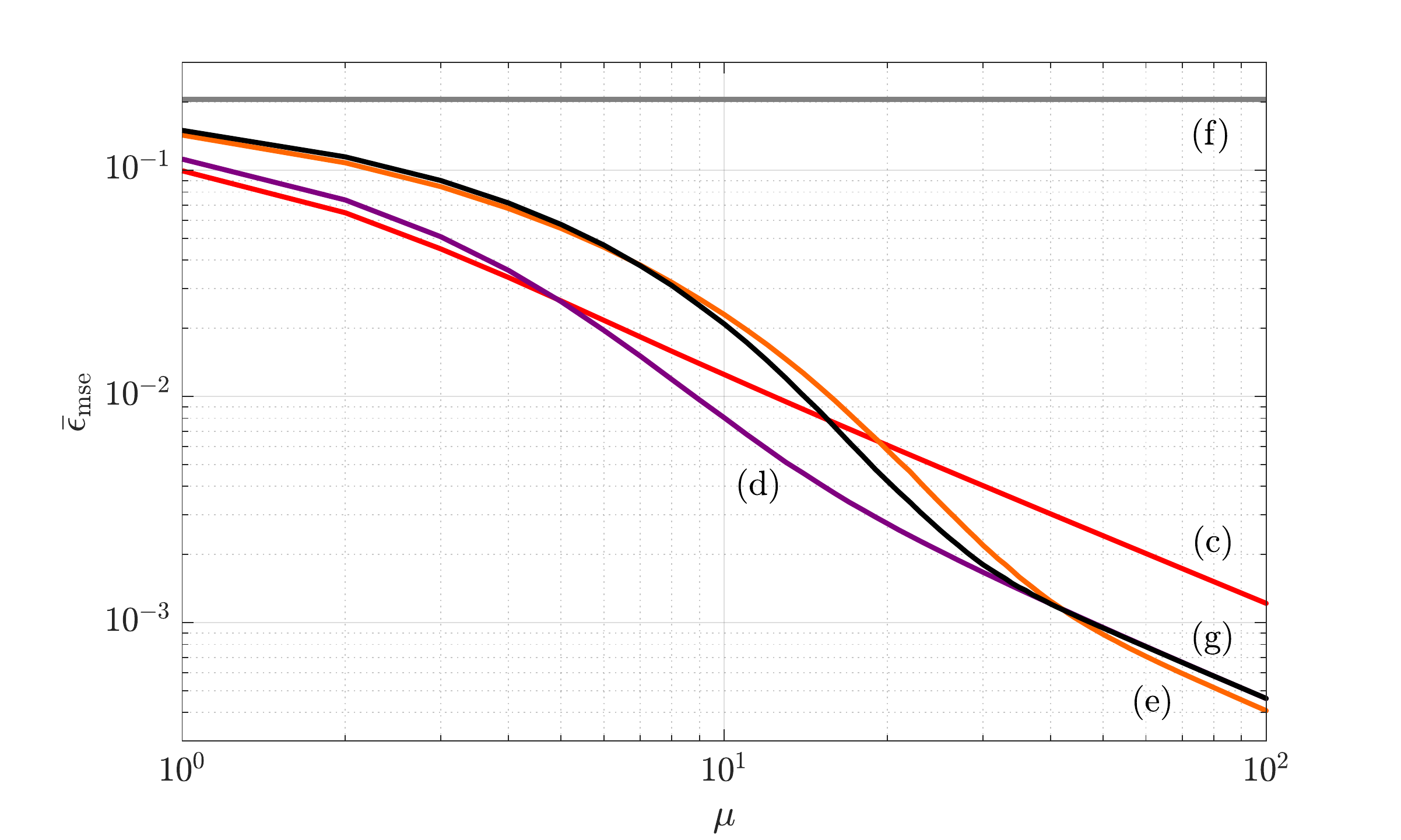}
	\caption[Role of photon correlations in the non-asymptotic regime]{Mean square error as a function of the number of repetitions using the optimal single-shot strategy for (c) the twin squeezed vacuum state with $\mathcal{Q}=3$ and $\mathcal{J} = 0$, (d) the squeezed entangled state with $\mathcal{Q}=9$ and $\mathcal{J} = -0.1$, (e) the twin squeezed cat state with $\mathcal{Q}=11.75$ and $\mathcal{J} = 0$, and (e) the twin squeezed cat state with $\mathcal{Q}=10.00$ and $\mathcal{J} = 0$, where $\mathcal{Q}$ and $\mathcal{J}$ quantify the intra-mode and inter-mode correlations, and having $\bar{n}=2$, $\bar{\theta} = 0$ and $W_0 = \pi/2$, while (f) represents the variance of the prior probability.}
\label{correlations_figure}
\end{figure}

Next we examine the mean square errors associated with the optimal case, the intermediate case and the twin squeezed vacuum from the previous family. Their graphs are represented in figure \ref{correlations_figure} and labelled respectively as (e), (g) and (c). If we compare the optimal and intermediate states first, we see that a larger amount of intra-mode correlations is associated with a larger number of repetitions needed to reach the asymptotic regime, since the former state requires $\mu_\tau=66$ and the latter $\mu_\tau=42$ (see table \ref{tablesummary}). Furthermore, by comparing the form of the graphs (e) and (g) in figure \ref{correlations_figure} for these two states we can observe that the transition from the non-asymptotic regime to the asymptotic regime is associated  with a larger uncertainty for the optimal twin squeezed cat state for which $\mathcal{Q}$ is also larger. Finally, the graph (c) shows that the twin squeezed vacuum state, which has the smallest $\mathcal{Q}$, performs worse than the two previous cases asymptotically, while its error is the lowest when $1\leqslant\mu \lesssim 10$. In other words, for this family of states there seems to be a trade-off between the performances in the asymptotic and non-asymptotic regimes that is associated with changes in $\mathcal{Q}$, which in practice would imply that increasing the amount of intra-mode correlations blindly can lead to high-uncertainty schemes in the regime of limited data. Moreover, we note that this conclusion is consistent with the related analysis by Tsang \cite{tsang2012} for the Rivas-Luis state \cite{rivas2012} based on the quantum Ziv-Zakai bound and our own analysis in section \ref{subsec:infiniteprecision}; both approaches demonstrate that if a certain parameter is modified such that the Fisher information increases arbitrarily, then the error cannot deviate substantially from the prior variance unless the number of trials is very large. 

Since increasing $\mathcal{Q}$ seems to be detrimental to the performance of our probes when the number of repetitions is  low, the next natural step is to investigate whether path entanglement could be useful in this regime. Including in our analysis the squeezed entangled state with $\mathcal{Q} = 9$ and $\mathcal{J}=-0.1$, which is labelled as (d) in figure \ref{correlations_figure}, we can see that this state converges asymptotically to the performance associated with the intermediate case of the twin squeezed cat family (g), that is, both probes have the same Fisher information. However, the graph of the squeezed entangled state presents a smaller curvature and a lower uncertainty when $\mu < 30$. The key aspect that distinguishes these two probes is that the squeezed entangled state has a lower amount of intra-mode correlations and a certain amount of beneficial path entanglement, which suggests that inter-mode correlations have helped to improve the precision in the non-asymptotic regime while keeping a large Fisher information. Hence, we conclude that path entanglement could be considerably more relevant in schemes that need to be optimised for a low number of trials than it is in the asymptotic regime. 

Despite these surprising results, we must acknowledge that our analysis is centred on a particular set of states, and that other schemes based on different states could show different properties\footnote{Furthermore, it is reasonable to expect that other schemes that allow other types of correlations behave differently too. For instance, in \cite{smirne2018} the authors showed that allowing entanglement between a finite number of probes in a frequency estimation protocol can lead to a less precise strategy.}. Therefore, the existence of a more general relationship between the number of trials and the usefulness of photon number correlations in interferometry for a given prior is an open question. 

\subsection{The effect of the prior information}
\label{prior_section}

In a wide set of inference problems that includes the scenarios presented here, the importance of the prior information depends on the number of shots. In particular, we know that the prior becomes less important as we increase the number of repetitions \cite{cox2000}, and this implies that, as we argued in chapter \ref{chap:methodology}, the prior probability will play an important role for making inferences if only a few experimental shots are possible. In that scenario it is crucial then to establish how different states of prior knowledge may affect the overall performance of a given metrology scheme.

\begin{figure}[t]
\centering
\includegraphics[trim={0.1cm 0.1cm 0.65cm 0.2cm},clip,width=7.7cm]{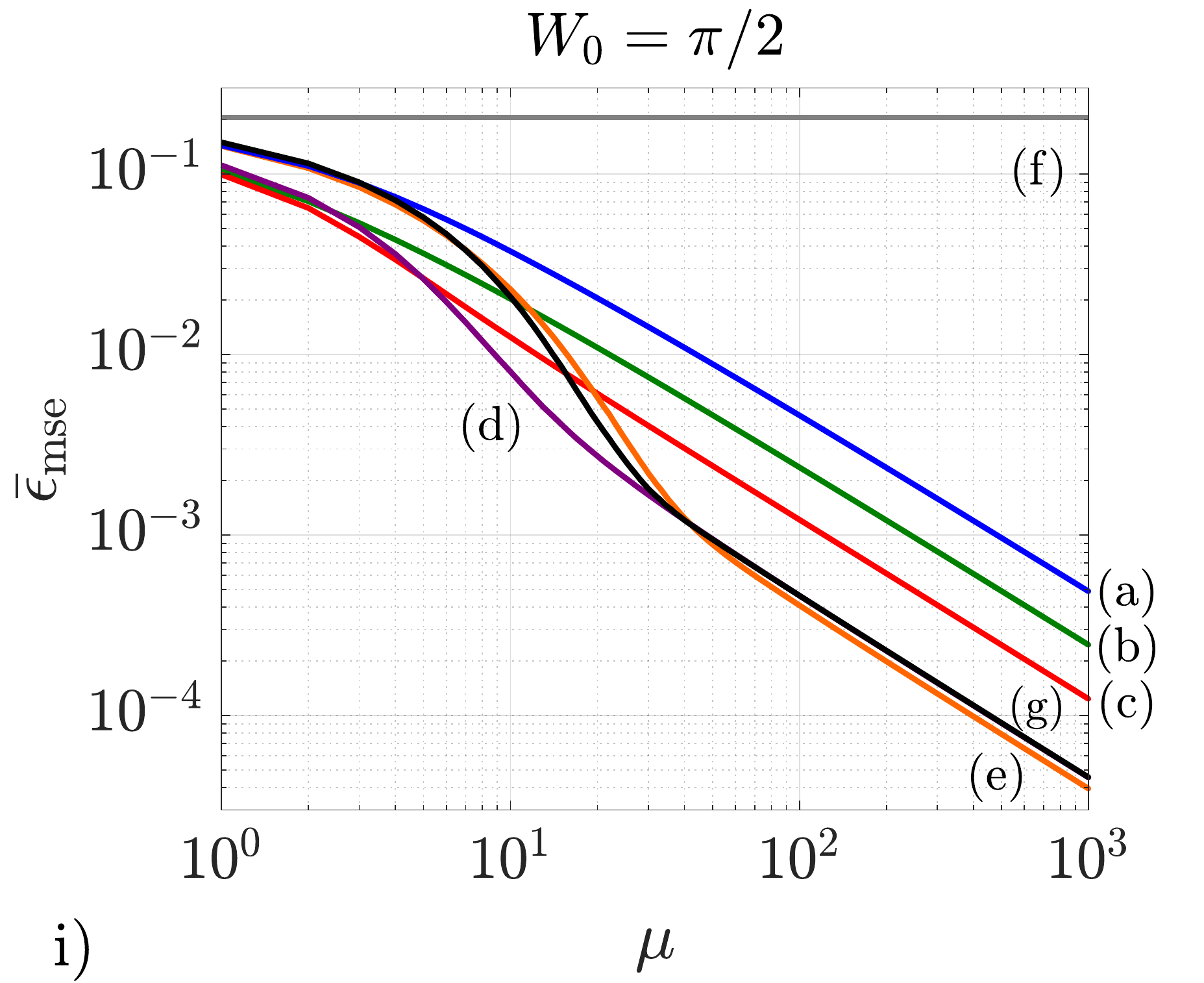}\includegraphics[trim={0.1cm 0.1cm 0.65cm 0.2cm},clip,width=7.7cm]{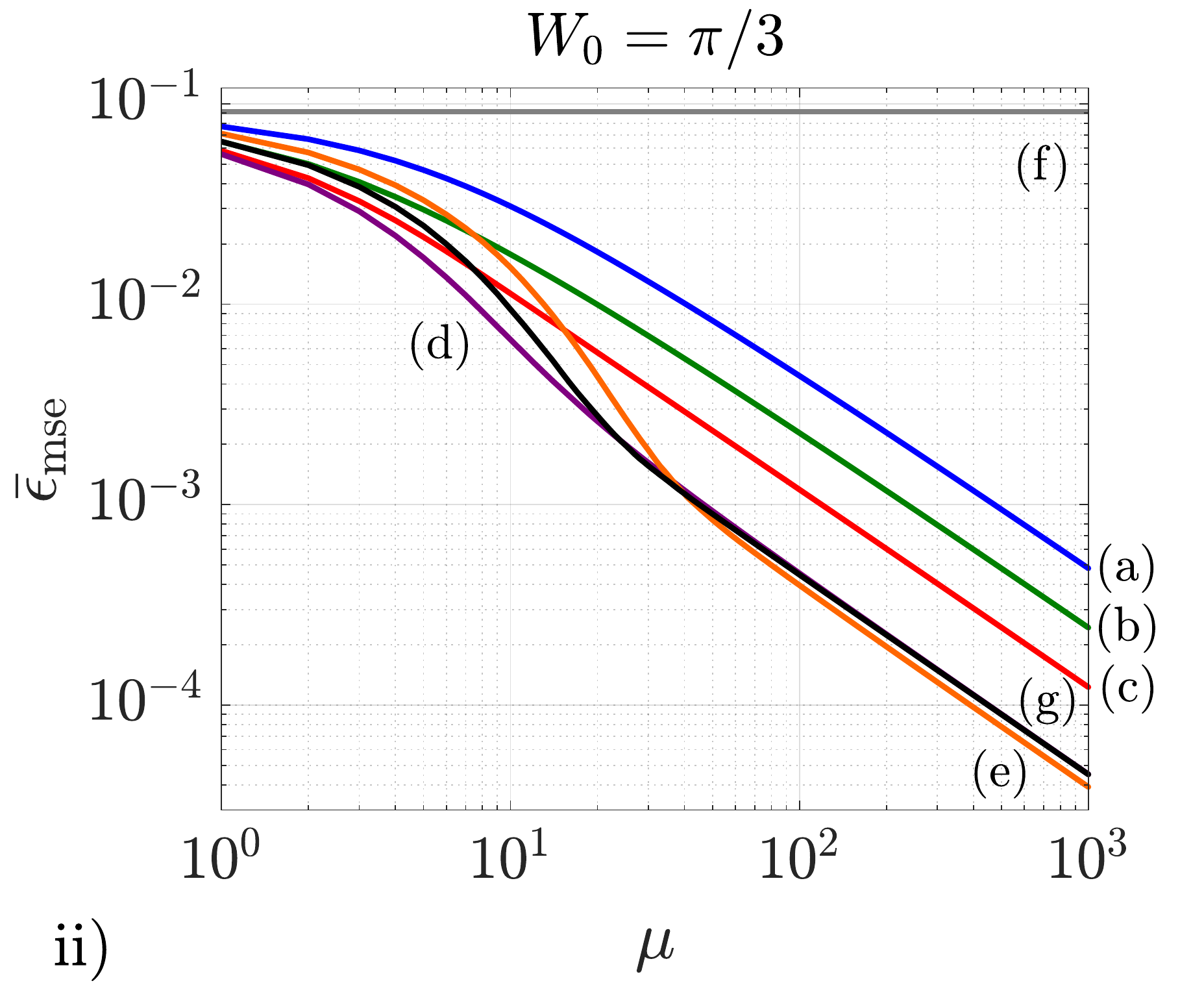}
\includegraphics[trim={0.1cm 0.1cm 0.65cm 0cm},clip,width=7.7cm]{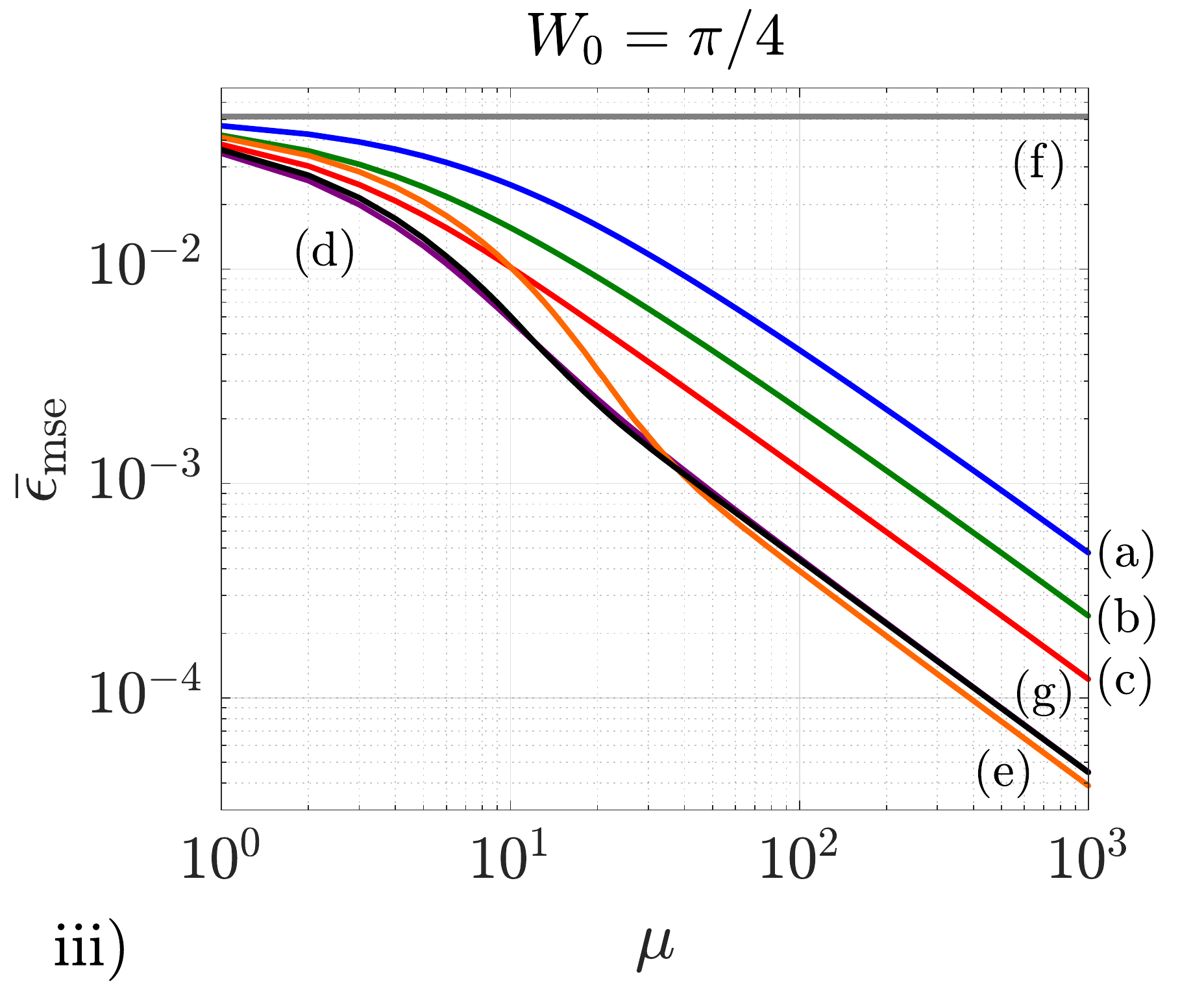}\includegraphics[trim={0.1cm 0.1cm 0.65cm 0.2cm},clip,width=7.7cm]{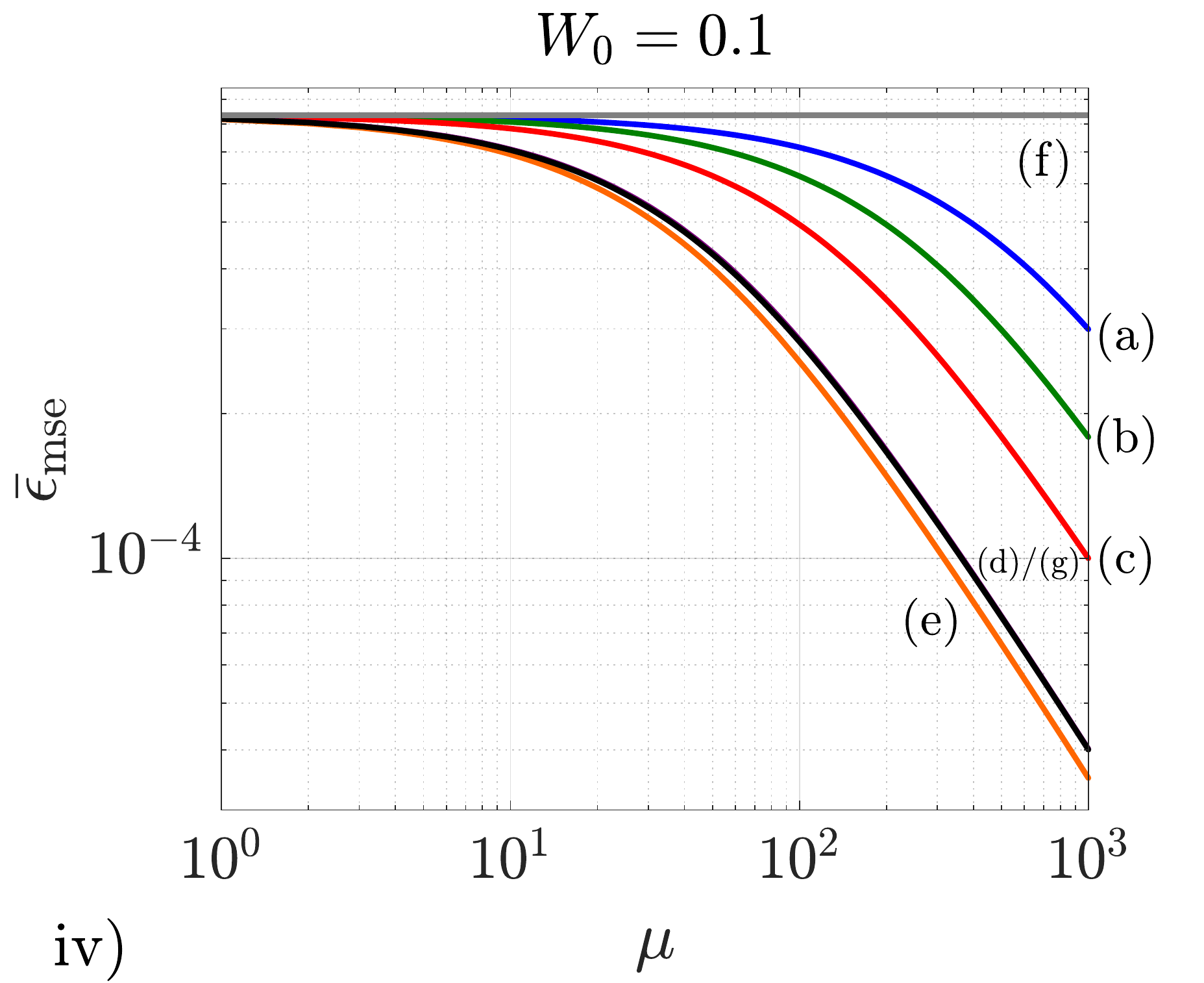}
	\caption[Prior information and the transition between local regimes]{i) Mean square error as a function of the number of repetitions using the optimal single-shot strategy for (a) the coherent state, (b) the NOON state, (c) the twin squeezed vacuum state, (d) the squeezed entangled state, (e) the (optimal) twin squeezed cat state, and (g) the (intermediate) twin squeezed cat state, with mean number of photons $\bar{n}=2$, prior mean $\bar{\theta} = 0$ and prior width $W_0 = \pi/2$, while (f) represents the variance of the prior probability; (ii) repetition of the calculation performed in (i) with prior width $W_0=\pi/3$, (iii) $W_0=\pi/4$, and (iv) $W_0=0.1$. The results in these figures represent the transition from the regime of intermediate prior knowledge and a low number of trials to the local regime of a narrow prior and a large number of measurements.}
\label{prior_figure}
\end{figure}

Taking the form of the uniform prior given in equation (\ref{prior_probability}), the parameters that we can alter are the prior width $W_0$ and the prior mean $\bar{\theta}$. In section \ref{main_results} we already mentioned that the bounds constructed in figure \ref{bounds_results}.i do not depend on $\bar{\theta}$, leaving $W_0$ as the only free parameter. In principle we should consider the possibility of having both $W_0>\pi/2$, which includes the intermediate and global regimes, and $W_0<\pi/2$, which encompasses the intermediate and local regimes. However, for large values of $W_0$ it is not possible to approximate the periodic deviation function in equation (\ref{sinerror}) to the square error. For that reason, we will only focus on the transition from the intermediate regime of prior knowledge to the local regime. 

To do this, let us start by calculating the optimal single-shot mean square error (equation (\ref{singleshot_bound}) or equation (\ref{shotbyshotmse}) with $\mu = 1$) for all the states with the prior widths $W_0 = \pi/2,~\pi/3,~\pi/4$ and $0.1$. The numerical results are shown in table \ref{prior_effect_summary}. While the best probe in the single-shot regime for $W_0 = \pi/2$ is the twin squeezed vacuum state, the squeezed entangled state becomes the preferable choice when $W_0 = \pi/3$ and $W_0 = \pi/4$, and we need to start with a prior with width $W_0 = 0.1$ in order to recover the twin squeezed cat state as the optimal state. Moreover, the ordering of probes in terms of their performance when $W_0 = 0.1$ is exactly the same as the ordering found in the asymptotic regime, which is also included in the last column of table \ref{prior_effect_summary}. Consequently, we can say that for our schemes the local regime due to a high amount of prior information is achieved when $W_0 \leqslant 0.1$.

An equivalent path to arrive to the same result relies on the approximation
\begin{equation}
\bar{\epsilon}_{\mathrm{mse}}\gtrsim\Delta \theta_p^2\left(1-\Delta\theta_p^2 F_q\right)
\label{bayes_bound_high_prior}
\end{equation}
employed in \cite{macieszczak2014bayesian, jarzyna2015true} for the single-shot mean square error. This relation was found in \cite{macieszczak2014bayesian} assuming a Gaussian prior with a narrow width but, in fact, it can be shown that it also holds for our flat prior if $W_0 \ll 1$. Assuming the latter condition, the Taylor expansion around $\bar{\theta}$ for the transformed state $\rho(\theta)$ is
\begin{equation} 
\rho(\theta) \approx \rho(\bar{\theta}) + \frac{\partial \rho(\bar{\theta})}{\partial \theta} (\theta - \bar{\theta}).
\label{rhoparameterapprox}
\end{equation}
Furthermore, recalling that $L(\bar{\theta})\rho(\bar{\theta}) + \rho(\bar{\theta}) L(\bar{\theta}) = 2\partial \rho(\bar{\theta})/\partial \theta$ for the symmetric logarithmic derivative $L(\bar{\theta})$ \cite{helstrom1976,paris2009,rafal2015}, equation (\ref{rhoparameterapprox}) can be rewritten as
\begin{equation}
\rho(\theta)  \approx \rho(\bar{\theta}) + \frac{1}{2}\left[ L(\bar{\theta})\rho(\bar{\theta}) + \rho(\bar{\theta}) L(\bar{\theta})\right] (\theta - \bar{\theta} ).
\label{density_matrix_approx}
\end{equation}
The next step is to introduce equation (\ref{density_matrix_approx}) in the expressions $\rho = \int d\theta p(\theta) \rho(\theta)$ and $\bar{\rho} = \int d\theta p(\theta) \rho(\theta) \theta$, finding that $\rho \approx \rho(\bar{\theta})$ and
\begin{equation}
\bar{\rho} \approx \bar{\theta}\rho(\bar{\theta}) +  \frac{\Delta \theta^2_p}{2}\left[ L(\bar{\theta})\rho(\bar{\theta}) + \rho(\theta) L(\bar{\theta})\right].
\end{equation}
Hence, from $S\rho + \rho S = 2\bar{\rho}$ and the previous approximations we can see that the equation to be solved in this regime is
\begin{equation}
\left[S - \theta \mathbb{I} - \Delta \theta^2_p L(\bar{\theta})\right]\rho(\bar{\theta}) + \rho(\bar{\theta}) \left[S - \theta \mathbb{I} - \Delta \theta^2_p L(\bar{\theta})\right] \approx 0, 
\end{equation} 
which means that the quantum estimator takes the form 
\begin{equation}
S \approx \bar{\theta} ~\mathbb{I} + \Delta \theta^2_p ~ L(\bar{\theta}).
\end{equation}
In turn, this implies that 
\begin{equation}
\mathrm{Tr}\left(\bar{\rho}S\right) \approx \bar{\theta}^2 + \Delta \theta^4_p~F_q(\bar{\theta}),
\label{approxbayesinfo}
\end{equation}
where $F_q(\bar{\theta}) = \mathrm{Tr}[\rho(\bar{\theta})L(\bar{\theta})^2]$ and we have used the fact that $\mathrm{Tr}[\rho(\bar{\theta})L(\bar{\theta})] = 0$, and by introducing equation (\ref{approxbayesinfo}) in the single-shot bound $\bar{\epsilon}_{\mathrm{mse}} \geqslant \int d\theta p(\theta)\theta^2 - \mathrm{Tr}(\bar{\rho}S) = \Delta\theta^2_p + \bar{\theta}^2 - \mathrm{Tr}(\bar{\rho}S)$ that was reviewed in sections \ref{subsec:singleshotparadigm} and \ref{subsec:originalderivation} we finally arrive at equation (\ref{bayes_bound_high_prior}).

That the Fisher information $F_q$ appears in equation (\ref{bayes_bound_high_prior}) as the key quantity to determine which scheme has the best performance for a given prior explains why the numerical results in table \ref{prior_effect_summary} for $W_0=0.1$ predict the same order of probes as the approximation $1/(\mu F_q)$ in the asymptotic regime of many repetitions. In both cases, the larger $F_q$, the better the performance. 

It is interesting to observe the similarity between the local regime of prior knowledge for a single shot and the local regime due to a large number of experiments. On the one hand, the best states for $W_0=\pi/2$ and $W_0=0.1$ have intra-mode correlations only, while for $W_0 = \pi/3$ and $W_0 = \pi/4$ the best state presents path entanglement too. On the other hand, figure \ref{prior_figure}.i shows that for $1\leqslant\mu<5$ and $\mu>40$ there is no inter-mode entanglement in the optimal probes, but it appears in the best state for $5<\mu<40$. One way of understanding this similar behaviour is to note that updating our posterior density via Bayes theorem after each new trial reduces the uncertainty in a way that is formally similar to making the prior narrower in a sequential way. Nevertheless, both processes are conceptually different. 

Finally, figures \ref{prior_figure}.i - \ref{prior_figure}.iv show the transition from the intermediate regime of prior knowledge and a low number of trials to a local regime with both high prior information and a large number of shots. This modifies the connection between the number of repetitions and the properties of different probes considerably, as can be seen by the change in the points where the graphs for different states cross each other as the prior width is reduced. As a result, establishing a pattern that helps us to understand what probes we need to use for different values of $\mu$ in the regime of limited data becomes more complicated than in the two previous sections. Fortunately, this is not a problem in real experiments because we typically know what our specific prior information is and we can always proceed on a case-by-case basis, but it constitutes an important obstacle to deriving more general conclusions.

\subsection{Performance of physical measurements}
\label{measurements_section}

Until now we have investigated the physical consequences of the bounds constructed following the procedure of section \ref{sec:methodb}. Nevertheless, in a real-world situation we also need to be able to generate concrete sequences of operations that can be implemented in the laboratory, study whether they saturate the theoretical bounds and, if they do not, determine how close to the fundamental minimum the associated uncertainty is. Since here we are using a fixed set of probe states, we need only consider sequences for implementing the measurement scheme.

States that can be generated using operations such as squeezing or displacement from the vacuum are generally easier to prepare than the abstract (and possibly entangled) states that arise in theoretical optimisations \cite{rafal2015,PaulProctor2016,schafermeier2018}; consequently, there is an intrinsically practical interest in exploring how close to the fundamental bounds this type of state can get. We already know that we can approach the quantum Cram\'{e}r-Rao bound  asymptotically for path-symmetric pure (but otherwise general) states when each individual measurement consists of counting photons after the action of a $50$:$50$ beam splitter \cite{HofmannHolger2009}. For instance, using that POM we have shown in section \ref{subsec:uncertaintynonasymptotic} that if $W_0 = \pi/2$ and we impose that the relative error in equation (\ref{saturation}) is $\varepsilon_{\tau}=0.05$, then this is true for the twin squeezed vacuum state for $\mu_\tau \geqslant 874$, although surpassing the $0.05$ threshold with the squeezed entangled state requires more than $\mu = 10^3$ repetitions because its convergence is slower.

\begin{figure}[t]
\centering
\includegraphics[trim={0.1cm 0.1cm 0.65cm 0.2cm},clip,width=7.7cm]{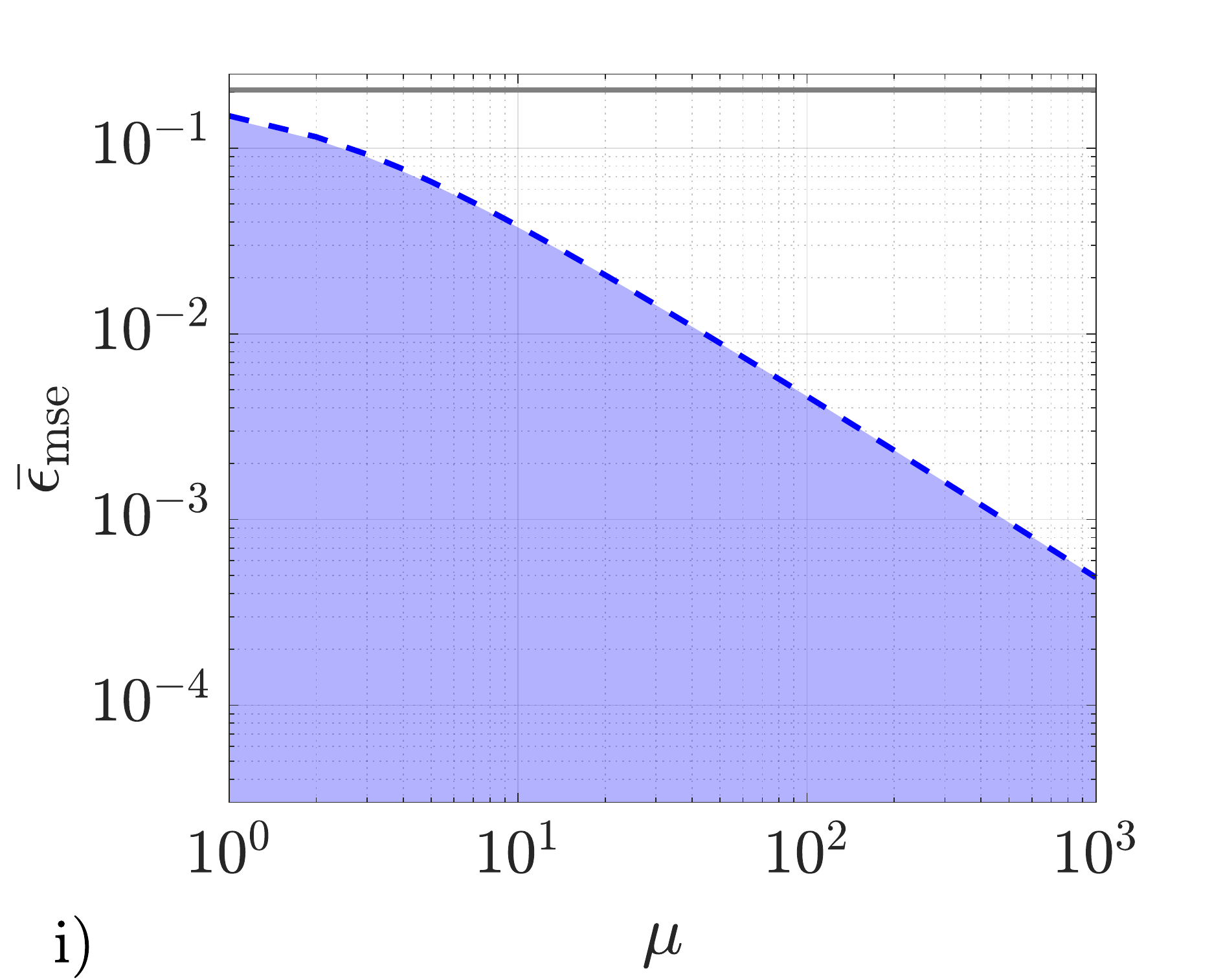}\includegraphics[trim={0.1cm 0.1cm 0.65cm 0.2cm},clip,width=7.7cm]{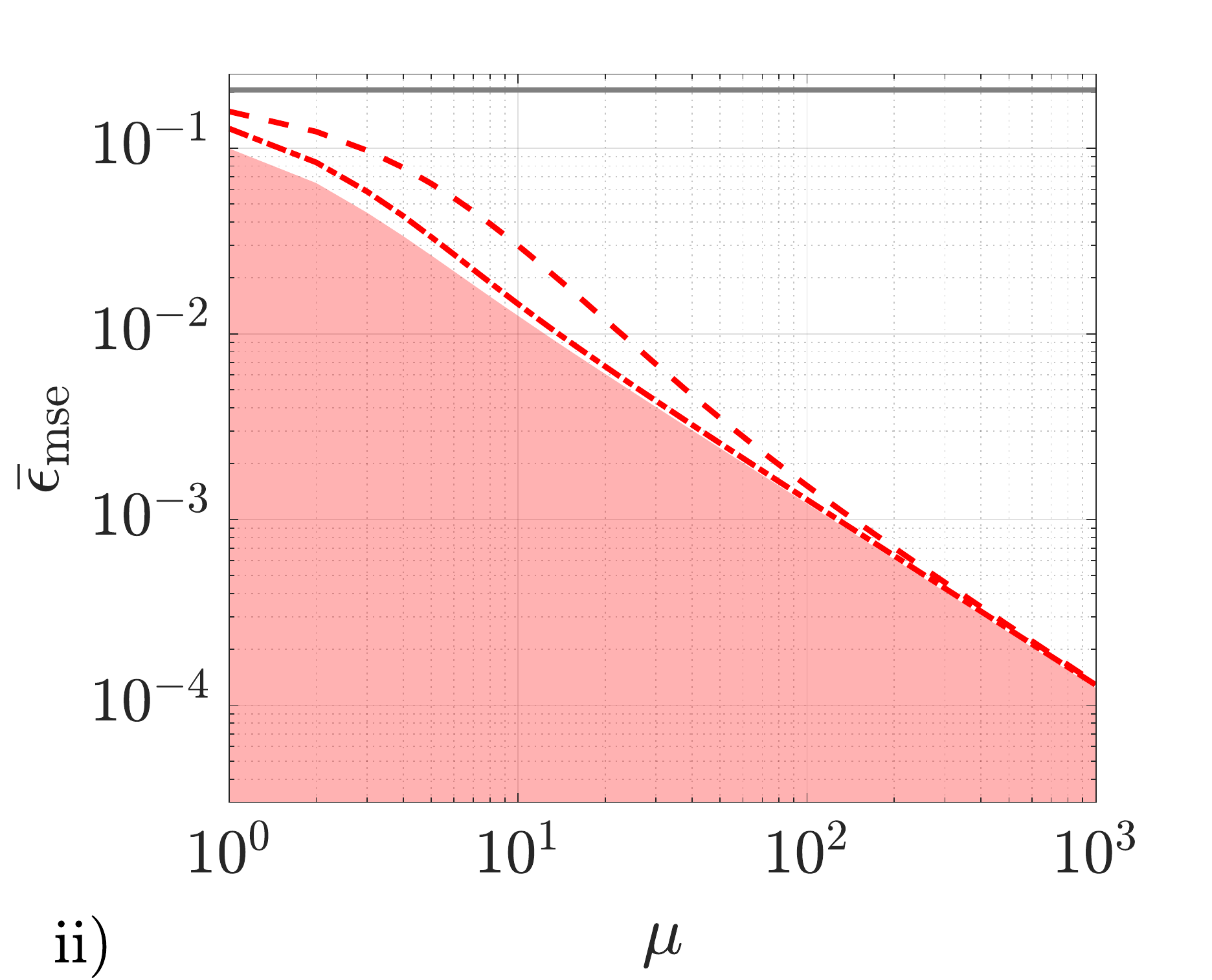}
\includegraphics[trim={0.1cm 0.1cm 0.65cm 0.2cm},clip,width=7.7cm]{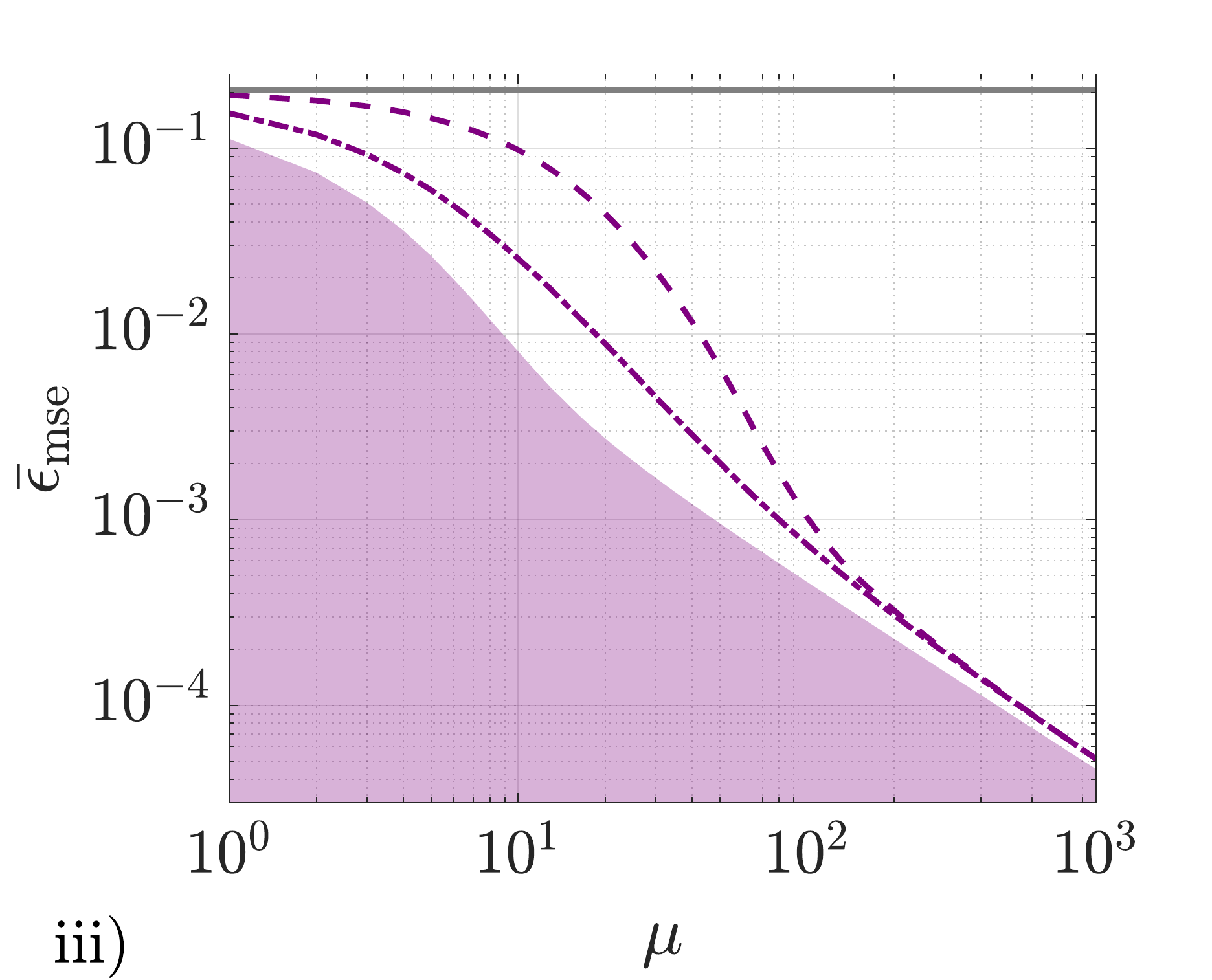}\includegraphics[trim={0.1cm 0.1cm 0.65cm 0.2cm},clip,width=7.7cm]{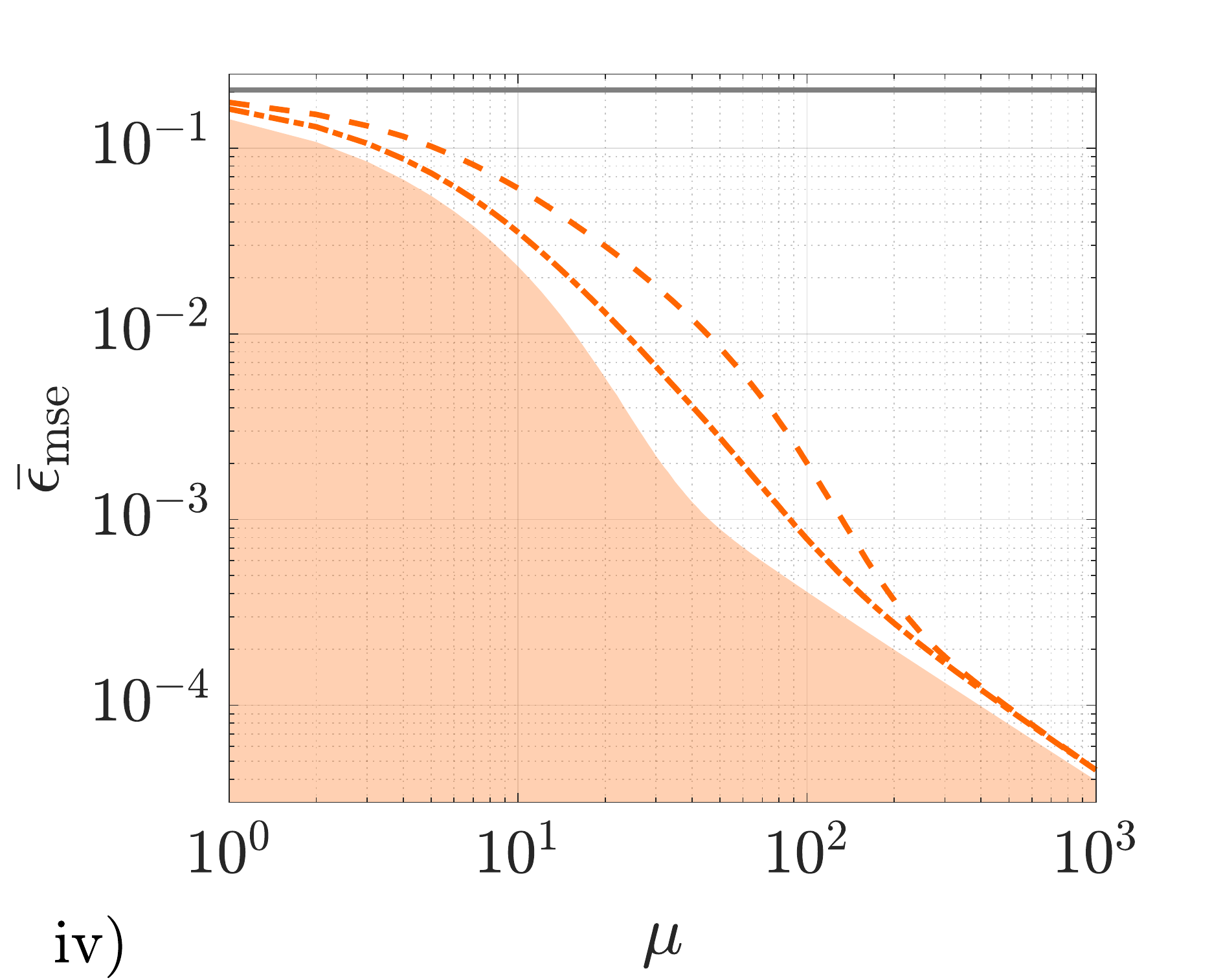}
	\caption[Performance of schemes with practical states and POMs]{i) Mean square error based on the optimal single-shot strategy (shaded area), error associated with the measurement of energy (dashed line) and prior variance (horizontal solid line) for the coherent state, (ii) the twin squeezed vacuum state, (iii) the squeezed entangled state, and (iv) the twin squeezed cat state, with mean number of photons $\bar{n}=2$, prior mean $\bar{\theta} = 0$ and prior width $W_0 = \pi/2$. Furthermore, the dash-dot graphs in (ii), (iii) and (iv) represents the uncertainty for the measurement of quadratures. The sequences of operations that implement the POMs that produce these results can be found in table \ref{POM_summary}.}
\label{continuousPOM}
\end{figure}

By using the bounds with $W_0=\pi/2$ and $\bar{\theta} = 0$ in section \ref{main_results} we can now answer this question in the regime of limited data too, both for the previous states and for the coherent and the twin squeezed cat states. As a preliminary step we have reproduced these bounds as shaded areas in figures \ref{continuousPOM}.i - \ref{continuousPOM}.iv for the coherent state, the twin squeezed vacuum state, the squeezed entangled state and the twin squeezed cat state, respectively. In addition, the dashed lines in those figures represent the mean square error associated with the measurement of the energy at each port of the interferometer (i.e., counting photons) after the action of a $50$:$50$ beam splitter. We draw attention to the fact that we have also introduced a known phase shift in the second port of the interferometer before this beam splitter is applied, the complete sequence of operations for each state being presented in table \ref{POM_summary}. The reason behind this choice is that we have found that the uncertainty of this POM depends on $\bar{\theta}$, and the extra phase shift allows us to achieve the optimal single shot precision when the prior is centred around $\bar{\theta} = 0$, which is our case. This dependence with $\bar{\theta}$ can be seen as a Bayesian analogue of those cases where the standard error propagation formula for a given observable depends on the unknown parameter $\theta$, which is not a problem in practice provided that the experiment is arranged close to an optimal operating point \cite{rafal2015}.

Importantly, the results in chapter \ref{chap:nonasymptotic} for $W_0=\pi/2$ were based on a prior centred around $\pi/4$ and did not include the extra phase shift discussed in the previous paragraph as part of the measurement scheme. However, we have found that the configuration in chapter \ref{chap:nonasymptotic} generates uncertainties that are numerically similar to those discussed here when the extra phase shifts are taken into account. Therefore, the comparison between both collections of results is meaningful for $W_0=\pi/2$.

\begin{table} [t]
\centering
{\renewcommand{\arraystretch}{1.45} 
\begin{tabular}{|l|c|c|c|}
\hline
Measurement & Observable & Projectors  & States \\
\hline
\hline
\begin{tabular}{@{}l@{}}$50$:$50$ splitter \& \\ counting (even) \end{tabular} & \begin{tabular}{@{}c@{}} $N_1 N_2 = \int dk~k \ketbra{k}{k}$, \\ with $N_i = a_i^{\dagger}a_i$ \end{tabular} & $\left\lbrace\mathrm{e}^{-i\frac{\pi}{4}N_2}\mathrm{e}^{-i\frac{\pi}{2}J_x}\ket{k}\right\rbrace_{k}$ & \begin{tabular}{@{}c@{}} All but \\ coherent \end{tabular} \\
\hline
\begin{tabular}{@{}l@{}}$50$:$50$ splitter \& \\ counting (odd) \end{tabular} & \begin{tabular}{@{}c@{}}$N_1 N_2 = \int dk~ k \ketbra{k}{k}$, \\ with $N_i = a_i^{\dagger}a_i$ \end{tabular}& $\left\lbrace\mathrm{e}^{-i\frac{\pi}{2}N_2}\mathrm{e}^{-i\frac{\pi}{2}J_x}\ket{k}\right\rbrace_{k}$ & Coherent \\
\hline
$\pi/8$ quadratures & \begin{tabular}{@{}c@{}}$X_1 X_2 = \int dq~ q \ketbra{q}{q}$, with \\ $X_i = [ \mathrm{e}^{i \frac{\pi}{8}} a_i^{\dagger} + \mathrm{e}^{-i\frac{\pi}{8}}a_i]/\sqrt{2}$\end{tabular}  & $\left\lbrace\mathrm{e}^{i\frac{\pi}{4}N_1}\mathrm{e}^{-i\frac{\pi}{2}J_x}\ket{q}\right\rbrace_{q}$ &  \begin{tabular}{@{}c@{}} All but \\ coherent \end{tabular} \\
\hline 
\begin{tabular}{@{}l@{}}Undoing \& \\  counting\end{tabular} & $N_1 N_2 = \int dk~ k \ketbra{k}{k}$ & $\left\lbrace\mathrm{e}^{i \pi J_z }\mathrm{e}^{i\frac{\pi}{2}J_x}D_1^{\dagger}\left(\alpha\right)\ket{k}\right\rbrace_{k}$ & Coherent \\
\hline
Parity POMs & \begin{tabular}{@{}c@{}} $\Pi_1 \Pi_2 = \int dp~ p \ketbra{p}{p}$, \\ with $\Pi_i = (-1)^{a_i^{\dagger}a_i}$ \end{tabular}& $\left\lbrace\mathrm{e}^{-i\frac{\pi}{4}N_2}\mathrm{e}^{-i\frac{\pi}{2}J_x}\ket{p}\right\rbrace_{p}$ & NOON \\ 
\hline
\end{tabular}}
\caption[Practical POMs to approach our non-asymptotic quantum bounds]{Sequences of quantum operations needed to implement the practical measurements discussed in section \ref{measurements_section}, whose uncertainty is represented in figures \ref{continuousPOM} and \ref{noonPOM}. Note that the observable column indicates the physical quantity that is being measured, and that the different combinations of phase shifts that appear in the third column have been chosen such that the schemes are optimal when the prior is centred around $\bar{\theta} = 0$ and $\bar{n} = 2$.}
\label{POM_summary}
\end{table}

To start our discussion of the low trial number regime with this POM, we first observe that, according to figure \ref{continuousPOM}.i, measuring energy with coherent states produces an uncertainty that is already very close to the associated bound for a low value of $\mu$. More concretely, the bound and the measurement error only differ in their second and third significant figures, as can be directly verified from the values in table \ref{practical_POM_summary}, where we provide the numerical uncertainties for the first ten shots of every scheme based on indefinite photon number strategies. Moreover, this can be further improved if instead we undo the preparation of the probe state before counting photons, that is, by reversing the $50$:$50$ beam splitter and the displacement from the vacuum operations that generated the coherent state in the first place. The extra known difference of phases showed in table \ref{POM_summary} is also needed for the case with $\bar{\theta}=0$ that we are considering. Nonetheless, taking into account the fact that both schemes produce an uncertainty whose first significant figure is that of the optimum (see table \ref{practical_POM_summary}), we conclude that, for most practical purposes, they are equally useful and optimal given any number of repetitions. 

\begin{table} [t]
\centering
{\renewcommand{\arraystretch}{1.2} 
\begin{tabular}{|c|c|c|c|c|c|}
\hline
\multicolumn{6}{|c|}{$\bar{\epsilon}_{\mathrm{mse}}\left(\mu = 1\right)$, $\dots$, $\bar{\epsilon}_{\mathrm{mse}}\left(\mu = 10 \right)$} \\
\hline
\hline
\multicolumn{3}{|c|}{Coherent state} & \multicolumn{3}{c|}{Twin squeezed vacuum state} \\
\hline
\begin{tabular}{@{}c@{}}Single-shot \\ POM\end{tabular} & \begin{tabular}{@{}c@{}}$50$:$50$ splitter \\ \& counting\end{tabular} & \begin{tabular}{@{}c@{}}Undoing \& \\ counting\end{tabular} & \begin{tabular}{@{}c@{}}Single-shot \\ POM\end{tabular} & \begin{tabular}{@{}c@{}}$50$:$50$ splitter \\ \& counting\end{tabular} & \begin{tabular}{@{}c@{}} $\pi/8$ \\ quadra. \end{tabular} \\
\hline
$1.44\cdot 10^{-1}$ & $1.49\cdot 10^{-1}$ & $1.47\cdot 10^{-1}$ & $9.94\cdot 10^{-2}$ & $1.57\cdot 10^{-1}$ & $1.27\cdot 10^{-1}$ \\
$1.11\cdot 10^{-1}$ & $1.15\cdot 10^{-1}$ & $1.13\cdot 10^{-1}$ & $6.48\cdot 10^{-2}$ & $1.23\cdot 10^{-1}$ & $8.37\cdot 10^{-2}$ \\
$8.94\cdot 10^{-2}$ & $9.25\cdot 10^{-2}$ & $9.07\cdot 10^{-2}$ & $4.49\cdot 10^{-2}$ & $9.71\cdot 10^{-2}$ & $5.83\cdot 10^{-2}$ \\
$7.47\cdot 10^{-2}$ & $7.70\cdot 10^{-2}$ & $7.56\cdot 10^{-2}$ & $3.36\cdot 10^{-2}$ & $7.85\cdot 10^{-2}$ & $4.31\cdot 10^{-2}$ \\
$6.40\cdot 10^{-2}$ & $6.59\cdot 10^{-2}$ & $6.47\cdot 10^{-2}$ & $2.64\cdot 10^{-2}$ & $6.44\cdot 10^{-2}$ & $3.32\cdot 10^{-2}$ \\
$5.60\cdot 10^{-2}$ & $5.74\cdot 10^{-2}$ & $5.66\cdot 10^{-2}$ & $2.17\cdot 10^{-2}$ & $5.38\cdot 10^{-2}$ & $2.67\cdot 10^{-2}$ \\
$4.98\cdot 10^{-2}$ & $5.10\cdot 10^{-2}$ & $5.02\cdot 10^{-2}$ & $1.83\cdot 10^{-2}$ & $4.56\cdot 10^{-2}$ & $2.22\cdot 10^{-2}$ \\
$4.48\cdot 10^{-2}$ & $4.58\cdot 10^{-2}$ & $4.51\cdot 10^{-2}$ & $1.58\cdot 10^{-2}$ & $3.91\cdot 10^{-2}$ & $1.89\cdot 10^{-2}$ \\
$4.07\cdot 10^{-2}$ & $4.15\cdot 10^{-2}$ & $4.10\cdot 10^{-2}$ & $1.40\cdot 10^{-2}$ & $3.39\cdot 10^{-2}$ & $1.64\cdot 10^{-2}$ \\
$3.74\cdot 10^{-2}$ & $3.80\cdot 10^{-2}$ & $3.76\cdot 10^{-2}$ & $1.25\cdot 10^{-2}$ & $2.98\cdot 10^{-2}$ & $1.45\cdot 10^{-2}$ \\
\hline
\hline
\multicolumn{3}{|c|}{Squeezed entangled state} & \multicolumn{3}{c|}{Twin squeezed cat state} \\
\hline
\begin{tabular}{@{}c@{}}Single-shot \\ POM\end{tabular} & \begin{tabular}{@{}c@{}}$50$:$50$ splitter \\ \& counting\end{tabular} & \begin{tabular}{@{}c@{}} $\pi/8$ \\ quadra. \end{tabular} & \begin{tabular}{@{}c@{}}Single-shot \\ POM\end{tabular} & \begin{tabular}{@{}c@{}}$50$:$50$ splitter \\ \& counting\end{tabular} & \begin{tabular}{@{}c@{}} $\pi/8$ \\ quadra. \end{tabular} \\
\hline 
$1.12\cdot 10^{-1}$ & $1.93\cdot 10^{-1}$ & $1.54\cdot 10^{-1}$ & $1.43\cdot 10^{-1}$ & $1.76\cdot 10^{-1}$ & $1.62\cdot 10^{-1}$ \\
$7.38\cdot 10^{-2}$ & $1.80\cdot 10^{-1}$ & $1.18\cdot 10^{-1}$ & $1.08\cdot 10^{-1}$ & $1.52\cdot 10^{-1}$ & $1.30\cdot 10^{-1}$ \\
$5.08\cdot 10^{-2}$ & $1.68\cdot 10^{-1}$ & $9.23\cdot 10^{-2}$ & $8.46\cdot 10^{-2}$ & $1.32\cdot 10^{-1}$ & $1.06\cdot 10^{-1}$ \\
$3.60\cdot 10^{-2}$ & $1.56\cdot 10^{-1}$ & $7.34\cdot 10^{-2}$ & $6.77\cdot 10^{-2}$ & $1.16\cdot 10^{-1}$ & $8.75\cdot 10^{-2}$ \\
$2.62\cdot 10^{-2}$ & $1.45\cdot 10^{-1}$ & $5.95\cdot 10^{-2}$ & $5.53\cdot 10^{-2}$ & $1.02\cdot 10^{-1}$ & $7.33\cdot 10^{-2}$ \\
$1.96\cdot 10^{-2}$ & $1.34\cdot 10^{-1}$ & $4.87\cdot 10^{-2}$ & $4.56\cdot 10^{-2}$ & $9.08\cdot 10^{-2}$ & $6.22\cdot 10^{-2}$ \\
$1.51\cdot 10^{-2}$ & $1.24\cdot 10^{-1}$ & $4.06\cdot 10^{-2}$ & $3.81\cdot 10^{-2}$ & $8.13\cdot 10^{-2}$ & $5.33\cdot 10^{-2}$ \\
$1.19\cdot 10^{-2}$ & $1.15\cdot 10^{-1}$ & $3.43\cdot 10^{-2}$ & $3.19\cdot 10^{-2}$ & $7.33\cdot 10^{-2}$ & $4.60\cdot 10^{-2}$ \\
$9.65\cdot 10^{-3}$ & $1.06\cdot 10^{-1}$ & $2.94\cdot 10^{-2}$ & $2.70\cdot 10^{-2}$ & $6.65\cdot 10^{-2}$ & $4.01\cdot 10^{-2}$ \\
$8.04\cdot 10^{-3}$ & $9.77\cdot 10^{-2}$ & $2.54\cdot 10^{-2}$ & $2.30\cdot 10^{-2}$ & $6.07\cdot 10^{-2}$ & $3.52\cdot 10^{-2}$ \\
\hline
\end{tabular}}
\caption[Numerical results for the indefinite photon number states]{Mean square error for the indefinite photon number states using the optimal single-shot POM and the physical measurement schemes described in the main text, with $1\leqslant\mu\leqslant 10$, $\bar{n}=2$, $\bar{\theta}=0$ and $W_0=\pi/2$.}
\label{practical_POM_summary}
\end{table}

The situation is very different when we consider the other three states in figures \ref{continuousPOM}.ii - \ref{continuousPOM}.iv, where the uncertainty of the energy measurement is now notably higher than each bound in the regime of limited data, the distance between the graphs of the measurement and those of the bounds being larger for a few repetitions than for a single shot. This measurement is particularly detrimental for the strategy based on the squeezed entangled state, since its error is very close to the prior variance (horizontal line in \ref{continuousPOM}.iii) when $\mu \sim 1$ and this indicates that almost no information is being gained there. Additionally, we can observe that the twin squeezed cat state in figure \ref{continuousPOM}.ii presents a slow convergence to the asymptotic Cram\'{e}r-Rao bound when we use this POM, compared with the twin squeezed vacuum probe state in figure \ref{continuousPOM}.ii or the coherent state in in figure \ref{continuousPOM}.i. Note that this is the same problem found in section \ref{subsec:uncertaintynonasymptotic} for the squeezed entangled state, which is also reproduced in our calculations here.

These results show that counting photons is not the best strategy to be followed when $\mu$ is low and the probes have been prepared in states with a large Fisher information such as the ones considered here, and this motivates the search for other practical alternatives. More concretely, instead of projecting onto the energy basis, we can consider the measurement of a different physical quantity. The dash-dot lines in figures \ref{continuousPOM}.ii - \ref{continuousPOM}.iv show the results where we have projected onto the eigenstates of the observable $X_1\otimes X_2$,
\begin{equation}
X_i =\frac{1}{\sqrt{2}}\left(a_i^\dagger \mathrm{e}^{i\pi/8} + a_i \mathrm{e}^{-i\pi/8}\right)
\label{quadrature}
\end{equation}
being a quadrature rotated by $\pi/8$ for the $i$-th mode \cite{barnett2002}, after having introduced the phase shift $\mathrm{exp}(i \frac{\pi}{4}a_1^\dagger a_1)$ and having applied a $50$:$50$ beam splitter (see table \ref{POM_summary})\footnote{Note that the eigenstates of the quadrature operator in equation (\ref{quadrature}) cannot be normalised \cite{barnett2002}, and that the eigenvectors mentioned in the main text refer to the numerical approximation associated with the truncated state that we are employing here.}. The error of this scheme also depends on $\bar{\theta}$.

By comparing the energy and quadrature POMs figures \ref{continuousPOM}.ii - \ref{continuousPOM}.iv we see that the graphs based on the latter measurement are substantially closer to the bounds than those for the former POM when the experiment is operating in the regime of limited data. In other words, we have found a physical measurement that improves over the results based on measuring the energy for the practical states under consideration and a low number of trials. Interestingly, the dash-dot lines still converge to the fundamental asymptotic bound, and this implies that in the asymptotic regime both schemes are, nevertheless, equivalent in practice and optimal.   

Although these results extend our findings in chapter \ref{chap:nonasymptotic}, figures \ref{continuousPOM}.ii - \ref{continuousPOM}.iv also show that it could still be possible to find other physical schemes with a better precision when $\mu$ is low, with a faster rate of convergence to the asymptotic minimum or even saturating the bound for any $\mu$. These are some of the questions that should be addressed for further progress in the design of experimentally feasible protocols that operate both in and out of the regime of limited data. 

\subsection{Optimality of NOON states}
\label{subsec:optnoon}

The fact that NOON states are conceptually simple makes them an excellent tool to understand metrology protocols, which is why we have chosen to study them separately. They emerge as the optimal probe that maximises the Fisher information over the definite photon number states \cite{demkowicz2011, jarzyna2016thesis}, and while they are unsuitable for a global estimation due to the multi-peak structure associated with the posterior probabilities that they generate (\cite{jarzyna2016thesis, kolodynski2014} and our analysis in section \ref{subsec:prioranalysis}), and they require that the scaling of the prior variance is already $\sim 1/\bar{n}^2$ in order to achieve the same scaling that the Cram\'{e}r-Rao bound predicts \cite{berry2012infinite, hall2012}, the results in section \ref{subsec:uncertaintynonasymptotic} have shown that they can still be useful to a certain extent in the intermediate regime of prior knowledge and limited data when the number of photons is low and the POM is based on measuring the energy at each port. In addition, this moderate usefulness also holds for the repetition of the single-shot optimal strategy according to our results in figure \ref{bounds_results}.i, since the NOON state performs better than the twin squeezed cat state for $1\leqslant\mu\leqslant10$. By studying the performance of this probe for different physical measurements with respect to the non-asymptotic bound we will see that NOON states are also optimal in another sense. 

\FloatBarrier

\begin{figure}[t]
\centering
\includegraphics[trim={0.1cm 0.1cm 0.65cm 0.5cm},clip,width=7.7cm]{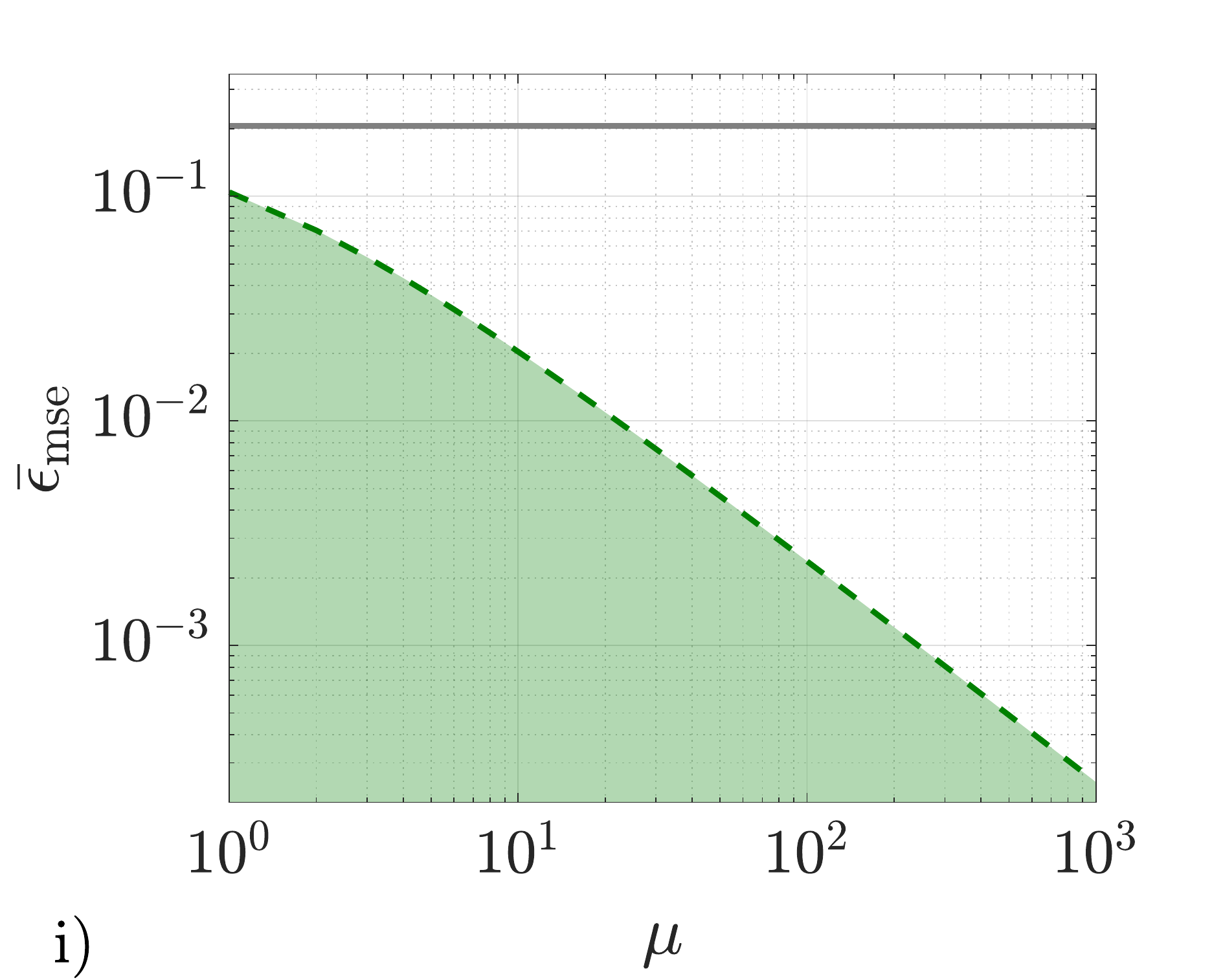}\includegraphics[trim={0.1cm 0.1cm 0.65cm 0.5cm},clip,width=7.7cm]{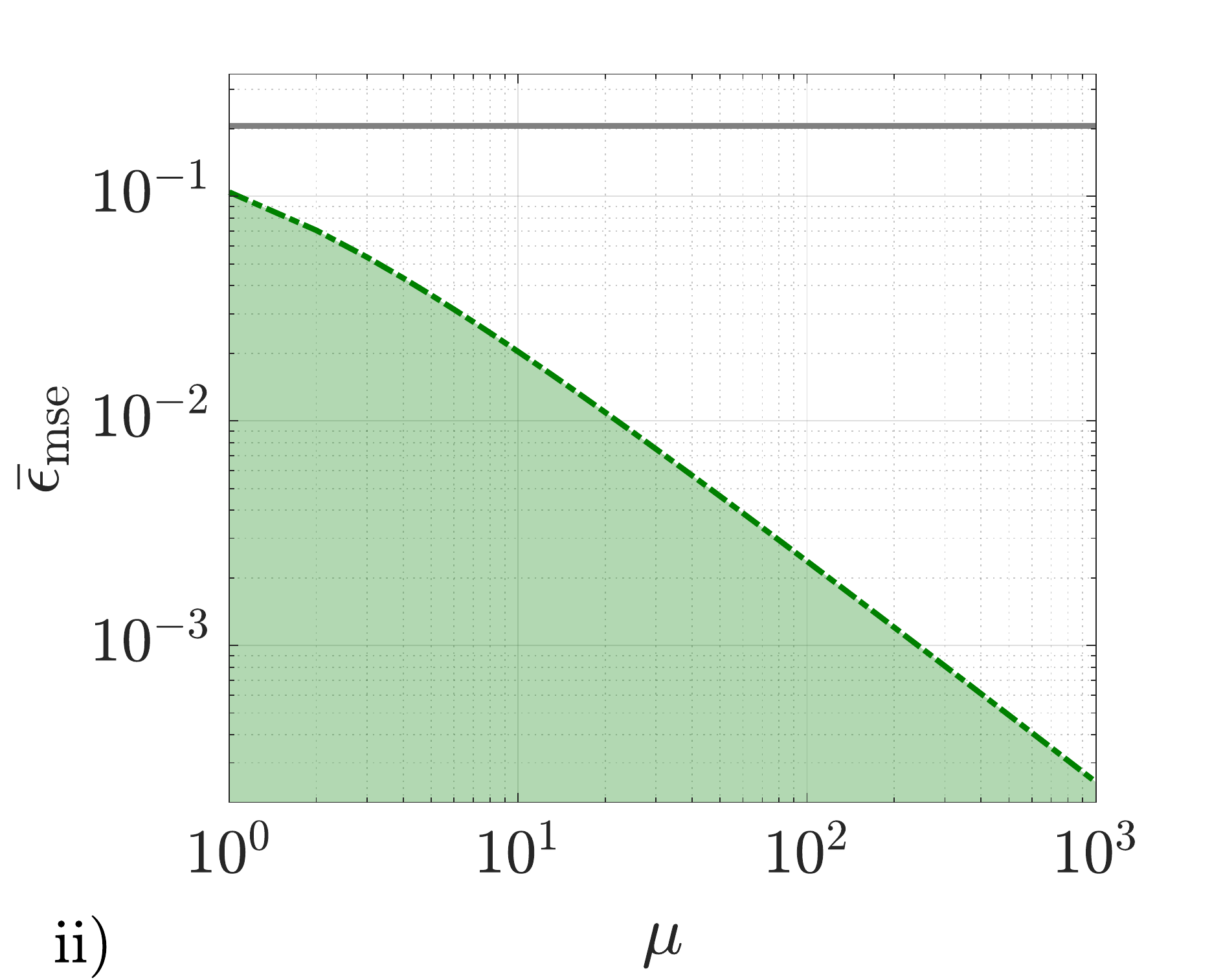}
\includegraphics[trim={0.1cm 0.1cm 0.65cm 0.5cm},clip,width=7.7cm]{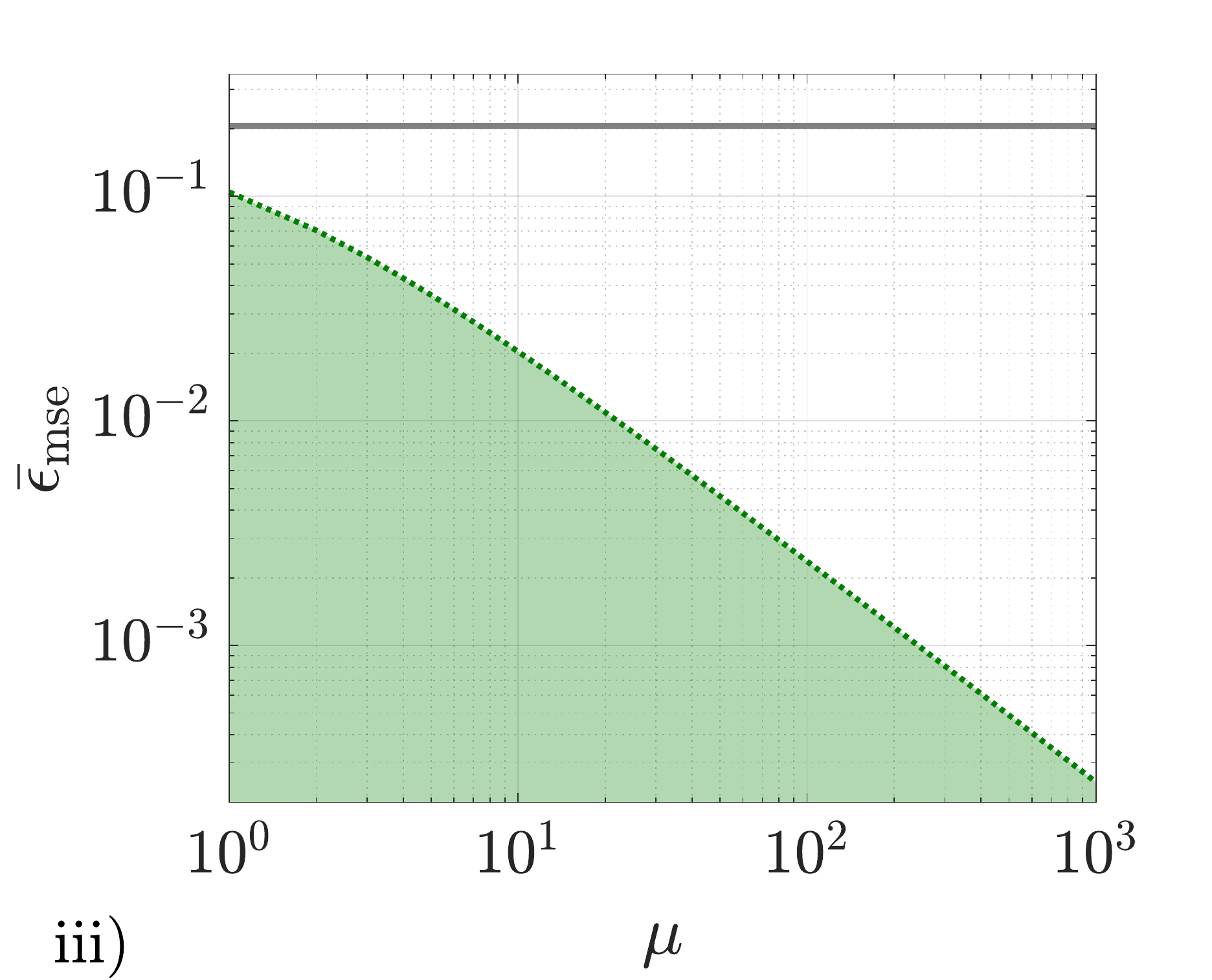}\includegraphics[trim={0.1cm 0.1cm 0.65cm 0.5cm},clip,width=7.7cm]{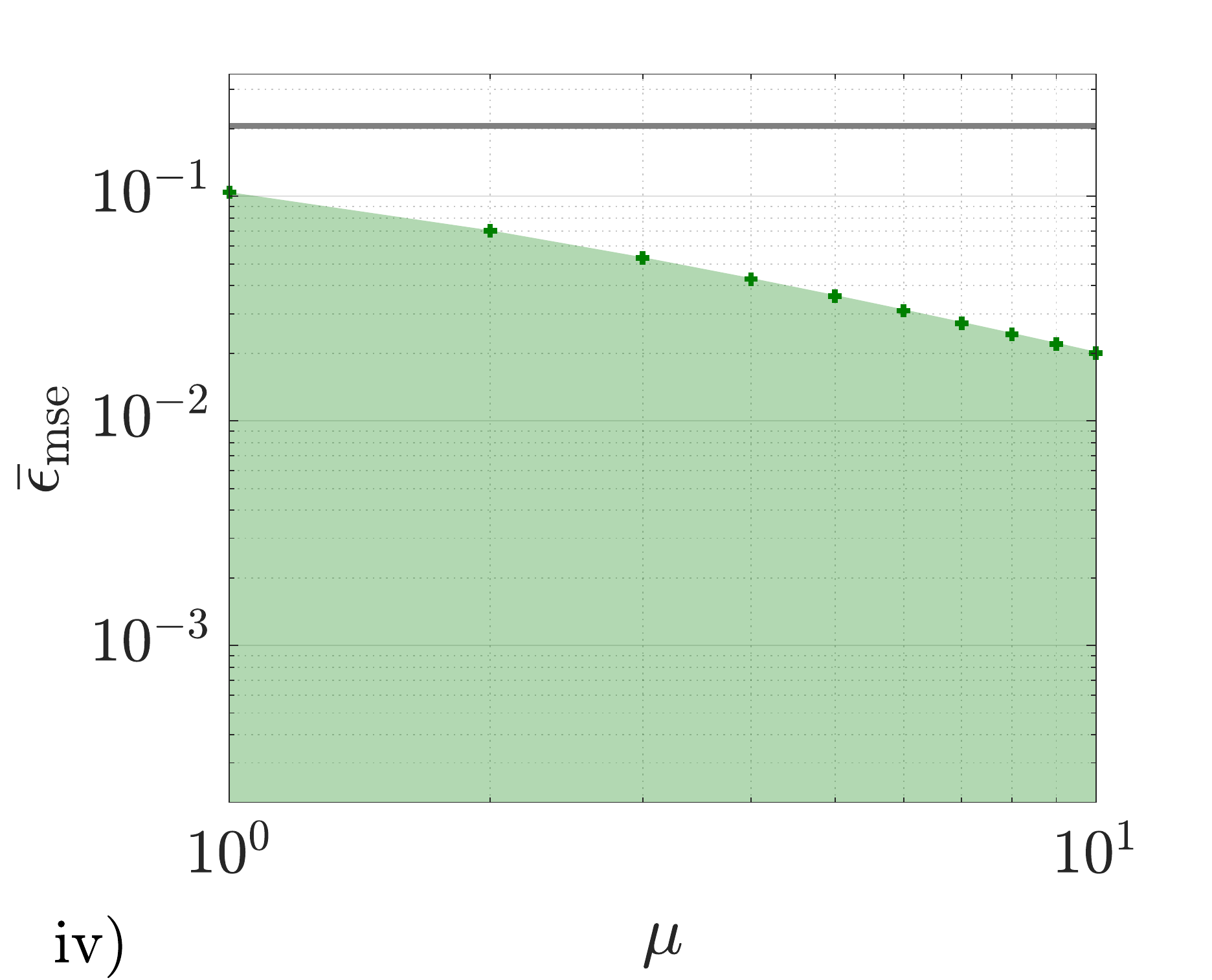}
	\caption[Performance of different POMs with a NOON probe state]{Mean square error based on the optimal single-shot strategy (shaded area), prior variance (horizontal solid line) and error associated with (i) the measurement of energy (dashed line), (ii) the measurement of quadratures (dash-dot line), (iii) parity measurements (dotted line), and (iv) the optimal collective measurement on $\mu$ copies of the probe (plus signs), for a NOON probe state with $\bar{n}=2$, $\bar{\theta} = 0$ and $W_0 = \pi/2$.}
\label{noonPOM}
\end{figure}

We start considering the two measurement schemes that we described in the previous section, that is, counting photons and measuring rotated quadratures after the introduction of some phase shifts that are indicated in table \ref{POM_summary}, and after the action of a $50$:$50$ beam splitter. The mean square errors generated by them for the NOON state, which are represented in figures \ref{noonPOM}.i and \ref{noonPOM}.ii, respectively, display a perfect agreement with the bounds for any number of repetitions. This can be further verified by observing that the numerical uncertainties for the first ten shots provided in table \ref{noon_POM_summary} are virtually identical. Additionally, for the photon counting measurement and a single shot it can be easily shown in an analytical fashion. To see it, first we calculate the likelihood function (see appendix \ref{sec:optrans}), finding that
\begin{eqnarray}
p(n,2-n|\theta) = \frac{\mathrm{cos}^2\left[\theta + (2n-3)\pi/4\right]}{ n! (2-n)!}.
\label{noonlikelihoodlimited}
\end{eqnarray}
Recalling that $n = 0, 1, 2$, equation (\ref{noonlikelihoodlimited}) can be expressed as
\begin{equation}
p(2,0|\theta) = p(0,2|\theta) = \frac{1}{2}~\mathrm{sin}^2\left(\theta - \frac{\pi}{4}\right), ~~ p(1,1|\theta) = \mathrm{cos}^2\left(\theta - \frac{\pi}{4}\right).
\label{likelihood_noon}
\end{equation}
Next we need to find the normalisation term of Bayes theorem, that is,
\begin{equation}
p(n,2-N) = \frac{2}{\pi} \int_{-\pi/4}^{\pi/4} d\theta p(n,2-n|\theta).
\label{normalisationbayes}
\end{equation}
Introducing equation (\ref{likelihood_noon}) in the formula for $p(n,2-n)$ we find that $p(2,0) = p(0,2) = 1/4$ and $p(1,1)=1/2$. At the same time, this gives us the last piece that we need to apply Bayes theorem and find the posterior probability $p(\theta|n,2-n) = p(\theta)p(n,2-n|\theta)/p(n,2-n)$, which in our case is
\begin{equation}
p(\theta|2,0) = p(\theta|0,2) = \frac{4}{\pi}\hspace{0.1em}\mathrm{sin}^2\left(\theta - \frac{\pi}{4}\right), ~~ p(\theta|1,1) = \frac{4}{\pi}\hspace{0.1em}\mathrm{cos}^2\left(\theta - \frac{\pi}{4}\right).
\label{posterior_noon}
\end{equation}
Now we observe that it is possible to rewrite the classically-optimal single-shot mean square error in equation (\ref{classicalbound}) as
\begin{equation}
\bar{\epsilon}_{\mathrm{mse}}(\mu=1) = \int d\theta p(\theta) \theta^2 - \int dn~ p(n,2-n) g_{\mathrm{opt}}(n,2-n)^2,
\end{equation}
where
\begin{equation}
g_{\mathrm{opt}}(n,2-n) = \int_{-\pi/4}^{\pi/4} d\theta p(\theta|n,2-n)\theta
\end{equation}
is the optimal estimator. Taking into account that $g_{\mathrm{opt}}(2,0) = g_{\mathrm{opt}}(0,2) = -1/\pi$ and $g_{\mathrm{opt}}(1,1) = 1/\pi$, the error associated to this POM is
\begin{equation}
\bar{\epsilon}_{\mathrm{mse}}(\mu=1) = \frac{2}{\pi}\int_{-\pi/4}^{\pi/4} d\theta \theta^2 - \frac{1}{\pi^2} = \frac{\pi^2}{48} - \frac{1}{\pi^2}.
\label{msesingleshotoptnoon}
\end{equation}
On the other hand, the single shot quantum bound is, in this case,
\begin{equation}
\bar{\epsilon}_{\mathrm{mse}}(\mu=1) \geqslant \frac{2}{\pi}\int_{-\pi/4}^{\pi/4} d\theta \theta^2 - \mathrm{Tr}(\bar{\rho}S) =
\frac{\pi^2}{48} - \frac{1}{\pi^2},
\label{mse_analytical_noon}
\end{equation}
which is the exact value found in equation (\ref{msesingleshotoptnoon}) (see the calculation of $\bar{\rho}$ and $S$ for the NOON state in section \ref{main_results}). Hence, the measurement under consideration saturates the single-shot bound, as we expected\footnote{We also notice the agreement of $\pi^2/48 - 1/\pi^2 \approx 0.104$ with the numerical results showed in table \ref{noon_POM_summary}.}.

\begin{table} [t]
\centering
{\renewcommand{\arraystretch}{1.2} 
\begin{tabular}{|c|c|c|c|c|}
\hline 
\multicolumn{5}{|c|}{$\bar{\epsilon}_{\mathrm{mse}}\left(\mu = 1\right)$, $\dots$, $\bar{\epsilon}_{\mathrm{mse}}\left(\mu = 10 \right)$} \\
\hline
\hline
\multicolumn{5}{|c|}{NOON state} \\
\hline
\begin{tabular}{@{}c@{}}Single-shot \\ POM \end{tabular} & \begin{tabular}{@{}c@{}}$50$:$50$ splitter \\ \& counting\end{tabular} & \begin{tabular}{@{}c@{}}$\pi/8$  \\ quadra. \end{tabular}  &  \begin{tabular}{@{}c@{}}Parity \\ POMs \end{tabular} & \begin{tabular}{@{}c@{}}Collective \\ POMs \end{tabular} \\
\hline 
$1.04\cdot 10^{-1}$ & $1.04\cdot 10^{-1}$ & $1.04\cdot 10^{-1}$ & $1.04\cdot 10^{-1}$ & $1.04\cdot 10^{-1}$ \\
$7.06\cdot 10^{-2}$ & $7.06\cdot 10^{-2}$ & $7.06\cdot 10^{-2}$ & $7.06\cdot 10^{-2}$ & $7.02\cdot 10^{-2}$ \\
$5.36\cdot 10^{-2}$ & $5.36\cdot 10^{-2}$ & $5.36\cdot 10^{-2}$ & $5.35\cdot 10^{-2}$ & $5.31\cdot 10^{-2}$ \\
$4.33\cdot 10^{-2}$ & $4.33\cdot 10^{-2}$ & $4.33\cdot 10^{-2}$ & $4.32\cdot 10^{-2}$ & $4.28\cdot 10^{-2}$ \\
$3.63\cdot 10^{-2}$ & $3.63\cdot 10^{-2}$ & $3.63\cdot 10^{-2}$ & $3.63\cdot 10^{-2}$ & $3.59\cdot 10^{-2}$ \\
$3.14\cdot 10^{-2}$ & $3.13\cdot 10^{-2}$ & $3.13\cdot 10^{-2}$ & $3.13\cdot 10^{-2}$ & $3.09\cdot 10^{-2}$ \\
$2.76\cdot 10^{-2}$ & $2.76\cdot 10^{-2}$ & $2.76\cdot 10^{-2}$ & $2.76\cdot 10^{-2}$ & $2.72\cdot 10^{-2}$ \\
$2.46\cdot 10^{-2}$ & $2.46\cdot 10^{-2}$ & $2.46\cdot 10^{-2}$ & $2.46\cdot 10^{-2}$ & $2.43\cdot 10^{-2}$ \\
$2.23\cdot 10^{-2}$ & $2.23\cdot 10^{-2}$ & $2.23\cdot 10^{-2}$ & $2.23\cdot 10^{-2}$ & $2.20\cdot 10^{-2}$ \\
$2.03\cdot 10^{-2}$ & $2.03\cdot 10^{-2}$ & $2.03\cdot 10^{-2}$ & $2.03\cdot 10^{-2}$ & $2.00\cdot 10^{-2}$ \\ 
\hline
\end{tabular}}
\caption[Numerical results for NOON states]{Mean square error for the NOON state using the optimal single-shot POM, the physical measurement schemes described in the main text and collective measurements, with $1\leqslant\mu\leqslant 10$, $\bar{n}=2$, $\bar{\theta}=0$ and $W_0=\pi/2$. We note that the calculation for collective measurements has been performed with a different numerical algorithm (see the Mathematica code in appendix \ref{sec:singleshotalgorithm}).}
\label{noon_POM_summary}
\end{table}

Similarly, a parity measurement based on the projectors of the observable $\Pi_1\otimes \Pi_2 =  (-1)^{a_1^{\dagger}a_1}\otimes (-1)^{a_2^{\dagger}a_2}$ \cite{gerry2010, chiruvelli2011}, and performed after introducing an extra phase shift and the action of a beam splitter (see table \ref{POM_summary}), also saturates the bound for all $\mu$, as it can be observed in figure \ref{noonPOM}.iii. This is consistent with the fact that the information about the phase is actually contained in the parity of the number of photons \cite{kolodynski2014, gerry2010, chiruvelli2011}. Interestingly, we have verified that counting photons and checking the parity at each port produces the same non-asymptotic results for the indefinite photon number states too. 

That different physical schemes are able to saturate the same quantum bound can be explained by recalling that the optimal quantum estimator $S$ is only defined on the support of $\rho$. In particular, for NOON states $\rho$ can be represented by a non-singular ($2 \times 2$) matrix in the number basis (see section \ref{main_results}), which is only a part of the full space including all the sectors with any number of photons. As a consequence, any measurement that coincides with the projectors $\ket{s_1}$ and $\ket{s_2}$ given in equation (\ref{noon_projectors}) in the part of the space that corresponds to the support of $\rho$ is going to be optimal, independently of the particular form of the POM elements. 

Furthermore, the same intuition can be used to understand why it is more difficult to saturate the bounds for indefinite photon number states in the non-asymptotic regime. For these states there is a non-zero probability of detecting any number of photons at each port of the interferometer, which implies that the optimal quantum estimator $S$ can be constrained in all the sectors of the operator space, and these constraints need to be fully satisfied to saturate the single-shot bound. However, as we accumulate more data we start to approach the quantum Cram\'{e}r-Rao bound, which is based on the equation $L(\theta) \rho(\theta) + \rho(\theta) L(\theta) = 2 \partial \rho(\theta)/\partial\theta$, and this equation only has a unique solution on the support of $\rho(\theta)$ \cite{genoni2008}, which in our case is simply a pure state. That is, finding physical measurements that saturate the asymptotic bounds is generally less demanding and, in fact, the errors of the physical measurements in figures \ref{continuousPOM}.i - \ref{continuousPOM}.iv converge to the fundamental bound. 

This state of affairs gives rise to an interesting situation. The Bayesian bounds in figure \ref{bounds_results}.i show that, in principle, the NOON state is not the best option among the probes that we are examining for any number of repetitions. In spite of this fact, if we compare the uncertainty associated with counting photons after undoing the preparation of a coherent state, the measurement of quadratures for the states based on the squeezing operator,  and any of the physical measurement previously discussed for the NOON state, then it can be shown that, in this case, the NOON state is the best probe when $1 \leqslant \mu \leqslant 3$. In particular, this conclusion can be extracted by inspection from tables \ref{practical_POM_summary} and \ref{noon_POM_summary}. This analysis highlights the importance of studying the possibility of saturating the theoretical bounds using realistic implementations in a particularly transparent way.

On the other hand, the mathematical simplicity of NOON states allows us to go one step further and study collective measurements \cite{jarzyna2015true, jarzyna2016thesis}. Until now this work has followed the model for repetitions that we introduced in section \ref{sec:problem}. However, we also saw that a more general possibility is to prepare $\mu$ identical copies of some probe and perform a single measurement on all of them at once. Given that repeating an experiment is generally easier than implementing collective techniques, it would interesting to find out whether collective POMs produce better uncertainties in those schemes whose associated calculations are tractable.

If we upgrade the optimal single-shot bound in equation (\ref{singleshot_bound}) to cover the collective case we find that
\begin{equation}
\bar{\epsilon}_\mathrm{mse} \geqslant \int d\theta p(\theta) \theta^2 - \mathrm{Tr}\left(\bar{\rho}_\mu S_\mu\right),
\label{singleshot_collective}
\end{equation} 
where now $S_\mu$ is given by $S_\mu \rho_\mu + \rho_\mu S_\mu = 2 \bar{\rho}_\mu$ with 
\begin{equation}
\rho_\mu = \int d\theta p(\theta) \smash[b]{ \underbrace{\rho(\theta) \otimes \cdots \otimes \rho(\theta)\,}_\text{$\mu$ times}}
\label{zerothmoment_collective}
\end{equation}
and
\begin{equation}
\bar{\rho}_\mu = \int d\theta p(\theta) \smash[b]{ \underbrace{\rho(\theta) \otimes \cdots \otimes \rho(\theta)\,}_\text{$\mu$ times}}\theta.
\label{firstmoment_collective}
\end{equation}
~\\[-10pt]

An algorithm to calculate equation (\ref{singleshot_collective}) for NOON states is proposed in appendix \ref{sec:singleshotalgorithm}, and its application for $1\leqslant \mu \leqslant 10$ results in the graph of figure \ref{noonPOM}.iv, which coincides with the bound generated by repeating the optimal strategy for a single probe\footnote{Unfortunately, it becomes numerically challenging to increase the number of copies, which is why we only consider $1\leqslant \mu \leqslant 10$ for this calculation.}. Numerically, this agreement occurs at least for the first significant figure, as it can be verified in table \ref{noon_POM_summary}. Thus we conclude that collective measurements do not provide a better performance than the practical measurements previously studied when we are working in the low-$\mu$ regime, each probe is prepared in a NOON state with $\bar{n} = 2$ and the prior width is $W_0=\pi/2$.

In summary, we have shown that there are measurements that can saturate the bound for the NOON state for all $\mu$ simultaneously. Consequently, NOON states do not only have a special status in the local regime, but also in the regime of limited data and moderate prior knowledge\footnote{In \cite{jarzyna2016thesis} it is argued that NOON states emerge as the optimal probe with a definite number of photons in the limit where the prior information dominates, something that is shown in \cite{demkowicz2011}, and it is concluded that, for that reason, using NOON states is almost useless in a practical Bayesian context. Although it is true that NOON states are limited due to the ambiguity that they introduce in the estimation and, more importantly, because of the difficulties to use them in real experiments \cite{schafermeier2018}, we draw attention to the fact that the regime where the prior knowledge may play a substantial role is relevant and useful whenever we need to make inferences from a practical scenario where only a low number of experiments can be performed.}. This can be explained by noticing that the optimal projectors for a single shot in equation (\ref{noon_projectors}) are the same that the projectors predicted by the symmetric logarithmic derivative that defines the quantum Fisher information \cite{kolodynski2014}. While this probe state is fragile and difficult to prepare in more realistic scenarios \cite{schafermeier2018}, these results are still interesting from a fundamental perspective, and they have helped us to understand the problems associated with saturating the bounds of more practical states that we need to overcome in the future. 

\section{Comparing our method with the alternative quantum bounds}
\label{sec:alternativevssingleshot}

The shot-by-shot strategy that we have constructed generates valid lower bounds for repetitive experiments. The fact that they are based on the single-shot optimum implies that, in a sense, they are fundamental in scenarios where the experiment is repeated, and we have seen that there is a constructive way of finding the theoretical POM that reaches them. 

A crucial aspect of our tool is that the calculations associated with it can be performed in an efficient way for practical schemes, having provided an algorithm with analytical and numerical components to achieve that goal. However, one could argue that some of the quantum bounds for low $\mu$ that we reviewed in section (\ref{subsec:alternativebounds}) are still computationally simpler, even when in general they also require a numerical treatment. In this section we analyse the relative merit of employing our strategy when the results generated by the latter are compared with the predictions of two alternative tools: the quantum Ziv-Zakai and Weiss-Weinstein bounds \cite{tsang2012, tsang2016}.

The Ziv-Zakai bound in equation (\ref{qzzb}) is already in a form that we can apply to our optical configuration, and the numerical algorithm that implements this operation can be found in appendix \ref{subsec:qzzbnum}. On the other hand, the Weiss-Weinstein bound in equation (\ref{qwwbpriorterm}) is expressed in terms of the quantity $f_c(s,\theta) = \int d\theta' p(\theta' + \theta)^s p(\theta')^{1-s}$, for $\lbrace \theta', p(\theta')\neq 0\rbrace$. A choice for $s$ that tends to produce tighter bounds is $s=1/2$ \cite{tsang2016, weinstein1988}, and we will also use it here. In that case, the previous quantity is simplified as $f_c(1/2,\theta) = \int d\theta' \sqrt{p(\theta' + \theta) p(\theta')}$, for $\lbrace \theta', p(\theta')\neq 0\rbrace$, which measures the overlap between the prior probability and a displaced version of it. For the flat prior of width $W_0$ that we are employing this overlap is simply $f_c(1/2,\theta) = 1-\abs{\theta}/W_0$. Consequently, the Weiss-Weinstein bound in equation (\ref{qwwbpriorterm}) becomes
\begin{equation}
\bar{\epsilon}_{\mathrm{mse}} \geqslant \sup_{\theta} \frac{\theta^2 \left(1-\frac{\theta}{W_0}\right)^2 \abs{f(\theta)}^{4\mu}/2}{\abs{f(\theta)}^{2\mu} - \left(1-\frac{2\theta}{W_0}\right)\mathrm{Re}\left\lbrace \left[f(\theta)^{2} {f(2\theta)}^{*}\right ]^\mu \right\rbrace},
\label{qwwbsimpler}
\end{equation}
with $0 \leqslant \theta < W_0 $. The algorithm to find this bound is provided in appendix \ref{subsec:qwwbnum}. 

\begin{figure}[t]
\centering
\includegraphics[trim={0.1cm 0.1cm 0.65cm 0.5cm},clip,width=7.7cm]{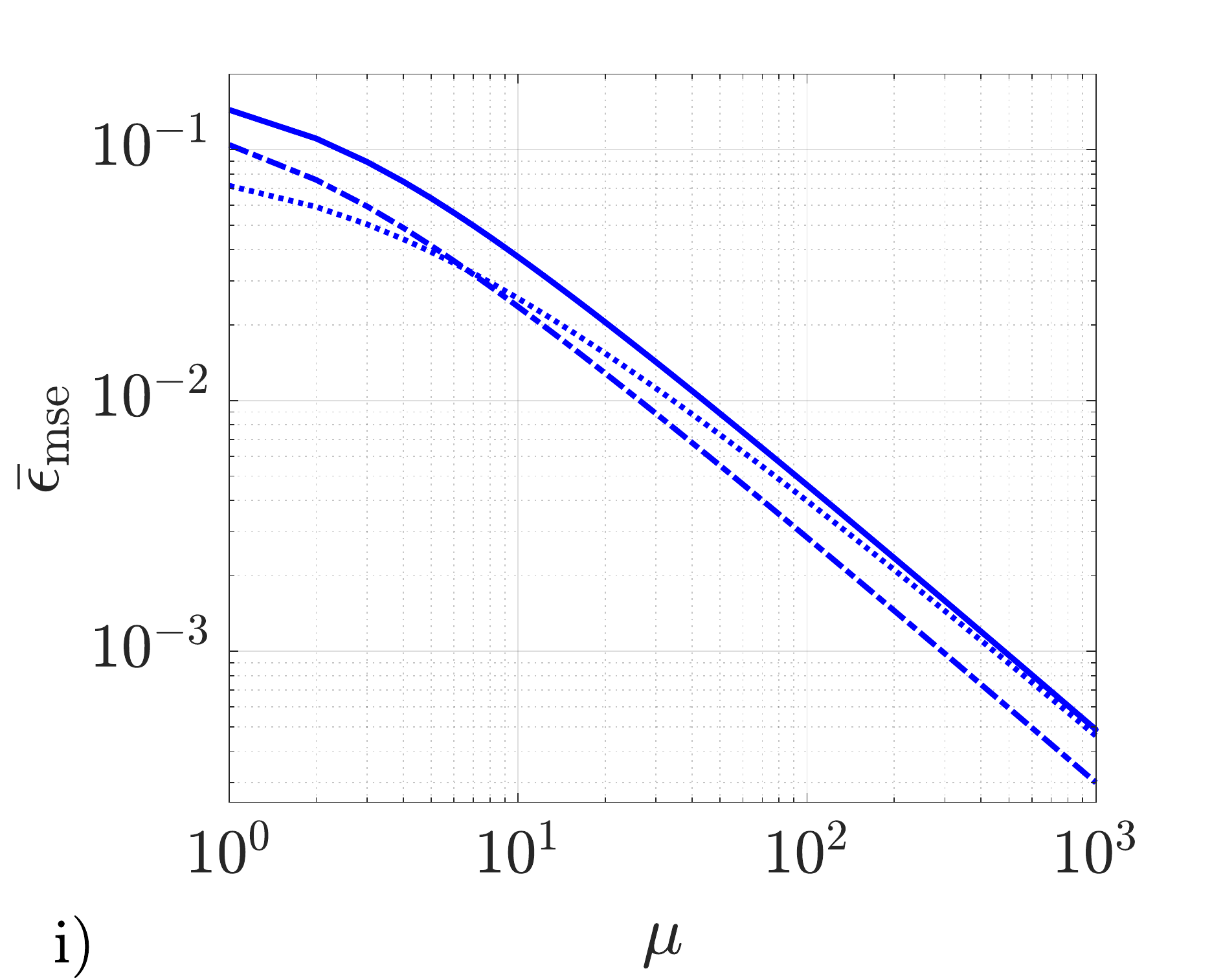}\includegraphics[trim={0.1cm 0.1cm 0.65cm 0.5cm},clip,width=7.7cm]{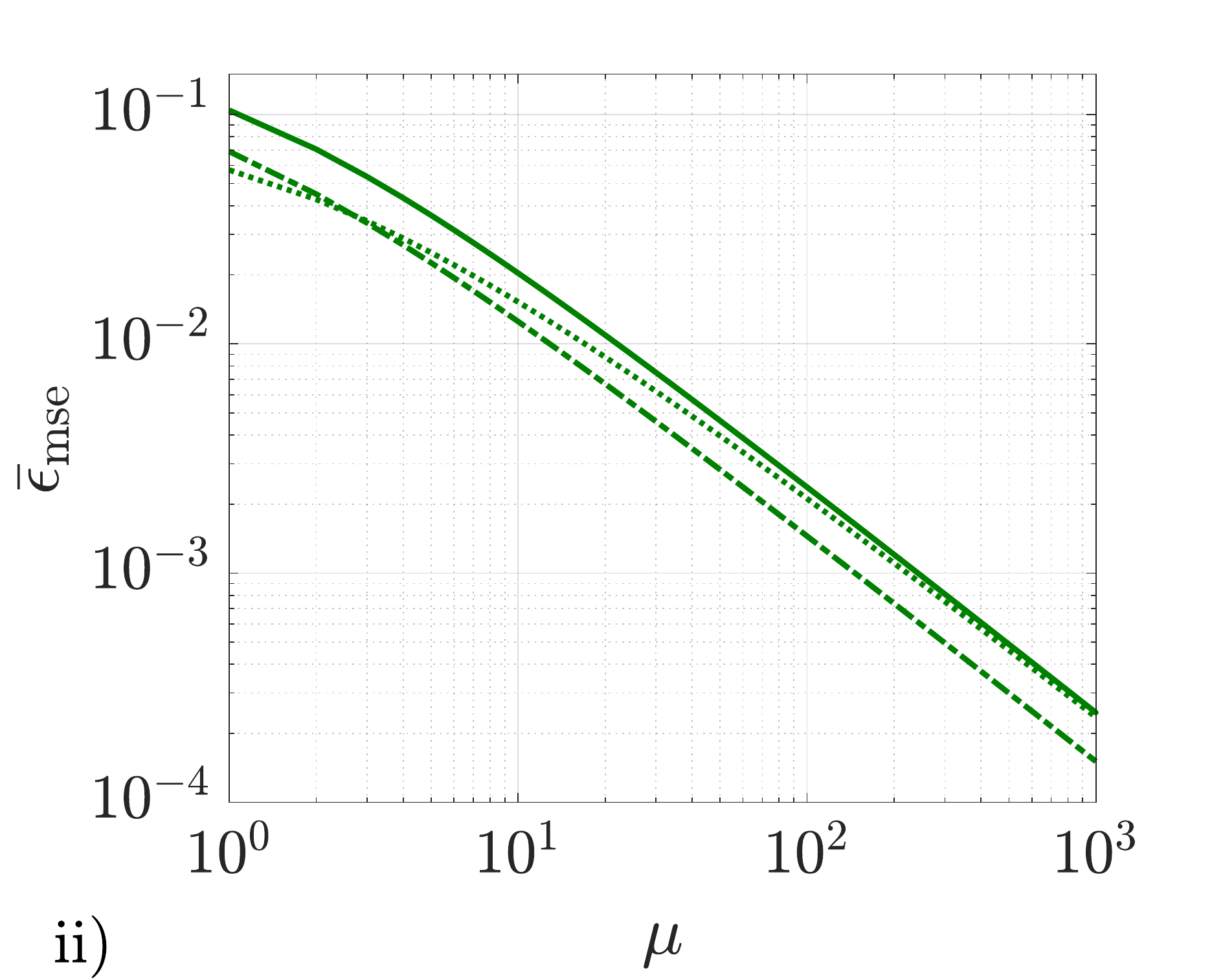}
\includegraphics[trim={0.1cm 0.1cm 0.65cm 0.5cm},clip,width=7.7cm]{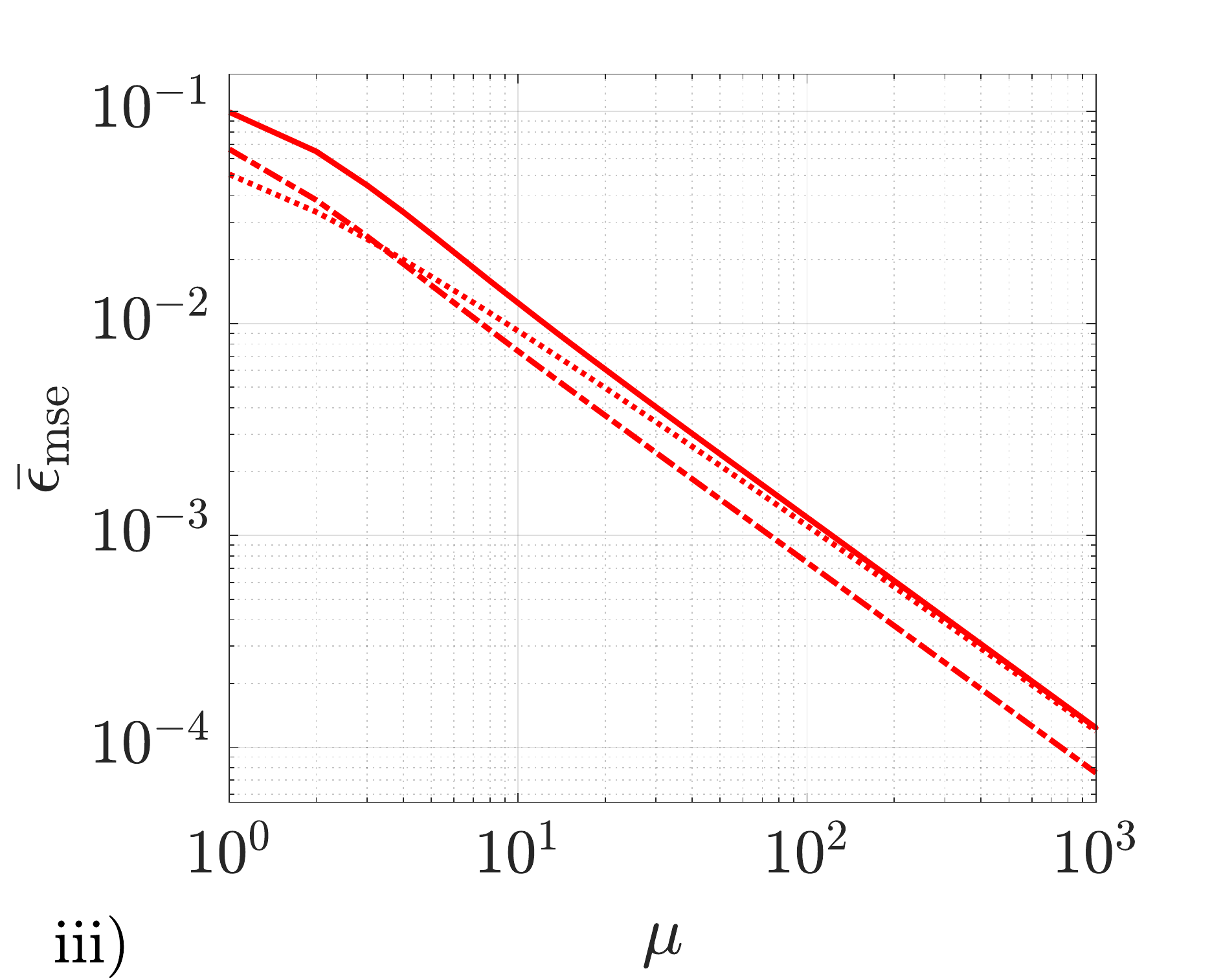}\includegraphics[trim={0.1cm 0.1cm 0.65cm 0.5cm},clip,width=7.7cm]{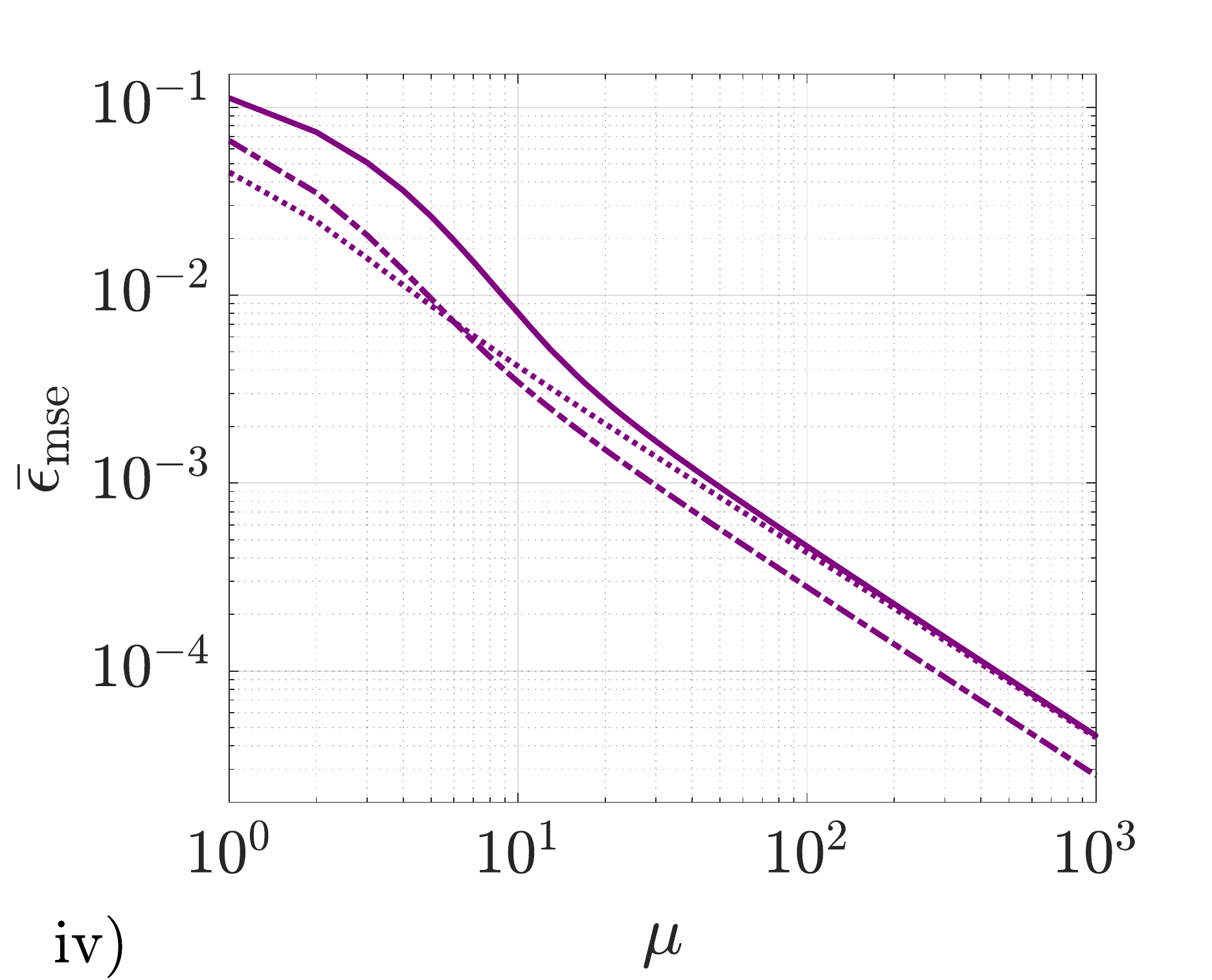}
\caption[Shot-by-shot bound versus Ziv-Zakai and Weiss-Weinstein]{Shot-by-shot strategy (solid line), quantum Ziv-Zakai bound (dash-dotted line) and quantum Weiss-Weinstein bound (dotted line) for (i) the coherent state, (ii) the NOON state, (iii) the twin squeezed vacuum state, and (iv) the squeezed entangled state, with $\bar{n} = 2$ and $W_{\mathrm{int}} = \pi/2$.}
\label{consistency}
\end{figure}

Figure \ref{consistency} shows the results of performing the previous calculations for (i) the coherent state, (ii) the NOON state, (iii) the twin squeezed entangled state and (iv) the twin squeezed cat state, where we have also included the bounds generated by repeating the single-shot optimal strategy (solid lines). As we can observe, neither the Ziv-Zakai bound (dash-dotted lines) nor the Weiss-Weinstein bound (dotted lines) coincides with with the single-shot optimum at $\mu = 1$, which means that these bounds are not tight in the single-shot regime. The Weiss-Weinstein bound is actually tight when $\mu \gg 1$, as it was proven in \cite{tsang2016}, while the Ziv-Zakai bound presents the correct scaling in this regime but it is not tight. Despite this, the latter bound is the tighter option when $\mu \sim 1$. 

The comparison for a low $\mu$ such that $\mu > 1$ is more subtle, since the bounds in equations (\ref{qzzb}) and (\ref{qwwbsimpler}) are defined for $\mu$ copies of the probe state, which in principle allows for collective measurements, while our bounds in section (\ref{sec:methodlimited}) are specifically designed for repetitions. Fortunately, we have been able to study collective techniques for NOON states, finding that both collective and repetitive measurements have the same precision for this configuration when $\mu$ is low. For that reason, the results for the NOON state in figure \ref{consistency}.ii demonstrate that the alternative bounds are loose in the intermediate regime. This implies that, in general, the quantum Ziv-Zakai and Weiss-Weinstein bounds are not tight in the non-asymptotic regime of optical metrology, in consistency with the examples in \cite{tsang2016}, and thus we conclude that our technique is preferred whenever we study repetitive experiments\footnote{Nonetheless, we observe that the quantum Ziv-Zakai and Weiss-Weinstein bounds correctly lower-bound the uncertainty for low values of $\mu$, in contrast to the Cram\'{e}r-Rao bound, since these bounds are valid for both biased and unbiased estimators \cite{tsang2012, tsang2016,  bayesbounds2007}.}. 

\section{Practical application: designing quantum experiments with genetic algorithms}
\label{sec:genetic}

The results in the previous sections, together with those in chapter \ref{chap:nonasymptotic}, complete our non-asymptotic methodology for single-parameter estimation problems. Now we would like to go a step further and conclude this chapter by exploiting our methods to design quantum states. In particular, we seek states that not only provide precision enhancements in the regime of limited data, but whose preparation is also associated with a concrete sequence of operations that can be implemented with current technology. Therefore, the findings in this section complement those in sections \ref{measurements_section} and \ref{subsec:optnoon}, where the sequences were only provided for POMs.

This problem belongs to the broader field of quantum state engineering, where ideas from machine learning and artificial intelligence are often utilised \cite{knott2016, jesus2018dec, driscoll2019}. In this context, Knott and his collaborators at Nottingham and Bristol \cite{jesus2018dec} proposed a genetic algorithm to design states in optical experiments. In what follows we show how the combination of these techniques with our methodology for single-parameter schemes provides a solution to the problem in the previous paragraph.

Let us first summarise the key ideas of the algorithm developed by Knott \emph{et al.} \cite{jesus2018dec}, which is called \emph{AdaQuantum}. One starts by creating a numerical toolbox with states, operations and measurements that can be performed with an optical arrangement. For example, the work \cite{jesus2018dec} includes a general splitter 
\begin{equation}
U_{ij}=\mathrm{exp}\left[\phi \left( a_i a_j^\dagger + a_i^\dagger a_j \right)\right]
\end{equation}
with transmissivity $T = \mathrm{cos}^2(\phi)$, the two-mode squeezing operator
\begin{equation}
S_{ij}=\mathrm{exp}\left(\zeta^{*}a_i a_j - \zeta a_i^\dagger a_j^\dagger\right)
\end{equation}
and the single-mode photon counting measurement $\ketbra{n}$, among others. 

The next step is to construct combinations of states, operations and measurements that together generate single-mode states. This is achieved by working in the space of $M$ modes and performing heralding measurements on $(M-1)$ of them, such that the state of the remaining mode is selected on the basis of the measurement outcome \cite{knott2016}. The results in this section have been obtained with $M = 2$. Then the final goal is to find arrangements of these experimental elements that give output states that are good for certain tasks, which is a search problem \cite{jesus2018dec}. 

Each of these combinations is said to be a \emph{genome}, and a collection of genomes is the \emph{population}. In addition, a \emph{fitness function} encoding the properties that we wish the final state to have is defined. This allows the algorithm to examine the values of such function for each of the genomes in the population, and to select those that are the fittest according to the criterion of the fitness function. The latter are then combined or modified, producing a new \emph{generation}, and the process is repeated until the probability of producing new improvements is very small.  

AdaQuantum, which was made available by its authors on GitHub \cite{adaquantum2019}, is flexible enough to accommodate different fitness functions, and this is precisely where our contribution starts. In particular, we prepared a MATLAB module including the numerical algorithms in appendices \ref{sec:singleshotalgorithm}, \ref{sec:pomnum} and \ref{sec:msematlab}, such that the Bayesian mean square error employed here could play the role of a fitness function in the regime of limited data and a realistic amount of prior knowledge. 

\begin{table} [t]
\centering
{\renewcommand{\arraystretch}{1.5} 
\begin{tabular}{|l|c|c|c|c|c|}
\hline
Setting & $\ket{\psi_{in}}$ & $\mathcal{O}_1$ & $\mathcal{O}_2$ & $\mathcal{O}_3$ & $\ketbra{n}$ \\
\hline
Ada, $\mu = 8$ & $|0,0\rangle$ & $S_{12}(\zeta = 0.89~\mathrm{e}^{i 0.031})$ & $U_{12}(T=0.69)$ &  $\mathrm{e}^{i N_1 0.32}$ & $\ketbra{4}$  \\
Ada, $\mu = 4$, $12$ & $|0,0\rangle$ & $S_{12}(\zeta = 0.91~\mathrm{e}^{i 0.040})$ & $U_{12}(T=0.66)$ & --- & $\ketbra{6}$ \\
Ada, $\mu = 1$ & $|0,0\rangle$ & $S_{12}(\zeta = 0.95~\mathrm{e}^{i 6.1})$ & $U_{12}(T=0.72)$ & --- & $\ketbra{2}$\\
\hline
\end{tabular}}
\caption[Sequences of quantum operations generated by AdaQuantum]{Details of the sequences generated by AdaQuantum using the Bayesian framework. The first two schemes are for a photon counting measurement after a $50$:$50$ beam splitter, and the last one is for the optimal single-shot POM. Note that $\ketbra{n}$ implements the heralding measurement that produces the single-mode state.}
\label{toolbox_summary}
\end{table}

Since AdaQuantum produces single-mode states and these cannot be used on its own because in experiments we can only access the information about the difference of phase shifts, one approach is to take a pair of such states as the input of our interferometer, that is, $\ket{\psi_0} = \ket{\psi}\otimes \ket{\psi}$, where $\ket{\psi}$ is the outcome of the genetic algorithm. The task given to AdaQuantum is thus to find the state $\ket{\psi}$ that minimises $\bar{\epsilon}_{\mathrm{mse}}$ for a given number of repetitions and measurement scheme. 

While the algorithm for the Bayesian error in appendix \ref{sec:msematlab} is relatively efficient when we only require the performance of a few schemes (section \ref{subsec:numalgorithm}), the genetic algorithm needs to calculate this error a large number of times in order to successfully evolve the optimal strategy through different generations, and this is numerically demanding. For that reason, we have chosen a narrow flat prior with $W_0 = \pi/12$ and $\bar{\theta} = 0$, so that the calculations associated with the integrals in equations (\ref{erropt}) and (\ref{shotbyshotmse}) for the mean square error are simplified\footnote{Notice that, according to our discussion in sections \ref{subsec:shotbyshot} and \ref{prior_section}, $W_0 = \pi/12 \approx 0.3$ is still a moderate amount of prior information.}. In addition, AdaQuantum assumes that $\bar{n} = 1$ for our configuration. As a consequence, the strategies in this section cannot be directly compared with those in the first part of this chapter.

\begin{figure}[t]
\centering
\includegraphics[trim={0.1cm 0.1cm 0.65cm 0.2cm},clip,width=7.7cm]{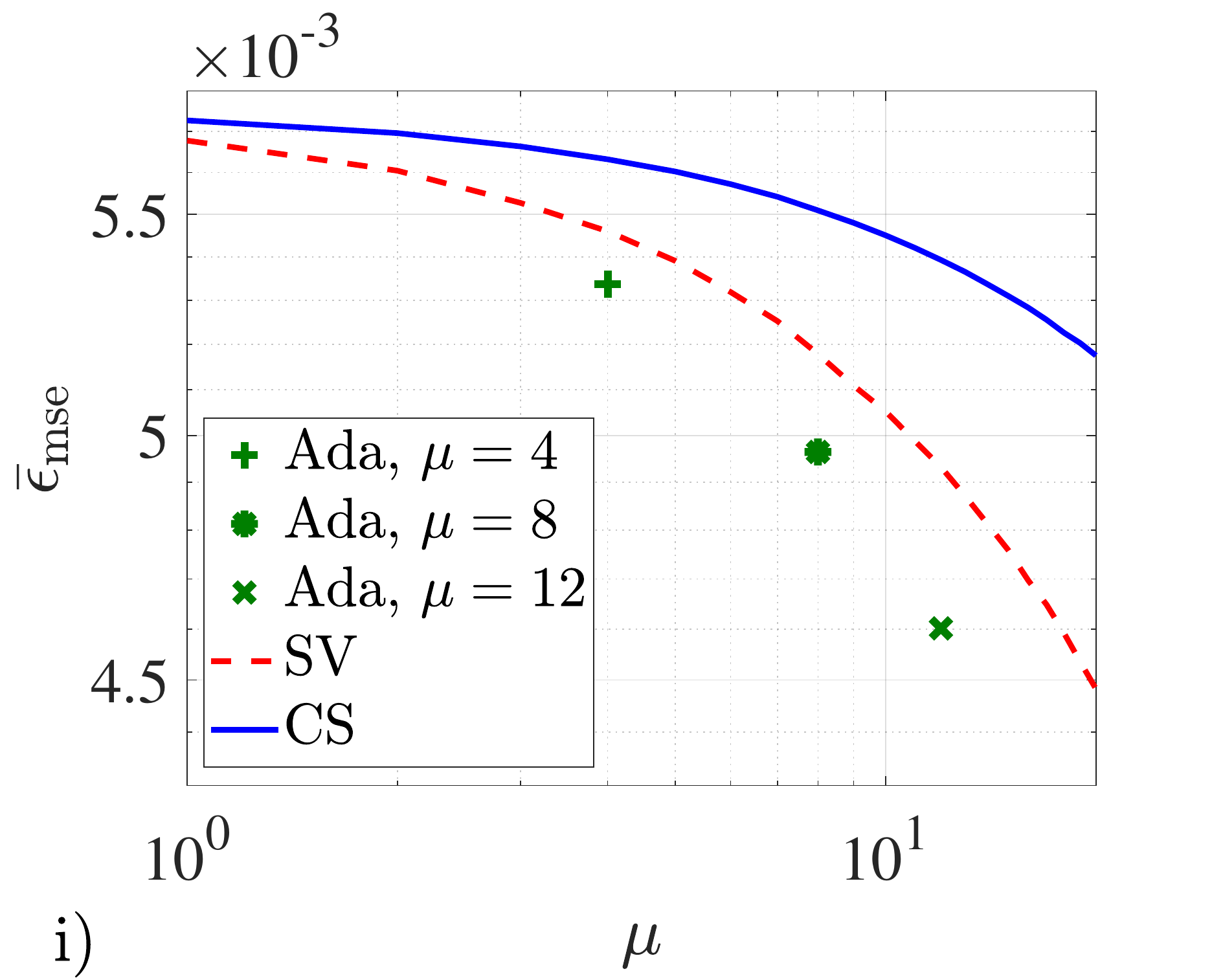}\includegraphics[trim={0.1cm 0.1cm 0.65cm 0.2cm},clip,width=7.7cm]{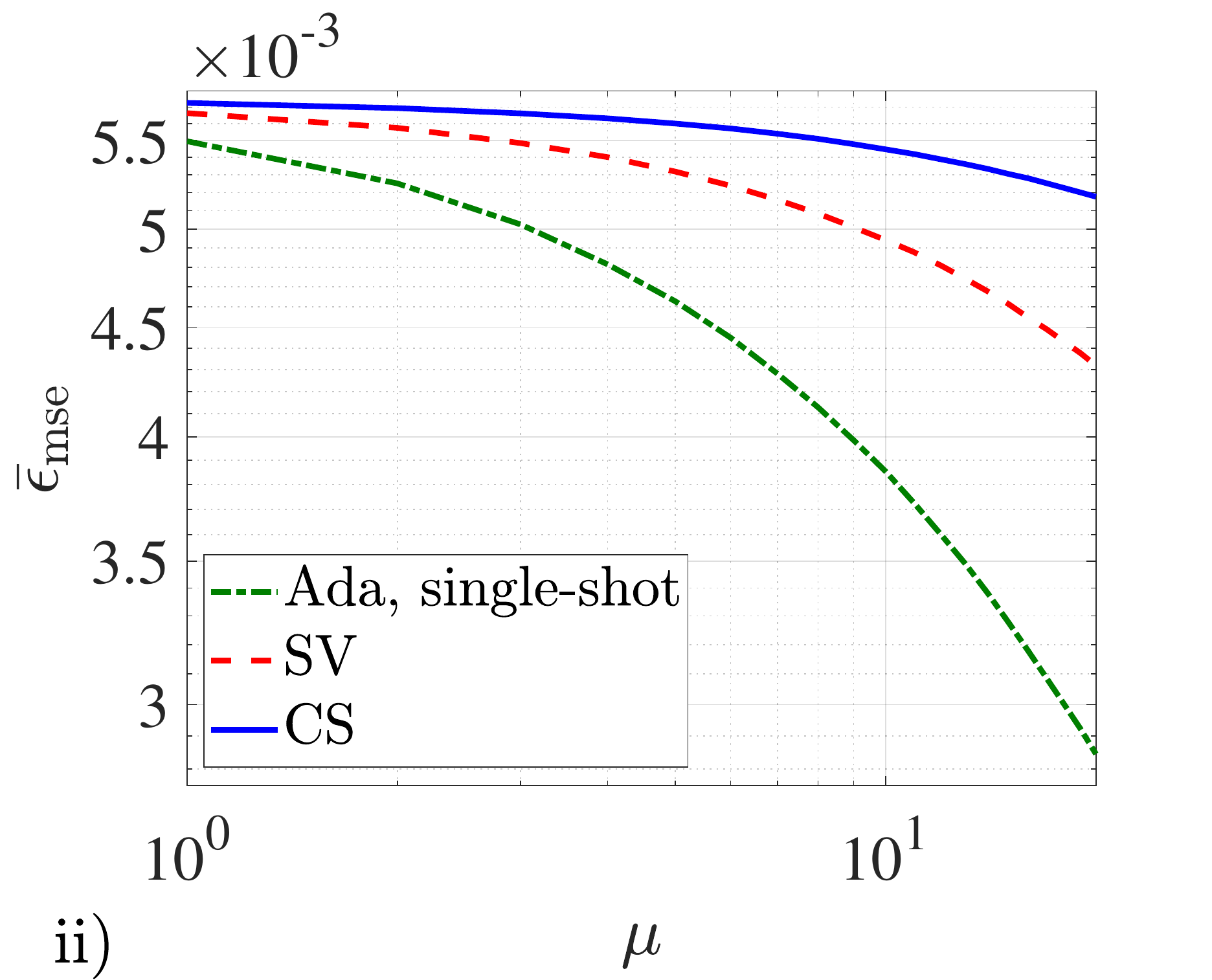}
\caption[Performance of the schemes found by AdaQuantum]{Mean square error as a function of the number of repetitions for (i) the coherent state (solid line, CS), the twin squeezed vacuum (dashed line, SV) and the states found by AdaQuantum for $\mu = 4$ (plus sign), $\mu = 8$ (asterisk) and $\mu = 12$ (cross) repetitions. Here the measurement scheme is based on counting photons after the action of a $50$:$50$ beam splitter. In (ii) we again consider the coherent and twin squeezed vacuum states, but now measured by their respective optimal single-shot POMs. The state found by AdaQuantum (dash-dot line) is then based on our shot-by-shot strategy, which already takes into account the optimal single-shot POM for the given state. All the configurations are based on a two-mode interferometer with $1$ photon on average, $W_0 = \pi/12$ and $\bar{\theta} = 0$.}
\label{bmse_ada_results}
\end{figure}

We will optimise the error using two different strategies. First we focus on one of the practically-motivated POMs that we have studied: counting photons after the action of a $50$:$50$ beam splitter, including an extra phase shift that is known and that takes into account the fact that the prior is centred around zero\footnote{In particular, we have chosen its even version in table \ref{POM_summary}.}, and we set the algorithm to optimise the error for $\mu = 4$, $\mu = 8$ and $\mu = 12$ repetitions. This search produces a state that takes the form $\ket{\psi} = \mathcal{N}\langle n|\hat{U}_{12}\hat{S}_{12}|0,0\rangle$ for $\mu = 4$ and $\mu=12$, where $\mathcal{N}$ is the normalisation, while for $\mu = 8$ we find $\ket{\psi} = \mathcal{N}\langle n|\hat{P}_1\hat{U}_{12}\hat{S}_{12}|0,0\rangle$, where $P_1$ is a phase shift in the first mode. Table \ref{toolbox_summary} provides the numerical parameters for the sequences of operations that would prepare these states in practice, and the uncertainty associated with two copies of the previous probes has been represented in figure \ref{bmse_ada_results}.i (individual points). Furthermore, we have included the errors for the coherent state (solid line, CS) and the twin squeezed vacuum state (dashed line, SV), as references. The motivation to use the latter as a benchmark is that while it illustrates the potential enhancement that quantum resources can provide, it is also a common and well-understood probe state, in contrast with more exotic choices such as those studied in section \ref{sec:methodlimited}. 

Figure \ref{bmse_ada_results}.i shows that the states found by AdaQuantum perform better than the chosen references, which demonstrates that AdaQuantum is able to optimise a Bayesian figure of merit in a regime where the Fisher information is often unsuitable. More concretely, we can quantify this improvement by introducing the quantity
\begin{equation}
I_{\mathrm{r}}=\frac{\bar{\epsilon}_{\mathrm{r}}-\bar{\epsilon}_{\mathrm{ada}}}{\bar{\epsilon}_{\mathrm{r}}},
\label{improvement_factor}
\end{equation}
where $\bar{\epsilon}_{\mathrm{r}}$ is the uncertainty of any of the reference states and a positive $I_{\mathrm{r}}$ indicates that there has been an improvement. Its calculation, whose results are summarised in table \ref{bayesian_ada_improvement}, shows an enhancement of between $2\%$ and $7\%$ with respect to the twin squeezed vacuum, and between $5\%$ and $15\%$ with respect to the coherent state.

\begin{table} [t]
\centering
{\renewcommand{\arraystretch}{1.2} 
\begin{tabular}{|c|c|c|c|c|c|}
\hline
\multicolumn{6}{|c|}{AdaQuantum's relative enhancement} \\
\hline
\hline
\multicolumn{4}{|c|}{$50$:$50$ splitter \& counting } & \multicolumn{2}{c|}{Single-shot POM} \\
\hline
Ref. & $I_{\mathrm{r}}(\mu=4)$ & $I_{\mathrm{r}}(\mu=8)$ & $I_{\mathrm{r}}(\mu=12)$ & Ref. & $I_{\mathrm{r}}(\mu=1)$ \\
\hline
SV & $0.02$ & $0.04$ & $0.07$ & SV & $0.03$ \\
CS & $0.05$ & $0.10$ & $0.15$ & CS & $0.04$ \\
\hline
\end{tabular}}
\caption[AdaQuantum's relative enhancement]{Improvement factor as defined in equation (\ref{improvement_factor}) to quantify the enhancement of the states found by AdaQuantum with respect to the twin squeezed vacuum state (SV) and the coherent (CS). The details of the experimental configuration are those indicated in the caption of figure \ref{bmse_ada_results} and in the main text.}
\label{bayesian_ada_improvement}
\end{table}

The second strategy is to select the optimal single-shot POM given by the eigenstates of the quantum estimator $S$ before the search of AdaQuantum starts, such that the fitness function is the uncertainty in equation (\ref{shotbyshotmse}). For this configuration we find another state with the form  $\ket{\psi} = \mathcal{N}\langle n|\hat{U}_{12}\hat{S}_{12}|0,0\rangle$, but with different parameters (see table \ref{toolbox_summary}). As table \ref{bayesian_ada_improvement} shows, this state is $3\%$ better than the twin squeezed vacuum measured by its correspondent single-shot POM, and $4\%$ better than the coherent state. In addition, we notice that the performance of a scheme where this probe is repeated $20$ times, which has been represented in figure \ref{bmse_ada_results}.ii, shows that the state found by AdaQuantum using the optimal single-shot POM is better than the benchmarks even when the number of repetitions grows. 

To summarise, we can say that the combination of AdaQuantum and the methods introduced in both this chapter and chapter \ref{chap:nonasymptotic} provides a robust method to find practical probe states with a strong performance for those systems that operate in the regime of limited data. Moreover, this may have important consequences for quantum metrology in a more general sense. We recall that, according to our discussion of the work by Macieszczak \emph{et al.} \cite{macieszczak2014bayesian} in section \ref{subsec:fundeq}, a way of finding quantum strategies that approach the optimum is to construct an algorithm where the state and the POM are sequentially optimised, which can be achieved by combining the fundamental equations (\ref{optequations}) found by Helstrom and Holevo \cite{helstrom1976, helstrom1974, holevo1973b, holevo1973} with the minimisation of equation (\ref{errmacieszczak}). In view of the promising results of this section, the possibility of using genetic algorithms with experimentally realisable operations to upgrade the strategy in \cite{macieszczak2014bayesian} suggests itself.

\section{Summary of results and conclusions}

We have proposed to use the strategy that is optimal after minimising the single-shot mean square error over all the possible POMs in a sequence of $\mu$ repeated experiments, completing in this way the part of our non-asymptotic methodology that is dedicated to single-parameter estimation problems. 

Given a state, a generator and a prior probability, we have seen that the bounds that arise from this technique are optimal for the first shot by construction, and that they also start to converge to the quantum Cram\'{e}r-Rao bound when $\mu\sim 10^2$. In addition, we have argued that they can be saturated using measurements that are equivalent to the projectors of the optimal quantum estimator $S$ for each repetition, and that this strategy is optimal for those experiments based on identical and independent trials where adaptive techniques or more general measurements are excluded. Furthermore, the comparison of our method with alternative tools such as the quantum Ziv-Zakai and Weiss-Weinstein bounds has revealed that our shot-by-shot strategy is preferred even when its calculation is not always as simple as that of more standard bounds, since the latter have been shown to be generally loose for optical protocols in the non-asymptotic regime. The calculation of the quantum Ziv-Zakai and Weiss-Weinstein bounds appeared in 
\cite{jesus2017}
\begin{displayquote}
\emph{Non-asymptotic analysis of quantum metrology protocols beyond the Cram\'{e}r-Rao bound}, \underline{Jes\'{u}s Rubio}, Paul Knott and Jacob Dunningham, J. Phys. Commun. 2 015027 (2018).
\end{displayquote}

The usefulness of this method in the context of quantum metrology has been demonstrated through the analysis of a Mach-Zehnder interferometer, and we have focused our study on three indefinite photon number states that have been proposed in the literature due to their large Fisher information: the twin squeezed vacuum state, the squeezed entangled state and the twin squeezed cat state. We have found that the twin squeezed vacuum state is the best option when $1\leqslant\mu <5$, $W_0=\pi/2$, and for $\mu = 1$, $W_0=\pi/3$; that the squeezed entangled state is the preferred choice if $5<\mu <40$, $W_0=\pi/2$
and when $\mu = 1$, $W_0=\pi/3$ or $W_0=\pi/4$; and that the twin squeezed cat state recovers its status of best probe due to its largest Fisher information when $\mu > 40$, $W_0=\pi/2$ and $\mu=1$, $W_0 = 0.1$. To the best of our knowledge, a fully Bayesian analysis in the terms explored in this work had not been done before for these probes. 

Using the twin squeezed cat state as a family of probes whose parameters can be modified for given mean number of photons and prior width, we have provided evidence that suggests that increasing the amount of intra-mode correlations, that is, the correlations within each arm of the interferometer, could be detrimental when the number of repetitions is low, which contrasts with the fact that the same type of correlations are actually beneficial in the asymptotic regime. Moreover, we have shown that using a state with less intra-mode correlations and a certain amount of path entanglement such as the squeezed entangled state appears to help to enhance the precision in the non-asymptotic regime without damaging the asymptotic performance in a dramatic way. Therefore, we conjecture that there might exist a more general relationship between the number of trials and the amount of  intra-mode and inter-mode correlations that could indicate how to reduce the uncertainty of the protocols in the regime of limited data.

It has been shown that, for a low number of trials, the usual strategy of counting photons after the action of a beam splitter is optimal for most practical purposes when the probe is prepared in a coherent state, although it does not saturate the non-asymptotic bounds for the other indefinite photon number states. However, we have found that in the latter case the situation can be improved if instead we measure quadratures rotated by $\pi/8$, since this scheme is closer to our bounds for low $\mu$. This result is particularly relevant because states prepared with operations such as squeezing or displacement from the vacuum and quadrature measurements are easier to implement in real-world situations. In addition, our calculations indicate that counting photons, measuring quadratures and implementing parity measurements are optimal strategies for any number of repetitions if the probe is in a NOON state, and that collective measurements on the first ten copies of this probe do not provide an advantage over the schemes based on identical and independent experiments.

Furthermore, we have addressed the inverse problem, such that the POM is fixed and the initial state is optimised over a set of experimentally feasible quantum operations. To achieve this, we have combined our single-parameter methodology developed in both this chapter and chapter \ref{chap:nonasymptotic} with the genetic algorithm \emph{AdaQuantum} proposed by Knott \emph{et al.} \cite{jesus2018dec}, finding that AdaQuantum is able to select probe states with precision enhancements over the chosen benchmarks for a low number of repetitions, and we have provided the specific sequences of operations that would allow us to prepare such states in the laboratory. This contribution has appeared in one of the sections of \cite{jesus2018dec}
\begin{displayquote}
\emph{Designing quantum experiments with a genetic algorithm}, Rosanna Nichols, Lana Mineh, \underline{Jes\'{u}s Rubio}, Jonathan C. F. Matthews and Paul A. Knott, Quantum Sci. Technol. 4 045012 (2019).
\end{displayquote}

It is important to note that we have not considered what happens in the presence of noise because our aim was to identify the novel effects that emerge directly from having a low number of trials without the interference of other features, which justifies our focus on ideal schemes. However, a comprehensive study of the effect of noise when the available data is limited is also crucial to model realistic scenarios. Although we leave this analysis for the future, in chapter \ref{chap:future}, which is where we will explore several potential ideas for future research, we provide an initial test to demonstrate the application of our method to a scheme where photon losses are present, finding that the qualitative behaviour of our practical results does not seem to change substantially for a reasonable amount of loss. 

We believe that these results constitute an important advance towards the establishment of a practical and useful methodology that will help us to design optimal metrology experiments taking the finite number of trials into account, and that they could play a crucial role in the design of realistic inference schemes once our approach has been combined with other features such as the presence of noise, larger numbers of photons, adaptive techniques or multi-parameter systems. The next two chapters of this thesis are precisely dedicated to extend our single-parameter methodology to the multi-parameter regime. 

The results of this chapter (other than those in sections \ref{sec:alternativevssingleshot} and \ref{sec:genetic}) have been published in \cite{jesus2018}
\begin{displayquote}
\emph{Quantum metrology in the presence of limited data}, \underline{Jes\'{u}s Rubio} and Jacob Dunningham, New J. Phys. 21 043037 (2019).
\end{displayquote}
\chapter{Quantum sensing networks and the role of correlations}
\label{chap:networks}

\section{Goals for the third stage of our methodology}
\label{sec:goals6}

While the methods in the two previous chapters can be applied to a wide range of single-parameter problems, in section \ref{subsec:multischemes} we argued that more realistic scenarios typically involve several pieces of unknown information; consequently, the next step is to accommodate our ideas to this type of schemes, and in this chapter we will generalise the hybrid estimation approach in chapter \ref{chap:nonasymptotic} to the multi-parameter regime. Given that the transition from single-parameter metrology to multi-parameter schemes opens the door to a vast set of new ways of enhancing our estimation protocols \cite{chiara2003, spagnolo2012, chiribella2012, humphreys2013, zhang2014, berry2015, baumgratz2016, knott2016local, gagatos2016gaussian, Szczykulska2016, jasminder2016, proctor2017networked, proctor2017networkedshort, vidrighin2014, szczykulska2017, zhang_lu2017, altenburg2017, zhuang2017, altenburg2018, hall2018, sekatski2019, qian2019, polino2018, roccia2018, jasminder2018, ge2018, eldredge2018, li2019, gatto2019}, it is important that we first identify the subset of estimation problems that we intend to investigate here. 

Let us recall our discussion in section \ref{subsec:multischemes}, where we introduced the quantum network model for distributed sensing proposed by Proctor \emph{et al.} \cite{proctor2017networked, proctor2017networkedshort}. We will focus our attention on a collection of sensors arranged such that a single parameter $\theta_i$ is encoded in the $i$-th sensor, and we will explore whether allowing for correlations between different sensors enhances the overall precision of their estimation. In the context of this configuration, Proctor \emph{et al.} \cite{proctor2017networked, proctor2017networkedshort} have shown that, assuming that the associated generators commute, such correlations are not needed to achieve the uncertainty that is optimal when the inverse of the Fisher information matrix is employed as the figure of merit \cite{proctor2017networked, proctor2017networkedshort}. Crucially, using this measure of uncertainty could be a potential caveat to the results of their framework, since we can always achieve any value for the local variances with the class of \emph{infinite-precision} states studied in section \ref{subsec:infiniteprecision}, while, at the same time, this type of probe can require a high amount of prior knowledge to work and a large number of trials to reach the Cram\'{e}r-Rao bound. Moreover, in chapter \ref{chap:limited} we have provided compelling evidence of the existence of a potential trade-off between the performances in the asymptotic and non-asymptotic regimes. Therefore, a study of the role of entanglement for the estimation of the parameters encoded in each sensor when the network operates in the non-asymptotic regime is needed, and this is one of the multi-parameter problems that both this chapter and chapter \ref{chap:multibayes} will address.

On the other hand, if we consider that the parameters in each sensor are local properties of the network, then a collection of functions of the parameters encoded in different sensors represents global properties. In that case, entangled strategies can provide a notable advantage with respect to local schemes \cite{proctor2017networked, proctor2017networkedshort} and, as a consequence, we can say that whether local or global strategies are preferable crucially depends on the type of information that we wish to extract\footnote{A related conclusion was reached by Altenburg and W\"{o}lk in \cite{altenburg2018}, where the authors further studied sequential strategies, that is, strategies that use entangled states to estimate one global property at a time, and concluded that no type of strategy  local, global or sequential - could be claimed to be optimal in general.}. The estimation of global properties or functions is particularly relevant in problems such as the interpolation of non-linear functions \cite{qian2019} and the determination of the coefficients in Taylor or Fourier expansions of some field \cite{sekatski2019}, and, from a fundamental perspective, it is important to understand the connection between the form of the functions to be estimated and the form of the quantum strategy that yields an optimal precision. For instance, using the model for quantum sensing networks under consideration it has been established that one can find entangled states that beat the best separable probe for the optimal estimation of a single function $f(\boldsymbol{\theta})$ that is linear \cite{proctor2017networked, proctor2017networkedshort, eldredge2018, altenburg2018, qian2019, gatto2019}, while a collection of linear functions that generate an orthogonal transformation (that is, $\boldsymbol{f}(\boldsymbol{\theta}) = V^{\transpose} \boldsymbol{\theta}$ with $V V^{-1}=\mathbb{I}$) can be estimated optimally with a completely local strategy \cite{proctor2017networked}. 

This state of affairs motivates the second and main multi-parameter problem that we wish to study in this chapter. More concretely, the problem of estimating several functions will be formulated using the asymptotic theory first, and then we will take those solutions as a guide to perform a Bayesian analysis of the non-asymptotic regime for some of these schemes. While a general answer is beyond the scope of our methods, we will see that it is possible to arrive to definite conclusions by focusing on a subclass of schemes with sensor-symmetric states, and by modelling its global properties with linear but otherwise general functions. Given that configuration, we will establish and exploit the link between the geometry of the linear functions and the strength of the inter-sensor correlations of the network, and we will explore how this may change as the number of repetitions varies.

Importantly, unlike with the schemes of chapter \ref{chap:nonasymptotic}, whose asymptotically optimal quantum strategies where available in the literature, here we will solve the asymptotic estimation problem explicitly before we perform its associated Bayesian analysis. In particular, suppose we denote the number of functions by $l$, and recall that $d$ is the number of unknown parameters. If the functions can be written in terms of an orthogonal transformation, then we have that $l = d$. Hence, from the previous discussion we see that the next natural step is to search for and examine the connection between global properties and the optimal strategy for their estimation when $l\neq 1$ and $l\neq d$, or when $l = d$ and the functions are linear but not orthogonal. This new intermediate regime is precisely the case that we will consider\footnote{Note that the estimation of several functions that are not orthogonal to each other has already arisen in specific applications. An example of this can be found in \cite{li2019}, where the authors studied a protocol for Ramsey interferometry. The importance of the results in the first half of this chapter rests on the partial generality and the fundamental focus of our approach, which is based on the framework provided by the quantum sensing network model in \cite{proctor2017networked}.}.  

In summary, the third stage of our methodology, which rests on combining the multi-parameter estimator that is exactly optimal with the asymptotically optimal quantum strategy, will serve as a basis to investigate how to harness correlations in multi-parameter problems when different amounts of data are available. Note that, as in chapter \ref{chap:nonasymptotic}, our approach not only provides a characterisation of our schemes in the non-asymptotic regime, but also estimates the number of repetitions and prior knowledge needed to recover the predictions of the Cram\'{e}r-Rao bound. Since the construction and maintenance of entangled networks are likely to be difficult in practice, our proposal may prove to be crucial in the study and implementation of sensing networks that operate with a realistic amount of data. 
 
\section{Methodology (part C)}

\subsection{Relevant information in multi-parameter schemes}
\label{subsec:relevantinfo}

Let us formulate the problem of estimating several functions in terms of the information content that we wish to extract using the network\footnote{We recall that this perspective was also exploited in chapter \ref{chap:methodology} to justify the suitability of different measures of uncertainty.}.

We will assign the terminology \emph{natural or primary properties} to those parameters that appear in the physical characterisation of the scheme. These have been denoted by $\boldsymbol{\theta}$ in the first part of this work. In addition, any collection of functions of them, which we can generally denote by $\boldsymbol{f}(\boldsymbol{\theta}) = (f_1(\boldsymbol{\theta}), \dots, f_l(\boldsymbol{\theta}))$, can be seen as  a set of \emph{derived or secondary properties} that we might wish to find. Importantly, there is a degree of arbitrariness in deciding which quantities are primary and which ones are secondary, and this should be fixed by the concrete application under analysis. For example, in previous chapters we have considered that the difference of optical phase shifts in a Mach-Zehnder interferometer is the natural parameter, instead of considering each phase shift independently and the difference as a function of them. 

The information that characterises the network is complete once we know all the natural parameters. In that case, we could also calculate any function of them. However, if the functions of interest only depend on some parameters, or they only require a particular combination of the primary properties, then it is not necessary nor optimal to spend resources in gathering all the information about the original parameters \cite{proctor2017networked}, since only certain pieces of information are relevant to us. In other words, studying functions of the original parameters is effectively equivalent to selecting which information the network should be focused on, so that we can optimise it accordingly.  For that reason, the specific form of the functions, i.e., $\boldsymbol{f}(\cdot)$, will be assumed to be known. 

In turn, this motivates the use of $\boldsymbol{f}[\boldsymbol{g}(\boldsymbol{m})] = (f_1[\boldsymbol{g}(\boldsymbol{m})], \dots, f_l[\boldsymbol{g}(\boldsymbol{m})])$ as the vector estimator for the functions, where we recall that $\boldsymbol{g}(\boldsymbol{m})$ are the estimators for the natural parameters. Although in principle we could consider more general estimators for the derived properties \cite{gill2011}, this choice is appropriate to highlight that the functions are part of the specification of the problem, and that the information is partial only due to the lack of knowledge about the parameters. Then, given some reasonable deviation function $\mathcal{D}\left[\boldsymbol{g}(\boldsymbol{m}), \boldsymbol{\theta}, \boldsymbol{f}(\cdot), \mathcal{W}_f \right]$ for the estimation of derived or secondary properties, we can approximate it with the square error as
\begin{eqnarray}
\mathcal{D}\left[\boldsymbol{g}(\boldsymbol{m}), \boldsymbol{\theta}, \boldsymbol{f}(\cdot), \mathcal{W}_f \right] \approx \mathrm{Tr}\left( \mathcal{W}_f \left\lbrace\boldsymbol{f}[\boldsymbol{g}(\boldsymbol{m})] - \boldsymbol{f}(\boldsymbol{\theta}) \right\rbrace \left\lbrace\boldsymbol{f}[\boldsymbol{g}(\boldsymbol{m})] - \boldsymbol{f}(\boldsymbol{\theta}) \right\rbrace^\transpose \right),
\label{errfunctions}
\end{eqnarray}
for a moderate amount of prior knowledge about the primary properties, where the weighting matrix for the functions is $\mathcal{W}_f = \mathrm{diag}(w_1, \cdots, w_l)$. Note that if the nature of the functions is such that the square error generates an appropriate measure of uncertainty, then the approximation in equation (\ref{errfunctions}) becomes an equality.  

To make the problem more tractable, we will assume that the secondary properties are linear, that is, $\boldsymbol{f}(\boldsymbol{\theta}) = V^{\transpose} \boldsymbol{\theta} + \boldsymbol{a}$, where $V$ is a $(d\times l)$ matrix and $\boldsymbol{a}$ is a column vector with $l$ components. In that case, the error in equation (\ref{errfunctions}) becomes
\begin{equation}
\mathcal{D}\left[\boldsymbol{g}(\boldsymbol{m}), \boldsymbol{\theta}, \boldsymbol{f}(\cdot), \mathcal{W}_f \right] \approx \mathrm{Tr}\left\lbrace \mathcal{W}_f  V^{\transpose} \left[\boldsymbol{g}(\boldsymbol{m}) - \boldsymbol{\theta}\right] \left[\boldsymbol{g}(\boldsymbol{m}) - \boldsymbol{\theta}\right]^\transpose  V \right\rbrace.
\label{errlinfun}
\end{equation}
The fact that it does not depend on $\boldsymbol{a}$ allows us to set $\boldsymbol{a} = \boldsymbol{0}$ without loss of generality. Thus the functions are simply $\boldsymbol{f}(\boldsymbol{\theta}) = V^{\transpose}\boldsymbol{\theta}$, such that the coefficients are encoded in the columns of $V$. 

The previous formalism is reduced to some of the extreme cases that have been studied by other authors (e.g., in \cite{proctor2017networked, proctor2017networkedshort, altenburg2018}). In particular, equation (\ref{errlinfun}) includes the estimation of a single function when $V$ is chosen as a column vector, and a collection of $l = d$ orthogonal functions is equivalent to imposing that $V$ is an orthogonal matrix. Note that, in the latter case, $\mathcal{W}_f = \mathrm{diag}(w_1, \dots, w_d) = \mathcal{W}$. Moreover, if the orthogonal transformation happens to be the identity, i.e., $V = \mathbb{I}$, then we recover the deviation function for the natural parameters in equation (\ref{multideviation}). However, this framework also covers a rich spectrum of possibilities where any number of linear but otherwise general functions are allowed. Therefore, it is clear that, despite our linearity assumption, this approach will allow us to explore the new regime described in section \ref{sec:goals6}.

Once we have selected a suitable deviation function for the multi-parameter problem of estimating functions, we can finally construct the measure of uncertainty 
\begin{eqnarray}
\bar{\epsilon}_{\mathrm{mse}} = \int d\boldsymbol{\theta} d\boldsymbol{m} ~p(\boldsymbol{\theta}, \boldsymbol{m})~\mathrm{Tr}\left\lbrace \mathcal{W}_f V^\transpose \left[\boldsymbol{g}(\boldsymbol{m}) - \boldsymbol{\theta}\right] \left[\boldsymbol{g}(\boldsymbol{m}) - \boldsymbol{\theta}\right]^\transpose V \right\rbrace
\label{msefunctions}
\end{eqnarray}
that is to be optimised. 

\subsection{The asymptotic regime for many parameters}
\label{subsec:multiasymp}

We start by optimising equation (\ref{msefunctions}) over all the possible vector estimators. First we rewrite it as 
\begin{align}
\bar{\epsilon}_\mathrm{mse} &= \mathrm{Tr}\left(\mathcal{W}_f V^\transpose \Sigma_\mathrm{mse} V\right) = \sum_{i=1}^l \sum_{j=1}^d \left(\mathcal{W}_f V^\transpose \Sigma_\mathrm{mse}\right)_{ij}V_{ji}
\nonumber \\
&= \sum_{j=1}^d \sum_{i=1}^l V_{ji} \left(\mathcal{W}_f V^\transpose \Sigma_\mathrm{mse}\right)_{ij} = \mathrm{Tr}\left(V \mathcal{W}_f V^\transpose \Sigma_\mathrm{mse}\right),
\end{align}
where we have defined the matrix square error
\begin{equation}
\Sigma_\mathrm{mse} = \int d\boldsymbol{\theta} d\boldsymbol{m} ~p(\boldsymbol{\theta}, \boldsymbol{m}) \left[\boldsymbol{g}(\boldsymbol{m}) - \boldsymbol{\theta}\right] \left[\boldsymbol{g}(\boldsymbol{m}) - \boldsymbol{\theta}\right]^\transpose.
\label{matrixmse}
\end{equation}
Given that both $V \mathcal{W}_f V^\transpose$ and $\Sigma_\mathrm{mse}$ are symmetric positive semi-definite matrices\footnote{That $V \mathcal{W}_f V^\transpose$ is symmetric can be easily verified as
\begin{align}
\left(V \mathcal{W}_f V^\transpose \right)^\transpose &= \left(\sum_{i,m=1}^d \sum_{k,j=1}^l V_{ij} (\mathcal{W}_f)_{jk} V_{mk} \boldsymbol{e}_i \boldsymbol{e}_m^\transpose\right)^\transpose = \sum_{i,m=1}^d \sum_{k,j=1}^l V_{ik} (\mathcal{W}_f)_{jk} V_{mj} \boldsymbol{e}_i \boldsymbol{e}_m^\transpose 
\nonumber \\
&= V \mathcal{W}_f^\transpose V^\transpose = V \mathcal{W}_f V^\transpose,
\nonumber
\end{align}
where $\boldsymbol{e}_i$ is the $i$-th element of the basis, while its positive semi-definiteness arises from the fact that, for any $\boldsymbol{u}$,
\begin{align}
\boldsymbol{u}^\transpose V \mathcal{W}_f V^\transpose \boldsymbol{u} &= \sum_{i, m = 1}^d \sum_{j, k = 1}^l u_i V_{ij} (\mathcal{W}_f)_{jk} V_{mk} u_m = \sum_{i, m = 1}^d \sum_{j = 1}^l u_i V_{ij} w_j V_{mj} u_m 
\nonumber \\
&= \sum_{j = 1}^l w_j \left(\sum_{i=1}^d u_i V_{ij} \right)^2 \geqslant 0.
\nonumber
\end{align}}, we can find the minimum value for $\bar{\epsilon}_\mathrm{mse}$ simply by searching for the estimators that make $\Sigma_{\mathrm{mse}}$ minimal in the matrix sense. In other words, we need to show that $\boldsymbol{u}^\transpose \Sigma_{\mathrm{mse}} \boldsymbol{u} \geqslant \boldsymbol{u}^\transpose C \boldsymbol{u}$ for an arbitrary real vector $\boldsymbol{u}$ and some symmetric positive semi-definite matrix $C$ that arises after selecting the optimal estimators\footnote{To see why this method works, note that, if $(\Sigma_{\mathrm{mse}} - C) \geqslant 0$, then
\begin{align}
\mathrm{Tr}\left[V \mathcal{W}_f V^\transpose \left(\Sigma_{\mathrm{mse}} - C \right)\right] &= \mathrm{Tr}\left(\sqrt{V \mathcal{W}_f V^\transpose}\sqrt{V \mathcal{W}_f V^\transpose} \sqrt{\Sigma_{\mathrm{mse}} - C }\sqrt{\Sigma_{\mathrm{mse}} - C }\right)
\nonumber \\
&= \mathrm{Tr}\left(\sqrt{V \mathcal{W}_f V^\transpose} \sqrt{\Sigma_{\mathrm{mse}} - C }\sqrt{\Sigma_{\mathrm{mse}} - C }\sqrt{V \mathcal{W}_f V^\transpose} \right)
\nonumber \\
&= \mathrm{Tr}\left[\sqrt{V \mathcal{W}_f V^\transpose} \sqrt{\Sigma_{\mathrm{mse}} - C }\left(\sqrt{V \mathcal{W}_f V^\transpose} \sqrt{\Sigma_{\mathrm{mse}} - C }\right)^\transpose \right]
\nonumber \\
&= \sum_{i,j = 1}^d \left(\sqrt{V \mathcal{W}_f V^\transpose} \sqrt{\Sigma_{\mathrm{mse}} - C }\right)_{ij}^2 \geqslant 0,
\nonumber
\end{align}
which implies that 
\begin{equation}
\bar{\epsilon}_{\mathrm{mse}} = \mathrm{Tr}\left(V \mathcal{W}_f V^\transpose \Sigma_{\mathrm{mse}}\right) \geqslant \mathrm{Tr}\left(V \mathcal{W}_f V^\transpose C\right).
\nonumber
\end{equation}
}.

Our task is then to minimise the scalar quantity
\begin{equation}
\boldsymbol{u}^\transpose \Sigma_{\mathrm{mse}} \boldsymbol{u} = \int d\boldsymbol{\theta} d\boldsymbol{m} ~p(\boldsymbol{\theta}, \boldsymbol{m}) \left[g_u(\boldsymbol{m}) - \theta_u  \right]^2,
\label{scalarquantitynetworks}
\end{equation}
with $g_u(\boldsymbol{m}) = \boldsymbol{u}^\transpose \boldsymbol{g}(\boldsymbol{m}) = \boldsymbol{g}^\transpose(\boldsymbol{m})  \boldsymbol{u}$, $\theta_u = \boldsymbol{u}^\transpose \boldsymbol{\theta} = \boldsymbol{\theta}^\transpose \boldsymbol{u}$ and arbitrary $\boldsymbol{u}$. Since $\boldsymbol{u}^\transpose \Sigma_{\mathrm{mse}} \boldsymbol{u}$ is a functional of $g_u(\boldsymbol{m})$, we can formulate another variational problem as
\begin{equation}
\delta \epsilon\left[g_u(\boldsymbol{m})\right] = \delta \int d\boldsymbol{m}~\mathcal{L}\left[\boldsymbol{m}, g_u(\boldsymbol{m}) \right] = 0,
\end{equation} 
where $\epsilon\left[g_u(\boldsymbol{m})\right] = \boldsymbol{u}^\transpose \Sigma_{\mathrm{mse}} \boldsymbol{u}$ and $\mathcal{L}\left[m, g_u(\boldsymbol{m}) \right] = \int d\boldsymbol{\theta} p(\boldsymbol{\theta}, \boldsymbol{m}) \left[g_u(\boldsymbol{m}) - \theta_u  \right]^2 $. Formally, this is the same type of calculation found in sections \ref{subsec:originalderivation} and \ref{theory}. As such, we know that the solution that makes $ \epsilon\left[g_u(\boldsymbol{m})\right]$ extremal is $g_u(\boldsymbol{m}) = \int d\boldsymbol{\theta} p(\boldsymbol{\theta}|\boldsymbol{m})\theta_u$, with $p(\boldsymbol{\theta}|\boldsymbol{m}) \propto p(\boldsymbol{\theta})p(\boldsymbol{m}|\boldsymbol{\theta})$, and we can expand it as 
\begin{equation}
\sum_{i=1}^d u_i \left[g_i\left(\boldsymbol{m}\boldsymbol\right) - \int d\boldsymbol{\theta} p(\boldsymbol{\theta}|\boldsymbol{m}) \theta_i \right] = 0.
\label{optmultiestimator}
\end{equation}
We also know that equation (\ref{optmultiestimator}) gives rise to a minimum; consequently, by inserting this solution in equation (\ref{scalarquantitynetworks}) we find that $\boldsymbol{u}^\transpose \Sigma_{\mathrm{mse}} \boldsymbol{u}$ is lower bounded by
\begin{equation}
\boldsymbol{u}^\transpose \int d\boldsymbol{m}~ p(\boldsymbol{m}) \left\lbrace \int d\boldsymbol{\theta} p(\boldsymbol{\theta}|\boldsymbol{m}) \boldsymbol{\theta}\boldsymbol{\theta}^\transpose - \left[\int d\boldsymbol{\theta} p(\boldsymbol{\theta}|\boldsymbol{m}) \boldsymbol{\theta}\right] \left[\int d\boldsymbol{\theta} p(\boldsymbol{\theta}|\boldsymbol{m}) \boldsymbol{\theta}\right]^\transpose \right\rbrace \boldsymbol{u},
\label{multiclassicalcondition}
\end{equation}
where $p(\boldsymbol{m}) = \int d\boldsymbol{\theta} p(\boldsymbol{\theta})p(\boldsymbol{m}|\boldsymbol{\theta})$. 

Equation (\ref{multiclassicalcondition}) must be less than or equal to $\boldsymbol{u}^\transpose \Sigma_{\mathrm{mse}} \boldsymbol{u}$ for any $\boldsymbol{u}$. As such, we conclude that the minimum matrix error is
\begin{equation}
\Sigma_\mathrm{mse} =  \int d\boldsymbol{m}~ p(\boldsymbol{m}) \Sigma(\boldsymbol{m}),
\label{matrixmsemin}
\end{equation}
with 
\begin{equation}
\Sigma(\boldsymbol{m}) = \int d\boldsymbol{\theta} p(\boldsymbol{\theta}|\boldsymbol{m}) \boldsymbol{\theta}\boldsymbol{\theta}^\transpose - \left[\int d\boldsymbol{\theta} p(\boldsymbol{\theta}|\boldsymbol{m}) \boldsymbol{\theta}\right] \left[\int d\boldsymbol{\theta} p(\boldsymbol{\theta}|\boldsymbol{m}) \boldsymbol{\theta}\right]^\transpose,
\end{equation}
and that this is achieved for the optimal vector estimator $\boldsymbol{g}\left(\boldsymbol{m}\right) = \int d\boldsymbol{\theta} p(\boldsymbol{\theta}|\boldsymbol{m}) \boldsymbol{\theta}$, in agreement with what is known in the literature \cite{kay1993}. Furthermore, combining equation (\ref{matrixmsemin}) with the original uncertainty in equation (\ref{msefunctions}), and expanding the result of that operation in components, we find that the minimum error for the estimation of the functions is 
\begin{eqnarray}
\bar{\epsilon}_{\mathrm{mse}} = \sum_{i=1}^l w_i \int d\boldsymbol{m}~ p(\boldsymbol{m}) \left\lbrace \int d\boldsymbol{\theta} p(\boldsymbol{\theta}|\boldsymbol{m}) f_i^2(\boldsymbol{\theta}) - \left[\int d\boldsymbol{\theta} p(\boldsymbol{\theta}|\boldsymbol{m}) f_i(\boldsymbol{\theta})\right]^2 \right\rbrace,
\label{msefunctionsmin}
\end{eqnarray}
where $f_i(\boldsymbol{\theta}) = \sum_{j=1}^d V_{ji} \theta_j$. This is the central quantity of this chapter.

Next we wish to select the quantum strategy that is asymptotically optimal, so that we need to examine the asymptotic behaviour of equation (\ref{msefunctionsmin}). This study mimics the strategy that we employed in section \ref{theory} for the scalar case, for which we followed the works \cite{jaynes2003, cox2000, bernardo1994}, and we also do it here. 

Suppose there is a region of the multi-parameter domain with hypervolume $\Delta$ where the likelihood $p(\boldsymbol{m}|\boldsymbol{\theta})$ becomes concentrated around an absolute maximum $\boldsymbol{\theta}_{\boldsymbol{m}}$ as $\mu \rightarrow \infty$, and that the prior for the primary properties is approximately flat in such region. Furthermore, the true vector parameter is $\boldsymbol{\theta}'$. In this regime we can then approximate $\mathrm{log}[p(\boldsymbol{m}|\boldsymbol{\theta})]$ formally as 
\begin{align}
\mathrm{log}[p(\boldsymbol{m}|\boldsymbol{\theta})] \approx  &~\mathrm{log}[p(\boldsymbol{m}|\boldsymbol{\theta}_{\boldsymbol{m}})]
\nonumber \\
&+ \frac{1}{2}\sum_{i, j = 1}^d \frac{\partial^2 \mathrm{log}[p(\boldsymbol{m}|\boldsymbol{\theta}_{\boldsymbol{m}})]}{\partial\theta_i \partial \theta_j}\left(\theta_i - \theta_{\boldsymbol{m},i} \right)\left(\theta_j - \theta_{\boldsymbol{m},j} \right). 
\end{align}
In addition, using the law of large numbers (section \ref{subsec:lln}) and the consistency of the maximum of the likelihood \cite{kay1993}, we can see that
\begin{equation}
\frac{\partial^2 \mathrm{log}[p(\boldsymbol{m}|\boldsymbol{\theta}_{\boldsymbol{m}})]}{\partial\theta_i \partial \theta_j} = \sum_{k=1}^\mu \frac{\partial^2 \mathrm{log}[p(m_i|\boldsymbol{\theta}_{\boldsymbol{m}})]}{\partial\theta_i \partial \theta_j} \approx \mu \int dm~p(m|\boldsymbol{\theta}')\frac{\partial^2 \mathrm{log}[p(m|\boldsymbol{\theta}')]}{\partial\theta_i \partial \theta_j},
\end{equation}
and since by expanding the derivative the latter term becomes the negative of the Fisher information matrix $F(\boldsymbol{\theta})$ in equation (\ref{fim}), we can approximate $p(\boldsymbol{m}|\boldsymbol{\theta})$ as
\begin{equation}
p(\boldsymbol{m}|\boldsymbol{\theta}) \approx p(\boldsymbol{m}|\boldsymbol{\theta}') \mathrm{exp}\left[-\frac{\mu}{2}\left(\boldsymbol{\theta}-\boldsymbol{\theta}' \right)^\transpose F(\boldsymbol{\theta}')\left(\boldsymbol{\theta}-\boldsymbol{\theta}' \right)\right].
\end{equation}

The Fisher information matrix is positive semi-definite in general, and as such it does not always have an inverse. In the context of the asymptotic theory \cite{sammy2016compatibility, proctor2017networked, pezze2017simultaneous}, a singular $F(\boldsymbol{\theta})$ would imply that one or more parameters cannot be estimated with a finite precision \cite{sammy2016compatibility}, and to extract the information about the other parameters one would need to resort to techniques such as the \emph{reduction method} proposed in \cite{proctor2017networked}, or simply to work in the support of $F(\boldsymbol{\theta})$. 

Crucially, that $F(\boldsymbol{\theta})^{-1}$ might not exist does not introduce any fundamental difficulty for the Bayesian estimation based on equation (\ref{msefunctionsmin}), which can always be performed, and this will be explicitly demonstrated in section \ref{subsec:multibayesnetworks} with an example. However, some care is still needed in order to optimise the quantum strategy using the asymptotic theory as a guide. For our purposes it suffices to only attempt the latter when the information matrix can be inverted, leaving for future work the extension of our procedure to cases where $F(\boldsymbol{\theta})$ is singular.

Having assumed that $F(\boldsymbol{\theta})^{-1}$ exists, we can now proceed to calculate three multivariate Gaussian integrals that will allow us to find the final form of the asymptotic error\footnote{The details of such calculations can be found in appendix \ref{sec:multigaussian}.}. By noting that
\begin{align}
\int d\boldsymbol{\theta} p(\boldsymbol{\theta}) p(\boldsymbol{m}|\boldsymbol{\theta}) &\approx  \frac{p(\boldsymbol{m}|\boldsymbol{\theta}')}{\Delta}\int_{-\boldsymbol{\infty}}^{\boldsymbol{\infty}} d\boldsymbol{\theta}\hspace{0.15em} \mathrm{e}^{-\frac{\mu}{2}\left(\boldsymbol{\theta}-\boldsymbol{\theta}' \right)^\transpose F(\boldsymbol{\theta}')\left(\boldsymbol{\theta}-\boldsymbol{\theta}' \right)} 
\nonumber \\
&= \frac{p(\boldsymbol{m}|\boldsymbol{\theta}')}{\Delta}\left\lbrace\frac{(2\pi)^d}{\mathrm{det}[\mu F(\boldsymbol{\theta}')]}\right\rbrace^{\frac{1}{2}},
\label{multigaussian0}
\end{align}
we have that the posterior can be approximated as 
\begin{equation}
p(\boldsymbol{\theta}|\boldsymbol{m}) = \frac{p(\boldsymbol{\theta})p(\boldsymbol{m}|\boldsymbol{\theta})}{p(\boldsymbol{m})} \approx \left\lbrace\frac{\mathrm{det}[\mu F(\boldsymbol{\theta}')]}{(2\pi)^d}\right\rbrace^{\frac{1}{2}} \mathrm{e}^{-\frac{\mu}{2}\left(\boldsymbol{\theta}-\boldsymbol{\theta}' \right)^\transpose F(\boldsymbol{\theta}')\left(\boldsymbol{\theta}-\boldsymbol{\theta}' \right)}.
\label{multiposterior}
\end{equation}
Furthermore, using equation (\ref{multiposterior}) we see that 
\begin{align}
\int d\boldsymbol{\theta} p(\boldsymbol{\theta}|\boldsymbol{m})\boldsymbol{\theta} &\approx  \left\lbrace\frac{\mathrm{det}[\mu F(\boldsymbol{\theta}')]}{(2\pi)^d}\right\rbrace^{\frac{1}{2}}\int_{-\boldsymbol{\infty}}^{\boldsymbol{\infty}} d\boldsymbol{\theta}\hspace{0.15em} \mathrm{e}^{-\frac{\mu}{2}\left(\boldsymbol{\theta}-\boldsymbol{\theta}' \right)^\transpose F(\boldsymbol{\theta}')\left(\boldsymbol{\theta}-\boldsymbol{\theta}' \right)} \boldsymbol{\theta} = \boldsymbol{\theta}',
\nonumber \\
\int d\boldsymbol{\theta} p(\boldsymbol{\theta}|\boldsymbol{m})\boldsymbol{\theta}\boldsymbol{\theta}^\transpose &\approx   \left\lbrace\frac{\mathrm{det}[\mu F(\boldsymbol{\theta}')]}{(2\pi)^d}\right\rbrace^{\frac{1}{2}}\int_{-\boldsymbol{\infty}}^{\boldsymbol{\infty}} d\boldsymbol{\theta}\hspace{0.15em} \mathrm{e}^{-\frac{\mu}{2}\left(\boldsymbol{\theta}-\boldsymbol{\theta}' \right)^\transpose F(\boldsymbol{\theta}')\left(\boldsymbol{\theta}-\boldsymbol{\theta}' \right)} \boldsymbol{\theta}\boldsymbol{\theta}^\transpose
\nonumber \\
&= \boldsymbol{\theta}'\left(\boldsymbol{\theta}'\right)^\transpose + \frac{F(\boldsymbol{\theta}')^{-1}}{\mu},
\label{multigaussian12}
\end{align}
and by inserting the previous results in equation (\ref{matrixmsemin}) we arrive at the asymptotic matrix error
\begin{equation}
\Sigma_{\mathrm{mse}} \approx \frac{1}{\mu}\int d\boldsymbol{\theta}' p(\boldsymbol{\theta}') \int d\boldsymbol{m}\hspace{0.15em}p(\boldsymbol{m}|\boldsymbol{\theta}') F(\boldsymbol{\theta}')^{-1} = \frac{1}{\mu}\int d\boldsymbol{\theta}' p(\boldsymbol{\theta}') F(\boldsymbol{\theta}')^{-1}.
\end{equation}

Following sections \ref{subsec:qapp} and \ref{sec:problem}, the natural parameters of the network will be encoded in the initial state $\rho_0$ as $\rho(\boldsymbol{\theta}) = \mathrm{e}^{-i\boldsymbol{K}\cdot \boldsymbol{\theta}}\rho_0 \mathrm{e}^{i\boldsymbol{K}\cdot \boldsymbol{\theta}}$, with $[K_i, K_j] = 0$. Since we will also employ pure states $\rho_0 = \ketbra{\psi_0}$, we have that 
\begin{equation}
\langle \psi(\boldsymbol{\theta})|[L_i(\boldsymbol{\theta}), L_j(\boldsymbol{\theta})]|\psi(\boldsymbol{\theta})\rangle = 4\langle \psi_0|[K_i, K_j]|\psi_0\rangle = 0,
\end{equation}
where we recall that $L_i(\boldsymbol{\theta})$ is the symmetric logarithmic derivative for the $i$-th natural parameter and that, for pure sates, $L_i(\boldsymbol{\theta})= 2 \partial \rho(\boldsymbol{\theta})/\partial \theta_i$. This means that we may find a POM such that $[F(\boldsymbol{\theta})]_{ij} = (F_q)_{ij}$ for a single copy, where $(F_q)_{ij} =  4\left( \langle  \psi_0 | K_i K_j |  \psi_0 \rangle - \langle  \psi_0 | K_i |  \psi_0 \rangle \langle  \psi_0 | K_j |  \psi_0 \rangle \right)$ is the quantum Fisher information matrix (\cite{sammy2016compatibility, pezze2017simultaneous} and section \ref{subsec:crb}). Therefore, the Bayesian uncertainty in equation (\ref{msefunctionsmin}) can be approximated as
\begin{equation}
\bar{\epsilon}_{\mathrm{mse}} \approx \bar{\epsilon}_{\mathrm{cr}} = \frac{1}{\mu} \mathrm{Tr}\left(\mathcal{W}_f V^\transpose F_q^{-1} V \right)
\label{multiqcrbfun}
\end{equation}
when the prior information is enough to identify the relevant region of the parameter domain. The right hand side is the quantum Cram\'{e}r-Rao bound for functions.

While both the previous discussion and our review in section \ref{subsec:crb} have led us to the same mathematical result, we would like to highlight that these two approaches are different from a conceptual point of view; the goal in section \ref{subsec:crb} was to find the conditions for the saturation of a multi-parameter bound, but here we are simply studying a limiting case (under certain assumptions) of the theory that arises when we employ the error in equation (\ref{msefunctionsmin}) as the measure of uncertainty. The crucial role of the theory in section \ref{subsec:crb} is that it allows us to see that the asymptotic expansion in this section is indeed optimal. 

Equation (\ref{multiqcrbfun}) enables us to study the asymptotic performance of quantum sensing networks. Once we have completed this step, we will search for a combination of $\rho_0$ and POM for which $F(\boldsymbol{\theta}) = F_q$, and we will perform a non-asymptotic analysis of that strategy using equation (\ref{msefunctionsmin}), where the optimal Bayes estimators $\boldsymbol{g}(\boldsymbol{\theta}) = \int d\boldsymbol{\theta}~ p(\boldsymbol{\theta}|\boldsymbol{m})\boldsymbol{\theta}$ have been selected. The latter step demands an extension of our techniques in chapter \ref{chap:nonasymptotic} to the multi-parameter regime, a generalisation that is developed in the next section. 

\subsection{Non-asymptotic analysis of multi-parameter protocols}
\label{subsec:multinonasym}

Let us consider that, according to our prior information, the natural parameters $\boldsymbol{\theta} = (\theta_1, \dots, \theta_d)$ can be initially thought of as if they were independent in the statistical sense, and that a displacement by an arbitrary real vector $\boldsymbol{c}$ does not change our state of information. That is, $\boldsymbol{\theta}$ and $\boldsymbol{\theta}' = \boldsymbol{\theta} + \boldsymbol{c}$ generate equivalent estimation problems. Note that if we knew that the parameters were optical phases, the displacement would be modulo $2\pi$. 

This invariance is equivalent to imposing that $p(\boldsymbol{\theta})d\boldsymbol{\theta} = p(\boldsymbol{\theta}')d\boldsymbol{\theta}'=p(\boldsymbol{\theta}+\boldsymbol{c})d\boldsymbol{\theta}$, which gives rise to the functional equation 
\begin{equation}
p(\boldsymbol{\theta}) = p(\boldsymbol{\theta}+\boldsymbol{c}).
\label{multipriorcondition}
\end{equation}
A way of searching for a solution is to expand the right hand side as
\begin{equation}
p(\boldsymbol{\theta}+\boldsymbol{c}) = \sum_{k=0}^\infty \frac{1}{k!} \left(\boldsymbol{c} \cdot \boldsymbol{\nabla} \right)^k p(\boldsymbol{\theta}),
\label{multitaylor}
\end{equation}
which is a multivariate Taylor expansion \cite{mathematics2004}, so that by introducing equation (\ref{multitaylor}) in (\ref{multipriorcondition}) we arrive at
\begin{equation}
\sum_{k=1}^\infty \frac{1}{k!} \left(\boldsymbol{c} \cdot \boldsymbol{\nabla} \right)^k p(\boldsymbol{\theta}) =0.
\end{equation}
Recalling that this must be fulfilled by an arbitrary vector $\boldsymbol{c}$, we see that this is satisfied when $p(\boldsymbol{\theta}) \propto 1$.

Since we are interested in the intermediate prior information regime, we further imagine that the previous argument is only approximately fulfilled in a portion of the parameter domain with hypervolume $\Delta_0$ that is centred around $\boldsymbol{\bar{\theta}}=(\bar{\theta}_1, \dots, \bar{\theta}_d)$. Moreover, given that a priori the parameters are thought of as independent, we may express the hypervolume $\Delta_0$ as $\Delta_0 = \prod_{i=1}^d W_{0,i}$, where $W_{0,i}$ is the prior width for the $i$-th parameter. Therefore, our multi-parameter prior probability will be
\begin{equation}
p(\boldsymbol{\theta})=1/\Delta_0 = 1/\left(\prod_{i=1}^d W_{0,i}\right), 
\label{multiprior}
\end{equation}
for $\boldsymbol{\theta}\in [\bar{\theta}_1 - W_{0, 1}/2, \bar{\theta}_1 + W_{0, 1}/2]\times \cdots \times [\bar{\theta}_d - W_{0, d}/2, \bar{\theta}_d + W_{0,d}/2]$, and zero otherwise. We notice that this argument generalises the analogous derivation in section \ref{experiment_prior} for a single parameter. 

On the other hand, in section \ref{subsec:multiasymp} we have learned that, to exploit the multi-parameter Cram\'{e}r-Rao bound as an asymptotic guide, $\Delta_0$ must be sufficiently small for the likelihood $p(\boldsymbol{m}|\boldsymbol{\theta})$ to develop a unique absolute maximum as $\mu$ grows. We will denote the largest hypervolume where this is the case by $\Delta_{\mathrm{int}}$, and we will call it \emph{intrinsic hypervolume}, in analogy with the notion of intrinsic width $W_{\mathrm{int}}$ introduced in chapter \ref{chap:nonasymptotic} and utilised in the context of a Mach-Zehnder interferometer.

A method to extract $W_{\mathrm{int}}$ by a visual inspection of the posterior probability\footnote{We recall that the posterior has the same extrema as the likelihood when the prior is flat.} was proposed and applied to single-parameter problems in sections \ref{experiment_prior} and \ref{subsec:prioranalysis}. In section \ref{subsec:multiprioranalysis} we will see that the same idea can be adapted in a straightforward way for a network with two primary properties, and the algorithm to implement it can be found in appendix \ref{sec:multiprior}. 

Unfortunately, in general it is less clear how to implement the previous method for large $d$. Moreover, the challenges associated with a large $d$ are even more serious when we face the calculation of $\bar{\epsilon}_\mathrm{mse}$ in equation (\ref{msefunctionsmin}) as a function of $\mu$. The algorithm that we constructed in section \ref{subsec:numalgorithm} was able to integrate a large amount of outcomes because it exploited how the information is updated within the expression for the Bayesian mean square error (see appendix \ref{sec:msematlab}). However, the integration associated with the parameter was implemented by a standard deterministic method, which is the type of approach that is known to become less efficient as the dimension grows. Given these difficulties, we will focus our multi-parameter non-asymptotic analysis on quantum sensing networks with two natural parameters, and we will only consider the general case with arbitrary $d$ for the preliminary asymptotic study in sections \ref{sec:networksasym} and \ref{subsec:intersensorasymp} and some of the single-shot scenarios in chapter \ref{chap:multibayes}. 

The algorithm that we will utilise to calculate $\bar{\epsilon}_\mathrm{mse}(\mu, d = 2)$ is:
\begin{enumerate}
\item A collection of $\mu$ experimental outcomes $\boldsymbol{m}$ is sampled from $p(m|\theta_1', \theta_2')$. These are used to construct the posterior $p(\theta_1, \theta_2|\boldsymbol{m})$ via Bayes theorem as
\begin{equation}
p(\theta_1, \theta_2|\boldsymbol{m}) \propto p(\theta_1, \theta_2) \prod_{i=1}^\mu p(m_i|\theta_1, \theta_2), 
\end{equation}
and we calculate the components of the experimental covariance matrix as
\begin{align}
\left[\Sigma(\boldsymbol{m})\right]_{ij} = &~ \int d\theta_1 d\theta_2 p(\theta_1, \theta_2|\boldsymbol{m})\theta_i \theta_j 
\nonumber \\
&- \left[\int d\theta_1 d\theta_2 p(\theta_1, \theta_2|\boldsymbol{m})\theta_i\right] \left[\int d\theta_1 d\theta_2 p(\theta_1, \theta_2|\boldsymbol{m})\theta_j\right].
\end{align}
Then we define the matrix $G \equiv V \mathcal{W}_f V^T$ and obtain the experimental error
\begin{equation}
\epsilon(\boldsymbol{m})=\mathrm{Tr}\left[G \Sigma(\boldsymbol{m})\right] = G_{11} [\Sigma(\boldsymbol{m})]_{11} + G_{22} [\Sigma(\boldsymbol{m})]_{22} + 2 G_{12} [\Sigma(\boldsymbol{m})]_{12}.
\label{linearexperrmatlab}
\end{equation}

\item The previous step is repeated a large number of times, so that the average of all the experimental uncertainties can be used as an approximation for
\begin{equation}
\epsilon(\theta_1', \theta_2') = \int d\boldsymbol{m}p(\boldsymbol{m}|\theta_1', \theta_2')\epsilon(\boldsymbol{m})
\end{equation}
due to the law of large numbers. 

\item Finally, this process is implemented for all the pairs $(\theta_1', \theta_2')$ in a discrete approximation to the parameter domain, and the resultant errors are averaged over the prior probability $p(\theta_1', \theta_1')$, arriving at
\begin{equation}
\int d\theta_1' d\theta_2' p(\theta_1', \theta_1') \epsilon(\theta_1', \theta_2') = \bar{\epsilon}_{\mathrm{mse}},
\end{equation}
which is the uncertainty for the linear functions in equation (\ref{msefunctionsmin}).
\end{enumerate}
Hence, this is an extension of the proposal in section \ref{subsec:numalgorithm}, and its numerical implementation in MATLAB can be found in appendix \ref{sec:multimsematlab}.

It is interesting to examine the nature of the matrix $G$ in equation (\ref{linearexperrmatlab}). The measure of uncertainty introduced in section \ref{sec:uncertainty} for the natural parameters had the form $\mathrm{Tr}(\mathcal{W} \Sigma)$, with $\mathcal{W}=\mathrm{diag}(w_1, \dots, w_d)$ and $\Sigma$ being the matrix error for a general deviation function. On the other hand, the uncertainty for linear functions in this chapter can be recast as $\mathrm{Tr}(V \mathcal{W}_f V^\transpose \Sigma)$, with $\mathcal{W}_f=\mathrm{diag}(w_1, \dots, w_l)$. Both of them are particular cases of the more general expression $\mathrm{Tr}(G \Sigma)$, where $G$ is a symmetric positive semi-definite matrix, and in fact this is how multi-parameter metrology is often presented in more general treatments (see, e.g., \cite{sammy2016compatibility}). 

The crucial observation is that different combinations of linear functions and weights can produce the same $G$. We may then say that using $G$ is more economical, in the sense that it throws away irrelevant details and it only keeps the part of the linear transformation introduced by $V \mathcal{W}_f V^\transpose$ that effectively affects the overall calculation of the uncertainty.  However, while this is indeed useful to perform calculations, it also implies that the structure of the specific linear functions that we wish to estimate in a particular problem is lost. This is why we have chosen to show the objects $(V, \mathcal{W}_f)$, or, equivalently, $(f_i(\cdot), w_i)$, explicitly in our derivations of sections \ref{subsec:relevantinfo} and \ref{subsec:multiasymp}, while we only consider $G$ for more pragmatic tasks. 

\section{Our methodology in action: results and discussion}
\label{subsec:multibayesnetworks}

\subsection{Sensor-symmetric states for quantum sensing networks}
\label{subsec:sensorsymmetric}

Suppose we have a collection of $d$ quantum sensors that are distributed in a portion of the physical space, and that we assume that each sensor is modelled by a qubit. Furthermore, since, in general, the spatial separation between sensors can be large, it is natural to choose a model with the following property: when one or more parameters are encoded in one of the sensors, the rest of them are not affected by this operation. To capture this idea we will employ the separable unitary encoding
\begin{equation}
U(\boldsymbol{\theta}) = \mathrm{e}^{-i \sigma_z \theta_1/2}\otimes \cdots \otimes \mathrm{e}^{-i \sigma_z \theta_d/2},
\label{networksunitary}
\end{equation}
where a single parameter is locally encoded in each sensor. Therefore, the natural properties of the system, represented by $\boldsymbol{\theta}$, are local, while the secondary properties that depend non-trivially on more than one natural parameter will be global. 

The unitary in equation (\ref{networksunitary}) is of the form seen in section \ref{subsec:multischemes} when the ancillary system is omitted, and it constitutes a particular case of the more general sensing model proposed by the authors of \cite{proctor2017networked}. Moreover, by rewriting it as
\begin{eqnarray}
U(\boldsymbol{\theta}) = \mathrm{exp}\left(-i\sum_{i=1}^d \sigma_{z, i}\theta_i/2\right) = \mathrm{exp}\left(-i \boldsymbol{K}\cdot\boldsymbol{\theta}\right),
\end{eqnarray}
where we have introduced the notation
\begin{equation}
2 K_i = \sigma_{z, i} \equiv \mathbb{I}_1\otimes\cdots\otimes\mathbb{I}_{i-1}\otimes \sigma_z\otimes\mathbb{I}_{i+1}\otimes\cdots\otimes\mathbb{I}_d,
\end{equation}
we can explicitly see that $[K_i, K_j] = [\sigma_z^{(i)},\sigma_z^{(j)}]/4 = 0$. The most general pure state for this system can be written in the $\sigma_z$ basis as
\begin{equation}
\ket{\psi_0} = \sum_{i_1 \dots i_d=0}^1 a_{i_1 \dots i_d} \ket{i_0, i_1 \dots i_d}, \hspace{0.2em} \text{with} \hspace{0.2em} \sum_{i_1 \dots i_d=0}^1 |a_{i_1 \dots i_d}|^2 = 1,
\label{genericprobe}
\end{equation}
and in principle we allow for any general measurement with POM elements $\lbrace E(m_j)\rbrace$ and outcomes $\lbrace m_j \rbrace$, which is to be performed on all the sensors at once during $j$-th repetition of the experiment. Finally, for this scenario we choose the resource operator to be trivial, that is, $R = \mathbb{I}$, so that $\langle R \rangle = 1$, with $\langle \Box \rangle = \langle \psi_0 | \Box  | \psi_0 \rangle$. This is unlike in the optical case, where each mode admitted different numbers of quanta, while  here we have that each sensor is a quantum entity on its own. 

We would like to see this configuration as a \emph{quantum network}. It is clear that each sensor can be regarded as a physical node of such network, but the nature of the link between different nodes is more subtle. In general, what is networked is the information related to each node, a part of which is the information collected by the network about each of the primary properties $\boldsymbol{\theta}$. More concretely, the links between nodes, if they exist, are to be understood as correlations associated with the initial probe $\ket{\psi_0}$, the POM $E(m)$ and the prior probability. 

In section \ref{subsec:multinonasym} we have assumed that, a priori, the primary parameters are to be thought of as independent, and this means that prior correlations do not play a role in our study. Furthermore, in section \ref{subsec:crb} we saw that the quantum Cram\'{e}r-Rao bound, which is the tool that we will use first, is a function of $\ket{\psi(\boldsymbol{\theta})} = U(\boldsymbol{\theta})\ket{\psi_0}$ alone. Hence, as a first step it is sufficient to examine the correlations associated with the preparation of the network. In particular, we will exploit the concept of \emph{inter-sensor correlations}, whose strength can be defined as \cite{knott2016local, proctor2017networked}
\begin{equation}
\mathcal{J}_{ij} = \frac{\langle K_i K_j \rangle - \langle K_i \rangle\langle K_j \rangle}{\Delta K_i \Delta K_j},
\label{multicorrelations}
\end{equation}
for $i \neq j$, where $\Delta K_i^2 = \langle K_i^2 \rangle - \langle K_i \rangle^2$ and $-1 \leqslant \mathcal{J}_{ij} \leqslant 1$. We can see that only the initial state and the generators that give rise to $\rho(\boldsymbol{\theta})$ are involved in this definition, and that the nature of these correlations is pairwise.

This formalism allows us to further introduce the concept of \emph{sensor-symmetric states}, which were defined by Proctor \emph{et al.} \cite{proctor2017networked} as those satisfying that
\begin{align}
v &= \langle K_i^2 \rangle - \langle K_i \rangle^2,
\nonumber \\
c &=\langle K_i K_j \rangle - \langle K_i \rangle\langle K_j \rangle,
\label{sensorsymcon}
\end{align}
for all $i$, $j$, where $c$ and $v$ are fixed values that characterise the preparation of the network. In turn, equation (\ref{multicorrelations}) becomes $\mathcal{J}_{ij}=\mathcal{J} = c/v$, also for all $i\neq j$, and for our qubit model we see that
\begin{align}
4v &= \langle \sigma_{z, i}^2 \rangle - \langle \sigma_{z, i} \rangle^2 = 1 - \langle \sigma_{z, i} \rangle^2,
\nonumber \\
4c &= \langle \sigma_{z, i} \sigma_{z, j} \rangle - \langle \sigma_{z, i} \rangle\langle \sigma_{z, j} \rangle,
\label{sensorsymconqubit}
\end{align}
where, in addition, $0 \leqslant 4v \leqslant 1$. This inequality for the variances stems from the fact that the eigenvalues for the $\sigma_z$ Pauli matrix are $\pm 1$, so that $\abs{\langle \sigma_{z, i} \rangle} \leqslant 1$.

Formally, sensor-symmetric configurations can be seen as a generalisation of those optical schemes based on path-symmetric states that we exploited in chapters \ref{chap:conceptual}, \ref{chap:nonasymptotic} and \ref{chap:limited}. Thus from our experience in previous chapters we can expect this arrangement to be very useful to simplify the calculations. Additionally, this type of configuration can be relevant in certain physical scenarios. For instance, they could be a reasonable choice for sensing a portion of space that either is largely homogeneous or we expect it to be, and they would also be appropriate if all the unknown parameters represent the same type of physical quantity, which is the case in imaging problems \cite{humphreys2013, zhang2014}. 

\subsection{Asymptotic estimation of arbitrary linear functions}
\label{sec:networksasym}

From our discussion in section \ref{subsec:multiasymp} we know that, under the appropriate conditions, we can expect the mean square error $\bar{\epsilon}_{\mathrm{mse}}$ to converge to the quantum Cram\'{e}r-Rao bound $\bar{\epsilon}_{\mathrm{cr}} = \mathrm{Tr}(\mathcal{W}_f V^\transpose F_q^{-1} V)/\mu$ as the number of trials $\mu$ grows. Let us thus examine the quantum strategies that are useful in this regime as the first step to apply our semiclassical methodology (section \ref{subsec:constructingmethod}).

If we denote by $\lbrace \boldsymbol{e}_i \rbrace$ the basis components of the real space where $\mathcal{W}_f$, $V$ and $F_q$ are defined, with $\boldsymbol{e}_i^\transpose\boldsymbol{e}_j = \delta_{ij}$, then from equation (\ref{sensorsymconqubit}) we have that
\begin{align}
F_q &= \sum_{i,j = 1}^d\left(\langle \sigma_{z, i} \sigma_{z, j} \rangle - \langle \sigma_{z, i}\rangle \langle \sigma_{z, j}\rangle \right) \boldsymbol{e}_i\boldsymbol{e}_j^\transpose = 4 \left(v \sum_{i=1}^d\boldsymbol{e}_i\boldsymbol{e}_i^\transpose + c \sum_{\underset{i\neq j}{i,j=1}}^d\boldsymbol{e}_i\boldsymbol{e}_j^\transpose  \right)
\nonumber \\
&= 4\left[(v-c)\mathbb{I} + c \mathcal{I}\right] = 4v\left[(1-\mathcal{J})\mathbb{I} + \mathcal{J} \mathcal{I}\right],
\label{fimsym}
\end{align}
where $\mathcal{I}$ is a $(d\times d)$ matrix of ones. This is the quantum Fisher information matrix for sensor-symmetric states.

To invert $F_q$, we need to impose the condition of positive definiteness, which is equivalent to require that its eigenvalues are strictly positive. Expressing $\mathcal{I}$ as $\mathcal{I} = \boldsymbol{1}\boldsymbol{1}^\transpose$, where $\boldsymbol{1}$ is the column vector of ones, the information matrix becomes $F_q = 4v\left[(1-\mathcal{J})\mathbb{I} + \mathcal{J} \boldsymbol{1}\boldsymbol{1}^\transpose\right]$. In that case, the characteristic equation for the eigenvalues $\lbrace \lambda \rbrace$ is
\begin{equation}
\mathrm{det} \left\lbrace 4v\left[\left(1-\mathcal{J}-\frac{\lambda}{4v}\right)\mathbb{I} + \mathcal{J}\boldsymbol{1}\boldsymbol{1}^\transpose\right] \right\rbrace = 0,
\label{characteristicfim}
\end{equation}
which upon using the identity $\mathrm{det}(X + \boldsymbol{y}\boldsymbol{z}^\transpose) = (1 + \boldsymbol{z}^\transpose X^{-1}\boldsymbol{y})\hspace{0.15em}\mathrm{det}(X)$, with $X = [4v(1-\mathcal{J}) - \lambda]\mathbb{I}$, $\boldsymbol{y} = 4v\mathcal{J}\boldsymbol{1}$ and $\boldsymbol{z}=\boldsymbol{1}$, implies that
\begin{equation}
\left\lbrace 4v\left[1+(d-1)\mathcal{J}\right]-\lambda\right\rbrace \left[4v\left(1-\mathcal{J}\right) - \lambda \right]^{d-1} = 0. 
\end{equation}
As a result, the eigenvalues of $F_q$ are $\lambda_1 = 4v[1+(d-1)\mathcal{J}]$, with multiplicity $1$, and $\lambda_2 = 4v(1-\mathcal{J})$, with multiplicity $d-1$, and by imposing that they are positive we conclude that $F_q$ is invertible when $1/(1-d)<\mathcal{J} < 1$. The rest of the calculations in this section assume that $\mathcal{J}$ lies in such open interval.

In \cite{proctor2017networked} the inverse of $F_q$ was found to be 
\begin{equation}
F_q^{-1} = \frac{\left[1 + (d-1)\mathcal{J}\right]\mathbb{I} - \mathcal{J}\mathcal{I}}{4v(1 -\mathcal{J})\left[1+(d-1)\mathcal{J}\right]}
\label{fimsyminv}
\end{equation}
for our configuration, and utilising this result we find that the asymptotic uncertainty for the estimation of linear functions is given by
\begin{equation}
\bar{\epsilon}_{\mathrm{cr}} = \frac{\left[1 + (d-2)\mathcal{J}\right]\mathrm{Tr}\left(\mathcal{W}_f V^\transpose V \right) - \mathcal{J}\mathrm{Tr}\left(\mathcal{W}_f V^\transpose \mathcal{X} V \right)}{4\mu v (1 -\mathcal{J})[1+(d-1)\mathcal{J}]},
\label{symmetricfunctionsfirst}
\end{equation}
where we have introduced the $(d\times d)$ matrix $\mathcal{X} \equiv \mathcal{I} - \mathbb{I}$ to separate the contribution to the uncertainty due to the diagonal elements of $F_q^{-1}$, which are the errors for each of the primary parameters, from that of the rest of the matrix.

Equation (\ref{symmetricfunctionsfirst}) shows that the uncertainty depends on three types of quantities: i) the number of repetitions $\mu$ and the number of parameters $d$, (ii) the combined properties of state and generators through the inter-sensor correlations $\mathcal{J}$ and the variance $v$, and (iii) two new quantities, $\mathrm{Tr}\left(\mathcal{W}_f V^\transpose V \right)$ and $\mathrm{Tr}\left(\mathcal{W}_f V^\transpose \mathcal{X} V \right)$, that are defined in terms of the functions encoded in $V$ and the weighting matrix $\mathcal{W}_f$. The next step is to investigate the physical meaning of the latter factors.

By relabelling the vector formed by the components of the $j$-th linear function as $\boldsymbol{f}_j$ (i.e., $f_j(\boldsymbol{\theta}) = \sum_{i=1}^d V_{ij} \theta_i \equiv \boldsymbol{f}_j^\transpose\boldsymbol{\theta}$), we can rewrite the first quantity in a more suggestive form as
\begin{align}
\mathrm{Tr}\left(\mathcal{W}_f V^T V \right) &= \sum_{i, j = 1}^l \sum_{k=1}^d \left(\mathcal{W}_f\right)_{ij} V_{kj} V_{ki} = \sum_{j = 1}^l w_j \sum_{k=1}^d V_{kj} V_{kj} 
\nonumber \\
&= \sum_{j=1}^l w_j \boldsymbol{f}_j^\transpose \boldsymbol{f}_j = \sum_{j=1}^l w_j \abs{\boldsymbol{f}_j}^2.
\end{align}
We observe that this is simply the weighted sum of the squared magnitudes of the vectors associated with the linear functions. Since $V \mathcal{W}_f V^T$ is positive semi-definitive, and excluding the degenerate case where all the coefficients vanish, we have that $\mathrm{Tr}(\mathcal{W}_f V^T V) = \mathrm{Tr}(V \mathcal{W}_f V^T) > 0$. In addition, when the functions are normalised, that is, $|\boldsymbol{f}_i| = 1$ for $1 \leqslant i \leqslant l$, and recalling that $\mathrm{Tr}(\mathcal{W}_f) = \sum_{i=1}^l w_i= 1$, we have that $\mathrm{Tr}(\mathcal{W}_f V^T V) = 1$. Hence, we define the \emph{normalisation term}
\begin{equation}
\mathcal{N}\equiv \mathrm{Tr}(\mathcal{W}_f V^T V) = \sum_{j=1}^l w_j \abs{\boldsymbol{f}_j}^2
\label{normterm}
\end{equation} 
satisfying that $\mathcal{N} > 0$, with $\mathcal{N} = 1$ for normalised linear functions.

On the other hand, the second term associated with the functions can be expressed as
\begin{align}
\mathrm{Tr}\left(\mathcal{W}_f V^T \mathcal{X}V\right) &= \mathrm{Tr}\left[\mathcal{W}_f V^T \left(\mathcal{I}-\mathbb{I}\right)V\right] = 
- \mathcal{N} + \sum_{i, j = 1}^l\sum_{k, m =1}^d \left(\mathcal{W}_f\right)_{ij} V_{kj} \mathcal{I}_{km} V_{mi}
\nonumber \\
&= - \mathcal{N} + \sum_{j = 1}^l w_j\sum_{k, m =1}^d V_{kj} 1_k 1_m V_{mj} = - \mathcal{N} + \sum_{j = 1}^l w_j \left(\sum_{k =1}^d V_{kj} 1_k \right)^2 
\nonumber \\
&= - \mathcal{N} + \sum_{j = 1}^l w_j \left(\boldsymbol{f}_j^\transpose\boldsymbol{1}\right)^2 = - \mathcal{N} + d \sum_{j = 1}^l w_j \abs{\boldsymbol{f}_j}^2 \mathrm{cos}^2\left(\varphi_{\boldsymbol{1},j}\right)
\nonumber \\
&= \sum_{j = 1}^l w_j \abs{\boldsymbol{f}_j}^2 \left[d\hspace{0.2em}\mathrm{cos}^2\left(\varphi_{\boldsymbol{1},j}\right) -1\right],
\label{geometrycalc}
\end{align}
where $\varphi_{\boldsymbol{1},j}$ is the angle between the vector associated with the $j$-th function and the direction defined by the vector of ones $\boldsymbol{1}$, and having used the fact that $|\boldsymbol{1}| = \sqrt{d}$. 

Recalling that $|\mathrm{cos}\left(\varphi_{\boldsymbol{1},j}\right)| \leqslant 1$ and using the result in equation (\ref{geometrycalc}), we see that $\mathrm{Tr}\left(\mathcal{W}_f V^T \mathcal{X}V\right)$ is bounded as
\begin{equation}
-\mathcal{N} \leqslant \mathrm{Tr}\left(\mathcal{W}_f V^T \mathcal{X}V\right) \leqslant \mathcal{N}(d-1),
\label{geometryinterval}
\end{equation}
and that the extremes are realised when either the functions are aligned with the direction of the vector of ones $\boldsymbol{1}$, or they lie in a subspace with $(l-1)$ dimensions orthogonal to it. In other words, we have shown that, for sensor-symmetric networks whose secondary properties are modelled by linear functions, there are two kinds of global properties that play a special role: the sum of all the natural parameters with equal weights, and any linear combination of them such that the sum of its coefficients vanishes. Any other set of global properties will produce some value for $\mathrm{Tr}\left(\mathcal{W}_f V^T \mathcal{X}V\right)$ lying within the interval defined in equation (\ref{geometryinterval}), and this value will be given by the geometry of the transformation defined by $V \mathcal{W}_f V^T$. This motivates the introduction of the \emph{geometry parameter}
\begin{equation}
\mathcal{G} \equiv \frac{1}{\mathcal{N}}\mathrm{Tr}\left(\mathcal{W}_f V^T \mathcal{X}V\right) = \frac{1}{\mathcal{N}}\sum_{j = 1}^l w_j \abs{\boldsymbol{f}_j}^2 \left[d\hspace{0.2em}\mathrm{cos}^2\left(\varphi_{\boldsymbol{1},j}\right) -1\right],
\label{geometryparameter}
\end{equation}
which satisfies that $-1 \leqslant \mathcal{G} \leqslant (d-1)$. 

\begin{figure}[t]
\centering
\includegraphics[trim={0cm 0.1cm 1.2cm 0cm},clip,width=7.75cm]{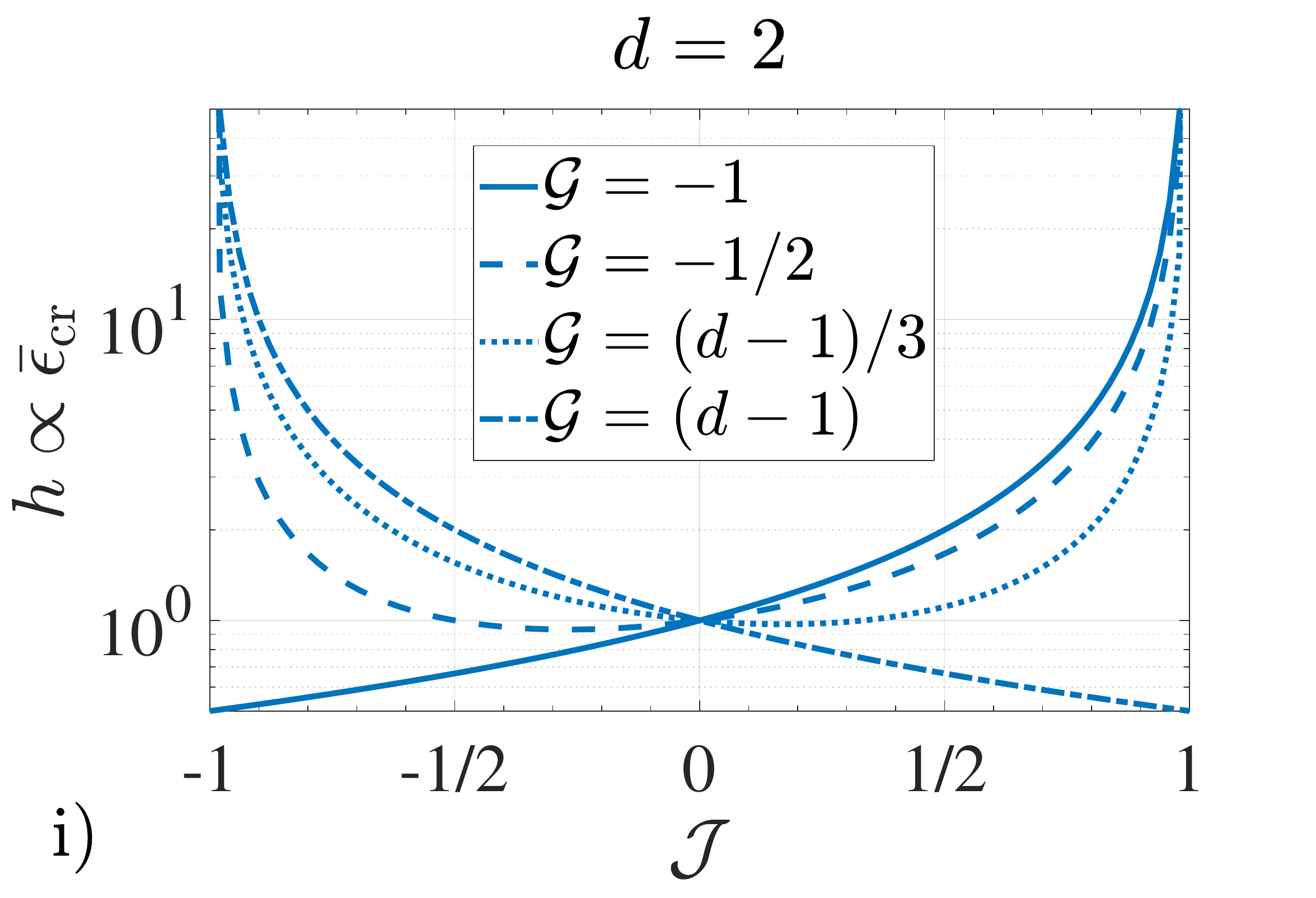}\includegraphics[trim={0cm 0.1cm 1.2cm 0cm},clip,width=7.75cm]{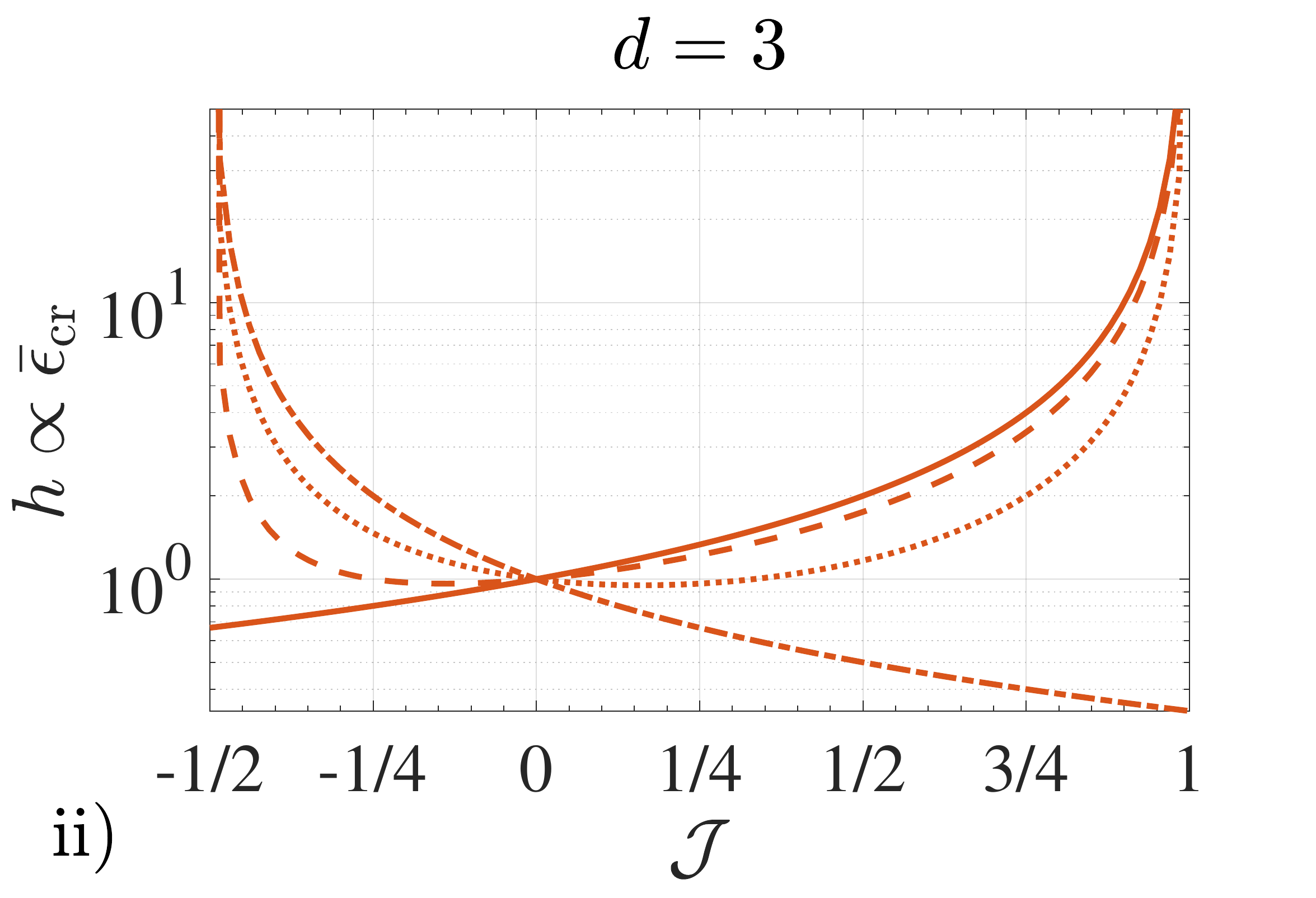}
\includegraphics[trim={0cm 0.1cm 1.2cm 0cm},clip,width=7.75cm]{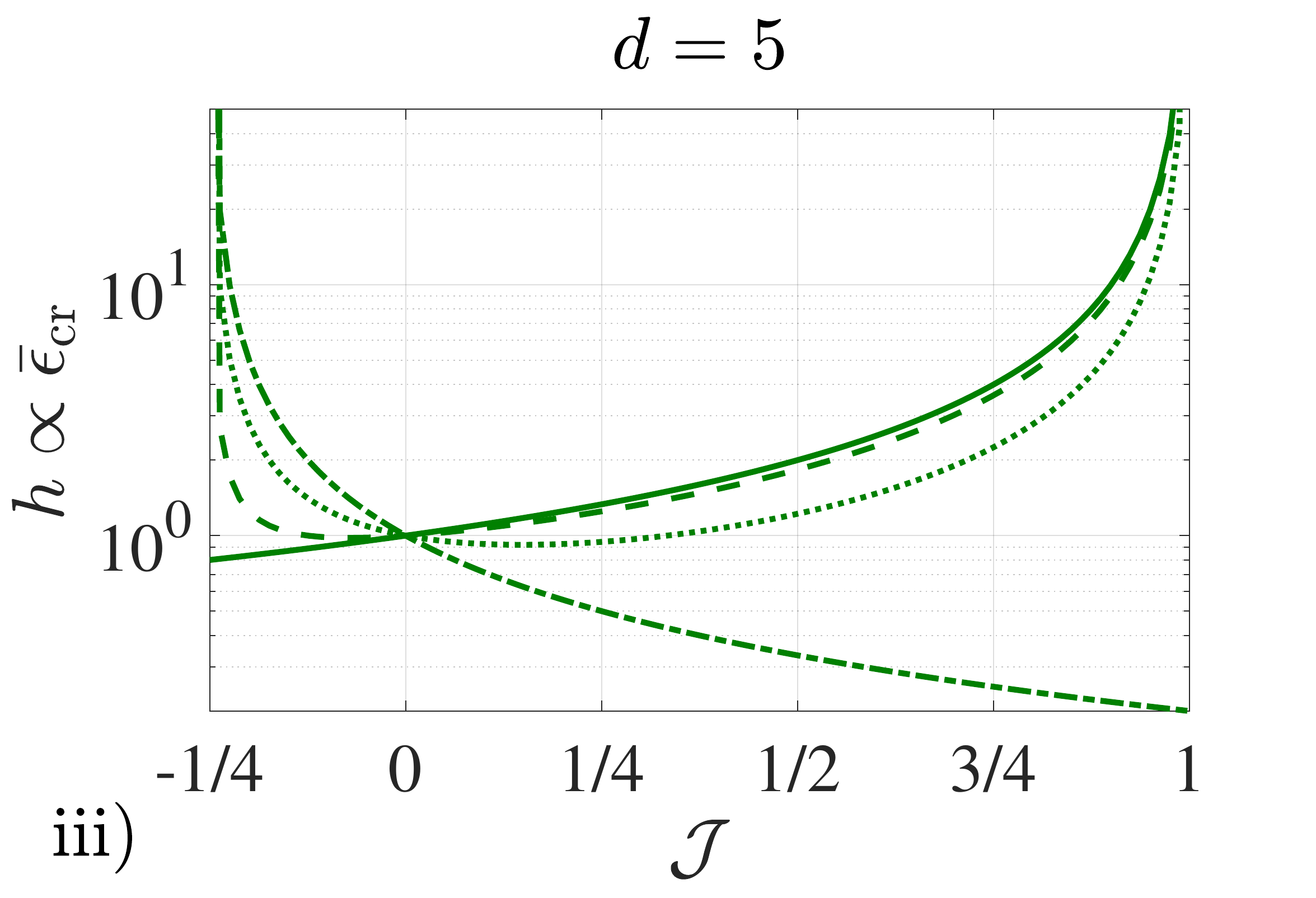}\includegraphics[trim={0cm 0.1cm 1.2cm 0cm},clip,width=7.75cm]{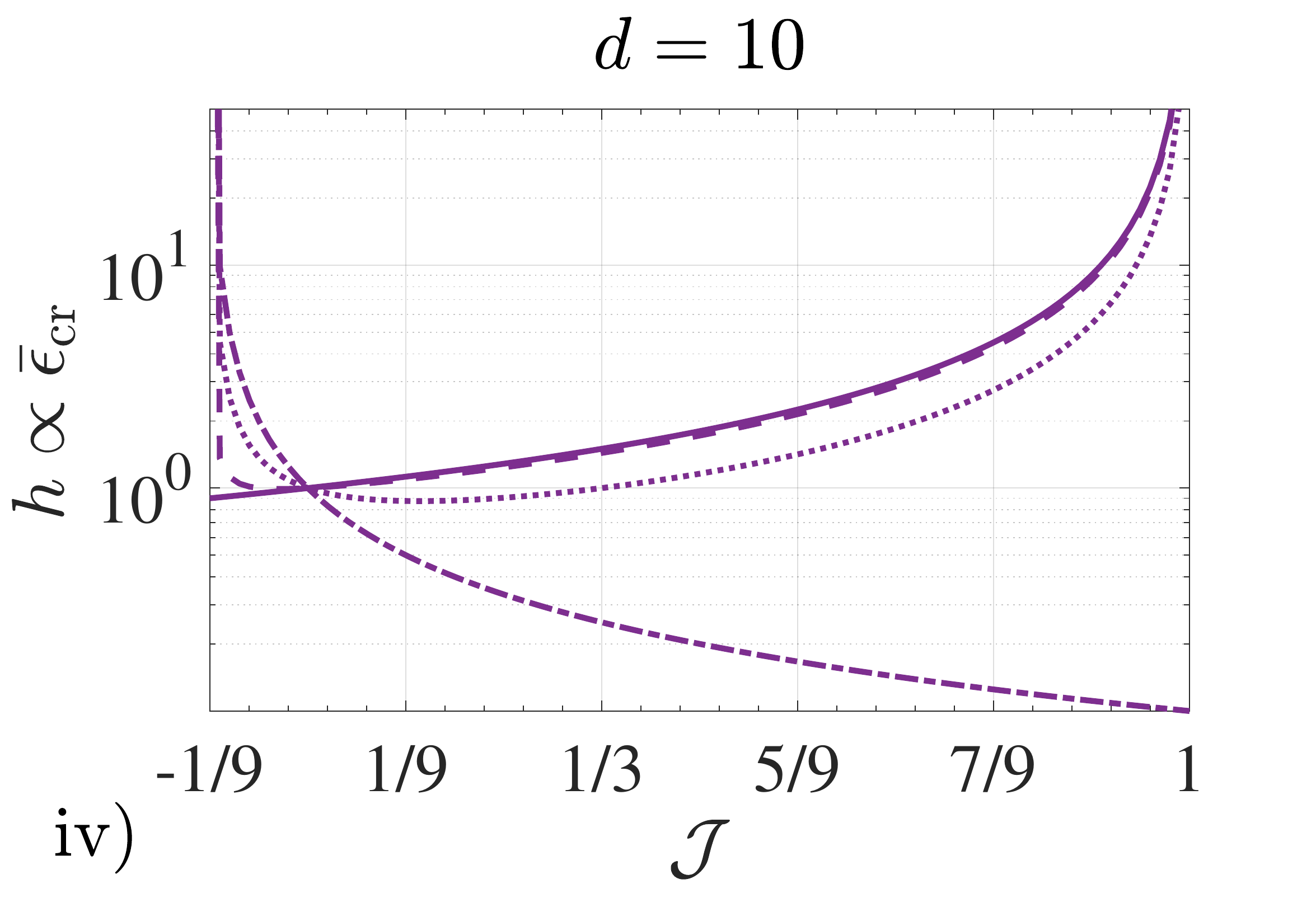}
	\caption[Asymptotic uncertainty, correlations and geometry of the functions]{Representation of the interplay between the inter-sensor correlations $\mathcal{J}$ and the geometry parameter $\mathcal{G}$ in equation (\ref{geometrylinkfactor}) for a quantum sensing network with (i) $d=2$, (ii) $d=3$, (iii) $d=5$ and (iv) $d=10$ natural parameters. We observe that, given $\mathcal{G} \in (-1, (d-1))$, the minimum uncertainty is achieved using a scheme with inter-sensor correlations of strength $\mathcal{J} \in (1/(1-d),1)$. The quantitative characterisation of these minima is provided in section \ref{subsec:intersensorasymp}.}
\label{geolinkplot}
\end{figure}

Inserting equations (\ref{normterm}) and (\ref{geometryparameter}) in equation (\ref{symmetricfunctionsfirst}) we find that the asymptotic uncertainty finally becomes
\begin{equation}
\bar{\epsilon}_{\mathrm{cr}} = \frac{\mathcal{N}}{4\mu v}h\left(\mathcal{J}, \mathcal{G}, d\right),
\label{symmetricfunctionssecond}
\end{equation}
where
\begin{equation}
h\left(\mathcal{J}, \mathcal{G}, d\right) = \frac{\left[1 + (d-2 - \mathcal{G})\mathcal{J}\right]}{(1 -\mathcal{J})[1+(d-1)\mathcal{J}]}.
\label{geometrylinkfactor}
\end{equation}
Given a sensor-symmetric network with $d$ local properties, the factor in equation (\ref{geometrylinkfactor}) codifies the interplay between the inter-sensor correlations $\mathcal{J}$ and the geometry parameter $\mathcal{G}$ for any linear property, which may be local or global. A representation of this interplay can be found in figure \ref{geolinkplot}. Furthermore, note that the formulas in equations (\ref{symmetricfunctionssecond}) and (\ref{geometrylinkfactor}) have been obtained without imposing further restrictions on the functions, which implies that this formalism can be applied to any number of arbitrary linear functions whose coefficients generate vectors that can form any angle and have any length. 

\subsection{The role of inter-sensor correlations I: asymptotic case}
\label{subsec:intersensorasymp}

Let us exploit the previous result to address the problem of selecting an arrangement that is optimal to estimate a specific set of linear functions, which was introduced in section \ref{subsec:relevantinfo}. Mathematically, we need to find the values for $v$ and $\mathcal{J}$ that are optimal for a given $\mathcal{G}$. One approach is to use the fact that, for qubits, $0\leqslant 4v\leqslant 1$, which allows us to lower bound equation (\ref{symmetricfunctionssecond}) as $\bar{\epsilon}_{\mathrm{cr}} \geqslant \bar{\epsilon}_{\mathrm{f}} = \mathcal{N}h\left(\mathcal{J}, \mathcal{G}, d\right)/\mu$ and focus on searching for the amount of correlations $\mathcal{J}$ that minimises this bound after having fixed $\mathcal{G}$, $d$ and $\mu$. In principle, there is no guarantee that the pairs $(4v = 1, \mathcal{J})$ generated by this method will correspond to any physical state, although the bounds on the asymptotic error constructed in this way would still be valid. Nevertheless, later in this section we will study an example that can realise a large portion of the pairs $(4v = 1, \mathcal{J})$ that we are going to predict.

\begin{figure}[t]
\centering
\includegraphics[trim={1cm 0.1cm 1.5cm 0.5cm},clip,width=14.75cm]{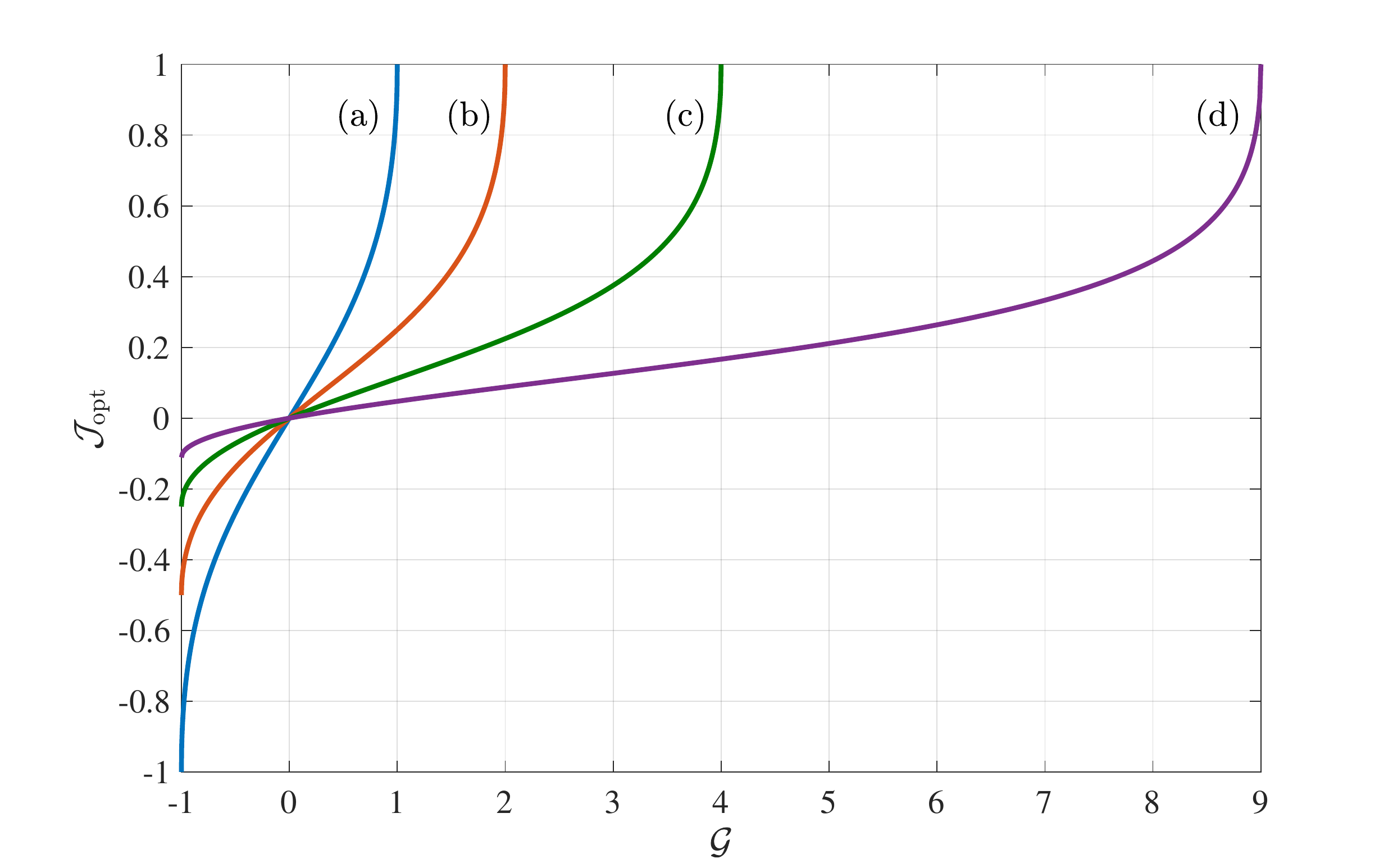}
	\caption[Optimal link between correlations and geometry of the functions]{Optimal link between the inter-sensor correlations $\mathcal{J}$ and the geometry $\mathcal{G}$ of a set of arbitrary linear functions, for $d=2$, $3$, $5$ and $10$. The analytical formula of this result has been provided in equation (\ref{optgeolinkanalytical}).}
\label{linkgeoent}
\end{figure}

The minimisation of $\bar{\epsilon}_{\mathrm{f}}$ reveals that, if $4v = 1$, and restricting our attention to the range $1/(1-d)<\mathcal{J} < 1$, the optimal strength for the inter-sensor correlations of the network is
\begin{equation}
\mathcal{J}_{\mathrm{opt}} = \frac{1}{\mathcal{G}+2-d}\left[1- \sqrt{\frac{(\mathcal{G}+1)(d-1-\mathcal{G})}{d-1}}\right],
\label{optgeolinkanalytical}
\end{equation}
for $-1 < \mathcal{G} < d-1$, which is determined by the structure of the functions alone via $\mathcal{G}$ once $d$ has been fixed\footnote{This result can be found as follows. If we look at $\bar{\epsilon}_{\mathrm{f}}$ as a function of $\mathcal{J}$, then the equation for its extrema is
\begin{equation}
\frac{\mathcal{N}}{\mu}\frac{\partial h\left(\mathcal{J}, \mathcal{G}, d\right)}{\partial \mathcal{J}} = \frac{\mathcal{N}}{\mu} \frac{(d-1)(d-2-\mathcal{G})\mathcal{J}^2+ 2(d-1)\mathcal{J} - \mathcal{G}}{(1 -\mathcal{J})^2[1+(d-1)\mathcal{J}]^2} = 0,
\nonumber
\label{slope}
\end{equation}
whose solutions are
\begin{equation}
\mathcal{J}_{\pm} = \frac{1}{\mathcal{G}+2-d}\left[1\mp \sqrt{\frac{(\mathcal{G}+1)(d-1-\mathcal{G})}{d-1}}\right].
\nonumber
\end{equation}
Since we need to restrict our study to the range $1/(1-d)<\mathcal{J} < 1$ for $F_q$ to be invertible, only $\mathcal{J_{+}}$ is a valid candidate to find a minimum. Next we examine the sign of the slope in the left hand side of equation (\ref{slope}) for some values of $\mathcal{J}$ around $\mathcal{J_{+}}$. By noticing that $\mathcal{N}/\mu > 0$ and using the endpoints of the domain for $\mathcal{J}$ we find that
\begin{equation}
\frac{\partial h\left(1 - \varepsilon, \mathcal{G}, d\right)}{\partial \mathcal{J}} > 0, ~~\frac{\partial h\left(1/(1-d) + \varepsilon, \mathcal{G}, d\right)}{\partial \mathcal{J}} < 0
\nonumber
\end{equation}
for an arbitrarily small $\varepsilon > 0$ when $\mathcal{G}\neq -1$, $\mathcal{G}\neq d-1$, which we exclude to guarantee that $\mathcal{J}\neq 1/(1-d)$, $\mathcal{J}\neq 1$. Consequently, $\mathcal{J_{+}}$ gives rise to the minimum that we were looking for.}. Crucially, equation (\ref{optgeolinkanalytical}) provides a map between correlations and geometry with one-to-one correspondence\footnote{Note that $\mathcal{J}_{\mathrm{opt}} \rightarrow (d-2)/[2(d-1)] $ when $\mathcal{G} \rightarrow d - 2$.}, as it can be directly verified in the representation of equation (\ref{optgeolinkanalytical}) in figure \ref{linkgeoent}. This is the central result of our asymptotic analysis.

Equation (\ref{optgeolinkanalytical}) reveals that, the more a collection of functions is clustered around the vector of ones $\boldsymbol{1}$, the larger the amount of positive correlations is required to be in order to perform the estimation optimally, provided that $4v=1$. Similarly, the amount of correlations with negative strength needs to be large if the functions are instead clustered around the subspace orthogonal to $\boldsymbol{1}$. The potential existence of this type of connection between geometry and correlations was precisely one of the general open questions identified by the authors of \cite{proctor2017networked}, which here has been answered in a definite way for the case of sensor-symmetric states. 

Furthermore, both equation (\ref{optgeolinkanalytical}) and figure \ref{linkgeoent} show that any amount of pairwise correlations would be detrimental whenever the geometry parameter vanishes. We need then to find out which kind of linear functions imply that $\mathcal{G} = 0$. To achieve this goal, let us recall our original definition for $\mathcal{G}$ in equation (\ref{geometryparameter}), that is, $\mathcal{G} = \mathrm{Tr}(\mathcal{W}_f V^\transpose \mathcal{X} V)/\mathcal{N}$. If we choose the uniform weighting matrix $\mathcal{W}_f = \mathbb{I}/l$ and $V$ is an orthogonal transformation, such that $VV^\transpose = V^\transpose V =  \mathbb{I}$, then 
\begin{equation}
\mathcal{G} = \frac{1}{\mathcal{N} l }\mathrm{Tr}(V V^\transpose \mathcal{X}) = \frac{1}{\mathcal{N} l}\mathrm{Tr}(\mathcal{X})=\frac{1}{\mathcal{N} l}  \mathrm{Tr}(\mathcal{I} - \mathbb{I})= 0.
\end{equation}
Now we observe that $\mathcal{J} = 0$, which is the optimal choice for the previous scenario, is always achieved by a separable qubit state $\ket{\psi_0} = (\sqrt{a}\ket{0} + \sqrt{1 - a}\ket{1})^{\otimes d}$, and by selecting $a = 1/2$ we have that $4v = 1$. Thus we can say that the estimation of a set of $l = d$ linear functions that are equally relevant and orthonormal can be carried out optimally by preparing our scheme with separable states. Moreover, since the estimation of the original parameters is a particular case of this type of transformation, our result implies that separable states are also sufficient for the optimal estimation of the primary properties. In other words, our formalism is consistent with the results available in the literature \cite{proctor2017networked, proctor2017networkedshort, altenburg2018, kok2017}. 

According to our definitions of \emph{local} and \emph{global} properties in section \ref{subsec:relevantinfo}, the previous conclusion is technically sufficient to affirm that while entangled pure states are generally useful for the optimal estimation of global properties, it is not true that we always need entangled probes in such case. However, a transformation that is orthogonal preserves angles and lengths, and for that reason one may argue that, in a sense, the information encoded by a set of functions that gives rise to an orthogonal transformation is equivalent to the information content of the original parameters, provided that the weighting matrices are uniform. Hence, it is perhaps not surprising that a local estimation strategy is preferred here, since Proctor \emph{et al.} \cite{proctor2017networked, proctor2017networkedshort} had already shown that the estimation of local properties associated with commuting generators can be performed optimally with a local strategy. 

In view of this, it is desirable to establish whether there are global properties with $\mathcal{G} = 0$ that select information that is not equivalent to estimate all the original parameters. First we observe that the eigendecomposition of $\mathcal{X}$, which is a symmetric matrix, is
\begin{equation}
\mathcal{X}_D = U_{\mathcal{X}}^\transpose \mathcal{X} U_{\mathcal{X}} = \mathrm{diag}\left[(d-1), -1, \dots, -1 \right], 
\end{equation} 
where the eigenvector for the first eigenvalue is $\boldsymbol{1}$ and those for the other eigenvalues belong to the orthogonal subspace\footnote{The characteristic equation for $\mathcal{X}$ is 
\begin{equation}
\mathrm{det}\left(\mathcal{X} - \lambda\mathbb{I}\right) = \mathrm{det}\left[\boldsymbol{1}\boldsymbol{1}^\transpose - (1 + \lambda)\mathbb{I}\right] \propto \left(1 - d + \lambda\right)\left(1+\lambda\right)^{d-1} = 0, 
\nonumber
\end{equation}
giving the eigenvalues $\lambda_1 = d-1$, with multiplicity $1$, and $\lambda_2 = -1$, with multiplicity $d-1$ (see the calculation for equation (\ref{characteristicfim}), which is formally similar to the one here). By inspection we see that $\boldsymbol{1}$ is one of the eigenvectors. Since the latter satisfies that $\mathcal{X}\boldsymbol{1} = (\boldsymbol{1}\boldsymbol{1}^\transpose - \mathbb{I})\boldsymbol{1} = (d-1)\boldsymbol{1}$, the rest of the eigenvalues must be associated with the subspace orthogonal to $\boldsymbol{1}$, and this concludes the eigendecomposition of $\mathcal{X}$.}. That implies that if we choose a single linear function as $V = \boldsymbol{f} = U_{\mathcal{X}}\boldsymbol{1}$, then we will have that $\mathcal{G} = \boldsymbol{1}^\transpose U_{\mathcal{X}}^\transpose \mathcal{X} U_{\mathcal{X}} \boldsymbol{1}/d = \boldsymbol{1}^\transpose \mathcal{X}_D \boldsymbol{1}/d=0$. Imagine that we consider a three-parameter network, so that 
\begin{eqnarray}
\boldsymbol{f} = U_{\mathcal{X}}\boldsymbol{1} = \frac{1}{\sqrt{6}}
\begin{pmatrix}
\sqrt{2} & \sqrt{3} & 1\\
\sqrt{2} & -\sqrt{3} & 1\\
\sqrt{2} & 0 & -2\\
\end{pmatrix} 
\begin{pmatrix}
1 \\
1 \\
1 \\
\end{pmatrix} 
= \frac{1}{\sqrt{6}}
\begin{pmatrix}
\sqrt{2} + \sqrt{3} + 1 \\
\sqrt{2} - \sqrt{3} + 1 \\
\sqrt{2} -2 \\
\end{pmatrix}.
\end{eqnarray}
Clearly, this is a global property, and, as a consequence, this result strengthens the idea that entanglement is sometimes not needed in scenarios where we are estimating global properties. Interestingly, the same argument fails for $d=2$, since in that case
\begin{equation}
\boldsymbol{f} = U_{\mathcal{X}}\boldsymbol{1} = \frac{1}{\sqrt{2}}
\begin{pmatrix}
1 & 1 \\
1 & -1
\end{pmatrix}
\begin{pmatrix}
1 \\
1
\end{pmatrix} =
\begin{pmatrix}
\sqrt{2} \\
0
\end{pmatrix},
\end{equation}
which simply rescales the first parameter, and this is a local property. Nonetheless, our conclusion above is still valid in general.

For the link between geometry and correlations in equation (\ref{optgeolinkanalytical}) to be truly useful, it is necessary that there are physical states with the properties that such link predicts as optimal. Consider first a network where $d=2$. Proctor \emph{et al.} (appendix E of \cite{proctor2017networked}) studied the estimation of $1 \leqslant l \leqslant d$ linear and normalised but otherwise arbitrary functions using the sensor-symmetric state 
\begin{equation}
\ket{\psi_0} = \frac{1}{\sqrt{2\left(1+\gamma^2\right)}}\left[\ket{0 0} + \gamma\left(\ket{01} + \ket{10}  \right) + \ket{1 1}\right],
\label{gammastatetwo}
\end{equation}
with $-\infty < \gamma < \infty$, which is a particular case of the more general formalism that we are developing in this chapter. The fact that the authors of \cite{proctor2017networked} succeeded in solving their problem completely with this state suggests that the latter may realise all the pairs $(4v = 1, \mathcal{J})$ that are optimal according to our results.

Recalling that $\sigma_z \ket{i} = (-1)^{i}\ket{i}$, we can see that, for the state in equation (\ref{gammastatetwo}), $\langle \sigma_{z,1} \rangle = \langle \sigma_{z,2} \rangle = 0$ and $\langle \sigma_{z,1} \sigma_{z,2} \rangle = \langle \sigma_{z,1} \sigma_{z,2} \rangle = (1-\gamma^2)/(1+\gamma^2)$, so that the variance is $4v = 4v_1 = 4v_2 = 1$ and the quantifier for the inter-sensor correlations can be written as a function of $\gamma$ as $\mathcal{J}=(1-\gamma^2)/(1+\gamma^2)$. This function reaches the maximum $\mathcal{J}=1$ at $\gamma = 0$, while it tends monotonically from such point to $\mathcal{J} = - 1$ when $\gamma \rightarrow \pm \infty$. In other words, for $d=2$ there is always a physical state that satisfies the condition imposed in equation (\ref{optgeolinkanalytical}) when $4v = 1$.

It is interesting to observe that $\gamma$ splits the state in a part where the sum of the parameters is encoded and a part that can encode the difference. More concretely\footnote{For this calculation we have used that
\begin{equation}
\mathrm{e}^{i\varphi\sigma_z}\ket{j} = \sum_{k=1}^\infty \frac{\left(i\varphi\right)^k}{k!} \sigma_z^k\ket{j} = \sum_{k=1}^\infty \frac{\left(i\varphi\right)^k}{k!} (-1)^{jk}\ket{j} = \mathrm{e}^{i(-1)^j \varphi}\ket{j}.
\nonumber
\end{equation}},
\begin{align}
\mathrm{e}^{-\frac{i}{2}(\sigma_{z,1}\theta_1+\sigma_{z,2}\theta_2)}\ket{\psi_0} = &~ \frac{1}{\sqrt{2\left(1+\gamma^2\right)}}\left[\mathrm{e}^{-\frac{i}{2}(\theta_1 + \theta_2)}\ket{0 0} + \mathrm{e}^{\frac{i}{2}(\theta_1 + \theta_2)}\ket{1 1} \right] 
\nonumber \\
&+ \frac{\gamma}{\sqrt{2\left(1+\gamma^2\right)}}\left[\mathrm{e}^{-\frac{i}{2}(\theta_1 - \theta_2)}\ket{01} + \mathrm{e}^{\frac{i}{2}(\theta_1 - \theta_2)}\ket{10} \right].
\label{twonetworktransformed}
\end{align}
A partial extension of this idea to the $d$-parameter case can be achieved by constructing a state where the part that encodes functions aligned with the direction of $\boldsymbol{1}$ is separated in an analogous fashion, i.e., 
\begin{eqnarray}
\ket{\psi_0} &=& \frac{1}{\sqrt{2\left[1 + \left( 2^{d-1}-1 \right)\gamma^2 \right]}} \left[\left(1-\gamma\right)\left(\ket{0}^{\otimes d} + \ket{1}^{\otimes d}\right) + \gamma \left(\ket{0} + \ket{1} \right)^{\otimes d} \right].
\nonumber \\
&\propto& \ket{0 0 \dots 0} + \ket{1 1 \dots 1} + \gamma \left(\text{the rest of the terms}\right).
\label{gammastategen}
\end{eqnarray}
For this probe, $4v_i = 1 - \langle \sigma_{z,i} \rangle^2 = 1  = 4v$ for all $i$, and $4c_{ij} = \langle \sigma_{z,i} \sigma_{z,j} \rangle - \langle \sigma_{z,i} \rangle\langle \sigma_{z,j} \rangle = \langle \sigma_{z,i} \sigma_{z,j} \rangle = (1-\gamma^2)/[1 + (2^{d-1}-1)\gamma^2] = 4c$ for all $i\neq j$, verifying in this way that the state in equation (\ref{gammastategen}) is also sensor symmetric. As a result, we can see that its inter-sensor correlations are given by
\begin{equation}
\mathcal{J} = \frac{1-\gamma^2}{1 + \left(2^{d-1}-1\right)\gamma^2}.
\label{gengammacorrelations}
\end{equation} 
If $0 \leqslant \abs{\gamma} \leqslant 1$, then we have that $1 \geqslant \mathcal{J} \geqslant 0$. This implies that there always exists a physical state associated with all the results in this section that require either positive inter-sensor correlations, or the absence of them. On the other hand, the amount of negative correlations that this state can cover lies in $ 0 > \mathcal{J} > - 1/(2^{d-1}-1)$, which corresponds to $1 < \abs{\gamma} < \infty$. Unfortunately, the amount of negative correlations that equation (\ref{optgeolinkanalytical}) might predict can lie in $ 0 > \mathcal{J} > 1/(1-d)$, where $1/(1-d) \leqslant - 1/(2^{d-1}-1)$ for $d\geqslant 2$ and the inequality is only saturated when $d=2$. Thus there is a subinterval not covered by equation (\ref{gammastategen}). Whether there are other physical states that may realise the missing values is an open question. 

Finally, we draw attention to the fact that the only entangled pure probes that may be asymptotically relevant for sensor-symmetric networks are those that give rise to inter-sensor correlations, while any other form of entanglement will be irrelevant in this type of scenario. To illustrate this idea, let us consider the state in equation (\ref{gammastategen}) for $d = 3$, and suppose that the functions to be estimated are associated with $\mathcal{G} = 0$. We have seen that, in that case, no inter-sensor correlations are needed to perform the estimation optimally, which implies that, according to equation (\ref{gengammacorrelations}), $\gamma = \pm 1$ . By inserting these parameters in equation (\ref{gammastategen}) we find that the optimal states are
\begin{equation}
\ket{\psi_{+}} = \frac{1}{2\sqrt{2}}\left(\ket{0}+\ket{1}\right)^{\otimes 3}
\end{equation}
and
\begin{equation}
\ket{\psi_{-}} = \frac{1}{2\sqrt{2}} \left[2\left(\ket{0}^{\otimes 3} + \ket{1}^{\otimes 3}\right) - \left(\ket{0} + \ket{1} \right)^{\otimes 3} \right].
\end{equation}
The first state is separable, but it can be shown that $\ket{\psi_{-}}$ is not. If we tried to write the latter as
$\ket{\psi_{-}} = (x_0\ket{0}+x_1\ket{1})(y_0\ket{0}+y_1\ket{1})(z_0\ket{0}+z_1\ket{1})$, with $|x_0|^2+|x_1|^2 = |y_0|^2+|y_1|^2  = |z_0|^2+|z_1|^2 = 1$, we would find contradictions such as 
\begin{equation}
\left[(x_0 = x_1) \land (x_0 = - x_1)\right]\land (|x_0|^2+|x_1|^2 = 1),
\end{equation}
which by \emph{reductio ad absurdum} allows us to conclude that the state with $\gamma = -1$ and $d = 3$ is entangled. Hence, while here entanglement is not required to reach the asymptotic optimum, neither is it necessarily detrimental. The only requirement imposed by our formalism is the absence of pairwise correlations, and the presence or absence of any other kind of correlation does not affect the uncertainty. 

\subsection{Multi-parameter prior information analysis}
\label{subsec:multiprioranalysis}

Now we focus on two-parameter networks and we turn to the more general problem of estimating linear functions when different amounts of data are available, using our findings about the properties of the asymptotically optimal quantum probes as a guide. To do this, our prior knowledge needs to be consistent with the idea that, if we were to keep repeating the experiment, our scheme would continue being useful (section \ref{theory}). In other words, our prior must allow for the asymptotic regime to be reached, which requires a multi-parameter analysis of the prior information.

In section \ref{subsec:multinonasym} we concluded that a suitable prior for this problem in the regime of moderate prior knowledge is the multi-parameter flat density in equation (\ref{multiprior}). Since our network is highly symmetric, it is appropriate to imagine that our prior information is similar for both parameters, and this justifies assuming that $W_{0,i} = W_0$ and $\bar{\theta}_i = \bar{\theta}$ for $i = 1 , 2$, where we recall that $W_{0, i}$ and $\bar{\theta}_i$ were the prior width and the prior mean for the $i$-th primary parameter. Hence, the prior area is simply $\Delta_0 = W_0^2$, and we will choose $\bar{\theta} = W_0/2$ for the Bayesian calculations in this chapter. 

Next we need to choose $W_0$ (and thus the prior area $\Delta_0$) such that the likelihood does not contain ambiguous information in the region where the original parameters can lie, and to construct the likelihood we have to select a measurement. As with the state, we wish to select an asymptotically optimal POM, which is achieved by requiring that $F(\boldsymbol{\theta})=F_q$. We know that a POM fulfilling this condition always exists for the scenario with pure states and commuting generators considered here (\cite{sammy2016compatibility, pezze2017simultaneous} and sections \ref{subsec:crb} and \ref{subsec:multiasymp}), since, in such case, the symmetric logarithmic derivatives are not unique and we may find some pair of logarithmic derivatives that commute, which would allow us to construct the POM \cite{sammy2016compatibility}. Alternatively, Humphreys \emph{et al.} \cite{humphreys2013} proposed a set of projectors such that one of the elements is the original state, while the rest are orthogonal to it, and this idea was refined and extended in \cite{pezze2017simultaneous} by identifying conditions that projective POMs need to fulfil to have that $F(\boldsymbol{\theta})=F_q$. However, for us it suffices to follow a simpler approach; first we will provide a qualitative argument suggesting a potential measurement scheme, and then we will verify that it is indeed optimal by a direct calculation.  

Although we wish to estimate functions, the condition $F(\boldsymbol{\theta})=F_q$ refers only to the original parameters, and we know that these can be estimated optimally using a local strategy (\cite{proctor2017networked, proctor2017networkedshort} and section \ref{subsec:intersensorasymp}). In view of this, a local POM might be sufficient to make the classical and quantum information matrices equal, and, in fact, this would be very useful for our analysis, since in that case we could associate any enhancement derived from the presence of correlations with the initial state. 

Consider then the local POM $\ket{n, k} = \left[\ket{0} + (-1)^n \ket{1}\right]\otimes[\ket{0} + (-1)^k \ket{1}]/2$, for $n, k = 0, 1$. In addition, we have seen that, if $d=2$, then the state in equation (\ref{gammastatetwo}) is sufficient to realise all the asymptotic results predicted by our theory. As such, we will use this probe for our Bayesian calculation. Combining this POM with the transformed state $\ket{\psi(\theta_1, \theta_1)} = \mathrm{e}^{-\frac{i}{2}(\sigma_{z,1}\theta_1+\sigma_{z,2}\theta_2)}\ket{\psi_0}$ in equation (\ref{twonetworktransformed}), the probability amplitude is
\begin{align}
\braket{n,k}{\psi(\theta_1, \theta_2)}  \propto &~ \mathrm{e}^{-\frac{i}{2}(\theta_1+\theta_2)} + (-1)^{n+k}\mathrm{e}^{\frac{i}{2}(\theta_1+\theta_2)}
\nonumber \\
&+\gamma\left[(-1)^k \mathrm{e}^{-\frac{i}{2}(\theta_1-\theta_2)} + (-1)^n \mathrm{e}^{\frac{i}{2}(\theta_1-\theta_2)}\right]
\nonumber \\
\propto &~ \mathrm{cos}\left\lbrace\left[\theta_1 + \theta_2 + \pi(k+n)\right]/2\right\rbrace
\nonumber \\
&+ \gamma\hspace{0.15em}\mathrm{cos}\left\lbrace\left[\theta_1 - \theta_2 - \pi(k-n)\right]/2\right\rbrace,
\end{align}
the modulus of the proportionality factor being $1/\sqrt{2(1+\gamma^2)}$. This allows us to find the likelihood function
\begin{equation}
p(n, k | \theta_1, \theta_2) = ||\braket{n,k}{\psi(\theta_1, \theta_2)}||^2 = \left[\mathrm{cos}(x_{+}) + \gamma \mathrm{cos}(x_{-})\right]^2/[2(1+\gamma^2)],
\label{multilikelihood}
\end{equation}
where we have introduced the notation $x_{\pm} \equiv \left[\theta_1 \pm \theta_2 \pm \pi(k\pm n)\right]/2$.

The elements of the classical Fisher information matrix for the probability in equation (\ref{multilikelihood}) are
\begin{align}
[F(\boldsymbol{\theta})]_{11} &= \sum_{n,k = 0}^1 \frac{1}{p(n, k | \theta_1, \theta_2)}\left[\frac{\partial p(n, k | \theta_1, \theta_2)}{\partial \theta_1} \right]^2 
\nonumber \\
&= \frac{1}{2\left(1+\gamma^2 \right)} \sum_{n,k = 0}^1  \left[ \mathrm{sin}(x_{+}) + \gamma \mathrm{sin}(x_{-}) \right]^2 = 1,
\label{classfimqubit1}
\end{align}
\begin{align}
[F(\boldsymbol{\theta})]_{22} &= \sum_{n,k = 0}^1 \frac{1}{p(n, k | \theta_1, \theta_2)}\left[\frac{\partial p(n, k | \theta_1, \theta_2)}{\partial \theta_2} \right]^2 
\nonumber \\
&= \frac{1}{2\left(1+\gamma^2 \right)} \sum_{n,k = 0}^1 \left[ \mathrm{sin}(x_{+}) - \gamma \mathrm{sin}(x_{-}) \right]^2 = 1,
\end{align}
and
\begin{align}
[F(\boldsymbol{\theta})]_{12} &= \sum_{n,k = 0}^1 \frac{1}{p(n, k | \theta_1, \theta_2)}\frac{\partial p(n, k | \theta_1, \theta_2)}{\partial \theta_1}\frac{\partial p(n, k | \theta_1, \theta_2)}{\partial \theta_2} 
\nonumber \\
&= \frac{1}{2\left(1+\gamma^2 \right)} \sum_{n,k = 0}^1 \left[ \mathrm{sin}^2(x_{+}) - \gamma^2 \mathrm{sin}^2(x_{-})  \right] = \frac{1-\gamma^2}{1+\gamma^2},
\label{classfimqubit3}
\end{align}
with $[F(\boldsymbol{\theta})]_{21}= [F(\boldsymbol{\theta})]_{12}$. On the other hand, in sections \ref{sec:networksasym} and \ref{subsec:intersensorasymp} (see also appendix E of \cite{proctor2017networked}) we have seen that, for this configuration,
\begin{equation}
F_q = 
\begin{pmatrix}
1 & \mathcal{J} \\
\mathcal{J} & 1
\end{pmatrix} =
\begin{pmatrix}
1 & (1-\gamma^2)/(1+\gamma^2) \\
(1-\gamma^2)/(1+\gamma^2) & 1
\end{pmatrix},
\end{equation}
which is identical to the classical Fisher information matrix in equations (\ref{classfimqubit1} - \ref{classfimqubit3}). Therefore, we conclude that the quantum strategy formed by the previous local POM and the state in equation (\ref{gammastatetwo}) is asymptotically optimal. 

Following section \ref{subsec:multinonasym}, one way of identifying the size of the region where the likelihood function of this strategy is free of ambiguities is to represent the posterior probability $p(\theta_1, \theta_2|\boldsymbol{n}, \boldsymbol{k}) \propto p(\boldsymbol{n}, \boldsymbol{k} |\theta_1, \theta_2)$, where $\boldsymbol{n} = (n_1, \dots, n_\mu)$ and $\boldsymbol{k} = (k_1, \dots, k_\mu)$, so that we can visualise the regions with an asymptotically unique absolute maximum in a direct fashion. The result of this operation, which is based on the algorithm in appendix \ref{sec:multiprior}, is shown in figure \ref{priornetwork} for several values of $\gamma$. 

\begin{figure}[t]
\centering
\includegraphics[trim={0.2cm 0cm 0.5cm 0cm},clip,width=5.1cm]{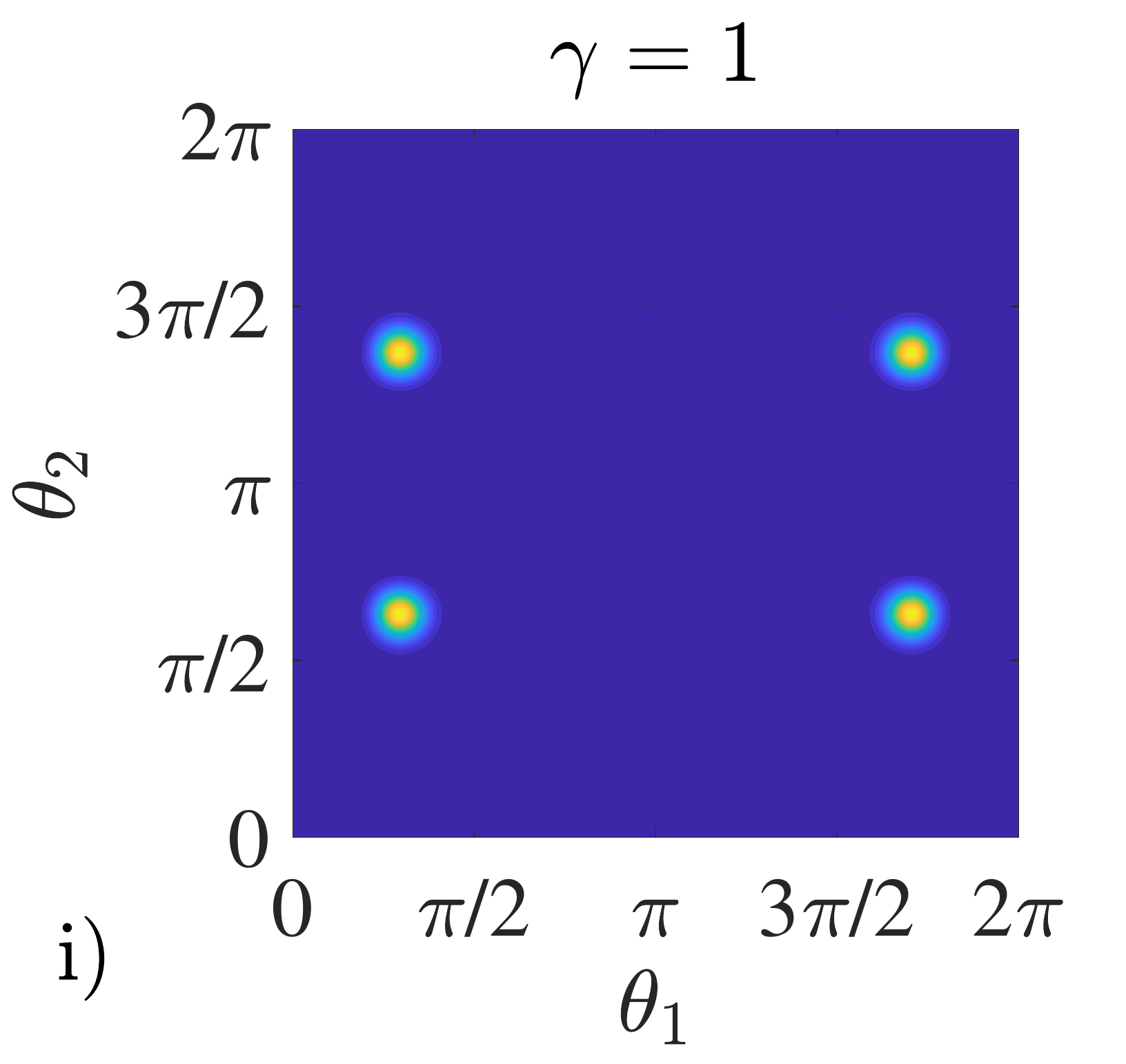}\includegraphics[trim={0.2cm 0cm 0.5cm 0cm},clip,width=5.1cm]{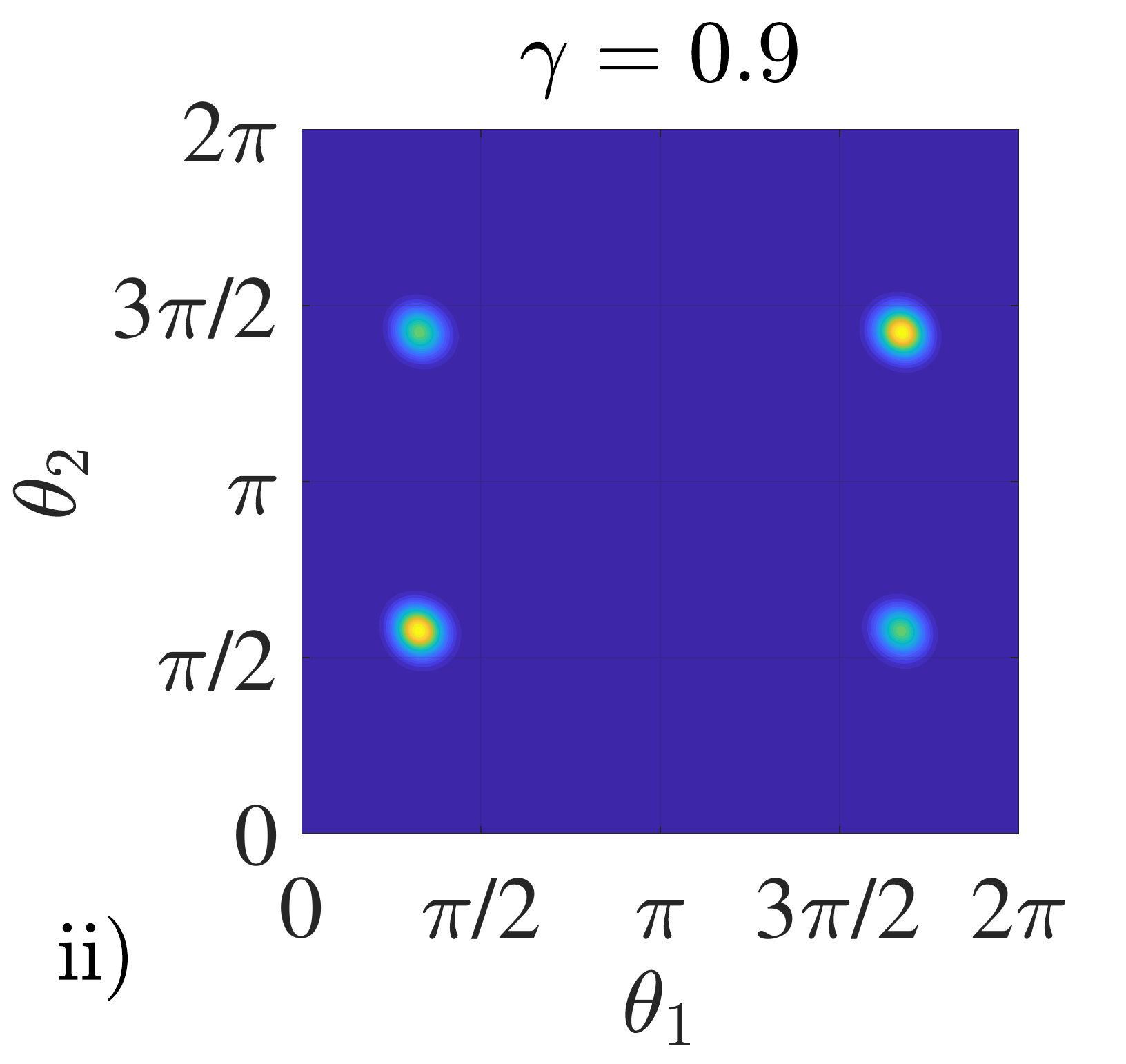}\includegraphics[trim={0.2cm 0cm 0.5cm 0cm},clip,width=5.1cm]{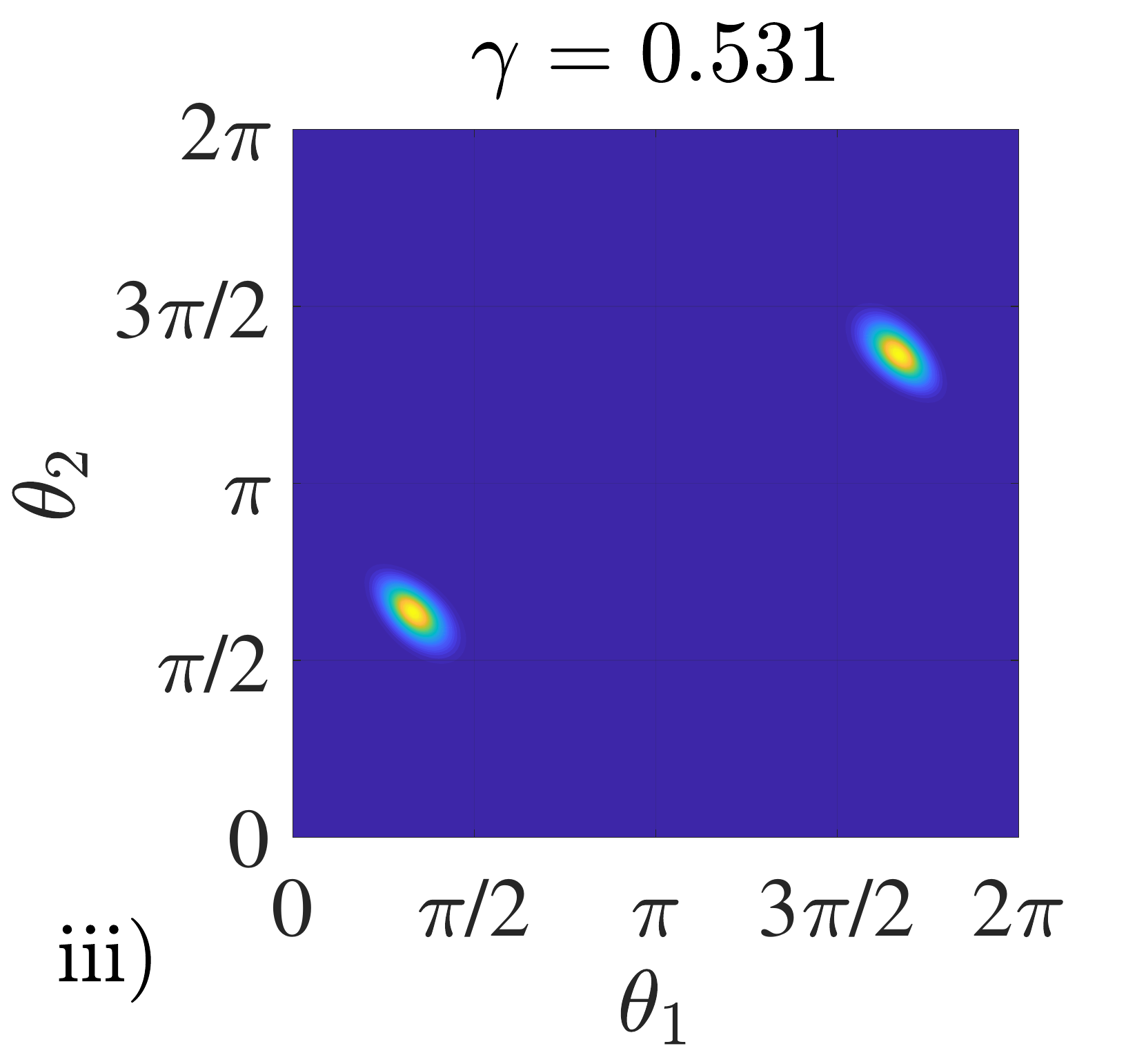}
\includegraphics[trim={0.2cm 0cm 0.5cm 0cm},clip,width=5.1cm]{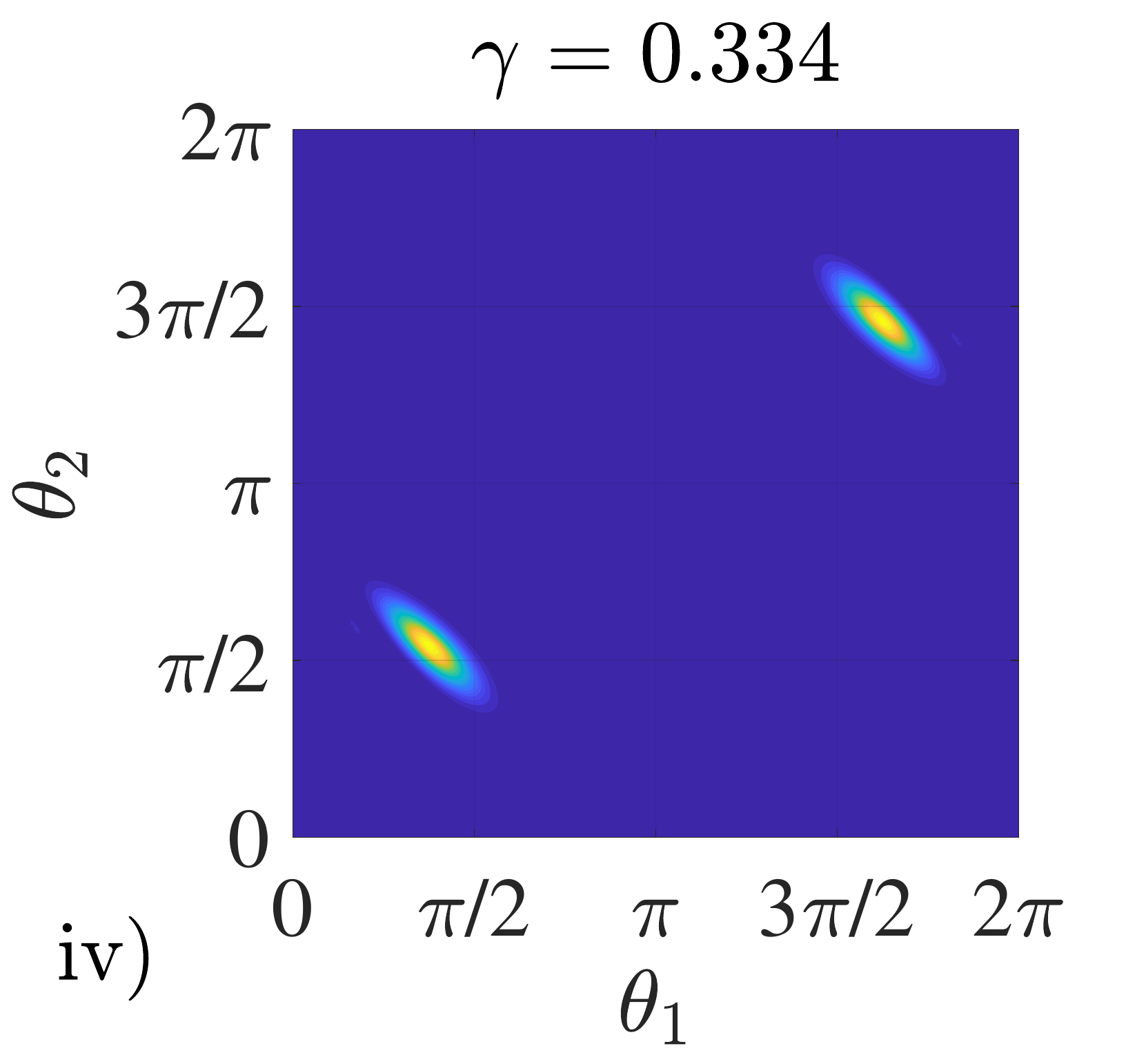}\includegraphics[trim={0.1cm 0cm 0.5cm 0cm},clip,width=5.1cm]{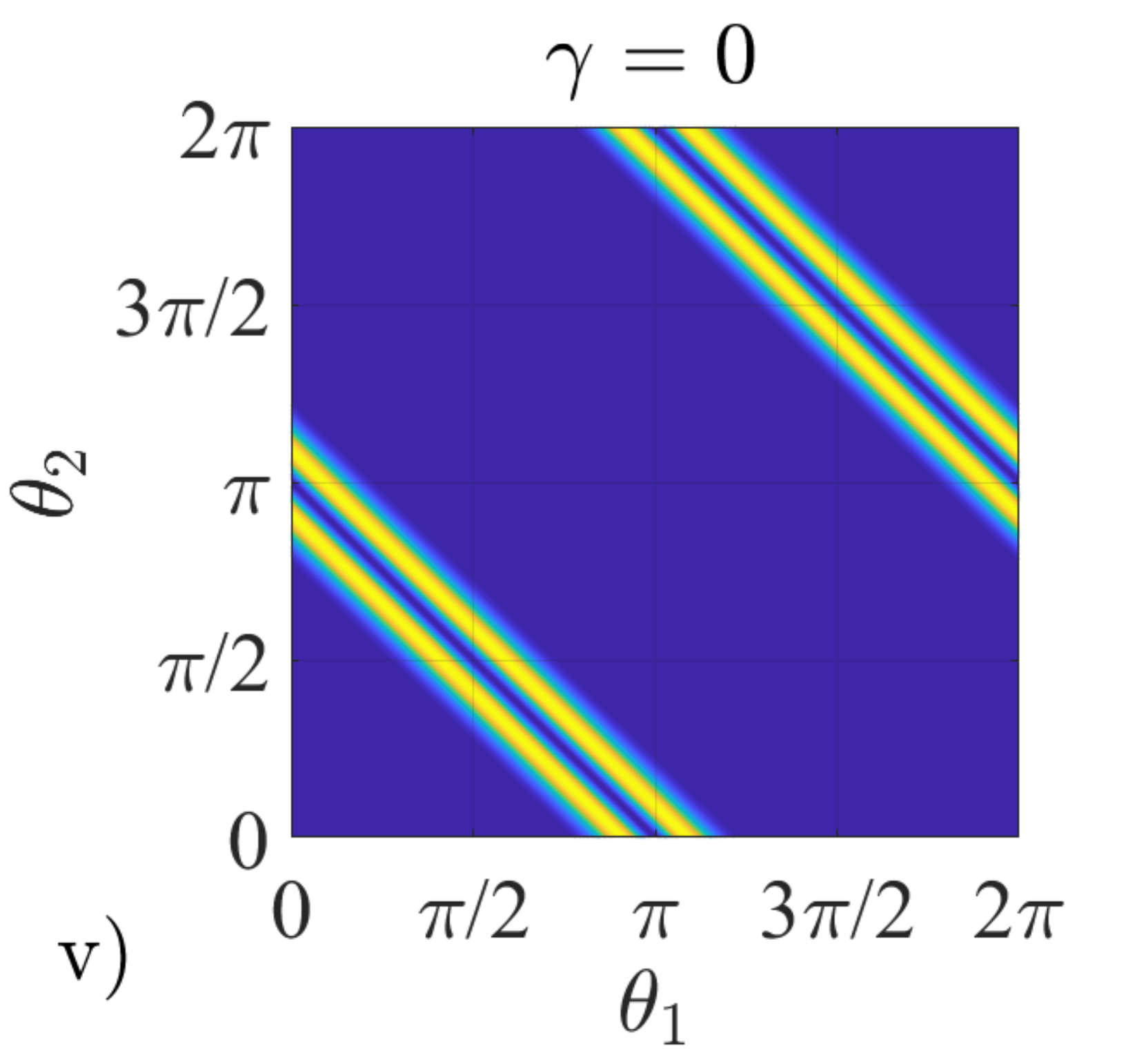}
\caption[Prior information analysis of a two-parameter scheme]{Posterior density functions for random simulations of $\mu = 100$ trials, a flat prior and the quantum strategy represented by the likelihood in equation (\ref{multilikelihood}), with (i) $\gamma = 1$, (ii) $\gamma = 0.9$, (iii) $\gamma = 0.531$, (iv) $\gamma = 0.334$ and (v) $\gamma = 0$. The simulated true values of the original parameters are $\theta'_1=1$ and $\theta'_2=2$. We draw attention to the fact that the representation for $1 < \gamma < \infty $ follows the same pattern but with the posterior peaks tending to the direction orthogonal to that in (v).}
\label{priornetwork}
\end{figure}

First we note that the simulations in figure \ref{priornetwork} have been restricted to the area $(\theta_1, \theta_2) \in [0, 2\pi]\times [0, 2\pi]$ because it is clear that the single-shot likelihood in equation (\ref{multilikelihood}) is invariant under   $\theta_i \rightarrow \theta_i + 2\pi m$, with $m = 0, \pm 1, \pm 2, \dots$ and $i=1,2$\footnote{The calculations required to arrive at this conclusion are analogous to those in section \ref{subsec:prioranalysis} for the NOON state.}, and thus it suffices to examine the symmetries within one period. While the number of maxima changes with $\gamma$, we can observe that all the ambiguities in figures \ref{priornetwork}.i - \ref{priornetwork}.iv can be avoided if the prior area satisfies that $\Delta_0 = W_0^2 \leqslant \pi^2$. 

The situation for $\gamma = 0$ in figure \ref{priornetwork}.v is, however, different. In that case, no single peak can be selected even after a large number of repetitions, which implies that such scheme does not have an asymptotic approximation. This is consistent with the fact that, if $\gamma = 0$, then $\mathcal{J} = 1$, and according to our results in section \ref{sec:networksasym}, this case needs to be excluded for the Fisher information matrix to be invertible. Furthermore, the same type of behaviour would have been observed if we had examined the limit $|\gamma| \rightarrow \infty$, for which $\mathcal{J}\rightarrow -1$. Hence, we only need to impose the existence of a unique maximum for $0 < |\gamma| < \infty$. As such, we conclude that, given our configuration, the intrinsic area is $\Delta_\mathrm{int} = W_\mathrm{int}^2 = \pi^2$, provided that the parameters are thought of as independent before the experiment is performed.

Crucially, the previous discussion does not imply that the scheme with $\gamma = 0$ is useless. Figure \ref{priornetwork}.v shows   that this scheme is giving information about the combination $\theta_2 + \theta_1 = \pi m$, with $m = 0, \pm 1, \pm 2, \dots$, that is, about the sum of the parameters. In fact, this can be seen in a very transparent way by inserting $\gamma = 0$ in equation (\ref{multilikelihood}), since then the likelihood for a single shot is only sensitive to the sum of the primary parameters. The calculations in the next section will reveal that while the performance of this scheme is generally poor in the asymptotic regime, it can be useful when $\mu$ is low.

\subsection{The role of inter-sensor correlations II: non-asymptotic case}
\label{subsec:correlationsmultibayes}

Using the quantum strategy in section \ref{subsec:multiprioranalysis} for a two-sensor qubit network  and the optimal estimator found in section \ref{subsec:multiasymp}, we wish to estimate two global properties of such network when the experiment operates both in and out of the regime of limited data. In particular, we consider the linear functions $f_1(\boldsymbol{\theta}) = (2\theta_1 + \pi \theta_2)/\sqrt{4+\pi^2}$ and $f_2(\boldsymbol{\theta}) = (2\theta_1 + \theta_2)/\sqrt{5}$, which we encode in the columns of $V$ as
\begin{equation}
V = \frac{1}{\sqrt{20 + 5\pi^2}}
\begin{pmatrix}
2\sqrt{5} & 2\sqrt{4+\pi^2}\\
\pi \sqrt{5} & \sqrt{4+\pi^2}
\end{pmatrix}.
\label{funlimiteddata}
\end{equation}
To complete this task, we assume that both functions are equally relevant, so that $\mathcal{W}_f = \mathbb{I}/2$, and that our prior knowledge is represented by the prior probability $p(\theta_1, \theta_2) = 4/\pi^2$, when $(\theta_1, \theta_2) \in [0, \pi/2]\times[0, \pi/2]$, and zero otherwise. Since $\Delta_0 = \pi^2/4 < \pi^2 = \Delta_\mathrm{int}$, our analysis in section \ref{subsec:multiprioranalysis} implies that this prior assignment will allow us to reach the asymptotic regime.

Let us start by comparing a local strategy with an entangled scheme that is asymptotically optimal. The former assumes that the experiment is arranged such that $\gamma = 1$, $\mathcal{J} = 0$, while to find the properties of the latter we need to recall our results in section \ref{subsec:intersensorasymp} for the asymptotic role of inter-sensor correlations. Equation (\ref{optgeolinkanalytical}) indicates that, for $d = 2$, 
\begin{equation}
\mathcal{J}_{\mathrm{opt}} = \left(1- \sqrt{1 - \mathcal{G}^2}\right)/\mathcal{G},
\label{twoparametergeolink}
\end{equation}
when $\mathcal{G}\neq 0$, and $\mathcal{J}_{\mathrm{opt}} = 0$ if $\mathcal{G} = 0$. In addition, $\mathcal{J} = (1-\gamma^2)/(1+\gamma^2)$, and by combining the latter expression with equation (\ref{twoparametergeolink}) we find that
\begin{equation}
\gamma_{\mathrm{opt}} = \pm \left(\frac{\mathcal{G}-1+\sqrt{1-\mathcal{G}^2}}{\mathcal{G} + 1 - \sqrt{1-\mathcal{G}^2}}\right)^{\frac{1}{2}},
\label{gammaoptlink}
\end{equation}
when $\mathcal{G}\neq 0$, and $\mathcal{\gamma}_{\mathrm{opt}} = 1$ if $\mathcal{G} = 0$. The normalisation term for the functions in equation (\ref{funlimiteddata}) is simply $\mathcal{N} = \mathrm{Tr}(\mathcal{W}_f V^\transpose V) = 1$, while the geometry parameter is $\mathcal{G} = \mathrm{Tr}\left(\mathcal{W}_f V^\transpose \mathcal{X} V \right)/\mathcal{N} = (8 + 10\pi + 2\pi^2)/(20 + 5\pi^2) \approx 0.853$, and by inserting this result in equations (\ref{twoparametergeolink}) and (\ref{gammaoptlink}) we have that $\gamma_\mathrm{opt} \approx \pm 0.531$ (we can choose the positive solution without loss of generality) and that $\mathcal{J} = 0.561$, where the latter verifies that this state is indeed entangled\footnote{This is because the two-sensor state in equation (\ref{gammastatetwo}) is only separable when $\gamma^2 = 1$. To show it, we just need to impose that
\begin{equation}
\left(x_0 \ket{0} + x_1\ket{1}\right)\left(y_0 \ket{0} + y_1\ket{1}\right) \propto \ket{00} + \gamma \left(\ket{01} + \ket{10} \right) + \ket{11},
\nonumber 
\end{equation}
from where the previous statement follows.}.   

\begin{figure}[t]
\centering
\includegraphics[trim={0.2cm 0cm 0.5cm 0cm},clip,width=14.75cm]{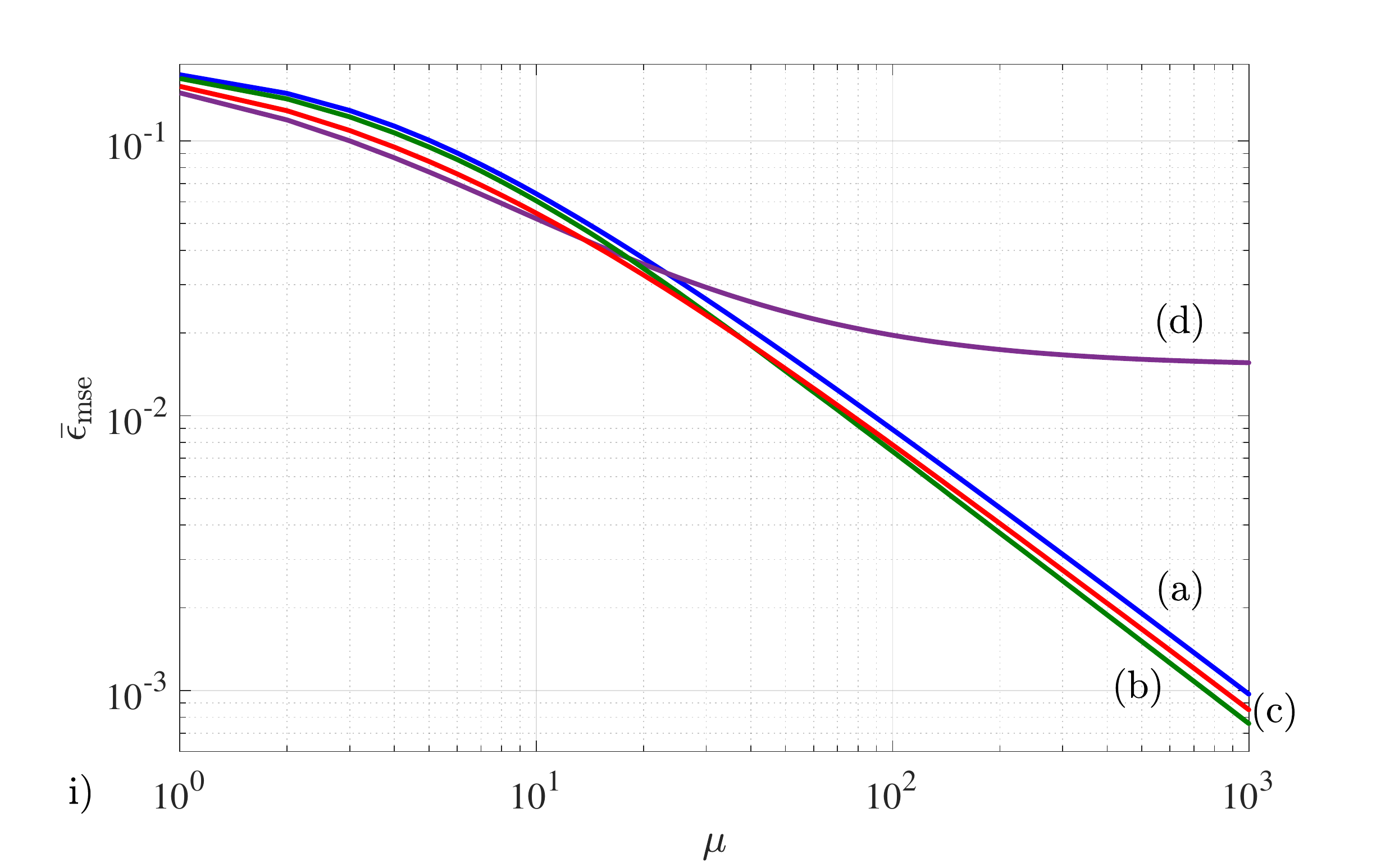}
\includegraphics[trim={0.2cm 0cm 0.5cm 0cm},clip,width=5.1cm]{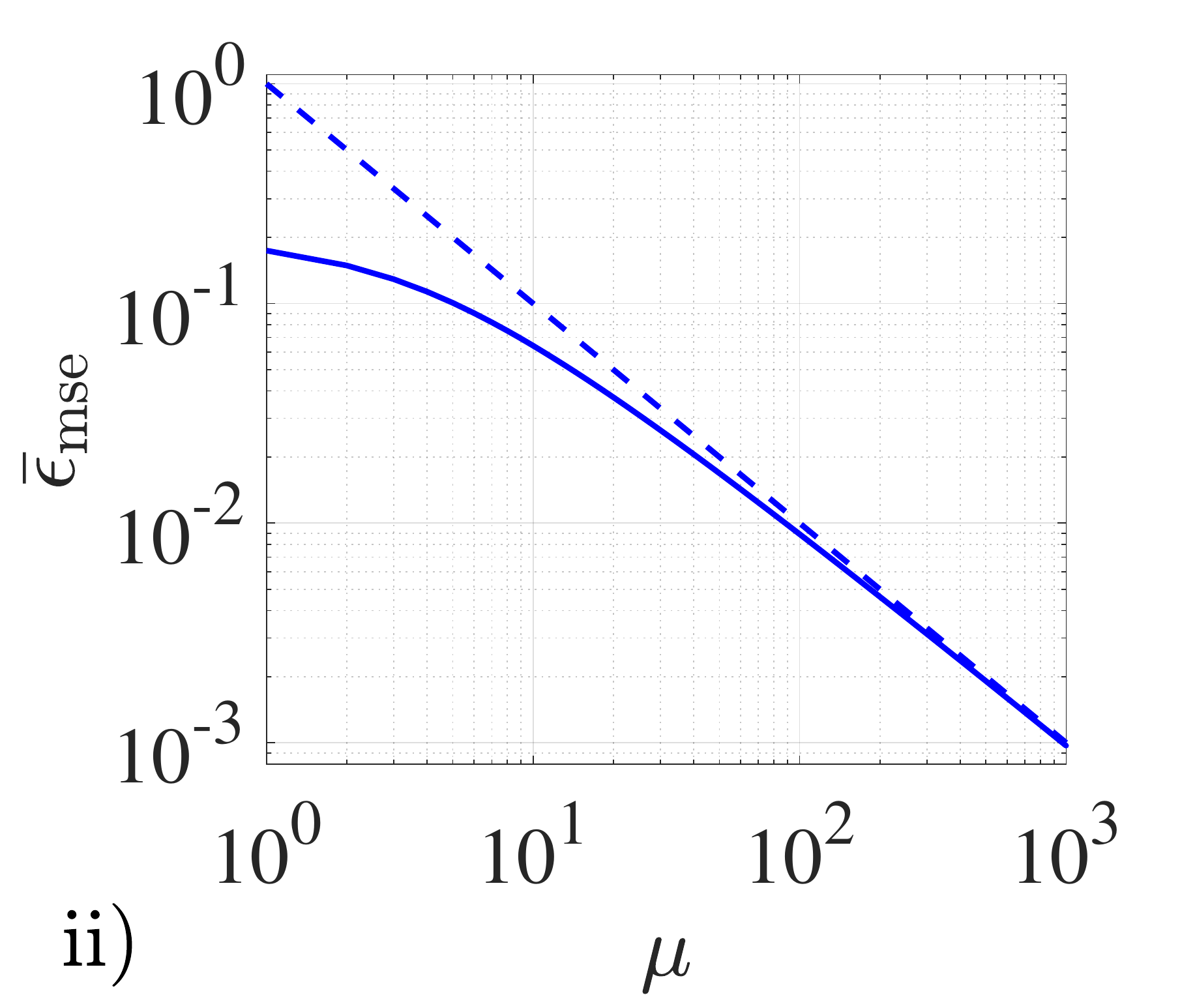}\includegraphics[trim={0.2cm 0cm 0.5cm 0cm},clip,width=5.1cm]{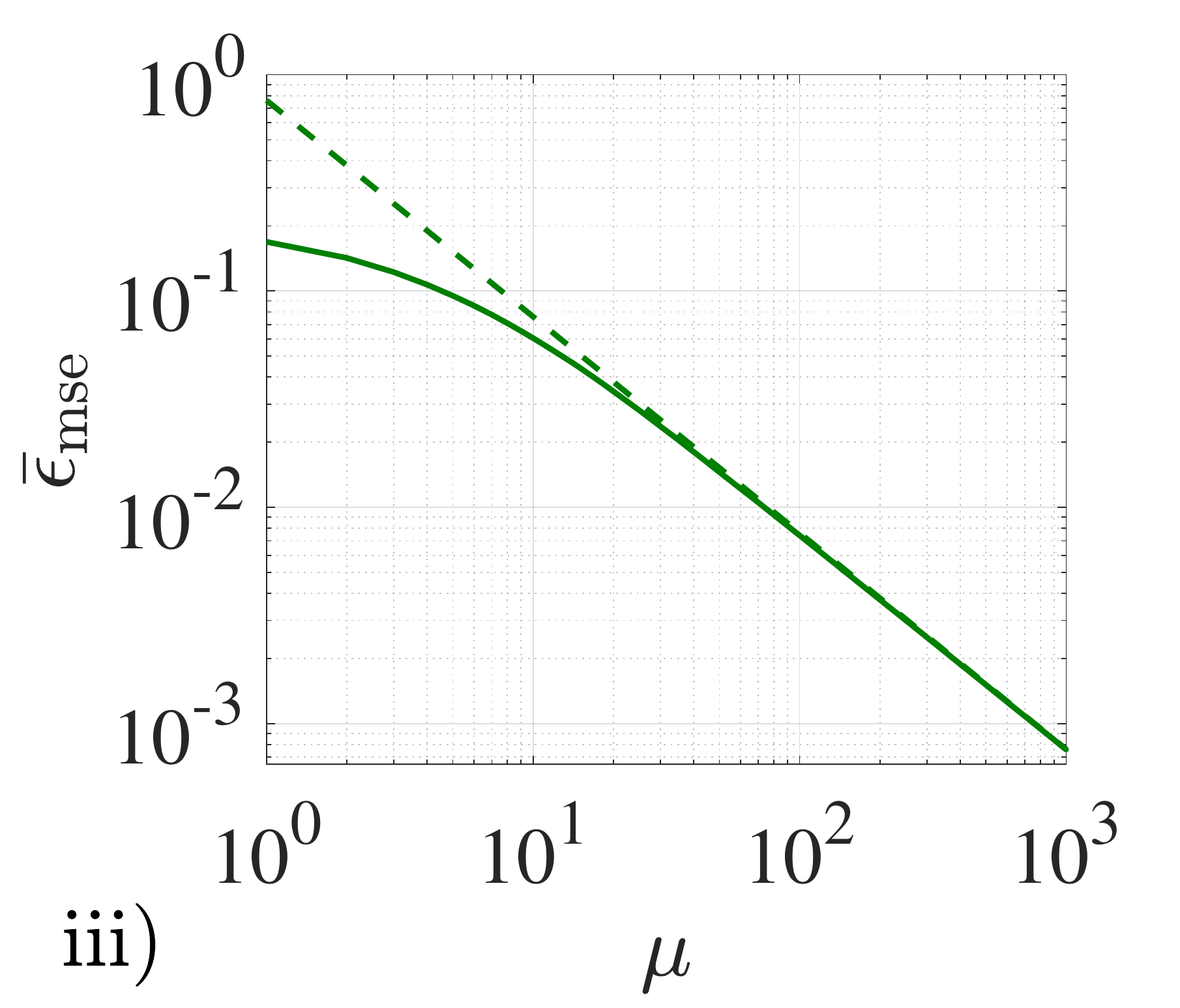}\includegraphics[trim={0.2cm 0cm 0.5cm 0cm},clip,width=5.1cm]{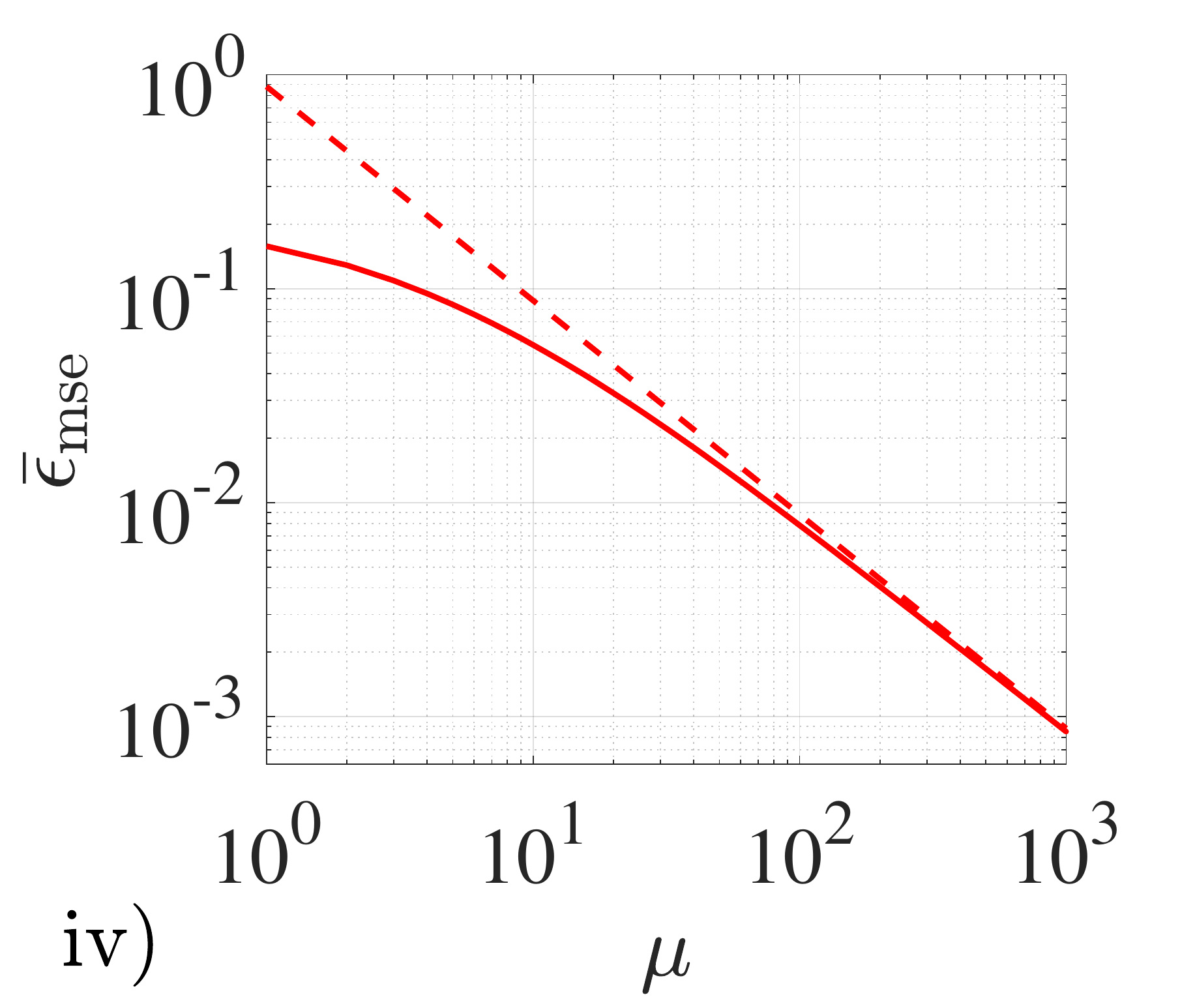}
\caption[Estimation of two linear functions in the non-asymptotic regime]{i) Mean square error for the estimation of the linear functions $f_1(\boldsymbol{\theta}) = (2\theta_1 + \pi \theta_2)/\sqrt{4+\pi^2}$ and $f_2(\boldsymbol{\theta}) = (2\theta_1 + \theta_2)/\sqrt{5}$ by means of the two-sensor qubit network introduced in section \ref{subsec:multiprioranalysis}, where (a) is a local strategy, with $\gamma = 1$, $\mathcal{J} = 0$; (b) is the asymptotically optimal entangled strategy, with $\gamma = 0.531$, $\mathcal{J} = 0.561$; (c) is a strategy whose enhancement has been balanced between the asymptotic and non-asymptotic regimes, with $\gamma = 0.334$, $\mathcal{J} = 0.799$; and (d) is a maximally entangled state, with $\gamma = 0$, $\mathcal{J} = 1$, while figures (ii - iv) compare the mean square error (solid lines) and the multi-parameter quantum Cram\'{e}r-Rao bound (dashed lines) for the strategies in (a - c). All the calculations assume the weighting matrix $\mathcal{W}_f = \mathbb{I}/2$ and a flat prior  of area $\Delta_0 = \pi^2/4$ and centred around $(\pi/4, \pi/4)$.}
\label{nonasymptoticnetwork}
\end{figure}

The numerical calculation of $\bar{\epsilon}_{\mathrm{mse}}$ in equation (\ref{msefunctionsmin}) for these two strategies can be performed with the algorithm in appendix \ref{sec:multimsematlab}, and the results have been represented in figure \ref{nonasymptoticnetwork}.i as graphs (a) for the local scheme and (b) for the optimal entangled strategy. We can observe that the local strategy performs worse than the entangled one for any number of repetitions. Therefore, in this case we have that the prediction made by the asymptotic theory is qualitatively preserved in the non-asymptotic regime. However, a closer analysis reveals that the distance between both graphs is considerably shorter when $1 \leqslant \mu \lesssim 20$ than when $\mu \gg 1$. This behaviour is reminiscent of what we found for a Mach-Zehnder interferometer in figure \ref{bounds_results}.i, where some of the probes with a large Fisher information (and thus with a good asymptotic performance) had an error very close to that of a coherent laser beam in the regime of limited data, the latter being an optical analogue of the notion of local strategy in this chapter. Moreover, from the optical study we learned that a better asymptotic error was sometimes associated with a worse performance in the regime of low $\mu$. As a consequence, a natural question is whether we could obtain an uncertainty that is lower than the error for the asymptotically optimal entangled state when the network operates in the non-asymptotic regime. 

\begin{table} [t]
\centering
{\renewcommand{\arraystretch}{1.2}
\begin{tabular}{|l|c|c|c|}
\hline
Strategy & $\gamma$ & $\mathcal{J}$ & $\mu_{\tau} (\Delta_0=\pi^2/4)$ \\
\hline
\hline
Local & $1$ & $0$ & $4.58\cdot 10^2$ \\
Asymptotically optimal & $0.531$ &$0.561$ & $4.3\cdot 10$ \\
Balanced enhancement & $0.334$ & $0.799$ & $5.37\cdot 10^2$  \\
Maximally entangled & $0$ & $1$ & $-$ \\
\hline
\end{tabular}}
\caption[Bayesian study of a quantum sensing network]{Properties of different strategies based on a two-parameter qubit network, where $\gamma$ selects the state and $\mathcal{J}$ is the amount of inter-sensor correlations. Furthermore, the third column provides the number of repetitions needed such that the relative error between the Bayesian uncertainty and the Cram\'{e}r-Rao bound is equal to or less than $5\%$ (that is, $\varepsilon_\tau = 0.05$ in equation (\ref{saturation})). These results demonstrate the state-dependent nature of the conditions required to approach the Cram\'{e}r-Rao in multi-parameter systems.}
\label{multinetworkstable}
\end{table}

To test this idea, let us select a third arrangement with an asymptotic error that lies between those of the local scheme and the asymptotically optimal strategy. The asymptotic error for our network can be written in terms of $\gamma$ as (see equations (\ref{symmetricfunctionssecond}) and (\ref{geometrylinkfactor}))
\begin{equation}
\bar{\epsilon}_\mathrm{mse} \approx \bar{\epsilon}_{\mathrm{cr}} = \frac{\left(1+\gamma^2\right)\left[\left(1-\mathcal{G}\right)+\left(1+\mathcal{G}\right)\gamma^2\right]}{4\mu \gamma^2} \equiv \bar{\epsilon}_{\mathrm{qbit}}\left(\gamma\right),
\label{crmsetwoqubitnetwork}
\end{equation}
and using this we can find the $\gamma$ of the strategy satisfying our desideratum above by imposing that
\begin{equation}
\bar{\epsilon}_{\mathrm{qbit}}\left(\gamma\right) = \frac{1}{2}\left[\bar{\epsilon}_{\mathrm{qbit}}\left(\gamma_{\mathrm{loc}}=0\right) + \bar{\epsilon}_{\mathrm{qbit}}\left(\gamma_{\mathrm{ent}}=0.531\right)\right].
\end{equation}
The solutions of this equation are $\gamma \approx \pm 0.334 , \pm 0.842$, and we take our third strategy to be prepared such that $\gamma = 0.334$, $\mathcal{J} = 0.799$, since this is the option with the lowest uncertainty for a single shot\footnote{In particular, $\bar{\epsilon}_{\mathrm{mse}}(\mu = 1, \gamma = 0.334) \approx 0.158$ and $\bar{\epsilon}_{\mathrm{mse}}(\mu = 1, \gamma = 0.842) \approx 0.173$.}. 

The uncertainty $\bar{\epsilon}_{\mathrm{mse}}$ for the third scheme has been represented as a function of the number of trials in figure \ref{nonasymptoticnetwork}.i, where it is labelled as (c). As expected, this error lies between the local and the asymptotically optimal strategies when $\mu \gg 1$, but this is no longer the case in the regime of limited data. More concretely, the graphs for the asymptotically optimal strategy and the new scheme cross each other when $\mu \approx 40$, so that the former is optimal when $\mu > 40$ and the latter is the preferred choice if $1 \leqslant \mu \lesssim 40$. Consequently, we may say that trading a part of the asymptotic enhancement is sometimes associated with an improved performance in the non-asymptotic regime, which is the same phenomenon that we uncovered in chapter \ref{chap:limited} for highly sensitive optical probes.

Interestingly, the balanced strategy ($\gamma = 0.334$, $\mathcal{J} = 0.799$) is associated with a larger amount of inter-sensor correlations, and it can be argued that this is consistent with the fact that this scheme provides a better precision in the non-asymptotic regime. To see why, let us first recall that, when $\mu$ is large, the information about the global properties is essentially provided by the experimental data that we are accumulating, so that the strength of the correlations predicted by the asymptotic theory is assuming a large amount of information. On the contrary, the information in the regime of limited data is a mixture of prior knowledge and experimental data, and given that we are employing a moderately vague prior, it is reasonable to expect the amount of entanglement that is optimal when we have an abundance of measurement data to be generally inappropriate in the non-asymptotic regime. By noting that the geometry parameter $\mathcal{G} \approx 0.853$ is relatively close to $1$, which was precisely the geometry value for the direction of the vector of ones $\boldsymbol{1}$ (i.e., our functions are clustered around the equally weighted sum of the parameters), we can compensate the low amount of information with a $\mathcal{J}$ that is closer to that associated with $\boldsymbol{1}$, which is $\mathcal{J} = 1$, in order to enhance the precision when $\mu$ is low. This is what (b) and (c) in figure \ref{nonasymptoticnetwork}.i show. 

We may push this intuition further and consider a network with $\gamma = 0$, $\mathcal{J} = 1$, which is a maximally entangled state. Its graph has been labelled as (d) in figure \ref{nonasymptoticnetwork}.i, and upon comparing it with the three previous strategies we see that the maximally entangled state is the best option when $1 \leqslant \mu \lesssim 10$. The price that we pay for this low-$\mu$ enhancement is that the scheme ceases to be useful after $\mu \approx 20$ trials, and it is asymptotically beaten by the rest of schemes, including the local strategy. We notice that this result is consistent with our analysis in section \ref{subsec:multiprioranalysis}, where we established that this probe is only sensitive to the equally weighted sum of the original parameters. 

The maximally entangled state is also a good example to illustrate that the main consequence of a non-invertible Fisher information matrix is the lack of the asymptotic approximation provided by the Cram\'{e}r-Rao bound, without this implying that we cannot perform the estimation using such strategy. On the contrary, for the local, asymptotically optimal and balanced strategies we have that the Bayesian mean square errors converge to their respective Cram\'{e}r-Rao bounds, as it may be verified by observing figures \ref{nonasymptoticnetwork}.ii - \ref{nonasymptoticnetwork}.iv. The number of repetitions required for the relative error between these Bayesian uncertainties and their asymptotic bounds to be equal to or less than $5\%$ runs from $\mu \sim 10$ to $\mu \sim 10^2$ (see table \ref{multinetworkstable}). 

In summary, we have demonstrated that the strength of the inter-sensor correlations that is useful to estimate a given collection of global properties changes substantially for different amounts of data, i.e., for different values of $\mu$. Since this is the same type of behaviour that we have established for single-parameter schemes in previous chapters, we conjecture that the novel effects associated with a limited amount of data that here have been uncovered using specific examples may actually be a more general feature of non-asymptotic quantum metrology and be generally present in a wide range of experiments operating in the regime of limited data.

\section{Summary of results and conclusions}

In this chapter we have made the transition from single-parameter problems to scenarios with several unknown pieces of information. One of the crucial advantages of exploiting multi-parameter schemes is the possibility of harnessing correlations between different sensors in an array of them, and to study this question we have built our work on the quantum sensing network model that Proctor \emph{et al.} \cite{proctor2017networked, proctor2017networkedshort} proposed as a framework for problems of distributed sensing. 

We have seen that previous results in the literature had established that the presence of correlations between sensors is particularly useful when we wish to estimate properties that can be seen as \emph{global} with respect to a partition in terms of spatially separated sensors. In the context of the model in \cite{proctor2017networked, proctor2017networkedshort}, a property is said to be \emph{local} if it can be represented by a locally encoded parameter, while a global property is modelled by a non-trivial function of two or more local parameters. Given these basic notions, our first step has been to introduce the concepts of \emph{natural or primary} and \emph{derived or secondary} properties for a quantum sensing network, where the former are the physical parameters that characterise the system and the latter are functions of them. Crucially, it has been argued that, to some extent, we are free to decide which parameters are natural and which ones are secondary, and for the purposes of this chapter we have taken the primary parameters to be local. 

Next we have carried out an analysis to determine the measure of uncertainty that is suitable for the estimation of functions in general, and of linear functions in particular, in analogy with our related discussion in section \ref{sec:uncertainty}, and this has provided us with a framework suitable to extend our methodology in chapter \ref{chap:nonasymptotic} to the multi-parameter regime. More concretely, we have selected the multi-parameter estimator that is optimal for any number of trials, and we have examined the asymptotic regime of the multi-parameter Bayesian error as a potentially useful guide for choosing the quantum strategy. Apart from the conditions on the prior information and the number of repetitions that we had already encountered in chapter \ref{chap:nonasymptotic}, the fact that the Fisher information matrix is sometimes singular introduces here a new potential difficulty. We have generalised our methods in previous chapters to find out the minimum amount of prior knowledge and trials that are needed for the multi-parameter Cram\'{e}r-Rao bound to be valid, and we have restricted our proposal of exploiting the asymptotic theory as a guide to cases where the information matrix is invertible.  Nonetheless, we have conjectured that it might be possible to adapt our approach to singular Fisher information matrices (e.g., working in the support of such matrix). 

The central question that we have addressed with this formalism is that of the role of inter-sensor correlations for the estimation of arbitrary linear functions using sensor-symmetric networks and different amounts of data. First we have centred our attention on the asymptotic part of the problem, and we have derived an analytical expression that provides us with a link between the geometry of the vectors formed by the components of the linear functions and the amount of inter-sensor correlations, such that the asymptotic uncertainty is optimal. Furthermore, we have shown that there exists a physical state for most of the optimal configurations that our result predicts. Crucially, this relationship between the amount of entanglement in a pure state and how much the vectors associated with the functions are clustered around certain directions was precisely one of the open questions that Proctor \emph{et al.} \cite{proctor2017networked} identified when they proposed their network model, and here we have provided a definite and complete answer for the case of sensor-symmetric states. Additionally, our results are applicable to any number of linear functions, while other approaches in the literature have generally focused on estimating either a single function or an orthonormal collection of them. 

Using these results we have been able to show that the largest amounts of correlations are associated for sensor-symmetric states with two special subspaces: the direction indicated by the vector of ones, and the subspace orthogonal to it. Furthermore, we have recovered the known result that orthogonal transformations, which include the estimation of the original parameters as a trivial case, can be estimated optimally without inter-sensor correlations.  

While orthogonal transformations are generally a form of global properties, we have given arguments suggesting that the information captured by an orthogonal transformation is, in a sense, equivalent to that encoded by all the original parameters, which in our model are local. In view of this, it was crucial to establish whether there exist other global properties that require no correlations to be estimated optimally. The answer to this question has been in the affirmative, and we have constructed an example demonstrating this idea explicitly. As a consequence, our results have strengthened the idea that entanglement is sometimes not needed even when we are estimating global properties. 

Moreover, another example with a three-sensor network has revealed that entanglement might not only be detrimental, but that it might also be irrelevant. The key idea is to observe that the asymptotic uncertainty only depends on the inter-sensor correlations, which are of a pairwise nature; consequently, other forms of entanglement that do not produce this specific type of correlations do not affect the estimation error in the asymptotic regime. 

On the other hand, the application of our link between geometry and correlations has allowed us to select an asymptotically optimal quantum strategy for our Bayesian analysis of sensor-symmetric networks. After finding a POM for which the classical and quantum Fisher information matrices coincide, we have determined the prior information that such scheme would require for the Cram\'{e}r-Rao bound to be a valid approximation, establishing in this way the size of the region where the estimation can be performed without ambiguities. Remarkably, we have succeeded in applying Jaynes's principle of transformation groups to our multi-parameter problem, and we have justified the use of a multivariate flat prior from first principles. 

From the study of the non-asymptotic uncertainty of this strategy we have learned that the amount of correlations that are needed to enhance the performance of the network crucially depends on the amount of data that has been collected. While our Bayesian analysis is still limited (we have only considered the case $d = 2$ for this part of the problem), the fact that we have found important results with such a low-dimensional estimation problem invites optimism and suggests that there is still a vast set of unexplored possibilities to be uncovered. For instance, it would be interesting to examine whether the irrelevancy of forms of entanglement other than those that generate inter-sensor correlations is also true for a low number of trials, which is a question that requires simulations where $d \geqslant 3$. Therefore, we conclude that our proposal provides a solid methodology to investigate the design of quantum sensing networks that operate in a regime with realistic amounts of data. 

The results of this chapter will appear in \cite{jesus2019a}
\begin{displayquote}
\emph{Quantum sensing networks for the estimation of linear functions}, \underline{Jes\'{u}s} \underline{Rubio}, Paul A. Knott, Timothy J. Proctor and Jacob A. Dunningham, in preparation (2020).
\end{displayquote}
\chapter{Bayesian multi-parameter quantum metrology}
\label{chap:multibayes}

\section{Goals for the final stage of our methodology}

Our study of quantum sensing networks has uncovered a wealth of new results associated with the interplay between correlations and different amounts of data. However, this is only a small part of the rich variety of novel effects that we expect to be relevant for multi-parameter non-asymptotic metrology. While the practical usefulness of our hybrid estimation method (optimal estimator plus asymptotically optimal quantum strategy) will certainly play an important role in exploring this line of thought, it is clear that a limited amount of data demands multi-parameter tools that are specifically designed to take into account the intrinsically Bayesian nature of this type of scenarios.  

In chapter \ref{chap:methodology} we saw that the fundamental equations for the optimal Bayesian quantum strategy have been known since the works of Helstrom, Holevo and others \cite{personick1971, helstrom1976, helstrom1974, holevo1973b, holevo1973, yuen1973}, and in chapter \ref{chap:limited} we exploited this formalism for single-parameter schemes in a way that takes into account the reality of experimental practice, where the resources are always finite and possibly limited. In particular, we have proposed to first calculate the single-shot optimal quantum strategy and then repeat it as many times as the application at hand demands or allows for, and we demonstrated that this procedure generates uncertainties that are optimised in a shot-by-shot fashion and that sometimes recover the Cram\'{e}r-Rao bound asymptotically.

The main task of this chapter is to extend the previous idea to the multi-parameter regime, a goal that will complete the construction of our non-asymptotic methodology. To achieve it, first we will derive a new single-shot lower bound on the multi-parameter uncertainty on the basis of the single-parameter optimum for the square error, so that our result will always be applicable to scenarios where there is a moderate amount of prior information. Then we will discuss how and under which circumstances we can employ our new tool in strategies where the same experiment is repeated several times, and we will illustrate the application of these ideas with two important examples: the two-parameter qubit network that we studied in chapter \ref{chap:networks}, and a discrete model of phase imaging. 

Our findings will provide new insights to understand the role of correlations when we wish to estimate the natural properties (i.e., the original parameters) of some experiment where the data is limited and the prior information is moderate. A comprehensive study of this question was, to the best of our knowledge, missing, since the literature has mainly addressed it using asymptotic tools \cite{proctor2017networked, proctor2017networkedshort, knott2016local, altenburg2018} and the Bayesian part of our analysis in chapter \ref{chap:networks} has been primarily dedicated to the estimation of functions of the original parameters. In addition, we will demonstrate that the multi-parameter Cram\'{e}r-Rao bound is sometimes recovered as a limiting case within our approach. While our results will not be as general as if we could solve the fundamental equations for the multi-parameter optimal strategy in an exact fashion (section \ref{subsec:fundeq}), they will be shown to constitute a reasonable alternative that not only can be applied to real-world problems, but that also relies on calculations that are tractable, both from a numerical and an analytical point of view.

\section{Methodology (part D)}

\subsection{A new multi-parameter single-shot quantum bound}
\label{subsec:mybound}

Suppose we have a probe state $\rho_0$ that is employed to encode the unknown parameters $\boldsymbol{\theta} = (\theta_1, \cdots, \theta_d)$, so that the transformed state is $\rho(\boldsymbol{\theta})$, and that we perform a single measurement $E(m)$ with outcome $m$. Then the likelihood function will be $p(m|\boldsymbol{\theta}) = \mathrm{Tr}[E(m) \rho(\boldsymbol{\theta})]$, and by combining it with the prior $p(\boldsymbol{\theta})$ into the joint density $p(\boldsymbol{\theta}, m) = p(\boldsymbol{\theta}) p(m|\boldsymbol{\theta})$ we can construct the uncertainty
\begin{equation}
\bar{\epsilon}_{\mathrm{mse}} = \sum_{i=1}^d w_i \int d\boldsymbol{\theta} dm ~p(\boldsymbol{\theta}, m) \left[g_i(m) - \theta_i  \right]^2,
\label{msegen}
\end{equation}
where we recall that $g_i(m)$ is the estimator for the $i$-th parameter and $w_i \geqslant 0$ its relative importance. 

In chapter \ref{chap:networks} we saw that the uncertainty in equation (\ref{msegen}) for the parameters $\boldsymbol{\theta}$ arises as a particular case of that in equation (\ref{msefunctions}) for linear functions when we choose the trivial transformation $V = \mathbb{I}$, and provided that we restrict the estimation to a single shot. For that reason, we can now complete the first step of our derivation here, which is to perform a classical optimisation over all the possible estimators, by simply applying the result of such optimisation in section \ref{subsec:multiasymp} for $V = \mathbb{I}$. 

More concretely, let us rewrite equation (\ref{msegen}) as $\bar{\epsilon}_{\mathrm{mse}} = \mathrm{Tr}[\mathcal{W} \Sigma_{\mathrm{mse}}]$, with $\mathcal{W} = \mathrm{diag}(w_1, \dots, w_d)$,
\begin{equation}
\Sigma_{\mathrm{mse}} =  \int d\boldsymbol{\theta} dm ~p(\boldsymbol{\theta}, m) \left[\boldsymbol{g}(m) - \boldsymbol{\theta} \right] \left[\boldsymbol{g}(m) - \boldsymbol{\theta} \right]^\transpose,
\end{equation}
and $\boldsymbol{g}(m) = (g_1(m), \dots, g_d(m))$, and let us recall that, since $\mathcal{W}$ is a positive semi-definite matrix, to minimise $\bar{\epsilon}_{\mathrm{mse}}$ it suffices to lower bound $\Sigma_{\mathrm{mse}}$ in the matrix sense. According to our discussion in section \ref{subsec:multiasymp}, this operation gives us that
\begin{equation}
\Sigma_{\mathrm{mse}} \geqslant \Sigma_{\mathrm{opt}}^c = \int dm~ p(m) \Sigma(m),
\label{optclasmse}
\end{equation}
where $p(m) = \int d\boldsymbol{\theta} p(\boldsymbol{\theta})p(m|\boldsymbol{\theta})$ and
\begin{equation}
\Sigma(m) = \int d\boldsymbol{\theta} p(\boldsymbol{\theta}|m) \boldsymbol{\theta}\boldsymbol{\theta}^\transpose - \left[\int d\boldsymbol{\theta} p(\boldsymbol{\theta}|m) \boldsymbol{\theta}\right] \left[\int d\boldsymbol{\theta} p(\boldsymbol{\theta}|m) \boldsymbol{\theta}\right]^\transpose.
\label{multiexperror}
\end{equation}

Now we observe that by integrating the outcomes in the first term of equation (\ref{multiexperror}), and expanding the posterior  as $p(\boldsymbol{\theta}|m) = p(\boldsymbol{\theta})p(m|\boldsymbol{\theta})/p(m)$ in its second term, we can express $\Sigma_{\mathrm{opt}}^c$ as
\begin{equation}
\Sigma_{\mathrm{opt}}^c = \int d\boldsymbol{\theta} p(\boldsymbol{\theta}) \boldsymbol{\theta}\boldsymbol{\theta}^\transpose - \int \frac{dm}{p(m)}\left[\int d\boldsymbol{\theta} p(\boldsymbol{\theta})p(m|\boldsymbol{\theta}) \boldsymbol{\theta}\right] \left[\int d\boldsymbol{\theta} p(\boldsymbol{\theta})p(m|\boldsymbol{\theta}) \boldsymbol{\theta}\right]^\transpose.
\label{mmsematrix}
\end{equation}
As we can see, the second term of this expression is reminiscent of the definition for the classical Fisher information matrix in equation (\ref{fim}). This is the same type of formal connection between Bayesian quantities and those belonging to the asymptotic theory that we studied in section \ref{subsec:originalderivation} for the single-parameter case, and upon making the quantum part of the problem explicit we can exploit this formal similarity as we did in that section to further lower bound $\Sigma_\mathrm{opt}^c$.

Using equation (\ref{mmsematrix}) and $p(m) = \int d\boldsymbol{\theta} p(\boldsymbol{\theta})p(m|\boldsymbol{\theta})$ we can construct the scalar quantity 
\begin{equation}
\boldsymbol{u}^\transpose \Sigma_{\mathrm{opt}}^c \boldsymbol{u} = \int d\boldsymbol{\theta} p(\boldsymbol{\theta})\theta_u^2 - \int dm  \frac{\left[ \int d\boldsymbol{\theta} p(\boldsymbol{\theta})p(m|\boldsymbol{\theta}) \theta_u \right]^2}{\int d\boldsymbol{\theta} p(\boldsymbol{\theta})p(m|\boldsymbol{\theta})},
\label{scalarquantity}
\end{equation}
where $\theta_u = \boldsymbol{u}^\transpose \boldsymbol{\theta} = \boldsymbol{\theta}^\transpose \boldsymbol{u}$ and $\boldsymbol{u}$ is some real vector, and by inserting $p(m|\boldsymbol{\theta}) = \mathrm{Tr}[E(m) \rho(\boldsymbol{\theta})]$ in equation (\ref{scalarquantity}) we find that
\begin{equation}
\boldsymbol{u}^\transpose \Sigma_{\mathrm{opt}}^c \boldsymbol{u} = \int d\boldsymbol{\theta} p(\boldsymbol{\theta})\theta_u^2 - \int dm  \frac{\mathrm{Tr}\left[ E(m) \bar{\rho}_u \right]^2}{\mathrm{Tr}\left[ E(m) \rho \right]},
\label{bayesanalogy}
\end{equation}
with $\rho = \int d\boldsymbol{\theta} p(\boldsymbol{\theta}) \rho(\boldsymbol{\theta})$ and $\bar{\rho}_u = \int d\boldsymbol{\theta} p(\boldsymbol{\theta}) \rho(\boldsymbol{\theta}) \theta_u$. We have thus arrived at an expression where the second term has taken an analogous form to that of the single-parameter classical Fisher information after having inserted the Born rule, and from our derivation in section \ref{subsec:originalderivation} we know that it is possible to bound this term following the same steps of the proof that gives rise to the Braunstein-Caves inequality for the Fisher information (\cite{BraunsteinCaves1994, genoni2008} and section \ref{subsec:crb}).  

If we follow this analogy and we introduce the Bayesian counterpart of the equation for the symmetric logarithmic derivative, that is, $S_u \rho + \rho S_u = 2\bar{\rho}_u$, then we have that
\begin{align}
\int dm  \frac{\mathrm{Tr}\left[ E(m) \bar{\rho}_u \right]^2}{\mathrm{Tr}\left[ E(m) \rho \right]} &= \int dm\left(\frac{\mathrm{Re}\left\lbrace\mathrm{Tr}\left[E(m) S_u\rho\right]\right\rbrace}{\sqrt{\mathrm{Tr}\left[ E(m) \rho\right]}}\right)^2
\nonumber \\
&\leqslant \int dm~\abs{\frac{\mathrm{Tr}\left[E(m) S_u\rho\right]}{\sqrt{\mathrm{Tr}\left[ E(m) \rho\right]}}}^2
\nonumber \\
&=\int dm~ \abs{\mathrm{Tr}\left[\frac{\rho^{\frac{1}{2}}E(m)^{\frac{1}{2}}}{\sqrt{\mathrm{Tr}\left[ E(m) \rho \right]}} E(m)^{\frac{1}{2}}S_u \rho^{\frac{1}{2}}\right]}^2
\nonumber \\
&\leqslant  \int dm~ \mathrm{Tr}\left[ E(m) S_u \rho S_u \right] = \mathrm{Tr}\left( \rho S_u^2 \right) \equiv \mathcal{K}_u,
\label{quantuminequalities}
\end{align}
where, as in section \ref{subsec:originalderivation}, we have used the Cauchy-Schwarz inequality $|\mathrm{Tr}[X^\dagger Y ]|^2 \leqslant \mathrm{Tr}[ X^\dagger X ] \mathrm{Tr}[Y^\dagger Y ]$ with 
\begin{eqnarray}
X = \frac{E(m)^{\frac{1}{2}} \rho^{\frac{1}{2}}}{\sqrt{\mathrm{Tr}\left[ E(m) \rho \right]}}, ~~Y = E(m)^{\frac{1}{2}} S_u \rho^{\frac{1}{2}}.
\label{csinequality}
\end{eqnarray}

On the other hand, recalling that $\theta_u = \sum_{i=1}^d u_i \theta_i$ we can see that $\bar{\rho}_u = \sum_{i=1}^d u_i \bar{\rho}_i$, with $\bar{\rho}_i = \int d\boldsymbol{\theta} p(\boldsymbol{\theta}) \rho(\boldsymbol{\theta}) \theta_i$. In turn, this allows us to express $S_u$ as $S_u = \sum_{i=1}^d u_i S_i$, with $S_i \rho + \rho S_i = 2\bar{\rho}_i$ and $S_i$ being a Hermitian operator. Since our aim is to derive a matrix inequality, we need to find a way of using the previous definitions to rewrite $\mathcal{K}_u$ as $\mathcal{K}_u = \boldsymbol{u}^\transpose \mathcal{K} \boldsymbol{u}$, where $\mathcal{K}_{ij} = \mathrm{Tr}(\rho A_{ij})$ is a matrix and $A_{ij}$ is some operator associated with the product of $S_i$ and $S_j$. Given that the operators $S_i$ and $S_j$ might not commute, let us first decompose $A_{ij}$ as
\begin{equation}
2 A_{ij} = \left(A_{ij}+A_{ij}^\dagger\right) + \left(A_{ij} - A_{ij}^\dagger\right),
\end{equation}
so that
\begin{equation}
\boldsymbol{u}^\transpose \mathcal{K} \boldsymbol{u} = \frac{1}{2}\left\lbrace\sum_{i, j=1}^d u_i u_j\mathrm{Tr}\left[\rho\left(A_{ij} +  A_{ij}^\dagger\right)\right] + \sum_{i, j=1}^d u_i u_j\mathrm{Tr}\left[\rho\left( A_{ij} - A_{ij}^\dagger\right)\right]\right\rbrace.
\end{equation}
If we were to take $A_{ij} = S_i S_j$, then we would find that 
\begin{align}
\boldsymbol{u}^\transpose \mathcal{K} \boldsymbol{u} &= \frac{1}{2}\left\lbrace\sum_{i, j=1}^d u_i u_j\mathrm{Tr}\left[\rho\left(S_i S_j +  S_j S_i\right)\right] + \sum_{i, j=1}^d u_i u_j\mathrm{Tr}\left[\rho\left( S_i S_j - S_j S_i\right)\right]\right\rbrace
\nonumber \\
&= \frac{1}{2}\left\lbrace\sum_{i, j=1}^d u_i u_j\mathrm{Tr}\left[\rho\left(S_i S_j +  S_j S_i\right)\right] + \mathrm{Tr}\left[\rho\left(S_u^2 - S_u^2\right)\right]\right\rbrace
\nonumber \\
&= \frac{1}{2}\sum_{i, j=1}^d u_i u_j\mathrm{Tr}\left[\rho\left(S_i S_j +  S_j S_i\right)\right] = \mathrm{Tr}\left(\rho S_u^2\right) = \mathcal{K}_u,
\end{align}
and the same result would have been obtained should we had chosen $A_{ij} = S_j S_i$ or the Hermitian version $A_{ij} = (S_i S_j + S_j S_i)/2$ instead. Therefore, we can take $\mathcal{K}$ to be a symmetric matrix with elements\footnote{See \cite{helstrom1968multiparameter} for the analogous operation in the original derivation of the multi-parameter quantum Cram\'{e}r-Rao bound.}
\begin{equation}
\mathcal{K}_{ij} = \mathrm{Tr}\left[\rho \left(S_i S_j + S_j S_i \right) \right]/2.
\label{bayesinfmatrix}
\end{equation}

The combination of equations (\ref{optclasmse}), (\ref{bayesanalogy}), (\ref{quantuminequalities}) and (\ref{bayesinfmatrix}), which must be valid for any $\boldsymbol{u}$, finally gives us the chain of matrix inequalities 
\begin{equation}
\Sigma_{\mathrm{mse}} \geqslant \Sigma_{\mathrm{opt}}^c \geqslant \Sigma_q = \int d\boldsymbol{\theta} p(\boldsymbol{\theta}) \boldsymbol{\theta}\boldsymbol{\theta}^\transpose - \mathcal{K}.
\label{myquantumbound}
\end{equation}
The quantum inequality in equation (\ref{myquantumbound}) is the central result of this chapter.

Applying this result to the original measure of uncertainty in equation (\ref{msegen}) we find that the scalar version of our new bound in equation (\ref{myquantumbound}) is
\begin{equation}
\bar{\epsilon}_\mathrm{mse} \geqslant \sum_{i=1}^d w_i \left[\int d\boldsymbol{\theta} p(\boldsymbol{\theta})\theta_i^2 - \mathrm{Tr}\left(\rho S_i^2\right) \right]. 
\label{multibayesbound}
\end{equation}
Furthermore, by noticing that $\mathrm{Tr}(\rho S_i) = \int d\boldsymbol{\theta} p(\boldsymbol{\theta})\theta_i$ and defining the uncertainties
\begin{equation}
\Delta \theta_{p,i}^2 = \int d\boldsymbol{\theta} p(\boldsymbol{\theta})\theta_i^2 - \left[ \int d\boldsymbol{\theta} p(\boldsymbol{\theta})\theta_i \right]^2
\end{equation}
and $\Delta S_{\rho, i}^2 = \mathrm{Tr}\left(\rho S_i^2 \right) - \mathrm{Tr}\left(\rho S_i \right)^2$ we may rewrite the bound in equation (\ref{multibayesbound}) as
\begin{equation}
\bar{\epsilon}_\mathrm{mse} \geqslant \sum_{i=1}^d w_i \left(\Delta \theta_{p,i}^2 - \Delta S_{\rho, i}^2 \right),
\end{equation}
which is the multi-parameter version of our equivalent expression for a single parameter found in section \ref{subsec:originalderivation}. 

\subsection{Towards a shot-by-shot strategy for many parameters}
\label{subsec:multibayessaturation}

In section \ref{subsec:multiasymp} we saw that the classical inequality in equations (\ref{optclasmse}) and (\ref{myquantumbound}) can be saturated when the estimators are given by the averages over the posterior probability, that is, $\boldsymbol{g}_{\mathrm{opt}}(m) = \int d\boldsymbol{\theta} p(\boldsymbol{\theta}|m)\boldsymbol{\theta}$. On the other hand, since the derivation in equation (\ref{quantuminequalities}) is formally identical to that in section \ref{subsec:originalderivation} for the single-parameter case, from our discussion there we know that the condition for the saturation of the first inequality in equation (\ref{quantuminequalities}) is that $\mathrm{Tr}[E(m)S_u\rho]$ is real, while the second inequality is saturated if and only if
\begin{equation}
\frac{E(m)^{\frac{1}{2}}\rho^{\frac{1}{2}}}{\mathrm{Tr}\left[E(m) \rho \right]} = \frac{E(m)^{\frac{1}{2}}S_u\rho^{\frac{1}{2}}}{\mathrm{Tr}\left[E(m) S_u \rho \right]}.
\end{equation}

If $[S_i, S_j] = 0$ for all $i$, $j$, then we may fulfil such conditions by constructing the measurement scheme with the projections onto the common eigenstates of this set of commuting operators. To verify it, let us first observe that 
\begin{align}
S_u &= \sum_{i=1}^d u_i S_i = \sum_{i=1}^d  u_i \int dm~ c_i(m) \ketbra{\psi(m)} 
\nonumber \\
&= \int dm~c_u(m) \ketbra{\psi(m)},
\end{align}
with $c_u(m) = \sum_{i=1}^d u_i c_i(m)$ and $\lbrace \ketbra{\psi(m)} \rbrace$ being the common eigenstates of $\lbrace S_i \rbrace$. Then, by using $E(m) =  \ketbra{\psi(m)}$ we find that
\begin{align}
\int dm  \frac{\mathrm{Tr}\left[ E(m) \bar{\rho}_u \right]^2}{\mathrm{Tr}\left[ E(m) \rho \right]} &= \int dm\left(\frac{\mathrm{Re}\left\lbrace\mathrm{Tr}\left[\ketbra{\psi(m)} S_u\rho\right]\right\rbrace}{\sqrt{\mathrm{Tr}\left[ \ketbra{\psi(m)} \rho\right]}}\right)^2
\nonumber \\
&= \int dm~ c_u^2(m) \mathrm{Tr}\left[\ketbra{\psi(m)} \rho\right]
\nonumber \\
&= \mathrm{Tr}\left(\rho S_u^2\right) = \mathcal{K}_u = \boldsymbol{u}^\transpose\mathcal{K}\boldsymbol{u},
\end{align}
and, as a consequence, our matrix quantum bound in equation (\ref{myquantumbound}) is achieved. 

The projective strategy based on $\lbrace \ketbra{\psi(m)} \rbrace$ can be thought of as if we were employing the projective measurements that are optimal to estimate each parameter in an independent fashion. Unfortunately, it is known that the optimal strategy for Bayesian multi-parameter estimation is not necessarily based on those projectors \cite{helstrom1974, personick1969thesis}, which is a manifestation of the fact that the quantum estimators $\lbrace S_i \rbrace$ do not need to commute. When $[S_i, S_j] \neq 0$, one possibility would be to search for the projective measurement that is simultaneously optimal for all the parameters. This is precisely what Personick did for $d = 2$ in \cite{personick1969thesis}, where he derived a set of equations for a new set of quantum estimators associated with the simultaneous strategy, and indeed it may be verified that these equations are generally different from those satisfied by $\lbrace S_i \rbrace$. Nevertheless, even if we can find the optimal projective strategy, this might not be optimal in a global sense, since it can be shown \cite{helstrom1974} that to achieve the optimal uncertainty predicted by Helstrom and Holevo's fundamental equations in section \ref{subsec:fundeq} we may need a multi-parameter strategy based on general POMs. In summary, we conclude that we cannot always saturate our bound.  

Despite these difficulties, our new tool can still be useful and informative. On the one hand, the results based on it will be tight and fundamental whenever the operators $\lbrace S_i \rbrace$ commute. If that happens, then the optimal single-shot measurement can be calculated using our bound, and in that case it is possible to generalise our shot-by-shot methodology in chapter \ref{chap:limited} to the multi-parameter regime. In particular, given $\mu$ identical and independent trials and the multi-parameter optimal single-shot POM $\ketbra{\psi(m_i)} \equiv \ketbra{\psi(s_i)}$ with outcome $m_i \equiv s_i$ in the $i$-th repetition, the optimal estimators that take into account the information from all the repetitions of this quantum strategy are $\boldsymbol{g}(\boldsymbol{s}) = \int d\boldsymbol{\theta} p(\boldsymbol{\theta}|\boldsymbol{s}) \boldsymbol{\theta}$, with $\boldsymbol{s} = (s_1, \dots, s_\mu)$, and the associated uncertainty can be expressed as 
\begin{equation}
\bar{\epsilon}_{\mathrm{mse}} = \sum_{i=1}^d w_i \int d\boldsymbol{s}~p(\boldsymbol{s}) \left\lbrace \int d\boldsymbol{\theta} p(\boldsymbol{\theta}|\boldsymbol{s}) \theta_i^2 - \left[\int d\boldsymbol{\theta} p(\boldsymbol{\theta}|\boldsymbol{s}) \theta_i \right]^2 \right\rbrace, 
\label{msegenmany}
\end{equation}
where $p(\boldsymbol{\theta}|\boldsymbol{s}) \propto p(\boldsymbol{\theta}) \prod_{j=1}^\mu \langle \psi(s_j) | \rho(\boldsymbol{\theta}) | \psi(s_j)\rangle$. For $d=2$, which is the case for one of the scenarios that we will study, this error can be numerically calculated as a function of $\mu$ using the two-parameter algorithm presented in section \ref{subsec:multinonasym} and appendix \ref{sec:multimsematlab}.

On the other hand, even if we cannot saturate our bound, it can be argued that it is still better than any other multi-parameter bound for the error in equation (\ref{msegen}) that also ignores the potential non-commutativity of $\lbrace S_i \rbrace$. This is because the latter type of bound will necessarily be equal to or lower than our equation (\ref{multibayesbound}), since the quantity $\int d\boldsymbol{\theta} p(\boldsymbol{\theta})\theta_i^2 - \mathrm{Tr}\left(\rho S_i^2\right)$ is the optimum for the estimation of $\theta_i$ (\cite{personick1971, yuen1973, helstrom1976} and sections \ref{subsec:singleshotparadigm} and \ref{subsec:originalderivation}). The practical consequence of this is that our result will produce bounds that can be tighter than proposals such as the multi-parameter version of the Ziv-Zakai bound in \cite{zhang2014}\footnote{In fact, this has already been demonstrated at the single-parameter level in section \ref{sec:alternativevssingleshot}, where we have found that the Ziv-Zakai and Weiss-Weinstein bounds \cite{tsang2012, tsang2016} are generally loose in optical scenarios with a finite number of repetitions.}. We leave for future work to examine the relative tightness of our bound with respect to other alternatives such as the multi-parameter Weiss-Weinstein bound \cite{tsang2016} or the bound for complex quantities that was derived by Yuen and Lax \cite{yuen1973} when applied to real parameters. 
 
Since the complexity of the calculations associated with our bound is similar to that of the Fisher information matrix for general density operators\footnote{In the single-parameter case this was first observed by Macieszczak \emph{et al.} \cite{macieszczak2014bayesian}.}, with the extra advantage of not having to invert $\mathcal{K}$, we conjecture that the tool that we have introduced may end playing a crucial role in analyses of multi-parameter metrology whenever Helstrom and Holevo's fundamental equations cannot be solved exactly in problems with several parameters. The rest of this chapter is dedicated to demonstrate its usefulness with concrete examples. 

\section{Our methodology in action: results and discussion}

\subsection{Qubit sensing network}
\label{subsec:multibayesqubit}

Our first example is the two-parameter qubit network that we studied in sections \ref{subsec:intersensorasymp} - \ref{subsec:correlationsmultibayes}, which was prepared in the probe state $\ket{\psi_0} = [\ket{00}+\gamma (\ket{01}+\ket{10})+\ket{11}]/\sqrt{2(1+\gamma^2)}$, with real $\gamma$, and which upon interacting with the object that we wish to study was transformed as $\ket{\psi(\theta_1, \theta_2)} = U(\theta_1, \theta_2) \ket{\psi_0}$ by the unitary operator $U(\theta_1, \theta_2) = \mathrm{exp}(-i\sigma_z\theta_1/2)\otimes \mathrm{exp}(-i\sigma_z\theta_1/2) = \mathrm{exp}[-i(\sigma_{z,1}\theta_1 + \sigma_{z,2}\theta_2)/2]$. 

Let us start by performing the single-shot Bayesian analysis. Assuming that we are working in the regime of moderate prior knowledge, so that we can use the flat prior $p(\theta_1, \theta_2) = 4/\pi^2$, when $(\theta_1, \theta_2) \in [-\pi/4, \pi/4]\times[-\pi/4, \pi/4]$, and zero otherwise, we have that\footnote{Unless otherwise indicated, all the analytical calculations of this chapter have been performed using Mathematica.}
\begin{align}
\rho &= \frac{4}{\pi^2}\int_{-\pi/4}^{\pi/4} d\theta_1 \int_{-\pi/4}^{\pi/4} d\theta_2\hspace{0.15em} \mathrm{e}^{-\frac{i}{2}(\sigma_{z,1}\theta_1 + \sigma_{z,2}\theta_2)}\ketbra{\psi_0}\mathrm{e}^{\frac{i}{2}(\sigma_{z,1}\theta_1 + \sigma_{z,2}\theta_2)}
\nonumber \\
&= \frac{1}{2\pi^2(1+\gamma^2)}
\left(
\begin{array}{cccc}
 \pi ^2 & 2 \sqrt{2} \pi  \gamma  & 2 \sqrt{2} \pi  \gamma  & 8 \\
 2 \sqrt{2} \pi  \gamma  & \pi ^2 \gamma ^2 & 8 \gamma ^2 & 2 \sqrt{2} \pi  \gamma  \\
 2 \sqrt{2} \pi  \gamma  & 8 \gamma ^2 & \pi ^2 \gamma ^2 & 2 \sqrt{2} \pi  \gamma  \\
 8 & 2 \sqrt{2} \pi  \gamma  & 2 \sqrt{2} \pi  \gamma  & \pi ^2 \\
\end{array}
\right),
\label{labeleffstate}
\end{align}
\begin{align}
\bar{\rho}_1 &= \frac{4}{\pi^2}\int_{-\pi/4}^{\pi/4} d\theta_1 \int_{-\pi/4}^{\pi/4} d\theta_2\hspace{0.15em} \mathrm{e}^{-\frac{i}{2}(\sigma_{z,1}\theta_1 + \sigma_{z,2}\theta_2)}\ketbra{\psi_0}\mathrm{e}^{\frac{i}{2}(\sigma_{z,1}\theta_1 + \sigma_{z,2}\theta_2)}\theta_1
\nonumber \\
&= \frac{i(4-\pi)}{2\sqrt{2}\pi^2(1+\gamma^2)}
\begin{pmatrix}
0 & 0 & -\pi\gamma & -2\sqrt{2} \\
0 & 0 & -2\sqrt{2}\gamma^2 & -\pi\gamma \\
\pi\gamma & 2\sqrt{2}\gamma^2 & 0 & 0 \\
2\sqrt{2} & \pi\gamma & 0 & 0 
\end{pmatrix},
\label{labeleffmean1}
\end{align}
and
\begin{align}
\bar{\rho}_2 &= \frac{4}{\pi^2}\int_{-\pi/4}^{\pi/4} d\theta_1 \int_{-\pi/4}^{\pi/4} d\theta_2\hspace{0.15em} \mathrm{e}^{-\frac{i}{2}(\sigma_{z,1}\theta_1 + \sigma_{z,2}\theta_2)}\ketbra{\psi_0}\mathrm{e}^{\frac{i}{2}(\sigma_{z,1}\theta_1 + \sigma_{z,2}\theta_2)}\theta_2
\nonumber \\
&= \frac{i(4-\pi)}{2\sqrt{2}\pi^2(1+\gamma^2)}
\begin{pmatrix}
0 & -\pi\gamma & 0 & -2\sqrt{2} \\
\pi\gamma & 0 & 2\sqrt{2}\gamma^2 & 0 \\
0 & -2\sqrt{2}\gamma^2 & 0 & -\pi\gamma \\
2\sqrt{2} & 0 & \pi\gamma & 0 
\end{pmatrix},
\label{labeleffmean2}
\end{align}
where the columns are labelled as $\ket{00}$, $\ket{01}$, $\ket{10}$ and $\ket{11}$. In addition, by inserting equations (\ref{labeleffstate} - \ref{labeleffmean2}) in $S_i\rho + \rho S_i = 2\bar{\rho}_i$, and using a two-parameter extension of the Mathematica algorithm in appendix \ref{sec:singleshotalgorithm}, we find that the independently optimal quantum estimators are
\begin{eqnarray}
S_1 = \frac{2 \left(4-\pi\right)}{\pi\left(1+\gamma^2\right)}\left( \frac{\gamma}{\sqrt{2}} \sigma_y\otimes\mathbb{I} + \frac{1-\gamma^2}{\pi}\sigma_x\otimes\sigma_y \right),
\label{qest1}
\end{eqnarray}
\begin{eqnarray}
S_2 = \frac{2 \left(4-\pi\right)}{\pi\left(1+\gamma^2\right)}\left( \frac{\gamma}{\sqrt{2}} \mathbb{I}\otimes\sigma_y + \frac{1-\gamma^2}{\pi}\sigma_y\otimes\sigma_x \right), 
\label{qest2}
\end{eqnarray}
which have been rewritten in terms of Pauli matrices to better visualise their structure. As a result, the single-shot bound in equation (\ref{multibayesbound}) is
\begin{eqnarray}
\bar{\epsilon}_{\mathrm{mse}} \geqslant \frac{\pi^2}{48} - \frac{2\left(4-\pi\right)^2\left[2-\left(4-\pi^2\right)\gamma^2 + 2\gamma^4 \right]}{\pi^4 \left(1+\gamma^2\right)^2},
\label{qubitoptmse}
\end{eqnarray}
having chosen both parameters to be equally important (i.e., $\mathcal{W} = \mathbb{I}/2$). 

As a first observation we note that equation (\ref{qubitoptmse}) achieves its minimum value at $\gamma = \pm 1$, so that $\bar{\epsilon}_{\mathrm{mse}} \geqslant \pi^2/48 - (4-\pi)^2/(2\pi^2) \approx 0.168$. Given that $\bar{\epsilon}_{\mathrm{prior}} = \pi^2/48 \approx 0.206$, we conclude that a single shot can improve our knowledge about $(\theta_1, \theta_2)$ by $18\%$ with respect to the prior uncertainty\footnote{The improvement is defined as $(\bar{\epsilon}_{\mathrm{prior}}-\bar{\epsilon}_{\mathrm{mse}})/\bar{\epsilon}_{\mathrm{prior}}$ multiplied by $100\%$ (see section \ref{sec:genetic}).}.

\begin{figure}[t]
\centering
\includegraphics[trim={1cm 0.1cm 1.3cm 1cm},clip,width=14.75cm]{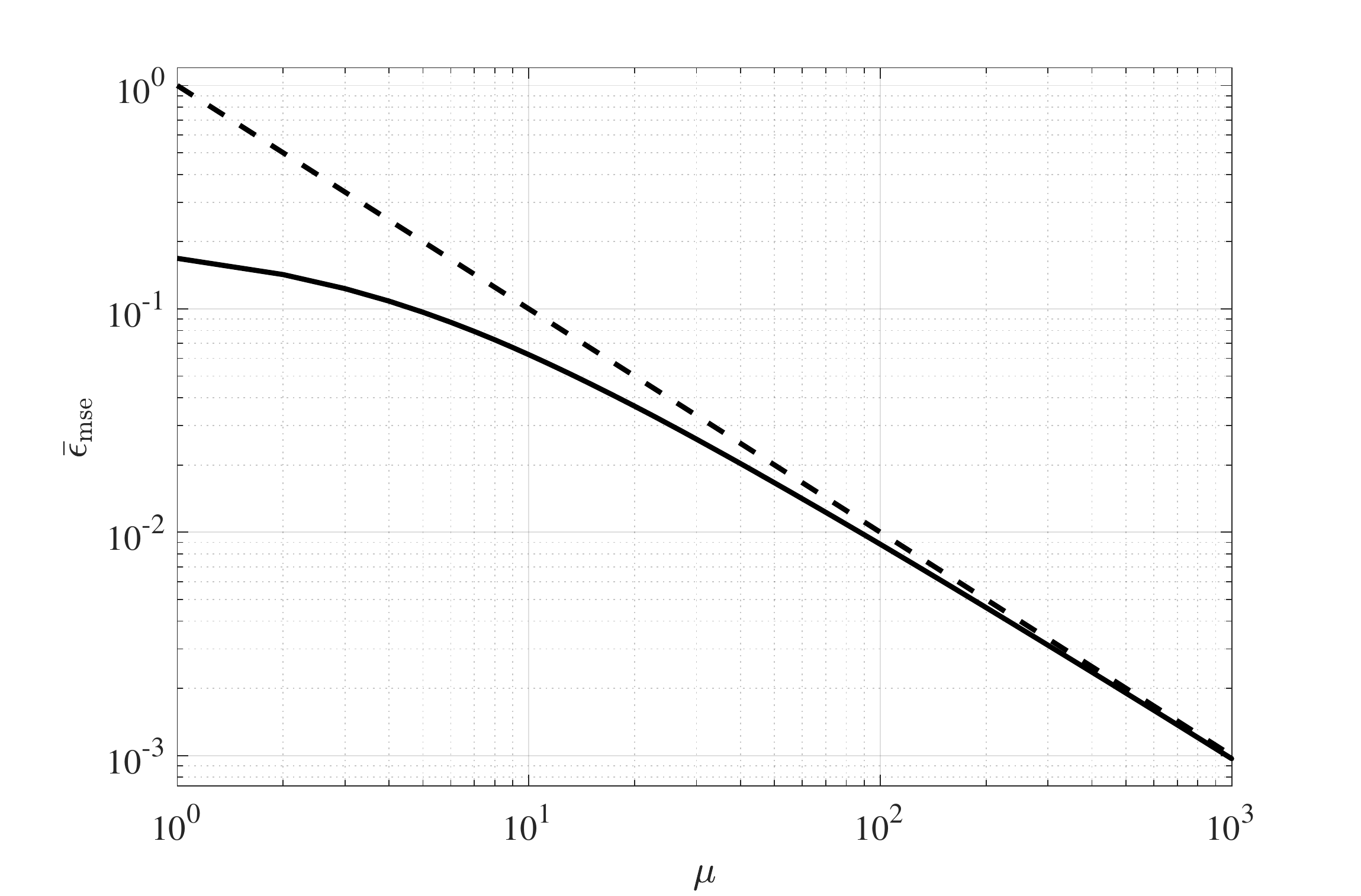}
	\caption[Shot-by-shot quantum bound for a two-parameter qubit network]{Mean square error in equation (\ref{msegenmany}) based on a single-shot optimal measurement (solid line) and quantum Cram\'{e}r-Rao bound (dashed line) for the two-parameter qubit network in the main text, with $\gamma = 1$ and a prior squared area $\pi^2/4$ centred around $(0, 0)$. The solid line is the result of optimising the scheme in a shot-by-shot fashion, and it is optimal at least for a single shot and for a large number of them. In addition, note that $\bar{\epsilon}_{\mathrm{cr}} = 1/\mu$ when $\gamma = 1$.}
\label{multibayes_plot}
\end{figure}

Furthermore, since $S_1$ and $S_2$ commute, in this case there is a measurement that achieves our single-shot bound. If we choose $\gamma = 1$, then 
\begin{eqnarray}
S_1 = \frac{ \left(4-\pi\right)}{\pi\sqrt{2}} \sigma_y\otimes\mathbb{I}, ~~S_2 = \frac{ \left(4-\pi\right)}{\pi\sqrt{2}} \mathbb{I}\otimes \sigma_y,
\label{qestloc}
\end{eqnarray}
and thus we can construct an optimal strategy given by the common projectors $\ket{s_+, s_+}$, $\ket{s_-, s_-}$, $\ket{s_+, s_-}$, $\ket{s_-, s_+}$, where $\ket{s_\pm} = (\ket{0}\pm i\ket{1})/\sqrt{2}$. We may then calculate the uncertainty for $\mu$ trials in equation (\ref{msegenmany}) using this measurement in each shot, and the result of this operation has been represented in figure \ref{multibayes_plot} with a solid line. The quantum Cram\'{e}r-Rao bound, which in this case is simply\footnote{To find this result, we first recall that the estimation of the original parameters is equivalent to estimate a set of linear functions with the trivial transformation $V=\mathbb{I}$, for which the geometry parameter is $\mathcal{G} = 0$ (see section \ref{sec:networksasym}), and equation (\ref{crmsetwoqubitnetwork}) indicates that, in this case, the Cram\'{e}r-Rao bound is the expression given in the main text.} $\bar{\epsilon}_{\mathrm{cr}}= \mathrm{Tr}(\mathcal{W} F_q^{-1})/\mu = (1+\gamma^2)^2/(4\mu\gamma^2)$, has also been included in the same figure as the dashed line, and, as we can observe, the latter is approached by the Bayesian error as $\mu$ grows. More concretely, the deviation of the asymptotic bound with respect to the exact calculation reaches the threshold of $\varepsilon_\tau = 0.05$ after $\mu = 5.05 \cdot 10^2$ repetitions (see \ref{subsec:asymsatu} for the definition the relative error $\varepsilon_\tau$), and it further decreases afer that point. Hence, our multi-parameter Bayesian strategy is optimal both for a single shot and for a large number of trials, which is the same behaviour that we found in the single-parameter protocols of chapter \ref{chap:limited}.

Remarkably, our result shows that this scheme does not require entanglement in order to approach the optimal single-shot uncertainty, since the strategy presented above (state plus POM) is local. That a local version of this scheme is optimal to estimate the original parameters was also concluded in \cite{proctor2017networked} from the analysis of its asymptotic performance, and such result may also be recovered from our asymptotic formalism in chapter \ref{chap:networks}. In other words, we have demonstrated that the fact that a global strategy is not needed for this protocol is not only true asymptotically, but also in the non-asymptotic regime when the scheme is implemented in a shot-by-shot fashion with the optimal single-shot measurement.

\subsection{Quantum imaging}

\begin{figure}[t]
\centering
\includegraphics[trim={4.5cm 4cm 4.75cm 3cm},clip,width=12cm]{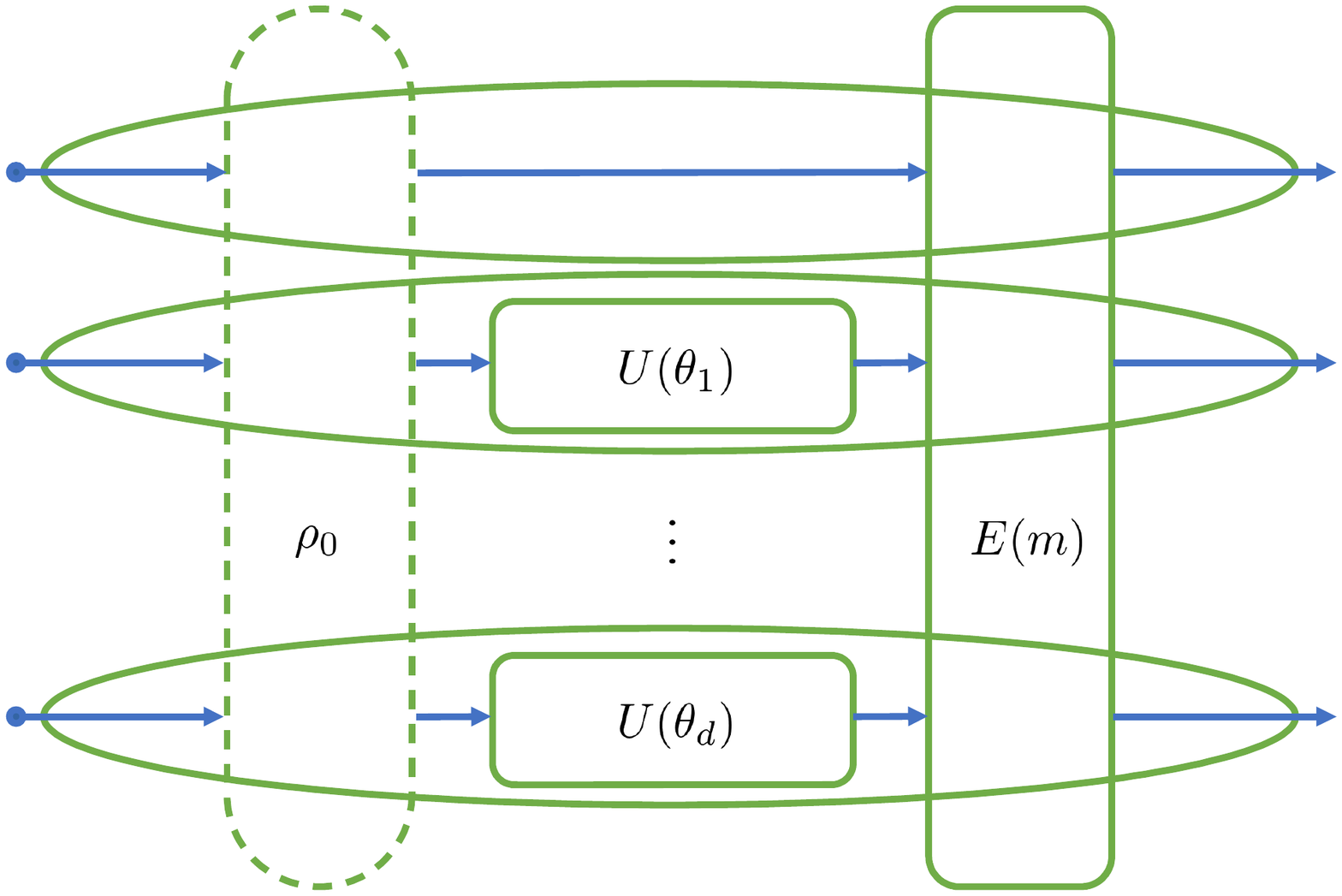}
\caption[Discrete model for phase imaging]{Discrete model for phase imaging, where a collection of $(d+1)$ optical modes are prepared in a (potentially entangled) state $\rho_0$, an unknown parameter $\theta_j$ is encoded in the $j$-th mode via the local unitary operator $U(\theta_j) = \mathrm{exp}(-i a_j^\dagger a_j \theta_j)$ and this operation is repeated for $d$ of them, and a (potentially global) measurement scheme $E(m)$ is implemented, with outcome $m$. As we saw in section \ref{subsec:multischemes}, this scheme is a particular case of the quantum sensing network model proposed by Proctor \emph{et al.} \cite{proctor2017networked} and exploited in chapter \ref{chap:networks}.}
\label{imagingmodel}
\end{figure}

The second scheme that we wish to examine is the discrete model of phase imaging explored by Humphreys \emph{et al.} \cite{humphreys2013} with the Cram\'{e}r-Rao bound, and by Macchiavello \cite{chiara2003} using covariant measurements\footnote{See chapter $4$ of \cite{holevo2011} for an introduction to the concept of \emph{covariant measurement}.}. In the former the scheme is assumed to operate in the asymptotic regime, while in the latter the calculation is carried out for a single shot but in the absence of prior knowledge. On the contrary, our calculations in this section assume an intermediate amount of prior information. 

Consider a system with $(d+1)$ optical modes, such that we encode a phase shift $\theta_j$ with a local unitary $U(\theta_j) = \mathrm{exp}(-i a_j^\dagger a_j \theta_j) = \mathrm{exp}(-i N_j \theta_j)$ in the $j$-th mode, for $1\leqslant j\leqslant d$, while the remaining mode $j=0$ is employed as a reference that has been calibrated in advance \cite{proctor2017networked}. The creation and annihilation operators of the $j$-th mode are $a_j^\dagger$ and $a_j$, respectively, and a schematic representation of this configuration can be found in figure \ref{imagingmodel}. Given this arrangement, a possible strategy is to follow a global approach and prepare the probe as
\begin{equation}
\ket{\psi_0} = \frac{1}{\sqrt{d+\alpha^2}}\left(\alpha\ket{\bar{n}~0 \cdots 0} + \cdots + \ket{0 \cdots 0~\bar{n}} \right),
\label{gnoons}
\end{equation}
which is a generalised NOON state \cite{knott2016local, humphreys2013} with a free parameter $\alpha$ that we take to be real. In this context the resource operator is $R = \sum_{j=0}^{d} N_j$, so that the total amount of resources per trial is given by the mean number of quanta, that is, $\langle \psi_0 |  R | \psi_0 \rangle = \bar{n}$. 

Let us first calculate the single-shot bound on the uncertainty associated with an estimation problem based on this scheme and where $d = 2$, $\bar{n} = 2$, $\mathcal{W} = \mathbb{I}/2$ and the prior probability is the same employed in the qubit case, and let us choose $\alpha = 1$, which is the balanced version of equation (\ref{gnoons}) \cite{knott2016local}. With this configuration we find that 
\begin{eqnarray}
\rho = \frac{1}{3}\left[\mathbb{I} + \frac{2(\lambda_1 + \lambda_4)}{\pi} + \frac{4\lambda_6}{\pi^2} \right],
\end{eqnarray}
and
\begin{eqnarray}
\bar{\rho}_1 = \frac{1}{3\pi}\left(\frac{2\lambda_7}{\pi}-\lambda_2\right),~~\bar{\rho}_2 = -\frac{1}{3\pi}\left(\frac{2\lambda_7}{\pi}+\lambda_5\right),
\end{eqnarray}
where $\lambda_i$ are Gell-Mann matrices\footnote{We recall that the Gell-Mann matrices are defined as \cite{gellmann1962}
\begin{align}
\lambda_1 &=
\begin{pmatrix}
0 & 1 & 0 \\
1 & 0 & 0 \\
0 & 0 & 0
\end{pmatrix},~
\lambda_2 = 
\begin{pmatrix}
0 & -i & 0 \\
i & 0 & 0 \\
0 & 0 & 0
\end{pmatrix},~
\lambda_3 = 
\begin{pmatrix}
1 & 0 & 0 \\
0 & -1 & 0 \\
0 & 0 & 0
\end{pmatrix},~
\lambda_4 =
\begin{pmatrix}
0 & 0 & 1 \\
0 & 0 & 0 \\
1 & 0 & 0
\end{pmatrix},
\nonumber \\
\lambda_5 &= 
\begin{pmatrix}
0 & 0 & -i \\
0 & 0 & 0 \\
i & 0 & 0
\end{pmatrix},~
\lambda_6 = 
\begin{pmatrix}
0 & 0 & 0 \\
0 & 0 & 1 \\
0 & 1 & 0
\end{pmatrix},~
\lambda_7 = 
\begin{pmatrix}
0 & 0 & 0 \\
0 & 0 & -i \\
0 & i & 0
\end{pmatrix},~
\lambda_8 = 
\begin{pmatrix}
1/\sqrt{3} & 0 & 0 \\
0 & 1/\sqrt{3} & 0 \\
0 & 0 & -2/\sqrt{3}
\end{pmatrix}.
\nonumber
\end{align}} \cite{gellmann1962}. Furthermore, introducing these results in $S_k\rho + \rho S_k = 2\bar{\rho}_k$ we find that the quantum estimators are
\begin{equation}
S_1 =  \frac{1}{\pi}\left[ \frac{\lambda_5-(1+\pi^2)\lambda_2}{2+\pi^2}  +\frac{\lambda_7}{\pi} \right],~~S_2 = \frac{1}{\pi}\left[ \frac{\lambda_2 -(1+\pi^2) \lambda_5}{2+\pi^2}  -\frac{\lambda_7}{\pi} \right],
\end{equation}
and the single-shot error is bounded as
\begin{equation}
\bar{\epsilon}_{\mathrm{mse}} \geqslant \frac{\pi^2}{48} - \frac{2\left(4 + 3\pi^2 + \pi^4\right)}{3\pi^4\left(2 + \pi^2\right)} \approx 0.130.
\label{photonbound}
\end{equation}

Unlike in the previous scenario, here $[S_1, S_2] \neq 0$, which implies that the bound does not provide a measurement to apply the shot-by-shot method in an optimal way. However, it can still provide useful information. On the one hand, we can study how close a given measurement can get. A numerical search by trial and error has revealed an approximated set of projectors with a precision almost as good as that given in equation (\ref{photonbound}). In particular, if we use 
\begin{align}
\bra{\varphi_a}&= (0.485 + 0.131 i, 0.441 - 0.070 i, -0.223 + 0.706 i),
\nonumber \\
\bra{\varphi_b} &= (0.688, - 0.208 - 0.432 i, -0.270 - 0.472 i),\nonumber \\
\bra{\varphi_c}&= (0.509 + 0.118 i, -0.284 + 0.700 i, 0.396 )
\end{align}
as the measurement scheme, where the components are labelled as $\ket{2,0,0}$, $\ket{0,2,0}$, $\ket{0,0,2}$, then we have that\footnote{The uncertainty for this POM can be numerically calculated using the MATLAB algorithm in appendix \ref{sec:multimsematlab}.} $\bar{\epsilon}_{\mathrm{mse}} \approx 0.142$. 

On the other hand, we may also explore the precision scaling that the bound is able to predict. In fact, recalling that the scaling associated with the global strategy in equation (\ref{gnoons}) can also be achieved with a local strategy when we work in the asymptotic regime \cite{knott2016local}, it would be desirable to establish whether the same phenomenon can be observed for $\mu = 1$ and a moderate prior.

To study this possibility, suppose we now have $d$ parameters, $\mathcal{W} = \mathbb{I}/d$ and a flat prior of hypervolume $(2\pi/\bar{n})^d$ with $\bar{n} \geqslant 4$, so that the prior knowledge is moderate and sufficient to avoid the periodicities associated with NOON states (see chapters \ref{chap:nonasymptotic} and \ref{chap:limited} and, e.g., \cite{alfredo2017, hall2012, friis2017}). In addition, to simplify the calculation of the bound in equation (\ref{multibayesbound}) let us relabel the components of the state in equation (\ref{gnoons}) as $\beta \equiv 1/\sqrt{d+\alpha^2}$ and $\beta' \equiv \alpha/\sqrt{d+\alpha^2}$, so that $\beta' = \sqrt{1-d\beta^2}$, and the basis kets as
\begin{equation}
\ket{0\dots 0~\bar{n}~0 \dots 0} = \ket{0}_0 \otimes \cdots \otimes \ket{0}_{j-1} \otimes \ket{\bar{n}}_j \otimes \ket{0}_{j+1}\otimes \cdots \ket{0}_d \equiv \ket{u_j}.
\end{equation}
Using these definitions and the fact that
\begin{equation}
\int_{-\frac{\pi}{\bar{n}}}^{\frac{\pi}{\bar{n}}} d\theta_j = \frac{2\pi}{\bar{n}},~~\int_{-\frac{\pi}{\bar{n}}}^{\frac{\pi}{\bar{n}}} d\theta_j \hspace{0.15em} \mathrm{e}^{\pm i \bar{n}\theta_j} \theta_j = \pm \frac{2i \pi}{\bar{n}^2} ,~~\int_{-\frac{\pi}{\bar{n}}}^{\frac{\pi}{\bar{n}}} d\theta_j\hspace{0.15em} \theta_j = \int_{-\pi/\bar{n}}^{\pi/\bar{n}} d\theta_j \hspace{0.15em} \mathrm{e}^{\pm i \bar{n}\theta_j} = 0
\end{equation}
we find that
\begin{align}
\rho &= \left(\frac{\bar{n}}{2\pi}\right)^d \int_{-\frac{\pi}{\bar{n}}}^{\frac{\pi}{\bar{n}}} d\theta_1 \cdots \int_{-\frac{\pi}{\bar{n}}}^{\frac{\pi}{\bar{n}}} d\theta_d \hspace{0.15em}\mathrm{e}^{-i \boldsymbol{N}\cdot\boldsymbol{\theta}} \ketbra{\psi_0}\mathrm{e}^{i \boldsymbol{N}\cdot\boldsymbol{\theta}}
\nonumber \\
&= \left(1 - d\beta^2\right)\ketbra{u_0} + \beta^2 \sum_{k=1}^d \ketbra{u_k},
\label{labeleffstatenoon}
\end{align}
and
\begin{align}
\bar{\rho}_k &= \left(\frac{\bar{n}}{2\pi}\right)^d \int_{-\frac{\pi}{\bar{n}}}^{\frac{\pi}{\bar{n}}} d\theta_1 \cdots \int_{-\frac{\pi}{\bar{n}}}^{\frac{\pi}{\bar{n}}} d\theta_d \hspace{0.15em} \mathrm{e}^{-i \boldsymbol{N}\cdot\boldsymbol{\theta}} \ketbra{\psi_0}\mathrm{e}^{i \boldsymbol{N}\cdot\boldsymbol{\theta}}\theta_k
\nonumber \\
&= \frac{-i\beta \sqrt{1-d\beta^2}}{\bar{n}} \left(\ketbra{u_k}{u_0} - \ketbra{u_0}{u_k}\right).
\end{align}

Next we need to solve $S_k \rho + \rho S_k = 2\bar{\rho_k}$. In section \ref{numcal} we saw that if we decompose $\rho$ as $\rho = \sum_i p_i \ketbra{\phi_i}$, then we can rewrite $S_k$ as
\begin{equation}
S_k = 2 \sum_{i j} \frac{\bra{\phi_i}\bar{\rho_k}\ket{\phi_j}}{p_i + p_j}\ketbra{\phi_i}{\phi_j},
\label{multiqestdiag}
\end{equation}
and by observing that $\rho$ in equation (\ref{labeleffstatenoon}) is already diagonal, equation (\ref{multiqestdiag}) simply becomes
\begin{equation}
S_k =  \frac{- 2i\beta \sqrt{1-d\beta^2}}{\bar{n}\left[1+\beta^2(1-d)\right]} \left(\ketbra{u_k}{u_0} - \ketbra{u_0}{u_k}\right).
\end{equation}

Inserting now the results for $\rho$ and the quantum estimators $S_k$ in equation (\ref{multibayesbound}) we find the bound
\begin{equation}
\bar{\epsilon}_{\mathrm{mse}} \geqslant \frac{1}{\bar{n}^2} \left[\frac{\pi^2}{3} - \frac{4\beta^2 (1-d\beta^2)}{1+\beta^2(1-d)} \right],
\end{equation}
which achieves its minimum at $\beta = 1/\sqrt{d+\sqrt{d}}$ (i.e., at $\alpha = d^{1/4}$). Thus
\begin{equation}
\bar{\epsilon}_{\mathrm{mse}} \geqslant \frac{1}{\bar{n}^2} \left[\frac{\pi^2}{3} - \frac{4}{(1+\sqrt{d})^2} \right] ~\underset{d\gg 1}{\longrightarrow}~ \frac{1}{\bar{n}^2}\left(\frac{\pi^2}{3} - \frac{4}{d}\right)
\label{globalscaling}
\end{equation}
for the global strategy. 

The bound in equation (\ref{globalscaling}) is to be compared to a local protocol such as $\rho_0^{\mathrm{ref}}\otimes \rho_0^{(1)}\otimes \cdots \otimes \rho_0^{(d)}$, with $\rho_0^{(i)} = |\phi_0^{(i)}\rangle \langle \phi_0^{(i)}|$ in the pure case. A choice for $\ket{\phi_0}$ capable of achieving the same asymptotic precision than the generalised NOON state is \cite{knott2016local}
\begin{equation}
\ket{\phi_0} = \left[\sqrt{1- \frac{\bar{n}}{N(d+1)}}\ket{0}+\sqrt{\frac{\bar{n}}{N(d+1)}}\ket{N}\right],
\label{localstrategy}
\end{equation}
where $N$ is a free parameter that can be varied while the total mean number of quanta $\bar{n}$ remains constant. The key idea is that this state can have arbitrarily large local variances as $N$ grows \cite{rivas2012, knott2016local, tsang2012}, so that it belongs to the family of \emph{infinite-precision} states that we examined in section \ref{subsec:infiniteprecision}. As a consequence, if we only used asymptotic tools, then it would appear to be possible not only to equate the performance of the global strategy, but to also supersede this and any other protocol. Nevertheless, the following calculation shows that our Bayesian bound produces a more physical result. 

Given the local strategy in equation (\ref{localstrategy}) and the flat prior of hypervolume $(2\pi/\bar{n})^d$ that we are using, let us express the single-shot bound in equation (\ref{multibayesbound}) as 
\begin{equation}
\bar{\epsilon}_{\mathrm{mse}} \geqslant  \frac{1}{\bar{n}^2} \left[\frac{\pi^2}{3} - f\left(N, \bar{n}, d\right) \right],
\end{equation}
where 
\begin{equation}
f\left(N, \bar{n}, d\right) \equiv \frac{\bar{n}^2}{d} \sum_{k=1}^d \mathrm{Tr}(\rho S_k^2). 
\end{equation}
Since the prior under consideration is separable (that is, $p(\boldsymbol{\theta})=p(\theta_1)\cdots p(\theta_d)$), in this case we have that $\rho = \rho_0^{\mathrm{ref}}\otimes \rho^{(1)}\otimes  \cdots \otimes \rho^{(d)}$ and $\bar{\rho}_k = \rho_0^{\mathrm{ref}}\otimes \rho^{(1)}\otimes  \cdots \otimes \bar{\rho}^{(k)} \otimes  \cdots \otimes \rho^{(d)}$. In turn, the individual quantum estimators take the form $S_k = \mathbb{I}_{\mathrm{ref}}\otimes\mathbb{I}\otimes \cdots \otimes S^{(k)} \otimes\cdots \otimes \mathbb{I}$, and the calculation of the optimal single-shot uncertainty for the local estimation of several phases is effectively reduced to the single-parameter calculation 
\begin{equation}
f\left(N, \bar{n}, d \right) = \bar{n}^2 \hspace{0.15em} \mathrm{Tr}(\varrho S^2),
\label{localfunction}
\end{equation}
where $\rho^{(k)} \equiv \varrho$ and $S^{(k)}\equiv S$ are single-mode operators and $\rho^{(k)}$ and $S^{(k)}$ are identical for all the modes. Performing calculations analogous to those in previous examples (and also similar to those in chapter \ref{chap:limited} for single-parameter NOON states), we find that\footnote{The interested reader can find further details of this calculation in \cite{jesus2019b}.}
\begin{equation}
f\left(N, \bar{n}, d \right) = \frac{4\bar{n}^3\left[\left(1 + d\right) N - \bar{n} \right]\left[N \pi \hspace{0.15em}\mathrm{cos}\left(N\pi/\bar{n}\right) - \bar{n}\hspace{0.15em}\mathrm{sin}\left(N\pi/\bar{n}\right) \right]^2}{\pi^2 N^6 \left(1 + d\right)^2},
\end{equation}
which presents two crucial properties:  
\begin{enumerate}
\item[a)] if $N \rightarrow \infty$, then $f(N,\bar{n},d) \rightarrow 0$, so that
\begin{equation}
\bar{\epsilon}_\mathrm{mse}  ~~\underset{N \rightarrow \infty}{\longrightarrow}~~  \frac{\pi^2}{3 \bar{n}^2} = \frac{1}{d} \sum_{i=1}^d \Delta \theta_{p,i}^2;
\end{equation}
\item[b)] if $N = \bar{n}$, then $f(N,\bar{n},d) = 4d/(1+d)^2$, and
\begin{equation}
\bar{\epsilon}_{\mathrm{mse}} \geqslant \frac{1}{\bar{n}^2}\left[\frac{\pi^2}{3} - \frac{4d}{(1+d)^2}\right] ~~\underset{d\gg 1}{\longrightarrow}~~ \frac{1}{\bar{n}^2}\left(\frac{\pi^2}{3} - \frac{4}{d}\right).
\label{localscaling}
\end{equation}
\end{enumerate}

From the first property it is clear that the local strategy in equation (\ref{localstrategy}) cannot produce an arbitrarily good precision by simply increasing $N$, which contrasts with the performance of these states when one attemps to use the asymptotic theory directly. An intuitive way of understanding this is to observe that the periodicity associated with equation (\ref{localstrategy}) is $2\pi/N$; consequently, the width where the value of a given phase may lie needs to be smaller as $N$ grows to avoid ambiguities, and thus the limit $N \rightarrow \infty$ is essentially equivalent to require that the unknown parameters are practically localised before we perform the estimation. Since the prior knowledge modelled by $p(\boldsymbol{\theta})$ is fixed by the situation under analysis, the high amount of prior information required as $N$ grows is not being provided, and the scheme is eventually unable to extract more information beyond what we knew to start with. This type of behaviour is well understood in single-parameter schemes \cite{tsang2012, berry2012infinite, hall2012, jesus2017, giovannetti2012subheisenberg, pezze2013, rafal2015}, and it complements our discussion about \emph{infinite-precision} states in section \ref{subsec:infiniteprecision}. 

The second property suggests that the global strategy is not required to get the scaling that appears in equations (\ref{globalscaling}) and (\ref{localscaling}), and that this is indeed the case can be shown by verifying that it is possible to reach the bound associated with the local strategy. Recalling that the form of the quantum estimators in the latter case is $S_k =  \mathbb{I}_{\mathrm{ref}}\otimes\mathbb{I}\otimes \cdots \otimes S^{(k)} \otimes\cdots \otimes \mathbb{I}$, we see that this implies that each operator $S_k$ commutes trivially with the rest; consequently, we can always construct an optimal strategy with local states and measurements as we did with the qubit network in section \ref{subsec:multibayesqubit}. This means that the local imaging scheme can be employed to achieve the scaling in equation (\ref{globalscaling}) provided that we choose the prior judiciously and that $N$ is finite, and that a global strategy is not necessary in such case, just as the work in \cite{knott2016local} demonstrated for schemes operating in the asymptotic regime. 

Importantly, note that while we know how to construct a measurement scheme to implement the local strategy for any number of parameters, a strategy whose uncertainty is close to the bound for the global scheme has been found only when $d=2$. We leave for future work to determine whether the scaling in equation (\ref{globalscaling}) can be also recovered by a global protocol such that $\mu = 1$ and $d>2$ and that operates with a moderate amount of prior knowledge.

\section{Summary of results and conclusions}

The method proposed in this chapter provides a framework to study realistic multi-parameter schemes where the empirical data is limited and the prior knowledge is moderate, extending in this way the approach introduced in chapter \ref{chap:limited} for single-parameter scenarios and completing our non-asymptotic methodology for quantum metrology. Taking into account that we are starting to witness the experimental implementation of multi-parameter protocols \cite{roccia2018, polino2018}, our proposal could play a crucial role in the design of future experiments once other realistic effects such as the presence of losses are included.

The application of our method to physical schemes such as a sensing network of qubits or a phase imaging protocol has revealed, in addition, important information about the role of entanglement for the estimation of several parameters, which complements our study of functions of those parameters in chapter \ref{chap:networks}. On the one hand, we have demonstrated that the simultaneous estimation of two parameters using a qubit network can be performed optimally with a local strategy when the number of trials is low and we are working in the intermediate prior information regime. On the other hand, we have seen that the scaling provided by the generalised NOON state can be recovered using a local scheme when $\mu = 1$ and the prior knowledge is moderate. That is, we have shown that the fact that entanglement is not needed to achieve the optimal uncertainty using these schemes is not only true in the asymptotic regime, but also in more realistic configurations. We expect this result to have important consequences in future developments of multi-parameter schemes. 

From a theoretical perspective, the most important result of this chapter is the derivation of a multi-parameter Bayesian bound on the single-shot estimation error. The Bayesian nature of this new tool guarantees that the prior information will be correctly taken into account, and we have demonstrated that our bound can be saturated in some circumstances. Moreover, if that is the case, then we can implement our proposal of performing theoretical metrology analyses by repeating the single-shot optimal strategy, which is one of the central ideas of this thesis.  

To derive our bound we have separated the classical optimisation from the manipulations associated with the quantum part of the problem, as we did in section \ref{subsec:originalderivation} for the single-parameter case. Alternatively, we could have started by constructing the scalar quantity $\boldsymbol{u}^\transpose \Sigma_{\mathrm{mse}}\boldsymbol{u}$, and then we could have instead employed any of the alternative single-parameter proofs available in the literature (see \cite{personick1971, helstrom1976, macieszczak2014bayesian} and our review in section \ref{subsec:singleshotparadigm}) to show that $\boldsymbol{u}^\transpose \Sigma_{\mathrm{mse}}\boldsymbol{u} \geqslant \int d\boldsymbol{\theta}p(\boldsymbol{\theta})\theta_u^2 - \mathrm{Tr}(\rho S_u^2)$, from where equation (\ref{multibayesbound}) follows. Note, however, that in that case the classical and quantum optimisations would be performed simultaneously. 

Among all the bounds that neglect the interference between optimal quantum strategies for different parameters due to their lack of commutativity, our result is arguably the preferred option, since it recovers the true optimum in the limit of a single parameter and gives the true multi-parameter optimum when $\lbrace S_i \rbrace$ commute. Furthermore, our analysis of the qubit network have revealed that the multi-parameter Cram\'{e}r-Rao bound can be recovered as an asymptotic limiting case of our bound, a transition that we have characterised using the method in section \ref{subsec:asymsatu}. Combining these observations with the fact that its calculation is relatively simple, we may conclude that our approach provides a reasonable balance between approaching the exact result and having a tractable problem, and while some care is needed when we use this tool to enquire about fundamental limits, it may be sufficient in many practical cases, as our examples with qubits and optical modes demonstrate. 

The results of this chapter have appeared in \cite{jesus2019b}
\begin{displayquote}
\emph{Bayesian multi-parameter quantum metrology with limited data}, \underline{Jes\'{u}s Rubio} and Jacob Dunningham, arXiv:1906.04123 (2019).
\end{displayquote}
\chapter{A look to the future}
\label{chap:future}

\section{Current limitations and the future of non-asymptotic metrology}

Non-asymptotic quantum metrology, as we have defined it in this thesis, was born out of the necessity of applying metrology techniques to situations with a limited amount of experimental data and a potentially moderate amount of prior knowledge, both of which generally lie outside of the scope of the theory based on the Fisher information and the Cram\'{e}r-Rao bound. At the heart of our approach lies the idea of developing an alternative way of doing quantum metrology by relying less on formal approximations and more on the identification of the physically relevant quantities. Nevertheless, our methodology is only the first iteration towards the completion of this task, and despite the wealth of new results that we have uncovered using our formalism, there are still potentially important upgrades for our methods.

One of the crucial tasks that we identified in chapters \ref{chap:limited} and \ref{chap:multibayes} as a potential upgrade is to extend our methods to cover experiments that not only operate in the regime of limited data, but that are also affected by the presence of noise. In the next section we will carry out a first simple analysis of an optical scheme with photon losses, and we will highlight important features to be explored in future work when these two realistic effects are taken into account in a combined fashion. 

Another interesting possibility for future work would be to implement in the laboratory those schemes that have been optimised using our shot-by-shot approach in chapters \ref{chap:limited} and \ref{chap:multibayes}. Let us illustrate how we would proceed with a single-parameter example. Given an experimental arrangement whose information is summarised in the quantum probability $p(m|\theta) = \mathrm{Tr}[E(m) \rho(\theta)]$, and given a moderate amount of prior knowledge encoded in $p(\theta)$, the first step is to find the optimal single-shot strategy that reaches the minimum of the square error $\bar{\epsilon}_{\mathrm{mse}} = \int d\theta dm\hspace{0.15em} p(\theta) p(m|\theta) [g(m) - \theta]^2$, which can be achieved by means of the single-shot quantum optimisation reviewed and exploited in chapter \ref{chap:limited} and in sections \ref{subsec:singleshotparadigm}, \ref{subsec:originalderivation} and \ref{loss}. This process will provide us with either the optimal POM $E(m) \equiv e_m^\mathrm{opt}$ for a given state, the optimal state $\rho_0^\mathrm{opt}$ for a given measurement, or a state and measurement that are both optimal. Although this is the same step that initiated our theoretical study of chapter \ref{chap:limited}, it is also the point where theory and experiment diverge. Our aim was to study the fundamental behaviour of schemes that operate with a limited amount of data, and as such the uncertainty that we have calculated has been averaged over both the parameter and the measurement outcomes. However, in a real-world experiment we will have a concrete string of outcomes, and according to our discussion in section \ref{sec:uncertainty}, the experimental error needs to be based on a figure of merit that depends on such outcomes, that is, in equation (\ref{errexp}) after having chosen the square error. Whether we know how to implement $e^\mathrm{opt}_m$ or how to prepare $\rho_0^\mathrm{opt}$ is a question beyond the scope of our method, although we note that our study with genetic algorithms in section \ref{sec:genetic} demonstrates that the shot-by-shot strategy may be made feasible with current technology. 

It is also important to note that the amount of resources per trial, given by $\langle R \rangle$ with resource operator $R$ (section \ref{sec:problem}), has been assumed to be small. While this assumption guarantees that our results are relevant for and applicable to sensing fragile systems \cite{eckert2007, pototschnig2011, carlton2010, taylor2013, taylor2015, taylor2016, PaulProctor2016}, it excludes other applications where a still finite but larger number of resources per trial may be allowed even if the data is still limited. As we saw in chapter \ref{chap:intro}, this is the case, in particular, for remote sensing \cite{shabir2015, kebei2013, lanzagorta2012, wang2016,zhuang2017}. Fortunately, increasing $\langle R \rangle$ does not alter the foundations of our methodology, and our methods can also be applied in those cases by simply using matrices with larger dimensions for the numerical simulation of the physical system under consideration. 

In terms of theoretical progress, our hybrid method based on selecting the optimal estimator and the asymptotically optimal quantum strategy opens the door to revisiting quantum metrology protocols that have been optimised using the Fisher information and the Cram\'{e}r-Rao bound, which are the majority. More concretely, we could perform a non-asymptotic analysis such as those in chapters \ref{chap:nonasymptotic} and \ref{chap:networks} to determine which of the results found by other authors in the context of the asymptotic theory could be carried over to and exploited in the non-asymptotic regime. 

Importantly, the hybrid method relies on the existence of an asymptotic approximation that coincides with the quantum Cram\'{e}r-Rao bound. A weaker possibility would be returning to $\int d\boldsymbol{\theta}p(\boldsymbol{\theta})F(\boldsymbol{\theta})^{-1}/\mu$ as a more general approximation for the matrix error $\Sigma_{\mathrm{mse}}$, and attempting to use such expression as a guide to select the quantum strategy by comparing different protocols, provided that $F(\boldsymbol{\theta})$ is never singular. Although the generality associated with the quantum Cram\'{e}r-Rao bound is lost in this way, by renouncing to such generality and considering the measurement scheme explicitly we might no longer need to restrict our attention to pure states and commuting generators, since these assumptions were precisely introduced as a simple way of having that $F(\boldsymbol{\theta}) = F_q$ for a single copy \cite{sammy2016compatibility}. Nonetheless, we recall that the problems associated with the existence of some useful asymptotic approximation do not affect any other form of Bayesian estimation where the quantum strategy is selected in a different way, including schemes without an asymptotic expansion at all (this was the case, for instance, of the qubit network with a maximally entangled state studied in chapter \ref{chap:networks}).

From a technical point of view, a current limitation is that associated with the calculation of the Bayesian uncertainty for quantum sensing networks. Due to the numerical difficulties discussed in section \ref{subsec:multinonasym}, our non-asymptotic analyses of multi-parameter schemes with $\mu > 1$ (chapters \ref{chap:networks} and \ref{chap:multibayes}) have been restricted to configurations with two natural parameters, i.e., $d = 2$. Therefore, developing methods to overcome this challenge may have a major impact in the long run, since we expect a plethora of new effects arising from the interplay between different amounts of data and the richer set of possibilities for inter-sensor correlations that emerges when $d \geqslant 3$. Some ideas in this direction include the modification of our algorithm in appendix \ref{sec:multimsematlab} such that the integrals associated with the unknown parameters are also performed with Monte Carlo techniques, or perhaps employing some other quantum bound whose calculation is simple enough to study cases where both $\mu$ and $d$ are unrestricted. One potential candidate fulfilling the latter requirement is the multi-parameter quantum Ziv-Zakai bound in \cite{zhang2014}, although, according to our discussions in sections \ref{subsec:alternativebounds}, \ref{sec:alternativevssingleshot} and \ref{subsec:multibayessaturation}, we cannot expect the results derived using this type of tool to be fundamental in general.  

To conclude this brief exploration of what the future of our methodology might look like, let us recall that the most general approach to the problem of quantum parameter estimation is that based on the equations for the optimal strategy discovered by Helstrom and Holevo \cite{helstrom1976, helstrom1974, holevo1973b, holevo1973}, which we reviewed in section \ref{subsec:fundeq}. That method, which amounts to optimising the uncertainty in a direct fashion, is arguably more fundamental than using bounds that only work in certain regimes, and, in a way, we might see our contribution in this thesis as a bridge between both worlds that has been carefully built by focusing on the physical aspects of the problem, as opposed to following a more abstract approach. As a consequence, any future refinements of our methods should move us closer to the true optima predicted by Helstrom and Holevo's Bayesian theory.

\section{The effect of photon losses}
\label{loss}

Following our previous discussion, let us perform an initial test of the application of our method to noisy scenarios. Dorner \emph{et al.} \cite{dorner2009} studied and solved the problem of photon losses in interferometry using the Fisher information, and here we follow the configuration described in that work. Suppose we consider another Mach-Zehnder interferometer with initial state $\ket{\psi_0} = \sum_{k=0}^2 c_k \ket{k,2-k}$ and where the unknown phase shift $\phi$ is now encoded in the first arm with the unitary transformation $\mathrm{exp}(-i N_1 \phi)$, where $N_i=a_1^\dagger a_1$. In addition, the photon losses in such arm are modelled using a fictitious beam splitter with transmissivity $\eta$. In that case, the transformed state is \cite{dorner2009}
\begin{equation}
\rho(\phi) = \mathrm{e}^{-i N_1 \phi}\left(\sum_{l=0}^2 K_{l,a_1}\ketbra{\psi_0}K_{l,a_1}^\dagger\right) \mathrm{e}^{i N_1 \phi},
\label{lossy_state}
\end{equation}
where $K_{k,a_1}=(1-\eta)^{l/2}\eta^{N_1/2}a_1^l/\sqrt{l!}$ are Kraus operators.

We need to find the state $\ket{\psi_0}$ that is optimal for a given amount of loss. Since for this initial test we are interested in analysing the specific proposal in \cite{dorner2009} and this work is based on the Fisher information, we will simply select the initial probe that has the largest $F_q$, and we will follow the methodology in chapter \ref{chap:limited} to find the Bayesian bound based on repeating the optimal single-shot strategy of this state. However, note a potentially better result could be found by optimising the single-shot bound instead. We leave this possibility for future work.

To represent a realistic amount of loss we can choose $\eta = 9/10$, and the components of the state with the largest $F_q$ for this value are $c_0 = 3/\sqrt{19}$, $c_1 = 0$ and $c_2 = \sqrt{10/19}$. Hence, equation (\ref{lossy_state}) becomes
\begin{equation}
\rho(\phi) = \frac{1}{190} 
\begin{pmatrix} 
1 & 0 & 0 & 0 \\
0 & 90 & 0 & 27 \sqrt{10}~\mathrm{e}^{i 2 \phi} \\
0 & 0 & 18 & 0 \\
0 & 27 \sqrt{10}~\mathrm{e}^{-i 2 \phi} & 0 & 81 
\end{pmatrix},
\end{equation}
where the columns are labelled as $\ket{0,0}$, $\ket{0,2}$, $\ket{1,0}$ and $\ket{2,0}$, respectively.

\begin{figure}[t]
\centering
\includegraphics[trim={0.75cm 0.1cm 1.3cm 1cm},clip,width=14.75cm]{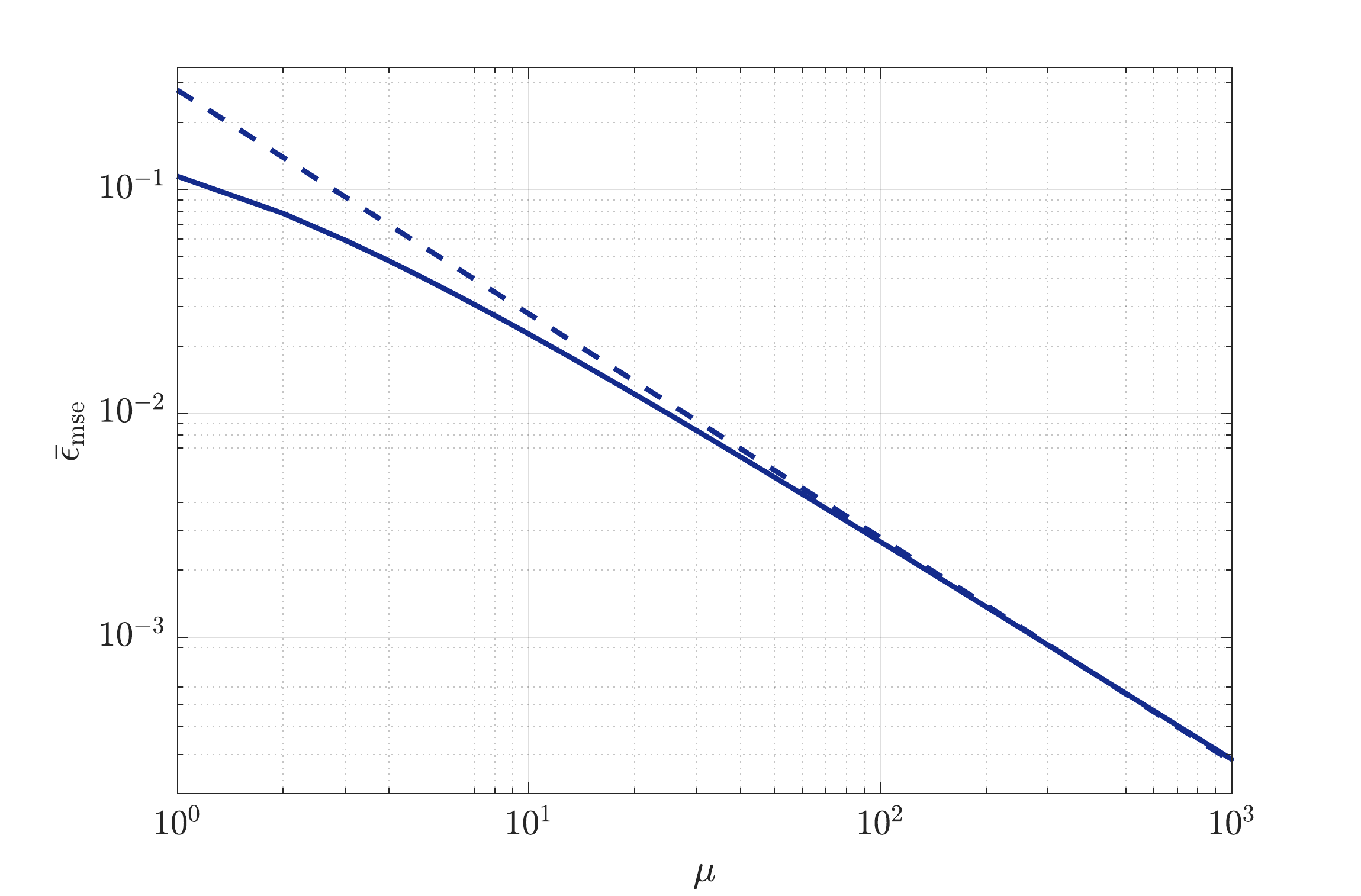}
	\caption[Shot-by-shot quantum bound with photon losses]{Mean square error based on the optimal single-shot strategy (solid line) and quantum Cram\'{e}r-Rao bound (dashed line) for a two-photon state whose Fisher information is optimal (see \cite{dorner2009}) that is fed to a Mach-Zehnder interferometer with photon losses in its first arm, with $\eta=0.9$, $\bar{\phi} = \pi/4$ and $W_0 = \pi/2$.}
\label{lossy_plot}
\end{figure}

The next step is to calculate the optimal single-shot strategy. Assuming that the prior $p(\phi)$ is a flat density of width $W_0 = \pi/2$ and centred around $\bar{\phi} = \pi/4$, we can calculate $\rho = \int d\phi p(\phi)\rho(\phi)$, $\bar{\rho} = \int d\phi p(\phi)\rho(\phi)\phi$ and insert the results into $S\rho + \rho S = 2\bar{\rho}$ to find the optimal quantum estimator
\begin{equation}
S=\frac{1}{76\pi}
\begin{pmatrix} 
19\pi^2 & 0 & 0 & 0 \\
0 & 19\pi^2 & 0 & -24 \sqrt{10} \\
0 & 0 & 19\pi^2 & 0 \\
0 & -24 \sqrt{10} & 0 & 19\pi^2
\end{pmatrix}.
\end{equation}
Using the eigenspaces of $S$ we may construct the projective measurement $\ket{s_1}=(-\ket{0,2}+\ket{2,0})/\sqrt{2}$, $\ket{s_2}=(\ket{0,2}+\ket{2,0})/\sqrt{2}$, $\ket{s_3}=\ket{1,0}$ and $\ket{s_4}=\ket{0,0}$. However, note that, in this case, the optimal single-shot POM is not unique due to the degeneracy of one of the eigenvalues of $S$.

Finally, we calculate the mean square error in equation (\ref{shotbyshotmse}) using this optimal single-shot measurement. The result has been represented in figure \ref{lossy_plot} as a solid line, which also includes the quantum Cram\'{e}r-Rao bound as a dashed line (the latter can be obtained using the expression for the Fisher information $F_q$ provided in \cite{dorner2009}). As we can see, the Bayesian error is very close to the quantum Cram\'{e}r-Rao bound, although a perfect convergence cannot be observed because the mean square error crosses the bound when $\mu \approx 4 \cdot 10^2$. 

It may be verified that the reason for this discrepancy is that the classical Fisher information associated with the chosen POM is no longer parameter-independent, and it only achieves the quantum Cram\'{e}r-Rao bound for certain values of $\phi$. As such, and recalling our discussion about the asymptotic regime in chapter \ref{chap:nonasymptotic}, this implies that, in this case, the true asymptotic approximation is $\int d\phi \hspace{0.15em}p(\phi)/[\mu F(\phi)]$, and not $1/(\mu F_q)$. Remarkably, this is unlike for the ideal schemes in chapter \ref{chap:limited}. Thus this phenomenon sets the scene for a future study about the fundamental limits that we may expect when the data is scarce and there is a certain amount of noise\footnote{We draw attention to the fact that the explanation for the discrepancy in figure \ref{lossy_plot} provided in this section complements the initial test with photon losses in our publication \cite{jesus2018}, where such explanation was not explicitly identified.}. 

On the other hand, if we were to look at this result from a more practical point of view, then we could conclude that a reasonable amount of photon losses does not alter substantially our findings for ideal schemes, since we have verified that after $\mu = 10^3$ repetitions the relative error between the Bayesian uncertainty and the quantum Cram\'{e}r-Rao bound in figure \ref{lossy_plot} is just $\varepsilon = 0.02$. Nevertheless, a deeper investigation including other sources of noise, new probe states and realistic measurements is required in order to construct a complete picture of the effect of noise when the available data is limited. 

\section{A more fundamental perspective}
\label{sec:fundtime}

While quantum metrology and quantum estimation theory are, in a sense, frameworks with an eminently pragmatic purpose, it is well known that they can also be used in a more fundamental way to construct uncertainty relations of a generalised type \cite{braunstein1996, helstrom1976}. In fact, we have already encountered a manifestation of this connection in section \ref{subsec:optint}, where we reviewed a path to arrive at the quantum Cram\'{e}r-Rao bound for pure states from a Mandelstam-Tamm uncertainty relation (see \cite{HofmannHolger2009}). 

Suppose we look at the quantum Cram\'{e}r-Rao bound as a generalised uncertainty relation. The results in this thesis have demonstrated the advantages of treating this bound as a limiting case of a more general theory, such that the former only emerges when certain conditions are fulfilled. If we follow this logic, then it is natural to enquire whether it would be possible to construct some sort of uncertainty relation that incorporates both the effect of the prior information and a finite number of shots. That a generalised uncertainty relation can be constructed with Bayesian quantities was in fact shown by Helstrom \cite{helstrom1976} using the sine error in equation (\ref{sinerror}) for a single shot, while Braunstein \emph{et al.} \cite{braunstein1996} considered several copies of the probe state within the context of the Cram\'{e}r-Rao bound. In view of this, by constructing an uncertainty relation that combines both features we would be able to extend the scope of uncertainty relations in quantum mechanics to cover scenarios where the data is limited and only a moderate amount of prior information is available. 

Although we leave for future work the detailed exploration of this possibility and of its potential consequences, we would like to illustrate what our non-asymptotic methodology has to say about this line of thought. To achieve that goal, let us consider a scenario where the parameter that we wish to estimate is the elapsed time from the evolution of a two-level system, which we denote by $t$. If the system is prepared in the pure state $\rho_0 = (\mathbb{I}+\sigma_x)/2$, the parameter is encoded as $\rho(t) = \mathrm{e}^{-i K t} \rho_0 \mathrm{e}^{i K t}$ and the generator is $K = (E/\hbar)\sigma_z$, with energy $E$, then the quantum Fisher information is
\begin{equation}
F_q = 4\left(\langle \psi_0 | K^2 | \psi_0 \rangle - \langle \psi_0 | K | \psi_0 \rangle^2\right) = 4\left[\mathrm{Tr}(\rho_0 K^2) -\mathrm{Tr}\left(\rho_0 K\right)^2\right] = \frac{4 E^2}{\hbar^2},
\end{equation}
so that the value of the quantum Cram\'{e}r-Rao bound is
\begin{equation}
\bar{\epsilon}_\mathrm{cr} = \frac{1}{\mu F_q} = \frac{\hbar^2}{4\mu E^2}.
\end{equation}
If we are working in the asymptotic regime, then $\bar{\epsilon}_{\mathrm{mse}}\gtrsim \bar{\epsilon}_{\mathrm{cr}}$, so that we can write 
\begin{equation}
E^2 \hspace{0.1em}\bar{\epsilon}_{\mathrm{mse}} \equiv E^2 \Delta t^2 \gtrsim \frac{\hbar^2}{4\mu},
\end{equation}
which indeed has the form that we would expect for an uncertainty relation.

\begin{figure}[t]
\centering
\includegraphics[trim={0.1cm 0.1cm 0.5cm 0.5cm},clip,width=7.7cm]{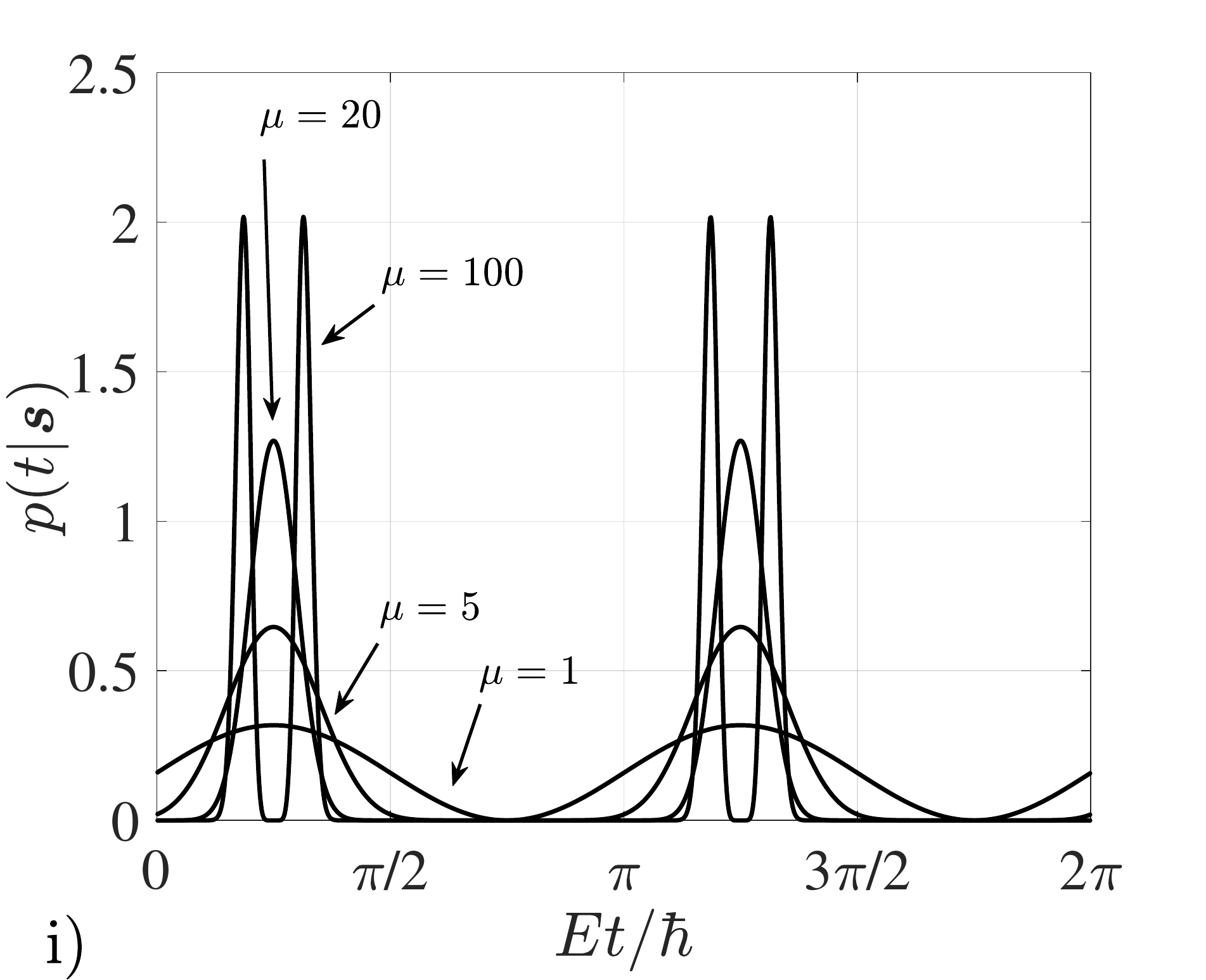}\includegraphics[trim={0cm 0cm 0.5cm 0.5cm},clip,width=7.7cm]{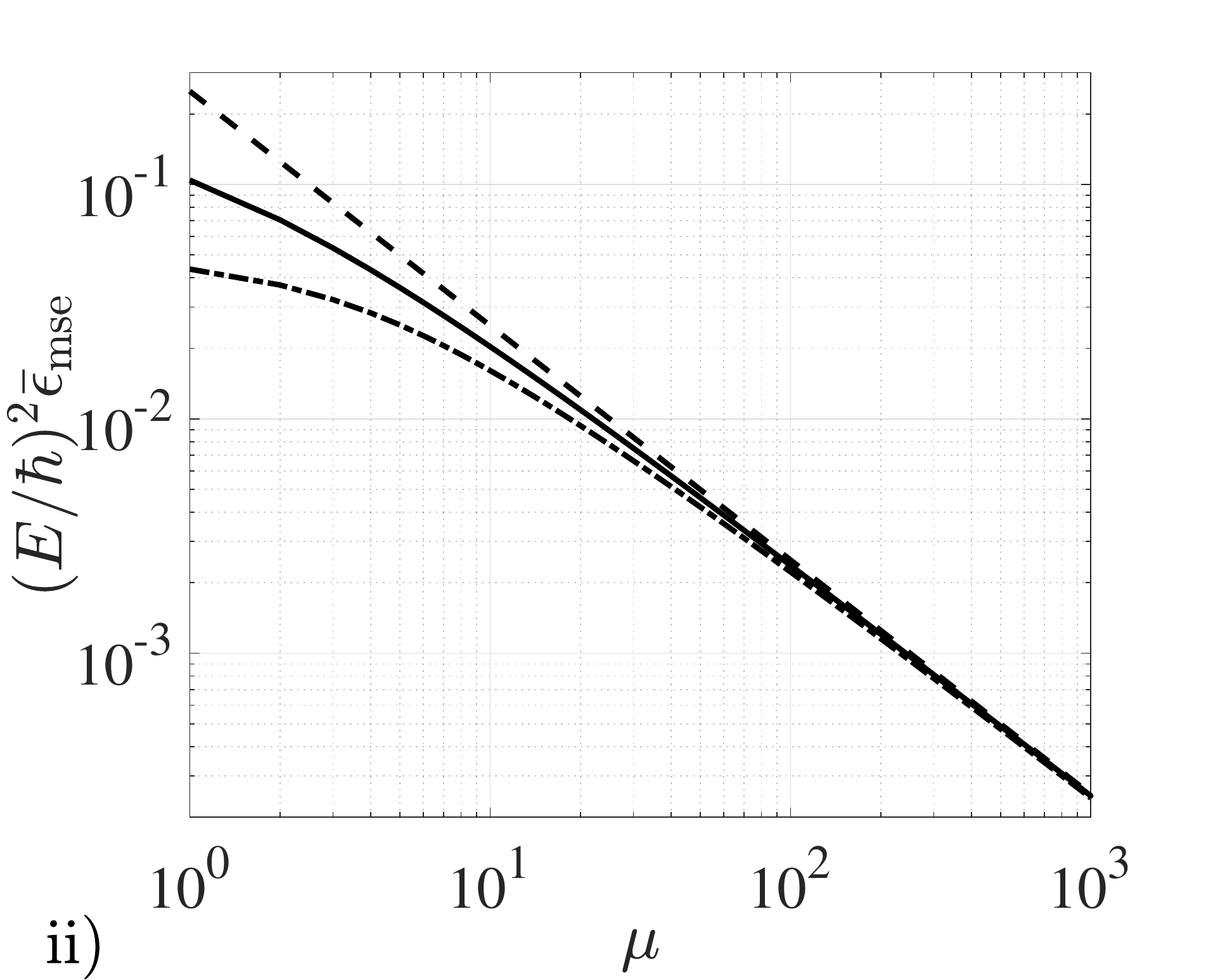} 
\caption[Bayesian time estimation]{i) Posterior probabilities for random simulations of 1, 5, 20 and 100 trials, a flat prior, the POM $\ketbra{s_\pm} = (\mathbb{I} \pm \sigma_y)/2$ and the state $\rho_0 = (\mathbb{I}+\sigma_x)/2$. In (ii) we have represented the mean square error for the previous configuration and prior widths $W_0=\pi\hbar/(2E)$ (solid line) and $W_0=\pi\hbar/(4E)$ (dash-dotted line), while the dashed line is the quantum Cram\'{e}r-Rao bound, which in this section plays the role of a generalised uncertainty relation. We draw attention to the fact that while the prior information alters the precision in the regime of limited data, both Bayesian schemes converge to the same asymptotic optimum.}
\label{timeestimation}
\end{figure}

To saturate this bound, first we need a measurement for which $F(t) = F_q$, where $F(t)$ is the classical Fisher information. We can verify by a direct calculation that a POM that satisfies such condition is $\ketbra{s_\pm} = (\mathbb{I} \pm \sigma_y)/2$. In particular, given that the transformed state is
\begin{align}
\rho(t) &= \mathrm{exp}\left(-i E t\sigma_z/\hbar\right)\rho_0\hspace{0.15em}\mathrm{exp}\left(i E t\sigma_z/\hbar\right)
\nonumber \\
&= \frac{1}{2}\left[\mathbb{I} + \mathrm{cos}\left(2 E t/\hbar\right)\sigma_x + \mathrm{sin}\left(2 E t/\hbar\right)\sigma_y\right],
\end{align}
where we have used the fact that $\mathrm{exp}(i A \sigma_z) = \mathrm{cos}(A)\mathbb{I} + i \mathrm{sin}(A)\sigma_z$, and that the single-shot likelihood function for the aforementioned POM is
\begin{equation}
p(s_\pm|t) = \langle s_\pm | \rho(t) | s_\pm \rangle = \frac{1}{2} \left[1 \pm \mathrm{sin}\left(2 E t/\hbar\right)\right],
\end{equation}
we find that
\begin{eqnarray}
F(t) = \frac{1}{p(s_{+}|t)}\left[\frac{\partial p(s_{+}|t)}{\partial t}\right]^2 + \frac{1}{p(s_{-}|t)}\left[\frac{\partial p(s_{-}|t)}{\partial t}\right]^2 = \frac{4 E^2}{\hbar^2} = F_q,
\end{eqnarray}
as desired. Furthermore, from our findings in chapter \ref{chap:nonasymptotic} we know that to reach the Cram\'{e}r-Rao bound we also need to select a region of the parameter domain where the likelihood does not contain ambiguous information. Assuming a flat prior in such region, we have seen that one way of identifying the intrinsic width\footnote{We recall that we have defined the intrinsic width as the largest width that a flat prior can have while the likelihood function still presents a unique absolute maximum after many repetitions (see chapter \ref{chap:nonasymptotic}).} $W_\mathrm{int}$ is to examine the maxima of the posterior probability $p(t|\boldsymbol{s}) \propto p(\boldsymbol{s}| t)$, where $\boldsymbol{s} = (s_1, \dots, s_\mu)$ are the outcomes of $\mu$ repetitions of the experiment. Figure \ref{timeestimation}.i shows the result of this operation, and upon its inspection we  conclude that $W_\mathrm{int} = \pi \hbar/(2E)$ if one of the boundaries of our prior probability is $(2k+1)\pi \hbar/(4E)$, where $k$ is an integer. The final requirement to achieve the Cram\'{e}r-Rao bound is to repeat the experiment a large number of times. 
 
We shall now compare this bound with our shot-by-shot method, which requires us to calculate the single-shot optimal POM. Suppose that we write our flat prior as\footnote{This form is more convenient to take into account the fact that the intrinsic width depends here on the origin of the prior.} $p(t) = 1/W_0$, for $t\in[a_0,a_0+W_0]$, and zero otherwise, where $a_0 = \pi\hbar/(4E)$ and $W_0 = W_{\mathrm{int}} = \pi\hbar/(2E)$. In that case, and recalling that the quantum estimator $S$ is given by $S\rho + \rho S = 2\bar{\rho}$, with $\rho = \int dt\hspace{0.15em} p(t)\rho(t)$ and $\bar{\rho} = \int dt\hspace{0.15em}p(t)\rho(t)t$, we find that 
\begin{equation}
S = \frac{\pi\hbar}{2E} \left(\mathbb{I}-\frac{2\sigma_y}{\pi^2}\right).
\end{equation}
The projectors of this operator are precisely the POM elements that we have examined in the previous paragraph, that is, $\ketbra{s_\pm} = (\mathbb{I} \pm \sigma_y)/2$. Hence, for this arrangement we have that the same measurement scheme that is optimal asymptotically is also optimal for a single-shot. Adapting the shot-by-shot uncertainty in section \ref{subsec:shotbyshot} to our present case we conclude that the error to be calculated is 
\begin{equation}
\bar{\epsilon}_{\mathrm{mse}} = \int d\boldsymbol{s}~p(\boldsymbol{s}) \left\lbrace \int dt\hspace{0.15em}p(t|\boldsymbol{s}) t^2 - \left[\int dt\hspace{0.15em} p(t|\boldsymbol{s}) t \right]^2 \right\rbrace,
\label{errortime}
\end{equation}
where the posterior is $p(t|\boldsymbol{s})= p(t) p(\boldsymbol{s}|t)/p(\boldsymbol{s})$, the likelihood for $\mu$ trials is $p(\boldsymbol{s}|t) = \prod_{i=1}^\mu \langle s_i | \rho(t) |s_i \rangle$ and $p(\boldsymbol{s}) = \int dt \hspace{0.15em}p(t)p(\boldsymbol{s}|t)$. 

The result of the previous calculation has been represented as the solid line of figure \ref{timeestimation}.ii, while the quantum Cram\'{e}r-Rao bound is the dashed line. As we can see, the asymptotic bound underestimates the precision when the number of repetitions is low, since our shot-by-shot uncertainty is lower in such regime. The reason for this discrepancy is that the latter is taking into account a certain amount of prior information, which is consistent with the qualitative picture that our results in previous chapters have revealed. The novelty here is that we are looking at the quantum Cram\'{e}r-Rao bound as a generalised uncertainty relation. Since the prior knowledge that goes into the mean square error is precisely taking into account the requirements to saturate the Cram\'{e}r-Rao bound, one way of interpreting this result is to conclude that an uncertainty relation that does not include the combined effect of the prior information and a finite number of trials might be losing important information about the fundamental limits of the scheme under analysis. This conjecture, should it be confirmed, might have important consequences for our basic understanding of uncertainty relations.  

\section{Summary of results and conclusions}

Our non-asymptotic methodology promises to open new and exciting lines of future research. On the practical side, the next natural step is to include the effect of noise within our formalism, and a first initial test of this possibility has been carried out using a lossy interferometer. This calculation has revealed that while our method may be applied to such scenario, there are also important differences with respect to the ideal case. Regarding the technical limitations that we have found for the numerical calculation of Bayesian quantities, we have identified the extension of our algorithms to cases with $d\geqslant 3 $ natural parameters as a key step, so that we can keep exploring the interesting interplay that we have uncovered between correlations and a limited amount of data. Finally, we have explored the possibility of using our methodology in a more fundamental context, and we have conjectured the potential existence of generalised uncertainty relations that include the combined effect of a limited amount of data and a moderate prior knowledge, illustrating this idea by applying our shot-by-shot method to the problem of estimating the elapsed time from the evolution of a quantum system. 
\chapter{Conclusions}
\label{chap:conclusions}

Every journey has a final destination, and it is time for us to reach ours. Our particular journey started with the realisation that, ultimately, the success of science as a method to understand the world stems from the solid foundation provided by the empirical facts that we are able to identify. Our ability to extract information from reality is thus crucial, and this is precisely where quantum metrology enters the scene as one of our best frameworks to enhance our ways of communicating with nature, which is achieved via measurements that in this case rely on the quantum properties of matter and light. 

Unfortunately, nature does not always provide us with as much information as we would like to have, and this has two fundamental consequences. On the  one hand, it means that the amount of empirical data that a certain experiment is allowed to extract might be very limited. On the other hand, the available prior knowledge will in many cases be moderate at best, since the prior information is essentially a manifestation of what we learned from either previous experiments or theories whose validity is grounded on empirical evidence.

Given that many quantum protocols are currently devised assuming either an abundance of measurement data, or a very good prior information, or perhaps both, it was crucial to revisit the techniques of quantum metrology and extend them such that they could be efficiently and reliably applied to scenarios where the previous limitations are present. This is exactly what the research in this thesis has achieved.  

The path that we have followed to solve this problem has led us to a new methodology that has opened the door to an alternative way of doing quantum metrology, and that promises to play a central role in any future study of non-asymptotic protocols designed for scenarios with a limited amount of data. That this is likely to be the case has in fact been demonstrated through many surprising new results that have emerged from the application of our methods to specific metrology schemes. 

One of the systems that we have studied in more depth is the Mach-Zehnder interferometer, which is a paradigmatic scheme in the context of optical interferometry. Here the goal is to provide a good estimate for a parameter that represents the difference of optical phase shifts between the arms of the interferometer. Focusing our attention on state-of-the-art probe states that can be constructed with operations such as displacements of the vacuum or squeezing, we have first shown that the number of repetitions and the minimum amount of prior knowledge that are needed for the asymptotic theory to be meaningful crucially depend on the specific properties that a given probe has. For example, given a fixed amount of prior knowledge, we have found that while common probes such as coherent states might require a small number of  trials to reach the performance predicted by the asymptotic theory, more exotic cases such as the squeezed entangled state present a much slower convergence. The crucial observation is that the latter promises a great precision-enhancement with respect to coherent states when we only look at the asymptotic theory, and yet the coherent state beats the squeezed entangled state when the data is limited and we perform a standard photon-counting measurement. The general conclusion is that the ordering of states in terms of their performance is dramatically affected by the number of times that the experiment is repeated, and thus maximising the Fisher information might not always be the best approach. 

The previous idea has been put on a more solid basis by means of our shot-by-shot optimisation method, that is, by repeating the quantum strategy that is optimal for a single shot. Remarkably, while this is a fully Bayesian approach that in principle does not rely on the Fisher information, our method sometimes recovers the predictions of the latter, either in the limit of a large number of repetitions, or in the limit of a narrow prior for a single shot. A crucial finding derived from this approach is the evidence for the existence of a trade-off between the asymptotic and non-asymptotic uncertainties. In particular, we have found that increasing the amount of photon correlations within each of the modes of the interferometer might be detrimental if the experiment is operating with a limited amount of data, despite the fact that these correlations are known to be extremely useful once we have reached the asymptotic regime. More surprisingly, the calculation of a state with less intra-mode correlations but a certain amount of mode entanglement has proven to be a better choice to keep the precision high in both the asymptotic and non-asymptotic regimes, even when the asymptotic theory indicates that mode entanglement only provides a limited advantage. As a consequence, our results indicate that we should pay more attention to the amount of data as a feature that might alter our assessment of the role of correlations in quantum metrology. 

On the other hand, we have shown that our shot-by-shot approach is a useful tool to generate precision bounds that, at least for repetitive experiments, have a certain fundamental character, and we have demonstrated that our bounds can be tighter than other proposals in the literature such as the quantum versions of the Ziv-Zakai and Weiss-Weinstein bounds. Interestingly, our bound for the NOON state is also its true fundamental limit, since we have shown that, in this case, general collective measurements are not better than simply repeating the single-shot optimal strategy. On a more practical note, our bounds have proven to be very useful to assess how close to the precision limits that we have calculated the uncertainty associated with practical measurements can be. For example, we have found that while measuring quadratures and a measurement based on counting photons are, for ideal schemes, equally precise when the scheme operates in the asymptotic regime, the former is closer to our bounds when the number of repetitions is small. In addition, by combining our method with a genetic algorithm we have provided sequences of operations to generate probes that not only have a good performance in the regime of limited data, but that may also be implemented in the laboratory with current technology. In other words, our non-asymptotic analysis of the Mach-Zehnder interferometer has revealed new theoretical properties about the interplay between amount of data, prior information and photon correlations that were previously unknown, and it has also provided specific procedures that may be relevant in real-world implementations of our protocols when these operate in the non-asymptotic regime.  

Given our aim of bringing quantum metrology techniques closer to the reality of experimental practice, our methodology would not have been complete if we had not addressed multi-parameter metrology problems, since many practical applications require the estimation of several pieces of information. In this context, we have chosen to focus on the design of quantum sensing networks, which is a model for distributed sensing. The implementation of this type of configuration might involve large distances between the quantum sensors that form the network (this is the case, e.g., in a network of satellites), and this makes the construction  and maintenance of these quantum networks potentially challenging. For that reason, it was crucial to identify strategies that can perform optimally even when the available resources are limited, including both the number of times that the protocol can be run and the amount of correlations between the sensors that we may have. The presence of several parameters provides, in addition, a set of possibilities to enhance the protocol that is larger than in the single-parameter case, and it is useful to split the problem in two parts. The first of them involves the estimation of properties that have been locally encoded in each sensor. Applying our shot-by-shot method we have shown that, in that case, entanglement between sensors is not required to achieve the optimal precision that a network of qubits could provide. In addition, we have found that neither is entanglement necessary to benefit from the precision-enhancement associated with a quantum imagining protocol when the latter is compared to the individual estimation based on Mach-Zehnder interferometers. Remarkably, this is the same conclusion that had been reached previously in the literature in the context of the asymptotic theory. Therefore, our result has effectively extended such conclusion to the regime of limited data and a moderate amount of prior knowledge.

The second part of the problem of quantum sensing networks involves the estimation of properties that are modelled by arbitrary functions of several locally-encoded parameters. For that reason, we may say that such functions represent global properties of the network. It was known that entanglement sometimes enhances the performance of the schemes designed for this specific problem notably, but the situations where this had been shown were mostly limited to considering a single function. Here we have been able to go a step further. In particular, we have considered linear but otherwise arbitrary functions, and we have solved the asymptotic estimation problem completely for the particular case of sensor-symmetric networks, which can be seen as a generalisation of the symmetric configurations that are typically utilised in optical interferometry. These asymptotic solutions were then employed as a guide to perform our non-asymptotic analysis, and we have shown that the amount of inter-sensor correlations that is optimal crucially depends on the number of repetitions and the prior information that it is being assumed, which is exactly the same type of phenomenon that we had uncovered for the Mach-Zehnder interferometer. For example, we have found that if the vectors formed with the components of functions are clustered around the direction associated with maximally entangled states, then these probes will be the best choice for a small number of trials and a vague prior, while only a moderate amount of positive correlations will be required to achieve the asymptotic optimum. Taking into account that the same type of behaviour has been established both for single-parameter and multi-parameter schemes, which were, in addition, based on physically different systems, our results suggest that the interplay between correlations, amount of data and prior knowledge is in fact a more general feature, so that we may expect it to also arise in other estimation problems. It appears that if we could learn how to control the aforementioned interplay in practice, then we would have at our disposition a remarkably large amount of unexplored possibilities to enhance non-asymptotic quantum protocols.

If we look at our results from a more fundamental point of view, then we can see that two very satisfactory features of the methodology that we have developed are its unified character and consistency. Indeed, the path that we have followed has effectively transformed an initial collection of techniques - many of them already known but often treated as if they were unrelated to each other - into a unified framework that offers a much broader perspective. We have seen that, for us, the first question to be asked before we start the optimisation of our protocols is what is the physically meaningful quantity that we should employ to assess the uncertainty, and whether we choose to rely on bounds or on any other technique is mostly related to which tools generate more tractable calculations. Our formalism then follows naturally from this point of view: given a measure of uncertainty that we wish to use with a class of protocols based on repetitive experiments, we can either optimise the system in a shot-by-shot fashion, which is arguably the most general and fundamental possibility for our particular case, or we can follow a weaker approach and only require that the protocol performs optimally as the data accumulate. We have shown that these simple but powerful ideas can be applied to both single-parameter and multi-parameter cases, and we have even derived a new multi-parameter quantum bound during the process of adapting our methods to the latter case. We can conclude that, as we announced in the introduction, we have proposed, constructed, explored and exploited a \emph{non-asymptotic quantum metrology}.

\clearpage
\phantomsection
\addcontentsline{toc}{chapter}{Bibliography}
\bibliography{phd_references_04012020}

\begin{thebibliography}{100}

\bibitem{englert2013}
B.-G. {Englert}.
\newblock {On quantum theory}.
\newblock {\em European Physical Journal D}, 67:238, 2013.

\bibitem{haag1996}
R.~Haag.
\newblock {\em Local quantum physics: fields, particles, algebras}.
\newblock Texts and monographs in physics. Springer, Berlin, New York, 2nd
  edition, 1996.

\bibitem{haag2016}
R.~Haag.
\newblock {On Quantum Theory}.
\newblock arXiv:1602.05426, 2016.

\bibitem{vim2012}
BIPM, IEC, IFCC, ILAC, ISO, IUPAC, IUPAP, and OIML.
\newblock {\em {International vocabulary of metrology - Basic and general
  concepts and associated terms (VIM)}}.
\newblock Joint Committee for Guides in Metrology, JCGM 200:2012, Geneva,
  International Organization for Standardization, 3rd edition, 2012.

\bibitem{spivak1980}
M.~Spivak.
\newblock {\em Calculus}.
\newblock Publish or Perish, Berkeley, Calif., 2nd edition, 1980.

\bibitem{helstrom1976}
C.~W. Helstrom.
\newblock {\em {Quantum Detection and Estimation Theory}}.
\newblock Academic Press, New York, 1976.

\bibitem{rafal2015}
R.~Demkowicz-Dobrza\ifmmode~\acute{n}\else \'{n}\fi{}ski, M.~Jarzyna, and
  J.~Ko\l{}ody\ifmmode~\acute{n}\else \'{n}\fi{}ski.
\newblock {Quantum Limits in Optical Interferometry}.
\newblock {\em Progress in Optics}, 60:345--435, 2015.

\bibitem{dowling2014}
J.~P. {Dowling} and K.~P. {Seshadreesan}.
\newblock Quantum optical technologies for metrology, sensing, and imaging.
\newblock {\em Journal of Lightwave Technology}, 33(12):2359--2370, 2015.

\bibitem{jaynes2003}
E.~T. Jaynes.
\newblock {\em {Probability Theory: The Logic of Science}}.
\newblock Cambridge University Press, 2003.

\bibitem{kay1993}
S.~M. Kay.
\newblock {\em {Fundamentals of Statistical Signal Processing: Estimation
  Theory}}.
\newblock Prentice Hall, 1993.

\bibitem{dowling2003}
J.~P. Dowling and G.~J. Milburn.
\newblock Quantum technology: the second quantum revolution.
\newblock {\em Philosophical Transactions of the Royal Society A: Mathematical,
  Physical and Engineering Sciences}, 361(1809):1655--1674, 2003.

\bibitem{degen2017}
C.~L. Degen, F.~Reinhard, and P.~Cappellaro.
\newblock Quantum sensing.
\newblock {\em {Reviews of Modern Physics}}, 89:035002, 2017.

\bibitem{browne2017}
D.~E. Browne, S.~Bose, F.~Mintert, and M.~S. Kim.
\newblock {From Quantum Optics to Quantum Technologies}.
\newblock {\em Progress in Quantum Electronics}, 54:2--18, 2017.

\bibitem{barnett2017}
S.~M. Barnett, A.~Beige, A.~Ekert, B.~M. Garraway, C.~H. Keitel, V.~Kendon,
  M.~Lein, G.~J. Milburn, H.~M. Moya-Cessa, M.~Murao, J.~K. Pachos, G.~M.
  Palma, E.~Paspalakis, S.~J.~D. Phoenix, B.~Piraux, M.~B. Plenio, B.~C.
  Sanders, J.~Twamley, A.~Vidiella-Barranco, and M.~S. Kim.
\newblock {Journeys from Quantum Optics to Quantum Technology}.
\newblock {\em Progress in Quantum Electronics}, 54:19--45, 2017.

\bibitem{acin2018}
A.~Ac{\'{\i}}n, I.~Bloch, H.~Buhrman, T.~Calarco, C.~Eichler, J.~Eisert,
  D.~Esteve, N.~Gisin, S.~J. Glaser, F.~Jelezko, S.~Kuhr, M.~Lewenstein, M.~F.
  Riedel, P.~O. Schmidt, R.~Thew, A.~Wallraff, I.~Walmsley, and F.~K. Wilhelm.
\newblock {The quantum technologies roadmap: a European community view}.
\newblock {\em New Journal of Physics}, 20(8):080201, 2018.

\bibitem{aasi2013}
J.~Aasi, J.~Abadie, B.~P. Abbott, et~al.
\newblock {Enhanced sensitivity of the LIGO gravitational wave detector by
  using squeezed states of light}.
\newblock {\em Nature Photonic}, 7:613--619, 2013.

\bibitem{pitkin2011}
M.~Pitkin, S.~Reid, S.~Rowan, and J.~Hough.
\newblock {Gravitational Wave Detection by Interferometry (Ground and Space)}.
\newblock {\em Living Reviews in Relativity}, 14(1):5, 2011.

\bibitem{taylor2013}
M.~A. Taylor, J.~Janousek, V.~Daria, J.~Knittel, B.~Hage, H.-A. Bachor, and
  W.~P. Bowen.
\newblock Biological measurement beyond the quantum limit.
\newblock {\em Nature Photonics}, 7:229--223, 2013.

\bibitem{taylor2015}
M.~Taylor.
\newblock {\em {Quantum Microscopy of Biological Systems}}.
\newblock Springer International Publishing Switzerland, 2015.

\bibitem{taylor2016}
M.~A. Taylor and W.~P. Bowen.
\newblock Quantum metrology and its application in biology.
\newblock {\em Physics Reports}, 615:1--59, 2016.

\bibitem{eckert2007}
K.~Eckert, O.~Romero-Isart, M.~Rodriguez, M.~Lewenstein, E.~S. Polzik, and
  A.~Sanpera.
\newblock Quantum non-demolition detection of strongly correlated systems.
\newblock {\em Nature Physics}, 4(1):50--54, 2007.

\bibitem{pototschnig2011}
M.~Pototschnig, Y.~Chassagneux, J.~Hwang, G.~Zumofen, A.~Renn, and
  V.~Sandoghdar.
\newblock Controlling the phase of a light beam with a single molecule.
\newblock {\em Physical Review Letters}, 107:063001, 2011.

\bibitem{carlton2010}
P.~M. Carlton, J.~Boulanger, C.~Kervrann, J.-B. Sibarita, J.~Salamero,
  S.~Gordon-Messer, D.~Bressan, J.~E. Haber, S.~Haase, L.~Shao, L.~Winoto,
  A.~Matsuda, P.~Kner, S.~Uzawa, M.~Gustafsson, Z.~Kam, D.~A. Agard, and J.~W.
  Sedat.
\newblock Fast live simultaneous multiwavelength four-dimensional optical
  microscopy.
\newblock {\em Proceedings of the National Academy of Sciences},
  107(37):16016--16022, 2010.

\bibitem{wolfgramm2013}
F.~Wolfgramm, C.~Vitelli, F.~A. Beduini, N.~Godbout, and M.~W. Mitchell.
\newblock Entanglement-enhanced probing of a delicate material system.
\newblock {\em Nature Photonics}, 7(1):28--32, 2013.

\bibitem{PaulProctor2016}
P.~A. Knott, T.~J. Proctor, A.~J. Hayes, J.~P. Cooling, and J.~A. Dunningham.
\newblock {Practical quantum metrology with large precision gains in the
  low-photon-number regime}.
\newblock {\em Physical Review A}, 93:033859, 2016.

\bibitem{baumgart2016}
I.~Baumgart, J.~M. Cai, A.~Retzker, M.~B. Plenio, and Ch. Wunderlich.
\newblock {Ultrasensitive Magnetometer using a Single Atom}.
\newblock {\em Physical Review Letters}, 116:240801, 2016.

\bibitem{shabir2015}
S.~Barzanjeh, S.~Guha, C.~Weedbrook, D.~Vitali, J.~H. Shapiro, and
  S.~Pirandola.
\newblock {Microwave Quantum Illumination}.
\newblock {\em Physical Review Letters}, 114:080503, 2015.

\bibitem{kebei2013}
K.~Jiang, H.~Lee, C.~C. Gerry, and J.~P. Dowling.
\newblock Super-resolving quantum radar: Coherent-state sources with homodyne
  detection suffice to beat the diffraction limit.
\newblock {\em Journal of Applied Physics}, 114(19):193102, 2013.

\bibitem{lanzagorta2012}
M.~Lanzagorta.
\newblock {\em Quantum radar}.
\newblock Synthesis digital library of engineering and computer science. Morgan
  \& Claypool, USA, 2012.

\bibitem{wang2016}
Q.~Wang, L.~Hao, Y.~Zhang, C.~Yang, X.~Yang, L.~Xu, and Y.~Zhao.
\newblock Optimal detection strategy for super-resolving quantum lidar.
\newblock {\em Journal of Applied Physics}, 119(2):023109, 2016.

\bibitem{zhuang2017}
Q.~Zhuang, Z.~Zhang, and J.~H. Shapiro.
\newblock Entanglement-enhanced lidars for simultaneous range and velocity
  measurements.
\newblock {\em Physical Review A}, 96:040304, 2017.

\bibitem{Dowling1998}
J.~P. Dowling.
\newblock {Correlated input-port, matter-wave interferometer: Quantum-noise
  limits to the atom-laser gyroscope}.
\newblock {\em Physical Review A}, 57:4736--4746, 1998.

\bibitem{proctor2017networked}
T.~J. {Proctor}, P.~A. {Knott}, and J.~A. {Dunningham}.
\newblock Networked quantum sensing.
\newblock arXiv:1702.04271, 2017.

\bibitem{proctor2017networkedshort}
T.~J. Proctor, P.~A. Knott, and J.~A. Dunningham.
\newblock Multiparameter estimation in networked quantum sensors.
\newblock {\em Physical Review Letters}, 120:080501, 2018.

\bibitem{ge2018}
W.~Ge, K.~Jacobs, Z.~Eldredge, A.~V. Gorshkov, and M.~Foss-Feig.
\newblock {Distributed Quantum Metrology with Linear Networks and Separable
  Inputs}.
\newblock {\em Physical Review Letters}, 121:043604, Jul 2018.

\bibitem{eldredge2018}
Z.~Eldredge, M.~Foss-Feig, J.~A. Gross, S.~L. Rolston, and A.~V. Gorshkov.
\newblock Optimal and secure measurement protocols for quantum sensor networks.
\newblock {\em Physical Review A}, 97:042337, 2018.

\bibitem{altenburg2018}
S.~Altenburg and S.~W\"{o}lk.
\newblock Multi-parameter estimation: global, local and sequential strategies.
\newblock {\em Physica Scripta}, 94(1):014001, 2018.

\bibitem{qian2019}
K.~Qian, Z.~Eldredge, W.~Ge, G.~Pagano, C.~Monroe, J.~V. Porto, and A.~V.
  Gorshkov.
\newblock Heisenberg-scaling measurement protocol for analytic functions with
  quantum sensor networks.
\newblock {\em Physical Review A}, 100:042304, 2019.

\bibitem{liao2017}
S.-K. {Liao}, W.-Q. {Cai}, W.-Y. {Liu}, L.~{Zhang}, Y.~{Li}, J.-G. {Ren},
  J.~{Yin}, Q.~{Shen}, Y.~{Cao}, Z.-P. {Li}, F.-Z. {Li}, X.-W. {Chen}, L.-H.
  {Sun}, J.-J. {Jia}, J.-J. {Wu}, X.-J. {Jiang}, J.-F. {Wang}, Y.-M. {Huang},
  Q.~{Wang}, Y.-L. {Zhou}, L.~{Deng}, T.~{Xi}, L.~{Ma}, T.~{Hu}, Q.~{Zhang},
  Y.-A. {Chen}, N.-L. {Liu}, X.-B. {Wang}, Z.-C. {Zhu}, C.-Y. {Lu}, R.~{Shu},
  C.-Z. {Peng}, J.-Y. {Wang}, and J.-W. {Pan}.
\newblock {Satellite-to-ground quantum key distribution}.
\newblock {\em Nature}, 549:43--47, 2017.

\bibitem{samuel2018}
S.~Fern\'{a}ndez-Lorenzo.
\newblock {\em Exploiting symmetry and criticality in quantum sensing and
  quantum simulation}.
\newblock PhD thesis, University of Sussex, 2018.

\bibitem{dunningham2006}
J.~A. Dunningham.
\newblock Using quantum theory to improve measurement precision.
\newblock {\em Contemporary Physics}, 47(5):257--267, 2006.

\bibitem{giovanetti2006review}
V.~Giovannetti, S.~Lloyd, and L.~Maccone.
\newblock Quantum metrology.
\newblock {\em Physical Review Letters}, 96:010401, 2006.

\bibitem{paris2009}
M.~G.~A. Paris.
\newblock Quantum estimation for quantum metrology.
\newblock {\em International Journal of Quantum Information},
  07(supp01):125--137, 2009.

\bibitem{braunstein_gaussian1992}
S.~L. Braunstein.
\newblock {How large a sample is needed for the maximum likelihood estimator to
  be approximately Gaussian?}
\newblock {\em Journal of Physics A: Mathematical and General}, 25(13):3813,
  1992.

\bibitem{braun2018}
D.~Braun, G.~Adesso, F.~Benatti, R.~Floreanini, U.~Marzolino, M.~W. Mitchell,
  and S.~Pirandola.
\newblock Quantum-enhanced measurements without entanglement.
\newblock {\em Reviews of Modern Physics}, 90:035006, 2018.

\bibitem{tsang2016}
X.-M. Lu and M.~Tsang.
\newblock {Quantum Weiss-Weinstein bounds for quantum metrology}.
\newblock {\em Quantum Science and Technology}, 1(1):015002, 2016.

\bibitem{liu2016}
J.~Liu and H.~Yuan.
\newblock Valid lower bound for all estimators in quantum parameter estimation.
\newblock {\em New Journal of Physics}, 18(9):093009, 2016.

\bibitem{lumino2017}
A.~Lumino, E.~Polino, A.~S. Rab, G.~Milani, N.~Spagnolo, N.~Wiebe, and
  F.~Sciarrino.
\newblock Experimental phase estimation enhanced by machine learning.
\newblock {\em Physical Review Applied}, 10:044033, 2018.

\bibitem{ariano1998}
G.~M. D'Ariano, C.~Macchiavello, and M.~F. Sacchi.
\newblock {On the general problem of quantum phase estimation }.
\newblock {\em Physics Letters A}, 248(2-4):103--108, 1998.

\bibitem{chiara2003}
C.~Macchiavello.
\newblock Optimal estimation of multiple phases.
\newblock {\em Physical Review A}, 67:062302, 2003.

\bibitem{chiribella2005}
G.~Chiribella, G.~M. D'Ariano, and M.~F. Sacchi.
\newblock Optimal estimation of group transformations using entanglement.
\newblock {\em Physical Review A}, 72:042338, 2005.

\bibitem{holevo2011}
A.~S. Holevo.
\newblock {\em {Probabilistic and Statistical Aspects of Quantum Theory}}.
\newblock Edizioni della Normale, Springer Basel, 2011.

\bibitem{helstrom1974}
C.~Helstrom and R.~Kennedy.
\newblock Noncommuting observables in quantum detection and estimation theory.
\newblock {\em IEEE Transactions on Information Theory}, 20(1):16--24, 1974.

\bibitem{holevo1973b}
A.~S. Holevo.
\newblock Statistical problems in quantum physics.
\newblock In G.~Maruyama and Y.~V. Prokhorov, editors, {\em Proceedings of the
  Second Japan-USSR Symposium on Probability Theory}, pages 104--119, Berlin,
  Heidelberg, 1973. Springer Berlin Heidelberg.

\bibitem{holevo1973}
A.~S. Holevo.
\newblock Statistical decision theory for quantum systems.
\newblock {\em Journal of Multivariate Analysis}, 3(4):337--394, 1973.

\bibitem{tsang2012}
M.~Tsang.
\newblock {Ziv-Zakai Error Bounds for Quantum Parameter Estimation}.
\newblock {\em Physical Review Letters}, 108:230401, 2012.

\bibitem{knott2016local}
P.~A. Knott, T.~J. Proctor, A.~J. Hayes, J.~F. Ralph, P.~Kok, and J.~A.
  Dunningham.
\newblock Local versus global strategies in multiparameter estimation.
\newblock {\em Physical Review A}, 94(6):062312, 2016.

\bibitem{definetti1990}
B.~de~Finetti.
\newblock {\em {Theory of Probability: a Critical introductory Treatment}}.
\newblock John Wiley \& Sons Ltd., {Wiley Classics Library} edition, 1990.

\bibitem{bernardo1994}
J.~M. Bernardo.
\newblock {\em Bayesian theory}.
\newblock Wiley, Chichester, 1994.

\bibitem{rosenthal2006}
J.~S. Rosenthal.
\newblock {\em A first look at rigorous probability theory}.
\newblock Singapore; Hackensack, N. J.: World Scientific, 2nd edition, 2006.

\bibitem{jeffreys1961}
H.~Jeffreys.
\newblock {\em Theory of Probability}.
\newblock Clarendon P., 3rd edition, 1961.

\bibitem{jiangwei2014}
J.~{Shang}, H.~{Khoon Ng}, and B.-G. {Englert}.
\newblock {Quantum state tomography: Mean squared error matters, bias does
  not}.
\newblock arXiv:1405.5350, 2014.

\bibitem{ballentine1998}
L.~E. Ballentine.
\newblock {\em Quantum mechanics: a modern development}.
\newblock World Scientific, Singapore, 1998.

\bibitem{vanhorn2003}
K.~S. Van~Horn.
\newblock {Constructing a logic of plausible inference: a guide to Cox's
  theorem}.
\newblock {\em International Journal of Approximate Reasoning}, 34(1):3--24,
  2003.

\bibitem{nidditch1962}
P.~H. Nidditch.
\newblock {\em Propositional calculus}.
\newblock London: Routledge \& Kegan Paul Ltd, 1962.

\bibitem{copi2016}
I.~M. Copi, C.~Cohen, and K.~McMahon.
\newblock {\em Introduction to logic}.
\newblock Essex, Pearson Education Limited, 14th edition, 2016.

\bibitem{shafer2004}
G.~Shafer.
\newblock {Comments on ``Constructing a logic of plausible inference: a guide
  to Cox's Theorem", by K. S. Van Horn}.
\newblock {\em {International Journal of Approximate Reasoning}},
  35(1):97--105, 2004.

\bibitem{vanhorn2004}
K.~S. Van~Horn.
\newblock {Response to Shafer's comments}.
\newblock {\em International Journal of Approximate Reasoning}, 35(1):107--110,
  2004.

\bibitem{cox1946}
R.~T. Cox.
\newblock {Probability, Frequency and Reasonable Expectation}.
\newblock {\em American Journal of Physics}, 14(1):1--13, 1946.

\bibitem{cox1961}
R.~T. Cox.
\newblock {\em The algebra of probable inference}.
\newblock Baltimore: The Johns Hopkins Press, 1961.

\bibitem{paris1994}
J.~B. Paris.
\newblock {\em The Uncertain Reasoner's Companion: A Mathematical Perspective}.
\newblock Cambridge University Press, 1994.

\bibitem{colyvan2004}
M.~Colyvan.
\newblock {The philosophical significance of Cox's theorem}.
\newblock {\em International Journal of Approximate Reasoning}, 37(1):71--85,
  2004.

\bibitem{clayton2017}
A.~Clayton and T.~Waddington.
\newblock {Bridging the intuition gap in Cox's theorem: A Jaynesian argument
  for universality}.
\newblock {\em International Journal of Approximate Reasoning}, 80:36--51,
  2017.

\bibitem{vanhorn2017}
K.~S. Van~Horn.
\newblock {From propositional logic to plausible reasoning: A uniqueness
  theorem}.
\newblock {\em International Journal of Approximate Reasoning}, 88:309--332,
  2017.

\bibitem{breuer2002}
H.-P. Breuer.
\newblock {\em The theory of open quantum systems}.
\newblock Oxford University Press, 2002.

\bibitem{ballentine2016}
L.~E. Ballentine.
\newblock Propensity, probability, and quantum theory.
\newblock {\em Foundations of Physics}, 46, 02 2016.

\bibitem{waerden1967}
B.~L. Waerden.
\newblock {\em Sources of Quantum Mechanics}.
\newblock North Holland, Amsterdam, 1967.

\bibitem{schwinger2001}
J.~Schwinger.
\newblock {\em {Quantum Mechanics: Symbolism of Atomic Measurements}}.
\newblock Edited by B.-G. Engler. Springer, New York, 2001.

\bibitem{barnett2014}
S.~M. Barnett.
\newblock {\em {Quantum Retrodiction}}.
\newblock {In: E. Andersson , P. \"{O}hberg (eds) Quantum Information and
  Coherence. Scottish Graduate Series. Springer, Cham}, 2014.

\bibitem{isham1995}
C.~J. Isham.
\newblock {\em Lectures on quantum theory: mathematical and structural
  foundations}.
\newblock Imperial College P., London, 1995.

\bibitem{ballentine1986}
L.~E. Ballentine.
\newblock Probability theory in quantum mechanics.
\newblock {\em American Journal of Physics}, 54(10):883--889, 1986.

\bibitem{fuchs2017}
C.~A. Fuchs.
\newblock {Notwithstanding Bohr, the Reasons for QBism}.
\newblock {\em Mind and Matter}, 2(15):245--300, 2017.

\bibitem{barnett2002}
S.~M. Barnett and P.~M. Radmore.
\newblock {\em {Methods in Theoretical Quantum Optics}}.
\newblock Oxford University Press, 2002.

\bibitem{nielsen2010}
M.~A. Nielsen and I.~L. Chuang.
\newblock {\em Quantum computation and quantum information}.
\newblock Cambridge University Press, Cambridge, 10th anniversary edition,
  2010.

\bibitem{dunningham2018}
J.~Dunningham and V.~Vedral.
\newblock {\em {Introductory Quantum Physics and Relativity}}.
\newblock World Scientific, 2nd edition, 2018.

\bibitem{kok2010}
P.~Kok.
\newblock {\em Introduction to optical quantum information processing}.
\newblock Cambridge University Press, Cambridge, 2010.

\bibitem{Szczykulska2016}
M.~Szczykulska, T.~Baumgratz, and A.~Datta.
\newblock Multi-parameter quantum metrology.
\newblock {\em Advances in Physics: X}, 1(4):621--639, 2016.

\bibitem{baumgratz2016}
T.~Baumgratz and A.~Datta.
\newblock {Quantum Enhanced Estimation of a Multidimensional Field}.
\newblock {\em Physical Review Letters}, 116:030801, 2016.

\bibitem{yurke1986}
B.~Yurke, S.~L. McCall, and J.~R. Klauder.
\newblock {SU(2) and SU(1,1) interferometers}.
\newblock {\em Physical Review A}, 33:4033--4054, 1986.

\bibitem{pezze2015}
L.~{Pezz{\`e}}, P.~Hyllus, and A.~Smerzi.
\newblock Phase-sensitivity bounds for two-mode interferometers.
\newblock {\em Physical Review A}, 91:032103, 2015.

\bibitem{HofmannHolger2009}
H.~F. Hofmann.
\newblock All path-symmetric pure states achieve their maximal phase
  sensitivity in conventional two-path interferometry.
\newblock {\em Physical Review A}, 79:033822, 2009.

\bibitem{sahota2015}
J.~Sahota and N.~Quesada.
\newblock {Quantum correlations in optical metrology: Heisenberg-limited phase
  estimation without mode entanglement}.
\newblock {\em Physical Review A}, 91:013808, 2015.

\bibitem{vidrighin2014}
M.~D. Vidrighin, G.~Donati, M.~G. Genoni, X.-M. Jin, W.~S. Kolthammer, M.~S.
  Kim, A.~Datta, M.~Barbieri, and I.~A. Walmsley.
\newblock Joint estimation of phase and phase diffusion for quantum metrology.
\newblock {\em Nature Communications}, 5, 2014.

\bibitem{szczykulska2017}
M.~Szczykulska, T.~Baumgratz, and A.~Datta.
\newblock Reaching for the quantum limits in the simultaneous estimation of
  phase and phase diffusion.
\newblock {\em Quantum Science and Technology}, 2(4):044004, 2017.

\bibitem{humphreys2013}
P.~C. Humphreys, M.~Barbieri, A.~Datta, and I.~A. Walmsley.
\newblock {Quantum Enhanced Multiple Phase Estimation}.
\newblock {\em Physical Review Letters}, 111:070403, 2013.

\bibitem{zhang_lu2017}
L.~Zhang and K.~W.~C. Chan.
\newblock {Quantum multiparameter estimation with generalized balanced
  multimode NOON-like states}.
\newblock {\em Physical Review A}, 95:032321, 2017.

\bibitem{jasminder2016}
J.~S. Sidhu and P.~Kok.
\newblock Quantum metrology of spatial deformation using arrays of classical
  and quantum light emitters.
\newblock {\em Physical Review A}, 95:063829, 2017.

\bibitem{jasminder2018}
J.~S. {Sidhu} and P.~{Kok}.
\newblock {Quantum Fisher information for general spatial deformations of
  quantum emitters}.
\newblock arXiv:1802.01601, 2018.

\bibitem{sammy2016compatibility}
S.~Ragy, M.~Jarzyna, and R.~Demkowicz-Dobrza\ifmmode~\acute{n}\else
  \'{n}\fi{}ski.
\newblock Compatibility in multiparameter quantum metrology.
\newblock {\em Physical Review A}, 94:052108, 2016.

\bibitem{jarzyna2012}
M.~Jarzyna and R.~Demkowicz-Dobrza\ifmmode~\acute{n}\else \'{n}\fi{}ski.
\newblock Quantum interferometry with and without an external phase reference.
\newblock {\em Physical Review A}, 85:011801, 2012.

\bibitem{spagnolo2012}
N.~Spagnolo, L.~Aparo, C.~Vitelli, A.~Crespi, R.~Ramponi, R.~Osellame,
  P.~Mataloni, and F.~Sciarrino.
\newblock Quantum interferometry with three-dimensional geometry.
\newblock {\em Scientific Reports}, 2(1), 2012.

\bibitem{berry2000}
D.~W. Berry and H.~M. Wiseman.
\newblock Optimal states and almost optimal adaptive measurements for quantum
  interferometry.
\newblock {\em Physical Review Letters}, 85:5098--5101, 2000.

\bibitem{esteban2017}
E.~Mart\'{\i}nez-Vargas, C.~Pineda, F.~Leyvraz, and P.~Barberis-Blostein.
\newblock Quantum estimation of unknown parameters.
\newblock {\em Physical Review A}, 95:012136, 2017.

\bibitem{kolodynski2014}
J.~Ko\l{}ody\ifmmode~\acute{n}\else \'{n}\fi{}ski.
\newblock {\em {Precision bounds in noisy quantum metrology}}.
\newblock PhD thesis, University of Warsaw, arXiv:1409.0535, 2014.

\bibitem{jarzyna2016thesis}
M.~{Jarzyna}.
\newblock {\em {Phase Coherence in Quantum Metrology and Communication}}.
\newblock PhD thesis, University of Warsaw, 2016.

\bibitem{berry2015}
D.~W. Berry, M.~Tsang, M.~J.~W. Hall, and H.~M. Wiseman.
\newblock {Quantum Bell-Ziv-Zakai Bounds and Heisenberg Limits for Waveform
  Estimation}.
\newblock {\em Physical Review X}, 5:031018, 2015.

\bibitem{demkowicz2011}
R.~Demkowicz-Dobrza\ifmmode~\acute{n}\else \'{n}\fi{}ski.
\newblock Optimal phase estimation with arbitrary a priori knowledge.
\newblock {\em Physical Review A}, 83:061802, 2011.

\bibitem{durkin2007}
G.~A. Durkin and J.~P. Dowling.
\newblock Local and global distinguishability in quantum interferometry.
\newblock {\em Physical Review Letters}, 99:070801, 2007.

\bibitem{li2018}
Y.~Li, L.~{Pezz{\`e}}, M.~Gessner, Z.~Ren, W.~Li, and A.~Smerzi.
\newblock {Frequentist and Bayesian Quantum Phase Estimation}.
\newblock {\em Entropy}, 20(9), 2018.

\bibitem{mathematics2004}
K.~F. Riley, M.~P. Hobson, and S.~J. Bence.
\newblock {\em {Mathematical methods for physics and engineering}}.
\newblock Cambridge University Press, 2004.

\bibitem{macieszczak2014bayesian}
K.~Macieszczak, M.~Fraas, and R.~Demkowicz-Dobrza\ifmmode~\acute{n}\else
  \'{n}\fi{}ski.
\newblock Bayesian quantum frequency estimation in presence of collective
  dephasing.
\newblock {\em New Journal of Physics}, 16(11):113002, 2014.

\bibitem{haase2018jul}
J.~F. {Haase}, A.~{Smirne}, S.~F. {Huelga}, J.~Ko\l{}ody\ifmmode~\acute{n}\else
  \'{n}\fi{}ski, and R.~Demkowicz-Dobrza\ifmmode~\acute{n}\else \'{n}\fi{}ski.
\newblock {Precision Limits in Quantum Metrology with Open Quantum Systems}.
\newblock {\em Quantum Measurements and Quantum Metrology}, 5:13--39, 2018.

\bibitem{jarzyna2015true}
M.~Jarzyna and R.~Demkowicz-Dobrza\ifmmode~\acute{n}\else \'{n}\fi{}ski.
\newblock True precision limits in quantum metrology.
\newblock {\em New Journal of Physics}, 17(1):013010, 2015.

\bibitem{rivas2012}
A.~Rivas and A.~Luis.
\newblock {Sub-Heisenberg estimation of non-random phase shifts}.
\newblock {\em New Journal of Physics}, 14(9):093052, 2012.

\bibitem{zhang2013}
Y.~R. Zhang, G.~R. Jin, J.~P. Cao, W.~M. Liu, and H.~Fan.
\newblock {Unbounded quantum Fisher information in two-path interferometry with
  finite photon number}.
\newblock {\em Journal of Physics A: Mathematical and Theoretical},
  46(3):035302, 2013.

\bibitem{giovannetti2012subheisenberg}
V.~Giovannetti and L.~Maccone.
\newblock {Sub-Heisenberg Estimation Strategies Are Ineffective}.
\newblock {\em Physical Review Letters}, 108:210404, 2012.

\bibitem{berry2012infinite}
D.~W. Berry, M.~J.~W. Hall, M.~Zwierz, and H.~M. Wiseman.
\newblock {Optimal Heisenberg-style bounds for the average performance of
  arbitrary phase estimates}.
\newblock {\em Physical Review A}, 86:053813, 2012.

\bibitem{pezze2013}
L.~{Pezz{\`e}}.
\newblock {Sub-Heisenberg phase uncertainties}.
\newblock {\em Physical Review A}, 88:060101, 2013.

\bibitem{alfredo2017}
A.~Luis.
\newblock {Breaking the weak Heisenberg limit}.
\newblock {\em Physical Review A}, 95:032113, 2017.

\bibitem{bayesbounds2007}
H.~L. van Trees and K.~L. Bell.
\newblock {\em {Bayesian Bounds for Parameter Estimation and Nonlinear
  Filtering/Tracking}}.
\newblock Jonh Wiley \& Sons, Inc., 2007.

\bibitem{vaart1998}
A.~W. van~der Vaart.
\newblock {\em Asymptotic statistics}.
\newblock Cambridge University Press, 1998.

\bibitem{BraunsteinCaves1994}
S.~L. Braunstein and C.~M. Caves.
\newblock Statistical distance and the geometry of quantum states.
\newblock {\em Physical Review Letters}, 72:3439--3443, 1994.

\bibitem{helstrom1967mmse}
C.~W. Helstrom.
\newblock Minimum mean-squared error of estimates in quantum statistics.
\newblock {\em Physics Letters A}, 25(2):101 -- 102, 1967.

\bibitem{genoni2008}
M.~G. Genoni, P.~Giorda, and M.~G.~A. Paris.
\newblock Optimal estimation of entanglement.
\newblock {\em Physical Review A}, 78:032303, 2008.

\bibitem{pezze2014}
L.~{Pezz{\`e}} and A.~Smerzi.
\newblock Quantum theory of phase estimation.
\newblock In G.M. Tino and M.A. Kasevich, editors, {\em Atom Interferometry,
  Proceedings of the International School of Physics Enrico Fermi, Course 188,
  Varenna}, page 691, IOS Press, Amsterdam, 2014. Springer Berlin Heidelberg.

\bibitem{pezze2017simultaneous}
L.~{Pezz{\`e}}, M.~A. Ciampini, N.~Spagnolo, P.~C. Humphreys, A.~Datta, I.~A.
  Walmsley, M.~Barbieri, F.~Sciarrino, and A.~Smerzi.
\newblock {Optimal Measurements for Simultaneous Quantum Estimation of Multiple
  Phases}.
\newblock {\em Physical Review Letters}, 119:130504, 2017.

\bibitem{albarelli2019}
F.~Albarelli, J.~F. Friel, and A.~Datta.
\newblock {Evaluating the Holevo Cram\'er-Rao Bound for Multiparameter Quantum
  Metrology}.
\newblock {\em Physical Review Letters}, 123:200503, 2019.

\bibitem{gill2011}
R.~D. Gill and M.~I. Gu\c{t}\u{a}.
\newblock {\em On asymptotic quantum statistical inference}, volume~9 of {\em
  Collections}, pages 105--127.
\newblock Institute of Mathematical Statistics, Beachwood, Ohio, USA, 2013.

\bibitem{fraser1964}
D.~A.~S. Fraser.
\newblock On local unbiased estimation.
\newblock {\em Journal of the Royal Statistical Society. Series B
  (Methodological)}, 26(1):46--51, 1964.

\bibitem{hall2012}
M.~J.~W. Hall and H.~M. Wiseman.
\newblock {Does Nonlinear Metrology Offer Improved Resolution? Answers from
  Quantum Information Theory}.
\newblock {\em Physical Review X}, 2:041006, 2012.

\bibitem{lecam1986}
L.~M. Le~Cam.
\newblock {\em Asymptotic methods in statistical decision theory}.
\newblock Springer, 1986.

\bibitem{gill1995}
R.~D. Gill and B.~Y. Levit.
\newblock {Applications of the van Trees inequality: A Bayesian Cram\'{e}r-Rao
  Bound}.
\newblock {\em Bernoulli}, 1(1/2):59--79, 1995.

\bibitem{zhang2014}
Y.-R. Zhang and H.~Fan.
\newblock Quantum metrological bounds for vector parameters.
\newblock {\em Physical Review A}, 90:043818, 2014.

\bibitem{personick1971}
S.~Personick.
\newblock Application of quantum estimation theory to analog communication over
  quantum channels.
\newblock {\em IEEE Transactions on Information Theory}, 17(3):240--246, 1971.

\bibitem{hansen2003}
F.~Hansen and G.~K. Pedersen.
\newblock {Jensen's Operator Inequality}.
\newblock {\em Bulletin of the London Mathematical Society}, 35(4):553--564,
  2003.

\bibitem{jesus2017}
Jes\'{u}s Rubio, P.~A. Knott, and J.~A. Dunningham.
\newblock {Non-asymptotic analysis of quantum metrology protocols beyond the
  Cram\'{e}r-Rao bound}.
\newblock {\em Journal of Physics Communications}, 2(1):015027, 2018.

\bibitem{personick1969thesis}
S.~Personick.
\newblock {\em Efficient analog communication over quantum channels}.
\newblock PhD thesis, Dep. Elec. Eng., M.I.T., Cambridge, Mass., 1969.

\bibitem{mashide2002}
M.~Sasaki, M.~Ban, and S.~M. Barnett.
\newblock {Optimal parameter estimation of a depolarizing channel}.
\newblock {\em Physical Review A}, 66:022308, 2002.

\bibitem{sekatski2017}
P.~Sekatski, M.~Skotiniotis, and W.~D\"ur.
\newblock {Improved Sensing with a Single Qubit}.
\newblock {\em Physical Review Letters}, 118:170801, 2017.

\bibitem{bernad2018}
J.~Z. Bern\'ad, C.~Sanavio, and A.~Xuereb.
\newblock Optimal estimation of the optomechanical coupling strength.
\newblock {\em Physical Review A}, 97:063821, 2018.

\bibitem{chabuda2016allanvariance}
K.~Chabuda, I.~D. Leroux, and R.~Demkowicz-Dobrza\ifmmode~\acute{n}\else
  \'{n}\fi{}ski.
\newblock {The quantum Allan variance}.
\newblock {\em New Journal of Physics}, 18(8):083035, 2016.

\bibitem{helstrom1968multiparameter}
C.~Helstrom.
\newblock The minimum variance of estimates in quantum signal detection.
\newblock {\em IEEE Transactions on Information Theory}, 14(2):234--242, 1968.

\bibitem{braunstein1996}
S.~L. Braunstein, C.~M. Caves, and G.~J. Milburn.
\newblock {Generalized Uncertainty Relations: Theory, Examples, and Lorentz
  Invariance}.
\newblock {\em Annals of Physics}, 247(1):135--173, 1996.

\bibitem{jesus2019b}
Jes\'{u}s Rubio and J.~Dunningham.
\newblock {Bayesian multi-parameter quantum metrology with limited data}.
\newblock arXiv:1906.04123, 2019.

\bibitem{dowling2008}
J.~P. Dowling.
\newblock {Quantum optical metrology - the lowdown on high-N00N states}.
\newblock {\em Contemporary Physics}, 49(2):125--143, 2008.

\bibitem{gerry2010}
C.~C. Gerry and J.~Mimih.
\newblock The parity operator in quantum optical metrology.
\newblock {\em Contemporary Physics}, 51(6):497--511, 2010.

\bibitem{chiruvelli2011}
A.~Chiruvelli and H.~Lee.
\newblock Parity measurements in quantum optical metrology.
\newblock {\em Journal of Modern Optics}, 58(11):945--953, 2011.

\bibitem{sahota2016}
J.~Sahota, N.~Quesada, and D.~F.~V. James.
\newblock Physical resources for optical phase estimation.
\newblock {\em Physical Review A}, 94:033817, 2016.

\bibitem{lee2016}
S.-Y. Lee, C.-W. Lee, J.~Lee, and H.~Nha.
\newblock Quantum phase estimation using path-symmetric entangled states.
\newblock {\em Scientific Reports}, 6:30306, 2016.

\bibitem{smirne2018}
A.~Smirne, A.~Lemmer, M.~B. Plenio, and S.~F. Huelga.
\newblock Improving the precision of frequency estimation via long-time
  coherences.
\newblock {\em Quantum Science and Technology}, 4(2):025004, 2019.

\bibitem{haase2018may}
J.~F. Haase, A.~Smirne, J.~Ko\l{}ody\ifmmode~\acute{n}\else \'{n}\fi{}ski,
  R.~Demkowicz-Dobrza\ifmmode~\acute{n}\else \'{n}\fi{}ski, and S.~F. Huelga.
\newblock Fundamental limits to frequency estimation: a comprehensive
  microscopic perspective.
\newblock {\em New Journal of Physics}, 20(5):053009, 2018.

\bibitem{braunstein_maxlikelihood1992}
S.~L. Braunstein, A.~S. Lane, and C.~M. Caves.
\newblock Maximum-likelihood analysis of multiple quantum phase measurements.
\newblock {\em Physical Review Letters}, 69:2153--2156, 1992.

\bibitem{cox2000}
D.~R. Cox and D.~V. Hinkley.
\newblock {\em {Theoretical Statistics}}.
\newblock Chapman \& Hall, 2000.

\bibitem{jaynes1968}
E.~T. Jaynes.
\newblock Prior probabilities.
\newblock {\em IEEE Transactions on Systems and Cybernetics}, 4(3):227--241,
  1968.

\bibitem{toussaint2011}
U.~von Toussaint.
\newblock Bayesian inference in physics.
\newblock {\em Reviews of Modern Physics}, 83:943--999, 2011.

\bibitem{friis2017}
N.~Friis, D.~Orsucci, M.~Skotiniotis, P.~Sekatski, V.~Dunjko, H.~J. Briegel,
  and W.~D\"{u}r.
\newblock Flexible resources for quantum metrology.
\newblock {\em New Journal of Physics}, 19(6):063044, 2017.

\bibitem{kass1996}
R.~E. Kass and L.~Wasserman.
\newblock {The Selection of Prior Distributions by Formal Rules}.
\newblock {\em Journal of the American Statistical Association},
  91(435):1343--1370, 1996.

\bibitem{numerics2014matlab}
A.~Quarteorni, F.~Saleri, and P.~Gervasio.
\newblock {\em {Scientific Computing with MATLAB and Octave}}.
\newblock Springer, 2014.

\bibitem{knott2016}
P.~A. Knott.
\newblock A search algorithm for quantum state engineering and metrology.
\newblock {\em New Journal of Physics}, 18(7):073033, 2016.

\bibitem{jesus2018dec}
R.~Nichols, L.~Mineh, Jes{\'{u}}s Rubio, J.~C.~F. Matthews, and P.~A. Knott.
\newblock Designing quantum experiments with a genetic algorithm.
\newblock {\em Quantum Science and Technology}, 4(4):045012, 2019.

\bibitem{bhatia1997}
R.~Bhatia.
\newblock {\em Matrix analysis}.
\newblock Graduate texts in mathematics; 169. Springer, New York, N.Y., 1997.

\bibitem{horn1985}
R.~A. Horn.
\newblock {\em Matrix analysis}.
\newblock Cambridge University Press, 1985.

\bibitem{alfredo1996}
A.~Luis and J.~Pe{\v{r}}ina.
\newblock Optimum phase-shift estimation and the quantum description of the
  phase difference.
\newblock {\em Physical Review A}, 54:4564--4570, 1996.

\bibitem{schafermeier2018}
C.~Sch\"{a}fermeier, M.~Je\v{z}ek, L.~S. Madsen, T.~Gehring, and U.~L.
  Andersen.
\newblock Deterministic phase measurements exhibiting super-sensitivity and
  super-resolution.
\newblock {\em Optica}, 5(1):60--64, Jan 2018.

\bibitem{weinstein1988}
E.~Weinstein and A.~J. Weiss.
\newblock A general class of lower bounds in parameter estimation.
\newblock {\em IEEE Transactions on Information Theory}, 34(2):338--342, 1988.

\bibitem{driscoll2019}
L.~O'Driscoll, R.~Nichols, and P.~A. Knott.
\newblock {A hybrid machine learning algorithm for designing quantum
  experiments}.
\newblock {\em Quantum Machine Intelligence}, 2019.

\bibitem{adaquantum2019}
AdaQuantum on~GitHub.
\newblock https://github.com/paulk444/AdaQuantum, 2019.

\bibitem{jesus2018}
Jes\'{u}s Rubio and J.~A. Dunningham.
\newblock Quantum metrology in the presence of limited data.
\newblock {\em New Journal of Physics}, 21(4):043037, 2019.

\bibitem{chiribella2012}
G.~Chiribella.
\newblock Optimal networks for quantum metrology: semidefinite programs and
  product rules.
\newblock {\em New Journal of Physics}, 14(12):125008, 2012.

\bibitem{gagatos2016gaussian}
C.~N. Gagatsos, D.~Branford, and A.~Datta.
\newblock Gaussian systems for quantum-enhanced multiple phase estimation.
\newblock {\em Physical Review A}, 94:042342, 2016.

\bibitem{altenburg2017}
S.~Altenburg, M.~Oszmaniec, S.~W\"{o}lk, and O.~G\"uhne.
\newblock Estimation of gradients in quantum metrology.
\newblock {\em Physical Review A}, 96:042319, 2017.

\bibitem{hall2018}
M.~J.~W. Hall.
\newblock {Entropic Heisenberg limits and uncertainty relations from the Holevo
  information bound}.
\newblock {\em Journal of Physics A: Mathematical and Theoretical},
  51(36):364001, 2018.

\bibitem{sekatski2019}
P.~{Sekatski}, S.~{W{\"o}lk}, and W.~{D{\"u}r}.
\newblock {Optimal distributed sensing in noisy environments}.
\newblock arXiv:1905.06765, 2019.

\bibitem{polino2018}
E.~Polino, M.~Riva, M.~Valeri, R.~Silvestri, G.~Corrielli, A.~Crespi,
  N.~Spagnolo, R.~Osellame, and F.~Sciarrino.
\newblock Experimental multiphase estimation on a chip.
\newblock {\em Optica}, 6(3):288--295, 2019.

\bibitem{roccia2018}
E.~Roccia, V.~Cimini, M.~Sbroscia, I.~Gianani, L.~Ruggiero, L.~Mancino, M.~G.
  Genoni, M.~Antonietta Ricci, and M.~Barbieri.
\newblock Multiparameter approach to quantum phase estimation with limited
  visibility.
\newblock {\em Optica}, 5(10):1171--1176, 2018.

\bibitem{li2019}
X.~Li, J.-H. Cao, Q.~Liu, M.~K. Tey, and L.~You.
\newblock {Multi-parameter Estimation with Multi-mode Ramsey Interferometry}.
\newblock arXiv:1905.01673, 2019.

\bibitem{gatto2019}
D.~Gatto, P.~Facchi, F.~A. Narducci, and V.~Tamma.
\newblock Distributed quantum metrology with a single squeezed-vacuum source.
\newblock {\em Physical Review Research}, 1:032024, 2019.

\bibitem{kok2017}
P.~Kok, J.~Dunningham, and J.~F. Ralph.
\newblock Role of entanglement in calibrating optical quantum gyroscopes.
\newblock {\em Physical Review A}, 95:012326, 2017.

\bibitem{jesus2019a}
Jes\'{u}s Rubio, P.~A. Knott, T.~J. Proctor, and J.~A. Dunningham.
\newblock {Quantum sensing networks for the estimation of linear functions, in
  preparation}, 2020.

\bibitem{yuen1973}
H.~Yuen and M.~Lax.
\newblock Multiple-parameter quantum estimation and measurement of
  nonselfadjoint observables.
\newblock {\em IEEE Transactions on Information Theory}, 19(6):740--750, 1973.

\bibitem{gellmann1962}
M.~Gell-Mann.
\newblock Symmetries of baryons and mesons.
\newblock {\em Physical Review}, 125:1067--1084, 1962.

\bibitem{dorner2009}
U.~Dorner, R.~Demkowicz-Dobrza\ifmmode~\acute{n}\else \'{n}\fi{}ski, B.~J.
  Smith, J.~S. Lundeen, W.~Wasilewski, K.~Banaszek, and I.~A. Walmsley.
\newblock {Optimal Quantum Phase Estimation}.
\newblock {\em Physical Review Letters}, 102:040403, 2009.

\bibitem{iwo1993}
I.~Bialynicki-Birula, M.~Freyberger, and W.~Schleich.
\newblock {Various measures of quantum phase uncertainty: a comparative study}.
\newblock {\em Physica Scripta}, 1993(T48):113, 1993.

\bibitem{jizba2016}
P.~Jizba, Y.~Ma, A.~Hayes, and J.~A. Dunningham.
\newblock One-parameter class of uncertainty relations based on entropy power.
\newblock {\em Physical Review E}, 93:060104, 2016.

\bibitem{jizba2017}
P.~Jizba, A.~Hayes, and J.~A. Dunningham.
\newblock New class of entropy-power-based uncertainty relations.
\newblock {\em Journal of Physics: Conference Series}, page 012054, 2017.

\bibitem{martinezvargas2019}
E.~{Mart{\'\i}nez-Vargas}, C.~{Pineda}, and P.~{Barberis-Blostein}.
\newblock {Quantum measurement optimization by decomposition of measurements
  into extremals}.
\newblock arXiv:1901.06179, 2019.

\end{thebibliography}

\appendix
\chapter{Supplemental material}
\label{app:supplemental}

\section{Other measures of uncertainty in estimation theory}
\label{sec:otheruncertainty}

The fact that different measures of uncertainty inform us about different aspects of some estimation problem is well known, both in classical \citep{jaynes2003} and quantum \cite{iwo1993} scenarios. Among all the available options, in the literature of quantum metrology one typically finds a clear distinction between frequentist and Bayesian uncertainties \cite{rafal2015, li2018}, which in practice are associated, respectively, with a high amount of prior information, in which case we say that they are \emph{local}, and with a low amount of prior knowledge, meaning that they are \emph{global} \cite{paris2009, haase2018jul}. However, we can also find studies including both local and global tools without introducing the former distinction \cite{hall2012}. In addition, it is common to associate the idea of fixed but unknown parameters with the frequentist approach, while Bayesian metrology is seen as if we were giving a random description of such parameters \cite{rafal2015}. Nevertheless, Bayesian uncertainties also admit parameters that are fixed but unknown, as our discussion in section \ref{sec:uncertainty} and the work in \cite{li2018} show.   

In view of this, a more transparent picture of the different types of uncertainty and how they should be used in metrology has a great potential in terms of establishing meaningful comparisons between optimal protocols. This is precisely one of the key advantages of the three-step method to construct uncertainties that we have proposed in the main text\footnote{Nonetheless, note that our selection of uncertainties in section \ref{sec:uncertainty} has been motivated by the practical requirements of our problem, and it does not include important alternatives such as the use of entropic uncertainty relations \cite{jizba2016, jizba2017, hall2018}.}. 

Interestingly, shortly after the development of our three-step construction (which we originally published in \cite{jesus2017} in the context of the algorithm in section \ref{subsec:numalgorithm}), a related approach was proposed by Li \emph{et al.} \cite{li2018}\footnote{However, both our work \cite{jesus2017} and \cite{li2018} are based on the square error and a single parameter, while the discussion in section \ref{sec:uncertainty} is more general.}. The authors of \cite{li2018} classified the uncertainties in terms of frequentist and Bayesian quantities, and, within each group, in terms of random and fixed parameters. This allowed them to analyse the mathematical relationships between uncertainties, and to establish which bounds are satisfied by different errors. 

The errors that Li \emph{et al.} identify for Bayesian metrology with fixed parameters are equivalent to those that we have presented in section \ref{sec:uncertainty} (and in our work \cite{jesus2017}). The other groups are useful to understand the origin of the paradoxes that can emerge when bounds that are only valid for certain quantities are misapplied \cite{li2018}. Still, notwithstanding the merits of this extended classification, it can be argued that our three fundamental categories provide a simpler perspective without a practical loss of generality.

For example, let us take the case of random parameters. In these schemes the experiment is repeated $\nu$ times, such that we make $\mu$ observations per repetition, and the random component appears because the unknown parameters can have different values in each repetition \cite{li2018}. The deviation function in this situation is \cite{martinezvargas2019}
\begin{equation}
\frac{1}{\nu} \sum_{i = 1}^\nu \mathcal{D}\left[\boldsymbol{g}(\boldsymbol{m}_i), \boldsymbol{\theta}_i \right].
\end{equation}
According to our discussion, a theorist that is designing the experiment needs a probability with information about outcomes and parameters. Suppose we have the distribution of the different values that the random parameters can acquire. The parameters may change when we rerun the experiment, but they remain fixed while we are generating the measurement outcomes $\boldsymbol{m}_i$ during the $i$-th repetition \cite{li2018}. If $\boldsymbol{\theta}_i$ represents the fixed but unknown values of that trial, then we know how likely is the appearance of each of the possible values that they could have acquired in that particular iteration, and thus we can encode the information about their random distribution in the prior $p(\boldsymbol{\theta}_i)$, provided that nothing else is known. Combining this with the likelihood $p(\boldsymbol{m}_i|\boldsymbol{\theta}_i)$, and noticing that the previous argument is identical for all the iterations of the experiment, we find the error
\begin{equation}
\frac{1}{\nu} \sum_{i = 1}^\nu \int d\boldsymbol{\theta}_i d\boldsymbol{m}_i ~ p(\boldsymbol{\theta}_i, \boldsymbol{m}_i) ~\mathcal{D}\left[\boldsymbol{g}(\boldsymbol{m}_i), \boldsymbol{\theta}_i \right].
\label{randomerr}
\end{equation}
Finally, by noting that the previous expression sums the same numerical uncertainty $\nu$ times, we conclude that equation (\ref{randomerr}) is formally identical to equation (\ref{errthe}). That is, we arrive to the known result that we can model either random or fixed but unknown parameters with the same mathematics in the context of a theoretical study, even when they are physically different situations. The crucial observation is that the previous analysis is simply a combination of $\nu$ scenarios, each of them belonging to our third type of uncertainty in section \ref{sec:uncertainty}.  

The case of frequentist uncertainties is more subtle. Frequentist metrology is based on the error \cite{paris2009, rafal2015, li2018}
\begin{equation}
\int d\boldsymbol{m}~p(\boldsymbol{m}|\boldsymbol{\theta}) ~\mathcal{D}[\boldsymbol{g}(\boldsymbol{m}),\boldsymbol{\theta}].
\label{freqerror}
\end{equation}
As noted in \cite{jaynes2003}, this is the error to be employed when we do not have access to a set of specific outcomes but we do know $\boldsymbol{\theta}$, which in principle is a different type of problem. However, it can be shown \cite{hall2012, rafal2015, kolodynski2014, haase2018jul} that if certain assumptions are fulfilled in a local region of the parameter domain, often combined with an asymptotic requirement of many repeats or copies, then the quantity in equation (\ref{freqerror}) can be made useful for parameter estimation in a wide range of practical cases. Its key advantage is that the related calculations are generally more tractable than the alternatives, but, at the same time, it can be argued that it is also physically unsatisfactory. To see why, note that $p(\boldsymbol{\theta})$ does not appear in equation (\ref{freqerror}), and yet knowledge about the local region of interest is precisely the type of prior information that is best represented with a prior probability. 

Furthermore, our calculations in previous chapters strongly suggest that, for metrology protocols, the local regime emerges naturally from the Bayesian error in equation (\ref{errthe}) whenever the appropriate conditions are fulfilled, which is in agreement with other studies that also connect Bayesian and non-Bayesian quantities in quantum metrology \cite{jarzyna2015true}. From a formal perspective, this behaviour is a more general feature of estimation problems, and it is not limited to metrology \cite{gill2011}. In other words, we can recover the same local simplicity without sacrificing conceptual consistency and rigour, and thus there is no need to switch frameworks and use equation (\ref{freqerror}) even if we only want to work in that regime. These are our reasons to exclude equation (\ref{freqerror}) from the list of basic measures of uncertainty, a choice that differs from the path generally followed \cite{paris2009, rafal2015, li2018} (with exceptions such as the work in \cite{hall2012}). 

A final question is whether some of these uncertainties could be associated with the errors that are directly measured in the laboratory \cite{li2018}. Since our quantities are constructed out of probabilities, by virtue of the law of large numbers we know that a necessary condition is to have access to a very large amount of measurements, provided that the probabilities describe repetitive experiments. We saw an example of this when we revisited the use of the error propagation formula for phase estimation in a Mach-Zehnder interferometer (see section \ref{subsec:optint}). Otherwise, the previous quantities cannot be experimentally accessed, and they merely summarise information based on either our theoretical analysis (equations (\ref{errsim}) and (\ref{errthe})) or on empirical outcomes (equation (\ref{errexp})). The regime of limited data involves, by definition, scenarios with a low number of measurements; consequently, while our results could be implemented in practice, the uncertainties involved in their design are of a theoretical nature.   

In conclusion, it may be argued that the uncertainties examined in the classification of section \ref{sec:uncertainty} can be sufficient to accommodate a wide range of practical scenarios, including not only the cases with random parameters or a low amount of prior information, but also those with a high amount of prior knowledge (i.e., that work in the local regime) or that involve fixed parameters.

\section{How large the prior width can be such that the use of a quadratic error is justified?}
\label{prior_sinapprox_appendix}

A good experiment should be arranged such that the uncertainty $\bar{\epsilon}$ decreases as a function of the number of observations $\mu$. If the parameter to be estimated is periodic and we use the sine error 
\begin{equation}
\bar{\epsilon} = 4\int d\theta d\boldsymbol{m} ~p(\theta,m)~\mathrm{sin}^2\left[\frac{g(\boldsymbol{m})-\theta}{2} \right]
\label{sinerrcomplete}
\end{equation} 
(see section \ref{sec:uncertainty}), then the former statement implies that the greatest value acquired by $\bar{\epsilon}$ in equation (\ref{sinerrcomplete}) is given by 
\begin{equation}
\bar{\epsilon}({\mu=0}) = 4\int d\theta ~p(\theta)~\mathrm{sin}^2\left(\frac{g - \theta}{2} \right).
\label{prior_sinerr}
\end{equation}
Furthermore, equation \ref{prior_sinerr} is simplified as
\begin{equation}
\bar{\epsilon}({\mu=0}) = \frac{4}{W_0}\int_0^{W_0} d\theta ~\mathrm{sin}^2\left(\frac{g - \theta}{2} \right)
\label{priorsinerrsimple}
\end{equation}
after using the uniform prior in equation (\ref{prior_probability}) with $\bar{\theta} = W_0/2$, and the minimum of equation (\ref{priorsinerrsimple}) is achieved when the estimator $g$ satisfies $\mathrm{cos}(g - W_0) = \mathrm{cos}(g)$, which for one period implies that $g = W_0 /2$. Hence,  
\begin{eqnarray}
\bar{\epsilon}(\mu=0) = \frac{4}{W_0}\int_0^{W_0} d\theta~ \mathrm{sin}^2\left(\frac{W_0}{4} - \frac{\theta}{2} \right)
= 2\left[1 - \frac{2}{W_0}\mathrm{sin}\left( \frac{W_0}{2}\right) \right].
\label{sin_muzero}
\end{eqnarray}
If we now expand equation (\ref{sin_muzero}) up to second order in $W_0$, we find that
\begin{equation}
\bar{\epsilon}(\mu=0) \approx \frac{{W_0}^2}{12},
\label{mse_muzero}
\end{equation}  
which is the prior that we would have found using the square error directly. 

\begin{figure}[t]
\centering
\includegraphics[trim={0.1cm 0.1cm 1.4cm 0.2cm},clip,width=7.45cm]{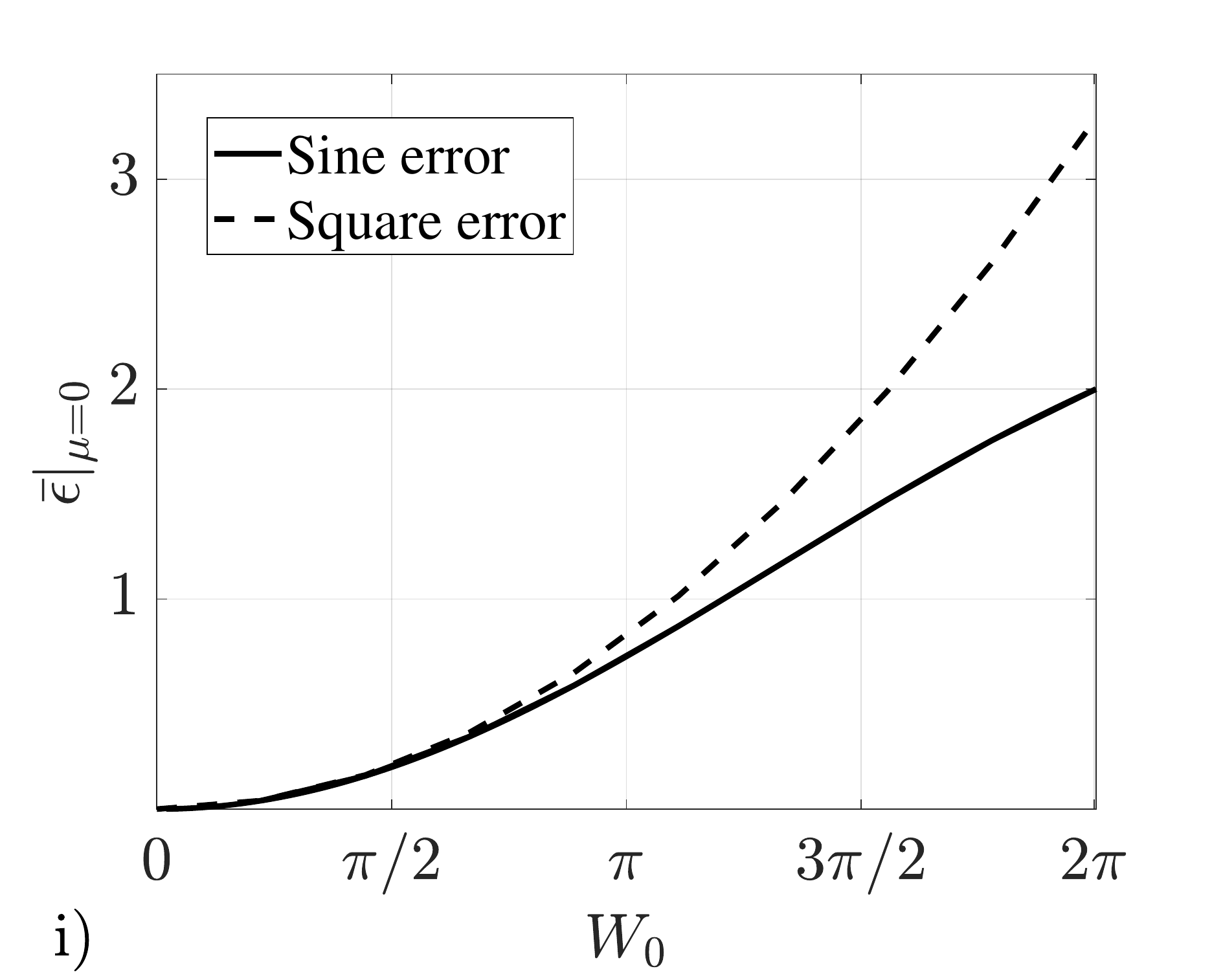}\includegraphics[trim={0.1cm 0.1cm 0.65cm 0.2cm},clip,width=7.75cm]{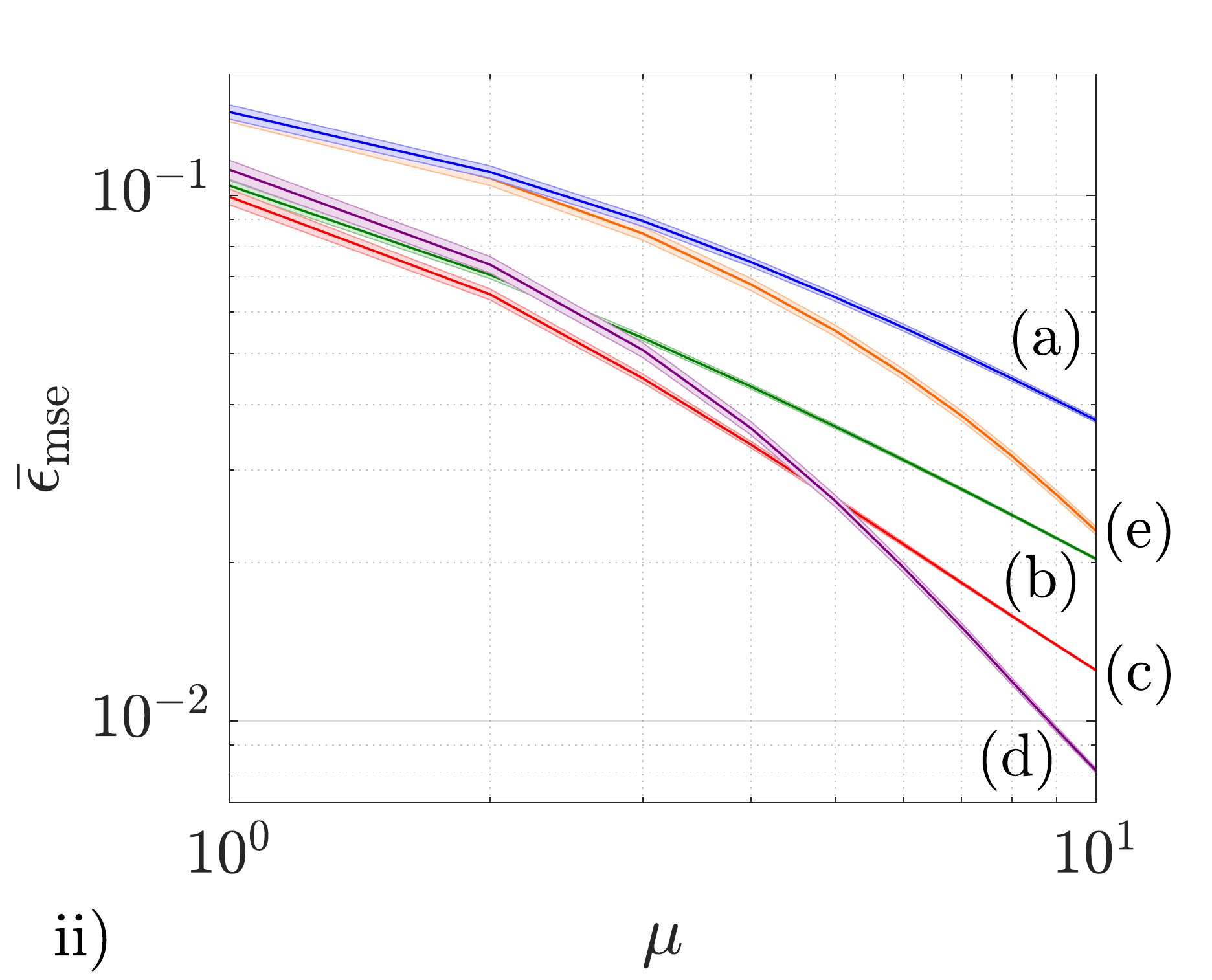}
	\caption[Square error as an approximation for a periodic uncertainty]{i) Comparison between the prior uncertainty ($\mu = 0$) given by a periodic error function and that associated to the mean square error as a function of $W_0$. Most of our results in Section \ref{results} are calculated using the values $W_0 = \pi/2$ and $W_0 = \pi/3$; (ii) Mean square error based on the optimal single-shot strategy (solid line) and bounds for the approximation error after having expanded the sine error up to second order (shaded area) for (a) the coherent state, (b) the NOON state, (c) the twin squeezed vacuum state, (d) the squeezed entangled state, and (e) the twin squeezed cat state, with $\bar{n} = 2$, $\bar{\theta}=0$ and $W_0 = \pi/2$. This figure shows that the mean square error is a suitable approximation for the mean sine error when we are in the regime of moderate prior knowledge.}
\label{circular_mse}
\end{figure}

According to figure \ref{circular_mse}.i, which compares equations (\ref{sin_muzero}) and (\ref{mse_muzero}) as a function of the width $W_0$, the approximation starts to fail in a notable way when $W_0 \approx \pi$. Given that in chapter \ref{chap:nonasymptotic} we calculated the mean square error for NOON, twin squeezed vacuum and squeezed entangled states with $W_0=\pi/2$, and that $W_0=\pi/3$ was employed with both the previous states and for a coherent beam, we can say that the approximation is reasonable for these configurations when $\mu = 0$. Moreover, $\abs{g(\boldsymbol{m}) - \theta}$ will not be greater than $W_0$ for $\mu > 0$, and thus a similar reasoning could be applied to the comparison of equations (\ref{sinerrcomplete}) and (\ref{errwork}). The only scheme for which this approximation is cruder is the coherent state with prior width $W_0 = \pi$. 

A more powerful argument to refine such threshold is also possible. First we observe that the approximation $4\hspace{0.15em}\mathrm{sin}^2\lbrace\left[g(\boldsymbol{m})-\theta\right]/2 \rbrace \approx [g(\boldsymbol{m})-\theta]^2$ relies on the quantity $|g(\boldsymbol{m})-\theta|/2$ being small, where, as we have seen, $|g(\boldsymbol{m})-\theta|/2 \leqslant W_0/2$. The minimum requirement that is natural to impose is that the variable for which the Taylor expansion is calculated (i.e., $|g(\boldsymbol{m})-\theta|/2$) is slightly smaller than $1$ at most, which is always the case if the width of our experiment satisfies that $W_0 \lesssim 2$. In principle this would still be a crude approximation if we were interested in the sine function itself. However, the sine error is then integrated over all the possible values for $\theta$ and $\boldsymbol{m} = (m_1, \dots , m_\mu)$. This implies that $|g(\boldsymbol{m})-\theta|/2 \sim 2$ when $W_0\sim2$ only for a few combinations of values, and the weight of those cases will decrease as the joint probability $p(\theta,\boldsymbol{m})$ accumulates more data. We conclude then that $W_0 \lesssim 2$ is a reasonable estimation for the range of validity of the mean square error in a problem with a periodic parameter. Note that this condition has the same order of magnitude than the estimation found in \cite{friis2017}, where the authors argued that the width of their Gaussian prior had to be $\pi/2$ or less, and it is a better estimation than the one obtained in the previous paragraphs. 

From the previous discussion we see that only the calculation of the first few shots could be potentially misleading if we use the mean square error. To show that this is not the case for the type of schemes analysed in the main text, let us estimate explicitly the error of the Taylor expansion for some of the schemes in chapter \ref{chap:limited}. First, using Taylor's theorem we have that $\mathrm{sin}^2(x) = x^2-x^4 \mathrm{cos}(2\varepsilon)/3$, where $\varepsilon \in [0,x]$  \cite{mathematics2004}. The first term is the approximation that we want to use, while the second term represents the error of this approximation. Using the fact that the cosine is bounded between $-1$ and $1$, the Taylor error can be estimated with 
\begin{equation}
\Delta \bar{\epsilon} = \frac{1}{12}\int d\theta d\boldsymbol{m}~p(\theta) p(\boldsymbol{m}|\theta) \left[g(\boldsymbol{m})-\theta\right]^4,
\label{taylorerror}
\end{equation}
and knowing that the optimal phase estimator is the average of the posterior probability $p(\theta|\boldsymbol{m}) \propto p(\theta)p(\boldsymbol{m}|\theta)$, we can rewrite equation (\ref{taylorerror}) as
\begin{equation}
\Delta \bar{\epsilon} = \frac{1}{12}\int d\theta' p(\theta') \int d\boldsymbol{m}~ p(\boldsymbol{m}|\theta') \Delta \bar{\epsilon}(\boldsymbol{m}),
\end{equation}
where
\begin{equation}
\Delta \bar{\epsilon}(\boldsymbol{m}) = \langle \theta^4 \rangle- 4 \langle \theta \rangle \langle \theta^3 \rangle + 6\langle \theta \rangle^2 \langle \theta^2 \rangle - 3\langle \theta \rangle^4
\end{equation}
and we have used the notation $\left\langle \Box \right\rangle = \int d\theta p(\theta|\boldsymbol{m}) \Box$. This is precisely the three-step decomposition introduced in section \ref{subsec:numalgorithm} to obtain the mean square error and, as such, we can compute $\Delta \bar{\epsilon}$ numerically in the same way.

This calculation is shown in figure \ref{circular_mse}.ii, where the graph in the middle of the shaded areas is $\bar{\epsilon}_{\mathrm{mse}}$ for $1\leqslant\mu \leqslant 10$ and $W_0 = \pi/2$ and the boundaries are given by $\pm \Delta \bar{\epsilon}$. We can see that the Taylor error bounds for the twin squeezed cat state, the squeezed entangled state and the twin squeezed cat state, which constitute the basis of our main results in chapter \ref{chap:limited}, do not overlap for any value of $\mu$. Therefore, all the comparisons made between these probes are valid. That the twin squeezed cat state and the coherent state overlap for $\mu = 1, 2, 3$ is not surprising, since their respective mean square errors also do (see figure \ref{bounds_results}.i), and the same observation hold for the NOON state and the squeezed entangled state when $\mu = 2$. On the other hand, the shaded area of the NOON state overlaps slightly with the top shaded area of the twin squeezed vacuum state when $\mu=1$. It is important to appreciate that the shaded areas are bounds for the Taylor error, and it is not guaranteed that the uncertainties for these two states actually coincide. However, even if they did, it would simply constitute another instance where the role of inter-mode and intra-mode correlations is altered in the regime of limited data, since a state with path entanglement that is beaten by a state with a large amount of intra-mode correlations in the asymptotic regime would reach the same uncertainty than the latter for a single shot. 

Finally, we also notice that the approximation will become even better as $W_0$ decreases, which is the case for the other prior widths that we have explored. Hence, we can conclude that the results that arise from the use of the mean square error as an approximation for the mean sine error in the regime of moderate prior knowledge are generally valid.

\section{Calculation details of some results in optical interferometry}
\label{sec:optrans}

This appendix presents the derivation of some auxiliary results employed in chapters \ref{chap:nonasymptotic} and \ref{chap:limited} for the non-asymptotic study of the Mach-Zehnder interferometer.

First we will verify that $U_{\mathrm{BS}} D_1(\alpha)\ket{0, 0} = |\alpha/\sqrt{2},-i\alpha/\sqrt{2}\rangle$. Using \cite{barnett2002}
\begin{equation}
\mathrm{exp}\left(X\right)f\left(Y\right)\mathrm{exp}\left(-X\right) = f\left[\mathrm{exp}\left(X\right)Y \mathrm{exp}\left(- X\right)\right],
\label{coherent1step}
\end{equation}
where $X$ and $Y$ are operators, we have that
\begin{align}
U_{\mathrm{BS}} D_1(\alpha) U_{\mathrm{BS}}^\dagger &= \mathrm{exp}\left(-i\frac{\pi}{2}J_x\right) \mathrm{exp}\left(\alpha a_1^{\dagger} - \alpha^{*}a_1\right) \mathrm{exp}\left(i\frac{\pi}{2}J_x\right) 
\nonumber \\
&= \mathrm{exp}\left[\alpha \left(\mathrm{e}^{-i\frac{\pi}{2}J_x}a_1^{\dagger}\mathrm{e}^{i\frac{\pi}{2}J_x}\right) - \alpha^{*}\left(\mathrm{e}^{-i\frac{\pi}{2}J_x} a_1 \mathrm{e}^{i\frac{\pi}{2}J_x}\right)\right].
\end{align}
In addition, 
\begin{align}
U_{\mathrm{BS}} a_1 U_{\mathrm{BS}}^\dagger &= a_1 + \left( -i \pi/2 \right) \comm*{J_x}{a_1} + \frac{\left( -i \pi/2 \right)^2}{2!} \comm*{J_x}{\comm*{J_x}{a_1}} + \ldots = 
\nonumber \\ 
&= a_1 + i \left(\pi/4 \right) a_2 - \frac{\left(\pi/4 \right)^2}{2!} a_1 + \ldots
\nonumber \\
&= \left[ 1 - \frac{\left(\pi/4 \right)^2}{2!} + \ldots \right] a_1 + i \left[\left(\pi/4 \right) + \ldots \right] a_2
\nonumber \\
&= \mathrm{cos}\left(\pi/4\right) a_1 + i\hspace{0.1em}\mathrm{sin}\left(\pi/4\right) a_2 = \left(a_1 + i a_2\right)/\sqrt{2}
\label{opid1}
\end{align}
after combining the Baker-Campbell-Hausdorff formula \cite{yurke1986}
\begin{equation}
\mathrm{e}^{z X}Y\mathrm{e}^{-z X} =  B + z \comm*{X}{Y} + \frac{z^2}{2!} \comm*{X}{\comm*{X}{Y}} + \ldots,
\label{BCH}
\end{equation}
with the commutation relations $\comm*{a_i}{a_j} = \comm*{a_i^\dagger}{a_j^\dagger}=0$, $\comm*{a_i}{a_j^\dagger}=\delta_{ij}$ (see section \ref{subsec:qapp}). Finally, by noticing that 
\begin{equation}
\mathrm{exp}\left(-i\frac{\pi}{2}J_x\right)\ket{0, 0} = \sum_{k=0}^\infty \frac{\left(-i\pi/4\right)^k}{k!} \left( a_1^\dagger a_2 + a_1 a_2^\dagger \right)^k \ket{0, 0} = \ket{0, 0},
\end{equation}
and taking into account equations (\ref{coherent1step}) and (\ref{opid1}), we arrive at
\begin{align}
U_{\mathrm{BS}} D_1(\alpha)\ket{0, 0} &= \left[U_{\mathrm{BS}} D_1(\alpha) U_{\mathrm{BS}}^\dagger\right] U_{\mathrm{BS}}\ket{0, 0}
\nonumber \\
&= \mathrm{exp}\left\lbrace\left(\frac{\alpha}{\sqrt{2}} a_1^\dagger - \frac{\alpha^{*}}{\sqrt{2}}a_1\right) + \left[\frac{\left(-i\alpha\right)}{\sqrt{2}} a_2^\dagger - \frac{\left(-i\alpha\right)^{*}}{\sqrt{2}}a_2 \right]  \right\rbrace \ket{0, 0}
\nonumber \\
&= D_1\left(\alpha/\sqrt{2}\right)D_2\left(-i\alpha/\sqrt{2}\right) \ket{0, 0}
\nonumber \\
&= |\alpha/\sqrt{2},-i\alpha/\sqrt{2}\rangle,
\end{align}
as we expected.

On the other hand, we need to find the likelihood function $p(n_1, n_2| \theta) = ||a(n_1, n_2| \theta)||^2$ associated with the probability amplitude
\begin{equation}
a(n_1, n_2| \theta) = \langle n_1, n_2| \mathrm{e}^{-i\frac{\pi}{2}J_x}\mathrm{e}^{i N_2 \phi}\mathrm{e}^{-iJ_z\theta}\ket{\psi_{\mathrm{NOON}}} = \langle n_1, n_2| \Phi(\theta) \rangle,
\end{equation}
where $| \Phi(\theta) \rangle =  \mathrm{e}^{-i\frac{\pi}{2}J_x}\mathrm{e}^{i N_2 \phi}\mathrm{e}^{-iJ_z\theta}\ket{\psi_{\mathrm{NOON}}}$, $\phi$ is a known phase shift and $\ket{\psi_{\mathrm{NOON}}} = (\ket{N, 0} + \ket{0, N})/\sqrt{2}$. The transformation associated with the phase shifts is
\begin{equation}
\mathrm{e}^{i N_2 \phi}\mathrm{e}^{-iJ_z\theta}\ket{\psi_{\mathrm{NOON}}} = \frac{1}{\sqrt{2}}\left[\mathrm{e}^{-iN\theta/2}\ket{N, 0} + \mathrm{e}^{iN\left(2\phi + \theta\right)/2}\ket{0, N}\right],
\label{phasenoon}
\end{equation}
since
\begin{equation}
\mathrm{e}^{i N_i x}\ket{n_i} = \sum_{k=0}^\infty \frac{\left(i x\right)^k}{k!} N_i^k \ket{n_i} = \sum_{k=0}^\infty \frac{\left(i x n_i\right)^k}{k!}\ket{n_i} = \mathrm{e}^{i n_i x}\ket{n_i}.
\end{equation}
Furthermore, from
\begin{align}
U_{\mathrm{BS}} a_2 U_{\mathrm{BS}}^\dagger &= a_2 + \left( -i \pi/2 \right) \comm*{J_x}{a_2} + \frac{\left( -i \pi/2 \right)^2}{2!} \comm*{J_x}{\comm*{J_x}{a_2}} + \ldots = 
\nonumber \\ 
&= a_2 + i \left(\pi/4 \right) a_1 - \frac{\left(\pi/4 \right)^2}{2!} a_2 + \ldots
\nonumber \\
&= i\left[\left(\pi/4 \right) + \ldots \right]a_1 + \left[ 1 - \frac{\left(\pi/4 \right)^2}{2!} + \ldots \right]a_2
\nonumber \\
&= i\hspace{0.1em}\mathrm{sin}\left(\pi/4\right)a_1 + \mathrm{cos}\left(\pi/4\right)a_2 = \left(i a_1 + a_2\right)/\sqrt{2} 
\label{opid2}
\end{align}
and equations (\ref{opid1}) and (\ref{phasenoon}) we find that 
\begin{align}
| \Phi(\theta) \rangle &=  \frac{1}{\sqrt{2 N!}}\left[\mathrm{e}^{-iN\theta/2}\left(\mathrm{e}^{-i\frac{\pi}{2}J_x} a_1^\dagger \mathrm{e}^{i\frac{\pi}{2}J_x}\right)^N + \frac{\mathrm{e}^{iN\left(2\phi + \theta\right)/2}}{\sqrt{2 N!}} \left(\mathrm{e}^{-i\frac{\pi}{2}J_x} a_2^\dagger \mathrm{e}^{i\frac{\pi}{2}J_x}\right)^N \right]\ket{0, 0}
\nonumber \\
&= \frac{1}{\sqrt{2^{N+1} N!}}\left[\mathrm{e}^{-iN\theta/2}\left(a_1^\dagger - i a_2^\dagger\right)^N + \mathrm{e}^{iN\left(2\phi + \theta\right)/2}\left(-i a_1^\dagger + a_2^\dagger\right)^N \right] \ket{0, 0}
\nonumber \\
&= \frac{1}{\sqrt{2^{N+1}}}\sum_{k=0}^N \sqrt{\frac{N!}{k! (N-k)!}}\left[ \left(-i\right)^{N-k}\mathrm{e}^{-iN\theta/2} + \left(-i\right)^k \mathrm{e}^{iN\left(2\phi + \theta\right)/2} \right] \ket{k, N-k}
\nonumber \\
&= \sqrt{\frac{2}{2^N}}\sum_{k=0}^N \sqrt{\frac{N!}{k! (N-k)!}} \hspace{0.1em}\mathrm{cos}\left[N\left(\theta+\phi\right)/2 + \left(2k - N\right)\pi/4 \right] \ket{k, N \hspace{-0.1em} - \hspace{-0.1em} k}\hspace{-0.1em}.\hspace{-0.5em}
\end{align}
Hence, 
\begin{equation}
a(n, N - n | \theta) = \sqrt{\frac{2N!}{2^N n! (N-n)!}} \hspace{0.1em}\mathrm{cos}\left[N\left(\theta+\phi\right)/2 + \left(2k - N\right)\pi/4 \right],
\label{noonamplitude}
\end{equation}
where we have used that NOON states have a definite number of photons. 

If $\phi = 0$, then equation (\ref{noonamplitude}) generates the probability density
\begin{equation}
p(n,N-n|\theta) = \frac{2 N! \hspace{0.2em} \mathrm{cos}^2\left[N\theta/2 + (2n-N)\pi/4\right]}{2^N n! (N-n)!} 
\end{equation}
that we employed in chapter \ref{chap:nonasymptotic}, while for $\phi = -\pi/4$ and $N = 2$ we have that
\begin{equation}
p(n,2-n|\theta) = \frac{\mathrm{cos}^2\left[\theta + (2n-3)\pi/4\right]}{ n! (2-n)!},
\end{equation}
which is the result examined in chapter \ref{chap:limited}.

\section{Multivariate Gaussian integrals}
\label{sec:multigaussian}

Here we summarise the standard calculation of the multivariate integrals in chapter \ref{chap:networks}, which also include those in chapter \ref{chap:nonasymptotic} when we take $d=1$. 

Let us denote the classical Fisher information matrix, which it is assumed to be positive definite, by $F(\boldsymbol{\theta}')\equiv F'$, and its eigendecomposition by $F' = Z F'_D Z^\transpose$, with $F'_D=\mathrm{diag}(z_1, \dots, z_d)$. In addition, define the transformation $\boldsymbol{\theta}-\boldsymbol{\theta}' = Z \boldsymbol{y}$ between $\boldsymbol{\theta}$ and $\boldsymbol{y}$, with Jacobian $\mathrm{det}(Z) = 1$. The first calculation is
\begin{align}
G_0 &= \int_{-\boldsymbol{\infty}}^{\boldsymbol{\infty}} d\boldsymbol{\theta}\hspace{0.15em}\mathrm{e} ^{-\frac{\mu}{2}\left(\boldsymbol{\theta}-\boldsymbol{\theta}'\right)^\transpose F' \left(\boldsymbol{\theta}-\boldsymbol{\theta}'\right)} 
= \int_{-\boldsymbol{\infty}}^{\boldsymbol{\infty}} d\boldsymbol{y}\hspace{0.15em}\mathrm{e} ^{-\frac{\mu}{2}\boldsymbol{y}^\transpose F'_D \boldsymbol{y}} 
\nonumber \\
&= \prod_{i=1}^d \int_{-\infty}^{\infty} dy_i\hspace{0.15em} \mathrm{e}^{-\frac{\mu}{2}y_i^2 z_i} = \left(\frac{2\pi}{\mu}\right)^{\frac{d}{2}}  \prod_{i=1}^d \frac{1}{\sqrt{z_i}} = \left[\frac{\left(2\pi\right)^d}{\mathrm{det}(\mu F')}\right]^{\frac{1}{2}},
\label{gaussian0}
\end{align}
since $\int_{-\infty}^{\infty} dy_i\hspace{0.15em} \mathrm{e}^{-\frac{\mu}{2}y_i^2 z_i} = [2\pi/(\mu z_i)]^{1/2}$. On the other hand, 
\begin{align}
G_{1,i} &= \int_{-\boldsymbol{\infty}}^{\boldsymbol{\infty}} d\boldsymbol{\theta}\hspace{0.15em}\mathrm{e} ^{-\frac{\mu}{2}\left(\boldsymbol{\theta}-\boldsymbol{\theta}'\right)^\transpose F' \left(\boldsymbol{\theta}-\boldsymbol{\theta}'\right)}\theta_i 
= \int_{-\boldsymbol{\infty}}^{\boldsymbol{\infty}} d\boldsymbol{y}\hspace{0.15em}\mathrm{e} ^{-\frac{\mu}{2}\boldsymbol{y}^\transpose F'_D \boldsymbol{y}}  \left(\theta_i' + \sum_{j=1}^d Z_{ij}y_j\right)
\nonumber \\
&=  G_0\hspace{0.15em}\theta'_i + \sum_{j=1}^d Z_{ji}\prod_{k=1}^d \int_{-\infty}^{\infty} dy_k\hspace{0.15em} \mathrm{e}^{-\frac{\mu}{2}y_k^2 z_k}y_j =  G_0\hspace{0.15em}\theta'_i
\label{gaussian1}
\end{align}
where we have used that $\int_{-\infty}^{\infty} dy_i\hspace{0.15em} \mathrm{e}^{-\frac{\mu}{2}y_i^2 z_i}y_i = 0$. Finally, 
\begin{align}
G_{2,ij} = &~ \int_{-\boldsymbol{\infty}}^{\boldsymbol{\infty}} d\boldsymbol{\theta}\hspace{0.15em}\mathrm{e} ^{-\frac{\mu}{2}\left(\boldsymbol{\theta}-\boldsymbol{\theta}'\right)^\transpose F' \left(\boldsymbol{\theta}-\boldsymbol{\theta}'\right)}\theta_i \theta_j
\nonumber \\
= &~ \int_{-\boldsymbol{\infty}}^{\boldsymbol{\infty}} d\boldsymbol{y}\hspace{0.15em}\mathrm{e} ^{-\frac{\mu}{2}\boldsymbol{y}^\transpose F'_D \boldsymbol{y}}  \left(\theta_i' + \sum_{k=1}^d Z_{ik}y_k\right)\left(\theta_j' + \sum_{l=1}^d Z_{jl}y_l\right)
\nonumber \\
= &~ G_0\hspace{0.15em}\theta_i' \theta_j'
+ \sum_{k=1}^d \left(\theta_i' Z_{jk} + \theta_j' Z_{ik} \right)\prod_{m=1}^d \int_{-\infty}^{\infty} dy_m\hspace{0.15em} \mathrm{e}^{-\frac{\mu}{2}y_m^2 z_m}y_k
\nonumber \\
&+ \sum_{k, l=1}^d Z_{ik} Z_{jl}\prod_{m=1}^d \int_{-\infty}^{\infty} dy_m\hspace{0.15em} \mathrm{e}^{-\frac{\mu}{2}y_m^2 z_m}y_k y_l
\nonumber \\
= &~ G_0\hspace{0.15em}\theta_i' \theta_j' + \left(\frac{2\pi}{\mu}\right)^{\frac{d}{2}}  \frac{1}{\mu} \sum_{l=1}^d \frac{Z_{il} Z_{jl}}{z_l} \frac{1}{\sqrt{z_l}} \prod_{\lbrace k=1,  \hspace{0.1em} k\neq l \rbrace}^d \frac{1}{\sqrt{z_k}}
\nonumber\\
= &~ G_0 \left\lbrace\theta_i' \theta_j' + \frac{\left[\left(F'\right)^{-1}\right]_{ij}}{\mu} \right\rbrace,
\label{gaussian2}
\end{align}
given that $\int_{-\infty}^{\infty} dy_i\hspace{0.15em} \mathrm{e}^{-\frac{\mu}{2}y_i^2 z_i}y_i^2 = [2\pi/(\mu z_i)^3]^{1/2}$.

Equations (\ref{gaussian0} - \ref{gaussian2}) lead us to the results in equations (\ref{norm_asy}), (\ref{gaussiansingle}), (\ref{multigaussian0}) and (\ref{multigaussian12}) in the main text. 
\chapter{Numerical toolbox for quantum interferometry}
\label{app:numsingle}

In this appendix we present a collection of algorithms that together allow to reproduce the numerical results for optical interferometry in chapters \ref{chap:nonasymptotic}, \ref{chap:limited} and \ref{chap:future}. Most of the codes are written in MATLAB, the only exception being the second algorithm in appendix \ref{sec:singleshotalgorithm}, which has been developed in Mathematica.

\section{Basic elements and initial states}
\label{sec:elementsnum}

Operators in the space of a single electromagnetic mode:
\begin{lstlisting}[language=Matlab]
function [creat] = creation(dimension)
% Matrix representation of the creation operator, where 'dimension' is
% the cutoff of the space.
creat=zeros(dimension,dimension);
for aa=1:dimension
    for bb=1:dimension
        if aa == (bb+1); creat(aa,bb)=sqrt(bb);
        else; creat(aa,bb)=0;
        end
    end
end
creat=sparse(creat); end
\end{lstlisting}

\begin{lstlisting}[language=Matlab]
function [id] = identity(dimension)
% Identity matrix
id=eye(dimension);
id=sparse(id); end
\end{lstlisting}

\justify{Optical elements:}
\begin{lstlisting}[language=Matlab, mathescape=true]
function [j1] = j1schwinger(dimension)
% Matrix representation of the J1 operator (Jordan-Schwinger map).
j1=0.5*(kron(creation(dimension),creation(dimension)') + kron(creation$\hspace{0.15em}\swarrow$
(dimension)',creation(dimension)));
j1=sparse(j1); end
\end{lstlisting}

\begin{lstlisting}[language=Matlab, mathescape=true]
function [j3] = j3schwinger(dimension)
% Matrix representation of the J3 operator (Jordan-Schwinger map).
j3=0.5*(kron(creation(dimension)*creation(dimension)',identity$\hspace{0.15em}\swarrow$
(dimension))-kron(identity(dimension),creation(dimension)*creation$\hspace{0.15em}\swarrow$
(dimension)'));
j3=sparse(j3); end
\end{lstlisting}

\begin{lstlisting}[language=Matlab]
function [v] = beam_splitter(dimension)
% Matrix representation of a 50:50 beam splitter. 
v=expm(-1i*0.5*pi*sparse(j1schwinger(dimension)));
v=sparse(v); end
\end{lstlisting}

\begin{lstlisting}[language=Matlab]
function [utheta] = phase_shift_diff(dimension, theta)
% Matrix representation of the unitary encoding of the unknown parameter 
% 'theta' (difference of phase shifts).
utheta=expm(-1i*theta*j3schwinger(dimension));
utheta=sparse(utheta); end
\end{lstlisting}

\justify{Initial probe states for a Mach-Zehnder interferometer:}
\begin{lstlisting}[language=Matlab]
function [zero] = vacuum(dimension)
% Vacuum state for a single mode.
temp=identity(dimension);
zero=temp(:,1);
zero=sparse(zero); end
\end{lstlisting}

\begin{lstlisting}[language=Matlab]
function [displ] = displacement(dimension,alpha)
% Matrix representation of the displacement operator, where 'alpha' is
% the amount of displacement. 
displ=expm(alpha*creation(dimension)-conj(alpha)*creation(dimension)');
displ=sparse(displ); end
\end{lstlisting}

\begin{lstlisting}[language=Matlab, mathescape=true]
function [squ] = squeeze(dimension,zeta)
% Matrix representation of the squeezing operator for a single mode, 
% where 'zeta' is the squeezing parameter.
squ=expm(0.5*(conj(zeta)*(creation(dimension)')^2-zeta*(creation$\hspace{0.15em}\swarrow$
(dimension))^2));
squ=sparse(squ); end
\end{lstlisting}

\begin{lstlisting}[language=Matlab, mathescape=true]
function  [initial_state] = initial_probe(state_sel)
% Common states in optical interferometry, where  'state_sel' is a number
% from 1 to 5 labelling the quantum probes
%
%   (1) Coherent state: |alpha/sqrt(2),-i*alpha/sqrt(2)>
%   (2) NOON state: (|N,0>+|0,N>)/sqrt(2)
%   (3) Twin squeezed vacuum state: S_a(z)S_b(z)|0,0>
%   (4) Squeezed entangled state: N (|z,0> + |0,z>)
%   (5) Twin squeezed cat state: [N S(z)(|alpha>+|-alpha>)]\otimes2
%
% whose componentes are generated in the number basis of a Mach-Zehnder 
% interferometer.
%
% The code is configured with the parameters
%
%   (1) alpha=sqrt(2)
%   (2) N=2
%   (3) z=asinh(1)
%   (4) z=log(2+sqrt(3))
%   (5) alpha=0.960149, z=1.2145
%
% so that the nbar number of quanta that enters the interferometer is 2.
%
% The cutoff for the vectors are: (1) 20, (2) 2, (3) 50, (4) 60 
% and (5) 50. These values are selected such that the numerical states 
% are a reasonable approximation to the analytical kets.

% State parameters
nbar=2;
number=nbar; % Mean number of photons
alpha=sqrt(nbar); % Displacement parameter
zeta=asinh(sqrt(nbar/2)); % Squeezing parameter
zent=log(2+sqrt(3));

alphacat=0.960149; % Maximum Fisher information for the twin squeezed 
zcat=1.2145;         % cat state

%alphacat=1.09048; % Same Fisher information for the twin squeezed cat 
%zcat=1.1025;        % state and the squeezed entangled state

if state_sel==1
    num_cutoff=20; % Cutoff for states
elseif state_sel==2
    num_cutoff=number;
elseif state_sel==3 || state_sel==5
    num_cutoff=50;
elseif state_sel==4
    num_cutoff=60;
end
op_cutoff=num_cutoff+1; % Cutoff for operators

% Initial state
if state_sel==1
    initial_temp=sparse(displacement(op_cutoff,alpha)*vacuum(op_cutoff));
    initial_state=sparse(kron(initial_temp,vacuum(op_cutoff)));
    initial_state=beam_splitter(op_cutoff)*initial_state;
elseif state_sel==2
    initial_temp=sparse((creation(op_cutoff)^number)*vacuum(op_cutoff));
    initial_state=sparse(kron(initial_temp,vacuum(op_cutoff))+kron(vacuum$\hspace{0.15em}\swarrow$
(op_cutoff),initial_temp));
elseif state_sel==3
    initial_temp1=sparse(squeeze(op_cutoff,zeta)*vacuum(op_cutoff));
    initial_temp2=sparse(squeeze(op_cutoff,zeta)*vacuum(op_cutoff));
    initial_state=sparse(kron(initial_temp1,initial_temp2));
elseif state_sel==4
    initial_state=(kron(squeeze(op_cutoff,zent),identity(op_cutoff))$\hspace{0.15em}\swarrow$         	
+kron(identity(op_cutoff),squeeze(op_cutoff,zent)))*kron(vacuum$\hspace{0.15em}\swarrow$
(op_cutoff),vacuum(op_cutoff));
elseif state_sel==5
    initial_state=squeeze(op_cutoff,zcat)*(displacement$\hspace{0.15em}\swarrow$
(op_cutoff,alphacat)+displacement(op_cutoff,-alphacat))*vacuum(op_cutoff);
    initial_state=kron(initial_state,initial_state);
    initial_state=initial_state/sqrt((initial_state'*initial_state));
end
initial_state=initial_state/sqrt((initial_state'*initial_state)); end
\end{lstlisting}

\section{Optimal quantum strategies for the square error criterion}
\label{sec:singleshotalgorithm}

The following algorithm has been utilised to calculate the optimal single-shot strategies in chapter \ref{chap:limited}:
\begin{enumerate}
\item The components $c_{nm}$ of $\ket{\psi_0}$ are numerically approximated in a finite space of dimension $d_c$ per mode. For the coherent state this dimension is $d_c=21$, and the number probability for this cutoff is $p_c \sim 10^{-19}$; for the twin squeezed vacuum state we have that $d_c = 51$ and $p_c \sim 10^{-17}$; $d_c = 61$ and $p_c \sim 10^{-5}$ for the squeezed entangled state; and $d_c = 51$ and $p_c \sim 10^{-10}$ for the twin squeezed cat state. This is achieved via the code for generating initial states in the previous appendix.
\item $\mathcal{K}$ and $\mathcal{L}$ are numerically generated using the formulas in equations (\ref{k_semianalytical_sol}) - (\ref{singularcases}). This allows us to calculate $\rho = \rho_0 \circ \mathcal{K}$ and $\rho = \rho_0 \circ \mathcal{L}$ in the number basis.
\item The basis of $\rho$ and $\bar{\rho}$ is changed as $\rho_D = \mathcal{V}^\dagger \rho \mathcal{V}$ and $\bar{\rho}_D = \mathcal{V}^\dagger \bar{\rho} \mathcal{V}$, where the columns of $\mathcal{V}$ are given by the eigenvectors $\ket{\phi_i}$ of $\rho$, $(\rho_D)_{ij} = p_i\delta_{ij}$ and $(\bar{\rho}_D)_{ij} = \bra{\phi_i} \bar{\rho} \ket{\phi_j}$.
Only the eigenvectors $\ket{\phi_i}$ whose eigenvalues $p_i$ satisfy that $p_i \gtrsim 10^{-12}$ are employed. 
\item Now we can calculate the elements $(S_D)_{ij} = \bra{\phi_i} S \ket{\phi_j}=2(\bar{\rho}_D)_{ij}/(p_i + p_j)$ directly.
\item We return to the original basis using $S=\mathcal{V} S_D \mathcal{V}^\dagger$.
\item Finally, we calculate the spectral decomposition of $S$, which gives us the estimates $\lbrace s\rbrace$ and the projectors $\lbrace \ket{s}\rbrace$.
\end{enumerate}
Its implementation in MATLAB is:

\begin{lstlisting}[language=Matlab, mathescape=true]
function [Sopt,sopt,soptvec_columns,bayes_bound,rho,pk,psik,rhobar,$\hspace{0.15em}\swarrow$
rhobarnew] = mz_optimal_1trial(initial_state,phase_width,phase_mean)
% Optimal single-shot strategy, where 'initial_state' is a pure state 
% for the Mach-Zehnder interferometer, 'phase_width' is the width of the 
% parameter domain and 'phase_mean' is its centre.
%
% This programme calculates:
%
%   a) the optimal quantum estimator 'Sopt'
%   b) the estimates 'sopt' for the unknown parameter given by the
%      spectrum of 'Sopt'
%   c) the optimal projective measurement for a single trial given by
%      the eigenvectors 'soptvec_columns' of 'Sopt'
%   d) the optimal single-shot mean square error 'bayes_bound'
%   e) the zero-th quantum moment of the transformed density matrix 'rho'
%   f) 'rho' in its diagonal basis, denoted by 'pk'
%   g) the matrix 'psik' whose columns are the eigenvectors of 'rho'
%   h) the first quantum moment of the transformed density matrix 'rhobar'
%   i) 'rhobar' in the eigenbasis of 'rho', denoted by 'rhobarnew'

% Calculation of 'rho' and 'rhobar'
index=1;
kvec=zeros(1,length(initial_state)^2);
lvec=zeros(1,length(initial_state)^2);
for x1=1:sqrt(length(initial_state))
    for y1=1:sqrt(length(initial_state))
        for z1=1:sqrt(length(initial_state))
            for t1=1:sqrt(length(initial_state))
                if (x1-1)-(y1-1)+(t1-1)-(z1-1)==0
                    K=phase_width;
                    L=phase_mean*phase_width;
                else
                    comp_temp=(x1-1)-(y1-1)+(t1-1)-(z1-1);
                    exp_temp=exp(-1i*comp_temp*phase_mean/2);
                    sin_temp=sin(comp_temp*phase_width/4);
                    cos_temp=cos(comp_temp*phase_width/4);
                    K=4*exp_temp*sin_temp/comp_temp;
                    L=exp_temp*(4*phase_mean*sin_temp/comp_temp$\hspace{0.15em}\swarrow$ 
+1i*2*phase_width*cos_temp/comp_temp - 1i*8*sin_temp/comp_temp^2);
                end
                kvec(index)=K/phase_width;
                lvec(index)=L/phase_width;
                index=index+1;
            end
        end
    end
end

kmat=sparse(vec2mat(kvec,sqrt(length(kvec))));
lmat=sparse(vec2mat(lvec,sqrt(length(lvec))));
initial_rho=kron(initial_state,initial_state');
rho=initial_rho.*kmat; rhobar=initial_rho.*lmat;
rho=full(rho); rhobar=full(rhobar);

% Eigenvalues and eigenvectors of 'rho'
[psik, pk] = eigs(rho,rank(rho));
psik=sparse(psik);
pk=sparse(pk);

pkvec=zeros(1,length(pk));
for x=1:length(pk)
    pkvec(x)=pk(x,x);
end

% 'rhobar' in the eigenbasis of 'rho'
rhobarnew=psik'*rhobar*psik;

% Optimal single-shot strategy: projectors and outcomes
Sopt_temp=zeros(length(pkvec),length(pkvec));
for a=1:length(pkvec)
    for b=1:length(pkvec)
        if pkvec(a)+ pkvec(b)>0
            Sopt_temp(a,b)=2*rhobarnew(a,b)/(pkvec(a)+pkvec(b));
        end
    end
end

Sopt_temp=sparse(Sopt_temp);
Sopt=psik*Sopt_temp*psik';
Sopt=full(Sopt);

[soptvec_columns, sopt_temp] = eigs(Sopt,rank(Sopt));
sopt=zeros(1,length(sopt_temp));
for x=1:length(sopt_temp)
    sopt(x)=sopt_temp(x,x);
end
soptvec_columns=sparse(soptvec_columns);
sopt=sparse(sopt);
if imag(sopt)<1e-5; sopt=real(sopt);
else
    error('The estimates of the unknown parameter must be real. Check$\hspace{0.15em}\swarrow$
the cutoff in the intermediate calculations.')
end

% Phase domain
phase=linspace(phase_mean-phase_width/2,phase_mean+phase_width/2,1000);

% Optimal single-shot mean square error
bayes_bound=trapz(phase,phase.*phase)/phase_width - trace(Sopt*Sopt*rho);
if imag(bayes_bound)<1e-10; bayes_bound=real(bayes_bound);
else
    error('The mean square error must be real. Check the cutoff$\hspace{0.15em}\swarrow$
in the intermediate calculations.')
end

end
\end{lstlisting}

The previous algorithm was extended in our work \cite{jesus2018} to calculate the collective measurement that is optimal on $\mu$ copies of a NOON state (see section \ref{subsec:optnoon}). However, a more economic alternative using Mathematica is:

\begin{mathematicathesis}
(* Optimal single-shot mean square error for collective POMs on 'mu'$\hspace{0.15em}\swarrow$
copies of a NOON state *)

(* Number of copies *)
Clear[mu]
mu = 4; (* 'mu' must be greater than or equal to 2 *)

(* Transformed density matrix *)
Clear[nbar, rhotemp, rhotheta]
nbar = 2;
rhotemp[theta_] := N[ {{1/2, Exp[-I*nbar*theta]/2},$\hspace{0.15em}\swarrow$
{Exp[I*nbar*theta]/2, 1/2}}]; 
rhotheta[theta] = KroneckerProduct[rhotemp[theta ], rhotemp[theta ]];
Do[rhotheta[theta ] = KroneckerProduct[rhotemp[theta ],$\hspace{0.15em}\swarrow$
rhotheta[theta ]], {j, mu - 2}]; (* 'j' is the index of repetition *)

(* Prior probability *)
Clear[priorwidth, priormean, a, b, prior]
priorwidth = Pi/2.;
priormean = 0;
a = priormean - priorwidth/2;
b = priormean + priorwidth/2;
prior[theta_] := 1/(b - a);

(* Calculation of 'rho' and its diagonal form*)
Clear[rho, rhodiag]
rho = Integrate[prior[theta]*rhotheta[theta], {theta, a, b}];
rhodiag = DiagonalMatrix[Eigenvalues[rho]];

(* Calculation of 'rhobar' and its form in the eigenbasis of rho*)
Clear[rhobar, change, rhobardiag]
rhobar = Integrate[prior[theta]*rhotheta[theta]*theta, {theta, a, b}];
change = Transpose[Eigenvectors[rho]];
rhobardiag = Inverse[change].rhobar.change;

(* Optimal quantum estimator *)
Clear[rhosupp, rhobarsupp, Sopt]
rhosupp = rhodiag[[1 ;; MatrixRank[rho], 1 ;; MatrixRank[rho]]];
rhobarsupp = rhobardiag[[1 ;; MatrixRank[rho], 1 ;; MatrixRank[rho]]];
Sopt = LyapunovSolve[rhosupp/2., rhosupp/2., rhobarsupp];

(* Optimal single-shot mean square error *)
Clear[emse]
emse = Round[NIntegrate[prior[theta]*theta^2, {theta, a, b}]$\hspace{0.15em}\swarrow$ 
- Tr[Sopt.Sopt.rhosupp], 10.^(-7)]

0.0428159 (* Result for mu = 4 *)
\end{mathematicathesis}

Furthermore, we notice that the single-shot calculation for a lossy interferometer in section \ref{loss} was carried out with a similar version of the Mathematica code above.

\section{Other quantum bounds}

\subsection{Quantum Cram\'{e}r-Rao bound}
\label{subsec:qcrbmatlab}

\begin{lstlisting}[language=Matlab]
function [qcrb] = mz_qcrb(initial_state,mu_max)
% Quantum Cramer-Rao bound as a function of the number of trials, where
% 'initial_state' is a pure state for the Mach-Zehnder interferometer
% and 'mu_max' is the maximum number of repetitions.

% Number of repetitions
observations=1:1:mu_max;

% Space cutoff (for a single mode)
op_cutoff=sqrt(length(initial_state));

% Quantum Fisher information (pure states and unitary encoding)
expectation_n=initial_state'*j3schwinger(op_cutoff)*initial_state;
expectation_n2=initial_state'*j3schwinger(op_cutoff)^2*initial_state;
qfi=4*(expectation_n2-expectation_n^2);

% Do we have information?
if qfi==0
    disp('The Quantum Fisher information is zero.')
    return
end

% Quantum Cramer-Rao Bound
qcrb=1./(observations*qfi);

end
\end{lstlisting}

\subsection{Quantum Ziv-Zakai bound}
\label{subsec:qzzbnum}

\begin{lstlisting}[language=Matlab, mathescape=true]
function [qzzb] = mz_qzzb(initial_state,phase_width,mu_max)
% Quantum Ziv-Zakai bound as a function of the number of trials, where
% 'initial_state' is a pure state for the Mach-Zehnder interferometer,
% 'phase_width' is the width of the parameter domain and 'mu_max' is 
% the maximum number of repetitions. 

% Space cutoff (for a single mode)
op_cutoff=sqrt(length(initial_state));

% Parameter domain
W=phase_width;
dim_theta=1000;
theta=linspace(0,W,dim_theta);

% Fidelity
fidelity=zeros(dim_theta,1);
for z=1:dim_theta
    after_phase_shift=sparse(phase_shift_diff(op_cutoff,theta(z))$\hspace{0.15em}\swarrow$
*initial_state);
    fidelity(z)=abs(initial_state'*after_phase_shift)^2;
end

% Quantum Ziv-Zakai Bound integrand
integrand=zeros(dim_theta,mu_max);
for runs=1:mu_max
    for z=1:dim_theta
        integrand(z,runs)=0.5.*theta(z).*(1-theta(z)./W)$\hspace{0.15em}\swarrow$
.*(1-sqrt(1-fidelity(z).^runs));    
    end
end
integrand=sparse(integrand);

% Quantum Ziv-Zakai Bound
qzzb=trapz(theta,integrand,1);

end
\end{lstlisting}

\subsection{Quantum Weiss-Weinstein bound}
\label{subsec:qwwbnum}

\begin{lstlisting}[language=Matlab, mathescape=true]
function [qwwb] = mz_qwwb(initial_state,phase_width,mu_max)
% Quantum Weiss-Weinstein bound as a function of the number of 
% trials, where 'initial_state' is a pure state for the Mach-Zehnder,
% interferometer 'phase_width' is the width of the parameter domain 
% and 'mu_max' is the maximum number of repetitions. 

% Space cutoff (for a single mode)
op_cutoff=sqrt(length(initial_state));

% Parameter domain
W=phase_width;
dim_theta=1000;
h=linspace(0,W,dim_theta);

% Quantum Weiss-Weinstein Bound
find_supremum=zeros(mu_max,length(h));
qwwb=zeros(1,mu_max);
for runs=1:mu_max
    for z=1:length(h)
        zeta=initial_state'*phase_shift_diff(op_cutoff,h(z))*initial_state;
        zeta2=initial_state'*phase_shift_diff(op_cutoff,2*h(z))$\hspace{0.15em}\swarrow$
*initial_state;
        fid_function=abs(zeta)^2;
        find_supremum(runs,z)=h(z)^2*(1-h(z)/W)^2*fid_function^(2*runs)$\hspace{0.15em}\swarrow$
/(2*fid_function^runs-2*(1-2*h(z)/W)*real(zeta^(2*runs)*conj(zeta2)^runs));
    end
    qwwb(runs)=max(find_supremum(runs,:));
end

end
\end{lstlisting}

\section{Measurement strategies}
\label{sec:pomnum}

\begin{lstlisting}[language=Matlab, mathescape=true]
function [outcomes,proj_columns] = mz_pom$\hspace{0.15em}\swarrow$
(state_choice,pom_choice,phase_width,phase_mean)
% Outcomes and POM elements of five projective measurement schemes:
%
%   1) Optimal single-shot POM
%   2) 50:50 beam splitter + photon counting
%   3) 50:50 beam splitter + measurement of quadratures rotated by pi/8
%   4) Undoing the preparation of the initial state + photon counting
%   5) 50:50 beam splitter + parity measurements
%
% where 'state_choice' labels the initial state, 'pom_choice' selects one
% of the previous measurement schemes, 'phase_width' is the width of the 
% phase domain and 'phase_mean' is its centre. 
%
% Some extra phase shifts that are assumed to be known have been added 
% to 2) - 5) in order to make the strategy optimal when the prior is 
% centred around zero.

% Space cutoff (for a single mode)
op_cutoff=sqrt(length(initial_probe(state_choice)));

if pom_choice==1
    % 1) Optimal single-shot POM
    [~,outcomes,proj_columns,~,~,~,~,~,~]=mz_optimal_1trial(initial_probe$\hspace{0.15em}\swarrow$
(state_choice),phase_width,phase_mean);
    
elseif pom_choice==2
    % 2) 50:50 beam splitter + photon counting
        
    % Observable quantity (number of photons at each port)
    observable=kron(creation(op_cutoff)*creation(op_cutoff)',creation$\hspace{0.15em}\swarrow$
(op_cutoff)*creation(op_cutoff)');
    [proj_columns,outcomes_temp]=eig(full(observable));
    outcomes=zeros(1,length(outcomes_temp));
    for x=1:length(outcomes_temp)
        outcomes(x)=outcomes_temp(x,x);
    end
    
    % Extra phase shift
    odd_shift=kron(identity(op_cutoff),expm(1i*(pi/2)*creation$\hspace{0.15em}\swarrow$
(op_cutoff)*creation(op_cutoff)'));
    even_shift=kron(identity(op_cutoff),expm(1i*(pi/4)*creation$\hspace{0.15em}\swarrow$
(op_cutoff)*creation(op_cutoff)'));
    
    if state_choice==1
        optimal_shift=odd_shift;
    else
        optimal_shift=even_shift;
    end
    
    % Effect of the 50:50 beam splitter
    proj_columns=optimal_shift'*beam_splitter(op_cutoff)*proj_columns;
    
elseif pom_choice==3
    % 3) 50:50 beam splitter + measurement of quadratures rotated by pi/8
        
    % Observable quantity
    if state_choice==1
        error('The quadrature POM is not available for coherent states.')
    else
        phasequad1=pi/8;
        phasequad2=phasequad1;
    end
    quad1=(creation(op_cutoff)*exp(1i*phasequad1)+creation$\hspace{0.15em}\swarrow$
(op_cutoff)'*exp(-1i*phasequad1))/sqrt(2);
    quad2=(creation(op_cutoff)*exp(1i*phasequad2)+creation$\hspace{0.15em}\swarrow$
(op_cutoff)'*exp(-1i*phasequad2))/sqrt(2);
    observable=kron(quad1,quad2);
    [proj_columns,outcomes_temp]=eig(full(observable));
    outcomes=zeros(1,length(outcomes_temp));
    for x=1:length(outcomes_temp)
        outcomes(x)=outcomes_temp(x,x);
    end
    
    % Extra phase shift
    optimal_shift=kron(expm(-1i*(pi/4)*creation(op_cutoff)*creation$\hspace{0.15em}\swarrow$
(op_cutoff)'),identity(op_cutoff));
            
    % Effect of the 50:50 beam splitter
    proj_columns=optimal_shift'*beam_splitter(op_cutoff)*proj_columns;
    
elseif pom_choice==4
    % 4) Undoing the preparation of the initial state + photon counting
    if state_choice==1
    else
        error('This POM is only available for coherent states.')
    end  
    
    % Observable quantity (number of photons at each port)
    observable=kron(creation(op_cutoff)*creation(op_cutoff)',creation$\hspace{0.15em}\swarrow$
(op_cutoff)*creation(op_cutoff)');
    [proj_columns,outcomes_temp]=eig(full(observable));
    outcomes=zeros(1,length(outcomes_temp));
    for x=1:length(outcomes_temp)
        outcomes(x)=outcomes_temp(x,x);
    end
    
    % Extra phase shifts
    optimal_shift=sparse(expm(-1i*pi*j3schwinger(op_cutoff)));
    
    % Unitary transformations to undo the preparation of the state
    bs=sparse(beam_splitter(op_cutoff)');
    cs_undo=sparse(kron(displacement(op_cutoff,sqrt(2)),identity$\hspace{0.15em}\swarrow$
(op_cutoff)));
    combined=cs_undo*bs*optimal_shift;
    proj_columns=combined'*proj_columns; 
        
elseif pom_choice==5
    % 5) 50:50 beam splitter + parity measurements
    
    % Observable quantity (parity of the number of photons at each port)
    paritya=sparse(kron(identity(op_cutoff),(-1)^(full(creation$\hspace{0.15em}\swarrow$
(op_cutoff)*creation(op_cutoff)'))));
    parityb=sparse(kron((-1)^(full(creation(op_cutoff)*creation$\hspace{0.15em}\swarrow$
(op_cutoff)')),identity(op_cutoff)));
    observable=full(paritya*parityb);
    [proj_columns,outcomes_temp]=eig(full(observable));
    outcomes=zeros(1,length(outcomes_temp));
    for x=1:length(outcomes_temp)
        outcomes(x)=outcomes_temp(x,x);
    end
    
    % Extra phase shift
    odd_shift=kron(identity(op_cutoff),expm(1i*(pi/2)*creation$\hspace{0.15em}\swarrow$
(op_cutoff)*creation(op_cutoff)'));
    even_shift=kron(identity(op_cutoff),expm(1i*(pi/4)*creation$\hspace{0.15em}\swarrow$
(op_cutoff)*creation(op_cutoff)'));
    
    if state_choice==1
        optimal_shift=odd_shift;
    else
        optimal_shift=even_shift;
    end
    
    % Effect of the 50:50 beam splitter
    proj_columns=optimal_shift'*beam_splitter(op_cutoff)*proj_columns;  
end

end
\end{lstlisting}

\section{Prior information analysis}
\label{sec:priormatlab}

This algorithm generates the graphs in chapter \ref{chap:nonasymptotic} for the prior information analysis of the Mach-Zehnder interferometer. A version of this code was also employed for time estimation in section \ref{sec:fundtime}.

\begin{lstlisting}[language=Matlab, mathescape=true]
% Prior information analysis for single-parameter schemes
%
% This programme uses Bayes theorem to generate the posterior probability
%
% p(theta|m_1, ..., m_mu)
%
% for a flat prior and the likelihood function given by the Born rule. 
% The initial state and the measurement scheme are those of a Mach-Zehnder 
% interferometer, and they can be selected from the respective MATLAB 
% functions in our interferometric toolbox by giving a value from 1 to 5
% for 'state_choice' and 'pom_choice'.
%
% Important observations:
%
% - The prior is defined over all the parameter domain, so that the 
% symmetries of the likelihood that enable us to find the intrinsic 
% width can be visualised. 
%
% - The variables 'prior_mean_1shot' and 'prior_width_1shot' are needed
% to specify the optimal single-shot POM, but they do not affect the
% other measurement schemes. 
%
% - The results for the prior information analysis in chapter 4 are 
% recovered when we remove the extra phase shifts in the second option
% of our MATLAB function mz_pom(.) and we select it. This is because
% the results in chapter 4 were obtained for a prior between 0 and W, 
% while the POMs included in our sample of codes are those associated
% with a prior centred around zero (see chapter 5). 
clear

% State and POM options (see the respective codes in previous sections)
state_choice=2;
pom_choice=2;

% Initial state
initial_state=initial_probe(state_choice);

% Space cutoff
op_cutoff=sqrt(length(initial_state));

% Parameter domain
prior_mean=pi;
prior_width=2*pi; % Complete parameter domain
a=prior_mean-prior_width/2;
b=prior_mean+prior_width/2;
dim_theta=1000;
theta=linspace(a,b,dim_theta);

% Simulation of the unknown true value
index_real=160;

% Measurement scheme
prior_mean_1shot=0;
prior_width_1shot=pi/2; 
[outcomes_space,proj_columns] = mz_pom(state_choice,pom_choice,$\hspace{0.15em}\swarrow$
prior_width_1shot,prior_mean_1shot);

% State after the phase shift, final state and amplitudes
amplitudes=zeros(length(outcomes_space),dim_theta);
for z=1:dim_theta
    after_phase_shift=sparse(phase_shift_diff(op_cutoff,theta(z))$\hspace{0.15em}\swarrow$
*initial_state);
    for x=1:length(outcomes_space)
        pom_element=proj_columns(:,x);
        amplitudes(x,z)=sparse(pom_element'*after_phase_shift);
    end
end

% Likelihood function (using the Born rule)
likelihood=amplitudes.*conj(amplitudes);

% Prior density function
prior=ones(1,dim_theta);
prior=prior/trapz(theta,prior);

% Updating via Bayes theorem
prob_temp=prior;
for runs=1:100
    
    % Simulation of an interferometric experiment
    prob_sim=likelihood(:,index_real);
    cumulative = cumsum(prob_sim); % Cumulative function
    prob_rand=rand; % Random selection
    auxiliar=cumulative-prob_rand;
    
    for x=1:length(outcomes_space)
        if auxiliar(x)>0
            index=x;
            break
        end
    end
    
    % Posterior density function
    prob_temp=sparse(prob_temp.*likelihood(index,:));
    if trapz(theta,prob_temp)>1e-16
        prob_temp=prob_temp./trapz(theta,prob_temp);
    else
        prob_temp=0;
    end
    
    % Posterior probability plots
    if runs==1
        plot(theta,prob_temp,'k-','LineWidth',2.5)
        hold on
    elseif runs==2; plot(theta,prob_temp,'k-','LineWidth',2.5)
    elseif runs==10; plot(theta,prob_temp,'k-','LineWidth',2.5)
    elseif runs==100; plot(theta,prob_temp,'k-','LineWidth',2.5)
        hold off
    end
    
end

% Plot specifications
grid
fontsize=21;
set(gcf,'units','points','position',[250,50,550,400])
xlabel('$\color{mauve}{\$}$\theta$\color{mauve}{\$}$','Interpreter','latex','FontSize',fontsize)
ylabel('$\color{mauve}{\$}$p(\theta | \textbf{\textit{m}})$\color{mauve}{\$}$','Interpreter','latex',$\hspace{0.15em}\swarrow$
'FontSize',fontsize)
xticks([0 pi/2 pi 3*pi/2 2*pi])
xticklabels({'0','\pi/2','\pi', '3\pi/2','2\pi'})
xlim([min(theta) max(theta)])
set(gca, 'FontSize', fontsize,'FontName','Times New Roman')
\end{lstlisting}

\section{Mean square error for any number of trials}
\label{sec:msematlab}

\begin{lstlisting}[language=Matlab, mathescape=true]
function [epsilon_trials]=mz_mse_trials(state_choice,pom_choice,$\hspace{0.15em}\swarrow$
prior_width,prior_mean,mu_max)
% Bayesian mean square error
%
% This programme calculates the mean square error as a function of the
% number of repetitions.
%
% To run it, we need to specify the variables 'state_choice', which 
% labels the initial state of a Mach-Zehnder interferometer; 'pom_choice',
% which selects the measurement scheme; 'phase_width', which is the 
% width of a flat prior probability; 'phase_mean', which is the centre
% of its domain; and 'mu_max', which is the maximum number of trials.
%
% Note that this code relies on other MATLAB functions of our numerical
% toolbox. The algorithm in this section has been exploited to calculate
% the mean square error for all the single-parameter cases treated in this 
% thesis, including the ideal schemes for optical interferometry studied 
% in chapters 4 and 5, the calculation of the Taylor error to verify the
% validity of our squared approximation in appendix A, our lossy analysis 
% in chapter 8 and our analysis of the elapsed time, also in chapter 8.

% Seed for the random generator
rng('shuffle') 

% Initial state
initial_state=initial_probe(state_choice);

% Space cutoff
op_cutoff=sqrt(length(initial_state));

% Parameter domain
a=prior_mean-prior_width/2;
b=prior_mean+prior_width/2;
dim_theta=1250;
theta=linspace(a,b,dim_theta);
num_steps=125;
step=round(dim_theta/num_steps);
if step-round(step)~=0
    disp('Error: dim_theta divided by num_steps must be an integer.')
    return
elseif num_steps<3
    disp('Error: the approximation for the external theta integral needs$\hspace{0.15em}\swarrow$
three rectangles at least.')
    return
end

% Monte Carlo sample size
tau_mc=1250;    

% Measurement scheme
[outcomes_space,proj_columns] = mz_pom(state_choice,pom_choice,$\hspace{0.15em}\swarrow$
prior_width, prior_mean);

% State after the phase shift, final state and amplitudes
amplitudes=zeros(length(outcomes_space),dim_theta);
for z=1:dim_theta
    after_phase_shift=sparse(phase_shift_diff(op_cutoff,theta(z))$\hspace{0.15em}\swarrow$
*initial_state);
    for x=1:length(outcomes_space)        
        pom_element=proj_columns(:,x);
        amplitudes(x,z)=sparse(pom_element'*after_phase_shift);
    end             
end

% Likelihood function
likelihood=amplitudes.*conj(amplitudes);
disp('The likelihood function has been created.')
if (1-sum(likelihood(:,1)))>1e-7
    error('The quantum probabilities do not sum to one.')
end

% Prior probability
prior=ones(1,dim_theta);
prior=prior/trapz(theta,prior);

% Bayesian inference
epsilon_bar=0;
for index_real=1:step:dim_theta
    epsilon_n=zeros(1,mu_max); % Preallocate vector
    epsilon_n_sum=zeros(1,mu_max);
    for times=1:tau_mc
        
        % Prior density function
        prob_temp=prior;
        for runs=1:mu_max
            
            % (Monte Carlo) Interferometric simulation
            prob_sim1=likelihood(:,index_real);
            cumulative1 = cumsum(prob_sim1); % Cumulative function
            prob_rand1=rand; % Random selection
            auxiliar1=cumulative1-prob_rand1;
           
            for x=1:length(outcomes_space)
                if auxiliar1(x)>0
                    index1=x;
                    break
                end
            end
            
            % Posterior density function
            prob_temp=sparse(prob_temp.*likelihood(index1,:));
            if trapz(theta,prob_temp)>1e-16
                prob_temp=prob_temp./trapz(theta,prob_temp);
            else
                prob_temp=0;
            end
            
            % Experimental square error
            theta_expe=trapz(theta,prob_temp.*theta);
            theta2_expe=trapz(theta,prob_temp.*theta.^2);
            epsilon_n(runs)=theta2_expe-theta_expe^2;    
        end
        
        % Monte Carlo sum
        epsilon_n_sum=epsilon_n_sum+epsilon_n;
    end
       
    % Monte Carlo approximation for the Bayesian error
    epsilon_average=epsilon_n_sum/(tau_mc);
    epsilon_bar=epsilon_bar+epsilon_average*prior(index_real)$\hspace{0.15em}\swarrow$
*(theta(2*step)-theta(step));
end
epsilon_trials=epsilon_bar;
\end{lstlisting}

The numerical precision for our calculation of $\bar{\epsilon}_{\mathrm{mse}}(\mu)$ can be estimated using the identity
\begin{equation}
\int d\theta p(\theta) \theta^2 = \int d\theta' p(\theta') \int d\boldsymbol{m}~ p(\boldsymbol{m}|\theta') \int d\theta p(\theta|\boldsymbol{m})\theta^2,
\end{equation}
where the right hand side is to be calculated numerically (with a code analogous to that presented above) and to be compared to the analytical solution for the left hand side. We have found that our numerical results for the Mach-Zehnder interferometer are valid up to the third significant figure.  
\chapter{Numerical toolbox for multi-parameter metrology}
\label{app:multinum}

\section{Multi-parameter prior information analysis}
\label{sec:multiprior}

\begin{lstlisting}[language=Matlab, mathescape=true]
% Two-parameter prior information analysis
%
% Prior information analysis for a qubit sensing network. The basic logic
% of the method parallels that for the single-parameter case (see appendix
% B.5 and chapter 6 for more details).

% Initial parameters
prior_mean1=pi; 
prior_mean2=prior_mean1;
prior_width1=2*pi;
prior_width2=2*pi;
mu_max=100;

% True values for the unknown parameters
theta1_real=1;
theta2_real=2;

% Initial state
gamma_par=1; % Local strategy
%gamma_par=0; % Maximally entangled strategy
%gamma_par=0.530696; % Asymptotically optimal strategy
%gamma_par=0.3343605926149827; % Balanced startegy
initial_state=sparse([1 gamma_par gamma_par 1])'/sqrt(2+2*gamma_par^2);

% Generators
sigmaz=sparse([1 0; 0 -1]);
g1=kron(sigmaz,identity(2))/2;
g2=kron(identity(2),sigmaz)/2;

% Asymptotically optimal local POM (F = F_q, chapter 6)
proj1=sparse([-1 -1 1 1])'/2;
proj2=sparse([1 1 1 1])'/2;
proj3=sparse([1 -1 -1 1])'/2;
proj4=sparse([-1 1 -1 1])'/2;
proj_columns=[proj1';proj2';proj3';proj4']';

% Optimal single-shot POM (chapter 7)
% proj1=sparse([1i 1 1 -1i])'/2;
% proj2=sparse([-1i 1 1 1i])'/2;
% proj3=sparse([1i -1 1 1i])'/2;
% proj4=sparse([-1i -1 1 -1i])'/2;
% proj_columns=[proj1';proj2';proj3';proj4']';

% Parameter domain
dim_theta=200;
a1=prior_mean1-prior_width1/2;
b1=prior_mean1+prior_width1/2;
theta1=linspace(a1,b1,dim_theta);
a2=prior_mean2-prior_width2/2;
b2=prior_mean2+prior_width2/2;
theta2=linspace(a2,b2,dim_theta);

% State after encoding the parameters, final state and amplitudes
amplitudes=zeros(size(proj_columns,2),dim_theta,dim_theta);
amplitudes_sparse=zeros(dim_theta,dim_theta,size(proj_columns,2));
for z1=1:dim_theta
    for z2=1:dim_theta
        after_encoding=sparse(expm(-1i*(g1*theta1(z1)+g2*theta1(z2))))$\hspace{0.15em}\swarrow$
*initial_state;
        for x=1:size(proj_columns,2)
            povm_element=proj_columns(:,x);     
            amplitudes_temp=sparse(povm_element)'*sparse(after_encoding);
            amplitudes(x,z1,z2)=amplitudes_temp;
            amplitudes_sparse(z1,z2,x)=amplitudes_temp;
        end

        % The second method of generating the amplitudes is included in 
        % order to use sparse later in the code.
    end
end

% Likelihood function
likelihood=amplitudes.*conj(amplitudes);
if (1-sum(likelihood(:,1,1)))>1e-7
    error('The quantum probabilities do not sum to one.')
end
likelihood_sparse=amplitudes_sparse.*conj(amplitudes_sparse);
if (1-sum(likelihood_sparse(1,1,:),3))>1e-7
    error('The quantum probabilities do not sum to one.')
end

% Prior probability
prior=ones(dim_theta,dim_theta);
prior=prior/trapz(theta2,trapz(theta1,prior));

% Simulation of the true values for the unkonwn parameters
for y=1:dim_theta 
    if theta1(y)>theta1_real || theta1(y)==theta1_real
        index_real1=y;
        break
    end
end

for y=1:dim_theta
    if theta2(y)>theta2_real || theta2(y)==theta2_real
        index_real2=y;
        break
    end
end

% Bayesian simulation
outcomes=zeros(1,mu_max);
for runs=1:mu_max
    
    % Simulation of the experimental outputs
    prob_sim=likelihood(:,index_real1,index_real2);
    cumulative1 = cumsum(prob_sim); % Cumulative function
    prob_rand=rand; % Random selection
    auxiliar=cumulative1-prob_rand;
    
    for x=1:size(proj_columns,2)
        if auxiliar(x)>0
            index=x;
            break
        end
    end
   
    outcomes(runs)=index;
end

% Prior density function
prob_temp=prior;
for runs=1:mu_max
    
    % Updated posterior density function
    ytemp=outcomes(runs);
    
    likesimulated=likelihood_sparse(:,:,ytemp);
    prob_temp=sparse(prob_temp.*likesimulated);
    prob_norm=sparse(trapz(theta2,trapz(theta1,prob_temp,1),2));
    if prob_norm>1e-16
        prob_temp=prob_temp/prob_norm;
    else
        prob_temp=0;
    end
    prob_temp=sparse(prob_temp);
end

% Plot of the posterior
contour(theta1',theta2',prob_temp,'LevelStep',0.1,'Fill','on')
xticks([0 pi/4 pi/2 3*pi/4 pi 5*pi/4 3*pi/2 7*pi/4 2*pi])
xticklabels({'0', '\pi/4', '\pi/2', '3\pi/4', '\pi', '5\pi/4', '3\pi/2',$\hspace{0.15em}\swarrow$
'7\pi/4','2\pi'})
yticks([0 pi/4 pi/2 3*pi/4 pi 5*pi/4 3*pi/2 7*pi/4 2*pi])
yticklabels({'0', '\pi/4', '\pi/2', '3\pi/4', '\pi', '5\pi/4', '3\pi/2',$\hspace{0.15em}\swarrow$
'7\pi/4','2\pi'})
xt = get(gca, 'XTick');
fontsize=32;
set(gca, 'FontSize', fontsize,'FontName','Times New Roman');
yt = get(gca, 'YTick');
set(gca, 'FontSize', fontsize,'FontName','Times New Roman');
grid
\end{lstlisting}

\section{Multi-parameter mean square error for any number of trials}
\label{sec:multimsematlab}

\begin{lstlisting}[language=Matlab, mathescape=true]
% Mean square error for the estimation of two linear functions
%
% The estimation scheme is a quantum sensing network with two qubits.
%
% Note that we use the trapezoidal rule 'trapz' for the inner parameter 
% integrals because these have peaked integrands, while Simpson's Rule 
% 'simps' is a better choice when this problem does not arise, which is
% the case for the outer parameter integrals.
clear

% Initial parameters
prior_mean1=pi/4;
prior_mean2=pi/4;
prior_width1=pi/2;
prior_width2=pi/2;
mu_max=1;

% Weighting matrix
WD=[1 0; 0 1]/2;

% Transformation representing the original parameters
%K=[1 0; 0 1];

% Transformation representing two linear functions
V=[2/sqrt(4+pi^2) 2/sqrt(5); pi/sqrt(4+pi^2) 1/sqrt(5)];

% Combination of linear transformation and weighting matrix
G=V*WD*V';

% Initial state
gamma_par=1; % Local strategy
%gamma_par=0; % Maximally entangled strategy
%gamma_par=0.530696; % Asymptotically optimal strategy
%gamma_par=0.3343605926149827; % Balanced startegy
initial_state=sparse([1 gamma_par gamma_par 1])'/sqrt(2+2*gamma_par^2);

% Generators
sigmaz=sparse([1 0; 0 -1]);
g1=kron(sigmaz,identity(2))/2;
g2=kron(identity(2),sigmaz)/2;

% Asymptotically optimal local POM (F = F_q, chapter 6)
proj1=sparse([-1 -1 1 1])'/2;
proj2=sparse([1 1 1 1])'/2;
proj3=sparse([1 -1 -1 1])'/2;
proj4=sparse([-1 1 -1 1])'/2;
proj_columns=[proj1';proj2';proj3';proj4']';

% Optimal single-shot POM (chapter 7)
% proj1=sparse([1i 1 1 -1i])'/2;
% proj2=sparse([-1i 1 1 1i])'/2;
% proj3=sparse([1i -1 1 1i])'/2;
% proj4=sparse([-1i -1 1 -1i])'/2;
% proj_columns=[proj1';proj2';proj3';proj4']';

% Parameter domain
dim_theta=100;
dim_theta_out=20;
a1=prior_mean1-prior_width1/2;
b1=prior_mean1+prior_width1/2;
theta1=linspace(a1,b1,dim_theta); % Inner parameter integrals
theta1_out=linspace(a1,b1,dim_theta_out); % Outer parameter integrals
a2=prior_mean2-prior_width2/2;
b2=prior_mean2+prior_width2/2;
theta2=linspace(a2,b2,dim_theta);
theta2_out=linspace(a2,b2,dim_theta_out);

% Monte Carlo sample size
tau_mc=200; 

% State after encoding the parameters, final state and amplitudes
amplitudes=zeros(size(proj_columns,2),dim_theta,dim_theta);
amplitudes_sparse=zeros(dim_theta,dim_theta,size(proj_columns,2));
for z1=1:dim_theta
    for z2=1:dim_theta
        after_encoding=sparse(expm(-1i*(g1*theta1(z1)+g2*theta1(z2))))$\hspace{0.15em}\swarrow$
*initial_state;
        for x=1:size(proj_columns,2)
            povm_element=proj_columns(:,x);     
            amplitudes_temp=sparse(povm_element)'*sparse(after_encoding);
            amplitudes(x,z1,z2)=amplitudes_temp;
            amplitudes_sparse(z1,z2,x)=amplitudes_temp;
        end

        % The second method of generating the amplitudes is included in 
        % order to use sparse later in the code.
    end
end

% Likelihood function
likelihood=amplitudes.*conj(amplitudes);
if (1-sum(likelihood(:,1,1)))>1e-7
    error('The quantum probabilities do not sum to one.')
end
likelihood_sparse=amplitudes_sparse.*conj(amplitudes_sparse);
if (1-sum(likelihood_sparse(1,1,:),3))>1e-7
    error('The quantum probabilities (sparse version) do not sum to one.')
end

% Prior probability
prior=ones(dim_theta,dim_theta);
prior=prior/trapz(theta2,trapz(theta1,prior));
prior_out=ones(dim_theta_out,dim_theta_out);
prior_out=prior_out/trapz(theta2_out,trapz(theta1_out,prior_out));

% Bayesian mean square error
epsilon_out=zeros(dim_theta_out,dim_theta_out);
for index_out1=1:dim_theta_out
    for index_out2=1:dim_theta_out
        
        % Matching outer and inner parameter indices       
        for y=1:dim_theta
            if theta1(y)>theta1_out(index_out1) ||$\hspace{0.15em}\swarrow$
theta1(y)==theta1_out(index_out1)
                index_real1=y;
                break
            end
        end
        
        for z=1:dim_theta
            if theta2(z)>theta2_out(index_out2) ||$\hspace{0.15em}\swarrow$
theta2(z)==theta2_out(index_out2)
                index_real2=z;
                break
            end
        end
        
        epsilon_n1=zeros(1,mu_max);
        epsilon_n2=zeros(1,mu_max);
        epsilon_n_offdia=zeros(1,mu_max);
        epsilon_n_sum=zeros(1,mu_max);
        for times=1:tau_mc
            
            % Prior density function
            prob_temp=sparse(prior);
            for runs=1:mu_max
                
                % (Monte Carlo) Outcome simulation
                prob_sim=likelihood(:,index_real1,index_real2);
                cumulative = cumsum(prob_sim); % Cumulative function
                prob_rand=rand; % Random selection
                auxiliar=cumulative-prob_rand;
                
                for x=1:size(proj_columns,2)
                    if auxiliar(x)>0
                        index_mc=x;
                        break
                    end
                end
                
                % Posterior density function
                likesimulated=likelihood_sparse(:,:,index_mc);
                prob_temp=sparse(prob_temp.*likesimulated);
                normalisation=sparse(trapz(theta2,trapz$\hspace{0.15em}\swarrow$
(theta1,prob_temp,1),2));
                if normalisation>1e-16
                    prob_temp=prob_temp/normalisation;
                else
                    prob_temp=0;
                end
                prob_temp=sparse(prob_temp);
                
                % Bayes estimator for the first parameter
                theta_expe1=trapz(theta1,trapz(theta2,prob_temp,2)$\hspace{0.15em}\swarrow$
.*theta1',1);
                theta2_expe1=trapz(theta1,trapz(theta2,prob_temp,2)$\hspace{0.15em}\swarrow$
.*theta1'.^2,1);
                epsilon_n1(runs)=theta2_expe1-theta_expe1^2;
                
                % Bayes estimator for the second parameter
                theta_expe2=trapz(theta2,trapz(theta1,prob_temp,1)$\hspace{0.15em}\swarrow$
.*theta2,2);
                theta2_expe2=trapz(theta2,trapz(theta1,prob_temp,1)$\hspace{0.15em}\swarrow$
.*theta2.^2,2);
                epsilon_n2(runs)=theta2_expe2-theta_expe2^2;
                
                % Off-diagonal terms (the covariance matrix is symmetric)
                theta2_offdia=trapz(theta1,trapz(theta2,prob_temp$\hspace{0.15em}\swarrow$
.*theta2,2).*theta1',1);
                epsilon_n_offdia(runs)=theta2_offdia-theta_expe1$\hspace{0.15em}\swarrow$
*theta_expe2;

            end
            
            % Monte Carlo sum with transformation and weighting matrices
            epsilon_n_sum=epsilon_n_sum+G(1,1)*epsilon_n1+G(2,2)$\hspace{0.15em}\swarrow$
*epsilon_n2+2*G(1,2)*epsilon_n_offdia;
        end
                  
        % Monte Carlo approximation
        epsilon_average=epsilon_n_sum/(tau_mc);
        for runs_out=1:mu_max
            epsilon_out(index_out1,index_out2,runs_out)$\hspace{0.15em}\swarrow$
=epsilon_average(runs_out);
        end
    end
end

% Outer integral
epsilon_trials=zeros(1,mu_max);
for runs_out=1:mu_max
    epsilon_temp=epsilon_out(:,:,runs_out);
    epsilon_trials(runs_out)=simps(theta2_out,simps(theta1_out,prior_out$\hspace{0.15em}\swarrow$
.*epsilon_temp));
end

% Observations
observations=1:1:mu_max;

% Fisher information matrix
F11=4*(initial_state'*g1^2*initial_state-(initial_state'$\hspace{0.15em}\swarrow$
*g1*initial_state)^2);
F12=4*(initial_state'*g1*g2*initial_state-(initial_state'$\hspace{0.15em}\swarrow$
*g1*initial_state)*(initial_state'*g2*initial_state));
F21=4*(initial_state'*g2*g1*initial_state-(initial_state'$\hspace{0.15em}\swarrow$
*g2*initial_state)*(initial_state'*g1*initial_state));
F22=4*(initial_state'*g2^2*initial_state-(initial_state'$\hspace{0.15em}\swarrow$
*g2*initial_state)^2);
F=[F11 F12; F21 F22];

% Quantum Cramer-Rao bound
qcrb=trace(G/F)./(observations);

% Save results
%save('qnetwork_results.txt','observations','epsilon_trials','qcrb',$\hspace{0.15em}\swarrow$
'-ascii')
\end{lstlisting}


\end{document}